\documentclass[amssymb,amsfont,12pt,oneside,raggedright,notitlepage]{book}
\DeclareMathAlphabet{\mathpzc}{OT1}{pzc}{m}{it} 
\usepackage[export]{adjustbox}
\usepackage{appendix}
\usepackage{color}
\usepackage{makeidx}
\usepackage{bm}
\usepackage{amsfonts}
\usepackage{amsmath}
\usepackage{amssymb}
\usepackage{mathrsfs}
\usepackage{subfig}
\RequirePackage{epsf}
\RequirePackage{longtable}
\usepackage{times}
\usepackage{graphicx,wrapfig}
\usepackage{epstopdf}
\usepackage{epsfig}
\usepackage{hyperref}
\usepackage{enumitem}
\usepackage{tikz-cd}
\usepackage{array}
\usepackage{float}
\usepackage{enumitem}
\usepackage{rotating}

\DeclareMathOperator{\sech}{sech}
\newcommand{\benr}{\begin{enumerate}[resume=ex,label= \sl Exercise \ \theenumi. ]\setlength\itemindent{1.2cm}}
\newcommand{\benalph}{\begin{enumerate}[label=\alph*.]}

\def\d{\delta}
\def\de{\delta}
\def\n{\nabla}

\def\ep{\epsilon}

\def\p{\partial}

\def\la{\lambda}

\def\p{\psi}
\def\a{\alpha}
\def\b{\beta}
\def\g{\gamma}

\def\nn{\nonumber}

\def\Lie{{\cal L}}
\def\be{\begin{equation}}
\def\ee{\end{equation}}
\def\ba{\begin{eqnarray}}
\def\ea{\end{eqnarray}}

\def\<{\noindent }
\def\cblue{\color{blue}}
\def\cred{\color{red}}
\def\cb{\color{black}}
\def\cvi{\color{violet}}
\definecolor{RedViolet}{cmyk}{0.07,0.90,0,0.34}
\def\crv{\color{RedViolet}}

\newcommand{\ben}{\begin{enumerate}}
\newcommand{\een}{\end{enumerate}}
\newcommand{\beq}{\begin{equation}}
\newcommand{\eeq}{\end{equation}}
\newcommand{\bea}{\begin{eqnarray}}
\newcommand{\eea}{\end{eqnarray}}
\newcommand{\beaa}{\begin{eqnarray*}}
\newcommand{\eeaa}{\end{eqnarray*}}
\newcommand{\dis}{\displaystyle}
\newcommand{\pa}{\partial}
\newcommand{\na}{\nabla}

\newcommand{\tr}{{}^3\!R}
\newcommand{\nas}{\nabla\!}
\def\2pi{2\pi}

\newcommand{\Gabu}{G^{\alpha\beta}}

\newcommand{\Lag}{{\mathscr L}}
\def\a{\alpha}
\def\b{\beta}
\def\c{\gamma}
\def\d{\delta}

\def\wt{\widetilde}
\def\MCA#1{\multicolumn{2}{c!{\quad\vline\quad}}{\text{#1}}}
\def\MCB#1{\multicolumn{2}{c}{\text{#1}}}
\newcommand{\bsube}{\begin{subequations}}
\newcommand{\esube}{\end{subequations}}
\newcommand{\upp}[2]{{\raise0.9ex\hbox{${}^{\ #2}$}}\!\!\!\!#1\,}
\newcommand{\up}[2]{{\raise0.4ex\hbox{${}^{\ #2}$}}\!\!\!#1\,}

\newcommand{\cancel}[2]{\underline{#1}\begin{picture}(0,0)\put(#2,-3){{}$_{{}_/}$}\end{picture}}
\newcommand{\cancell}[2]{\underline{#1}\begin{picture}(0,0)\put(#2,-3){{}$_{{}_/\!_/}$}\end{picture}}
\def\dflat{\partial}
\usepackage{tikz}
\usetikzlibrary{positioning}
\tikzset{>=stealth}
\newcommand{\tikzmark}[3][]{\tikz[remember picture,baseline] \node [anchor=base,#1](#2) {$#3$};}

\newcommand{\Ibar}{I\hspace{-1.8mm}\raisebox{.2ex}{-}}
\def\p{\sigma}
\def\q{\tau}
\def\DDelta{\hbox{$\Delta$\kern-.55em\hbox{\raise.20em\hbox{\tiny$\backslash$\ }}}}
\def\ov{\overline}
\def\barm{\overline m}
\def\barbm{\overline{\bm m}}
\def\scrL{\mathscr L}
\def\cB{{\cal B}}

\def\wh{\widehat}
\newcommand{\G}[2]{\ensuremath{\Gamma^{#1}{}_{#2}}}
\makeindex
\begin{document}
\centerline{\huge\crv Notes on Gravitational Physics}
\vspace{1cm}

\centerline{\large John L. Friedman}
\vspace{4mm}
\centerline{\large Department of Physics, University of Wisconsin-Milwaukee}
\vspace*{1\baselineskip}
\centerline{\today}
\vspace{2cm}

These notes are self-contained, with the first six chapters used for a one-semester course, with suggested supplementary texts by Wald\cite{waldbook}, Misner, Thorne, \& Wheeler (MTW)\cite{MTW}, and, particularly for gravitational waves, by Schutz\cite{schutzbook} and Thorne \& Blandford \cite{thorneblandford}.  In its treatment of topics covered 
in these standard texts, the presentation here typically 
includes steps between equations that are skipped in Wald or MTW. Treatments  
of gravitational waves, particle orbits in black-hole backgrounds, 
the Teukolsky equation, and the initial value equations are motivated in part 
by the dramatic discoveries of gravitational  waves from the inspiral and coalescence of binary black holes and neutron stars, advances in numerical relativity, and the expected launch of the LISA space-based observatory.  
Students are assumed to have encountered special relativity, 
but these notes give a detailed presentation with a geometrical orientation, 
starting with with time dilation and length contraction and including 
relativistic particles, fluids, electromagnetism, and curvilinear 
coordinates.  Chapters 2-5 cover curvature, the Einstein equation, relativistic stars, and black holes.  Chapter 6, on gravitational waves, includes 
a discussion of detection and of noise in interferometric detectors.  
Chapter 7 is a brief introduction to cosmology, deriving the metrics of 
homogeneous isotropic space, the equations governing a universe with matter, radiation and vacuum energy, and their solutions.  It includes sections 
on the cosmological redshift and on using gravitational waves to measure the 
Hubble constant.

Chapter \ref{c:iv}, on the initial value problem, has  
a section on the form of the equations used in numerical relativity.
Its notation is that used, for example, in Baumgarte \& Shapiro\cite{bsbook21} 
and Shibata\cite{shibatabook15}; the presentation here is taken in part 
from the text by Friedman and Stergioulas \cite{fsbook}.  
The notes also have a chapter on the Newman-Penrose 
formalism and the Teukolsky equation. Following that is a chapter on 
black-hole thermodynamics and a final chapter on the gravitational action 
and on conserved quantities for asymptotically flat spacetimes, 
using Noether's theorem.  An appendix presents forms, densities, integration, and Cartan calculus, the first parts 
taken from Ref. \cite{fsbook}. 

\vspace{1.5cm} Corrections and suggestions have been made by several students, particularly by Charalampos Markakis to an earlier version of these notes and 
recently by Nikolaus Prusinski and James McCrickard.  A number of the 
black-and-white illustrations were drawn by Michael Hero.\\

\newpage
\centerline{\bf Notation and conventions} 
\index{notation}\index{conventions}

\begin{center}
\begin{tabular}{*{4}{l}}
\\
\hline\hline

Signature & $-+++$ \\
Indices $\alpha, \beta, \ldots$ & spacetime abstract \\
Indices $\mu, \nu$ & spacetime concrete\\
Indices $a,b, \ldots$ & space and $n$-dimensional abstract \\
Indices $i,j, \ldots$ & space and $n$-dimensional concrete \\
Spacetime & $M$ \\
Metric on $M$ & $g_{\a\b}$\\
Spacelike hypersurface & $\Sigma$  \\
Unit normal to $\Sigma$ & $n^\a$ \\
Projection $\perp n^\a$ & $\g_\a^\b$  \\
Riemann tensor & $[\nabla_a, \nabla_b]v_c = R_{abc}{}^d v_d $  \\
Ricci tensor &$ R_{ab} = R_{acb}{}^c$\\

\hline\hline
\end{tabular}
\end{center}

 In these notes, hyperref links to articles and books available online are in blue, as in this link to 
a translation of \href{https://isidore.co/misc/Physics%20papers%20and%20books/St.%20John%27s%20College%27s,%20TAC%27s%20curricula%27s,%20et%20alii%20sci.%20papers/1916-%20The%20Foundation%20of%20the%20General%20Theory%20of%20Relativity%20(Einstein).pdf}{Einstein's 1916 paper} and the following links to other notes on general relativity and gravitational physics, listed alphabetically by author: \\  

\noindent
\href{https://physics.mcmaster.ca/~cburgess/Notes/GRnotes.pdf}{C. P. Burgess notes}, short, with a good section on lensing.\\
\href{https://www.preposterousuniverse.com/grnotes/}{Sean Carroll's Lecture Notes on GR} (subsequently  expanded to Carroll's widely used textbook).\cite{carroll19}\\
\href{https://community.wvu.edu/~stmcwilliams/Sean_McWilliams/SP20_PHYS_704_files/Lecture_slides.pdf}{Mike Guidry's Modern General Relativity Lecture Notes} \\
\href{https://home.uni-leipzig.de/~tet/wp-content/uploads/2016/04/GR2015_0416.pdf}{Hollands and Sanders, Lecture Notes on General Relativity}\\
\href{https://webspace.science.uu.nl/~hooft101/lectures/genrel_2010.pdf}{Gerard 't Hooft's Intro to GR}\\
\href{https://www.damtp.cam.ac.uk/user/hsr1000/lecturenotes_2012.pdf}{Harvey Reall's General Relativity} from a Cambridge course.\\

The first several chapters of Hartle's {\sl Gravity}\cite{hartlebook} give a ``physics-first'' presentation of special relativity and an 
introduction to curved space; Gourgoulhon's \href{https://dokumen.pub/special-relativity-in-general-frames-from-particles-to-astrophysics-9783642372759-9783642372766-3642372759.html}{\sl Special Relativity in General Frames}\cite{gourgoulhon13} is, in contrast, a recent presentation of special relativity with a mathematics-first perspective.
For a bibliography of GR books, look \href{https://ned.ipac.caltech.edu/level5/March01/Carroll3/Carroll001.html}{here}. More recent texts include some listed in the abstract as well as the longer version
{\sl Numerical Relativity}\cite{BS10} of the Baumgarte-Shapiro book, Eric Poisson's {\sl A Relativists Toolkit}\cite{poisson07}, {\sl Gravity} by Poisson and Cliff Will\cite{PW14}, and Eric Gourgoulhon's 
\href{https://arxiv.org/abs/gr-qc/0703035}{\sl 3+1 Formalism in General Relativity}\cite{Gourgoulhon2012}.  Two gravitational-wave texts are 
{\sl Gravitational-Wave Physics and Astronomy}, by Jolien Creighton \& Warren Anderson, \cite{creightonanderson} and {\sl Gravitational Waves}, by  
Michele Maggiore, \cite{maggiore08}.   

\tableofcontents

\renewcommand{\labelenumi}{\Roman{enumi}}
\renewcommand{\labelenumii}{\Alph{enumii}}
\renewcommand{\labelenumiii}{\arabic{enumiii}}
\newpage 
\vspace{-5cm}

\noindent

\color{white}
\chapter{Introduction: Physics in Flat Spacetime}\vspace{-6cm} \cb
{\bf \Huge Chapter 1.  Introduction:\\ \\ \phantom{xxxxxxxxx}Physics in a Flat Spacetime}\label{c:fst}frd
\index{flat spacetime, Chap.~\ref{c:fst}|(}
\index{spacetime!flat, Chap.~\ref{c:fst}|(}

\section{Concepts of Space and Time}
\index{concepts of space and time}

The history of the way Western cultures have viewed space and time can
-- in retrospect -- be described by a sequence in which most of what 
seemed to be the obvious features were gradually discarded.  The following
description is a retrospective caricature of the history, inaccurate in
that, for example, Galileo is said to think of spacetime as flat when
he had no precise notion that could properly be called ``spacetime.''\\
\vspace{3mm}

\begin{wrapfigure}[7]{r}{8cm}
 		\vspace{-1cm}
                \includegraphics[width=6cm]{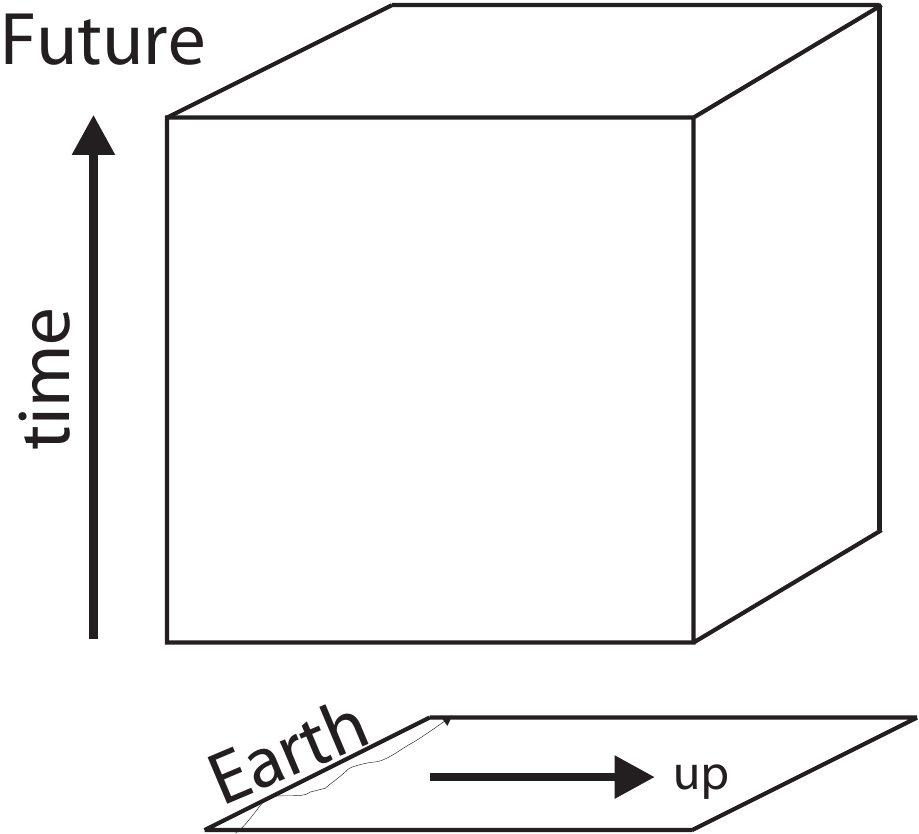}
                \label{IA1}
\end{wrapfigure}

\noindent Pre Greek  

\indent Earth flat

	Preferred spatial direction (up)

	Space absolute

	Time absolute

	Preferred time direction

	\hspace{.5in}(toward future)
	
	Space flat
\vspace{1.5cm}

\index{absolute space}
Space is {\sl absolute} if it makes sense to say that events at two different times occur at the same place.  All observers agree on what it means to be at rest -- to stay at the same point of space.  \\
\index{absolute time}
Time is {\sl absolute} if it makes sense to say that events at two different places occurred at the same time. All observers agree on what events are simultaneous, and synchronized watches 
at different places read the same elapsed time between pairs of simultaneous events.\\

\begin{wrapfigure}[4]{r}{8cm}
                \includegraphics[width=8cm]{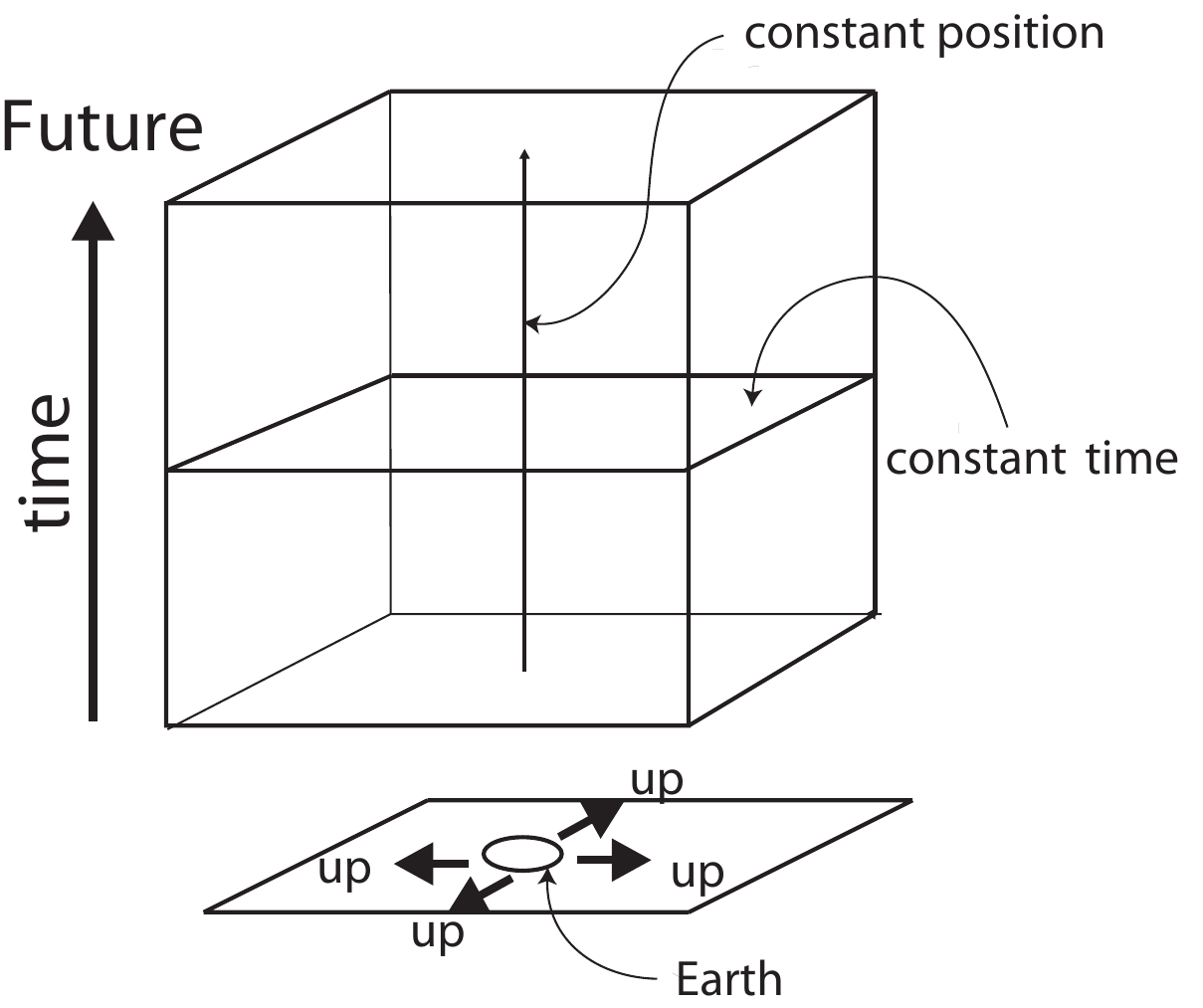}
                \label{IA2}
\end{wrapfigure}

\noindent Greek 

	{\em Earth curved} (Italics mark change)

	{\em Preferred set of spatial directions (up)}

	Space absolute

	Time absolute

	Preferred time direction 

	Space flat

\pagebreak 
\index{Galilean relativity|(}
\noindent{\sl The Copernican Revolution and Galilean relativity}\cb\\
\index{Copernican revolution}
The loss of absolute space begins with Copernicus.
His model (and presumably that of Aristarchus, 1700 years earlier) 
rested on a spectacular coincidence:  The retrograde motion of all five planets is 
simultaneously explained if one simply places the Sun, instead of the Earth, at the
center of the universe, regards the Earth as a planet, and places what
were then six planets in nearly circular orbits around the Sun. For each
outer planet --- Mars, say --- the explanation is simply this: Because
the period of the Earth is shorter than the period of Mars, as the Earth
orbits the Sun, it periodically passes Mars. When it does so,
Mars appears to go backward relative to the background
stars.  \\

The model explains an additional coincidence: When the 
outer planets -- e.g., Jupiter and Saturn -- are in the same place in 
the sky, their retrograde motion occurs at nearly the same time. (A time lapse video 
that illustrates this dance, showing Jupiter and Saturn moving by the Pleaides and Hyades, is here: \href{https://apod.nasa.gov/apod/image/0112/JuSa2000_tezel.gif}{pas de deux})\\

Finally, the Copernican model \index{Copernican model} gives a relation between the time a planet spends 
in its retrograde loop \index{retrograde motion of planets} and the angular size of the loop in the sky:  First, the known periods of 
the planet and the Earth, together with the time the planet spends moving retrograde, 
determine the radius of the planet's orbit in terms of the radius of the Earth's orbit.
Knowing this radius fixes the angular size of the retrograde loop. 
So one gets five numerical retrodictions, one number for each known planet, 
that have to coincide with observation for the model to hold. \\

That Copernicus found the distance to each of the planets in terms
of the Earth's distance from the Sun would have been one final, breathtaking 
coincidence -- and in this case a prediction, except that it was hundreds of years
before the distance from the Sun to the planets could finally be
measured. Textbooks emphasize the fact that Copernicus' model of retrograde 
motion was not significantly more accurate than Ptolemy's Earth-centered model.
Because he had to make corrections to circular orbits by hand to account for 
what turned out to be elliptical orbits, the overall number of parameters
Copernicus used for the same accuracy in orbital prediction was no smaller 
than Ptolemy's. 
\\

Once one understands that the Earth is not at rest and that it is one of
the planets, our picture of the universe changes dramatically. The other
planets, no longer heavenly, are ordinary objects like
the Earth. They are presumed to be made of the same kind of matter that
the Earth is made of and to obey the same laws. There is no longer a
natural rest frame, and Galileo's conclusion is that we cannot tell what 
is at rest and what is moving.  

\newpage

Here are two quotes, one from Galileo \index{Galileo} and one from Sir James Jeans that may be 
useful to your students for historical perspective if you are teaching physics or astronomy: 
\begin{figure}[H]
               \begin{center}
		\includegraphics[width=.5\textwidth]{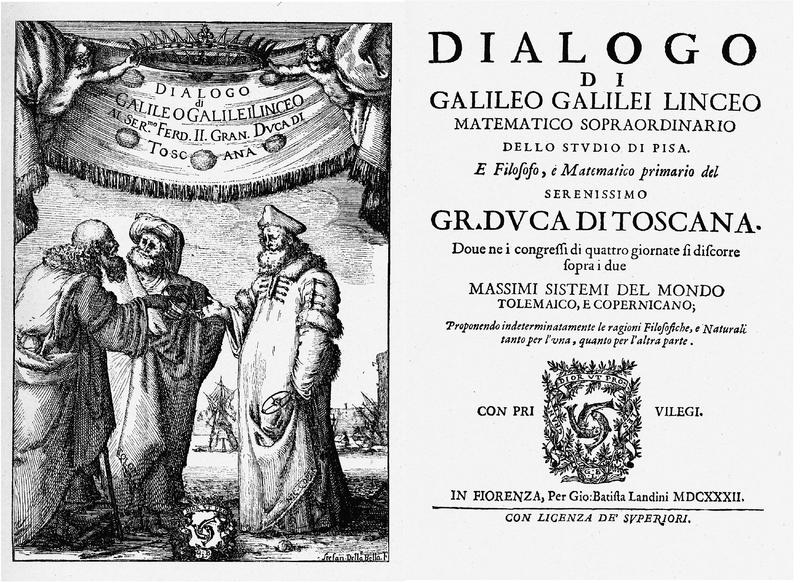}
		\end{center}
\end{figure}
 
{\sl Excerpt from Galileo's Dialogue Concerning the Two Chief World Systems, 
Ptolemaic and Copernican, 1632}\ \href{https://en.wikipedia.org/wiki/Galileo%27s_ship}{Dialogo} \\

\hspace{1cm}\begin{minipage}[h]{6.2in}\leftskip 0.2in 

[SALVATI] Shut yourself up with some friend in the largest room below decks of some large
ship and there procure gnats, flies, and other such small winged creatures. Also get a great
tub full of water and within it put certain fishes; let also a certain bottle be hung up,
which drop by drop lets forth its water into another narrow-necked bottle placed
underneath. Then, the ship lying still, observe how those small winged animals fly with
like velocity towards all parts of the room; how the fish swim indifferently towards all
sides; and how the distilling drops all fall into the bottle placed underneath. And casting
anything toward your friend, you need not throw it with more force one way than another,
provided the distances be equal; and leaping with your legs together, you will reach as far
one way as another. Having observed all these particulars, though no man doubts that, so
long as the vessel stands still, they ought to take place in this manner, make the ship move
with what velocity you please, so long as the motion is uniform and not fluctuating this
way and that. You will not be able to discern the least alteration in all the forenamed
effects, nor can you gather by any of them whether the ship moves or stands still. ...in
throwing something to your friend you do not need to throw harder if he is towards the
front of the ship from you... the drops from the upper bottle still fall into the lower bottle
even though the ship may have moved many feet while the drop is in the air ... Of this
correspondence of effects the cause is that the ship’s motion is common to all the things
contained in it and to the air also; I mean if those things be shut up in the room; but in
case those things were above the deck in the open air, and not obliged to follow the course
of the ship, differences would be observed, ... smoke would stay behind... .

[SAGREDO] Though it did not occur to me to try any of this out when I was at sea, I am
sure you are right. I remember being in my cabin wondering a hundred times whether the
ship was moving or not, and sometimes I imagined it to be moving one way when in fact it
was moving the other way. I am therefore satisfied that no experiment that can be done in
a closed cabin can determine the speed or direction of motion of a ship in steady motion.

\end{minipage}

\newpage

A description of Galileo's work, from Sir James Jeans, \href{https://archive.org/details/in.ernet.dli.2015.210060}{\sl The Growth of Physical Science}:\\

\hspace{1cm}\begin{minipage}[h]{6.2in}\leftskip 0.2in 
 
He set up a gently sloping plank, some 12 yards in length, and made polished
steel balls roll down a narrow groove cut into it.  With this simple apparatus,
he was able to verify his conjecture that the speed of fall increased uniformly
with time -- the law of uniform acceleration. It was one of the great moments
in the history of science. 
 
\crv{For now it became clear that the effect of force was not to {\em 
produce} motion, but to {\em change} motion - to produce acceleration; a body 
on which no force acts moves at uniform speed.}

\index{Galileo}
\cb Galileo's rolling steel balls, moving in a horizontal plane, continued 
their motion with undiminished speed until they were checked by friction and 
the resistance of the air.  This was not entirely new.  Plutarch, in 100 AD, 
had written, ``Everything is carried along by the motion natural to it, if it 
is not deflected by something else.'' But Galileo was the first to establish 
the principle experimentally.  Where others had conjectured, Galileo proved.\\
\end{minipage}

The idea that force is needed to produce motion is overturned by the
realization that we cannot tell what is at rest and what is moving.
After the Copernican revolution (after Galileo), it is clear that
a force is needed to change the velocity of an object, not to make it
move.  The understanding is made precise in Newton's first two laws, 
the first due to Galileo.  

\begin{enumerate}
\item[1.] With no force acting on it, an object moves at constant speed in a
straight line.
\item[2.] ${\bm F} = m\bm a$.
\end{enumerate}

A Galilean boost, changing each particle trajectory $\bm x(t)$ by the transformation 
\be
 \bm x(t) \rightarrow \bm x(t) +{\bf v}t  
\ee
leaves Newton's laws invariant: Because $\dis \frac{d^2}{dt^2}[\bm x(t)] =\frac{d^2}{dt^2}[\bm x(t)+{\bf v}t]$, the history of a system of particles for which each 
particle satisfies ${\bm F} = m\bm a$ is mapped to a boosted history satisfying ${\bm F} = m\bm a$.  The trajectory of an observer who says her frame is at rest, 
with trajectory $(t,\bm x(t)) = (t,\bm 0) $ is mapped to the trajectory 
$\cblue (t,\bm x(t)) = (t,{\bf v }t)$ of an observer who says that {\sl his} frame is at rest: 
$\cblue (t',\bm x'(t'))= (t,\bm 0)$. Time is still absolute, the same $t$ for both observers.\\

\begin{wrapfigure}[7]{r}{8cm}\vspace{-4mm}
                \includegraphics[width=7cm]{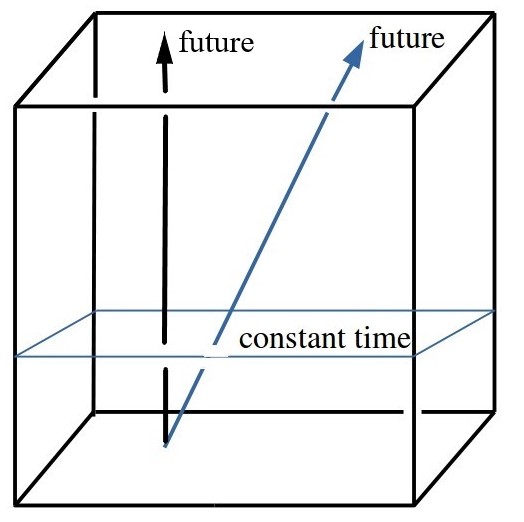}
                \label{IA3}
                \end{wrapfigure}
\noindent Galilean
	
	Earth curved

	{\em No preferred spatial direction}

	{\em No absolute space}

	Time absolute

	{\sl Preferred set of time directions}
	
	Spacetime flat
\index{Galilean relativity|)}
\newpage

Here's the abstract of Einstein's 1905 paper, introducing special relativity.\\
(Full text: \url{https://www.fourmilab.ch/etexts/einstein/specrel/www/})\\

\centerline{\bf ON THE ELECTRODYNAMICS OF MOVING BODIES}
\vspace{3mm}

\hspace{1cm}\begin{minipage}[h]{6.2in}\leftskip 0.2in 
It is known that Maxwell's electrodynamics--as usually understood at the present time--when applied to moving bodies, leads to asymmetries which do not appear to be inherent in the phenomena. Take, for example, the reciprocal electrodynamic action of a magnet and a conductor. The observable phenomenon here depends only on the relative motion of the conductor and the magnet, whereas the customary view draws a sharp distinction between the two cases in which either the one or the other of these bodies is in motion\ldots \\ 
Examples of this sort, together with the unsuccessful attempts to discover any motion of the Earth relatively to the “light medium,” suggest that the phenomena of electrodynamics as well as of mechanics possess no properties corresponding to the idea of absolute rest. They suggest rather that, as has already been shown to the first order of small quantities, the same laws of electrodynamics and optics will be valid for all frames of reference for which the equations of mechanics hold good.  We will raise this conjecture (the purport of which will hereafter be called the “Principle of Relativity”) to the status of a postulate, and also introduce another postulate, which is only apparently irreconcilable with the former, namely, that light is always propagated in empty space with a definite velocity $c$ which is independent of the state of motion of the emitting body. These two postulates suffice for the attainment of a simple and consistent theory of the electrodynamics of moving bodies based on Maxwell's theory for stationary bodies.   
\end{minipage}
\vspace{-8mm}

\begin{figure}[ht]
               \begin{center}
		\includegraphics[width=\textwidth]{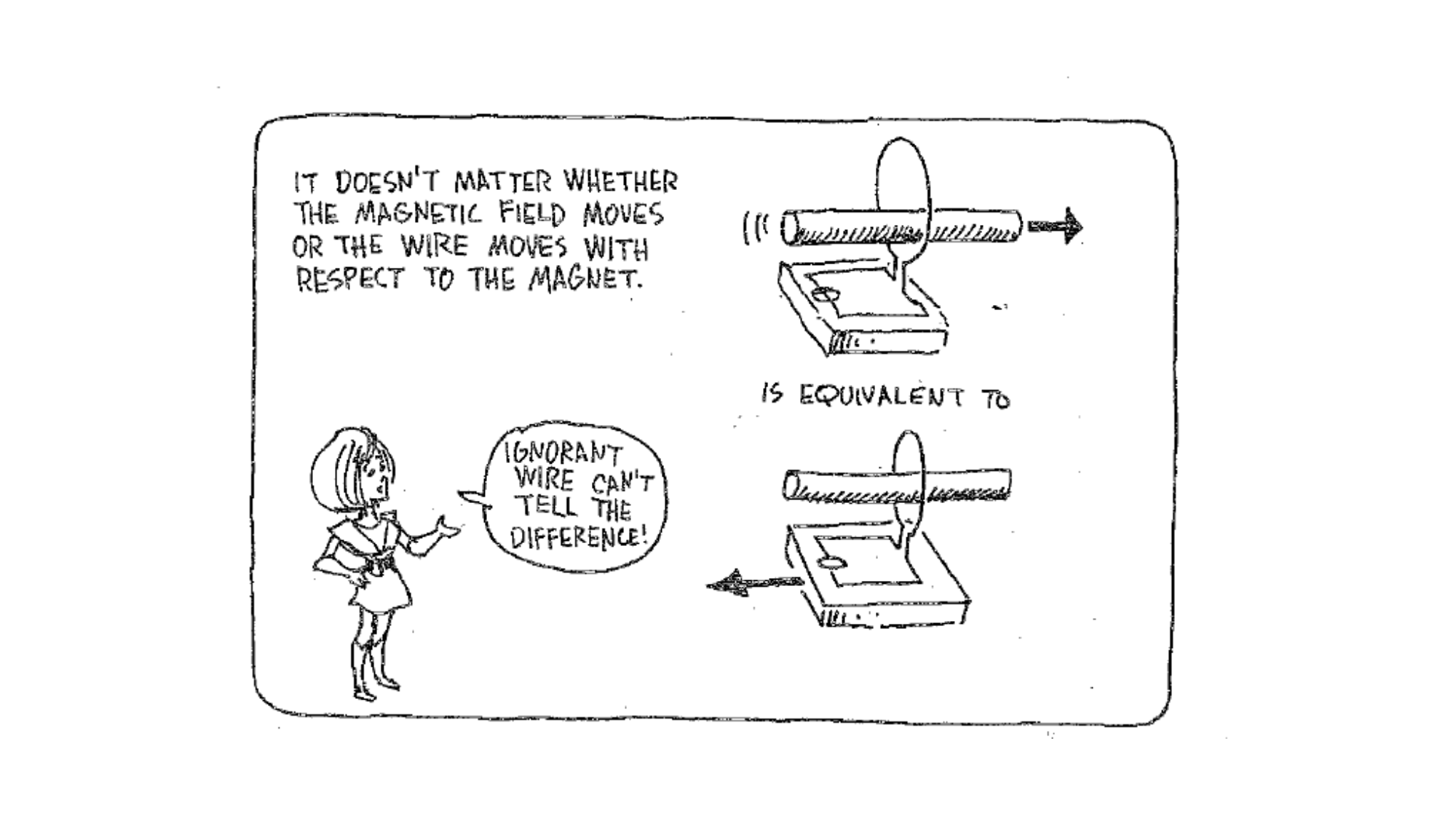}
		\end{center}\vspace{-18mm}
		\caption{From Gonick \& Huffman's Cartoon Guide to Physics\cite{gonickbook}}
\label{f:magnet}
\end{figure}

\mbox{Again a remarkable coincidence from the lack of absolute space underlies special relativity.}    
Hidden in Maxwell's equations \index{Maxwell's equations} in their original form -- and still in the form you first encounter -- is a structure that guarantees their invariance and that of the Lorentz force law: In a frame for which the wire is moving, the $q{\bf v}\times\bm B$ 
force on the moving electrons produces the current. In the frame in which the magnet moves, the equations conspire to give an electric field whose force $q\bm E$ produces the {\sl same} current, as measured by the ammeter in Fig. 1.1.    \\    
  
\newpage

\begin{wrapfigure}[10]{r}{10cm}
                 \includegraphics[width=8cm]{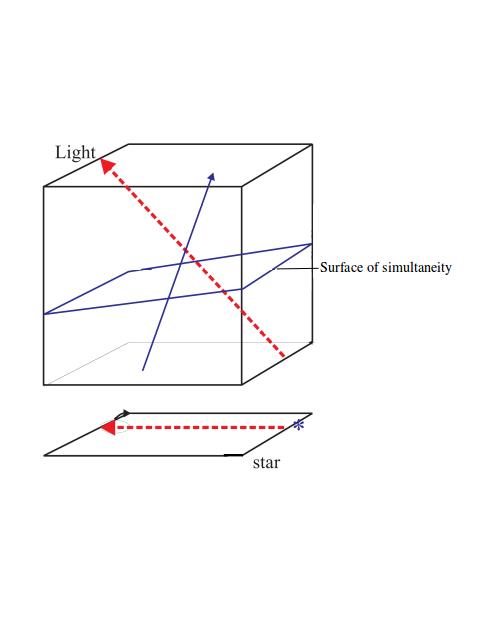}
                \label{IA4c}
                \end{wrapfigure}
\color{white}
.

.

.

.
\color{black}

\noindent Special Relativity \index{special relativity}\index{relativity!special relativity}

	Earth curved

	No preferred spatial direction

	No absolute space

	{\em No absolute time}

	Preferred set of time directions
	
	Spacetime flat

\vspace{2.5cm}

With the understanding that light is a wave, it appeared possible to regain absolute space: You are at rest relative to space (or to the ether that fills it) if light travels at the same speed $c$ in all directions.  But Maxwell's equations and hence the speed of light are invariant under a boost of the dynamical system to an identical system moving with a velocity $\bf v$.  
In order that the speed of light be unchanged by boosting the emitter or observer, however, 
one must discard not only absolute space but also absolute time.    
Observers moving relative to one another must disagree on what events are simultaneous. 

If you have taken special relativity, you encountered this historical logic, leading to time dilation, length  contraction, and, ultimately, the law that governs the geometry of flat space, the distance 
between two events: 
\[
   ds^2 = -c^2 dt^2 + dx^2+dy^2 + dz^2.
\]
The special case of this Minkowski metric is the Pythagorean metric $dx^2+dy^2+dz^2$ for the distance between simultaneous events as defined by a particular observer. 
\cblue If you have not followed the historical path to the Minkowski metric, 
the argument, inferring the Minkowski metric from the 
observer independence of the speed of light, is given as an appendix 
\ref{s:minkappendix} at the end of this section, and it would make sense 
to read that before going on to the next section. 
You may also want to read, for example, Chapter 4 of Hartle's {\sl Gravity}.  \cb
Beginning with the next section, \ref{s:minkowski}, we will follow the logic of the law itself, inferrring the physics from the geometry of a flat spacetime.  

Discarding the final assumption--that space is flat--begins with mathematicians in the 18th century, in particular with Gauss and his student Riemann: \\
{\sl \indent Therefore, either the reality on which our space is based must form a discrete manifold or else the\\ 
\indent  reason for the metric relationships must be sought for, externally, in the binding forces acting on it.}\\
\indent {\sl On the Hypotheses which lie at the Bases of Geometry.}, Bernhard Riemann, 1854\\
\indent\url{https://www.emis.de/classics/Riemann/WKCGeom.pdf}\\
Riemann sought but failed to find a relation between the geometry of space and the forces of gravity and electromagnetism.  He tried this before Maxwell's equations had shown that 
light was an electromagnetic wave. Identifying space with an ether comprised of particles, 
he sought to relate forces to changes in geometry, speaking of  \\
{\sl 1) the resistance with which a particle opposes a change of its volume,
and
2) the resistance with which a physical line element opposes a change
of length.
Gravity and electric attraction and repulsion are founded on the first
part, light, heat propagation, electrodynamic and magnetic attraction and
repulsion on the second part.} \cite{riemann1853} (Translation here from \href{https://hal.science/hal-01436900/document}{Bottazini-Tazzioli})

\newpage

Why should gravity in particular be related to geometry?  Here's a version from Robert Geroch, appropriate for general audiences and parties:  To decide what a straight line is, 
you could stretch a rope to find the shortest distance between two points.  
But gravity bends the rope. Stretching the rope more tightly decreases the bending, 
but there is a fundamental limit to the stretching:  As the tension increases, the speed of waves in the string increases, and the speed of the waves cannot be greater than the speed of light.  You could extend a solid rod between the two points and measure the length of the rod with a ruler, but gravity bends the rod and the ruler. Choosing increasingly rigid and incompressible rods decreases the bending, but the more rigid the rod, the greater the speed of sound, and the speed of sound must be less than the speed of light.  Finally, you could use light itself, shining a laser between the two points, its path defining a straight line in spacetime.  But gravity bends the light.  Anything you use to decide 
what a straight line is-- or, more generally, to determine the geometry of spacetime-- is curved by gravity. That fact suggests gravity is tied in a fundamental way to the geometry of spacetime. 
Einstein's motivation, based on the equivalence principle and perhaps already familiar to you, 
comes in the next chapter.

\begin{wrapfigure}[15]{r}{3cm}
              \includegraphics[width=4cm]{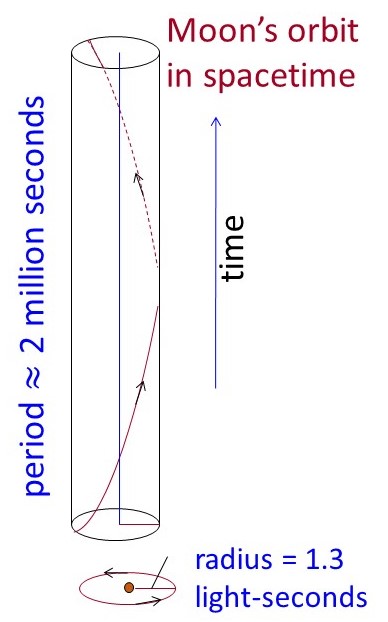}
                \label{moon_orbit}
\end{wrapfigure}

Two obvious obstacles blocked Riemann and later mathematicians (Sylvester, Clifford) from relating curvature to the curved trajectories of particles in a gravitational field:\\ 
\indent The geometers thought of the curvature of space, not spacetime. \\
\indent Space (and spacetime) is very nearly flat in the solar system.\\
Euclidean geometry is accurate to better than one 
part in a billion at the Earth's surface, but the spatial path of a ball or planet is 
nothing like a straight line.\\
 So why isn't it nonsense to say that particles move on 
geodesics, on paths of shortest length? 
\index{geodesic}
{\cblue The reason is that the {\sl spacetime} paths 
are very nearly straight:}  
If time is measured in seconds, distance is measured in light-seconds. 
The Earth-Moon distance is about 1.3 light-seconds, and the period of the Moon's orbit 
is $2.3\times 10^6$~s, so the spacetime trajectory of the Moon is a nearly straight spiral (see diagram on right).
Similarly, a ball that is tossed up and comes back down in 1 s reaches a height 
of $4\times 10^{-9}$ light-seconds (1.2 m), so it travels along a nearly straight path in 
spacetime, deviating from straight by a distance only $4\times 10^{-9}$ times the total 
distance it moves in spacetime.  The spacetime curvature of the path is comparable to the departure from Euclidean of the spacetime (and spatial) geometry.  \\
\vspace{-9mm} 

\begin{wrapfigure}[9]{r}{8cm}
               \includegraphics[width=8cm]{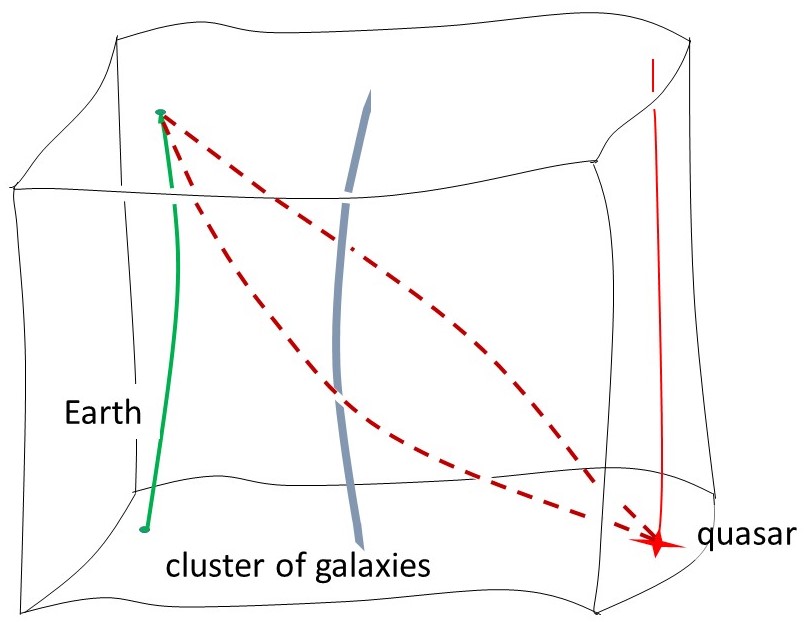}
                \label{IA5}
\end{wrapfigure}
\color{white}
.
.

\color{black}

\noindent General Relativity\index{general relativity}\index{relativity!general relativity}

	Earth curved

	No preferred spatial direction

	No absolute space

	No absolute time

	Preferred set of time directions

	\hspace{3mm}(if spacetime is time orientable)

	{\sl Spacetime curved}\hfill

\vspace{3cm}
\color{white}.\cb \hspace{7cm}
Light traveling from a quasar to Earth along two geodesics.\\
\color{white}.\cb \hspace{7.6cm} Gravitational lensing produces two images of the object.

\pagebreak

\noindent {\sl Spacetime diagrams}\\
\index{spacetime diagram}
Once there is no absolute time, histories can no longer be uniquely
described by a set of snapshots.  Instead, the usual   visualization is a
spacetime diagram.  Here are some examples.\\
\noindent
a.	The history of a string is a sheet.

\begin{figure}[h]
\begin{center}
\includegraphics[width=8cm]{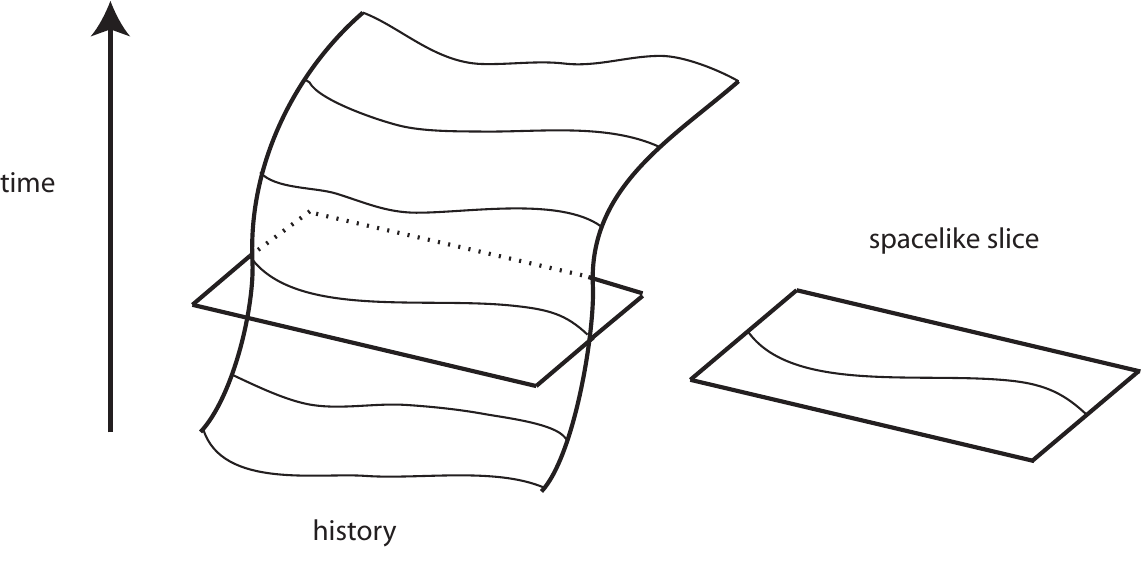}
\end{center}
\label{IIA6adobe}
\end{figure}

\noindent
b.	The history of a circle is a cylinder.

\begin{figure}[h]
\begin{center}
\includegraphics[width=8cm]{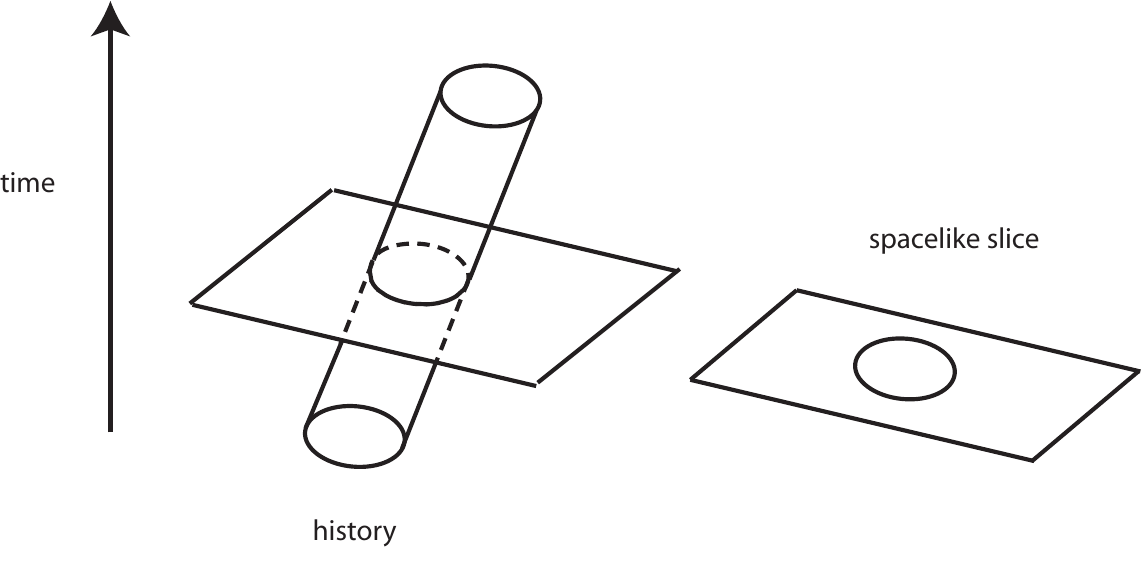}
\end{center}
\label{IIB7adobe}
\end{figure}

\noindent
c.	The history of a circle that starts as a point, expands, then ends as
a point is a sphere.

\begin{figure}[h]                                                              
\begin{center}
\includegraphics[width=10cm]{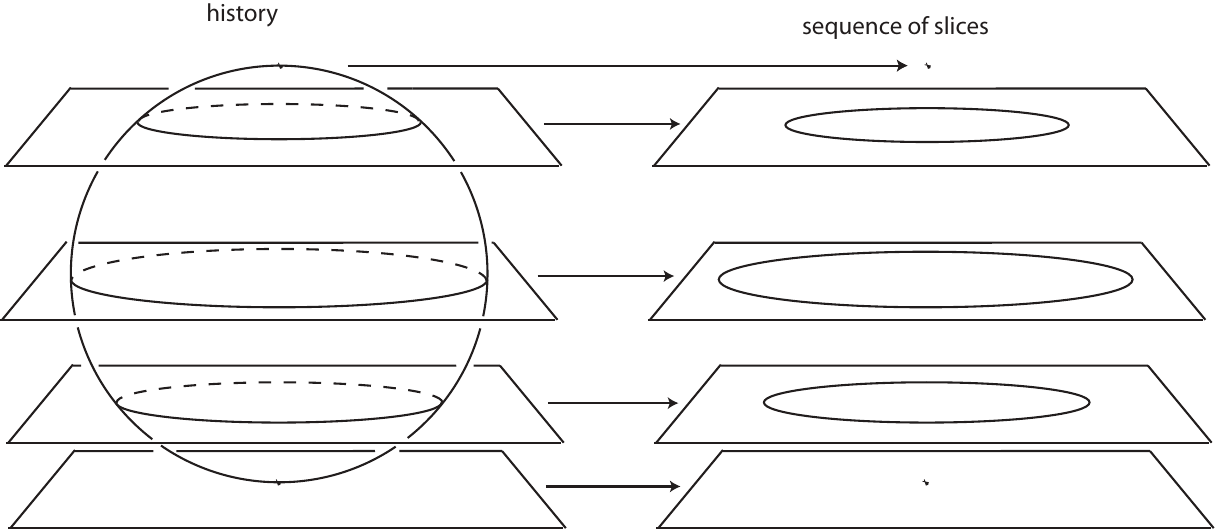}    
\end{center}                      
\label{IIC8adobe}                                                              
\end{figure}                                                                   

\pagebreak
\noindent
d.	The history of a balloon that starts from a point, expands,
contracts, and ends as a point is a 3-dimensional sphere $S^3$.
(This is the set of points in $\mathbb R^4$ satisfying $t^2+x^2+y^2+z^2=1$.)
\vspace{3mm}

\centerline{history:1 dimension suppressed \hspace{1.2cm} sequence of slices}
 
\begin{figure}[h]
\begin{center}
\includegraphics[width=.6\textwidth]{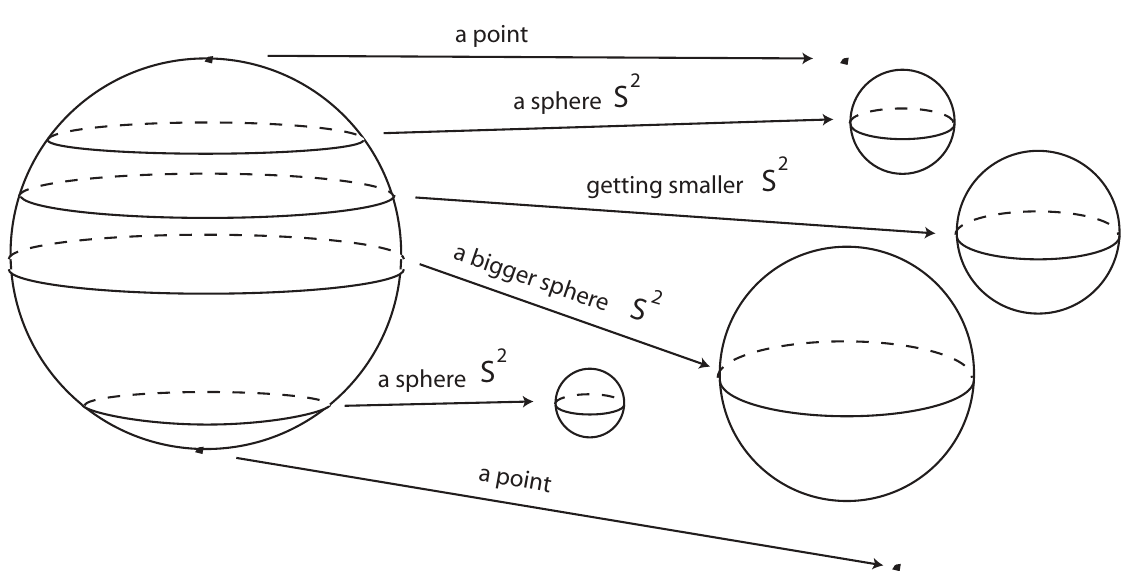}    
\end{center}                                                                             
\label{IID9adobe}                                                              
\end{figure}                                                                
\noindent   
e.	Closed universe:  
The history is a sequence of $S^3$'s starting from
a point, expanding, then contracting to a point.

\begin{figure}[h!] 
\begin{center}
\includegraphics[width=.6\textwidth]{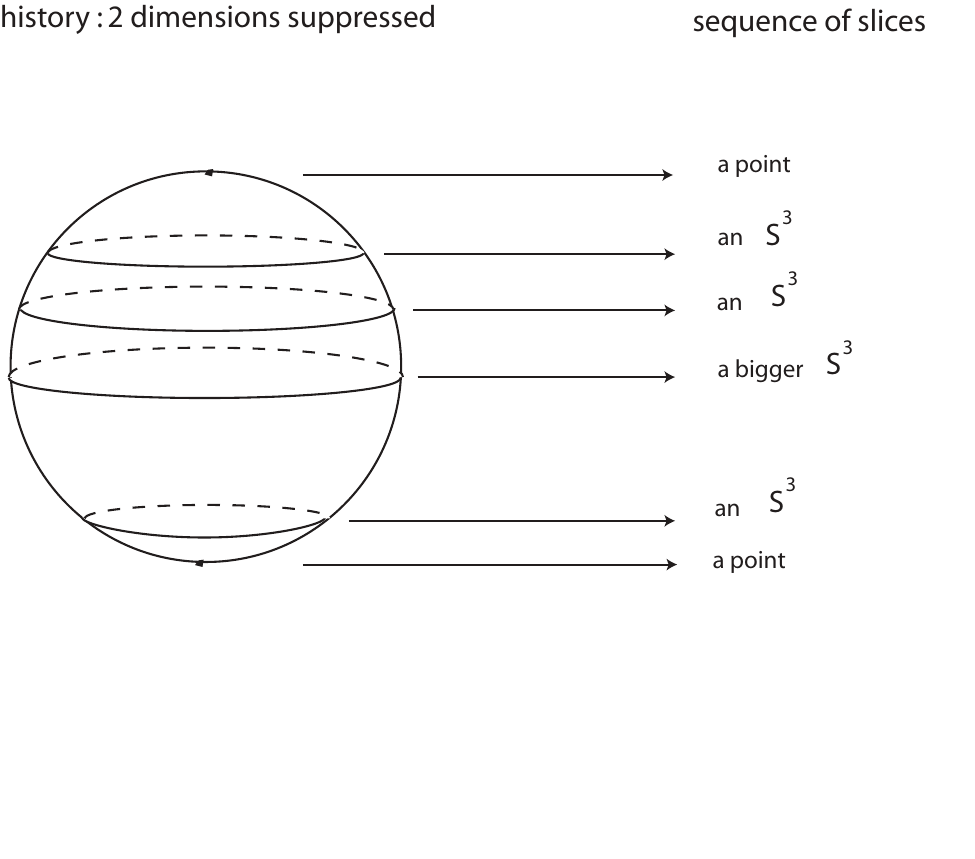}    
\end{center}                                                                                        
\label{IIE10adobe}                                                             
\end{figure}                                                                   

\newpage

\noindent Spacetime diagrams will soon be used to explicate some typical
paradoxes of special relativity.  In the meantime, it's sometimes useful in
understanding objects that can't be embedded in three dimensions - like
Klein bottles - to construct them as histories, with time the extra
dimension.  \\

\noindent A torus as a spacetime history

\begin{figure}[h] 
\begin{center}
\includegraphics[width=10cm]{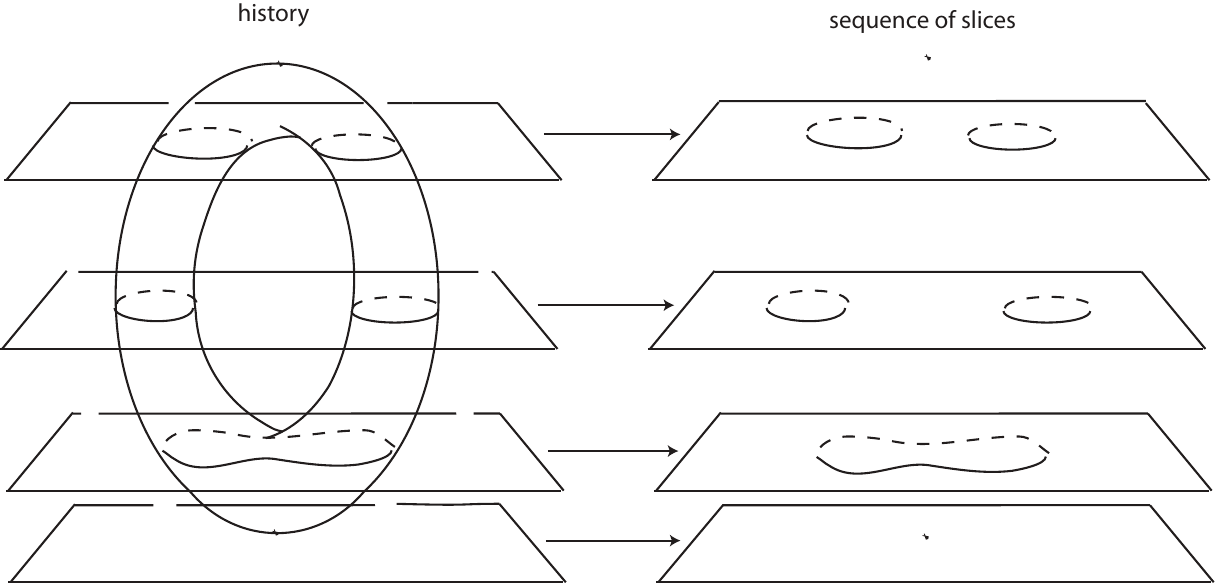}    
\end{center}                                                             
\label{IIE11adobe}                                                             
\end{figure}

\index{Klein bottle}
\noindent Klein bottle.  A Klein bottle can't be embedded in three dimensions without self-intersection.  It can, however, be embedded in four dimensions and so is a smooth non-intersecting history in spacetime:  
 Click on  \href{https://drive.google.com/file/d/1v1OIhDsdpfyoMTg8QJzQ76qW-vD-YRuk/view?usp=sharing}{torus-Klein bottle movie}.  If your browser doesn't 
play it, download the mp4 file to view it.  
As shown in this movie, to create a Klein bottle as a smooth submanifold of spacetime, take two circles (modeled as two wire loops) that coincide at $t=0$, separate them, rotate one by $\pi$, and then bring them together at $t=T$. \\

\centerline{Here is a somewhat different sequence of slices, similar now 
to the slices of the torus in the previous figure}

\begin{figure}[H] 
\begin{center}
\includegraphics[width=.2\textwidth]{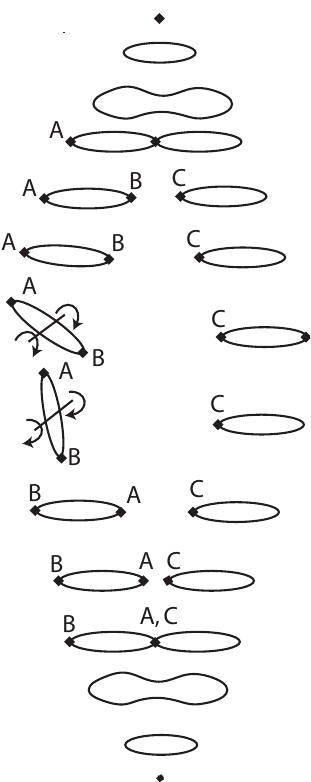}    
\end{center}                                                                                 
\label{f:klein}                                                             
\end{figure} 
\vspace{-4mm}                                                                  

\newpage

\section{Minkowski space}
\label{s:minkowski}\index{Minkowski space}
\index{special relativity|(}\index{relativity!special relativity|(}

{\sl The views of space and time which I wish to lay before you 
have sprung from the soil of experimental physics, and therein 
lies their strength. They are radical. Henceforth, space by itself, 
and time by itself, are doomed to fade away into mere shadows, 
and only a kind of union of the two will preserve an independent reality.} 
-- \href{https://archive.org/details/in.ernet.dli.2015.214561/page/n81/mode/2up}
{Hermann Minkowski, 1908}, 
translation above by Perrett and Jeffery.   \\ 

Readers are assumed to have encountered  some parts of special relativity: 
Lorentz transformations and the 4-velocity, 4-momentum, and rest mass of particles.
See, for example, Feynman's {\sl Lectures on Physics},  v. I, Chaps. 15-17 
\url{https://www.feynmanlectures.caltech.edu/I_toc.html}. 
(More intensive treatments are in Hartle's {\sl Gravity}\cite{hartlebook} 
with a ``physics-first'' presentation, and Gourgoulhon's comprehensive
{\sl Special Relativity in General Frames}\cite{gourgoulhon13},
with a mathematics-first perspective.)\\

We begin by deriving the usual results of special relativity, 
time dilation and length contraction, from the Minkowski metric.  
This is the reverse of the usual historical presentation, in which 
time dilation and length contraction are inferred from observer 
independence of the speed of light  and are then used to obtain the metric.  
Sect.~\ref{s:minkappendix}, beginning on p.~\pageref{s:minkappendix}
is a supplement for readers who are not familiar with that standard path 
to the Minkowski metric. If that includes you, you may want to look first at 
the introduction to the metric on this page and then work through the supplement, 
before going ahead to the rest of this chapter.

\subsection{Minkowski metric}\index{Minkowski metric}\index{metric!Minkowski}
\label{s:minkmetric}

The Euclidean geometry of flat 3-dimensional space is completely specified by the Pythagorean relation
\index{coordinates!Cartesian} 
\be
   ds^2 = dx^2 + dy^2 + dz^2.   
\label{e:flat1}\ee
Because the space is flat, the distance between finitely separated events is 
similarly given by 
\be
   (\Delta s)^2 = (\Delta x)^2 + (\Delta y)^2 + (\Delta z)^2,   
\ee  
and this is the distance $\Delta s$ between two events that are simultaneous for a given observer in terms of that observer's Cartesian coordinates.  
 
Underlying the unification of space and time is the universe's speed limit, the fact the speed of information $c$ is observer-independent. The Pythagorean relation is the 3-dimensional special case of the distance between any two events in flat 4-dimensional space, given in 
terms of the Cartesian coordinates of any inertial observer by  
\be
   (\Delta s)^2 = -c^2(\Delta t)^2 + (\Delta x)^2 + (\Delta y)^2 + (\Delta z)^2,   
\ee
with $\Delta s^2$ negative when the separation between events is timelike, when $c^2\Delta t^2>\Delta x^2 + \Delta y^2 + \Delta z^2$.  The proper time $\Delta \tau$ on a clock moving 
along a straight line between the events is then given by
\be
  (\Delta\tau)^2 = (\Delta t)^2 - [(\Delta x)^2 + (\Delta y)^2 + (\Delta z)^2]/c^2.
\ee
To complete the experimentally implied unification of space and time, we choose units with $c=1$ and write 
\be
   (\Delta s)^2 = -(\Delta t)^2 + (\Delta x)^2 + (\Delta y)^2 + (\Delta z)^2.    
\label{e:Deltas2}\ee
{\sl Units with $c=1$}.  If time and distance are measured in meters, 1 m in time is 
1 m$/c$ = 3.3336 ns;%
\footnote{Mermin \cite{mermin09}, noting that $c\approx$ 1 foot/nanosecond, defines a phoot to make this exact:1 phoot $ := c\times 1$ ns $\approx$ 0.984 ft.\\ \phantom{xxx} Then \mbox{$c=$ 1 phoot/ns.}} \\
if both are measured in seconds, 1 second of spatial distance is one 
light-second;\\
and if both are measured in years, 1 year of spatial distance is one light-year (ly). 

The Minkowski geometry of spacetime is now determined by the spacetime Pythagorean relation
\be\crv
  ds^2 =  -dt^2 + dx^2 + dy^2 + dz^2  \cb
\label{e:minkowski}\ee 
with the spacetime distance between two finitely separated points given by Eq.~\eqref{e:Deltas2}.
Notice that $\Delta s^2$ is called the squared {\sl distance} between events, whether it is positive or not. \\

We will use up indices to label coordinates in space, writing $\{x^i\}$, $i=1$-$3$. 
Similarly, up Greek indices will label coordinates in spacetime, $\{x^\mu\}$, $\mu=0$-$3$.
A vector $\bm v$ joining points in Euclidean space then has components denoted $v^i$.  
Its squared length $\bm v\cdot \bm v$, the squared distance between the points, is given in Cartesian coordinates by 
\be 
   \bm v\cdot \bm v = \sum_{ij} \delta_{ij} v^i v^j = (v^1)^2+(v^2)^2 + (v^3)^2,
\ee 
where $\delta_{ij}$ is the Kronecker $\delta$.  \\

\noindent{\bf Definition}.  The {\sl Euclidean metric} is the tensor whose components in Cartesian 
coordinates are $\delta_{ij}$. \\ 

From now on, we will use the {\sl Einstein summation convention}:  An expression with a repeated index implicitly denotes a sum over that index:    
\be
  \delta_{ij} v^i v^j :=\sum_{ij} \delta_{ij} v^i v^j.  
\label{e:Econvention}\ee
For reasons that will become clear in our discussion of vectors, dual vectors and tensors, 
the metric is written with down indices, and each implicit sum will ordinarily involve 
a pair of indices with one index up and one down, as in Eq.~\eqref{e:Econvention}.   

A vector $\bm v$ joining points in Minkowski space has components $v^\mu$.  Its squared length $\bm v\cdot \bm v$, the squared distance between the points, is then given in Cartesian coordinates by 
\be 
  \bm v\cdot\bm v = \eta_{\mu\nu} v^\mu v^\nu = -(v^0)^2+ (v^1)^2+(v^2)^2 + (v^3)^2,
\ee 
where 
\be
   \left\|\eta_{\mu\nu} \right\| = \left\| \begin{matrix} -1 &0&0&0\\
			    0& 1 & 0&0\\
			    0&0&  1 &0\\	
			    0&0&0&1  
	  \end{matrix} \right\|.  
\ee
\noindent{\bf Definition}.  The {\sl Minkowski metric} is the tensor whose components in Cartesian 
coordinates are $\eta_{\mu\nu}$. \\ 

\newpage
\noindent{\sl Spacelike, timelike, and null vectors} \index{spacelike!vector|textbf}\index{timelike!vector|textbf}\index{null vector|textbf}\\
Because of the minus sign in the Minkowski metric, the squared length of a vector, 
$\bm v\cdot \bm v$, can be positive negative or zero. \vspace{-5mm} 

\begin{wrapfigure}[0]{r}{10cm}
                \begin{center} \vspace{-10mm}
                \includegraphics[width=8cm]{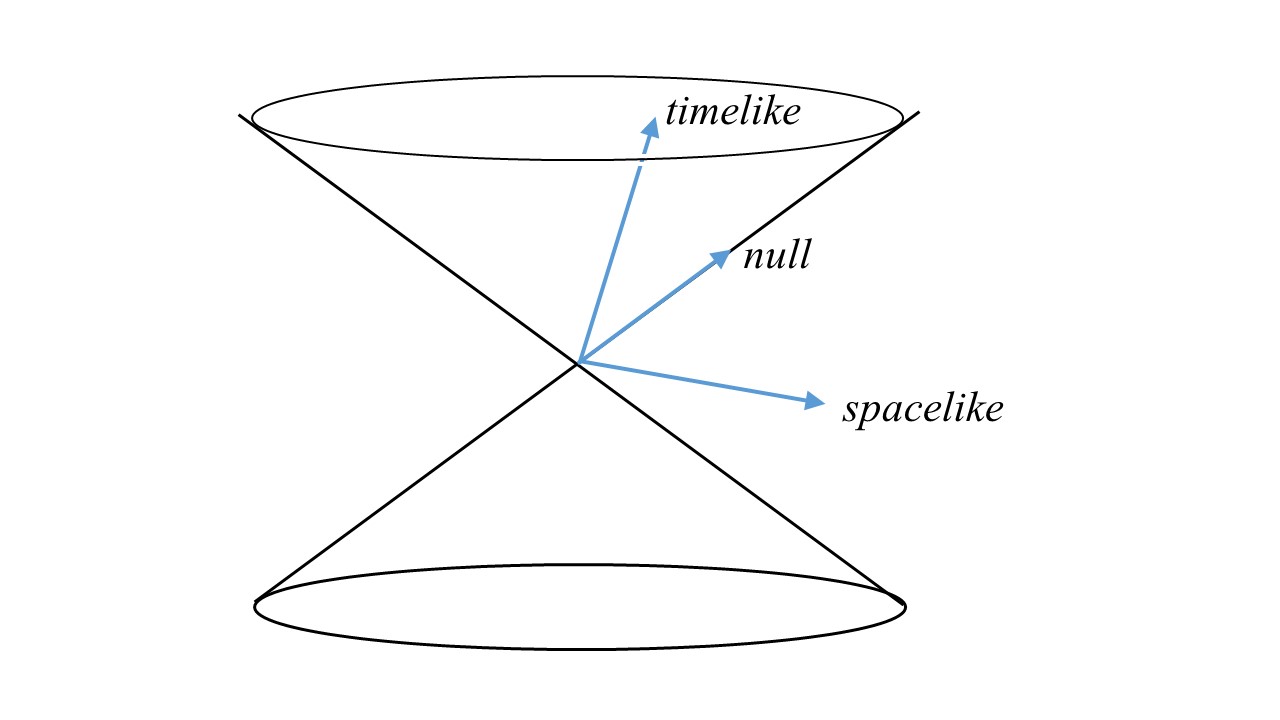}
		\end{center}
                \label{w21}
                \end{wrapfigure}
{\color{white} .}
      
If $\bm v\cdot \bm v < 0, \ \bm v$ is timelike, \index{vector!timelike}
\\
\vspace{4mm}

if $\bm v\cdot \bm v = 0, \ \bm v$ is null, and \index{vector!spacelike}
\\
\vspace{4mm}

if $\bm v\cdot \bm v > 0, \ \bm v$ is spacelike.\index{vector! null}\index{null vector}
\color{white}
.

.

.

\color{black}
\noindent  The null vectors from a given point lie on a 3-dimensional double cone, the light cone.\index{light cone} Timelike vectors fill the interior of the cone, and spacelike vectors lie outside it.  A choice of future and past picks out future and past cones generated by future and past pointing null vectors, 
respectively.  

The distance between two timelike separated events $P$ and $Q$ is the time read on a clock that moves in a straight line (no acceleration) from one event to the other.  
This is called the proper time of the clock or, for an observer with a clock, the 
observer's proper time.  
\begin{figure}[H]
                \begin{center} 
                \includegraphics[width=9cm]{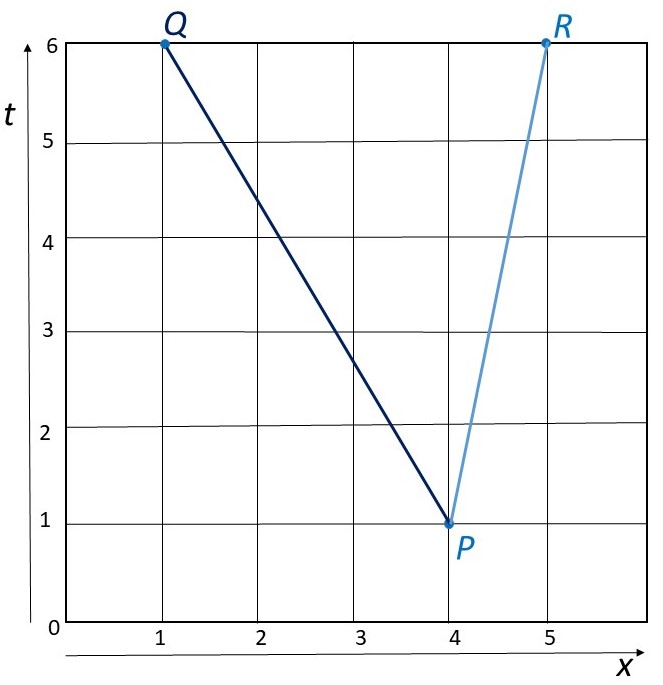}
		\end{center}
	\caption{The proper time $\tau$ read by clocks moving along the dark and light blue trajectories from $P$ to $Q$ and $P$ to $R$ is the spacetime distance between the respective events.}
                \label{time}
                \end{figure} 
In the figure, with $t$ and $x$ in seconds, the proper time between events $P$ and $Q$ is $\Delta\tau= \sqrt{5^2-3^2} = 4$ s; between $P$ and $R$ it is $\sqrt{5^2-1}=\sqrt{24}\approx 4.9$ s.  Notice that, although the path PQ looks longer in the spatial Euclidean geometry of your monitor (or paper), the length of the 
path in the actual Lorentzian geometry of spacetime is shorter.  

If spacetime diagrams are not familiar to you, read Chapter 4 of Hartle's {\sl Gravity}.  
Here's an excerpt (continuing until the exercises before the {\sl Surfaces of Simultaneity} section below):\\

\index{spacetime diagram}
To describe four-dimensional spacetime we first introduce a tool, which is so
simple it appears trivial, but so powerful it is indispensable. This is the idea of a
spacetime diagram. A spacetime diagram is a plot of two of the coordinate axes of
an inertial frame--two coordinate axes of spacetime. Since there are four axes and
only two dimensions on a piece of paper, two or at most three of these axes can
be drawn. Spacetime diagrams are slices or sections of spacetime in much the
same way as an $x$-$y$ plot is a two-dimensional slice of 3-dimensional space. A
typical example is shown in the figure above.  It is convenient to use $ct$ rather 
than $t$ as an axis, because then both have the same dimension. [Hartle will set 
$c=1$ later.]  

\index{event}
A point $P$ in spacetime can be called an {\sl event} because an event 
occurs at a particular place at a particular time, that is, at a point in spacetime. For example, a supernova explosion happened at the event in spacetime that occurred 
in A.~D.~1054 at the location of the Crab nebula. An event $P$ can be located in 
spacetime by giving its coordinates $(t_P , x_P , y_P , z_P )$ in an inertial frame.
A particle describes a curve in spacetime called a world line [In these notes, 
it will be called the particle's {\sl trajectory}]. It is the curve of 
positions of the particle at different instants, i.e., $x(t)$. The figure below shows a spacetime diagram with two sample world lines. 
\begin{figure}[H]
                \begin{center} 
                \includegraphics[width=11cm]{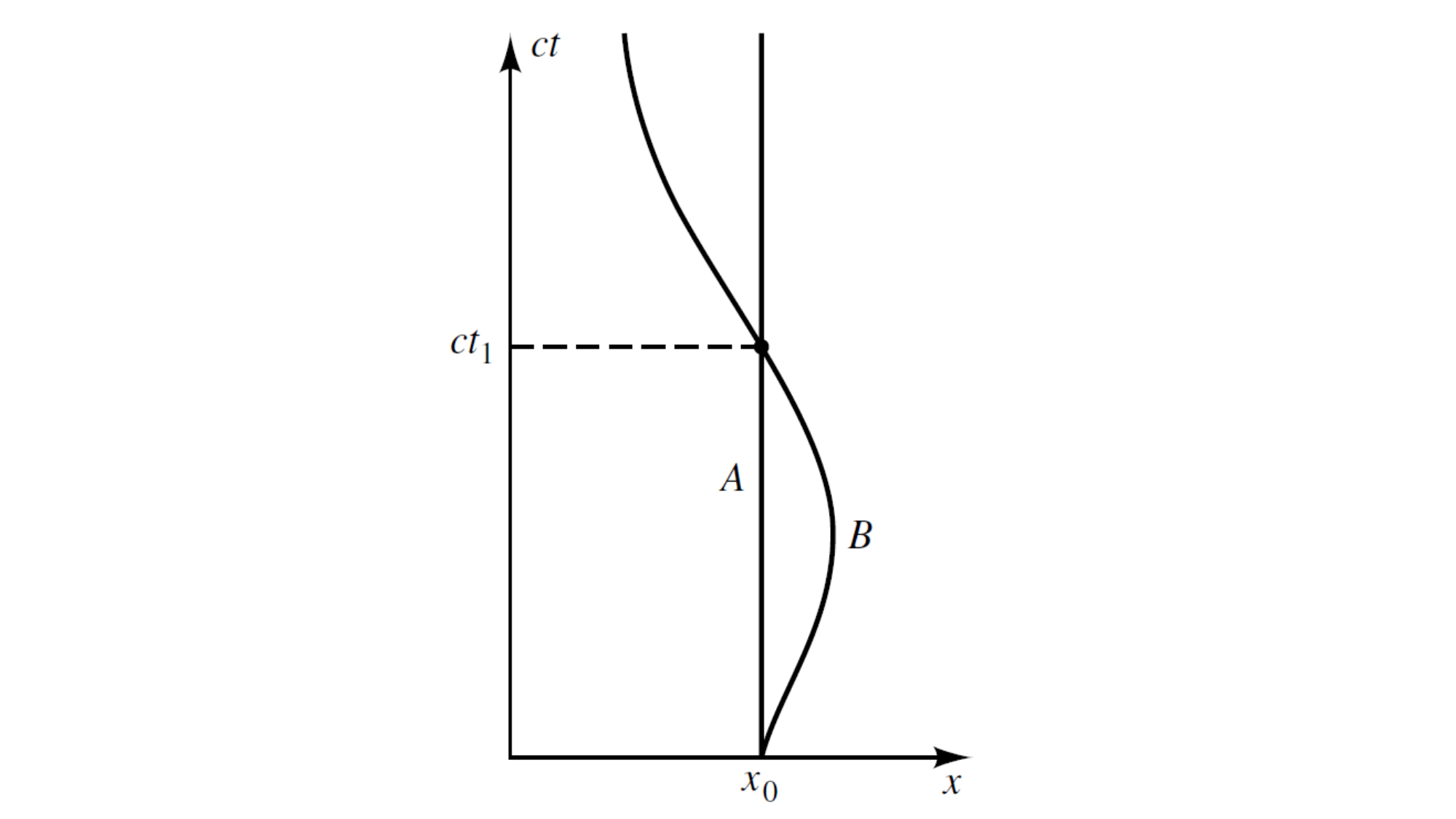}
		\end{center}\vspace{-5mm} 
	\caption*{Hartle Fig.~4.5. World lines in spacetime.  $A$ is the world line of a particle that sits at rest at $x_0$ for all time in the inertial frame $(ct,x)$.  World line $B$ represents an observer who accelerates away from $x_0$ at time $t=0$, decelerates, reverses direction, crosses $x_0$ at $t=t_1$, and heads off toward negative $x$. }
                \label{trajectorys}
                \end{figure} 
\noindent
The slope of the world line gives the
ratio $c/v^x$ since $d(ct)/d x = c dt/d x = c/v^x$ . Zero velocity corresponds to infinite slope. A velocity of $c$ corresponds to a slope of unity. Light rays therefore
move along the 45$^\circ$ lines in a spacetime diagram. \\

\noindent
[After introducing the Minkowski metric, Hartle continues with:]\\
Test your understanding of this by answering the following questions about the lengths between points in the figures
in the spacetime diagram in Figure 4.8 [copied on next page].  \\
(a) Which of the sides of triangle ABC is the longest? Which is the shortest?
What are the lengths in the units of the grid?\\
(b) Which is the shorter path between points A and C--the straight-line path
between A and C or the path through the other sides of ABC?\\
Then for (c) and (d), answer the same questions for triangle $A'B'C'$.\\

\begin{figure}[h!]
                \begin{center} 
                \includegraphics[width=15cm]{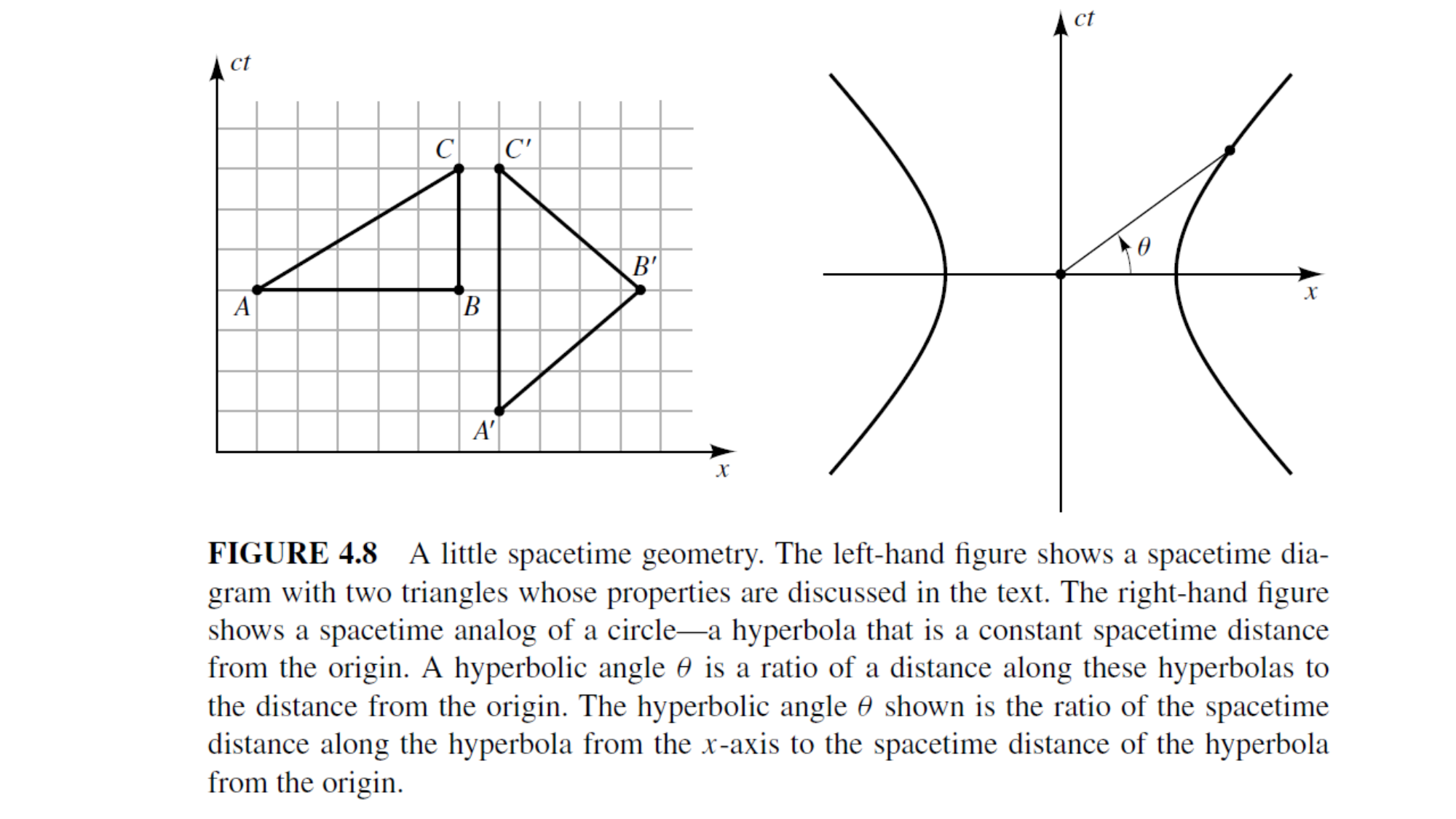}
		\end{center}
                \label{hartle4.8}
\end{figure}

\noindent{\sl Example: Spacetime Diagrams as Maps of Spacetime}.\\
\index{diagram, spacetime}\index{spacetime diagram}
No one would
think of confusing the relationships between lengths on a Mercator map of the
world with the relationships between true distances on the surface of the Earth.
A Mercator map is a projection of the geometry of the globe on a sheet of paper,
There are analogies between the elements of plane geometry and the geometry
of a spacetime diagram. One is illustrated in Figure 4.8. 

The analog of a circle of
radius $R$ centered on the origin is the locus of points a constant spacetime distance
from the origin. This consists of the hyperbolae $x^2-(ct)^2 = R^2$. The ratios of
arcs along a hyperbola to $R$ define hyperbolic angles, as shown in the figure, with
the relation
\be
     ct = R\sinh\theta, \qquad x = R\cosh\theta.  
\ee
It's useful to be able to understand these analogies, but it does not prove useful 
in relativity to pursue them too far. [We'll see later that a hyperbola of this 
kind is the trajectory of a uniformly accelerating particle.] \\

The following exercises have one- or two-line solutions. \\
 
\benr
\item $\tau$-Ceti, 12 ly away, hosts the closest potentially habitable planet.   
Draw a spacetime diagram for a spacecraft traveling to $\tau$-Ceti and back at speed 
$\frac{12}{13} c$, ignoring time spent accelerating.  
Use the lengths of the diagram in its Minkowski geometry to find the proper time elapsed on a spacecraft clock.  (If you are tempted to convert to m, cm, or s, stop and think.) 

\item What is the proper time $\tau$ read by a clock that moves with trajectory 
$(t, x(t), y(t))$ from \\
$t=t_1$ to $t=t_2$. Use the Minkowski metric \eqref{e:minkowski}.

\item What is $\tau(t)$ for a circular orbit with speed $v$? 

\een

\newpage

\subsection{Surfaces of simultaneity}
\index{surface of simultaneity}\index{simultaneity}
\label{s:simultaneity}

The Cartesian coordinates of the diagram in Fig.~\ref{time} above are those of an inertial 
observer who moves along a vertical line; the horizontal $t=$ constant lines are 
her surfaces of simultaneous events.  We can define surfaces of simultaneity in 
a democratic way, for any inertial observer as follows.  To construct a three dimensional space of events she calls simultaneous, she bounces light rays off events and calls the time $t$ of each event the average of the proper times measured on her clock when she sent and got back the signal.   Two events will be simultaneous if the times $t$ she computes in this way are the same.  
\begin{figure}[H]
\centering
\parbox{0.45\textwidth}{\includegraphics[width=\linewidth]{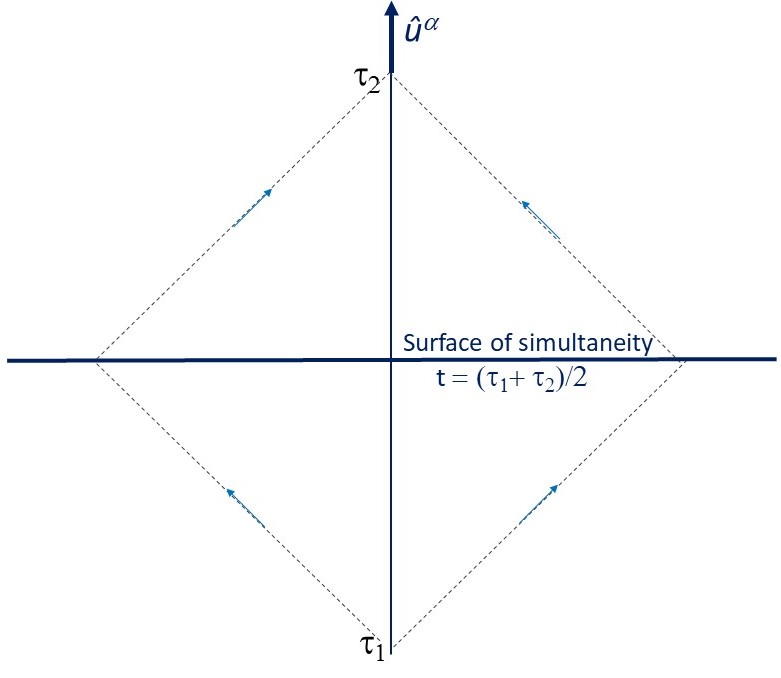}}%
\qquad
\parbox{0.45\textwidth}{\includegraphics[width=\linewidth]{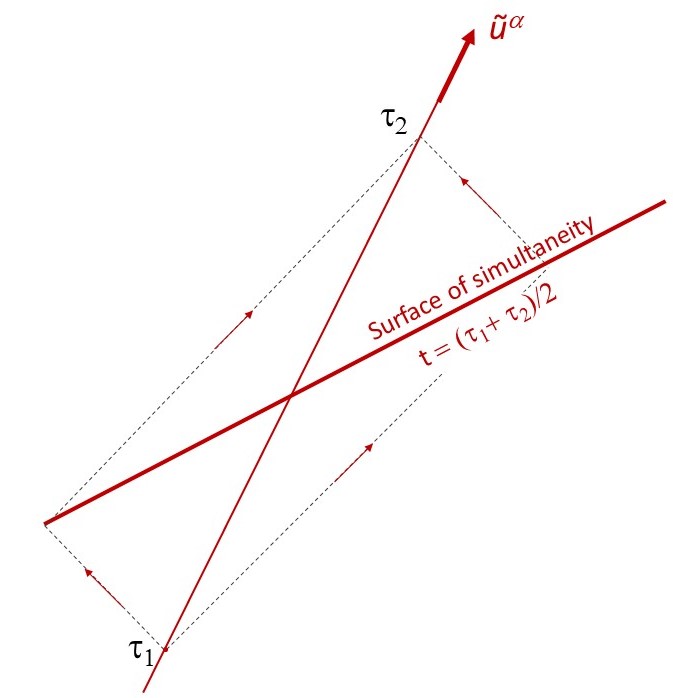}}
\caption{\hspace{27mm}(a)\hspace{60mm}(b) \\
\phantom{x}\hspace{28mm} Light rays are shown as dotted lines in each diagram.}%
\label{simultaneity1}\end{figure}
\vspace{-2mm} 

The set of all simultaneous events is called the surface of simultaneity of 
the observer, and two examples are shown in (a) and (b) above. Call observer 
(a) Alice and observer (b) Bob.  In a natural coordinate system of an inertial 
observer, the surfaces of simultaneity are the $t$ = constant hyperplanes.
\index{coordinates!natural coordinates of flat space}\index{coordinates!inertial}

It is clear from the figure that Alice's trajectory is 
orthogonal to her surfaces of simultaneity with respect to the Minkowski 
metric.  That is, a unit vector $\widehat{\bm u}$ tangent to Alice's trajectory 
has components $(1,0)$, while a vector $\widehat{\bm x}$ tangent to her 
surface of simultaneity 
has components $(0,1)$, and their dot product is 
$\eta_{\mu\nu}\widehat u^\mu \widehat x^\nu = -(1)(0) + (0)(1) = 0$.  
Because we don't know who is moving and who is at rest, it must be that Bob's 
trajectory is orthogonal to his surface of simultaneity. 

His trajectory is $(t,vt)$ in Alice's coordinates.  Its tangent $(1,v)$ is 
orthogonal to the line with coordinates $(t,t/v)$ and tangent $(1,1/v)$:  
\[ 
   (1,v)\cdot(1,1/v) = -1+1 = 0.  
\]
In more familiar language, Bob's 4-velocity is the unit tangent vector $\wt u$ with components $(\wt u^\mu) =\gamma(1,v)$ in Alice's frame, and the orthogonal unit vector 
proportional to $(1,1/v)$ has components $(\wt x^\mu) = \gamma(v,1)$. \\

\benr\item 
\label{ex:simultaneity} Check directly that Bob's surfaces of simultaneity, 
constructed by bouncing light-rays off events, as in Fig.~\ref{simultaneity1}(b), 
have the slope $dx/dt=1/v$ we have just found.  The easiest way is to use the fact 
that the light rays, at 45$^\circ$ angles, form a rectangle. \\
(Solution is Fig.~\ref{f:simangle} in the appendix to this Section.)
\een

\subsection{Time dilation and length contraction from the Minkowski metric}
\index{special relativity}\index{time dilation}

\noindent{\sl Time dilation} \hspace{1cm}   

Consider a slicing of spacetime by the surfaces of simultaneity 
of an inertial observer, together with a clock that moves with constant speed relative to that observer.  {\sl Time dilation} gives the proper time read by the clock as it moves from one of 
these $t=$ constant surfaces another, in terms of the difference $\Delta t$.  
For example, time dilation looks at the proper time of the clock carried by Bob as it passes a set of {\sl synchronized clocks} at rest relative to another inertial observer, Alice. Along 
Alice's surface of simultaneity, the synchronized clocks all read the same time $t$.   

\begin{figure}[h!]
  \centering
  \subfloat[]{\includegraphics[width=0.35\textwidth]{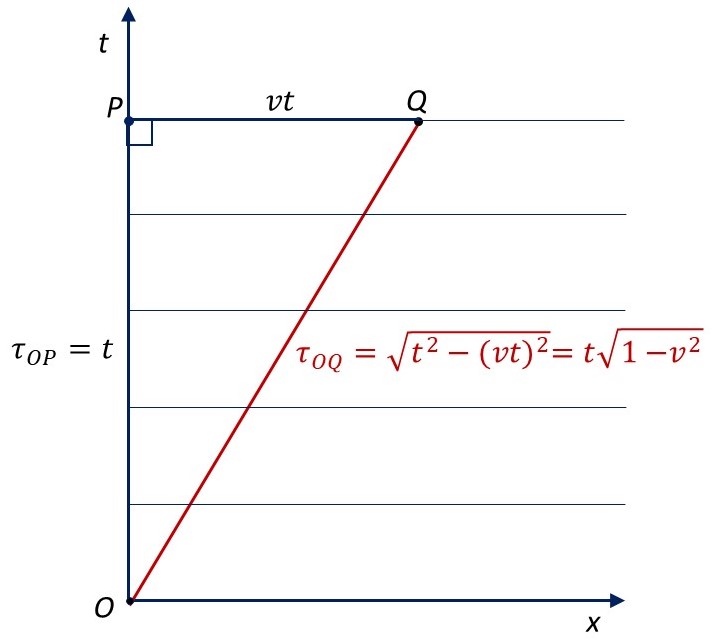}\label{fig:f1}}
  \hfill
  \subfloat[]{\includegraphics[width=0.65\textwidth]{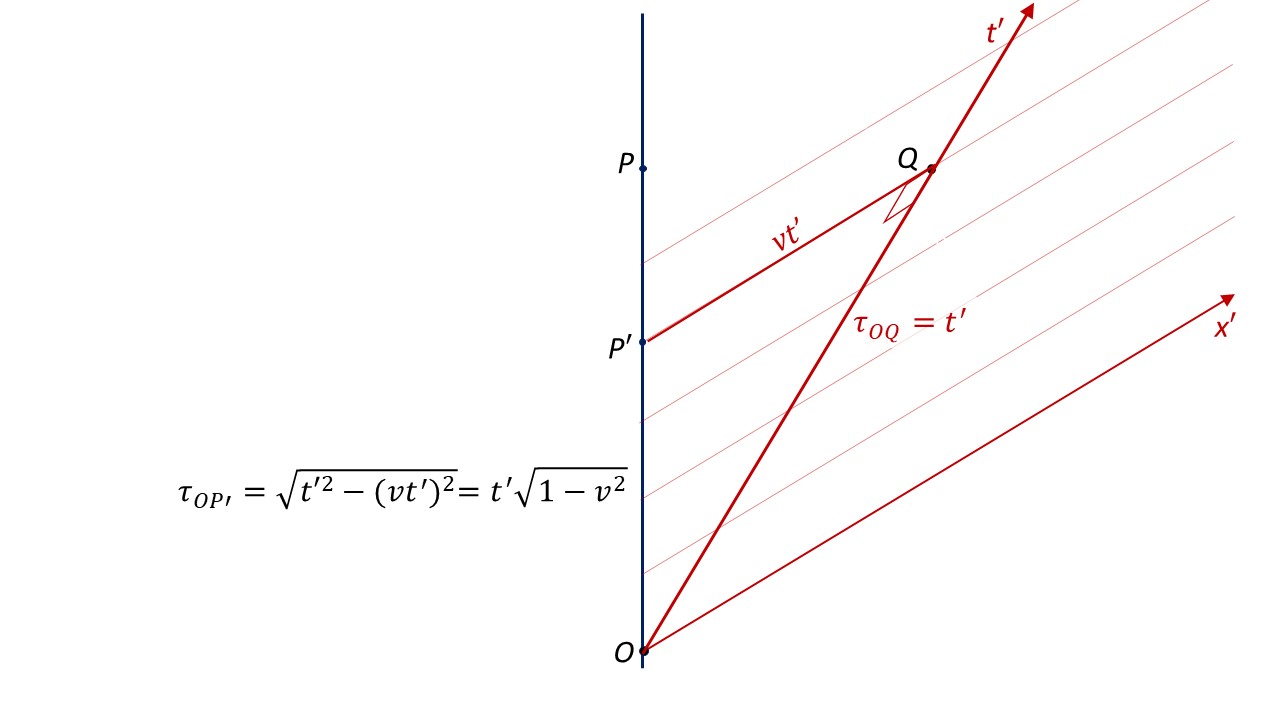}\label{fig:simultaneity3}}
  \caption{Bob moves from $O$ to $Q$ in proper time $\tau_{OQ}$.  Alice moves from $O$ to $P$ in 
proper time $\tau_{OP}$, where, as shown in (a), $P$ lies on Alice's surface of simultaneity 
through $Q$, with time coordinate $t=\tau_{OP}$, larger than $\tau_{OQ}$. From triangle 
OPQ, $\ \tau_{OQ} = t\sqrt{1-v^2}$. \\
In (b), Bob's surfaces of simultaneity are shown.  When Alice reaches point $P'$ in her proper time $\tau_{OP'}$, she is already crossing Bob's surface of simultaneity through $Q$, with $\dis\tau_{OP'} = t'\sqrt{1-v^2}$. }
\label{sim}
\end{figure}

In figure ~\ref{sim}, Alice and Bob, carrying clocks, move relative to one another at speed $v$. 
The same spacetime with the spacetime trajectories of Alice and Bob is shown in (a) and (b).  
In (a) the spacetime is sliced by Alice's surfaces of simultaneity, in (b) by Bob's.  

Diagram (a) uses the Minkowski metric to compare 
Bob's proper clock time $\tau_{OQ}$ to the larger change in the coordinate $t=\tau_{OP}$ labeling 
Alice's surface of simultaneity through $P$ and $Q$: From $\triangle$ OPQ,\\
\centerline{$\ \tau_{OQ} = \tau_{OP}\sqrt{1-v^2}$.}      
Diagram (b) shows the same clock trajectories, but with Bob's surfaces of simultaneity.
When Alice reaches $P'$ she is already crossing Bob's surface of simultaneity through $Q$. 
Diagram (b) uses the Minkowski metric to compare 
Alice's proper clock time $\tau_{OP'}$ to the larger change in the coordinate\\ \mbox{$t'=\tau_{OQ}$} labeling 
Bob's surface of simultaneity through $Q$ and $P'$: From triangle $OQP'$ (notice OQ$\perp$QP$'$) 
\centerline{$\ \tau_{OP'} = \tau_{OQ}\sqrt{1-v^2}$.}

The result, of course, is that you can't tell whose clock is at rest and whose is moving.
In each case the proper time elapsed on a clock is smaller 
than the difference in coordinate time between the first and last surface of simultaneity 
of an observer moving with speed $v$ relative to the clock. And it is smaller by the factor $\sqrt{1-v^2}$.\\  

\newpage

\noindent{\sl Length contraction}\index{length contraction}\\

What is the length of a stick?  In its own rest frame, it is the spacetime distance 
between two events, the simultaneous positions of the two ends of the stick as defined 
by an observer, Alice, moving with the stick. \vspace{-4mm}
\begin{figure}[h!]
		\begin{center}
                \includegraphics[width=.5\textwidth]{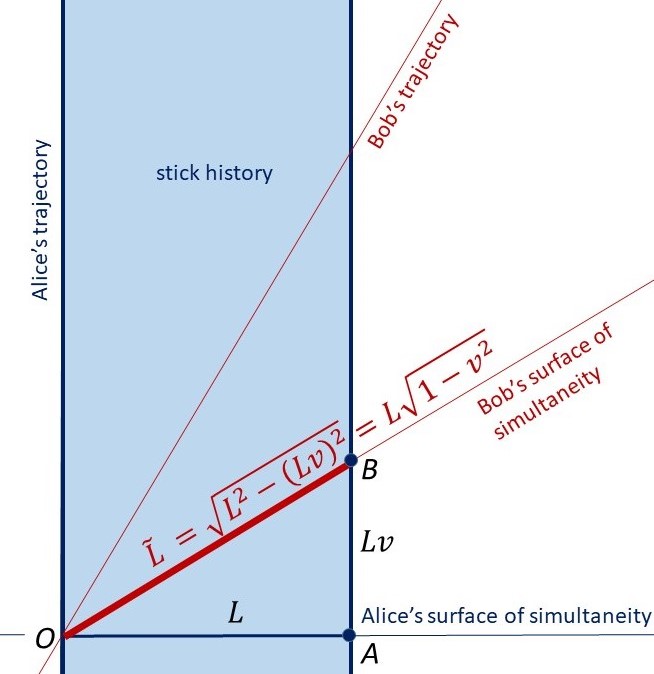}
		\end{center} 
\vspace{-6mm}
\caption{Alice's surface of simultaneity through $O$ intersects the stick's history along $OA$, with 
length $L$, the proper length of the stick.  Bob's surface of simultaneity intersects the history 
along $OB$, with length $\ \dis\wt L$. }
\label{simultaneity4} 
\end{figure} 

In Fig.~\ref{simultaneity4}, the blue shaded 
region is the spacetime stick, the history of the stick in spacetime.  Its length 
$L$ in its own frame is the length of the stick's intersection with 
Alice's surface of simultaneity.  The stick's length in Bob's
frame is the length of the stick's intersection with Bob's surface of 
simultaneity, namely the line $(t,t/v)$ in Alice's (and the stick's) coordinates.   
It intersects the left end of the stick at the origin and it intersects the 
right end of the stick when $(t,t/v) = (t,L)$, implying $t=Lv$.  
The length of the stick in Bob's frame is then the length $\wt L = OB$, the distance between events $O$ and $B$.
From triangle OAB,   
\be
   \wt L = \sqrt{-(Lv)^2 + L^2} = L\sqrt{1-v^2},  
\label{e:lcontract}\ee 
shorter than the proper length by $\sqrt{1-v^2}$. 

Because a boost tilts the surface of simultaneity only in the plane of the boost, 
lengths orthogonal to that plane are unchanged.  For an object whose proper 
volume $V$ is fixed, a slice boosted by a velocity $v$ then intersects the history of 
the object in a volume 
\be
   \wt V = V\sqrt{1-v^2}.
\label{e:vcontract}\ee 
To make this obvious for an object of arbitrary shape, consider a boost in the 
$t$-$x$ plane of the object ($t$ and $x$ the coordinates of an observer at rest relative to the object).  Divide the volume into small cubes 
aligned with the coordinate axes.  Then only the length of a cube in the $x$ direction 
changes with the change of slice, and the volume of each cube is smaller by $\sqrt{1-v^2}$
in the boosted slice of its history. 

\newpage
\subsection{Paradoxes}\index{paradoxes, special relativity}

	The usual paradoxes of special relativity arise because the notion of
simultaneity is observer dependent in reality but observer independent in
ordinary language and experience.  We have already considered the paradox of 
two relatively moving observers, each saying that the clock carried by the other 
is running slow.  One paradox will be treated here;
another will be a problem.  {\cblue They can all be similarly resolved, 
with the system described as a single consistent spacetime, 
sliced in different ways by the surfaces of simultaneity of different observers}. 

	Consider a box having length $L$ in its rest frame and a rod of length
$\ell > L$ in its (rod's) rest frame.  When the rod moves (rapidly) toward
the box, the box sees the rod's length contracted to such an extent that
when it reaches the box it can fit comfortably inside.  The rod, on the
other hand, sees the box contracted and thinks it can't come close to
making it inside.  But the box can actually catch the rod by lowering a
flap when the right end of the rod arrives and raising it when the left end of
the rod has passed.  Everyone must then agree that the closed box has a rod inside.  
How can this be consistent with what the rod sees?  (The flap of the box can be
as small as desired, and the time needed to raise and lower it negligible
compared to any other times in the problem, so the details of the flap can
be ignored).

	We will draw a spacetime diagram to describe the history of the rod and
box, and then look at the slicings of that history by the rod's and the
box's surfaces of simultaneity, showing that both rod and box are correct
\textit{and} consistent.
\index{spacetime diagram}
\vspace{-2mm}

\begin{figure}[h!]
\begin{center}
\includegraphics[width=14cm]{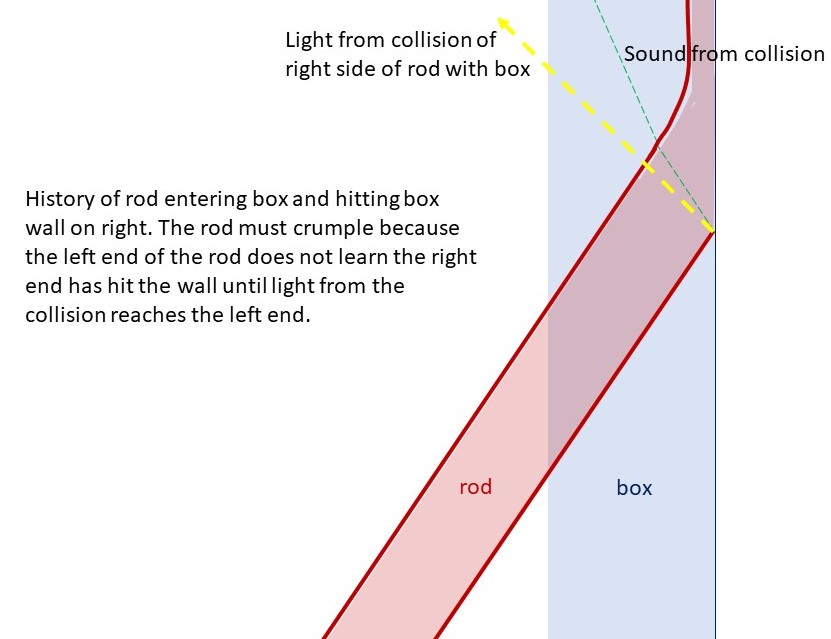}
\end{center}
\caption{Histories, like the history shown here, are observer-independent.  
What changes with different observers is the way the history is sliced by 
the observers' surfaces of simultaneity.}
\end{figure}

\newpage
\noindent In the slicing of history by the surfaces of simultaneity of 
an observer moving with the velocity of the initial rod, 
the left side of the rod is still outside the box when its right end hits the wall of the box.  

\begin{figure}[h!]
\begin{center}
\includegraphics[width=.9\textwidth]{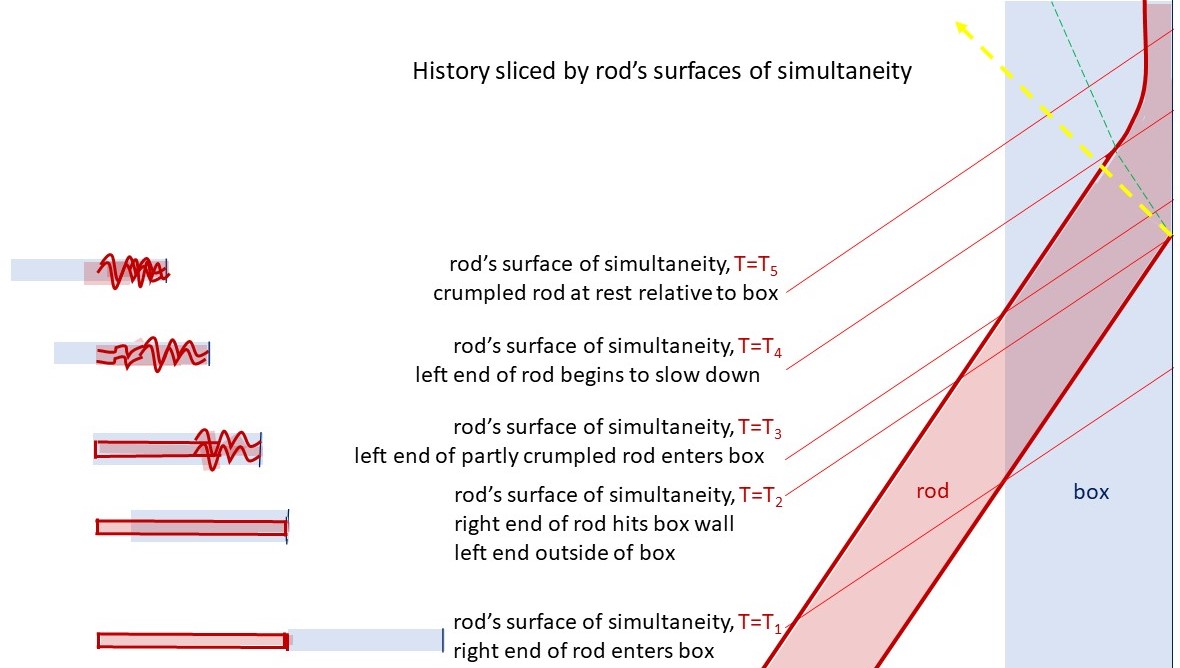}
\end{center}
\end{figure}
 
\noindent In the slicing of the history by the box's surfaces of simultaneity, the entire rod is inside the box before its right end hits the wall of the box. 

\begin{figure}[h!]
\begin{center}
\includegraphics[width=.8\textwidth]{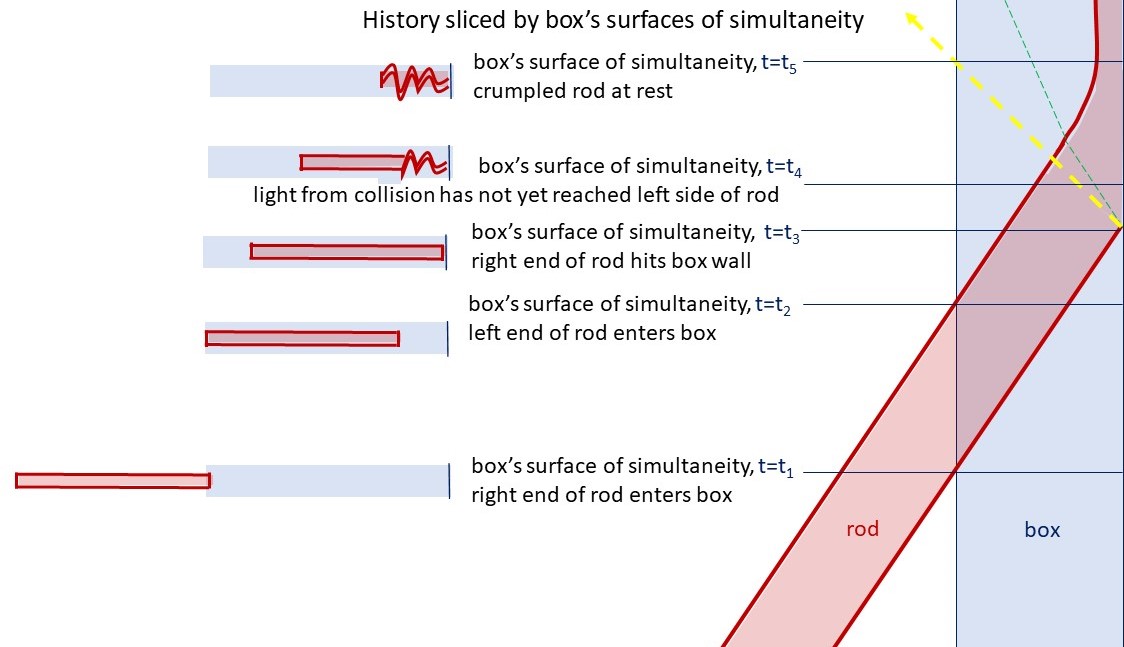}
\end{center}
\end{figure}

The lack of rigidity of the rod is required to resolve the paradox, 
and the lack of rigidity is implied by causality:  The left end of the rod does not change its 
speed until information from the collision has traveled from right to left across the rod. \\ 

\noindent {\em Exercise}:  (Lightman, Press, Price, Teukolsky \#
1.24 \cite{lppt})

Two giant frogs are captured, imprisoned in a large metal cylinder, and
placed on an airplane.  While in flight, the storage door accidentally opens
and the cylinder containing the frogs falls out.  Sensing something amiss,
the frogs decide to try to break out.  Centering themselves in the
cylinder, they push off from each other and slam simultaneously into the
ends of the cylinder.  They instantly push off from the ends and shoot
across the cylinder past each other into the opposite ends.  This continues
until the cylinder hits the ground.  Consider how this looks from some
other inertial frame falling at another speed.  In this frame the frogs do
\textit{not} hit the ends of the cylinder simultaneously, so the cylinder
jerks back and forth about its mean speed $v$.  The cylinder, however, was
at rest in one inertial frame.  Does that mean that one inertial frame can
jerk back and forth with respect to another?\\

\vspace{1cm}

As mentioned earlier, the next section is a supplement for readers who are not 
familiar with the path leading from the observer independence of the 
speed of light to the Minkowski metric. 
If the argument is familiar to you, skip to Sect.~\ref{s:tensors} on tensors 
and vector spaces. Later results will not refer to this supplementary section.

\begin{subappendices}  
\section{Supplement: Inferring the Minkowski metric from observer independence of the speed of light.}
\label{s:minkappendix}\index{Minkowski metric!from observer independence of $c$}

In this section, we'll go over the standard arguments giving time dilation and length contraction and then show how they lead to the Minkowski metric, 
\be
   \Delta s^2 = -\Delta t^2 + \Delta x^2 + \Delta y^2 + \Delta z^2.
\label{e:mink0}\ee
 
When the right side of Eq.~\ref{e:mink0} is positive, it gives the squared distance between two spacelike-related events, 
$P$ and $Q$.  That is, $PQ$ is the distance read on a ruler carried by an observer for whom the 
two events are simultaneous.  When the right side is negative, we can write 
\be
   \Delta \tau^2 = \Delta t^2 - \Delta x^2 - \Delta y^2 - \Delta z^2, 
\ee 
with $\Delta \tau$ the proper time elapsed on a clock that moves along an unaccelerated path 
from event $P$ to event $Q$ (where $P$ is the earlier event).

The argument will show that the spacetime version of the Pythagorean relation, 
\[
   \Delta s^2 = -\Delta t^2 + \Delta x^2 + \Delta y^2 + \Delta z^2,
\]
is implied by its special case when $\Delta t = 0$,
\[
  \Delta s^2 = \Delta x^2 + \Delta y^2 + \Delta z^2, 
\]  
together with the fact that $c=1$ for all observers.   

We begin with a quick review of the observer-dependence of simultaneity.\\

\noindent
{\sl Loss of absolute time}. \\

  The fact that the speed of light is the same to all observers-- implied by the observer-independence of Maxwell's equations and shown by the Michelson-Morley experiment-- requires us to abandon absolute time:
The argument that different observers must disagree on what events are simultaneous is simple.  It 
just uses the fact that Alice, on a moving train, and Bob, on the ground, watching the train pass him, 
each see light moving at speed $c$ in all directions regardless of the motion of the source:  Alice  
sets off a firecracker at the center of her train car and observes the light hit the two sides 
simultaneously.  Bob on the ground watches the train move to the right, so he sees the light hit the 
right side after it hits the left side. 
See, for example, minute 8 onward in this 
\href{https://www.youtube.com/watch?v=feBT0Anpg4A}{Mechanical Universe video} for a clear and 
entertaining presentation.  It goes on to show the tilted surfaces of simultaneity in spacetime that we now derive. \\

\begin{figure}[H]
\centering
\parbox{0.45\textwidth}{\includegraphics[width=\linewidth]{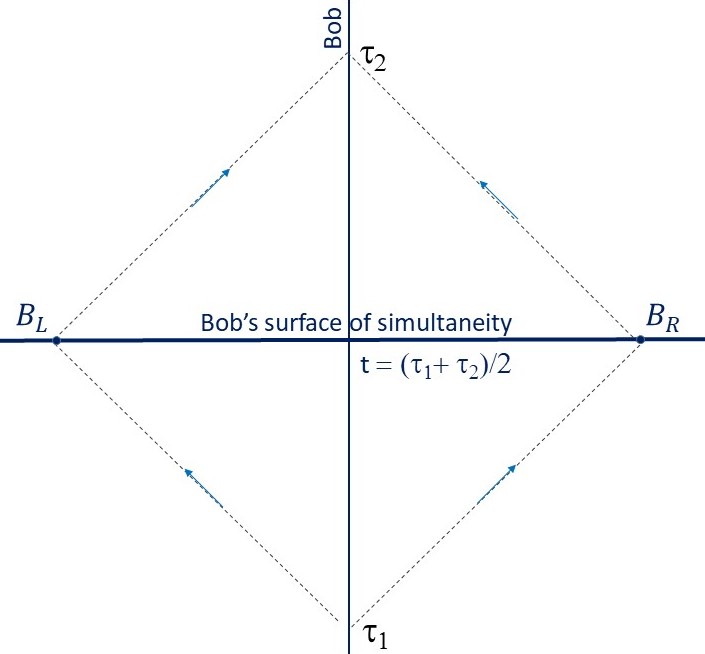}}%
\qquad
\parbox{0.45\textwidth}{\includegraphics[width=\linewidth]{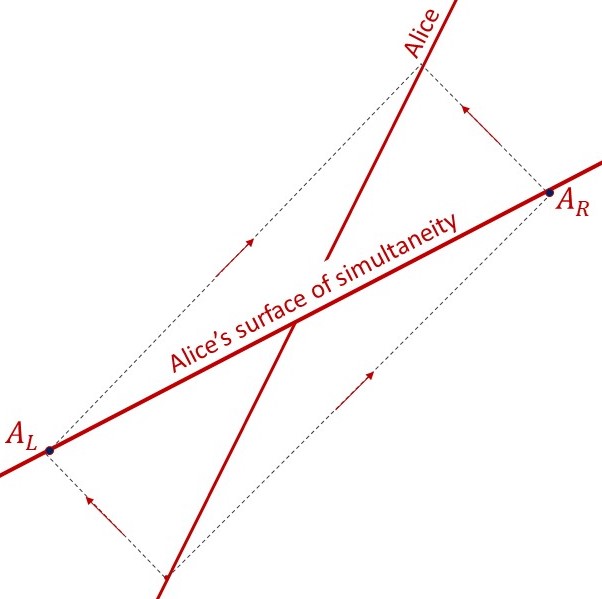}}
\caption{\hspace{27mm}(b)\hspace{60mm}(a) \\
Light rays are shown as dotted lines in each diagram.
The diagram now has Bob's trajectory vertical.}%
\label{f:simb}\end{figure}

To make the section self-contained we again define the surfaces of 
simultaneity of an observer. To be democratic, the spacetime diagram, Fig.~\ref{f:simb}, is drawn in Bob's frame this time, with his 
trajectory vertical,  
and Bob chooses his $x$-axis in the direction of Alice's motion. 
Alice moves at speed $v$ relative to Bob. Alice and Bob each determine 
their surfaces of simultaneity by sending light rays in opposite 
directions toward mirrors.  
\index{spacetime diagram}

\vspace{-2mm} 

The spacetime events $B_L$ and $B_R$ at 
which Bob's light rays bounce are simultaneous if the rays return 
to him at the same time. He assigns the same time $t$ to all events 
on his surface of simultaneity. Then if $\tau$ is time read on Bob's clock, 
the surface through $B_L$ and $B_R$ is a surface of constant 
$\dis t_B = \frac{\tau_1+\tau_2}2$, where $\tau_1$ is Bob's clock time 
when the light rays leave and $\tau_2$ the clock time when they 
return.  With this definition, Alice's trajectory satisfies $x=x_0 + vt$, 
with slope $\dis \frac{\Delta x_B}{\Delta t_B} = v$.   

Because the light rays Alice sends to right and left 
are at 45$^\circ$ to vertical in a spacetime diagram, the rays form 
a rectangle.  Her trajectory and surface of simultaneity are along 
the two diagonals of the rectangle.  In the diagram, that means 
the angle between Alice's surface of simultaneity and horizontal 
is the same as the angle between her trajectory and vertical.
The geometry is shown in the diagram below.  
\begin{figure}[H]
\centering
\includegraphics[width=0.5\linewidth]{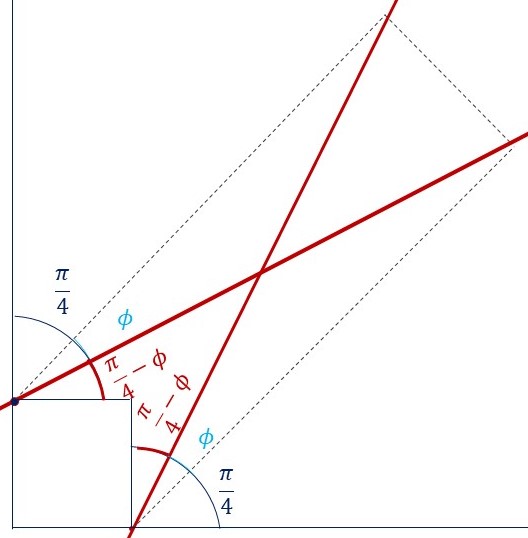}
\caption{Light rays form $\pi/4$ angles with vertical and horizontal. 
In the artificial Euclidean geometry of Bob's coordinate grid (and this paper or screen), the 
two angles labeled $\phi$ are the equal angles between the diagonals 
and the sides of the rectangle formed by the light rays.  
Then the angle between the trajectory and vertical and that 
between the surface of simultaneity and horizontal are each $\pi/4-\phi$.  }%
\label{f:simangle}\end{figure}
\vspace{-2mm} 

So, with respect to Bob's coordinates, her surface of simultaneity satisfies
\be
   \frac{\Delta t_B}{\Delta x_B} = v, \qquad  \frac{\Delta x_B}{\Delta t_B}=\frac1v.
\label{e:dxdt}\ee 
Note that the angles shown in Fig.~\ref{f:simangle} are angles in the 
artificial Euclidean metric $dt^2+dx^2$ of Bob's 
coordinates, which is also the metric of your screen or paper. They 
are not angles in the real Lorentzian geometry.  Using 
the Euclidean angles just makes it easy to find  
the coordinate ratio $\Delta x_B/\Delta t_B$.  

We do not yet need to relate Alice's coordinates to Bob's.

\subsection{Time dilation and length contraction}

\noindent {\sl Time dilation} \index{time dilation} \\
For those of you who have not seen the derivations before, they may be 
clearer with $c$ not set to 1, so $c$ will be kept until we go on to the 
derivation of the Minkowski metric.  

The experiment compares a time interval $\Delta t_B$ measured by Bob's synchronized clocks 
to proper time $\Delta \tau = \Delta t_A$ measured by Alice's single clock 
that moves past them at speed $v$.
Bob's clocks are at rest relative to the ground.  The experiment 
shows that the time measured on Bob's clocks is longer than the time measured 
on Alice's clock by the factor $\gamma = 1/\sqrt{1-v^2/c^2}$.  
 
In the experiment, Alice's clock is on the floor of a train of height $\ell$ 
that moves relative to Bob at speed $v$. A light signal from Alice's clock 
bounces off the ceiling and returns to her clock, traveling vertically 
as seen by Alice.  Alice's clock thus reads a roundtrip time $2\Delta \tau = 2\ell/c$.%
\footnote{We are looking at the roundtrip path of the photon instead of just the 
path from floor to ceiling because the comparison is between a single 
clock and a set of synchronized clocks, with the single clock moving 
relative to the synchronized clocks.  To directly measure 
the time for a photon to travel from floor to ceiling, Alice would need to 
synchronize a clock on the ceiling with her ground clock.}

\begin{figure}[h!]
  \centering
  \subfloat[]{\includegraphics[width=0.36\textwidth]{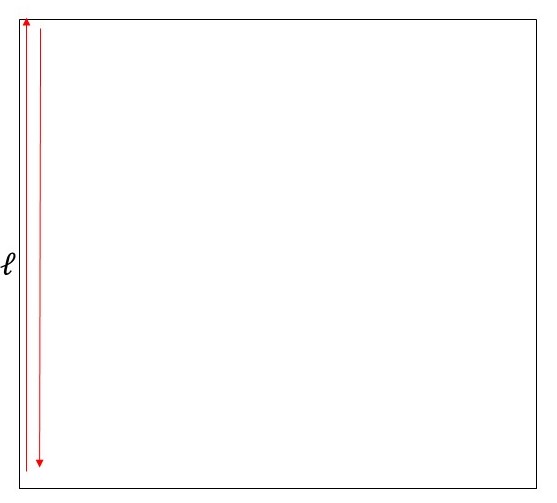}\label{fig:d1}}
  \hfill
  \subfloat[]{\includegraphics[width=0.6\textwidth]{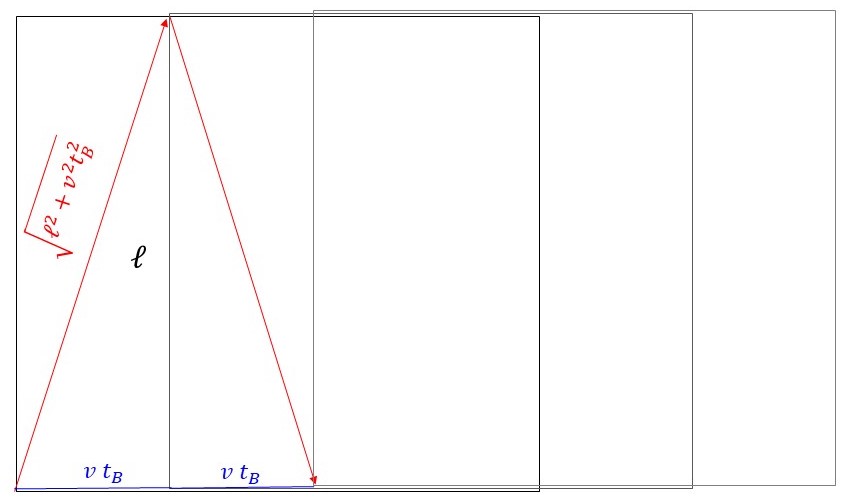}\label{fig:d2}}
  \caption{(a) In Alice's frame, a photon traverses a vertical path 
from floor to ceiling and back. \\
(b) In Bob's frame the photon moves a horizontal distance $2vt_B$ while 
it goes from floor to ceiling and back. Three successive positions of 
the train are shown as three overlapping rectangles, one in black, the others in gray.}
\label{f:dilation}
\end{figure}

Bob sees the light travel along diagonal paths, 
starting from his ground clock at $t_B$, bouncing off the ceiling, and meeting a 
synchronized clock when that clock reads time $t_B + 2\Delta t_B$. 
By the Pythagorean theorem, the total distance traveled is 
$2\sqrt{\ell^2+v^2 \Delta t_B^2} = 2\sqrt{c^2 \Delta \tau^2+ v^2 \Delta t_B^2}$. \  \ 
Because Bob must measure a speed $c$, that distance is $2c\Delta t_B$:  
\begin{align}
  2c \Delta t_B &= 2\sqrt{c^2\Delta \tau^2+ v^2 \Delta t_B^2} \ \Longrightarrow \nonumber\\
 \cblue \Delta t_B &= \cblue \frac{\Delta\tau}{\sqrt{1-v^2/c^2}}\cb.  
\label{e:dilation}\end{align} 

\noindent {\sl Length contraction}\index{length contraction} \\

   We next use use the time-dilation result of the last section to relate the 
the length of a ruler held by Bob to its length measured by Alice, who sees the 
ruler move past her at speed $v$.  The ruler's proper length $\ell$ is the 
length measured in its rest frame, the length measured by Bob.  
Because Bob sees Alice moving at speed $v$, his synchronized clocks measure a time $\Delta t_B=\ell/v$ for Alice to move from one side of the ruler to the other: 
$\Delta t_B$ is the time interval measured on Bob's synchronized clocks at each end of the ruler, between event $P$ when Alice is at the left end of the ruler and 
event $Q$ when she is at the right end. 

But we have just seen that the time elapsed on Alice's clock between two events on Bob's 
clocks as Alice's clock passes them is 
$\Delta t_A = \Delta t_B\sqrt{1-v^2/c^2}$. Now Alice watches the ruler move past her with 
speed $v$ in time $\Delta t_A$, and she thereby measures a length $v\Delta t_A$ of the ruler: 
\[
   \ell_A = v \Delta t_A = v \Delta t_B \sqrt{1-v^2/c^2} = \ell\sqrt{1-v^2/c^2}.  
\]
That is, if Alice moves relative to a ruler at speed $v$ along the line of the ruler, she will measure a contracted length 
\be
  \cblue \ell_A = \ell \sqrt{1-v^2/c^2}.\cb
\label{e:contraction}\ee

If two stars are at rest relative to one another, their proper distance $\ell$ is 
the distance measured by an observer at rest relative to them.  A spacecraft 
traveling from one star to another measures the smaller distance 
$\ell\sqrt{1-v^2/c^2}$ corresponding to a time elapsed on its clock that 
is smaller than $\ell/v$ by the factor $\dis\sqrt{1-v^2/c^2}$.  \\

From now on, we return to units with $c=1$.  

\subsection{Minkowski metric} 
\label{s:mink}
\index{Minkowski metric!inferring time dilation, length contraction}

We now relate time dilation and length contraction, Eqs.~\eqref{e:dilation} and \eqref{e:contraction}, to the geometry of flat space.  
This relation between physics and geometry comes from a set of 
identifications:
\ben[label=\arabic*.]
\item The set of events is spacetime; the path of an unaccelerated particle is a straight line in spacetime.  

\item The time $\Delta \tau$ elapsed on an unaccelerated clock is the spacetime distance between two events: That is, the distance $PQ$ between event $P$, the spacetime point at which a clock reads $\tau$, and event $Q$ at which the same clock reads $\tau+\Delta\tau$ is $PQ=\Delta\tau$. 

This is a physical interpretation of the distance between timelike separated events.
 
\item The distance between spacelike related events $P$ and $Q$ is the distance
measured by an observer for whom the events are simultaneous. It is the distance
measured by a ruler for which event $P$ is the left end of the ruler at a time $t$ 
and $Q$ is the right end at the same time $t$, according to an observer at 
rest relative to the ruler. That is, a ruler measures the distance between events 
that are simultaneous in the rest frame of the ruler. 
 
\een

The usual Pythagorean relation, together with assumption (3) means that 
on any surface of simultaneity $S$, the distance between points is given by 
\[
  \Delta s^2 = \Delta x^2 + \Delta y^2 + \Delta z^2, 
\] 
where the coordinates measure distances in orthogonal directions along the 
surface.  We take, as the fourth coordinate $t$, proper time measured by 
the inertial (unaccelerated) observers for whom $S$ is a surface of simultaneity. 
Then $t,x,y,z$ will be coordinates for spacetime, with $x,y,z$ constant and 
$t$ proper time along the trajectory of each of these observers.  

We first derive the spacetime Pythagorean relation for timelike related events 
from the time dilation equation.  We take Bob to be an inertial observer, whose 
coordinates are $t,x,y,z$ as described.  As in the time dilation experiment above, 
Alice's clock will move at speed $v$ past two of Bob's synchronized clocks.  
In the spacetime diagram below, the trajectory of Bob's clocks is vertical, 
and the horizontal line is one of Bob's surfaces of simultaneity, a set of 
events for which Bob's clocks all read the same time $t$.  

\begin{figure}[H]
  \centering
  \includegraphics[width=0.4\textwidth]{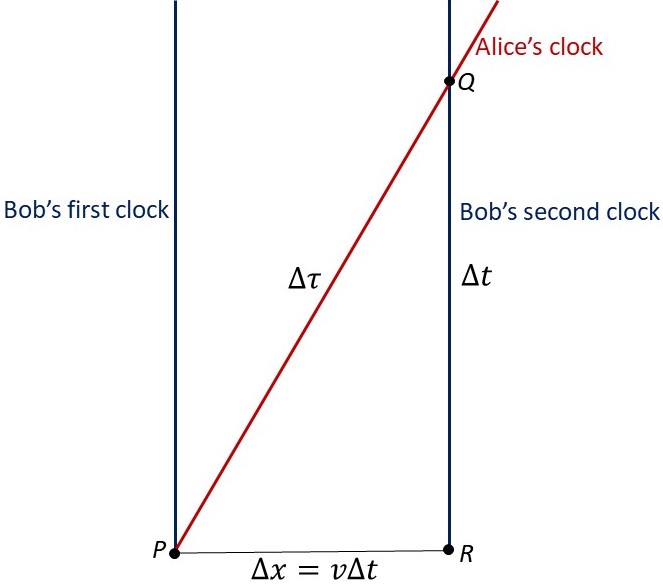}
  \caption{Bob's clocks move along vertical trajectories, 
and Alice's clock moves along the diagonal (red) trajectory. Alice's clock 
meeting Bob's first clock is event $P$; Alice's clock meeting Bob's second 
clock is event $Q$. }
\label{fig:contraction3}
\end{figure}
 
Bob's first clock at $P$ is synchronized with his second clock at $R$, each reading 
time $t$, with the two clocks separated in Bob's frame by a distance $\Delta x$: 
Then $PR = \Delta x$, because $P$ and $R$ lie in a surface of simultaneity.  
Alice's clock, which passes Bob's clock at $P$, reaches Bob's second 
clock at a point $Q$, when that clock reads a time $t+\Delta t$.  The statement  
"Bob measures a speed $v$ for Alice" means $\Delta x = v\Delta t$. That is, 
the proper time $QR$ elapsed on Bob's second clock is $QR = \Delta t = \Delta x/v$.   
Finally, the third side $PQ$ of triangle $PQR$ is the proper time elapsed on 
Alice's clock between events $P$ and $Q$.  From the time dilation 
equation~\eqref{e:dilation}, the length $PQ$ is  $\Delta\tau = \Delta t\sqrt{1-v^2}$.  

The lengths of the three sides of the triangle therefore satisfy the spacetime Pythagorean relation, 
\be 
   (PQ)^2 =  (QR)^2 - (PR)^2. 
\label{e:pyth2}\ee 
Check:
\[ 
(PQ)^2 = \Delta \tau^2 = (1-v^2) \Delta t^2 = \Delta t^2 - (v\Delta t)^2 = (QR)^2 - (PR)^2,
\]  
or 
\be 
  \Delta \tau^2 = \Delta t^2 - \Delta x^2 . 
\ee   
With a generic orientation of Bob's $x$-axis in his surface of simultaneity, $PR = \sqrt{\Delta x^2 + \Delta y^2 + \Delta z^2}$, 
and we have 
\be 
  \Delta \tau^2 = \Delta t^2 - \Delta x^2 - \Delta y^2 - \Delta z^2,  
\ee 
as claimed. \\
\newpage

Finally, we show that the spacetime Pythagorean relation holds for two spacelike related events.
We'll use length-contraction, Eq.~\eqref{e:contraction}, to relate the proper length $\Delta x$ of 
a ruler carried by Bob to its length measured by Alice.  Let points $P$ and $Q$ be the two ends of 
the ruler in one of Alice's surfaces of simultaneity - the left and right ends of the ruler at 
the same time according to Alice.  
In the diagram, the history of the ruler is the blue rectangle.  
Choose Bob's coordinates so that $P$ and $Q$ are in his $t$-$x$ plane with $\Delta t$ and 
$\Delta x$ positive.  
If $P$ and $Q$ are spacelike related, then $\Delta t < \Delta x$ in Bob's coordinates. From Eq.~\eqref{e:dxdt}, the surface of simultaneity of an observer (Alice) moving with speed $v$ along Bob's $x$ direction  has 
slope $\Delta t/\Delta x = v$, with $\Delta t$ and $\Delta x$ labeled 
in the diagram below. (With $c$ not set to 1, $v=c^2 \Delta t/\Delta x$.)   

\begin{figure}[ht!]
  \centering
  \includegraphics[width=0.4\textwidth]{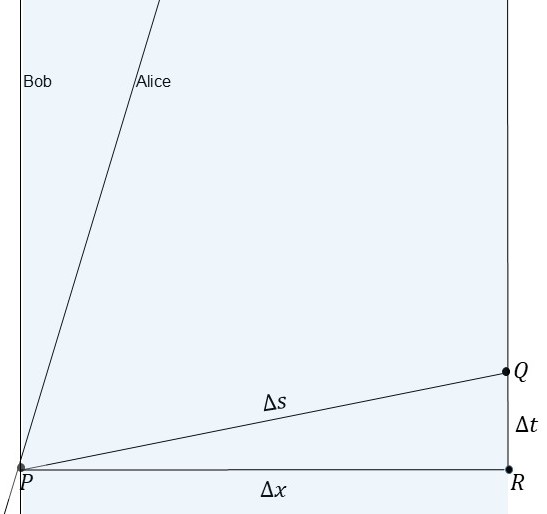}
  \caption{The light blue region is the history of a ruler at rest in Bob's frame. $PQ$ and $PR$ are, respectively, intersections of the ruler with Alice's and Bob's surfaces of simultaneity.  
Because events $P$ and $R$ are simultaneous in the rest frame of the ruler, the length  \mbox{$PR = \Delta x$} is the proper length of the ruler and is its length as measured 
by Bob.  The length $PQ = \Delta s$ is the ruler's length as measured by Alice. }
\label{fig:contraction4}
\end{figure}

The ruler at Bob's time $t$ is the line $PR$, the intersection of 
Bob's surface of simultaneity with the ruler history.  The length of the 
ruler according to Alice is the line $PQ$, the intersection of her 
surface of simultaneity with the ruler history.  Because Alice 
sees the ruler move past her with speed $v$, she measures the contracted 
length $PQ = \Delta s = \Delta x\,\sqrt{1-v^2}$, implying the spacetime
Pythagorean relation 
for the triangle $PQR$: 
\[
    (PQ)^2  = -(QR)^2 + (PR)^2.
    \]
Check:
\begin{align*}
    (PQ)^2 & = (\Delta x\sqrt{1-v^2})^2 
	    = \Delta x^2[1 - (\Delta t/\Delta x)^2] = \Delta x^2 -\Delta t^2 \\
          & = -(QR)^2 + (PR)^2,  \quad\mbox{or}\\
   \Delta s^2 &= -\Delta t^2 + \Delta x^2.
\end{align*} 
With a general direction of Bob's spatial axes, we then have 
\be
\Delta s^2 = -\Delta t^2 + \Delta x^2 + \Delta y^2 + \Delta z^2,
\ee
as claimed. \\ 

If $PQ$ is the trajectory of a photon, 
$\Delta t = \sqrt{\Delta x^2 + \Delta y^2 + \Delta z^2}$. Requiring the spacetime distance between $P$ and $Q$ to be continuous 
as a function of $P$ and $Q$ then implies the spacetime distance 
$\Delta s$ between lightlike related events is zero. \\      

With the supplement's derivation of the Minkowski metric in hand, the reader can return to 
Sect.~\ref{s:minkmetric}  

\setcounter{chapter}{1}

\end{subappendices}
 
\subsection{Lorentz transformations} \index{Lorentz transformation}

\noindent{\sl The metric as a dot product}\\
The dot product of two vectors $\bm u$ and $\bm v$ in Euclidean space 
has, in Cartesian coordinates, the form
\[
  \bm u \cdot \bm v = \delta_{ij}u^i u^j.
\]
Similarly, the dot product of two vectors $\bm u$ and $\bm v$ in 
$\mathbb R^4$ with the Minkowski metric is given by 
\be
  \bm u \cdot \bm v = \eta_{\mu\nu}u^\mu v^\nu. 
\label{e:Lorentzdot}\ee  

\noindent{\sl Symmetries of flat space} 

\index{symmetry}\index{symmetry!spacetime symmetry}\index{symmetry!Lorentz transformation} 
\index{symmetry!active} \index{symmetry!passive}\index{passive transformation}\index{passive transformation!Lorentz transformation}\index{coordinates!Lorentz transformation}
The equations of Newtonian physics are invariant under rotations and translations. The active form of the invariance is the statement that a constant rotation or translation takes an isolated system to a physically indistinguishable 
isolated system.  In its passive form, this is the invariance 
of the equations under a rotation or translation of the coordinates.  
The 
invariance is tied to the invariance of flat space under the Euclidean group, 
the group of transformations 
\[
   \bm x \mapsto R\bm x + \bm a
\] 
with $R$ a rotation or rotation-reflection.
\index{Lorentz transformation!active and passive}  

\index{rotation group}
The rotation group is the group of all linear transformations of flat space that 
preserve lengths and angles, or, equivalently, the dot product of any two vectors 
\[
  \bm u\cdot \bm v = \delta_{ij}u^i u^j = u^1 v^1 + u^2 v^2 + u^3 v^3.
\] 
\vspace{-2mm}
Let $\bar{\bm u} = R\bm u, \quad \bar{\bm v} = R\bm v$.  Preserving 
the dot product means $\bar{\bm u}\cdot \bar{\bm v} = \bm u\cdot\bm v$, or  
\be
  \delta_{ij} R^i{}_k u^k R^j{}_l v^l =\delta_{ij} u^i v^j , \mbox{ all } \bm u, \bm v 
\Longrightarrow     \delta_{ij} R^i{}_k R^j{}_l = \delta_{kl}.
\ee
That is, a rotation $R$ preserves the metric $\delta_{ij}$ of flat space.
The Euclidean group is then the group of length and angle preserving maps: These are 
{\sl isometries}, maps that preserve the metric.

Newton's laws are also invariant, as noted earlier, under Galilean transformations, 
boosts by an arbitrary velocity $\bm v$: 
\[
  t\mapsto t,\ \ \bm x\mapsto \bm x + \bm v t ,  
\] 
mixing space and time, but preserving $t=$ constant surfaces.  These are 
not isometries of spacetime, but we will see that {\sl they are 
the Newtonian limit of spacetime isometries\index{isometry}\index{symmetry!isometry}}, Lorentz boosts of 
Minkowski space.\\ 
\index{Minkowski space|textbf} 
\index{spacetime!flat spacetime|textbf}\index{flat spacetime|textbf} 
\hspace{-2mm}{\bf Definition}. {\sl Minkowski space}\index{Minkowski space} is $\mathbb R^4$ with the Minkowski 
metric.\\
 We will use the terms {\sl Minkowski space} and {\sl flat spacetime} 
interchangeably.   \\

\noindent{\sl Symmetries of flat spacetime} 

The {\sl Lorentz group} is the group of linear transformations preserving the 
dot product \eqref{e:Lorentzdot}.  Again, with $\bar u = \Lambda u, \quad \bar v = \Lambda v$, preserving dot products 
means $\bar u\cdot \bar v = u\cdot v$ or 
\be
  \eta_{\mu\nu} \Lambda^\mu{}_\sigma u^\sigma\Lambda^\nu{}_\tau v^\tau 
	=\eta_{\mu\nu}  u^\mu v^\nu , \mbox{ all } u, v \Longrightarrow 
  \eta_{\mu\nu} \Lambda^\mu{}_\sigma\Lambda^\nu{}_\tau  = \eta_{\sigma\tau}.
\ee
\index{Lorentz transformation|textbf}
That is, $\Lambda$ preserves the metric: A Lorentz transformation is a linear 
map of Minkowski space to itself that preserves the Minkowski metric.  

One typically describes the group by picking a timelike direction $\widehat{\bm u}$ 
and slicing the spacetime by spacelike hypersurfaces orthogonal to $\widehat{\bm u}$. 
These are the surfaces of simultaneity of an observer whose 4-velocity is  
$\widehat{\bm u}$.  The Lorentz group is generated by the rotations of these 
hypersurfaces and by boosts.\index{hypersurface}  \\  
In an active description, a boost\index{boost}\index{active transformation!boost}\index{Lorentz transformation!boost} $\Lambda$ maps a point $P$ of Minkowski 
space to the point $\wt P = \Lambda(P)$. If $P$ has 
coordinates $(t, x, y, z)$, the coordinates of $\wt P$, for a boost in the $t$-$x$ 
plane are are given by 
\begin{align*}
   \wt t &= \frac{t+vx}{\sqrt{1-v^2}} = \gamma (t+vx) \\
   \wt x &= \frac{x+v\rm t}{\sqrt{1-v^2}} = \gamma(x+vt), 
   \quad \rm \wt y=y, \ \wt z=z
\end{align*}
or 
\be
\Lambda = \begin{pmatrix} 
		\gamma& \g v & 0 &0\\
		 \g v&  \g & 0 & 0\\
		 0	  & 0		&1 &0\\
		 0	  & 0		&0 &1
			   \end{pmatrix}.
\label{e:boost}\ee  

A boost maps a 4-velocity $\widehat{\bm u}$ with components $\widehat u^\mu = (1,0,0,0)$ to a 4-velocity $\wt u$ with 
components $\wt u^\mu = \Lambda^\mu{}_\nu u^\nu$, or $(\wt u^\mu) = (\gamma, \gamma v,0,0)$.    
The surface of simultaneity of Alice with 4-velocity $\widehat{\bm u}$ is mapped 
to the surface of simultaneity of Bob with 4-velocity $\wt u$.  In particular, the unit 
vector with components $(\widehat x^\mu) = (0,1,0,0)$ is mapped to 
\[
  (\wt x^\mu) = (\Lambda^\mu{}_\nu \widehat x^\nu) = (\gamma v, \gamma, 0, 0),  
\]
orthogonal to $\wt u^\mu$.  These unit vectors are, or course, the vectors tangent to 
the trajectories of Alice and Bob and to their surfaces of simultaneity in Fig.~\ref{sim}. 

The Poincar\'e group is the group of Lorentz transformations and translations, 
\be
   P \mapsto \Lambda(P) + a,
\ee
the group of length and angle preserving maps, isometries of the Minkowski metric.

We can write a boost in a form resembling a rotation, by setting 
$\cosh\vartheta = \frac1{\sqrt{1-v^2}},\ \sinh\vartheta = \frac v{\sqrt{1-v^2}}$:  
\be
\begin{pmatrix}\wt t\\ \wt x\end{pmatrix} 
=  \Lambda \begin{pmatrix} t\\ x\end{pmatrix}, \mbox{ with }  
\Lambda = \begin{pmatrix} 
		\cosh\vartheta & \sinh\vartheta & 0 &0\\
		 \sinh\vartheta&  \cosh\vartheta & 0 & 0\\
		 0	  & 0		&1 &0\\
		 0	  & 0		&0 &1
			   \end{pmatrix}.\ \ 
\label{e:txboost}\ee
A formal relation of the boost $\Lambda$ to a rotation ${\cal R}$ comes from the replacement $t\rightarrow \tau = it$ that takes the Minkowski metric to 
the Euclidean metric $d\tau^2 + dx^2+dy^2 + dz^2$, analytically continued to real $\tau$. The boost of 
the $t$-$x$ plane becomes a rotation by $\theta$ in the $\tau$-$x$ plane, 
\be
\begin{pmatrix}\wt \tau\\ \wt x\end{pmatrix} 
=  {\cal R} \begin{pmatrix} \tau\\ x\end{pmatrix}, \mbox{ with }  
{\cal R} = \begin{pmatrix} 
		\cos\theta& -\sin\theta & 0 &0\\
		 \sin\theta&  \cos\theta & 0 & 0\\
		 0	  & 0		&1 &0\\
		 0	  & 0		&0 &1
			   \end{pmatrix}
\label{e:txboostr}\ee
with $\vartheta = i\theta$. The next exercise, with solution, is the derivation.  \\

\noindent{\sl Exercise}. Show that analytically continuing $\tau = it$ to 
real $\tau$ takes the boost \eqref{e:boost} to the rotation 
\eqref{e:txboostr}, with $\dis \theta = \frac1i\vartheta$ analytically 
continued to real $\theta$.    \\

\noindent{\sl Solution}. Start with the identities 
$\dis \cosh\vartheta = \cos\frac\vartheta i, \quad
\sinh\vartheta = i\sin\frac\vartheta i $,\ and with the definitions $\dis\theta:=\frac\vartheta i$, $\tau=it$, $\wt \tau = i\wt t$.    
From the boost \eqref{e:boost} taking $(t,x)$ to $(\wt t, \wt x)$, we have
\begin{align*}
\wt \tau = i\wt t& = i(t\cosh\vartheta + x\sinh\vartheta) 
		  = \tau\cos\frac\vartheta i - x\sin\frac\vartheta i
		  =\tau\cos\theta-x\sin\theta\\
\wt x &= x\cosh\vartheta + t\sinh\vartheta 
	= x\cos\frac\vartheta i + t\, i\sin\frac\vartheta i
	= x\cos\theta + \tau\sin\theta.\qquad\Box
\end{align*}
 \vspace{2mm}

You are more likely to have encountered the passive version of a boost\index{passive transformation!boost} in which the points of spacetime are fixed and the coordinates changed:  Again Bob moves with speed $v$ relative to Alice.  Alice and Bob assign to each point $P$ 
spacetime coordinates $t(P),x(P),y(P),z(P)$ and $t'(P),x'(P),y'(P),z'(P)$, related by
\begin{align*}
   t' &= \frac{t-vx}{\sqrt{1-v^2}} = \gamma (t-vx) \\
   x' &= \frac{x-v\rm t}{\sqrt{1-v^2}} = \gamma(x-vt)\\
   y' &=y \\
   z' &=z.
\end{align*}
Alice's and Bob's surfaces of simultaneity are of course the $t=$ constant and 
$t'=$ constant surfaces.  The origin is chosen with Alice's trajectory given by 
the line $x=y=z=0$, and Bob's trajectory is $x'=y'=z'=0$.      
 
\setcounter{section}{2}
\section{Tensors and Vector Spaces}
\label{s:tensors}
\subsection{Mathematical preliminaries}

We start with some notation: \\
$\mathbb R$ is the real line, $\mathbb C$ the complex plane. \\ 
${\mathbb R}^n$ is the set of $n$-tuples (ordered sets of $n$ real numbers).  \\
Then $\bm x$ = ($x^1, \dots , x^{n}\ $)  is in ${\mathbb R}^n$ if $x^1,\dots, x^n$ 
are in $\mathbb R$. We'll use $(x^0, \dots, x^3)$ for spacetime. 

Summation is as usual implied over a repeated index. 

Our conventions will be such 
that repeated indices will occur in pairs with one upper and one lower index.

\noindent A {\em group} is a set $G$ together with a binary
operation $\ \cdot{}\ $ , satisfying 
\ben
\item[(1)]	$a, b$ in $G$ $\Rightarrow $ $a$ $\cdot {}$ $b$ in $G$ 

\item[(2)]	$a, b, c$ in $G$ 
$\Rightarrow a \cdot (b\cdot c) = (a\cdot b) \cdot c$.

\item[(3)]	There is an identity element $e$ in $G$ such that 
$a\cdot e = e \cdot  a = a$, all $a$ in $G$. 

\item[(4)]	For every $a$ in $G$ there is an element $a^{-1}$ in $G$ with 
$a\cdot a^{-1} = a^{-1}\cdot a = e$. 
\een
An {\em abelian} group is a group in which the product $\cdot $ is commutative:
 $a\cdot b = b\cdot a$, all $a$, $b$ in $G$. \\

\noindent {\sl Example 1}: The integers are an abelian group under
addition (the ``product" is + and the identity is $0$). 

\noindent {\sl Example 2}: The reals are an abelian group under addition. 

\noindent {\sl Example 3}: The complex numbers are an abelian group under 
addition. 

\noindent {\sl Example 4}: ${\mathbb R}^n$ is an abelian group under addition
defined by 
\[
  (x^1,\dots,x^n) + (y^1,\dots,y^n) = (x^1+y^1,\dots, x^n + y^n).
\]
\noindent {\sl Example 5}:	The group GL${_4}$ is the group of all
linear maps $a: {\mathbb R}^4 \rightarrow  {\mathbb R}^4$ for which 
$a({\bm x}) \neq  0$ unless ${\bm x} = 0$. The element $a$ can be represented as an 
$n\times n$ matrix by writing $x \rightarrow  x'$ with $x'{}^i = a^i {}_j x^j $.  
Then the condition that $a({\bm x}) \neq 0$ for  $\mathbf x \neq  0$ is equivalent
to requiring that the matrix $a^i{}_j $ be nonsingular (that $\det
\|a^i{}_j \| \neq 0)$, and is also equivalent to the existence of an
inverse map $a^{-1}$ in $GL_4$.\\ 

\noindent {\sl Example 6}: The smooth invertible maps of $\mathbb R^4$ onto
$\mathbb R^4$ form a group with composition the product.  This is the group of
diffeomorphisms of $\mathbb R^4$. (Invertible here means having a smooth inverse.) 
\index{diffeomorphism}

\benr
\item Prove that $GL_n$, defined as the set of
all $n\times n$ matrices with nonzero determinant is a group.  In particular,
show that each element $\| a^i_j\|$ has an inverse.\\ 
\een

\subsection{Vector Space}

Writing laws of Newtonian physics in terms of vectors and tensors makes them manifestly invariant under rotations.  And numbers obtained by measurements can 
be written as dot products of tensors and vectors and are thus manifestly independent of coordinates or a choice of basis.  In relativity, the laws of 
physics are again written in terms of vectors and tensors, now with spacetime 
indices, and are thereby manifestly Lorentz invariant.  Vectors $\bm v$ and 
$\bm w$ can be added, with $\bm v+\bm w= \bm w+\bm v$, and you can multiply 
them by real numbers $r$, so one defines a vector space as an abelian group under addition, 
together with an operation of scalar multiplication:   

\index{vector space|textbf}
\noindent{\bf Definition}: A set $V$ is a real vector space if $V$ is an abelian
group under an operation (addition) that we denote by + and if there 
is a second operation (scalar multiplication) that associates with every
$r$ in $\mathbb R$ and ${\bm v}$ in $V$ an element $r{\bm v}$ 
in $V$ and that satisfies

(1)  $r(\bm u +{\bm v}) = r\bm u  + r{\bm v}$ \hspace{1in} (2)
$(r+s){\bm v} = r{\bm v} + s{\bm v}$\\

(3)	$r(s{\bm v}) = (rs){\bm v} $ \hspace{1.4in} 
(4)	$1{\bm v} = {\bm v}$,\hspace{.75in} 
all $r, s$ in $\mathbb R$, $\bm u ,{\bm v}$ in $V$\\	

\noindent A complex vector space is defined in exactly the same way, except
that $r$ and $s$ are in $\mathbb C$ instead of $\mathbb R$.\\	

Because $V$ is an abelian group, there is a zero vector $\bm 0$ for which 
$\bm v + \bm 0 = \bm 0 + \bm v = \bm v$, all $\bm v$ in $V$. \\

\noindent {\sl Example 1}: ${\mathbb R}^n$ is a vector space with scalar
multiplication defined by \\
\[ r(x^0, \dots, x^{n-1}) = (rx^0, \dots, rx^{n-1}) \]

\benr\item
Show that $0 \cdot \bm v = \bm 0$, all $\bm v$ in $V$. 

\item Show that ${\mathbb R}^n$ is a vector space with $\bm 0 = (0,\ldots,0)$, checking 
that it is an abelian group under addition and that conditions (1)-(4) are satisfied
\een

\noindent {\sl Example 2}:	The set of square integrable functions 
($\int  f^2 dx < \infty $) on the real line is a vector space:\\
\phantom{xxxxxxxxxx} Addition is addition and scalar multiplication is scalar multiplication.\\

\noindent {\sl Example 3}:	A Hilbert space is a complete complex vector 
space together with an inner product.\\

\noindent {\bf Definition}:	$L: V\rightarrow W$ is a {\em linear} map
between two vector spaces $V$ and $W$ if 
\[ 
L(r\bm u  + s{\bm v}) = rL(\bm u ) + sL({\bm v}),\ 
	\mbox{ all } r, s\in {\mathbb R}, \bm u , {\bm v} \in V.  
\] 

\noindent Two vector spaces $V$ and $W$ are \textit{isomorphic} if there is
a one to one and onto linear map $L: V \rightarrow W$.

\noindent {\sl Example}:  The group of isomorphisms from ${\mathbb R^4}$ to
itself is $GL_4$.\\

\par	A vector space $V$ is finite dimensional if there is a finite set of
vectors \{${\bm e}_0,\dots,{\bm e}_{n-1}$\} that span $V$, i.e., such that
every vector ${\bm v}$ in $V$ can be written in the form

\[ {\bm v} = v^i{\bm e}_i, \hspace{.5in} v^i \in \mathbb R . \]

\noindent The smallest number $n$ of vectors that span $V$ is called the
dimension of $V$ and a set of $n$ vectors that span $V$ is a \textit{basis}
\index{basis|textbf}\index{vector space!basis, frame}
or frame\index{frame|textbf} of $V$.  The numbers $v^i$ are the \textit{components} of ${\bm v}$
in the frame \{${\bm e}_0$,\dots,${\bm e}_{n-1}$\}.
\index{component!of a vector|textbf}  Note that the $n$ basis
vectors are linearly independent. That is, no one of them -- say ${\bm e}_0$
-- is a linear combination of the rest.  If it were, if $\bm e_0$, could be written 
in terms of the others,  
\[ 
{\bm e}_0 = r^1{\bm e}_1+\cdots +r^{n-1}{\bm e}_{n-1}, 
\]
then {\em any} vector ${\bm v}$ would be a linear combination of the remaining basis vectors, 
${\bm e}_1$, \dots, ${\bm e}_{n-1}$,\\
${\bm v} = v^0{\bm e}_0 + v^1{\bm e}_1+\cdots + v^{n-1}{\bm e}_{n-1} =
(v^1+r^1v^0){\bm e}_1 + \cdots + (v^{n-1} + r^{n-1}v^0) {\bm e}_{n-1}$. \\
The $n-1$ vectors ${\bm e}_1,\dots , {\bm e}_{n-1}$ would therefore span $V$,
contradicting the assumption that the smallest set spanning $V$ has $n$
vectors.

\noindent {\sl Example}:  ${\mathbb R}^n$ has the natural basis

\begin{align*} &{\ \ \bf E}_0 \,= (1, 0, \dots, 0)\\
	&\phantom{E_0\ \ \ \ \, .}\vdots \\
	&{\bf E}_{n-1} = (0, \dots, 0, 1),
\end{align*}
and the $i^{th}$ component of ${\bm x} = (x^0, \dots, x^{n-1})$ is the 
number $x^{i}$.\\  

	Every $n$-dimensional vector space is, in fact, isomorphic to ${\mathbb R}^n$, and
an isomorphism $L$ is easy to find:  Let $\{{\bm e}_0, \dots,
{\bm e}_{n-1}\}$ be a basis of $V$.  Then any ${\bm v}$ in $V$ is ${\bm v}
= v^i {\bm e}_i$ and the isomorphism is simply
\be
L({\bm v}) = (v^0, \dots, v^{n-1}) = v^i {\bf E}_i. 
\label{L}\ee
\benr
\item Show that $L$ \textit{is} an
isomorphism, that it's linear, onto, and one to one.  Show first 
that a linear map $L$ is one-to-one if and only if
\[L({\bm v}) = 0 \Longrightarrow {\bm v} = 0. \]
\een

 To have an isomorphism $V \rightarrow {\mathbb R}^n$ means
that if you give me a vector in $V$, I'll give you $n$ real numbers.  And a
basis of $V$ provides just that:  For each vector ${\bm v}$, it gives you
the $n$ numbers ($v^0, \dots, v^{n-1}$), the components of ${\bm v}$ along
the basis.\\

\noindent {\em Change of basis}:
\index{basis!change of basis} 

We'll use MTW notation, in which one writes the components of a 
vector $\bm v$ in two different bases as $v^i$ and $v^{i'}$, with 
the prime associated with the index.  This reflects the fact that 
the vector itself does not change under a change of basis. 

If  $({\bm e}_0, \dots, {\bm e}_{n-1})$ and $({\bm e}_{0'}, \dots,
{\bm e}_{n-1'})$ are any two bases of $V$, then each vector ${\bm e}_i$
is a linear combination of the primed basis vectors ${\bm e}_{i'}$:
\be {\bm e}_i = {\bm e}_{j'}\,a^{j'}{}_i. \ee
 
Analogously,
\be {\bm e}_{i'} = {\bm e}_j\,a^j{}_{i'} . \ee
Because $\|a^{j}{}_{i'} \|$ takes the unprimed basis to the primed basis, and $ \|a^{j'}{}_{i} \|$ 
takes the primed basis back to the unprimed basis, they must be inverses of one another. \\
{\sl Claim}:  The matrices $\|a^{j'}{}_{i} \|$ and $\|a^j{}_{i'} \|$ are
inverses:
\be a^i{}_{k'}\, a^{k'}{}_j  = \delta^i_j \quad \mbox{ and } \quad 
a^{i'}{}_k\,  a^k{}_{j'} = \delta_{j'}^{i'}. \ee
{\sl Proof}:  ${\bm e}_i a^i{}_{k'} a^{k'}{}_j  
	= {\bm e}_{k'} a^{k'}{}_j  = {\bm e}_j = \delta_j^i {\bm e}_i$.  
Equating coefficients of each ${\bm e}_i$ gives 
$  a^i{}_{k'}\, a^{k'}{}_j  = \delta_i^j$.\\  
To show the second equation, exchange primed and unprimed indices in the proof.

If a vector ${\bm v}$ has components $v^i$ with respect to
${{\bm e}_i}$, its components with respect to ${{\bm e}_i'}$ are
\be 
	v^{i'} = a^{i'}{}_j \, v^j~~.\label{vi'} 
\ee
Similarly,
\be 
	v^i = a^i{}_{j'}\,v^{j'}~~. 
\label{vi}\ee
Proof: We have ${\bm v} = v^i {\bm e}_i = v^{i'} {\bm e}_{i'}$ and
${\bm e}_{i'} = {\bm e}_j a^j{}_{i'}$.  
Then 
${\bm e}_i v^i = {\bm e}_j a^j{}_{i'}v^{i'} =   {\bm e}_i a^i{}_{j'}v^{j'}$.\\
Equating coefficients of ${\bm e}_i$ gives (\ref{vi}).\\

\benr
\item  In the natural basis ${\bm e}_1 = (1, 0,
0)$, ${\bm e}_2 = (0, 1, 0)$, ${\bm e}_3 = (0, 0, 1)$ for ${\mathbb R}^3$, a vector
has components $v^1 = 1$ \hspace{.6in} $v^2 = 2$ \hspace{.6in}    $v^3 = 3$.

\noindent (a)	What are the components of ${\bm v}$ relative to the bases

	$\qquad {\bm e}_{1'}  = (1, 1, 0),$ \hspace{.3in}  
${\bm e}_{2'}  = (0, 2, 0),$
\hspace{.3in} ${\bm e}_{3'}  = (1, 1, 1);$ \\
and 

	$\qquad {\bm e}_{1''} = (2, 0, 0),$ \hspace{.3in}  ${\bm e}_{2''} = (0, 2, 0),$
\hspace{.3in} ${\bm e}_{3''}  = (0, 0, 2)?$

\noindent (b)	Find the matrices $a^{i'}{}_j,\  a^i{}_{j'},\  a^i{}_{j''},
	\  a^{i''}{}_j,\  a^{i''}{}_{j'}$. 
\een

\subsection{Dual Space}\index{dual vector space}\index{vector space!dual space}
\label{sec:dual} 

In quantum mechanics, the wave function $\psi$ of a single particle, say,
can be regarded as a ``ket'' vector $|\psi\rangle$ in the Hilbert space $H$ of
square integrable functions on space (${\mathbb R}^3$). It can also be regarded as a
``bra'' vector $\langle\psi|$, a linear map from $H$ to $\mathbb C$ by
\[
\langle\psi |:\ \  | \widetilde{\psi}\rangle \ 
	\rightarrow \ \langle\psi | \widetilde{\psi}\rangle 
	= \int\psi^*\widetilde\psi d^3x , 
\]
\noindent where $\widetilde{\psi}$ is any other square integrable function
($|\wt \psi\rangle$ a vector in $H$).

The vector ${\bm x} = (x^0, \dots, x^3)$ in $\mathbb R^4$ can similarly be regarded
as a linear map from $\mathbb R^4$ to $\mathbb R$ by 
$\dis {\bm x}\!:\  {\bf y} \mapsto -x^0y^0 + x^1y^1 + x^2y^2 + x^3y^3$, 
where ${\bf y}$ is any vector in $\mathbb R^4$.  If
\[x_\mu = \eta_{\mu\nu}x^\nu, \]
we can write ${\bm x}\!:\  {\bf y} \longrightarrow x_\mu y^\mu$.

  The set of all linear maps from $V$ to $\mathbb R$ also is called the dual space, $V^*$.  
It is a vector space (Exercise below), and, if $V$ is finite-dimensional 
of dimension $n$, then $V^*$ also has 
dimension $n$.  If 
${\bm\sigma}$ and ${\bm\tau}$ are two linear maps from $V$ to $\mathbb R$, 
their sum is 
\[
({\bm\sigma} + {\bm\tau})({\bm v}) 
	= {\bm\sigma} ({\bm v}) + {\bm\tau} ({\bm v}) 
\]
\nopagebreak
\hspace{2.5in}  real number = real number + real number

Scalar multiplication is given by 
\[ 
	(r{\bm\sigma})({\bm v}) = r({\bm\sigma}({\bm v})) 
\]
\hspace{2.5in}real number = real number $r$ $ \times\ $real number $\bm\sigma(\bf v)$\\

The vectors ${\bm\sigma}$ in $V^*$ are called dual vectors or 1-forms or covectors.\\
\index{vector!dual vector, covector}\index{dual vector}\index{covector}

\benr \item Show that  $V^*$ is a vector space. 
\een

To see that $V^*$ is $n$ dimensional when $V$ is $n$ dimensional, 
we introduce a basis $\{{\bm e}_i\}$
for $V$.  Let $\bm\sigma $ be any linear map, and ${\bm v} = v^i
{\bm e}_i$ a vector in $V$.  Because $\bm\sigma$ is linear, 
 ${\bm\sigma}({\bm v}) = v^i{\bm\sigma}({\bm e}_i)$.  
So if we define the $n$ real numbers $\sigma_i $ by
\be  \sigma_i = {\bm\sigma}({\bm e}_i), \ee
we have ${\bm\sigma}({\bm v})  = \sigma_i v^i $.  Now comes the key
part:  We are going to define a basis for $V^*$ in which $\sigma_i$ will
be the $i^{th}$ component of ${\bm\sigma}$.  If we let $\bm\omega^i$ be the 
linear map
\[\bm\omega^i({\bm v}) = v^i \]
Then ${\bm\sigma}$ is a linear combination of the $\bm\omega^i$:
\[{\bm\sigma} = \sigma_i \bm\omega^i .\]
Check:  ${\bm\sigma}({\bm v}) = \sigma_i v^i  = \sigma_i \bm\omega^i({\bm v})$.
\index{basis!dual basis}

\index{component!of a dual vector|textbf}
\noindent Thus $\{\bm\omega^i\}$ is a basis for $V^*$,  and the numbers 
$\sigma_1, \dots, \sigma_n$ are the components of ${\bm\sigma}$ in the 
basis $\{{\bm \omega}^i \}$.\footnote{MTW and Schutz use this notation for the 
dual basis, following Cartan.  Wald writes $\bm e^i$ instead of $\bm\omega^i$.}
In particular, $V^*$ is $n$-dimensional.  A basis $\{\bm\omega^i\}$ for 
$V^*$ is called the basis {\em dual to} $\{{\bm e}_i\}$ if 
\[ \bm\omega^i ({\bm e}_j ) = \delta^i_j, \]
and the basis we have just introduced satisfies this relation.   
\\

\noindent{\em Change of basis}\index{basis!change of basis} \\

\noindent If ${\bm e}_{i'} = {\bm e}_j\, a^j_{\ i'} $, and ${\bm\sigma}$
is a dual vector, then
\be 
   \sigma_{i'} = a^j_{i'}  \sigma_j . 
\ee
\noindent {\sl Proof}.  $\sigma_{i'} = \sigma ({\bm e}_{i'}) 
= {\bm\sigma}({\bm e}_j a^j_{\ i'}) = a^j{}_{\ i'}{\bm\sigma}({\bm e}_j ) 
= a^j_{\ i'} \sigma_j $. \\

	The basis dual to $\{{\bm e}_{i'}\}$ is \{$\bm\omega^{i'}\}$, with 
\[ \bm\omega^{i'} = a^{i'}_{\ j}\, \bm\omega^j .\]
Check:
\be 
\bm\omega^{i'} ({\bm v}) 	= a^{i'}_{\ j}\, \bm\omega^j ({\bm v}) 
			= a^{i'}_{\ j}\, v^j = v^{i'} .
\ee
	Finally, note that $\sigma({\bm v})$ is a number, and it never heard of a
basis, so $\sigma_i v^i$ must be independent of the choice of basis.
 Check:\\
\[ \sigma_{i'} v^{i'} = (a^j_{\ i'} \sigma_j )(a^{i'}_{\ k} v^k) 
			= (a^j_{\ i'} a^{i'}_{\ k} )\sigma_j v^k 
			= \delta^j{}_k \sigma_j v^k = \sigma_j v^j .\]
\noindent
{\em Abstract index notation} (invented by Roger Penrose):
\index{abstract index|textbf}\index{index!abstract index|textbf}

Instead of using arrows or boldface, we will often write a vector in the manner 
$v^a$, and a
dual vector will be written $\sigma_a$ .  The symbol $v^a$ will $\textit{not}$ mean the
$a^{th}$ component of the vector -- it will be the same as writing
${\bm v}$.  
\be
   v^a\equiv \bm v\ .
\ee
Then $\sigma_a v^a$ and $v^a\sigma_a$ both mean
$\sigma({\bm v})$.  There is no reason to 
introduce a basis every time we want to talk about a vector, but the boldface 
(or arrow) notation is clumsy for tensor algebra. \\
{\em Index conventions}: 
\label{notation!index}
Latin indices will be used for any number of 
dimensions, Greek for spacetime only.  
Latin and Greek indices at the beginning of the alphabet, $a,b,c,\dots$ and 
$\alpha,\beta, \gamma, \dots$ will be abstract ($a$ for ``abstract''); 
Latin and Greek indices from $i$ ($\iota$) onward, 
$i,j,k,l,m,n,\dots$ and $\iota,\kappa,\lambda, \mu,\nu, \dots$, will  
be concrete.\index{index!concrete index|textbf} \index{concrete index|textbf} \\ 

\subsection{Tensors}
\label{ss:tensors}\index{tensor|(}

The definition of a connecting vector, as a directed line joining two points, does 
not need a coordinate system or a choice of basis. And the definition  
specifying magnitude and direction again does not involve a basis.   
Tensors similarly arise in physics in a way that 
does not involve the introduction of a basis or the components of the tensor.    
In particular, the tensor you are 
likely to have encountered first, the moment of inertial tensor\index{moment of inertia tensor} $I_{ab}$, relates 
the angular velocity vector $\Omega^a$\index{angular velocity} of a rigid body 
to the body's rotational kinetic energy and to its angular momentum.\index{angular momentum}  
\[ E = \frac12 I_{ab} \Omega^a \Omega^b, \qquad
   L_a = I_{ab}\Omega^b.
\] 
That is, $I_{ab}$ can be regarded as the map $\Omega^b \rightarrow L_a$ from angular 
velocity to angular momentum, or as the map  $(\Omega^a, \Omega^b) \rightarrow 2E$.
The map  $\Omega^b \rightarrow L_a$ is linear; and the map  $(\Omega^a, \Omega^b) \rightarrow 2E$ 
is bilinear (separately linear in each argument).  Wald uses abstract indices, while 
MTW use only concrete indices and write $\bm I(\bm u,\bm v)$ instead of $I_{ab}u^a u^b$.   

We already introduced a dual vector as a linear map from vectors $v^a$ to numbers;
$\sigma_a:  v^a \longrightarrow \sigma_av^a$. A vector can similarly be
regarded as a linear map from dual vectors $\sigma_a$ to numbers;\\
\centerline{$v^a: \sigma_a \longrightarrow \sigma_a v^a$.}

\cblue We define a covariant tensor $T_{ab}$ as a map from pairs
of vectors to numbers $(u^a , v^b) \longrightarrow T_{ab}u^av^b$, 
with $T_{ab}$ linear in $u^a$ and $v^b$ separately
\index{tensor|textbf}\cb:
\begin{eqnarray}
T_{ab}(ru^a+sw^a)v^b &=& rT_{ab}u^av^b + sT_{ab}w^av^b \\
T_{ab}u^a(rv^b+sw^b) &=& rT_{ab}u^av^b + sT_{ab}u^aw^b
\end{eqnarray}
The components of $T_{ab}$ in a basis $(e^a_0, e^a_1, e^a_2, e^a_3)$ are,
for example,
\index{component!of a tensor|textbf}
\[T_{12} = T_{ab}e_1^ae^b_2 \]
\benr 
\item a) Show that in a different basis ${\bm e}_{i'} = {\bm e}_i a^i_{\, i'} $
	\be T_{i'j'} = a^i_{\, i'} a^j_{\, j'}T_{ij} \ .
	\label{transf}\ee
\een
$T_{ab}$ is called a second rank covariant tensor. A (third rank, mixed)
tensor $S^{ab}{}_c$ is analogously a map that takes two dual vectors and
one vector to a number,  $(\sigma_a, \tau_b, v^c) \rightarrow
S^{ab}{}_c\sigma_a\tau_bv^c$, with the map linear in each argument.

It is worth reiterating that each component of a tensor is a number. Each 
component of a tensor field is then a scalar field. 

\noindent
{\sl Exercise 8 b)}. Prove that the components of
$S^{ab}{}_c $ transform under a change of basis in the manner 
\be 
	S^{i'j'}{}_{k'} = a^{\, i'}_{\, i}a^{\, j'}_{\, j} a^k_{\, k'}S^{ij}{}_k . 
\ee

\noindent \crv In general then, a tensor $T^{a\cdots b}{}_{c\cdots d}$ 
with $r$ up indices and $s$ down indices  
is a multilinear map from $s$ vectors and $r$ dual vectors to $\mathbb R$\cb.
 Its components transform under a change of basis in the manner
\be 
 T^{i'\cdots j'}{}_{k'\cdots l'} 
 = a^{i'}_{\, i}\cdots a^{j'}_{\,j} a^k_{\,k'}\cdots a^l_{\,l'} 
   T^{i\cdots j}{}_{k\cdots l}. 
\label{transf2}\ee
\noindent The tensors with $r$ up indices and $s$ down indices 
form a vector space:  Each linear combination of tensors of the same type is again a 
linear map:  \\
\hspace{.5in} $pS^{a\cdots b}{}_{c\cdots d} + qT^{a\cdots b}{}_{c\cdots d}\ \ $   maps
$\ \ (\sigma_a, \dots, \tau_b, u^c, \dots, v^d)\ \ $ to\\
\\
$(pS^{a\cdots b}{}_{c\cdots d} +
qT^{a\cdots b}{}_{c\cdots d})\sigma_a\dots\tau_bu^c\cdots .v^d =
pS^{a\cdots b}{}_{c\cdots d}\sigma_a \cdots \tau_b u^c\cdots v^d + qT^{a\cdots b}{}_{c\cdots d}\sigma_a \cdots \tau_bu^c\cdots v^d$.  \\

In the older mathematics texts, tensors are often defined by the
transformation law, Eq. (\ref{transf2})  
-- and, in physics texts, by the equivalent law with the $a^{j'}{}_i~'s$ restricted to Lorentz
transformations.   The transformation law (\ref{transf}) is equivalent
to saying\\
{\sl the number $T_{i j}u^i v^j $ depends only on the vectors $u^a$ and $v^b$; 
it is independent of the choice of basis}.\\
\noindent But this statement means that $(u^a,v^b)\rightarrow T_{i
j}u^i v^j $ is a well defined map from vectors to numbers; in
other words, giving a tensor in the sense of a set of numbers $T_{i
j}$ that transform by (\ref{transf}) is the same as giving a bilinear
map $T_{ab}$ from vectors to real numbers.  The components of $T_{ab}$ are
the numbers $T_{i j}$.\\

\noindent {\em Outer product}

If $S_{ab}$ and $T^{ab}{}_c$ are tensors, so is $S_{ab}T^{cd}{}_e$ defined
by $S_{ab}T^{cd}{}_e u^av^b\sigma_c\tau_dw^e =
(S_{ab}u^av^b)(T^{cd}{}_e\sigma_c \tau_dw^e)$.   $S_{ab}T^{cd}{}_e$ is the
outer product of $S_{ab}$ and $T^{ab}{}_c$, and its components in a basis
are, of course, $S_{i j}T^{k l}{}_m$. 

Any tensor $T^{ab}$, say, can be built up out of vectors (and dual vectors) by
taking outer products.  That is, if $(\widehat t^\alpha, \widehat x^\alpha, \widehat y^\alpha, \widehat z^\alpha)$ 
is a basis for
$V$ (think of $\widehat t^\alpha$ as the unit vector along a $t$-axis in Minkowski
space), any vector can be written in the form
\[ A^\alpha = A^t\widehat t^\alpha + A^x\widehat x^\alpha + A^y\widehat y^\alpha + A^z\widehat z^\alpha \] 
(where $A^t, A^x, A^y,$ and $A^z$ are just components of $A^\alpha$ in the basis
$(\widehat t^\alpha, \widehat x^\alpha, \widehat y^\alpha, \widehat z^\alpha)$, and any tensor $T^{\alpha\beta}$ can be written
\be T^{\alpha\beta} = T^{tt}\widehat t^\alpha \widehat t^\beta + T^{tx}\widehat t^\alpha \widehat x^\beta 
+ T^{xt}\widehat x^\alpha \widehat t^\beta + T^{ty}\widehat t^\alpha \widehat y^\beta +
\cdots \ee 
For example, the Minkowski metric is
\be \eta^{\alpha\beta} = -\widehat t^\alpha \widehat t^\beta + \widehat x^\alpha \widehat x^\beta 
		+ \widehat y^\alpha \widehat y^\beta + \widehat z^\alpha\widehat z^\beta . \ee
 
   The index-free notation for the outer product of tensors $\bm S$ and $\bm T$ is $\bm S\otimes \bm T$.
\vspace{2mm}
   
\noindent {\em Contraction}

Given the tensor $T^a{}_b{}^{cd}{}_e = u^a\sigma_bv^c w^d\tau_e$, we can
obtain a new tensor of lower rank by contracting on one lower and one upper
index.  That is, a tensor $\wt T_b^{cd} \equiv T^a{}_b{}^{cd}{}_a$, with 
one down and two up indices is defined by 
\be T^a{}_b{}^{cd}{}_a = u^a\sigma_b v^c w^d\tau_a \equiv
\sigma_b v^c w^d \!\!\underbrace{(u^a\tau_a)}_{\mbox{a number}} \ee

Since any tensor can be written as a sum of outer products of vectors, \\
\centerline{
$A^{ab}{}_{cde} = u^a v^b\sigma_c \tau_d\omega_e + w^ax^b\zeta_c\eta
_d\xi_e + \cdots,$}
 one can define contraction between any up and any down
index by 
\be A^{ae}{}_{cde} = u^av^e\sigma_c \tau_d\omega_e + w^ax^e\zeta_c\eta
_d\xi_e + \cdots \quad .
\label{contract1}\ee
The meaning of contraction is easy to understand in terms of components:
\be 
	A^{i m}{}_{k l m} \equiv \sum^{n}_{m=1} A^{i m}{}_{k l m}\quad . 
\label{contract2}\ee

\benr \item Show the definition (\ref{contract1}) implies 
(\ref{contract2}) by looking first at tensors of the form $u^a\sigma_b$, 
then of the form $u^a{}\cdots v^b\sigma_c{}\cdots \tau_d$, and finally of 
the general form:  linear combinations of these.
\een

\noindent
{\em Index Symmetries}
\index{index!symmetries|textbf}\index{symmetry!index symmetries|textbf}

A tensor $T^{abc}{}_d$ is symmetric in the two upper indices $a$ and $c$ if
$T^{abc}{}_d = T^{cba}{}_d$.  \\
This means that the map, $T^{abc}{}_d$ is symmetric under
interchange of its first and third arguments:
\[ 
	T^{{\cblue a}b{\cblue c}}{}_d{\cblue\sigma_a}\tau_b{\cblue\omega_c}v^d 
	= T^{{\cblue c}b{\cblue a}}{}_d{\cblue\sigma_a}\tau_b{\cblue\omega_c}v^d 
	= T^{{\cblue a}b{\cblue c}}{}_d{\cblue\omega_a}\tau_b{\cblue\sigma_c}v^d.
\]
(The second equality here is a change in the names of the dummy indices.)

\noindent A tensor $S_{a\ cde}^{\ \,b}$ is similarly said to be symmetric in
a pair of lower indices -- in $a$ and $c$, say -- if \\
\centerline{$S_{a\ cde}^{\ \,b} = S_{c\ ade}^{\ \,b}$}.   \\
A tensor $T^{abc}$ is antisymmetric under interchange of $a$ and
$b$ if $T^{abc} = -T^{bac}$.  \\
$T^{a\cdots b}$ is totally antisymmetric
(totally symmetric) if it is antisymmetric (symmetric) under interchange of 
any pair of indices.  If $T^{ab}$ is any tensor
\be T^{(ab)} := \frac{1}{2} (T^{ab} + T^{ba}) \ee
is a symmetric tensor.  Similarly for $r$ indices, $T^{a_1\cdots a_r}:$
\be T^{(a_1\cdots a_r)} := \frac{1}{r!} 
\sum_{\mbox{permutations $\pi$ of $r$ integers}}
T^{a_{\pi(1)}\cdots\, a_{\pi(r)} }\ee

\noindent (For example, if $r = 3$, one permutation $\pi$ is 
$\pi(1) = 2,\ \pi(2) = 3,\ \pi(3) = 1$; 
another is \\  $\pi(1) =3,\ \pi(2) = 2,\ \pi(3) = 1$. ) \\
Thus
\be T^{(abc)} = \frac{1}{3!}(T^{abc} + T^{bca} + T^{cab} + T^{acb} +
T^{bac} + T^{cba}) \ee
Analogously, given any $T^{a{}_1\cdots a_r}$, one constructs the totally
antisymmetric tensor
\be 
  T^{[a_1\cdots a_r]} = \frac{1}{r!} \sum_\pi{(\text{sign}~\pi)}T^{a_{\pi(1)}\cdots a_{\pi(r)}}~, 
\label{antisymm1}\ee
where (sign $\pi$) is $+1$ when $\pi$ is an even permutation, $-1$ when $\pi$ is
odd. Eq. (\ref{antisymm1}) is commonly written
\be T^{[a_1\cdots a_r]} = \frac{1}{r!} \sum_{\pi}
(-1)^\pi T^{a_{\pi(1)}\cdots a_{\pi(r)}}~. 
\label{antisymm2}\ee

For example,
\be T^{[ab]} = \frac{1}{2}(T^{ab} - T^{ba}),\ee
and
\be T^{[abc]} = \frac{1}{3!} (T^{abc} + T^{bca} + T^{cab} - T^{acb} -
T^{bac} - T^{cba}). \ee
Then $T^{a\cdots b}$ is totally symmetric if $T^{a\cdots b} =
T^{(a\cdots b)}$, totally antisymmetric if $T^{a\cdots b} = T^{[a\cdots b]}$.\\

\noindent{\sl Syntax}

 A repeated index is called a {\sl dummy index}; an index that is 
not repeated is a {\sl free index}.  In the expression 
$T^{abc}{}_{ac}$, $a$ and $c$ are dummy indices, and the index $b$ is free.  
Any letter can be used for a dummy index, but one cannot use the same 
letter for two different dummy indices in an expression: 
\[ 
v^{ac} T_{abc} = v^{ad} T_{abd}, \mbox{ but } v^{aa}T_{aba} \mbox{ is nonsense}.\]
In a tensor equation or in an expression involving a sum of tensors, the 
free indices must match: 
\[ 
   v^a = w^a \mbox{ and } v^a+w^a \mbox{ are equivalent to } \bm v = \bm w 
   \mbox{ and } \bm v+\bm w, 
\]
but you are not allowed to write
\[
 v^a = w^b \mbox{ or } v^a + w^b. 
\]
  
Finally, the expressions $\sigma_a \tau_b$ and $\tau_b \sigma_a$ denote the 
same tensor, $\sigma_a \tau_b \equiv \tau_b \sigma_a$, because, when acting on 
the arguments $\bm u,\bm v$, written as $u^a v^b$, they give the same number:
\[
  \sigma_a \tau_b\ u^a v^b = (\sigma_a u^a)(\tau_b u^b) = (\tau_b u^b)(\sigma_a u^a) = \tau_b \sigma_a\ u^a v^b. 
\]
Similarly, $\sigma_a \tau_b\ u^a v^b = \sigma_a\tau_b v^b u^a $.  More generally, 
\[
  S^{a\cdots b}{}_{c\cdots d} T^{e\cdots f}{}_{g\cdots h}\equiv 
  	 T^{e\cdots f}{}_{g\cdots h}S^{a\cdots b}{}_{c\cdots d} .  
\]  
 \index{tensor|)} 

\subsection{The tensors \texorpdfstring{$\bm{\eta_{\alpha\beta},\,\bm\delta^\alpha_\beta  }$}, 
and \texorpdfstring{$\bm{\epsilon_{\alpha\beta\gamma\delta} }$}.}
\label{s:eta_delta_ep}

{\bf Definition}. {\crv A Minkowski metric on a four dimensional vector space is a symmetric tensor
$\eta_{\alpha\beta} $ with {\em signature} \mbox{$-+++$}. }
\index{Minkowski metric} \\ 
Because $\|\eta_{\mu \nu} \|$ is a
symmetric matrix, there is a basis in which it is diagonal, and to say that
its signature is $-+++$ means that in such a basis it has one negative and
three positive diagonal elements (eigenvalues).  For example:
\[ \| \eta_{\mu \nu} \| = \left| \left| \begin{array}{cccc}
-\frac{1}{4} & & & \\
& 4 & & \\
& & 9 & \\
& & & 16 \end{array} \right| \right| \]
\noindent By redefining the basis ${\bm e}_{0'} = 2{\bm e}_0 ,
{\bm e}_{1'} = \frac{1}{2}{\bm e}_1, {\bm e}_{2'} =\frac{1}{3}{\bm e}_2, 
{\bm e}_{3'  } = \frac{1}{4} {\bm e}_3,\ \ $ we have\\
$\eta_{0'0'}  = \eta_{\alpha\beta} e_{0'}{}^\alpha\, e_{0'}{}^\beta
= \eta_{\alpha\beta} (2e_{0}{}^\alpha)(2e_{0}{}^\beta)
= 4\eta_{00}  = -1$ and $\eta_{1'1'}
= \eta_{2'2'} = \eta_{3'3'} = 1$. That is, 
\[ \| \eta_{\mu' \nu'} \| = \left| \left| \begin{array}{cccc}
-1 & & & \\
& 1 & & \\
& & 1 & \\
& & & 1 \end{array} \right| \right| \]

The tensor $\delta^\alpha_\beta  $ is defined by 
$\delta^\alpha_\beta  \sigma_\alpha  v^\beta  = \sigma_\alpha  v^\alpha $, 
and it has components $\delta^\mu_\nu $ in {\em every} basis.  Moreover
$\delta^\alpha_\beta  v^\beta  = v^\alpha $ and $\delta^\alpha_\beta  \sigma_\alpha   = \sigma_\beta  $,  all $v^\alpha ,
\sigma_\beta  $.\\

Given a Minkowski metric $\eta_{\alpha\beta} $, one defines a contravariant tensor
$\eta^{\alpha\beta} $ by \\
(1) $\eta^{\alpha\beta} $ symmetric\\
(2) $\eta^{\alpha\gamma}\eta_{\beta\gamma} = \delta^\alpha_\beta$. \\
Then $ \| \eta^{\mu \nu} \| $ is the matrix inverse of 
$ \| \eta_{\mu \nu}\| $.  In an orthonormal basis,\index{orthonormal basis}
\index{basis!orthonormal}\index{frame!orthonormal} a basis in which \\

$ \| \eta_{\mu \nu} \| = \left| \left| \begin{array}{cccc}
-1 & & & \\
& 1 & & \\
& & 1 & \\
& & & 1 \end{array} \right| \right|,$ \ 
the inverse matrix is again $ \| \eta^{\mu
\nu} \|  =  \left| \left| \begin{array}{cccc}
-1 & & & \\
& 1 & & \\
& & 1 & \\
& & & 1 \end{array} \right| \right|$.

\noindent
$\eta_{\alpha\beta} $ can be regarded as a map from vectors to dual vectors:  If $\xi^\alpha$
is a vector, the map is $\xi^\alpha  \rightarrow \eta_{\alpha\beta} \xi^\beta$.  So given a
metric $\eta_{\alpha\beta} $, we have a natural way to identify vectors and
dual vectors, and the same letter $\xi$ will be used for the dual vector and for
the identified vector:  
\be \xi_\alpha   = \eta_{\alpha\beta} \xi^\beta  .\ee
Then $\eta^{\alpha\beta} \xi_\beta   = \eta^{\alpha\beta} \eta_{\beta\gamma}\xi^\gamma  = \delta^\alpha {}_\gamma  \xi^\gamma  =
\xi^\alpha $.  One says that $\eta_{\a\b}$ {\sl lowers the index} of a vector $\xi^\a$ and 
that $\eta^{\a\b}$ {\sl raises the index} of a dual vector $\xi_\a$. 
\index{lowering index}\index{raising index}\index{index!raising and lowering} 

Similarly, if $T^{\alpha\beta} $ is a contravariant tensor, we write
\be T_{\alpha\beta}  = \eta_{\alpha\gamma}\eta_{\beta\delta}T^{\gamma\delta} ,\ee
and it follows that
\be T^{\alpha\beta}  =  \eta^{\alpha\gamma}\eta^{\beta\delta}T_{\gamma\delta}. \ee

	In a four dimensional vector space, there is -- up to a constant -- only
one totally antisymmetric tensor 
$\epsilon_{\alpha\beta\gamma\delta}  = \epsilon_{[\alpha\beta\gamma\delta]}$.
This is easy to see in terms of components: 
$\epsilon_{0122} = -\epsilon_{0122}$, interchanging 2~and~2,\\
whence $\epsilon_{0122} = 0$ and $\epsilon_{\mu\nu\sigma\tau}$ can have no 
nonvanishing components with repeated index.  The only nonzero components are
$\epsilon_{0123},\ \epsilon_{0132},\ \epsilon_{1203}$, etc. and these are
related by a permutation of $\{0,1,2,3\}$. 
\[
	\epsilon_{0123} = -\epsilon_{0132} = \epsilon_{1032} = \cdots ~.
\] 
Then $\epsilon_{\alpha\beta\gamma\delta} $ can be fixed by setting 
$\epsilon_{0123} = 1$ in a right handed orthonormal basis whose 
timelike basis vector points to the future. With this normalization, we have
\be \epsilon_{\alpha\beta\gamma\delta}\epsilon^{\alpha\beta\gamma\delta}  = -4!
\label{epep}\ee

\noindent
{\sl Exercise}. \vspace{-2mm}	
\ben[label=\alph*.]\item	Show that (\ref{epep}) implies that
$\epsilon_{0123} = \pm1$ in an orthonormal basis.
\item	Suppose that, in a right-handed, future-pointing basis, $\epsilon_{0123} = 1$.  
Show that if $\widehat{\epsilon}_{0123} =  1$ in a left handed future pointing basis, then
$\widehat{\epsilon}_{\alpha\beta\gamma\delta}  = -\epsilon_{\alpha\beta\gamma\delta}$.
\een
\vspace{5mm}

\subsection{Tensor Fields and \texorpdfstring{$\bm\nabla_\alpha$}.} 
\label{s:tensorfields}

	The arena of special relativity is Minkowski space, $M$, which we can now
define as a four-dimensional vector space together with a Minkowski metric
$\eta_{\alpha\beta} $.  This definition singles out an origin -- the zero vector {\bf 0},
but physics doesn't hand us a preferred event so we can equally well regard
another point $A$ as the origin of a vector space, identifying any point B
with the vector $V^\alpha_{AB} = B^\alpha -A^\alpha $.
\vspace{-5mm}
\begin{figure}[h] 
                \begin{center}
                \includegraphics[width=.7\textwidth]{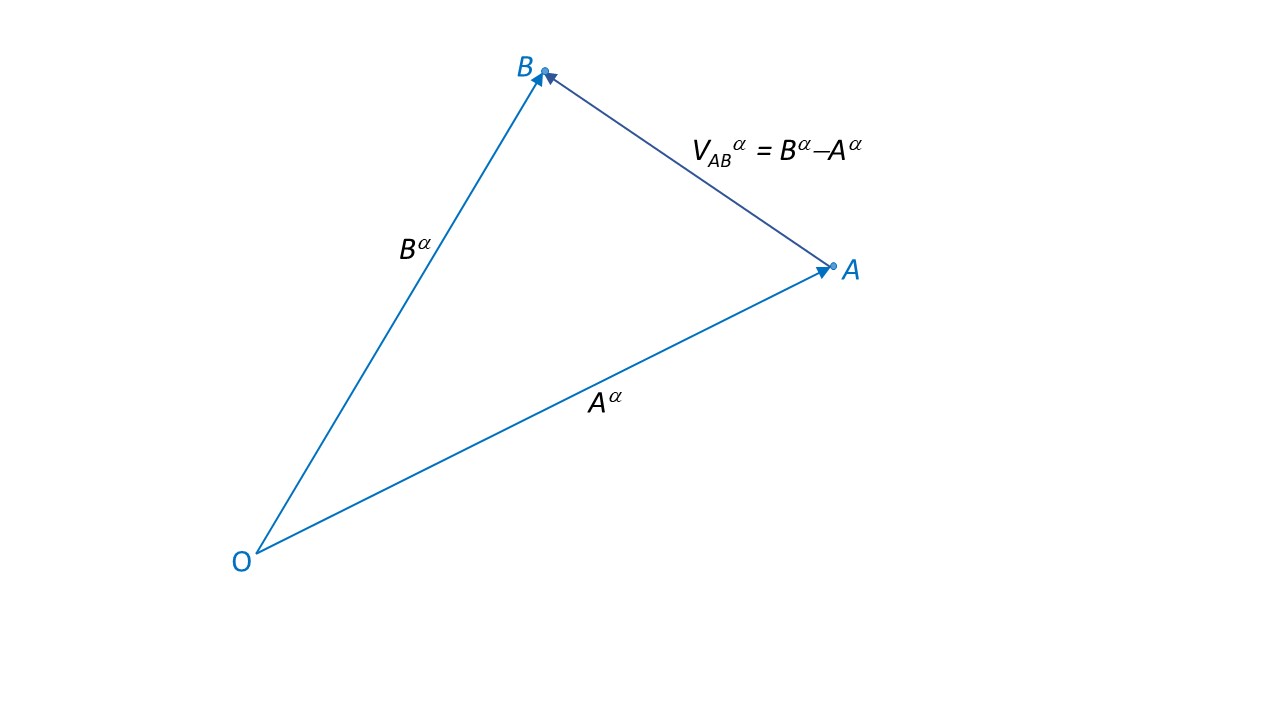}
		\end{center}
\label{vab} 
\end{figure} 
\vspace{-2cm}

	In this way, we regard the connecting vectors $V^\alpha_{AB}$ as vectors at
$A$, and $\eta_{\alpha\beta} $ may be regarded as a metric for the vector space at
each point A:
\be 
   \eta_{\alpha\beta} V^\alpha_{AB}V^\beta_{AC} 
	= \eta_{\alpha\beta} (B^\alpha -A^\alpha )(C^\beta -A^\beta ). 
\ee
If we choose $A$ as the origin of the vector space, then $\bf A = 0$, and $V^\alpha_{AB}= B^\alpha$. 
\newpage

The natural Cartesian coordinate systems with $A$ as origin assign as coordinates of the point $B$ the components of $B^\alpha$ with respect to an orthonormal basis,
$\wh{\bm t}, \wh{\bm x}, \wh{\bm y}, \wh{\bm z}$ at $A$: 
\index{coordinates!Cartesian}

\begin{figure}[H] 
                \begin{center}
                \hspace{2cm}\includegraphics[width=.5\textwidth]{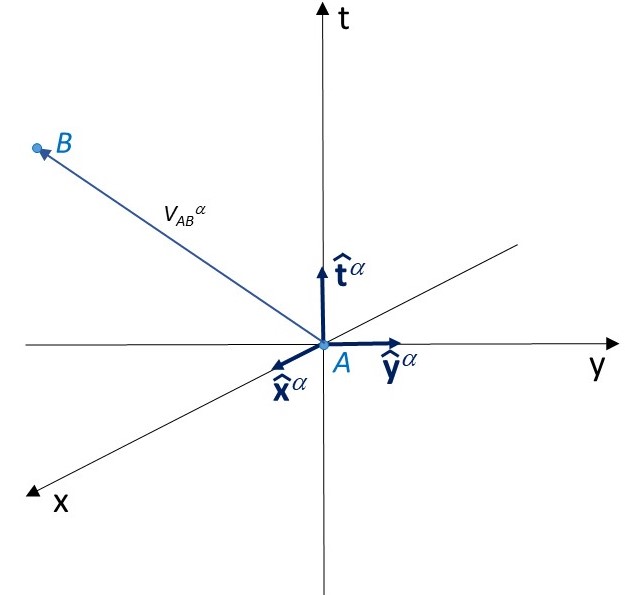}
		\end{center}
\label{w23a} 
\end{figure} 

\vspace{-7mm}

\centerline{$\bm B = {\bm V}_{AB} = B^t\wh{\bm t} + B^x \wh{\bm x}+ B^y\wh{\bm y}+ B^z\wh{\bm z}$}
\vspace{2mm}

\noindent
The $t,x,y,z$ coordinates of the point $B$ are then $B^t=t,B^x=x,B^y=y,B^z=z$.\\

\index{coordinates!curvilinear}\index{curvilinear coordinates}
In curvilinear coordinates, the 
identification is gone:  Suppose the point $B$ has coordinates $t,r,\phi,z$, with $x=r\cos\phi$, $y=r\sin\phi$,  
and let $\wh{\bm r}$ and $\wh{\bm \phi}$ be unit vectors in the $r$ and $\phi$ directions. 
Because the vector $\bm B$ from the origin to the point $B$ is radial, $B^{\hat\phi}=0$, not $\phi$.  
(Formally, $B^{\hat\phi} = - B^x\sin\phi + B^y\cos\phi = 0$.)  

We'll call the Cartesian coordinates associated with an orthonormal basis and a choice 
of origin {\sl natural} coordinates for $M$.  Any two natural
coordinate sytems for $M$ are related by a change of origin (translation)
together with a change of orthonormal basis (Lorentz transformation) -- i.e.,
two natural coordinate systems $\{x^\mu\}$ and $\{x'^\mu\} $ are related by a 
Poincar\'e transformation, $x'^\mu=\Lambda^\mu{}_\nu x^\nu + a^\mu$.\\

\noindent A {\sl scalar field} on $M$ is a map $f$:  $M \rightarrow\mathbb R$ assigning to each point $P$ in $M$ a number $f(P)$.\\
\index{scalar field}  
When the context gives the word ``function'' the unambiguous meaning 
$f:M\rightarrow\mathbb R$, 
it will be sometimes be used instead of the longer term ``scalar field.'' \\

\noindent A {\sl vector field} on $M$ assigns to each point $P$ a vector $v^\alpha (P)$
at $P$. \index{vector!vector field}\\

\noindent If $v^\alpha $ is a vector at $P$, $\eta_{\alpha\beta} v^\beta  = v_\alpha  $ is a dual vector
at $P$, and if $u^\alpha $ is another vector at $P,~u^\alpha v^\beta$ is a tensor at $P$.
In this way one constructs all tensors at $P$.  A tensor field $T^{\alpha\beta} {}_\gamma  $
on $M$ assigns to each point $P$ a tensor $T^{\alpha\beta} {}_\gamma  (P)$ at $P$.

In a curved space, one no longer has a natural way to define a connecting vector between an 
arbitrary pair of points.  But we will see that the tangent to a curve and the gradient of a function 
can naturally be identified with vectors and dual vectors, respectively.  Here is the flat-space 
description, written in a way that will later let us use most of it in curved space. 
\newpage   

A path through $P$ is defined as a map $c: {\mathbb R} \rightarrow M$, whose image intersects $P$.  Here we take $c(0) = P$.

\begin{figure}[h] 
                \begin{center}
                \includegraphics[width=.7\textwidth]{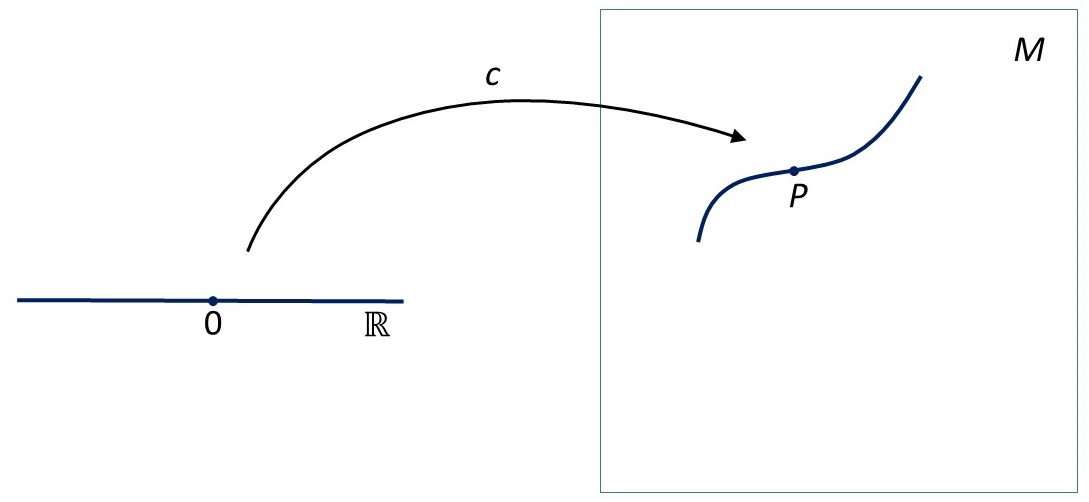}
		\end{center}
\label{w23b} 
\end{figure} 
\noindent{\sl Tangent vectors}. In flat space, we can regard each point 
$c(\lambda)$ as a vector at $P$ (the connecting vector from $P$ to $c(\lambda))$ and so
define the tangent to $c$ at $P$ as the vector
\be 
	v^\alpha  = \left.\frac{d}{d\lambda}c^\alpha (\lambda)\right|_{\lambda = 0}.   
\ee
Two curves are tangent at $P$ if they have the same tangent vector.

A vector $v^\alpha$ at $P$ maps functions on $M$ to real numbers by taking their
derivative in the direction of $v^\alpha $.  That is,
\be {\bm v}(f) = \left.\frac{d}{d\lambda} f(c(\lambda)) \right|_{\lambda = 0},
\label{e:vf}\ee
where $c(\lambda)$ is a curve through $P$ with tangent $v^\alpha $.  This will be 
used to {\sl define} vectors in a curved space.  

\begin{figure}[H] 
                \begin{center}
                \includegraphics[width=5cm]{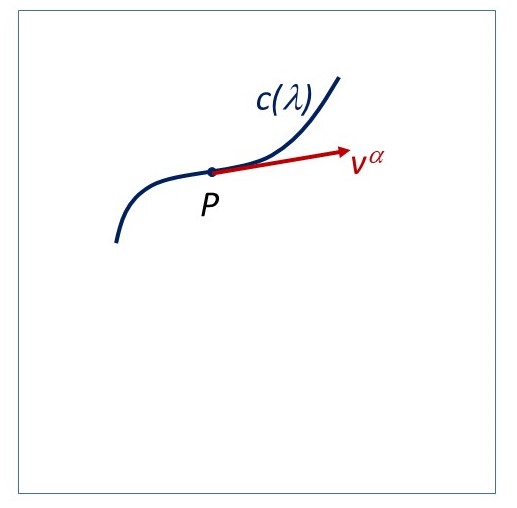}
		\end{center}\vspace{-8mm}
\label{w24a} 
\end{figure} 
\vspace{2mm}

\noindent {\sl Example}:  Let $t^\alpha $ be the tangent to the path $\lambda
\rightarrow (t+\lambda, x, y, z)$.\\
i.e. $t^\alpha $ is a unit vector in the $t$ direction at the point $(t,x, y,
z)$.\\

\noindent Then 
\be 
	{\bm t}(f) = \frac{d}{d\lambda}f(t+\lambda,x,y,z) 
		   =  \frac{\partial f}{\partial t} (t,x,y,z).
\label{e:tf}\ee
For this reason, the vector $t^\alpha $ is sometimes written ${\bm\partial}_t$.\\
\newpage

\noindent{\sl Gradients of functions are dual vectors}. \\
\index{gradient!of a scalar}
A function $f$ gives rise to a linear map from vectors at $P$ to real numbers:

\hspace{.7in}$\dis v^\alpha  \rightarrow 
	\left. \frac{d}{d\lambda} f(c(\lambda))\right|_{\lambda=0}$, 
where $c(\lambda)$ has tangent $v^\alpha  $. \\

\noindent
In a natural coordinate system at $P$
\begin{eqnarray}
 \left.\frac{d}{d\lambda} f(c(\lambda)) \right|_{\lambda=0}
 &=& \frac{\partial f}{\partial x^\mu}[c(\lambda)] 
		\left.\frac{dc^\mu}{d\lambda} \right|_{\lambda=0} \nonumber \\
&=& \frac{\partial f}{\partial x^{\mu}} (P)\ v^{\mu} , \end{eqnarray}
and so the map is clearly linear:
\begin{eqnarray*}r\,u + s\, v &\rightarrow&\frac{\partial f}{\partial
x^{\mu}} (P)(r\,u^{\mu} + s\, v^{\mu}) \\
&=& r\frac{\partial f}{\partial x^{\mu}} (P)u^{\mu} + s\frac{\partial
f}{\partial x^{\mu}} (P)\, v^{\mu}.\end{eqnarray*}
\noindent In other words, the map is a dual vector, and one calls it
$\nabla_\alpha  f$.  Thus $\nabla_\alpha  f:  v^\alpha  \rightarrow v^\alpha \nabla_\alpha  f$ has
components
\be \nabla_{\mu}f =\frac{\partial f}{\partial x^{\mu}} \ee 
Given a function $f$, one thereby acquires a dual vector field $\nabla_\alpha  f$,
the gradient of $f$.

\begin{wrapfigure}[5]{r}{6cm}
                \begin{center}\vspace{-10mm}
                \includegraphics[width=6cm]{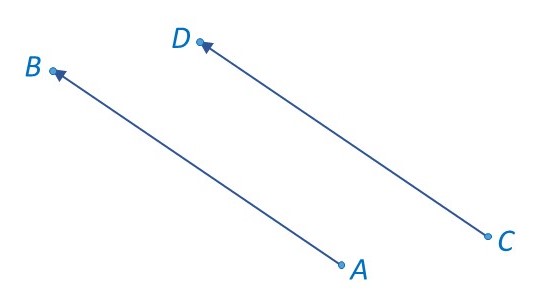}
		\end{center}
                \label{w25}
                \end{wrapfigure}
\color{white}
.

\color{black}
To extend the action of $\nabla_\alpha  $ to vectors, we need to compare 
vectors at two different points.  To do so, we use the  
natural parallel transport that identifies a vector $v^\alpha_{AB}$ at $A$ 
with the vector $v^\alpha_{CD}$ at $C$ when 
$v^\alpha_{AB} \equiv B^\alpha  - A^\alpha  = D^\alpha  - C^\alpha  
\equiv v^\alpha_{CD}$.
\color{white}
.

.

.

.
.
.
\color{black}
A vector field $v^\alpha$ is said to be constant if, given its value 
$v^\alpha(A)$ at any point $A$, one obtains the vector $v^\alpha (B)$ 
by parallel transport of $v^\alpha (A)$ from $A$ to $B$. Equivalently, 
the constant vector fields have constant components in a natural coordinate system:  
We can write 
\be 
	v^\alpha  = v^t \widehat t^\alpha  + v^x  \widehat x^\alpha  + v^y  \widehat y^\alpha  + v^z \widehat z^\alpha , 
\label{e:vcomponents}\ee
\noindent where \\
$\widehat t^\alpha $ is tangent to the curve $(t+\lambda, x, y, z) = c_0(\lambda)$\\
$\widehat x^\alpha $ is tangent to the curve $(t,x+\lambda, y, z) = c_1(\lambda)$,\\
$\ldots$.\\
Then the vector field $v^\alpha $ is a constant vector field when $v^t, v^x, v^y$, and $v^z$ are
constants.  We set
\be 
	\nabla_\alpha  v^\beta  = 0 
\ee
when $v^\alpha $ is a constant vector field and extend $\nabla_\alpha  $ to arbitrary
vector fields by the Leibnitz rule
\be \nabla_\alpha  (fv^\beta ) = (\nabla_\alpha  f)v^\beta  + f\nabla_\alpha  v^\beta,  \ee
$v^\alpha$ now not constant. 
Since any vector can be written in terms of constant vector fields $\widehat t^\alpha $,
$\widehat x^\alpha $, $\widehat y^\alpha $ and $\widehat z^\alpha $ in the form \eqref{e:vcomponents},
we have $\nabla_\alpha  v^\beta  = (\nabla_\alpha  v^t)\widehat t^\beta  
+ (\nabla_\alpha  v^x)\widehat x^\beta  + (\nabla_\alpha  v^y)\widehat y^\beta  
 + (\nabla_\alpha  v^z)\widehat z^\beta$.  $(v^t , v^x , v^y$ and $v^z$ are
four scalar fields,  the components of $v^\alpha $ in the basis 
$\widehat t^\alpha , \widehat x^\alpha  , \widehat y^\alpha  , \widehat z^\alpha )$.  
Thus in the basis 
$\widehat t^\alpha , \widehat x^\alpha  , \widehat y^\alpha  , \widehat z^\alpha $, 
the components of $\nabla_\alpha  v^\beta$ are
\begin{eqnarray} \nabla_\mu  v^\nu &=&\frac{\partial v^t}{\partial x^\mu} 
\delta^\nu_t + \frac{\partial v^x}{\partial x^\mu} \delta^\nu_x +
\frac{\partial v^y}{\partial x^\mu} \delta^\nu_y +
\frac{\partial v^z}{\partial x^\mu} \delta^\nu_z  \nonumber \\
&=& \frac{\partial v^\nu}{\partial x^\mu} \hspace{.5in} (t^\nu =
\delta^\nu_t,\ \widehat x^\nu = \delta^\nu_x,~~ \text{etc.}) \end{eqnarray}
The operator is extended to dual vectors via
\be \nabla_\alpha   \xi_\beta   \equiv \eta_{\beta\gamma} \nabla_\alpha   
\xi^\gamma  \ee
and to arbitrary tensor fields by saying what tensor (multilinear map) is denoted by 
$\nabla_\alpha  T^{\beta\cdots\gamma}  {}_{\delta\cdots\epsilon}  $:
\bea
(\nabla_\alpha  T^{\beta\cdots\gamma}  {}_{\delta\cdots\epsilon})
	w^\alpha \sigma_\beta   \cdots \tau_\gamma  u^\delta  \cdots v^\epsilon  
&\equiv &\nabla_\alpha  (T^{\beta\cdots\gamma}  {}_{\delta\cdots\epsilon}  
	\sigma_\beta   \cdots \tau_\gamma  u^\delta  {}\cdots v^\epsilon )w^\alpha  
								\nonumber\\
&& -T^{\beta\cdots\gamma}  {}_{\delta\cdots\epsilon}  
(\nabla_\alpha   \sigma_\beta  ) \cdots \tau_\gamma  u^\delta  {}\cdots
				v^\epsilon w^\alpha  \nonumber\\
&& - \cdots  -T^{\beta\cdots\gamma}  {}_{\delta\cdots\epsilon}  \sigma_\beta   
\cdots \tau_\gamma  u^\delta  {}\cdots (\nabla_\alpha  v^\epsilon )w^\alpha 
\label{nabla2}\eea 

\benr \item Prove that the components of
$\nabla_\alpha  T^{\beta\cdots\gamma}  {}_{\delta\cdots\epsilon}  $ 
in a natural coordinate system (i.e. in
the basis $\widehat t^\alpha , \widehat x^\alpha, \widehat y^\alpha , \widehat z^\alpha $ 
of vector fields tangent to the curves
$(t+\lambda, x, y, z)$, $(t, x+\lambda, y, z)$ etc. of a natural coordinate
system) are 
\be \nabla_\lambda T^{\mu\cdots\nu}{}_{\iota\cdots \kappa} =
\frac{\partial}{\partial x^\lambda } T^{\mu\cdots\nu}{}_{\iota\cdots\kappa} .\ee
\een
Note that in terms of the natural coordinate components, 
equation (\ref{nabla2}) is the Leibnitz rule 

\bea\frac{\partial}{\partial
x^\lambda }(T^{\mu\cdots\nu}{}_{\iota\cdots\kappa}\sigma_\mu
\cdots \tau_{\nu}u^{\iota}{}\cdots v^\kappa) 
&=& (\frac{\partial}{\partial
x^\lambda }T^{\mu\cdots\nu}{}_{\iota\cdots\kappa}) 
	\sigma{}_{\mu}\cdots \tau_{\nu}u^{\iota}{}\cdots v^\kappa\nonumber\\
&& + T^{\mu\cdots\nu}{}_{\iota\cdots\kappa}\frac{\partial \sigma
_\mu}{\partial x^\lambda}\cdots \tau_\nu u^{\iota}\cdots v^\kappa +
\cdots 
\label{nabla3}\eea
\noindent It follows immediately from (\ref{nabla3}) that
 \be\nabla_\alpha  \delta^\beta {}_\gamma   = 0,\  
 \nabla_\alpha  \eta_{\beta\gamma} = 0,\  
 \nabla_\alpha  \eta^{\beta\gamma} = 0,\ 
 \nabla_\alpha   \epsilon_{\beta\gamma\delta\epsilon} = 0. 
\ee
(If all the components of a tensor vanish, the tensor itself vanishes.)
\newpage

\section{Particles}
\label{s:particles}

\subsection{Particle Trajectories}
\index{particle|textbf}
	
	Our primary aim in the rather abstract treatment of tensors given in Sect.~\ref{ss:tensors}
was to develop a formalism that could be easily generalized to
curved spacetimes.  However, most of the standard
calculations of special relativity are more easily performed in a
coordinate-free framework - once one gets used to it and no longer feels
the need to translate back to a more familiar language.  Sect.~\ref{s:particles} 
is therefore designed largely to restate concepts and calculations you
already know in a somewhat different language.

	The motion of a particle can be described as a path in spacetime -- its
trajectory or worldline -- $c(\lambda)$.
\begin{wrapfigure}[10]{r}{5cm}
                \begin{center}\vspace{-6mm}
                \includegraphics[width=6cm]{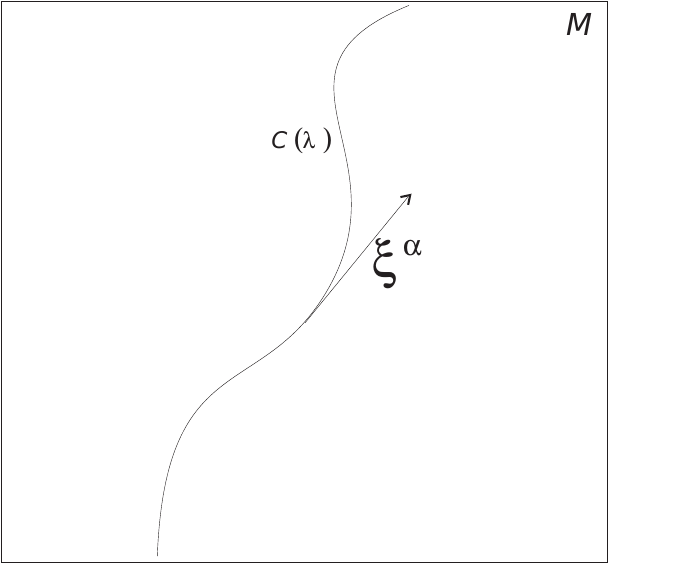}
		\end{center}
                 \label{w27}
                \end{wrapfigure}

The tangent vector to the path, 
\[ 
	\xi^\alpha  = \frac{dc^\alpha (\lambda)}{d\lambda}, 
\]
is timelike:  That's what it means to say the particle travels slower than
the speed of light. 
\color{white}
.

.

.

.

.

.

\color{black}

\noindent The set of all points (spacetime events) that can be reached by a
particle starting at a point $P$ is the interior of the future light cone
at $P$, and it can be identified with the set of all future pointing
timelike vectors at $P$.

\begin{figure}[h] 
                \begin{center}
                \includegraphics[width=6cm]{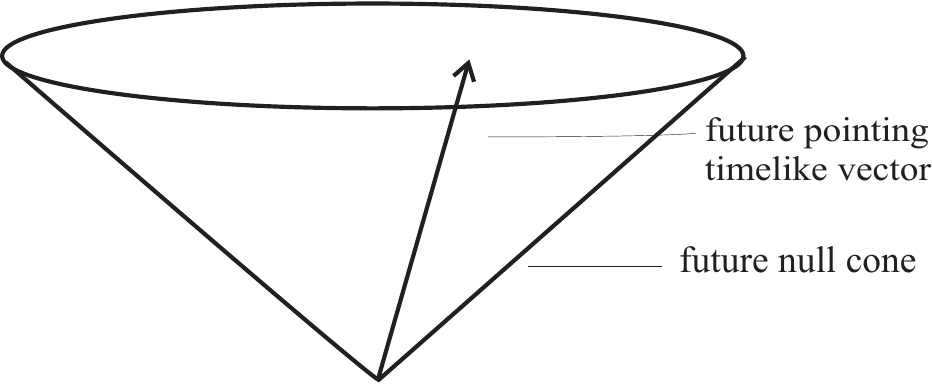}
		\end{center}
\label{w28a} 
\end{figure} 

  If the event $P$ is an explosion, so that an expanding spherical
pulse of light is sent out from it, then the light emerging from this event
forms a cone in spacetime with vertex $P$.  (A slicing of the light's
history is a sequence of spheres expanding from a point.)  Formally the
cone is the set of all future-pointing null vectors at $P$.

	The velocity (four-velocity) of a particle is the unit vector tangent to
its trajectory,
\be u^\alpha  = \xi^\alpha (-\xi^\beta \xi_\beta  )^{-1/2} ,\ee
with
\be 
u^\alpha u_\alpha  
=\xi^\alpha (-\xi^\beta\xi_\beta)^{-1/2}\xi_\alpha(-\xi^\gamma\xi_\gamma)^{-1/2}
=\xi^\alpha\xi_\alpha  (-\xi^\beta \xi_\beta)^{-1} = -1. 
\ee
\index{velocity!four-velocity}

\noindent A zero rest mass particle has no unit four-velocity;
the tangent to its trajectory is 
a null vector $k^\alpha$, $k^\alpha k_\alpha=0$.\\

\begin{wrapfigure}[12]{r}{5.5cm}
                \begin{center}\vspace{-13mm}
                \includegraphics[width=7cm]{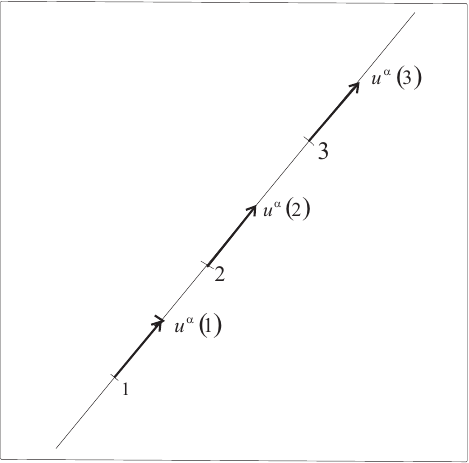}
		\end{center}
                \label{w28b}
                \end{wrapfigure}
                
	A free particle has a straight line trajectory - its tangent vector is
parallel propagated along itself, 
\be u^\beta \nabla_\beta  u^\alpha  = 0. \ee

A preferred observer in Minkowski space (an inertial observer) will mean
someone who moves along the trajectory of a free particle and who measures
components of tensors along a basis 
$(u^\alpha , \widehat x^\alpha , \widehat y^\alpha , \widehat z^\alpha )$  where $u^\alpha $ is
his velocity and $\widehat x^\alpha , \widehat y^\alpha , \widehat z^\alpha $ are unit 
vectors orthogonal to each other and to $u^\alpha $
and parallel propagated along $u^\alpha $.  Such {\em inertial frames} are the
preferred bases of constant vector fields introduced in Sect.~\ref{s:tensorfields}.\\

\vspace{4mm} 

In general, of course, particles and observers need not travel in straight
lines.  One can, for example, describe the measurements made by an
accelerating, rotating, observer: At each point along a timelike trajectory 
(the observer's worldline) one chooses a basis whose timelike 
basis vector is the observer's $u^\alpha$.\index{basis!basis of observer}  The spacelike vectors
are then unit, orthogonal to $u^\alpha $, and have orientation fixed by the
observer's rotation.  Think of an observer as someone with three
perpendicular rulers (thin rods) and a ticking watch.  
The connecting vector between ticks is the timelike basis vector; 
and the intersection of the slice orthogonal to $u^\alpha $ with the 
history of each ruler is a spacelike basis vector.

	History of an observer with watch and one ruler; the observer 
rotates by $\pi$ and then stops rotating. \\
\begin{figure}[h]
		\begin{center}
                \includegraphics[width=3cm]{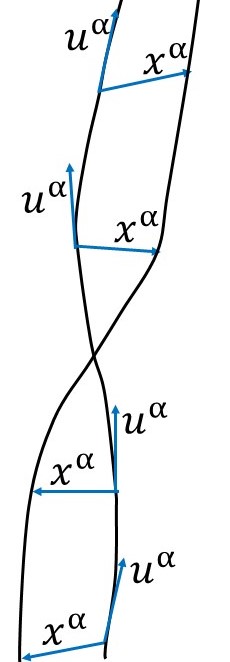}
		\end{center}
                \label{Page30}
\end{figure}	
\newpage

	The physical meaning of the metric $\eta_{\alpha\beta} $ is that it provides the
measured distance between events.  In particular, if a clock moves from $A$
to $B$ along a trajectory $c(\lambda), c(0) = A, c(\lambda) = B$, it will
tick off a time	
\be  
\tau_c  (\lambda) = \int^\lambda_0 [-\xi^\alpha (\lambda ) \xi^\beta (\lambda)\eta_{\alpha\beta} ]^{1/2}d\lambda , 
\quad \mbox{ where } 
\xi^\alpha (\lambda) = \frac{dc^\alpha (\lambda)}{d\lambda}.
\label{tau}\ee
$\tau_c $ is called the {\sl proper time} along the path $c$.\\

	Furthermore, there is a preferred time between two (timelike related)
events $A$ and $B$, namely \\
\centerline{$
	\tau = [-\eta_{\alpha\beta} (B^\alpha -A^\alpha )(B^\alpha -A^\beta)]^{1/2}$,}
and $\tau$ is called the proper time between the events.  We will see in a
minute that $\tau$ is the time measured by a freely moving clock whose
trajectory is the straight line from $A$ to $B$.  But first note that if
$c$ is parameterized by proper time $\tau$, then 
$\dis\xi^\alpha  = \frac{dc^\alpha (\tau)}{d\tau} = u^\alpha $, the unit vector tangent to $c$.\\

\noindent{\sl Proof}:  The proof is essentially immediate from the definition (\ref{tau}) of proper time. 
For any $\lambda(\tau)$, Eq.~(\ref{tau}) implies $\dis \frac{d\tau}{d\lambda} = [-\xi^\alpha\xi_\alpha]^{1/2}$.  When $\tau = \lambda$, we have $d\tau/d\lambda=1$ and $\xi^\alpha=dc^\alpha/d\tau$, implying  
\[
 1 = \left[ -\xi^\alpha \xi_\alpha\right]^{1/2},  \quad
 \ \xi^\alpha \xi_\alpha = \frac{dc^\alpha }{d\tau}\frac{dc_\alpha  }{d\tau} = -1. 
\]

	For a freely moving  clock -- the timepiece of an inertial observer -- the
trajectory will be 
\be c^\alpha (\tau) = \tau u^\alpha ~, \ee
parameterized by clock time, since $\dis\frac{dc^\alpha }{d\tau}$ is the constant
vector $u^\alpha $.  If $\tau_A$ is the clock's time at $A$ and $\tau_B$ its
time at $B$, 
\[ c^\alpha (\tau_A) = A^\alpha ~, \hspace{.3in} 
	c^\alpha (\tau_B) = B^\alpha ~,~ \text{and} \]
\begin{eqnarray*} 
(A^\alpha -B^\alpha )(A_\alpha  -B_\alpha  ) 
&= &(\tau_A u^\alpha - \tau_Bu^\alpha )(\tau_A u_\alpha  - \tau_Bu_\alpha) \\
&=& - (\tau_A-\tau_B)^2 . 
\end{eqnarray*}
That is, the proper time between the events $A$ and $B$ is the proper time
along the straight line trajectory from $A$ to $B$.\\

\subsection{Boosts and ``Addition of Velocity''}
\index{velocity!addition of velocity}

	We will use ``velocity" in this section to mean 4-velocity, and ``speed"
to mean the magnitude of the 3-velocity seen by an inertial observer.  

\noindent{\sl Addition of velocities}.\\
 
\noindent Suppose now that three observers, Alice, Bob and Cat,  move in straight lines in a plane in spacetime.  \\
Bob moves at speed $v_1$ relative to Alice.\\
Cat moves at speed $v_2$ relative to Bob.  \\
We are to find the speed $v$ of Cat relative to Alice.\\
Alice's coordinates are $t,x$,\\   
Bob's coordinates are $\ov t, \ov x$.

The familiar derivations are variants of the following: \\
In terms of Bob's coordinates, Cat's speed 
relative to Bob is $v_2 = d\overline x/d\overline t$.\\ 
Alice's coordinates $\overline t,\overline x$ are related to Bob's by 
\[
t = \gamma_1 (\overline t+v_1\overline x) \qquad x = \gamma_1(v_1\overline t + \overline x).  
\]
and Cat's speed relative to Alice is given by
\be
   \cblue v = \cb \frac{dx}{dt} = \frac{\g_1(d\overline x + v_1 d\overline t)}
   				     {\g_1(d\overline t + v_1 d\overline x)} 
   			     =  \frac{d\overline x/d\overline t + v_1}
   				     {1 + v_1 d\overline x/d\overline t}
   			    \cblue = \frac{v_1+v_2}{1+v_1 v_2}\ ;
\label{e:add_vel}\ee  

But it's instructive to see how the derivation goes as a product of the 
boosts that map Alice's frame to Bob's and Bob's frame to Cat's.\\ 
 
\noindent
Alice's frame $\{\bm u, \wh{\bm x}\}\equiv \{ \wh{\bm e}_\mu \}$,\\   
Bob's frame $\{\ov{\bm u}, \ov{\bm x}\}\equiv\{\ov{\bm e}_\mu,\}$, \\
Cat's frame $(\wt{\bm u}, \wt{\bm x})\equiv \{\wt{\bm e}_\mu\}$.
\vspace{-20mm} 
\begin{figure}[H]
               \begin{center}
		\includegraphics[width=.3\textwidth]{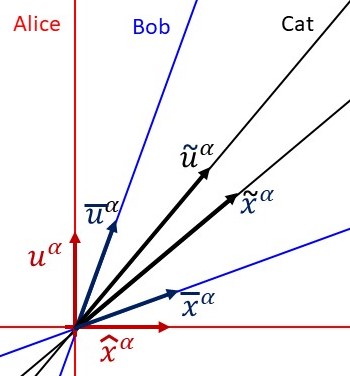}
		\end{center}
\end{figure}
Write the successive boosts and their product as in Eq.~\eqref{e:boost}.  
with indices suppressed and the $y,z$ dimensions omitted for brevity: 
\[
    \ov{\bm e} = \bm e \Lambda_1,\quad
    \wt{\bm e} = \ov{\bm e} \Lambda_2, \quad 
    \wt{\bm e}   = {\bm e} \Lambda,  \quad 
	\Lambda = \Lambda_1\Lambda_2, 
\]
where, with $\eta = \gamma v$, 
\begin{align*}
\Lambda_1  
	&=\begin{bmatrix}
		\gamma_1 & \eta_1  \\
		\eta_1  & \gamma_1 \\
	\end{bmatrix}, \qquad
\Lambda_2= 
	\begin{bmatrix}
		\gamma_2 & \eta_2  \\
		\eta_2  & \gamma_2 \\
	\end{bmatrix}.
\end{align*}
Then
\[
\Lambda = \begin{bmatrix}
 				\gamma & \eta \\
 				\eta & \gamma  \\
			\end{bmatrix}
= \Lambda_1 \Lambda_2 
 = \begin{bmatrix}
 \gamma_1\gamma_2 +\eta_1\eta_2 &\quad \g_1\eta_2+\g_2\eta_1\\
  \g_1\eta_2+\g_2\eta_1 
  &  \gamma_1\gamma_2 +\eta_1\eta_2 \\
			\end{bmatrix}, 
\]
\noindent
and
\be 
	\gamma =  \g_1\g_2 + \eta_1\eta_2
\qquad
	\eta = \g_1\eta_2+\g_2\eta_1. 
\label{addv}\ee 

\noindent Rewriting this solution in terms of $v, v_1, v_2$ 
gives the usual form \eqref{e:add_vel}
\[
	v = \frac\eta\gamma = \frac{v_1+v_2}{1+v_1 v_2} .
\]
But an equally useful form is obtained by noting that, by
(\ref{addv}),
\begin{eqnarray*}\gamma + \eta &=& \g_1\g_2 +
\eta_1\eta_2 + \g_1\eta_2 +
\eta_1\g_2 \\
&=& (\g_1 + \eta_1)(\g_2 + \eta_2) .\end{eqnarray*}
In other words $\log (\gamma + \eta)$ is additive!
\be
 \log(\gamma + \eta) 
	= \log(\g_2 + \eta_2) + \log(\g_1 + \eta_1).
\ee
Now
\[ \gamma + \eta = \frac{1+v}{\sqrt{1-v^2}} = \sqrt{\frac{1+v}{1-v}} ~,
~~\text{so}  \]
\be\cblue 
  \log\left(\frac{1+v}{1-v}\right) =  \log\left(\frac{1+v_1}{1-v_1}\right) 
			   + \log\left(\frac{1+v_2}{1-v_2}\right).
\label{e:additive}\ee 
The combination $\dis \vartheta=\log\sqrt{\frac{1+v}{1-v}}$ is called the {\sl velocity parameter}, and we encountered it in Eq.~\eqref{e:txboost}, 
the form of a boost that shows it as the Lorentz analog of a rotation.
\footnote{There is no standard notation.  Schutz calls the velocity parameter $u$, 
and these notes, like Hartle, use $\vartheta$, a reminder of its role as the 
parameter of a Lorentz transformation corresponding to the angle $\theta$ of a rotation. }
It's easy to check that $v~=~\tanh \vartheta$.\\

If you remember that the Doppler shift in the frequency of light is given by 
\be
  \widetilde\omega = \omega\sqrt{\frac{1+\wt v}{1-\wt v}} = \omega e^\vartheta ,    
\label{e:doppler}\ee
then the additivity of $\vartheta$ is not really new:  
Eq.~\eqref{e:doppler} implies that successive boosts by $v_1$ and $v_2$ change the frequency (and a photon's energy) by the product 
\mbox{$e^\vartheta = e^{\vartheta_1+\vartheta_2}$}, implying 
$\cblue\vartheta = \vartheta_1+\vartheta_2$.  \\
(If you haven't derived the Doppler shift, see \ref{ex:doppler} )

\benr 
\item  (From the problems to Feynman v. I) A cart rolls on a table with speed $v$.
A smaller cart rolls on the first in the same direction with speed $v$
relative to the first cart, and so on up to $n$ carts.   
What is the speed $v_n$ of the $n^{th}$ cart relative to the table? What is 
\( \lim_{n \rightarrow \infty}~ v_n \)?
\label{ex:cart}
\item[{\sl Exercise.\phantom{ $11'$}}] If you didn't do the last exercise using Eq.~\eqref{e:additive}, do so now.

\item Consider the curve $c(\lambda)$ given
in the natural coordinate system by 
\beaa
x(\lambda) &=& \int^\lambda r\cos \theta \cos \phi d\lambda,~~y(\lambda) =
\int^\lambda r\cos \theta \sin \phi d\lambda \\
z(\lambda) &=&\int^\lambda r\sin \theta d\lambda,\qquad ~~t(\lambda) =
\int^\lambda rd\lambda,
\eeaa
where $r$, $\theta$, and $\phi$ are arbitrary functions of $\lambda$.\\
\noindent(a) \  Show that $c(\lambda)$ is a null curve.\\
\noindent (b)\  Under what conditions on the functions $r(\lambda)$,
$\theta(\lambda)$ and $\phi(\lambda)$ will $c(\lambda)$ be a straight line?
\een
\newpage

\subsection{Surfaces of simultaneity and the 3+1 decomposition of spacetime}

	An observer having velocity $u^\alpha $ can separate $p^\alpha $ into energy and
three-momentum in a manner analogous to the decomposition of $u^\alpha $ in
the last section into vectors along and orthogonal to the observer's
velocity.  The decomposition is frequently encountered and deserves some
discussion.  The surfaces of simultaneity of an observer with
velocity $\widehat u^\alpha $ are orthogonal to $\widehat u^\alpha $, and so are spanned by three vectors
$\widehat x^\alpha , \widehat y^\alpha , \widehat z^\alpha $ perpendicular to $\widehat u^\alpha $.  
In a natural coordinate system of
an inertial observer (points are labeled by their components along the
basis $\widehat u^\alpha, \widehat x^\alpha, \widehat y^\alpha, \widehat z^\alpha $) 
the surfaces of simultaneity are, of course, the $t$ = constant hyperplanes.\\

	If $\wt u^\alpha $ is moving at speed $v$ relative to $\widehat u^\alpha $, there is an 
easy way to draw the surfaces of simultaneity of $\widehat u^\alpha $ and $\wt u^\alpha $. 
\vspace{-8mm}

\begin{figure}[h]
		\begin{center}
                \includegraphics[width=10cm]{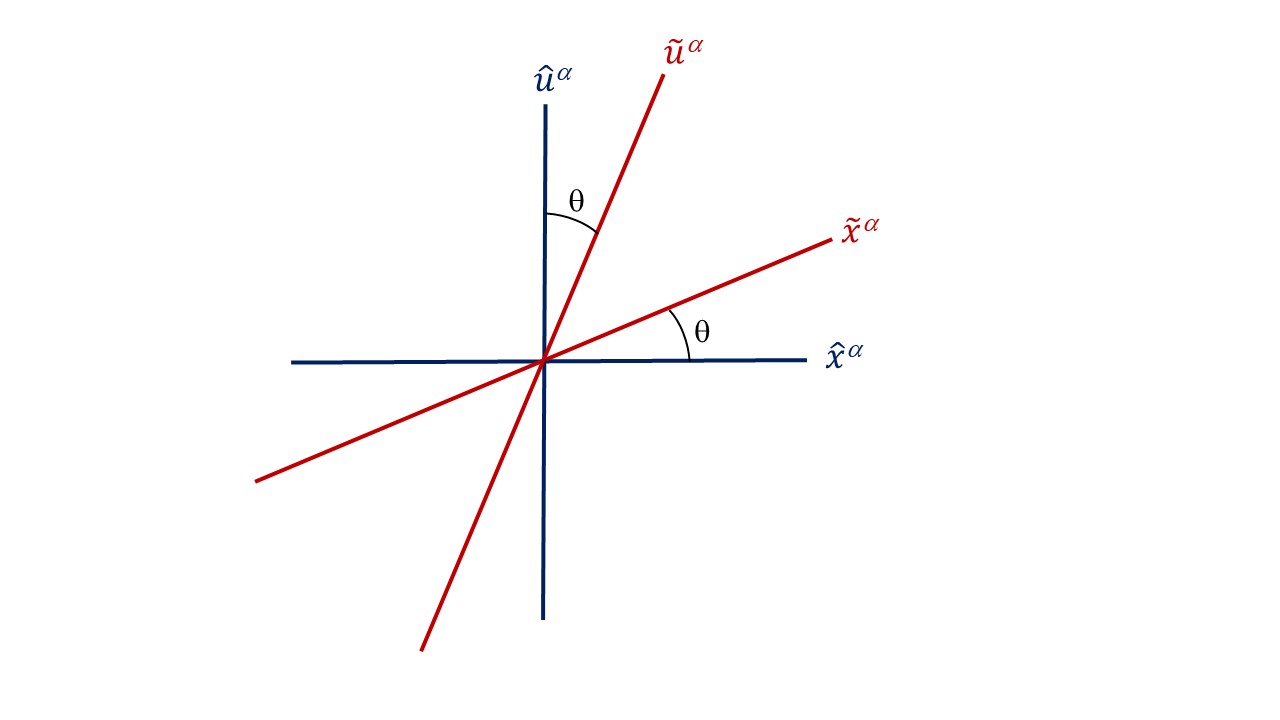}
		\end{center}  
\label{boost} 
\end{figure} 
\vspace{-8mm}

\noindent With $\tan\theta=v$, the 2-D drawing looks like the figure above. \\

\noindent
Here the angles labeled $\theta$ are angles of the Euclidean geometry of the screen (or paper), {\sl not of the actual Lorentzian geometry of spacetime.} \\
 
\noindent{\sl Check that $\tan\theta=v$}:  In the Lorentzian geometry of spacetime, $\widehat u^\alpha$ is perpendicular to $\widehat x^\alpha$ and $\wt u^\alpha$ is perpendicular to $\wt x^\alpha$.  By drawing the diagram with $\widehat u^\alpha$ vertical and  
with light-rays at $45^\circ$, we have made $\widehat x^\alpha$ perpendicular to 
$\widehat u^\alpha$ in the Euclidean spatial geometry of the diagram.  Now 
boosting the basis $(\wh u^\a,\wh x^\a)$ gives the basis $(\wt u^\a, \wt x^\a)$, 
with 
\[   
\wt u^\alpha  = \gamma\widehat u^\alpha  + \eta\widehat x^\alpha, \qquad 
\widetilde{x}^\alpha  = \eta \widehat u^\alpha  + \gamma\widehat x^\alpha.
\]
Then in the Euclidean geometry of the diagram the angle between 
$\wt u$ and $\widehat{u}$ is given by \\
\centerline{$\dis\tan\theta = \frac{\eta}{\gamma} = v$,}
and the angle between $\widehat{x}^\alpha $ and $\widehat x^\alpha $ 
is also given by\\
\centerline{ $\dis\tan\theta = \frac{\eta}{\gamma} = v$.  }
In the actual Lorentzian geometry, the dot products $\wt u^\a \wh u_\a$ and 
$\wt x^\a\wh x_\a$ are 
$-\wt u^\a \wh u_\a = \wt x^\a\wh x_\a = \gamma = \cosh\vartheta$, 
in agreement our previous expression,  Eq.~\eqref{e:boost}, for a boost.\\

\newpage
\noindent
{\sl 3+1 decomposition of tensors: Spacetime $\rightarrow$ space+time relative to an observer with velocity $u^\alpha$}.

\begin{figure}[H]
               \begin{center}
		\includegraphics[width=.7\textwidth]{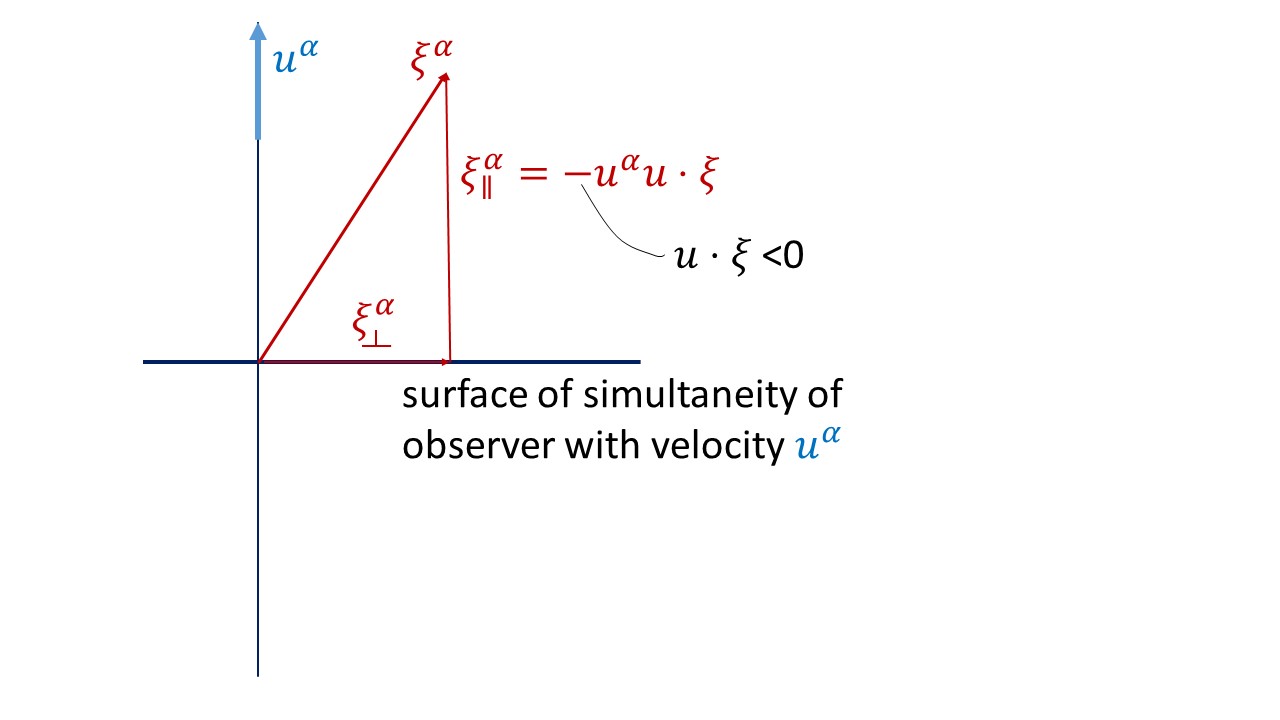}
		\end{center}
\end{figure}
\vspace{-15mm}

 The tensor
\be \g^\alpha {}_\beta   = \delta^\alpha {}_\beta   + u^\alpha u_\beta   \ee

\noindent projects vectors into the surface orthogonal to $u^\alpha $.  
In other words, if $\xi^\alpha $ is any vector,

\be 
	\xi^\alpha_\perp := \g^\alpha {}_\beta  \xi^\beta  
\ee
is a vector orthogonal to $u^\alpha $:
\be 
 \g^\alpha {}_\beta  u^\beta  = \delta^\alpha{}_\beta u^\beta  + u^\alpha u_\beta  u^\beta  
			     = u^\alpha  - u^\alpha  = 0;
\ee
similarly, $u_\alpha \g^\alpha_\beta=0$, implying
\[ 
	u_\alpha  \xi^\alpha_\perp = u_\alpha  \g^\alpha {}_\beta  \xi^\beta  = 0~.
\]
If $\xi^\alpha $ is already orthogonal to $u^\alpha $, then
\[ 
	\xi^\alpha_\perp = \xi^\alpha.
\]
{\sl Proof}: $\ \dis \xi_\perp^\alpha  = \g^\alpha {}_\beta   \xi^\beta  
			 = \delta^\alpha {}_\beta   \xi^\beta  + u^\alpha u_\beta  \xi^\beta  
			 = \xi^\alpha$, because $\ u_\beta  \xi^\beta  = 0$.  

Any vector can be written as the sum of a vector along $u^\alpha $ and one
orthogonal to $u^\alpha $:
\be 
	\xi^\alpha  = -(\xi^\beta {}u_\beta  )u^\alpha  + \xi^\alpha_\perp ~. 
\ee
Similarly any tensor can be decomposed in such a way that each
index is either along $u^\alpha $ or orthogonal to it:  e.g.
\begin{eqnarray}
	T^{\alpha\beta}   
		&=& \delta^\alpha {}_\gamma  \delta^\beta {}_\delta   T^{\gamma\delta} 
 		= (\g^\alpha {}_\gamma -u^\alpha u_\gamma) (\g^\beta {}_\delta -u^\beta u_\delta)
			T^{\gamma\delta} 
\nonumber \\
		&=& T_\perp^{\alpha\beta} - u^\alpha\, \g^\beta {}_\delta\,u_\gamma  T^{\gamma\delta} 
  				- u^\beta {}\, \gamma^\alpha {}_\gamma\,  u_\delta T^{\gamma\delta} 
  				+ u^\alpha u^\beta u_\gamma  u_\delta   T^{\gamma\delta}; 
\label{3+1}\end{eqnarray}

\noindent here 
$T^{\alpha\beta}_\perp = \g^\alpha {}_\gamma \g^\beta {}_\delta T^{\gamma\delta}$ 
is a tensor orthogonal to $u^\alpha$  
(that is, $T^{\alpha\beta}_\perp u_\beta = 0 = T^{\alpha\beta}_\perp u_\alpha$);
$u^\alpha \g^\beta{}_\delta  u_\gamma T^{\gamma\delta}$ has one index ($\alpha$) 
along $u^\alpha $ and one index ($\beta$) orthogonal to $u^\alpha$;
and $u^\alpha u^\beta u_\gamma  u_\delta   T^{\gamma\delta}$ has both 
indices along $u^\alpha$.

\subsection{Energy, momentum and acceleration of a particle}

A particle of rest mass $m$ and velocity $v^\alpha $ is described by a momentum
vector $p^\alpha  = mv^\alpha $.  Its rest mass is thus related to its momentum by
$p^\alpha p_\alpha   = -m^2$.

	The momentum $p^\alpha $ of a particle can be decomposed by an
observer $u^\alpha $ in the manner
\be 
	p^\alpha  = Eu^\alpha  + p^\alpha_\perp , 
\label{e:p3+1}\ee
where $E = -p^\alpha {}u_\alpha  $ and 
$p^\alpha_\perp = \g^\alpha{}_\beta\, p^\beta$ are the energy and
3-momentum of the particle.  
\index{energy!of particle|textbf} \index{momentum! of particle|textbf} 
Then 
\[
	p^\alpha {}p_\alpha   = -E^2 + p^2_\perp = -m^2.
\]
If the observer is at rest with respect to the particle, 
$u^\alpha  = v^\alpha $ so $p^\alpha  = mv^\alpha $ is already parallel to $u^\alpha $
and $E = m,\ p_\perp^\alpha  = 0$.  
When the relative speed of particle and observer is $v = (1-\gamma^{-2})^{1/2}$ with 
$\gamma = -u^\alpha v_\alpha$,
\be  E = m\gamma \hspace{.5in} p_\perp = m\eta~.\ee 

In calling the 4-vector $p_\perp^\a$ the 3-momentum, we are associating a 4-vector 
whose component along $\na_\a t$ vanishes 
with the 3-dimensional vector $p^a$, the usual 3-momentum seen by an observer 
with 4-velocity $t^\a=-\na^\a t$.  That is, $p_\perp^\a$ is the unique 
4-vector orthogonal to $t^\a$ whose spatial part is the 3-vector $p^a$:  
The components of $p^a$ are the nonzero components of $p_\perp^\a$, 
and they are the spatial components of $p^\a$ in a chart $\{t,x^i\}$:
\be
  p_\perp^i = p^i, \qquad p_\perp^t = 0.    
\ee 
\vspace{3mm}
   
     For an inertial observer, the $t$=constant surfaces with $-\na^\a t$ tangent to the observer's trajectory define a $3+1$ decomposition of this form for tensor fields on Minkowski space. Even when the observer is accelerating, such $3+1$ decompositions of tensors relative to observers make sense {\sl for tensors at the position of the observer}.  In describing tensor {\sl fields} on regions of spacetime far 
from an accelerated observer, however, they are not useful because ``later" and ``earlier" surfaces of simultaneity will intersect:
\begin{figure}[h]
		\begin{center}
                \includegraphics[width=.4\textwidth]{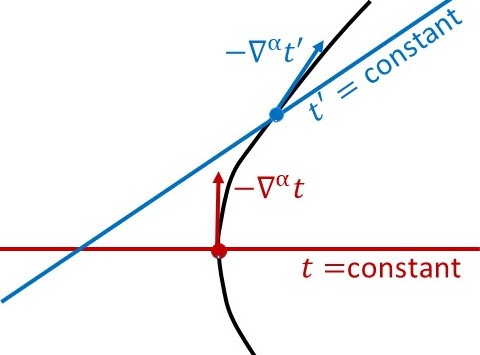}
		\end{center}
\caption{The black curve is the trajectory of an accelerated observer.  
Her 4-velocities at different points of her trajectory are the vectors $-\na^\a t$ and $\na^\a t'$, normal to hypersurfaces $t=$constant and $t'=$~constant that intersect.} 
\label{accelerated_observer}
\end{figure}

\newpage
\noindent {\sl Acceleration}: If $c(\tau)$ is a timelike path parameterized by
proper time, we found that $\dis u^\alpha~=~\frac{dc^\alpha (\tau)}{d\tau}$.  The
acceleration $a^\alpha $ is defined by

\be a^\alpha  = \frac{du^\alpha (\tau)}{d\tau} = \frac{d^2}{d\tau^2} c^\alpha (\tau)\ . \ee

\noindent We can write this as $a^\alpha  = u^\beta {}\nabla_\beta  u^\alpha $ if we regard $u^\alpha $
as a vector field defined along the path $c(\tau)$:

\be \frac{d}{d\tau} u^\alpha (c(\tau)) = \nabla_\beta  u^\alpha  (c(\tau)) \frac{dc^\beta }{d\tau}
= u^\beta {}\nabla_\beta  u^\alpha  \ee

\noindent (Note that $\nabla_\beta  u^\alpha $ alone is meaningless,  however, because 
$u^\alpha $ is not defined off of $c(\tau)$ -- only $u^\beta \nabla_\beta  u^\alpha $, the derivative along $c({\tau})$, has meaning.)  The acceleration $a^\alpha $ is a vector orthogonal
to $u^\alpha $:
\[ u_\alpha  a^\alpha  = u_\alpha  \frac{d}{d\tau}u^\alpha (\tau) = u^\alpha \frac{d}{d\tau}
u_\alpha  (\tau)\ \ \mbox{ (because $\eta_{\alpha\beta} $ is a constant tensor field)}.
\]
\[ \Rightarrow\ \  u_\alpha   \frac{d}{d\tau} u^\alpha  =
\frac{d}{d\tau}(u_\alpha  u^\alpha )\frac{1}{2} = \frac{1}{2}\frac{d}{d\tau}(-1) = 0.\]
That is, $a^\alpha $ lies in the surface of simultaneity of an inertial
observer at rest with respect to $u^\alpha $.  For such a {\em comoving} observer, the
components of $u^\alpha $ are (instantaneously)\\
\centerline
{$(u^\mu) = (1, 0, 0, 0)$. }
A short time later, \\
\centerline{
$(u^\mu) = (1,{\bf v}) + O(v^2)$}
(use $\gamma = (1-v^2)^{-1/2} = 1 + \frac{1}{2} v^2 + \cdots )$. Thus $a^\alpha=du^\alpha/dt$ 
has components
\[ 
	(a^\mu) = (0, {\bm a}_N) + O(v^2),
\]
where ${\bm a}_N$ is the Newtonian acceleration and $t$ the observer's
proper time. 

   At this time, call it $t=0$, when the particle is at rest relative to
the observer, their trajectories are tangent and so $\dis\frac{dt}{d\tau} = 1$,
where $\tau$ is the particle's proper time as above.  Thus
$\dis a^\mu=\frac{du^{\mu}}{d\tau} = (0, {\bm a}_N)$ at $t = 0$, when $v = 0$;
since $\dis \frac{du^\alpha}{d\tau} = a^\alpha$, the vector $a^\alpha $ is the Newtonian
acceleration measured by an inertial observer instantaneously at rest with
respect to the particle.  It can be measured either by using clocks and
rulers or by transporting an ``accelerometer" -- a box
with orthogonal springs supporting a mass -- along the world line of the
particle.  Then $a^\alpha $ will be proportional to a vector orthogonal to $u^\alpha $ that
joins the center of the box to the (in general) displaced position of he
mass.
\begin{figure}[h!]
\begin{center}
\includegraphics[width=10cm]{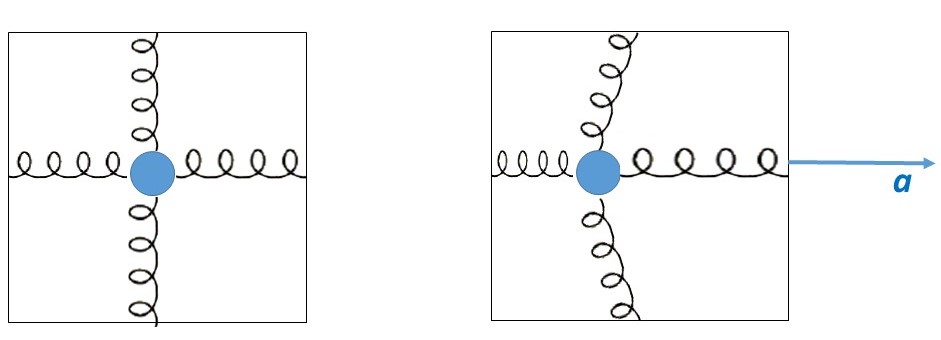}
\end{center}
\label{fig:accelerometer}
\end{figure}

A particle is said to be uniformly acelerating if
\[a^\alpha {}a_\alpha   = \text{constant} \]

\noindent and if $a^\alpha $ always lies in a plane (in spacetime). In other
words, the acceleration is always in the $x$-direction of some inertial
observer and its magnitude is constant as measured by an accelerometer  or
by an observer moving with the particle.  Such a particle will appear, for
example, to come in along the \mbox{$x$-axis} at nearly the speed of light, to slow
down, stop, reverse direction and exit at a speed that increases to 1 ($c$)
again.  The trajectory is most simply obtained using not an orthonormal
basis, but two null (lightlike) vectors along the ingoing and outgoing asymptotic null
directions;  these are constant vector fields, commonly written as $\ell^\alpha $ and $n^\alpha$.  Because 
they are null, 
\be 
	\ell^\alpha \ell_\alpha   = 0, \hspace{.5in} n^\alpha n_\alpha   = 0 .
\label{llnn}\ee 
We will normalize by
\be 
	\ell^\alpha  n_\alpha   = -\frac12\ .
\label{ln}\ee
For example, if $\widehat t^\alpha $ and $\widehat x^\alpha $
are unit orthogonal vectors in the plane of $\ell^\alpha,  n^\alpha $, 
one could take\\
\be \ell^\alpha  = \frac{\widehat t^\alpha +\widehat x^\alpha}2,\  n^\alpha  
	= \frac{\widehat t^\alpha -\widehat x^\alpha }{2}.
\label{e:tx} \ee
Because $u^\alpha $ is in the plane of $\ell^\alpha $,  $n^\alpha $, we have  
$u^\alpha  = \lambda \ell^\alpha  + \mu n^\alpha $, $u^\alpha {}u_\alpha   = -1$, 
implying $\dis\lambda = \frac{1}{\mu}$ and
\be 
 	u^\alpha  = \lambda \ell^\alpha  + \lambda^{-1} n^\alpha  .
\ee 
Then
\be a^\alpha  = \frac{du^\alpha }{d\tau} = \dot{\lambda} \ell^\alpha  -
\frac{\dot{\lambda}}{\lambda^2} n^\alpha ,\  \mbox{ where } \ (\,{}^\cdot\,):= \frac{d}{d\tau}.\ee
We have used the fact that $\ell^\alpha $ and $n^\alpha $ are constant vector fields 
to infer $\dis\frac{d}{d\tau} \ell^\alpha  = 0 = \frac{d}{d\tau}n^\alpha$.
\begin{figure}[h!]
\begin{center}
\includegraphics[width=4cm]{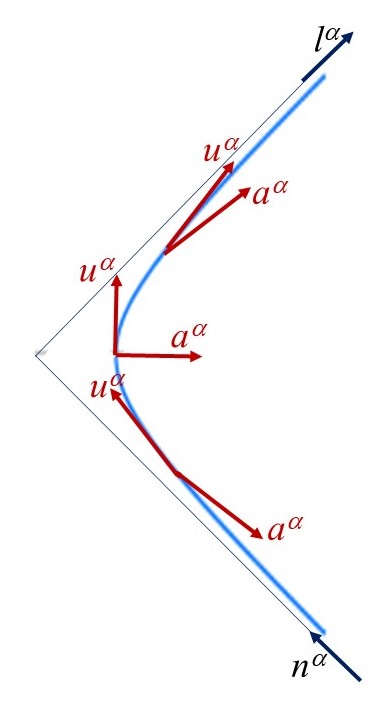}
\end{center}
\caption{As shown below, a particle with constant acceleration travels along a hyperbola, and it remains at a fixed spacetime distance from an origin (the origin is the intersection of the null asymptotes).}
\label{fig:acceleration}
\end{figure}

The relation $a^\alpha a_\alpha =$ constant $\equiv a^2$ has the form
\begin{align*}
a^2 &= \left(\dot{\lambda}\ell_\alpha - \frac{\dot{\lambda}}{\lambda^2}n_\alpha\right)
       \left(\dot{\lambda} \ell^\alpha - \frac{\dot{\lambda}}{\lambda^2}n^\alpha\right) 
     = -2\frac{\dot{\lambda^2}}{\lambda^2}\ell_\alpha n^\alpha \hspace{.5in} 
	\mbox{(use \eqref{llnn})} 
\\
    &= \frac{\dot{\lambda^2}}{\lambda^2}   \hspace{.5in} \mbox{(use \eqref{ln}).}
\end{align*}
Then
\[
	\frac{\dot{\lambda}}{\lambda} = \pm a, \ \mbox{ implying }  
	\lambda = \lambda_+\,e^{a\tau} + \lambda_-\, e^{-a\tau},
\]
for some constants $\lambda_+,\lambda_-$.  
We want $u^\alpha  \rightarrow k \ell^\alpha $ as $\tau \rightarrow \infty,\ \ \ \  u^\alpha 
\rightarrow k'n^\alpha$ as $\tau \rightarrow - \infty$, and so we pick $\lambda =
\lambda_+e^{a\tau}$, giving 
$u^\alpha  = \lambda_+e^{a\tau}\ell^\alpha+\lambda_+^{-1}e^{-a\tau}n^\alpha $.  
If, in addition, we fix the time $\tau$ at which $u^\alpha  = \ell^\alpha  + n^\alpha $ to be $\tau = 0$, then $\lambda_+ = 1$ and $u^\alpha(\tau)  = e^{a\tau}\ell^\alpha  + e^{-a\tau}n^\alpha $.  
After integrating and differentiating $u^\alpha(\tau))$, we have  
\be\cblue
 	c^\alpha(\tau) = \frac{1}{a} (e^{a\tau}\ell^\alpha  - e^{-a\tau}n^\alpha ), \qquad
	u^\alpha(\tau)  = e^{a\tau}\ell^\alpha  + e^{-a\tau}n^\alpha ~, \qquad
        a^\alpha(\tau) = a (e^{a\tau}\ell^\alpha  - e^{-a\tau}n^\alpha )\cb,
\ee
as shown above. Notice that the acceleration vector $a^\alpha$ at each point $c(\tau)$ of the trajectory is proportional to the spacelike vector $c^\alpha(\tau)$ from the origin to that point.

 In terms of the basis $\widehat t^\alpha ,\widehat x^\alpha $ with 
\[ \ell^\alpha  = \frac{\widehat t^\alpha +\widehat x^\alpha }{2} , \hspace{.5in} n^\alpha  = \frac{\widehat t^\alpha -\widehat x^\alpha }{2} ,\]
\be\cblue
  c^\alpha = \frac{1}{a}[\sinh(a\tau) ~\widehat t^\alpha 
			+ \cosh (a\tau)\widehat x^\alpha ], 
\qquad u^\alpha  = \cosh (a\tau)~ \widehat t^\alpha  + \sinh (a\tau) ~ \widehat x^\alpha,
\qquad a^\alpha = a^2\,c^\alpha\ \cb.
\ee
The path $c(\tau)$ is a hyperbola, whose $x$ and $t$ coordinates satisfy
\be 
 x^2 - t^2 = \frac{1}{a^2}(\cosh^2 a\tau - \sinh^2 a\tau) =
\frac{1}{a^2}. 
 \ee
That is, each point of the hyperbola in Fig.~\ref{fig:acceleration} is at the 
same spacetime distance $\dis s = \sqrt{c^\alpha c_\alpha}  = \frac1a$ from the origin!  The 
hyperbola is the Lorentzian analog of a circle. And  
the Euclidean analytic continuation to imaginary time $\mathpzc t = i\,t$ is  a circle whose Euclidean distance from the origin is $\mathpzc t^2+x^2= 1/a$.\\

\benr \item Assuming that people can't endure a
rocket acceleration much greater than $g$ for years at a time, the distance
a single generation of astronauts can travel is limited.  If your (proper)
acceleration is $g$ for half the trip, and you then decelerate at $g$ for
the second half, how far can you go in 20 years (on your clock)?  How long
does it take to reach Andromeda $(d = 2.2 \times 10^6$ ly)?

\item Prerequisite for the next section:\\
 (a) Verify the identity 
\[ 
\epsilon_{ijm}\epsilon_{klm} = \delta_{ik}\delta_{jl} - \delta_{il}\delta_{jk}
\]

(b) Using $({\bf A\times B})_i = \epsilon_{ijk}A_j B_k = \epsilon_{jki}A_j B_k$
and the identity of (a), show that \\
\phantom{xxxx} $\bf (A\times B)\cdot ( C\times D) 
			= A\!\cdot\! C\ B\!\cdot\! D - A\!\cdot\! D\ B\!\cdot\! C$.\\
(c) Similarly show that 
$\bf A\times (B\times C) = B\ A\!\cdot C\! - C\ A\!\cdot\! B$.
\een

\newpage

\section{Electromagnetism}
\label{s:em}\index{electromagnetism|(}\index{electromagnetism!flat spacetime|(}

\subsection{Gaussian-cgs units}\index{Gaussian units}\index{electromagnetism!Gaussian units}

With the electric and magnetic fields understood as parts of a single electromagnetic field, 
it is natural to choose units in which they have the same dimension:  These are Gaussian units.   
We first review Gaussian units in their usual form, keeping $c$, and then set $c$ to 1.  
In addition to giving $\bm E$ and $\bm B$ the same dimension, the units are chosen to 
give Coulomb's law the simple form,
\be
 F = \frac{q_1 q_2}{r^2}.
\label{e:Coulomb} \ee
\index{electromagnetism!Coulomb's law}\index{Coulomb's law}

To define the units, one just absorbs $\dis\frac1{4\pi\epsilon_0}$ into the definition of $q$: 
Using the subscripts $G$ and ${SI}$ for quantities in Gaussian and in SI units, we relate 
charge in Gaussian units to charge in $SI$ units by 
\be 
   q_G =  \frac{q_{SI}}{\sqrt {4\pi\epsilon_0}}. 
\ee 
\index{charge!Gaussian units}
The force between two charges then has the form \eqref{e:Coulomb} 
\[
   F = \frac{1}{4\pi\epsilon_0}\frac{q_{SI}^2}{r^2} \ 
     =  \frac{q_G^2}{r^2}. 
\]

With the above definition, charge no longer has an independent dimension: 
Instead, the force law implies a dimension in terms 
of powers of mass, length and time:
\begin{align*}
[F] &= [q^2]/L^2 \Rightarrow  [q]^2 = MLT^{-2} L^2 = ML^3 T^{-2}\\
\Rightarrow [q] &= M^{\frac12}L^{\frac32}T^{-1}.
\end{align*}

The Gaussian unit of charge is called an esu (electrostatic unit).  In cgs units, 
\[
   1\ \mbox{esu} \equiv 1 \ \mbox{ g}^{\frac12} \mbox{cm}^{\frac32}\mbox{s}^{-1}.
\] 
To relate esu to Coulombs, write 
\[
 \frac{1}{4\pi\ep_0} = 9\times10^9\ \rm Nm^2C^{-2} = 9\times10^9\ kg\ m^3 s^{-2} 
C^{-2},
\]
Then 
\begin{align*}
\rm \frac{1\ C}{\sqrt{4\pi\ep_0}}&=\rm  1\ C \ \sqrt{9\times10^9 kg\ m^3 s^{-2} C^{-2}}\\
      &=\rm  3\times10^9\ g^\frac12 cm^{\frac32} s^{-1} = 3\times10^9\ \mbox{esu}
\end{align*}
Then a charge of 1 C is a charge of $3\times10^9$ esu.

The electric field $E$ is as usual defined as force per unit 
charge, now with dimension 
\[
[E] = [F/q] = \frac{MLT^{-2}}{M^\frac12 L^{\frac32}T^{-1}} = M^\frac12 L^{-\frac12} 
T^{-1}.
\]
\index{electric field!Gaussian units}
In Gaussian units the unit of electric field is thus 1 g$^\frac12$ cm$^{-\frac12}$ 
s$^{-1}$.  
The unit of magnetic field (1 Gauss) is the same as that of the electric field:
\index{magnetic field!Gaussian units}
\[
   1 G = 1\ \mbox{g}^{1/2} \mbox{ cm}^{3/2} \mbox s^{-1}
\]

Finally, Maxwell's equations and the Lorentz force law have the form 
\bsube\begin{align}
\bm\nabla \cdot \bm B &= 0\qquad\quad\quad\ 
	\frac1c\partial_t{\bm B} + \bm\nabla\times{\bm E}  = 0,
\label{e:Maxwella}\\
\bm\nabla\cdot {\bm E}& = 4\pi\rho \qquad 
	- \frac1c\partial_t{\bm E} + \bm\nabla\times{\bm B}  = \frac{4\pi}c {\bm j},
\label{e:Maxwellb}\end{align}\label{e:Maxwell0}\esube
\index{electromagnetism!Maxwell's equations}\index{Maxwell's equations}
\be
{\bm F} = q({\bm E} + \frac{{\bm v}}{c}\times{\bm B}).
\ee
\index{Lorentz force}\index{force!Lorentz}

\subsection{The 4-dimensional electromagnetic (Faraday) tensor}
\index{electromagnetism! electromagnetic (Faraday) tensor} \index{Faraday tensor} 

A spinless charged particle at rest has only an electric field.  
Boosting the particle gives a current, so boosting $\bm E$ gives an 
electromagnetic field with nonzero $\bm B$.  
Similarly, Faraday's experiment with a moving magnet, illustrated in the cartoon of 
Fig.~\ref{f:magnet}, implies that the boosted magnetic field of a boosted magnet
is an electromagnetic field with nonzero $\bm E$.    
Equivalently, in passive terminology, as seen by a boosted observer, a pure electric field 
and a pure magnetic field are each superpositions of electric and magnetic fields.

There is, however, a difficulty in saying that $\bm B$ and $\bm E$ are parts of 
a single electromagnetic field: $\bm B$ is a pseudovector under reflection. That is,
its sign after reflection is opposite to the the sign of a vector under 
reflection. \index{magnetic field!as pseudovector and two-index tensor} 

\begin{figure}[H]
 \centerline{\includegraphics[width=0.3\textwidth]{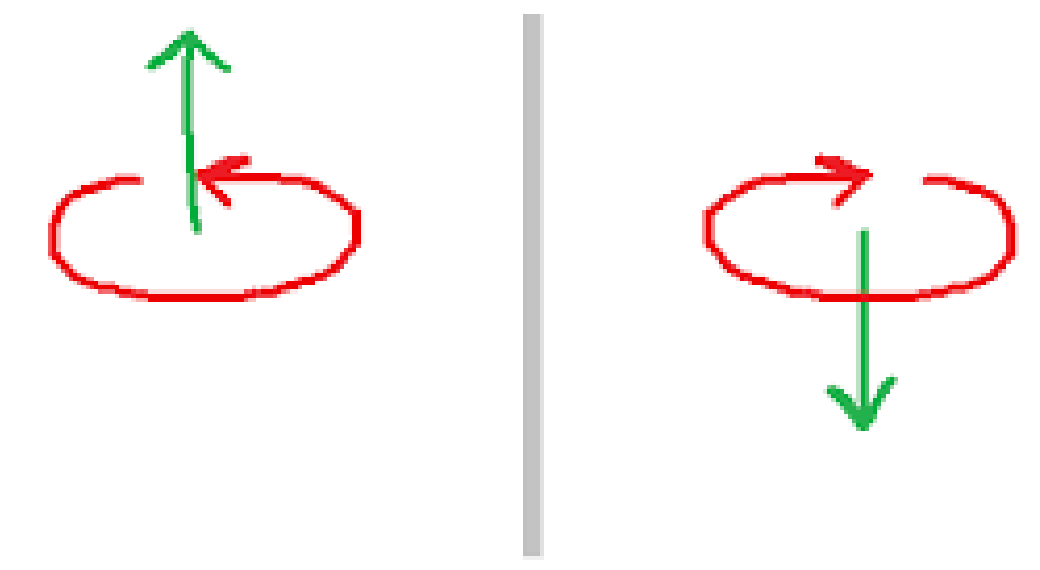}}
\caption{The magnetic fields of a current loop and of the reflected loop (both in 
red) are shown as green arrows parallel to a mirror. Because the magnetic fields 
point in opposite directions, $\bm B$ is a pseudovector.}  
 \end{figure}
How can $\bm E$ and $\bm B$ be two parts of a single field? The answer is that 
$\bm B$ is not naturally a vector but a two-index tensor:
 \be
 \cblue	B^a = \frac12 \epsilon^{abc} B_{bc}, \qquad B_{ab} = \epsilon_{abc} B^c\qquad B_{ab} = B_{[ab]}
\label{e:Bab}\ee
 The tensor $\epsilon_{abc}$ changes sign under reflection, so 
 $B_{ab}$ is an ordinary tensor, not a pseudotensor.

In an orthonormal frame, 
 \begin{align}
    B_{12} &= B^3 = B_3, \quad B_{23} = B_1,\quad B_{31}= B_2 \nonumber\\
    [B_{ij}] & = \begin{bmatrix}
    		0 & B_3 & -B_2 \\
    		-B_3 & 0 & B_1\\
    		B_2 & -B_1 & 0
 	    \end{bmatrix}
 \end{align}
The antisymmetric spatial tensor $B_{ab}$ must be part of an antisymmetric 
spacetime tensor $F_{\a\b} = F_{[\a\b]}$, and happily the number of components 
is right:  
\begin{eqnarray*}
\begin{array}{rl} \bm E &{\mbox{3 independent
components}}\vspace{1mm}\\
\bm B & 3 {\mbox{ independent components}}\vspace{1mm}\\
F_{\a\b}  {\mbox{ an antisymmetric tensor}} &6 {\mbox{ independent components}}
\end{array}
\end{eqnarray*} That is, a 4x4 antisymmetric matrix has 6 independent components.  
We have $F_{ij} = B_{ij}$ for the spatial components. 
The remaining nonzero components are $F_{0i}= -F_{i0}$, so the 3-vector with components
$F_{0i}$ must be proportional to $E_a$. \\

We fix the constant of proportionality by looking at the two sourcefree Maxwell equations. We have  
\[
0 = \na_a B^a = \ep^{abc}\na_a B_{bc}\ \mbox{ or }\ \na_{[a} B_{bc]} =0.
\]   
Because $F_{ij} = B_{ij}$, the equation $0 = \bm\nabla\cdot \bm B$ has components
$\na_{[i} F_{jk]} = 0$. The spacetime equation must then be 
\[
  0 = \na_{[\a} F_{\b\c]}, 
\]
whose $012$ component (multiplied by three) is 
\[
0 = \pa_0 F_{12} + \pa_1 F_{20} + \pa_2 F_{01} 
	= \pa_0 B_3 + \pa_1 {F_{20}} + \pa_2 F_{01} .
\]
The second sourcefree Maxwell equation is $\pa_0 \bm B + \bm\na\times\bm E = 0$, with 
component
$
  0 = \pa_0 B_3 + \pa_1 E_2 - \pa_2 E_1,  
$
implying $F_{20} = E_2$, $F_{01} = -E_1$.  We then have, for the components 
of $F_{\a\b}$ in an orthonormal frame, 
\be
\|F_{\mu\nu}\| =
\left\|\begin{matrix}0   &-E_1 & -E_2 &  -E_3  \\
	       E_1 & 0  & B_3 & -B_2  \\
	       E_2 & -B_3  & 0 & B_1  \\
	       E_3 &  B_2  & -B_1 & 0  \\
\end{matrix}\label{e:Fmunu}\right\| .
\ee
This is MTW (3.7).  We now return to our frame-independent formalism, with an arbitrary 
unit vector $t^\a$ used for $3+1$ decompositions.  

\subsection{\texorpdfstring{$F_{\alpha\beta} , j^\alpha  , E^\alpha  , B^\alpha,\ \rho,\ \ A_\alpha$} .}

{\em Conventions in these notes agree with those of MTW and Wald. 
Jackson \cite{jackson98}, however, uses a $+---$ signature, 
and his field tensor differs from ours by a sign: 
$F^{\alpha\beta} = - F_{\rm Jackson}^{\alpha\beta}$.}\\

With the electromagnetic tensor in hand, we summarize the spacetime form 
of electrodynamics and then check agreement with the 
Lorentz force law and with the two remaining
Maxwell equations, the Maxwell equations with source.  \\ 

Maxwell's equations are 
\index{charge!charged particle}\index{charge!charge current density}\index{current density}
\index{density!charge, current}
\bsube
\begin{align} \crv
	\nabla_{[\alpha}F_{\beta\gamma]} &\crv =0 
\label{e:dF}\\
\crv
\nabla_\beta   F^{\alpha\beta} &\crv=4\pi j^\alpha, 
\label{e:divF}\end{align}\label{e:Maxwell}\esube
\index{Maxwell's equations!in terms of $F_{\a\b}$}
\index{electromagnetism!Maxwell's equations in terms of $F_{\a\b}$}
where $j^\alpha $ is the current density, a vector field that depends 
only on the charged matter.  The motion of a charged particle is described 
by a timelike trajectory that satisfies the equation of motion 
(Lorentz force law)
\index{equation of motion!Lorentz force law}\index{Lorentz force law}
\be \crv
	u^\beta \nabla_\beta   p^\alpha  = q\,F^\alpha {}_\beta   u^\beta  , \cb
\label{lorentz}\ee
where $p^\alpha  = mu^\alpha $, and $q$ is the particle's charge.

	If an observer with velocity $t^\alpha $ measures the components of 
$F_{\alpha\beta} $ along vectors parallel or orthogonal to $t^\alpha $
(his natural basis vectors) he will be splitting $F_{\alpha\beta} $ in
the way described in the last section and by Eq.~(\ref{3+1}).   Because $F_{\alpha\beta}$ is antisymmetric, 
its diagonal components vanish:  In particular, \mbox{$F_{\alpha\beta} t^\alpha t^\beta=F_{tt}=0$}. 
(That is, $F_{tt} = -F_{tt} \Longrightarrow F_{tt} = 0$.)
\footnote{More generally, if $S^{\alpha\beta}$ is any symmetric tensor, and $A_{\alpha\beta}$ 
is antisymmetric, $A_{\alpha\beta}S^{\alpha\beta} =0$.  
To see this, use $A_{\alpha\beta}  =-A_{\beta\alpha}$ to write \\
\centerline{
$A_{\alpha\beta} S^{\alpha\beta} = -A_{\beta\alpha}S^{\alpha\beta} 
		=-A_{\beta\alpha} S^{\beta\alpha} 
	        =-A_{\alpha\beta} S^{\alpha\beta} 
 \Longrightarrow \ A_{\alpha\beta} S^{\alpha\beta}=0$.}  
The same manipulation shows that, in general,
$A^{\alpha\beta\gamma\cdots\delta}$ antisymmetric in $\alpha$ and $\beta$ and 
$S^{\alpha\beta\gamma\cdots\delta}$
symmetric in $\alpha$ and $\beta$ together imply 
$A^{\alpha\beta\gamma\cdots\delta}S_{\alpha\beta} {}^{\epsilon\cdots\zeta}=0$.}
  
Furthermore, 
$\g_\alpha{}^\beta t^\gamma  F_{\beta\gamma} 
	= (\delta^\alpha {}_\beta+t^\alpha  t^\beta)t^\gamma  F_{\beta\gamma} 
	= F_{\alpha\gamma}t^\gamma  $ 
		(again using $F_{\beta\gamma}t^\beta t^\gamma = 0$) and 
$\g_\alpha{}^\gamma  t^\beta   F_{\beta\gamma} =-F_{\alpha\gamma}t^\gamma $, 
so the only independent parts of the decomposed $F_{\alpha\beta} $ are
\be
 F_{\alpha\beta} t^\beta   \hspace{2mm} {\text{and}}\hspace{2mm} 
 \g_\alpha {}^\gamma  \g_\beta {}^\delta    F_{\gamma\delta} =: F^\perp_{\alpha\beta}  . 
\ee
The spatial 4-vector 
\index{electric field!from electromagnetic tensor $F_{\a\b}$} 
\be 
	E_\alpha := F_{\alpha\beta} t^\beta  
\label{electric}\ee
\index{electric field}
is called the electric field, and the antisymmetric spatial tensor $F^\perp_{\alpha\beta} $ corresponds to the magnetic field $B^\alpha $, also a spatial vector: \index{spatial vector!in 3+1 decomposition}
\begin{eqnarray}
B^\alpha  &:=& \frac{1}{2} \epsilon^{\delta\alpha\beta\gamma}t_\delta F^\perp_{\beta\gamma}\nonumber\\
&=& \frac{1}{2} \epsilon^{\delta\alpha\beta\gamma}t_\delta 
	(\delta_\beta^\epsilon +t_\beta t^\epsilon )(\delta_\gamma^\zeta  +t_\gamma  t^\zeta 
)F_{\epsilon\zeta}\nonumber\\
&=& \frac{1}{2} \epsilon^{\delta\alpha\beta\gamma}t_\delta   F_{\beta\gamma} , 
\label{magnetic}\end{eqnarray} 
where we have again used the index antisymmetry, in this case of  
$\epsilon^{\alpha\beta\gamma\delta} $ to infer $\epsilon^{\delta\alpha\beta\gamma}t_\delta t_\beta=0$.

	Note that
\be 
\epsilon_{\alpha\beta\gamma}  \equiv \epsilon_{\delta\alpha\beta\gamma}t^\delta   
\label{e:3epsilon}\ee
\index{magnetic field!from electromagnetic tensor $F_{\a\b}$}

\noindent is an antisymmetric spatial tensor (each index orthogonal to
$t^\alpha $) and $\epsilon_{\alpha\beta\gamma} \epsilon^{\alpha\beta\gamma} =3!$. In the orthonormal basis
$t^\alpha $, $\widehat x^\alpha $, $\widehat y^\alpha $, $\widehat z^\alpha $, $\epsilon _{123}=\epsilon_{0123}=1$ so the
spatial components of $\epsilon_{\alpha\beta\gamma} $ are
\[ \epsilon_{\lambda\mu\nu} = \left\{ \begin{array}{rll} 
	1,   &\lambda\mu\nu &{\mbox{an even permutation of 123}}\\ 
	-1,  &\lambda\mu\nu &{\mbox{an odd permutation of 123}} ,
	\end{array}\right.
\]
and $\epsilon_{\lambda\mu\nu} =0$ if any index is repeated or has the
value 0.

The equation defining $B^\alpha$ in terms of $F^{\alpha\beta}$ is then
\[ 
	B^\alpha = \frac12\epsilon^{\alpha\beta\gamma} F_{\beta\gamma}. 
\]

Some useful identities involving $\epsilon_{\alpha\beta\gamma} $ and $\epsilon
_{\alpha\beta\gamma\delta} $ are
\index{epsilon identities}
\begin{eqnarray}
\begin{array}{ll}\epsilon^{\alpha\beta\gamma\delta} \epsilon_{\epsilon\zeta\eta\theta} 
   = -4!0!\delta^{[\alpha}_{\;\epsilon}\delta^\beta_\zeta\delta^\gamma_\eta\delta_\theta^{\delta]} 
  &= -0! \sum\limits_{\pi\in{\mathbb P}_4} (-1)^\pi 
		\delta_\epsilon^{\pi(\alpha )}\ldots\delta_\g^{\pi(\delta)}\nonumber \\
\epsilon^{\alpha\beta\gamma\delta} \epsilon_{\epsilon\zeta\eta\delta} 
   = -3!1!\delta^{[\alpha}_{\;\epsilon}\delta^\beta_\zeta\delta^{\gamma]}_\eta	
  &= -1! \sum\limits_{\pi\in{\mathbb P}_3} (-1)^\pi  		
     \delta_\epsilon^{\pi(\alpha)} \ldots\delta_\eta^{\pi(\gamma)}\nonumber \\
\epsilon^{\alpha\beta\gamma\delta} \epsilon_{\epsilon\zeta\gamma\delta} 
	  = -2!2!\delta^{[\alpha}_{\;\epsilon}\delta_\zeta^{\beta]} 
	&=-2! \sum\limits_{\pi\in{\mathbb P}_2} (-1)^\pi
		\delta_\epsilon^{\pi(\alpha)} \delta_\zeta^{\pi(\beta)}\nonumber \\
\epsilon^{\alpha\beta\gamma\delta} \epsilon_{\epsilon\beta\gamma\delta} 
	=-1!3!\delta_\epsilon^\alpha  &=  -3!\delta_\epsilon^\alpha \nonumber \\
\epsilon^{\alpha\beta\gamma\delta} \epsilon_{\alpha\beta\gamma\delta}  
	= -0!4! &= -4! 
\end{array}\end{eqnarray}

\begin{eqnarray}
\begin{array}{ll} \epsilon^{\alpha\beta\gamma} \epsilon_{\delta\epsilon\zeta} 
	   = 3!0! \g^{[\alpha}_{\;\delta} \g^\beta_\epsilon \g_\zeta^{\gamma]} 
	&= 0! \sum\limits_\pi (-1)^\pi \g_\delta  ^{\pi(\alpha)} \g_\epsilon^{\pi(\beta )}\g_\zeta^{\pi(\gamma)}
									\vspace{1mm}\\
\epsilon^{\alpha\beta\gamma} \epsilon_{\delta\epsilon\gamma} = 2!1! \g_\delta  ^{[\alpha}\g_\epsilon^{\beta]} 
	&= 1!\sum\limits_\pi (-1)^\pi \g_\delta^{\pi(\alpha)} \g_\epsilon^{\pi(\beta )}\\
\epsilon^{\alpha\beta\gamma} \epsilon_{\delta\beta\gamma} = 1!2! \g_\delta  ^\alpha  
	&= 2! \g_\delta^\alpha \\
\epsilon^{\alpha\beta\gamma} \epsilon_{\alpha\beta\gamma} = 0!3! &= 3!\end{array} 
\end{eqnarray}
where $\pi\in {\mathbb P}_4$ is a permutation of the four letters $\alpha\beta\gamma\delta$, 
with $\pi\in {\mathbb P}_3$, $\pi\in {\mathbb P}_2$ defined analogously. (Here 
$\mathbb P_n$ is the group of permutation of $n$ objects.)

\vspace{2mm}	
	
  The relations (\ref{electric}) and (\ref{magnetic}) for $E^\alpha $ and 
$B^\alpha $ in terms of $F^{\alpha\beta} $ can be inverted to give
\be \cblue
F_{\alpha\beta}  = t_\alpha  E_\beta   -t_\beta  E_\alpha   
		  + \epsilon_{\alpha\beta\gamma} B^\gamma,\cb
\label{e:FEB}\ee
equivalent to the matrix equation \eqref{e:Fmunu}.
To show this, first note that
\be
	B^\alpha  = \frac{1}{2} \epsilon^{\alpha\beta\gamma}  F_{\beta\gamma}
\label{magnetic1}\ee
can be inverted as in Eq.\eqref{e:Bab}
\[
   B_{\a\b} = \epsilon_{\a\b\g} B^\g,
\] 
with $B_{\a\b}$ the spatial part of $F_{\a\b}$, 
\[
 B_{\a\b} := \g^\gamma_\alpha \g^\delta_\beta  F_{\gamma\delta}.
\] 
Check:   
\[ 
\epsilon_{\alpha\beta\gamma} B^\gamma 
	= \epsilon_{\alpha\beta\gamma}
	    \left(\frac{1}{2}\epsilon^{\gamma\delta\epsilon} F_{\delta\epsilon}\right) 
 =\frac{1}{2}(\g^\delta_\alpha \g^\epsilon_\beta 
  -\g^\epsilon_\alpha \g^\delta_\beta)F_{\delta\epsilon} 
  = \g^\gamma_\alpha \g^\delta_\beta  F_{\gamma\delta},
\]
where antisymmetry of $F_{\alpha\beta} $ was used to obtain the last equality. 
\be 
	F^\perp_{\alpha\beta}  = \epsilon_{\alpha\beta\gamma} B^\gamma .  
\ee
$F^\perp_{\alpha\beta}$ is the spatial tensor equivalent to $B_{ab}$ in the last section.  
Its nonzero components are $F^\perp_{ij} = B_{ij}$. \\  \\
Thus 
\begin{align*}
t_\alpha  E_\beta - t_\beta E_\alpha + \epsilon_{\alpha\beta\gamma} B^\gamma  
 &= t_\alpha F_{\beta\gamma}t^\gamma -t_\beta F_{\alpha\gamma}t^\gamma 
  + \g_\alpha{}^\gamma \g_\beta{}^\delta F_{\gamma\delta}\\
&= \cancel{t_\alpha  F_{\beta\gamma}t^\gamma }{-25} 
  - \cancell{t_\beta F_{\alpha\gamma}t^\gamma}{-25} + F_{\alpha\beta}  
  + \cancel{t_\alpha   t^\gamma  F_{\gamma\beta}}{-25} 
  + \cancell{t_\beta  t^\delta F_{\alpha\delta}}{-20} 
  + \underbrace{t_\alpha t^\gamma  t_\beta  t^\delta F_{\gamma\delta}}_0 \\ 
&= F_{\alpha\beta}  .
\end{align*} 

	Finally, before writing Maxwell's equations and the Lorentz force law in
terms of $E^\alpha $ and $B^\alpha $, we decompose $j^\alpha $:
\be 
	j^\alpha  = \rho_e t^\alpha  + j_\perp^\alpha  , 
\label{j3+1}\ee
\index{charge!charge density}\index{charge!charge current density}\index{current density}
where $\rho_e$ is the charge density and $j_\perp^\alpha $ the 3-current seen by the
observer with velocity $t^\alpha $.  

For example, the current $j^\alpha$ associated with charges of density $\rho_e$ in their own basis, 
whose velocity is $u^\alpha$, is $j^\alpha = \rho_e u^\alpha$.  In terms of a basis with 
timelike unit vector $\wt t^\alpha$, the 3+1 decomposition of $u^\alpha$ is 
 $ u^\alpha = \gamma \wt t^\alpha + u_\perp^\alpha$, or
\be 
	u^\alpha  = \gamma(\wt t^\alpha  + \wt v^\alpha  ) , \mbox{ implying } 
	\widetilde\rho_e := -j_\alpha \wt t^\alpha = \gamma \rho_e.
\label{utv}\ee

\subsection{Maxwell and Lorentz}
	Let us now regard $t^\alpha$ as the constant timelike vector field
obtained by parallel transporting the observer's velocity to all points of
$M$, so that $E^\alpha $ and $B^\alpha $ have meaning everywhere.

	The expression $F^\alpha{}_\beta   u^\beta$ occurring in the Lorentz force law can
now be written (using (\ref{e:FEB}) and (\ref{utv}) as
\begin{eqnarray*}
F^\alpha {}_\beta   u^\beta  
 &=& [t^\alpha  E_\beta   -t_\beta   E^\alpha  
		+ \epsilon^\alpha {}_{\beta\gamma} B^\gamma ]\gamma (t^\beta  +v^\beta  )\\
 &=& \gamma(E^\alpha  + \epsilon^\alpha {}_{\beta\gamma} v^\beta B^\gamma + v^\beta  E_\beta  t^\alpha )
\end{eqnarray*}
 
Project (\ref{lorentz}) along $t^\alpha $, denoting the particle's energy, $-p_\alpha t^\alpha$, by E to 
distinguish it from the magnitude $E$ of the electric field:
\[ t_\alpha   u^\beta \nabla_\beta   p^\alpha  = t_\alpha q F^\alpha{}_\beta  u^\beta \] 
\[ u^\beta \nabla_\beta   (-{\rm E}) =-q\gamma E_\beta   v^\beta  , \hspace{3mm} {\text{or}}\]
\be 
	\frac{d{\rm E}}{dt} = qE_\beta   v^\beta  , 
\label{dedt}\ee 
using $\dis u\cdot\nabla {\rm E} =\frac{d{\rm E}}{d\tau} 
	= \gamma \frac{d{\rm E}}{dt}$.  Eq. (\ref{dedt})
says that the particle's energy E changes by the work $q {\bf v} \cdot{\bm E} $
done per unit time by the field.  (In this last equation, $t$ has replaced 
$\tau$ as the path parameter along the particle's trajectory).  \\

Project (\ref{lorentz}) orthogonal to $t^\alpha $:
\[ 
\g^\alpha{}_\gamma\,  u^\beta \nabla_\beta   p^\gamma  
  = q\g^\alpha {}_\gamma  F^\gamma {}_\beta  u^\beta  
  = q\g^\alpha {}_\gamma\, \gamma 
    (E^\gamma +\epsilon^\gamma{}_{\beta\delta} v^\beta  B^\delta )
\] 
Using $\g^\alpha {}_\gamma  $ constant,  $\nabla_\beta   \g^\alpha {}_\gamma  =0$,
we have
\[ 
  u^\beta \nabla_\beta  p_\perp^\alpha  = q\gamma\,(E^\alpha  
	+\epsilon^\alpha {}_{\beta\gamma} v^\beta B^\gamma );
\]
using $d/d\tau = \gamma d/dt$ then gives
\be 
\frac{d}{dt} p_\perp^\alpha  
  = q(E^\alpha  +\epsilon^\alpha {}_{\beta\gamma} v^\beta B^\gamma );
\ee 
\index{force!Lorentz}\index{Lorentz force}
this is the usual Lorentz force law, 
\be
   \frac{d}{dt} {\bm p}_\perp 
  = q(\bm E  +\bm v\times\bm B).
\ee
\index{Lorentz force}\index{electromagnetism!Lorentz force} 

	Maxwell's equations in terms of $E^\alpha $ and $B^\alpha $ emerge in an analogous
way. 
The projection of $\nabla_\beta  F^{\alpha\beta}  = 4\pi j^\alpha $ along $t^\alpha$ is
\[ 
t_\alpha  \nabla_\beta  F^{\alpha\beta}  = 4\pi t_\alpha   j^\alpha  , 
	\hspace{3mm} \mbox{or, by (\ref{j3+1})}
\]
\be \nabla_\beta   E^\beta  = 4\pi\rho_e \ee 
\[
	\mbox{i.e., }\ \nabla\cdot{\bm E} = 4\pi\rho_e\,.
\]  
The projection orthogonal to $t^\alpha $ is, using (\ref{e:FEB}), 
\begin{align}
 \nabla_\beta [\g^\alpha{}_\gamma  
	  (t^\gamma E^\beta-t^\beta E^\gamma+ \epsilon^{\gamma\beta\delta}B_\delta)] 
	&= 4\pi j_\perp{}^\alpha \vspace{1mm}\\
-t^\beta  \nabla_\beta  E^\alpha + \epsilon^{\alpha\beta\gamma} \nabla_\beta  B_\gamma   
	&=  4\pi j_\perp{}^\alpha ;\vspace{1mm}\\
{\text{i.e.},}\; - \frac{\partial{\bm E} }{\partial t} +\bm \nabla\times\bm B
&=  4\pi\bm j.\nonumber
\end{align}
We have already seen that the equation $\nabla_{[\alpha}F_{\beta\gamma]} =0$ is equivalent to
\be 
\epsilon^{\alpha\beta\gamma\delta}  \nabla_\beta   F_{\gamma\delta} = 0 . 
\label{*df}\ee 
For completeness, here's a check using our covariant formalism.  
Projecting (\ref{*df}) along $t^\alpha  $ gives $\bm\nabla\cdot \bm B=0$:
\begin{eqnarray}
0 &=& \nabla_\beta   [t_\alpha   \epsilon^{\alpha\beta\gamma\delta}  (t_\gamma  E_\delta   -t_\delta    E_\gamma   + \epsilon_{\gamma\delta\epsilon}
B^\epsilon )]\nonumber\\
&=& \nabla_\beta  (\g^\beta {}_\epsilon B^\epsilon ) \hspace{15mm} (\epsilon^{\beta\gamma\delta}\epsilon_{\epsilon\gamma\delta} =
\g^\beta {}_\epsilon ) \nonumber\\
&=& \nabla_\beta  B^\beta  . 
\end{eqnarray} 
Projecting orthogonal to $t^\alpha$ gives $\dis \frac{\partial\bm B}{\partial t} + \bm\nabla\times \bm E = 0$: 
\begin{align*}
 0 & = \epsilon^{\zeta\beta\gamma\delta} \nabla_\beta
	   [t_\gamma E_\delta -t_\delta E_\gamma 
		+\epsilon_{\gamma\delta\epsilon}B^ \epsilon]\g^\alpha{}_\zeta  
\\
  &=  \epsilon^{\alpha\beta\delta}\nabla_\beta  E_\delta
    + \epsilon^{\alpha\beta\gamma} \nabla_\beta E_\gamma 
    - 2(\delta^\zeta _\eta\delta^\beta_\epsilon-\delta^\zeta_\epsilon\delta^\beta_\eta)     
        \nabla_\beta  B^\epsilon t^\eta \g^\alpha{}_\zeta 
\\
  &= 2\epsilon^{\alpha\beta\gamma} \nabla_\beta  E_\gamma   
    + 2t^\beta \nabla_\beta  B^\alpha
\end{align*}
or 
\be
t^\beta \nabla_\beta  B^\alpha  + \epsilon^{\alpha\beta\gamma} \nabla_\beta  E_\gamma  =0.
\ee
Collecting the equations, we have the equivalent, with spacetime indices, of Eqs.~\eqref{e:Maxwell0}
\index{Maxwell's equations|textbf}\index{electromagnetism!Maxwell's equations|textbf}
\cblue\begin{eqnarray}
\begin{array}{ll}  
\nabla_\beta  B^\beta &=0 \qquad -t^\beta \nabla_\beta  B^\alpha  + \epsilon^{\alpha\beta\gamma} \nabla_\beta  E_\gamma  =0\\
\nabla_\beta  E^\beta  &= 4\pi\rho_e \qquad t^\beta \nabla_\beta  E^\alpha  +\epsilon^{\alpha\beta\gamma} \nabla_\beta  B_\gamma  =4\pi j_\perp{}^\alpha 
 \end{array} 
\label{e:Maxwell4}\end{eqnarray}\cb 
\newpage

\noindent 
\benr\item
Let $^\ast F^{\alpha\beta}  = \frac{1}{2} \epsilon
^{\alpha\beta\gamma\delta}  F_{\gamma\delta}$.
\ben
\item[(a)] Show that if $F_{\alpha\beta} $ satisfies the sourcefree Maxwell
equations, so does \\
$\widehat F_{\alpha\beta}  = F_{\alpha\beta} \cos\eta 
+{}^*F_{\alpha\beta} \sin\eta$.
($F\rightarrow \widehat F$ is called a duality rotation).
\item[(b)]  If $\eta = \frac{\pi}{2}$, show that (for any $t^\alpha $) $\widehat
B^\alpha  = +E^\alpha $, $\widehat E^\alpha  =-B^\alpha $.
\item[(c)]  Show that, for any $\eta$, 
\[ 	\widehat B^\alpha  = B^\alpha\cos\eta + E^\alpha \sin\eta\qquad 
	\widehat E^\alpha  = E^\alpha\cos\eta - B^\alpha\sin\eta  \]
\index{duality!duality rotation of electromagnetic field}
\index{electromagnetism!duality rotation}
\een

\een

\subsection{Conservation of Charge}\index{charge!charge conservation}\index{conservation laws!charge}\index{electromagnetism!charge conservation}

Because the tensor $F^{\alpha\beta} $ is antisymmetric,
\[ \nabla_\alpha  \nabla_\beta   F^{\alpha\beta}  = \frac{1}{2} [\nabla_\alpha  \nabla_\beta  
-\nabla_\beta  \nabla_\alpha   ]F^{\alpha\beta} =0\]
That is, in natural coordinates on $M$,
\[ 
  \nabla_\mu  \nabla_\nu   F^{\mu\nu}  
	= \frac{\partial}{\partial x^\mu}\frac{\partial}{\partial x^\nu} F^{\mu \nu}, \hspace{3mm} {\text{and}}\hspace{3mm} 
  (\partial_\mu \partial_\nu -\partial_\nu \partial_\mu )F^{\mu \nu} = 0,
\]
for smooth $F^{\mu \nu}$.  But
\[ \nabla_\alpha  \nabla_\beta   F^{\alpha\beta}  = \nabla_\alpha   (4\pi j^\alpha  ), \hspace{3mm}
{\text{whence}}\]
\be \crv
	\nabla_\alpha   j^\alpha  = 0\cb. 
\label{e:chargecons0}\ee 
	The equation $\nabla_\alpha   j^\alpha  =0$ is called conservation of charge.  In terms
of the charge $\rho_e$ and current $j_\perp^\alpha $ of an observer with velocity 
$t^\alpha $, it can be written 
\be 
 \nabla_\alpha (\rho_e u^\alpha + j_\perp{}^\alpha) 
	= u^\alpha  \nabla_\alpha   \rho_e + \nabla_\alpha  j_\perp{}^\alpha  = 0,
\ee
or, more familiarly, 
\be 
	\partial_t\rho_e  + \bm \nabla\cdot\bm j_\perp = 0.
\label{e:chargecons}\ee
The corresponding integral conservation law is obtained by integrating \eqref{e:chargecons0} \ ($\equiv$ \eqref{e:chargecons}) 
over a 4-volume sandwiched between two $t$ = constant surfaces:
\begin{align*}
   0 &= \int_{t_1}^{t_2} dt\int dxdydz 
		\left(\partial_t\rho_e + \bm\nabla\cdot\bm j_\perp\right) \\
     &= \int_{t=t_2} dxdydz \rho_e -\int_{t=t_1} dxdydz\rho_e
       +\int_{t_1}^{t_2} dt 
		\int_{\substack{\rm spatial\\\rm infinity}} \bm j_\perp\cdot d{\bf A}\,.
\end{align*} 
Thus if $\bm j_\perp=0$ at spatial infinity (e.g., if the matter vanishes
outside some finite region) the total charge $Q$ is conserved from one
hypersurface $t=t_1$ to another $(t=t_2)$: 
\[
	Q = \int_{t=t_2} \rho_e  dV = \int_{t=t_1} \rho_e  dV.
\]
\subsection{Vector Potential}
\index{electromagnetism!vector potential}\index{vector potential}

The 3-dimensional vector potential $A_a$ for the magnetic field and the scalar potential $\Phi_e$ for the electric field come from the 3+1 decomposition of a 4-dimensional vector potential $A_\a$.  That is, the sourcefree Maxwell equation \eqref{e:dF}, $\nabla_{[\alpha}F_{\beta\gamma]}  =0 $, implies that the electromagnetic field $F_{\a\b}$ can be written in terms of a vector potential $A_\a$, 
\be
  F_{\a\b} = \na_\a A_\b - \na_\b A_\a.  
\label{e:FdA}\ee 
The 3+1 decomposition of $A_\a$ associated with the unit timelike vector 
$t^\a$ of an observer has the form 
\be
   A_\a = A^\perp_\a + \Phi_e t_\a, \quad \mbox{ where }\quad A^\perp_\a  := \gamma_\a{}^\b A_\b, \quad \Phi_e := - A_\alpha t^\a. 
\ee
Then Eqs.~\eqref{e:FEB} and \eqref{e:FdA} imply
\bsube\begin{align}
  B^\a &= \epsilon^{\a\b\c}\na_\b A^\perp_\c, \qquad E_\a = -\na_\a \Phi_e - \pa_t A^\perp_\a, 
  \quad \mbox{ or }\\
   \bm B &= \bm\na\times \bm A, \quad\qquad \bm E = - \bm \na \Phi_e -\pa_t \bm A,  
\end{align}\esube
where $\bm A$ here means the 3-dimensional vector potential $A_a$.  
\index{electromagnetism|)}\index{electromagnetism!flat spacetime|)}  
\newpage

\section{Continuous Matter} \index{continuous matter}
 
\subsection{The stress tensor and the stress-energy tensor}
\index{stress-energy tensor|(}
See Feynman, {\sl Lectures on Physics v. II}, Chap. 31, 40. \\
\url{https://www.feynmanlectures.caltech.edu/II_toc.html}; \\
Schutz Chap. 4;\\
Blandford and Thorne, {\sl Modern Classical Physics}\cite{thorneblandford} , Chap. 1 \\
\url{http://www.pmaweb.caltech.edu/Courses/ph136/yr2012/}.\\

We start with the stress tensor.  \index{stress tensor|(} 
Here's the introduction to it from Blandford-Thorne: \\
\begin{quote}
Press your hands together in the $y$-$z$ plane and feel the force that one hand exerts on the
other across a tiny area $A$ -- say, one square millimeter of your hands' palms [figure below].
That force, of course, is a vector $\bm F$. It has a normal component (along the $x$ direction). It
also has a tangential component: if you try to slide your hands past each other, you feel a
component of force along their surface, a “shear” force in the $y$ and $z$ directions. Not only is
the force $\bm F$ vectorial; so is the 2-surface across which it acts, $\bm\Sigma=A \bm e_x$. (Here $\bm e_x$ is the unit vector orthogonal to the tiny area $A$, and we have chosen the negative side of the surface to be the $-x$ side and the positive side to be $+x$. With this choice, the force $\bm F$ is that which
the negative hand, on the $-x$ side, exerts on the positive hand.)
\end{quote}\vspace{-5mm}

\begin{figure}[h!]
		\begin{center}
                \includegraphics[width=4cm]{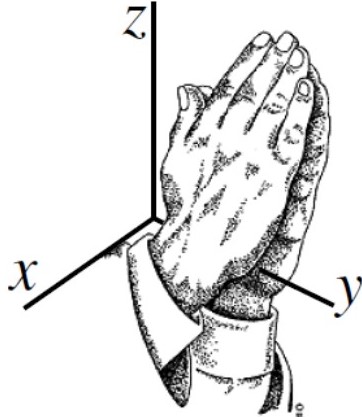}
		\end{center}
                 \label{bt1.5}
\end{figure}
\newpage

Suppose that inside a continuous material (solid or fluid) one makes a
small cut perpendicular to a unit vector $\bf n$.  The cut separates the
matter on one side of a small plane from the matter on the other side.
\vspace{-1.5cm}

\begin{wrapfigure}[10]{r}{5.5cm}
		\begin{center}
                \includegraphics[width=6cm]{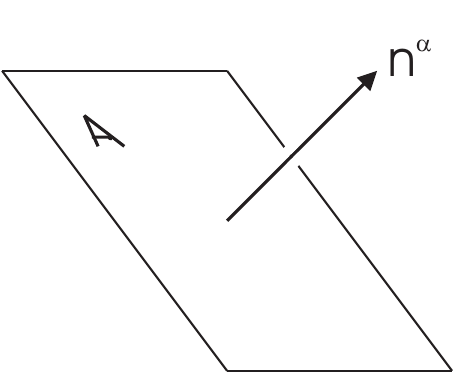}
		\end{center}
                 \label{IE1}
 \end{wrapfigure}	
\color{white}
.

.

.

\color{black}

Let $\bm F$ be the force needed to keep the matter on, say, the $+\bm n$
side in the state it would have been in had there been no cut. 
The force $\bm F$ is just the same force that had been exerted 
before the cut was made by the matter on the opposite side. And the force on 
the $-\bm n$ side is, of course, equal and opposite to the force on the $+\bm n$ side 
of the plane. Let ${\bm A} = {\bm n}A$, where $A$ is the area of the cut. 
\vspace{10mm}

Because the cut is small, $\bm F$ is linear in $\bm A$.  That is, any smooth 
function $\bm f(\bm x)$ has, about $\bm x=\bm 0$, the Taylor expansion 
\[
  f^i(\bm x) = f^i(\bm 0) + \partial_j f^i(\bm 0) x^j + O(x^2).
\]
Here the net force $\bm F$ on $\bm A$ vanishes for $\bm A=\bm 0$, so for small $A$,    
$F^i(\bm A) = S^{ij} A_j$, where $\dis S^{ij}:=\frac{\partial F^i}{\partial A_j}(\bm 0)$.
But a linear map from vectors to vectors is a tensor:  What we have shown is that 
that there is a tensor $S^{ab}$, the {\sl stress tensor}, for which 
\be 
    F^a = S^{ab} A_b.  
\label{e:flinear}\ee
Equivalently,  
\[ 
	S^{ab} n_b\; {\mbox{is the force across a unit area orthogonal to }} n_a .
\]
Since $\bm F  =\frac{d\bm p}{dt}$, one also says that
\begin{quote}
$S^{ab}n_b$ is the rate of flow of momentum across a unit area orthogonal
to $n_a$.
\end{quote}

\noindent{\sl Sign of the stress tensor}: 
Let's go back to the Blandford-Thorne description with two 
hands pressing each other. The hand on the $-x$ side exerts a 
force in the $+x$ direction, so $S_{xx} > 0$.  If your hands are stuck together with 
glue and you are trying to pull them apart, tension instead of pressure, then 
$S_{xx} < 0$.  For shear, if the hand on the $-x$ side is pulling up on the 
other hand, then $S_{zx}>0$:  The force on the hand on the $+x$ side is in the $+z$ 
direction.  \\

In a gas, if you make a cut, you need to exert a force $PA$ perpendicular to the 
cut, where $P$ is the gas pressure, to hold the gas in place.  That is, the pressure 
exerts a force  along $n^a$, with the same magnitude $PA$ for every orientation of the cut:  
\[ 
    S^a{}_bn^b = Pn^a, \quad \mbox{all}\ n^a.  
    \]
Then $S^a_b/P$ is the identity map $\delta^a_b$, and we have
\[ S^a{}_b = P\delta^a_b .\]
Note that in a gas, the molecules flowing in the positive $\bm n$ direction ($v^a n_a >0$) 
across a unit area orthogonal to $n^a$ carry total momentum $Pn^a$  across a 
unit area in unit time; so in this case the momentum flux $S^a{}_bn^b$ is tangible. 

In physics, the word {\sl fluid} refers both to liquids and gases. More 
generally, what distinguishes a fluid from a solid is that a fluid cannot 
maintain a shear stress: A fluid is continuous matter with stress tensor 
\be
 S^a{}_b = P\delta^a_b .
\ee
\index{stress tensor!perfect fluid}
(An {\sl imperfect fluid}, a fluid with viscosity,
is intermediate between a perfect fluid and a solid.  It has nonzero shear 
stress proportional to the gradient of its velocity.) 
\index{fluid!imperfect fluid}\index{imperfect fluid}
\newpage 

For continuous matter in general the stress tensor is symmetric:\\  
{\sl Claim}:  $S^{ab}$ is a symmetric tensor: $S^{ab} = S^{ba}$\,.\\
\\
\noindent{\sl Proof}:  We look at the torque on a small piece of the matter (called a {\sl fluid element} in the case of a fluid). The torque $\tau_z$ about the $z$-axis involves the $x$- and $y$-projections of the forces on four faces, shown below for a small unit cube.  
\begin{figure}[h!]
\begin{center}
\includegraphics[width=6cm]{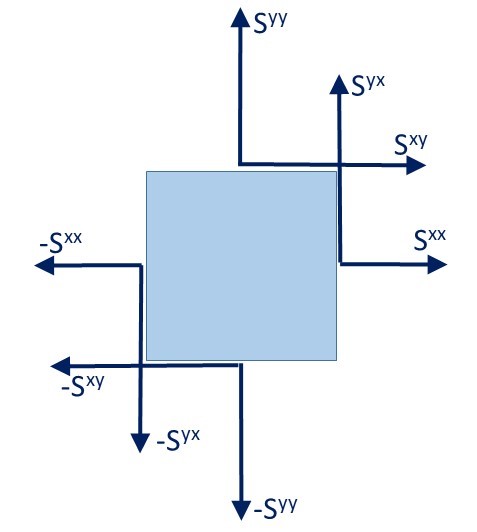}
\end{center}
\caption{Figure adapted from Feynman Lectures, v. II, Fig. 31-9.}
\label{fig:Sab}
\end{figure}

\noindent
For a cube with sides $\Delta x,\,\Delta y,\, \Delta z$, we have 
\begin{eqnarray*}
\tau_z &=& 2\left(- S^{xy}\Delta x\Delta z\right) \frac{\Delta y}{2} +
2\left( S^{yx}\Delta y\Delta z\right) \frac{\Delta x}{2}\\
&=& \left( -S^{xy}+S^{yx}\right)\Delta V\,\qquad \mbox{where\ } \Delta V = \Delta x\Delta y \Delta z.
\end{eqnarray*}
With $\rho, M, I$, and $\alpha$ the density, mass, moment of inertia, and angular 
acceleration of the cube, 
\begin{align*}
\tau &= I\alpha = \frac1{12} M(\Delta x^2+\Delta y^2)\alpha 
	= \frac1{12} \rho\Delta V (\Delta x^2+\Delta y^2)\alpha\\
\Rightarrow \alpha &= \frac{12}{\rho} \left( S^{yx}-S^{xy}\right)
\frac{1}{\Delta x^2+\Delta y^2} ,
\end{align*}

\noindent and tiny cubes rotate with arbitrarily large angular acceleration $\alpha$ unless
\[ S^{xy} = S^{yx} . \]
 \index{stress tensor|)} 

	The stress tensor is generalized to a four dimensional tensor
$T^{\alpha\beta}$ in the following way.  An observer $\widehat t^\alpha$ sees as her
stress tensor the spatial part of $T^{\alpha\beta}$
\[ T_\perp{}^{\alpha\beta} := \g^\alpha{}_\gamma \g^\beta{}_\delta
T^{\gamma\delta} = S^{\alpha\beta} ,\]
where $\g_{\alpha\beta}$ is, as usual, the projection onto the space 
orthogonal to $\widehat t^\alpha$.

Since the time component of momentum is energy, $T^{tx} \equiv
- T^{\alpha\beta}\ \widehat t_\alpha \widehat x_\beta = T^{\alpha\beta}\nabla_\alpha t\,\widehat x_\beta$ is the rate of flow of energy across a
unit spatial area orthogonal to $\widehat x_\alpha$.  And just as 
the 3-vector of flow (electric current $\bf j$, say) has as its time component 
the density (charge density $\rho_e$), here 
\begin{eqnarray*}
T^{xt} &=& {\mbox{density of $x$-momentum and}}\\
T^{tt} &=& {\mbox{energy density}}
\end{eqnarray*}
\index{energy!energy density}\index{density!energy density}\index{energy density}
In a basis moving with the material at a point $P$, $T^{it} = T^{ti} = 0$,
$i=1$-$3$, and $T^{\mu\nu}$ has the symmetric form
\[ \| T^{\mu\nu}\|
	= \left\| \begin{matrix}
 			T^{tt} &0&0&0\\
				 0&&&\\
			   0&&S^{ij}&\\
    				0&&&\\
	   \end{matrix} \right\|.
\]
Consequently $T^{\alpha\beta}$ is symmetric in any basis.

\benr
\item A rod has cross sectional area A and mass per unit length
$\mu$. 
\benalph\item What is its stress-energy tensor when it is under a tension
F? (assume F is uniformly distributed over the cross section)
\item What is its stress-energy tensor when it is compressed?  
\label{ex21}
\een

\item  \label{ex:drumhead} Find the stress-energy tensor at a point of a stretched drumhead, 
assuming mass per unit area $\sigma$ and uniform stress.  
The thickness of the drumhead is $h$ and the force per unit length 
along the edge of the drum is $\cal F$.  
Explain your choice of sign for components of the 
stress tensor.  (Solution below) 
\begin{figure}[h!]
\begin{center}
\includegraphics[width=.3\textwidth]{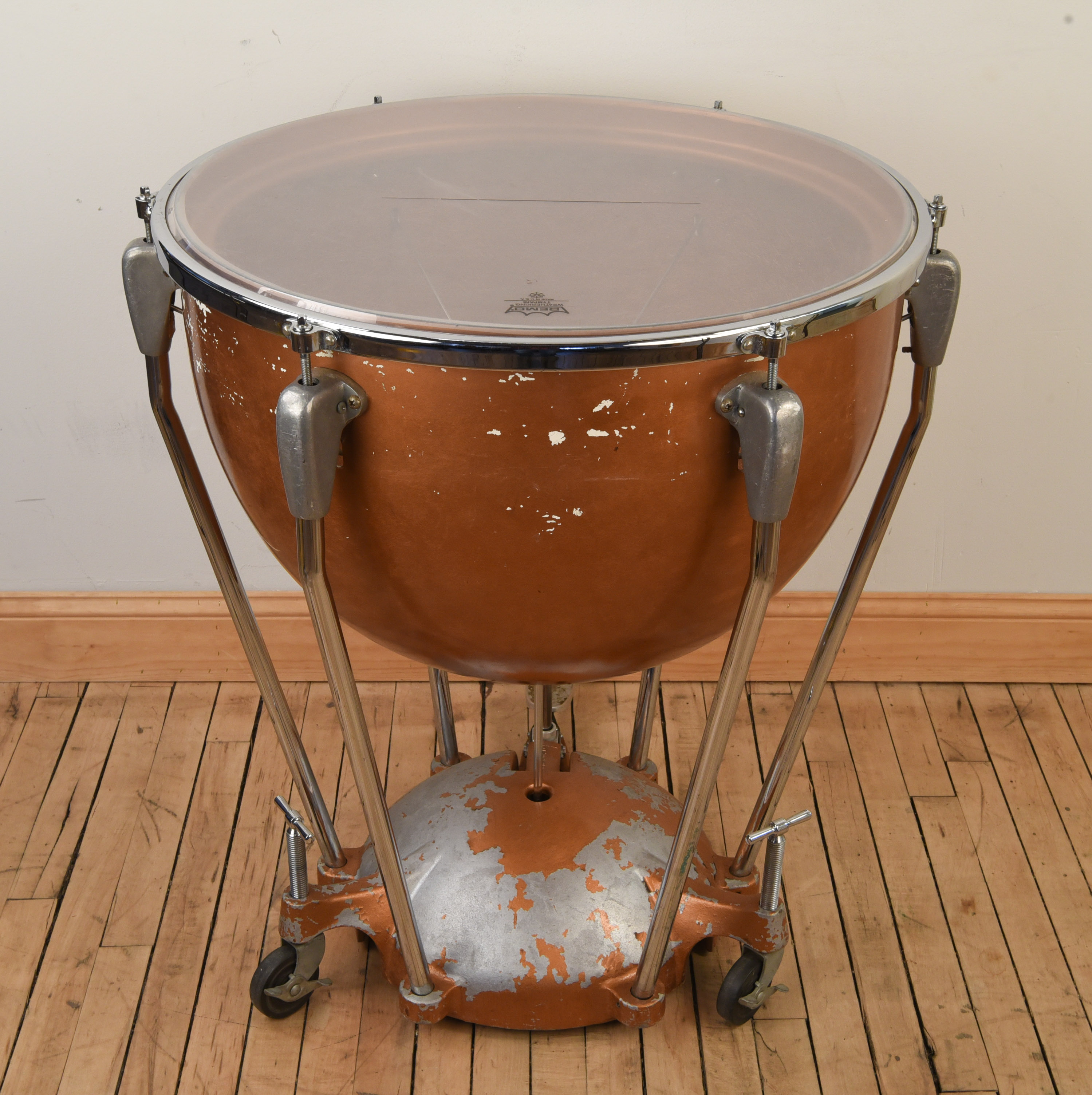}
\end{center}
\end{figure} 

\item A rope of mass per unit length $\mu$ has a static breaking
strength F.  What is the maximum F can be without violating the
``weak energy condition'' $T_{\alpha\beta}\hat t^\alpha \hat t^\beta>0$, all
timelike $\hat t^\alpha$? How close is a steel cable to this maximum strength?\\

Two of these exercises are problems 5.5 and 5.6 of Lightman, Press, Price and Teukolsky\cite{lppt}.\\
\label{ex22}
\een
\vspace{-2mm}

\noindent Solution to \ref{ex:drumhead} Note that the drumhead is uniformly stretched: Increasing the radial distance of points from the center increases the circumference 
of a circle about the center. Take $\hat x$ and $\hat y$ to be in the plane of the drumhead.  The force per unit area along the edge of the drum is ${\cal F}/h$. To keep the head in place on the $+x$ side of a small cut perpendicular to $\hat x$, you need to exert a force per unit area ${\cal F}/h$ in the $-\hat x$ direction.  That means that 
$S^{xx} = -{\cal F}/h$.  Similarly $S^{yy} =-{\cal F}/h$. The density is $\rho = \sigma/h$, implying 
\[
  [T^{\mu\nu}]= \begin{bmatrix} \sigma/h &0&0&0 \\ 0 &-{\cal F}/h &0&0\\ 0 & 0&-{\cal F}/h &0 \\ 0 &0 &0&0
		\end{bmatrix}
\] 
\vspace{3mm}

\noindent{\sl Electromagnetic stress-energy tensor} \\

 The stress-energy tensor of an electromagnetic field is
 \index{stress-energy tensor!electromagnetic field}
\index{electromagnetism!stress-energy tensor}
\be 
  T^{\alpha\beta}  
	= \frac{1}{4\pi} \left[ F^{\alpha\gamma} F^\beta {}_\gamma 
	 - \frac{1}{4} \eta^{\alpha\beta}  F^{\gamma\delta} F_{\gamma\delta}\right]. 
\label{e:Tem}\ee
Its $3+1$ decomposition associated with unit timelike vector $t^\a$ is 
\be
T^{\a\b} = \frac1{4\pi} \left[ \frac12(E^2+B^2) (t^\a t^\b + \g^{\a\b} )
		-E^\a E^\b - B^\a B^\b 
	+t^\a (E\times B)^\b + t^\b (E\times B)^\a\right].
\label{e:Tem3+1}\ee
\benr \item Lorentz force from divergence of $T^{\a\b}$.
\item
Prove that if $F^{\alpha\beta} $ is a
Maxwell field with source $j^\alpha $,
\[ \nabla_\beta  T^{\alpha\beta}  =-F^{\alpha\beta} j_\beta   .\]

\item Using the $3+1$ decomposition \eqref{e:FEB} of the Faraday tensor, show that the form \eqref{e:Tem3+1} of $T^{\alpha\beta} $ in terms of $E^\alpha $,
$B^\alpha $ and $t^\alpha $ follows from Eq.\eqref{e:Tem}.

\item 
Show that $T^{\alpha\beta} $ is invariant
under duality rotations:  $\widehat T^{\alpha\beta}  = T^{\alpha\beta} $.\\

Hint:
\vspace{-4mm}

\benalph
\item  Write $^\ast F^{\alpha\beta} $ in terms of $E^\alpha $, $B^\alpha $, and $t^\alpha $.
\item Show that $^\ast F^{\alpha\gamma} F^\beta {}_\gamma  = \eta^{\alpha\beta}  E_\gamma  B^\gamma $.
\item  Infer that $^\ast F^{\alpha\gamma} F^\beta {}_\gamma  + {^\ast}F^{\beta\gamma} F^\alpha {}_\gamma  -
\frac{1}{2} \eta^{\alpha\beta}  {^\ast}F^{\gamma\delta} F_{\gamma\delta} =0$.
\end{enumerate}
\een
 \index{stress-energy tensor|)} 

\subsection{Perfect fluids in a Newtonian context}

Continuous matter is a model for
a large assembly of particles in which a continuous energy density $\rho$ can
reasonably describe the macroscopic distribution of mass. The statement that the density varies
smoothly is the thermodynamic assumption that one can divide the dust into
boxes large enough that many particles are in each box (so that the density
$\rho = \frac{mN}{V}$ can be regarded as a continuous function of the
number $N$ of particles of mass $m$ in a box of volume $V$) and small
enough that the change in $\rho$ from one box to an adjacent box is
small:  $\rho$ can be regarded as a continuous function of position. 

In fluids, one assumes that the microscopic particles are interacting and that 
they collide frequently enough that the mean free path is small
compared with the scale on which the density $\rho$ changes.  One further
assumes that a mean 3-velocity $\bf v$ can be defined in boxes small
compared to the macroscopic length scale but large compared to the mean
free path and that it will be continuous over the matter.  An observer moving with this 
velocity $\bf v$ of the fluid, will see the collisions randomly
distribute the nearby particle velocities so that the particles will look
locally isotropic and will exhibit a macroscopic pressure $P$.\\

We'll begin in the Newtonian approximation to make things 
intuitively clear.  Formally, however, the relativistic equations are simpler.  \\

\noindent{\sl Euler equation: ${\bm F} = m{\bm a}$}\\
The Euler equation is the Newtonian equation of motion, ${\bm F} = m{\bm a}$
for a fluid element, a small piece of fluid.   
Consider a fluid element with density $\rho$ and velocity $\bf v$.  
\begin{figure}[h!]
\begin{center}
\includegraphics[width=7cm]{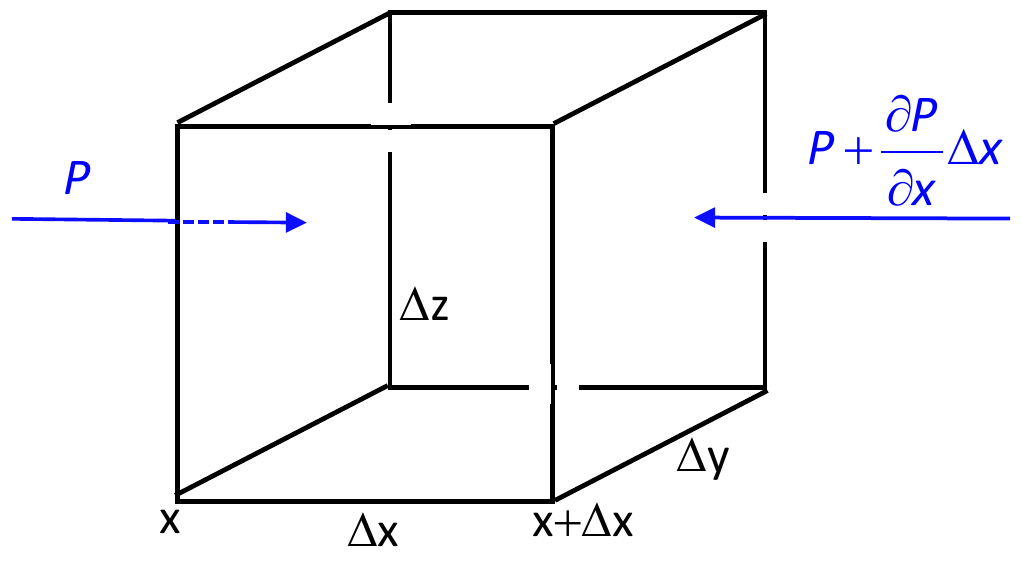}
\end{center}
\label{fig:euler}
\end{figure}

\vspace{-4mm}

\noindent
The pressure on the left face is $P(x)$; the pressure on the right face is 
$\dis P(x+\Delta x)=P(x)+\frac{\partial P}{\partial x}\Delta x$.\\
With $A=\Delta y\Delta z$ the area of the left and right faces of the box, 
the net force in the $x$-direction is
\begin{eqnarray*}
F_x &=& P(x)  A - P(x+\Delta x)  A\\
&=& -\frac{\partial P}{\partial x} V .
\end{eqnarray*}
where $ V = \Delta x\Delta y \Delta z$ is the volume of the fluid
element.  Replacing the index $x$ by $y$ and $z$, we have 
\[ 
	{\bm F}  = -{\bm\nabla} P\; V .
\]
We want to write ${\bm F}  = m{\bm a} $ or
\[ -\bm\nabla P\; V = \rho V\; {\bm a}  ,\]
and we need to find ${\bm a} $ in terms of the velocity field ${\bf v}(t,\bm x)$.
The vector field ${\bf v}(t,{\bm x})$ has the meaning that at time $t$ the fluid
element at $\bm x$ has velocity ${\bf v}(t,{\bm x})$.  Thus at time $t+\Delta t$
that same fluid element is at ${\bm x} +{\bf v} ({\bm x},t)\Delta t$ and has
velocity ${\bf v}(t+\Delta t,{\bm x}+{\bf v}({\bm x})\Delta t)$.  The fluid
element has changed its velocity by
\begin{eqnarray*}
  \Delta {\bf v} &=& {\bf v} (t+\Delta t, {\bm x} +{\bf v}({\bm x},t)\Delta t) 
			- {\bf v}(t, {\bm x})\\
		&=& \left(\frac{\partial{\bf v}}{\partial t}
			   +{\bf v}\cdot\nabla {\bf v}\right)\Delta t,
\end{eqnarray*}
in time $\Delta t$, and its acceleration is therefore
\beq 
	{\bm a} = \left( \partial_t +{\bf v}\cdot\nabla \right) {\bf v} . 
\eeq
In this way we obtain Euler's equation of motion 
\href{https://scholarlycommons.pacific.edu/euler-works/226/}{(``Principes g\'en\'eraux du mouvement des fluides,'' M\'emoires de l'Acad\'emie des Sciences de Berlin, 1757)}
\beq 
	\rho\, (\partial_t+{\bf v}\cdot\bm\nabla )\,{\bf v} 
	=-\bm\nabla P\ .
\label{104}\eeq
\index{equation of motion!Euler equation}

	In the presence of a gravitational field, with potential $\Phi$ satisfying
$\nabla^2\Phi = 4\pi G\rho$, there is an additional force 
$-\rho V\ \bm\nabla\Phi$ on each fluid element; and the Euler equation becomes
\beq \cblue
    \rho\, (\partial_t+{\bf v}\cdot\bm\nabla )\,{\bf v} 
	=-\bm\nabla P -\rho\bm\nabla\Phi .
\label{e:euler}
\eeq
  
\noindent{\sl Conservation of mass: The continuity equation}\index{conservation laws!mass}
\index{continuity equation}

As a fluid element moves its volume changes. Because its mass is conserved 
(in a relativistic context use baryon mass) a fractional increase $\Delta V/V$ 
in its volume is equal to the fractional decrease $\Delta\rho/\rho$ in its 
density.  
\beq
   \rho V = \mbox{ constant } \Longrightarrow 
\frac{\Delta\rho}{\rho} = - \frac{\Delta V}{V} \quad\mbox{or}\quad 
\frac{d\rho/dt}{\rho} = - \frac{dV/dt}{V},
\label{e:cmass0}\eeq 
where $\rho = \rho(t,{\bm x}(t))$, $V= V(t,{\bm x}(t))$. \\

We'll begin with the change in the volume of the fluid element as it 
moves.  It is helpful first to recall or notice the geometrical meaning of 
the divergence of a vector field.  If each point in a volume $V$ moves by 
a small amount $\bm\xi(x)$, from an initial position $\bm x$ to a final position
$\bar{\bm x} = \bm x+\bm\xi$, the volume of the box changes by 
$\Delta V = \overline V-V = \nabla\cdot\bm\xi\,V$: That is, $\bm\nabla\cdot\bm\xi$ 
is the fractional change in volume.  
This is really Gauss's theorem:  
As illustrated by the figure below, moving $V$ to $\overline V$ moves
each point of the surface $S$ of a volume $V$ along $\bm\xi$, changing the 
the volume of the box by 
\[
\dis \Delta V = \int_S \bm\xi\cdot d\bm S\,, 
\]
to lowest order in $\bm\xi$.
\begin{figure}[ht!]
\begin{center}
\includegraphics[width=.4\textwidth]{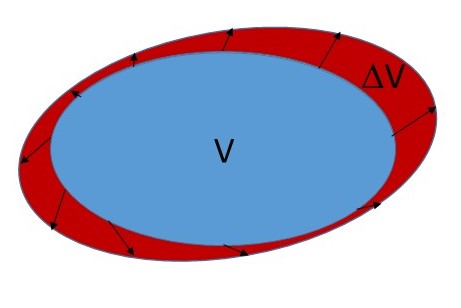}
\end{center}
\label{fig:gauss1}
\end{figure}

\noindent
Gauss's theorem now implies    
\beq
  \Delta V = \int_V \bm\nabla\cdot \bm\xi \,dV. 
\eeq

For a small volume $V$ the volume then changes by   
$\dis 
  \Delta V = V \bm\nabla\cdot\bm\xi,   
$
or
\beq{\color{blue}
 \frac{\Delta V}V = \bm\nabla\cdot\bm\xi},
\label{e:gauss00}\eeq
to lowest order in $V$ and $\bm\xi$.  That is, as claimed,  
$\nabla\cdot\bm\xi$ is the fractional change in volume.

Go back now to a fluid with a velocity field ${\bf v}(x,t)$.  In a 
time $\Delta t$ the fluid at $\bm x$ moves to ${\bm x} + \bm\xi$, where 
$\bm\xi = {\bf v}\Delta t$. 
Then, writing $\nabla\cdot\bm\xi = \nabla\cdot({\bf v}\,\Delta t)$, we have
\beq
   \frac{dV/dt}V = \bm\nabla\cdot\bf v.   
\label{e:divv}\eeq
The change in density is given by 
\beq
  \frac d{dt}\rho(t,{\bm x}(t)) = \partial_t \rho + \partial_i\rho \frac {dx_i}{dt} 
	= (\partial_t +{\bf v}\cdot \bm\nabla)\rho. 
\eeq
Finally, with these expressions for $dV/dt$ and $d\rho/dt$, conservation of mass 
\eqref{e:cmass0} is  
\[ 
  \frac1\rho(\partial_t +{\bf v}\cdot \bm\nabla)\rho = -\bm\nabla\cdot\bf v, 
\]
or 
\beq 
	\partial_t\rho + \nabla\cdot (\rho{\bf v}) = 0. 
\label{e:consm1}\eeq
This is commonly called the {\sl continuity equation}. 
\index{continuity equation|textbf} 

To summarize:  A fluid is characterized by its pressure $P$, 
density $\rho$ and 3-velocity ${\bf v}$.  In the Newtonian approximation, its motion is governed by the equations 
\bsube\crv\begin{align}
  	\nabla^2\Phi &= 4\pi G\rho,\qquad \lim_{r\rightarrow\infty}\Phi = 0\cb,
\label{e:poisson}\\ 
	\partial_t\,\rho + \nabla\cdot (\rho{\bf v}) &= 0\cb, 
\label{e:masscons}\\
\nonumber\\
 	\crv \rho\, (\partial_t+{\bf v}\cdot\bm\nabla )\,{\bf v} 
			&= -\bm\nabla P -\rho\bm\nabla\Phi \cb.
\label{110}\end{align}\esube
\cb

\subsection{Relativistic equation of motion and conservation laws.}\index{conservation laws}
\label{s:conservation_laws} 

Mass conservation is the Newtonian limit of energy conservation, and 
the Newtonian equation of motion, $\bm F = m\bm a$, is conservation of 
momentum, when the field responsible for the force is included in the 
stress-energy tensor $T^{\alpha\beta}$.  In terms of $T^{\alpha\beta}$, 
energy conservation and the equation of motion (momentum conservation) 
are projections of the single equation, 
\be
   \nabla_\beta T^{\alpha\beta} = 0, 
\label{e:divT}\ee
along and orthogonal to the velocity $\widehat t^\alpha$ of an observer. 

We begin by showing that Eq.~\eqref{e:divT}, the statement that $T^{\alpha\beta}$ 
is divergencefree, has the meaning of energy-momentum conservation.\index{conservation laws!energy-momentum}\index{energy!energy conservation in flat space}  

\begin{figure}[ht!]
\begin{center}
\includegraphics[width=.75\textwidth]{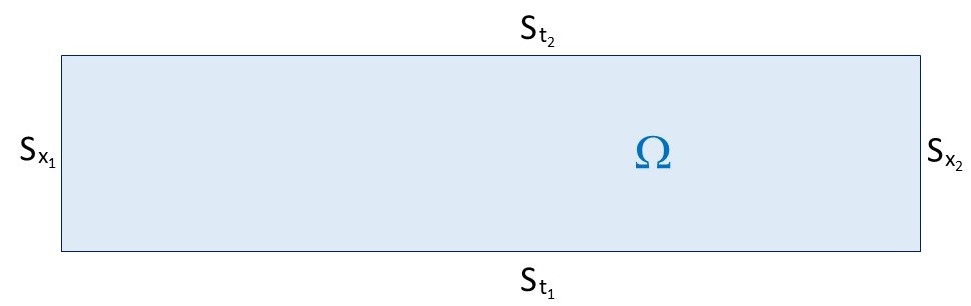}
\end{center}
\end{figure}

\noindent Consider a four-dimensional box $\Omega$ whose boundary
$\partial\Omega$ consists of the top and bottom volumes $S_{t_{1}}$ and
$S_{t_{2}}$ together with six timelike sides, $S_{x_{1}},\ldots
,S_{z_{2}}$.  (A timelike hypersurface is one that has timelike vectors at each point.  
A spacelike hypersurface contains only spacelike vectors.)
\index{hypersurface!spacelike|textbf}\index{hypersurface!timelike|textbf} 
\index{spacelike!hypersurface|textbf}\index{timelike!hypersurface|textbf}
The total $x$-momentum in the volume $S_{t_{1}}$ (i.e.\ in
the box at $t=t_1$) is
\[ 
	p_1{}^x = \int_{S_{t_1}} {\mbox{ (density of $x$-momentum)}}\ dxdydz
		=\int_{S_{t_1}} T^{\alpha\beta}\widehat x_\alpha \nabla_\beta t\ dxdydz.
\]
(Using $\nabla_\beta t$ instead of $\widehat t_\beta$ avoids a minus sign.) Similarly, the total $x$-momentum in $S_{t_2}$ is
\[ 
	p_2{}^x =\int_{S_{t_{2}}} T^{\alpha\beta}\widehat x_\alpha \nabla_\beta t dxdydz .
\]
The total $x$-momentum that left through the timelike sides is
\begin{eqnarray*}
\Delta p^x &=& \int\; {\mbox{(rate of $x$-momentum flow per unit area) }}
dAdt\\
&=& \int_{S_{x_{1}}} T^{\alpha\beta}\widehat x_\alpha (-\nabla_\beta x)dydzdt +
\int_{S_{x_{2}}} T^{\alpha\beta}\widehat x_\alpha \nabla_\beta x dydzdt +\cdots +
\int_{S_{z_{2}}} T^{\alpha\beta} \widehat x_\alpha \nabla_\beta z dxdydt
\end{eqnarray*}
If we now denote by $n_\alpha$ the unit outward normal at each point of the
boundary $\partial\Omega$, and by $dS$ each 3-dimensional volume element of
the surfaces $(dxdydz, dtdxdy$, etc.), conservation of $x$-momentum,
\[ p_1{}^x -p_2{}^x + \Delta p^x = 0,\]
takes the form \index{conservation laws!momentum}
\be \int_{\partial\Omega} T^{\alpha\beta} \widehat x_\alpha n_\beta dS = 0.
\label{93}\ee

Integrating over each coordinate $x^\mu$ establishes Gauss'
theorem for our  4-dimensional box:  For any vector field $j^\alpha$
\be 
	\int_\Omega \nabla_\alpha j^\alpha d^4V
		= \int_{\partial\Omega} j^\alpha n_\alpha dS.
\label{94}\ee
($\partial \Omega$ means the boundary of $\Omega$.)  
Thus if $\int\limits_{\partial\Omega} j^\alpha n_\alpha dS=0$ for
arbitrarily small boxes $\Omega$ about a point $\cal P$, (\ref{94})
$\Rightarrow \nabla_\alpha j^\alpha =0$ at $\cal P$.  And (\ref{93}) means
that for any constant vector field $\widehat x^\alpha$,
\[ 
	\nabla_\b (T^{\alpha\beta}\widehat x_\a )=0 ,
\]
($x$-momentum is conserved), implying Eq.~\eqref{e:divT}
\[ 
	\nabla_\b T^{\alpha\beta} = 0 .
\]

In curved spacetime, this equation is unchanged, but there will in general be no constant vector fields 
like $\widehat x^\a$, and, more generally, no vector fields $v^\a$ for which $\nabla_\a(T^{\a\b}v_\b) = 0$.  In flat space, $\widehat x^\a$ is along a translation symmetry of the geometry, and the current 
$j^\a = T^{\a\b} \widehat x_\b$ is conserved because of that symmetry. 
\index{symmetry!relation to conserved current}   

\subsection{Dust}
\index{dust}\index{perfect fluid!dust}

By dust is meant a collection of a large number of very small, very light
particles which do not interact directly with each other, whose density
varies smoothly, and whose velocity field (the unit tangent vectors to the
particle trajectories) is also smooth. Because of its simplicity, dust is 
often used as a first, highly idealized, model for baryonic matter in 
the universe (the particles are galaxies); for stellar collapse (historically, for 
early studies of spherical and slightly nonspherical collapse); for the disk of spiral 
galaxies (the particles include the stars in the disk, as well as the interstellar medium 
of actual dust particles and gas); and for thin accretion 
disks around black holes and neutron stars.  In the last example, a disk 
is thin if its pressure is small compared to the rotational energy of the disk; 
in general, ignoring pressure in comparison with the kinetic energy of the 
macroscopic fluid motion is the dust approximation.  

Requiring the velocity field $u^\alpha$ to be smooth means that
nearby dust particles travel along nearly parallel world lines.\\
\begin{figure}[ht!]
\begin{center}
\includegraphics[width=.4\textwidth]{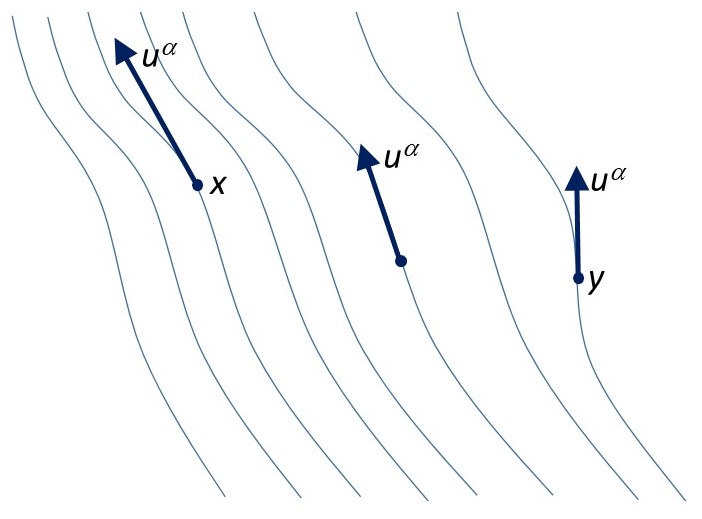}
\end{center}
\end{figure}

In order to define density, one needs to know what surface of simultaneity is being used. The natural choice is to take at each point a small volume orthogonal to the fluid 
velocity at that point.
\begin{figure}[ht!]
\begin{center}
\includegraphics[width=.4\textwidth]{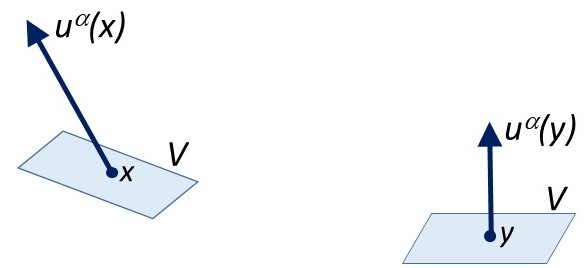}
\end{center}
\end{figure}
\vspace{-3mm}
\noindent Then the rest-mass density $\rho$ at a point $\bm x$ is the density measured at
$\bm x$ by an observer moving with the dust at $\bm x$---an observer with velocity
$u^\alpha(\bm x)$.  Such an observer, for whom the dust appears to be locally
at rest, is called {\em comoving}.  

	If there are no forces acting on the dust (if, for example, one can
neglect not only the two-particle gravitational interactions, but also the
interaction of each particle with the gravitational field of the dust as a
whole) then each particle will travel on a straight line:\\
\centerline{$ u^\beta\nabla_\beta u^\alpha = 0.$}
We will see that the same equation holds in curved spacetime, with $\nabla$ the 
covariant derivative associated with a metric whose curvature 
encompasses gravity.  

	This equation, together with conservation of mass completely describes
freely moving dust, and the equation describing conservation of mass can be
obtained in the following way. 
Consider a 4-volume $\Omega$ with sides tangent to the trajectories and
whose top and bottom faces are volumes $V_1$ and $V_2$: Think of   
$V_1$ and $V_2$ as spacelike slices of the dust's history.  
Conservation of particles is the statement that if $N$ particles ``enter''
(are in the volume) $V_1$, $N$ particles ``leave'' $V_2$.  If $m$ 
is the mass of a particle (or the average mass per particle, if there 
are several species), mass conservation has the form  \vspace{-3mm}
\be mN_1 = m N_2.\ee
\begin{figure}[ht!]
\begin{center}
\includegraphics[width=.2\textwidth]{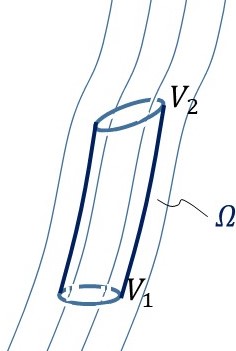}
\end{center}
\end{figure}
\vspace{-8mm}

Our goal is to write this equation in terms of the density $\rho$.  
Because $u^\alpha$ varies from point to point, however, 
even if we take $V_1$ and $V_2$ to be slices of constant time $t$, 
they will not, in general, be orthogonal to the velocity.  
The rest-mass of the particles entering a small 
comoving volume $\Delta V$ that is orthogonal to $u^\alpha$ is just $M=\rho \Delta V$. 
So we need to find the mass of a small slice $\Delta\wt V$ of the dust that is not orthogonal to $u^\alpha$.
\vspace{-3mm}
\begin{figure}[ht!]
	\begin{center}
		\includegraphics[width=.4\textwidth]{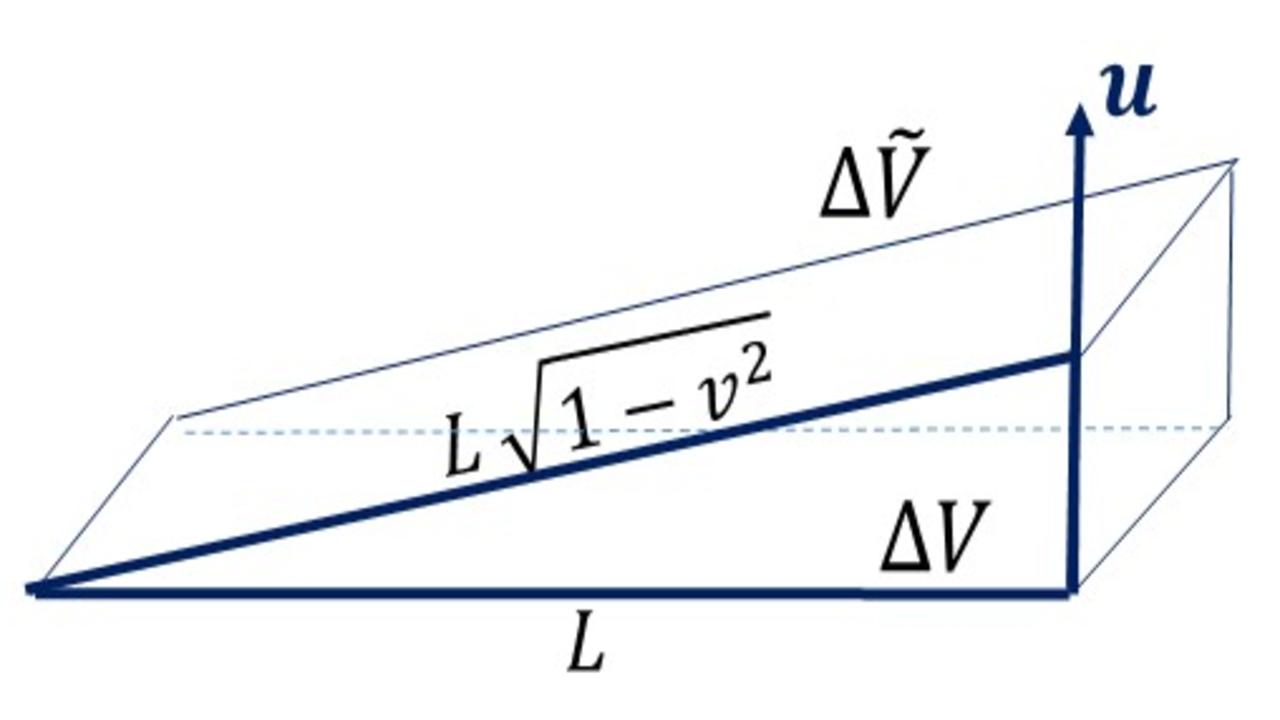}
	\end{center}
\label{dust3}
\end{figure}
Boosting the small slice $\Delta V$ of the fluid by a velocity $v$ gives a slice with volume 
\be
	\Delta \wt V = \Delta V\sqrt{1-v^2} \equiv \Delta V \widetilde\gamma^{-1},
\ee
by Eq.~\eqref{e:vcontract} -- just the fact that lengths orthogonal to the boost 
are unchanged and the length along the boost is smaller than the proper length by the factor 
$\widetilde\gamma^{-1}$.   We choose the slices to be small enough that the difference 
in the density and velocity from $\Delta V$ to $\Delta\wt V$ is negligible: 
$\wt u^\a = u^\a$. Because every dust particle that passes through $\Delta V$ also passes through $\Delta\wt V$, 
the rest mass $\Delta M$  is the same in volume $\Delta V$ and in $\Delta\wt V$, 
\[
    \Delta M = \rho \Delta V = \rho\wt \gamma \Delta\wt V.
\]  
Let $n_\alpha=-u_\alpha$ be the normal to $\Delta V$ with sign 
chosen to make $u^\alpha n_\alpha = 1$, and let $\wt n_\alpha$ be the corresponding 
unit normal to $\Delta \wt V$. Then $\wt n_\alpha$ is related to $n_\a$ by a boost 
with velocity $v$, implying
\index{normal to a spacelike hypersurface}
\index{normal to a spacelike hypersurface!unit normal} 

\[
   u^\alpha \wt n_\alpha  = -n^\a \wt n_\a = \widetilde\gamma.   
\]
We now have
\[ 
 \Delta M = \rho u^\alpha n_\alpha \Delta V 
	=  \rho \wt u^\alpha \wt n_\alpha\Delta \wt V,
\]
and the mass of dust in an arbitrary spacelike volume $V$ is 
\[
       M = \int_V \rho u^\alpha n_\alpha dV.   
\]   
If, as in our previous integral conservation equations, we now take $n_\alpha$ to be the outward normal to the boundary of $\Omega$, conservation of mass from $V_1$ to $V_2$ is 
\be 
	M_2-M_1 = \int_{V_2} \rho u^\alpha n_\alpha dV - \int_{V_1} \rho u^\alpha n_\alpha dV = 0.
\label{96}\ee
For example, if $V_1$ and $V_2$ are $t=$ constant surfaces, $n_\a$ is $\nabla_\a t$ on $V_2$ and 
$-\nabla_\a t$ on $V_1$.      
This is the integral version of mass conservation.  The corresponding
differential equation again follows from Gauss' theorem,
\[ 
\int_\Omega \nabla_\alpha (\rho u^\alpha )d\Omega 
= \int_{V_2} \rho u^\alpha n_\alpha dV + \int_{V_1} \rho u^\alpha n_\alpha dV .
\]
By our construction, because our spatial volume of dust moves with the dust, no particles 
leave it and there is no contribution on the right side of the equation 
from the timelike part of the boundary of $\Omega$.  
Formally, because $u^\alpha$ is tangent to the timelike sides of the boundary, 
$u^\alpha n_\alpha = 0$ there.  
Because we can choose an arbitrarily small volume of the form $\Omega$ 
about any point, (\ref{96}) implies 
that $\nabla_\alpha (\rho u^\alpha )=0$.

	To summarize:  Freely moving dust is described by a scalar field $\rho$
and a unit timelike vector field $u^\alpha$ satisfying the equations
\be \cblue
	\nabla_\beta (\rho u^\beta )=0, \qquad
	u^\beta\nabla_\beta u^\alpha = 0 .\cb
\label{e:dust}\ee
{\em Stress-energy tensor (energy-momentum tensor)}:  A comoving observer sees 
the dust at rest
and non-interacting, so there is no stress and $T^{ij} = 0$, $i,j = 1$-$3$.
Furthermore, there is no momentum density (since $u^i=0$), so $T^{0i} = 0$.
 Finally, he sees an energy density $\rho$:  $T^{00}=\rho$.
\[ \| T^{\mu\nu}\| = \left\| \begin{matrix}
 			\rho &&&\cr &0&&\cr &&0&\cr
&&&0\cr
			\end{matrix} \right\|.
\]
\index{density!energy density}\index{energy!energy density}\index{energy density}
Now for a comoving observer whose velocity has components $(u^\mu) = (1,0,0,0)$, the stress-energy tensor has components 
$T^{\mu\nu} = \rho u^\mu u^\nu$; and that implies the tensor equation
\be 
	T^{\alpha\beta} = \rho u^\alpha u^\beta.
\label{98}\ee
\index{stress-energy tensor!dust} \index{dust!stress-energy tensor}

 From the fact that the Newtonian equations describing a fluid with 
no external force express conservation of mass and momentum of each 
fluid element, one should expect that $\nabla_\beta T^{\alpha\beta} = 0$ will provide the equations that describe the motion of the fluid. \\ 
First the equations describing dust:  
Projecting along $u_\alpha$ gives conservation of energy (mass):
\index{conservation laws!energy}
\begin{eqnarray*}
0 &=& u_\alpha\nabla_\beta T^{\alpha\beta} = u_\alpha\left[
u^\alpha\nabla_\beta (\rho u^\beta )+(\nabla_\beta u^\alpha )\rho
u^\beta\right]\\
&=& -\nabla_\beta (\rho u^\beta), 
\end{eqnarray*}
where we used 
$u_\alpha\nabla_\beta u^\alpha = \frac{1}{2} \nabla_\beta (u^\alpha u_\alpha
) = \frac{1}{2} \nabla_\beta (-1) = 0$ to eliminate the second term.  

Projecting orthogonal to $u^\alpha$ gives the equation of motion:  We define the  projection operator $q^\a_\b$ orthogonal to $u^\a$ by
\be \cblue
	q^\alpha_\beta = \delta^\a_\b  + u^\alpha u_\beta;
\label{e:qab}\ee
In other words, $q^\a_\b$ is the tensor that maps $u^\a$ to $0$ and leaves vectors orthogonal to $u^\a$ unchanged, \\
\centerline{$q^\a_\b u^\b = 0, \qquad q^\a_\b v^\b = v^\a, \mbox{ for } v^\a \perp u^a$.}
Then the projection of the equation $\nabla_\beta T^{\alpha\beta}=$ orthogonal to $u^\a$ is 
\[ 0=q^\alpha{}_\gamma \nabla_\beta T^{\beta\gamma} =
q^\alpha{}_\gamma\nabla_\beta (\rho u^\beta u^\gamma ) = \rho u^\beta
q^\alpha{}_\gamma\nabla_\beta u^\gamma =\rho u^\beta\nabla_\beta u^\alpha ,\]
again using $u_\gamma\nabla_\beta u^\gamma =0$ to write 
$q^\alpha{}_\gamma\nabla_\beta u^\gamma = (\delta^\alpha{}_\gamma +u^\alpha
u_\gamma )\nabla_\beta u^\gamma = \nabla_\beta u^\alpha$.\\

\noindent {\em Charged dust}:\index{charge!charged dust}\index{dust!charged dust}  If the dust is charged, but the
interparticle separation is large enough that we can ignore the
electromagnetic interaction between any two dust particles (so that
$F^{\alpha\beta}$ can be computed from an average current density 
$j^\alpha = \rho_e u^\alpha$, or from $j^\alpha = \rho_e u^\alpha +$ external
sources), then each dust particle will travel on a trajectory
\[ 
	u^\beta\nabla_\beta u^\alpha = \frac q{m} F^{\alpha\beta}u_\beta ,
\]
where $q$ and $m$ are the charge and mass of the dust particle.  In the
continuum limit, $q\rightarrow 0$ and $m\rightarrow 0$ but $\frac{eN}{V}$
and $\frac{mN}{V}$ remain finite and 
$\dis \rho_e = \lim\limits_{\stackrel{\scriptstyle{q\rightarrow 0}}{N\rightarrow\infty}}
	\frac{qN}{V}$,\ \ \  
$\dis \rho = \rho_m 
	= \lim\limits_{\stackrel{\scriptstyle{m\rightarrow 0}}{N\rightarrow\infty}}
		\frac{mN}{V}$ 
are the scalar fields that describe the dust.  Thus the equation of motion is
\be \rho u^\beta \nabla_\beta u^\alpha = \rho_e F^{\alpha\beta} u_\beta .
\label{99}\ee
When the dust is the only source of $F^{\alpha\beta}$, so that
\be \nabla_\beta F^{\alpha\beta} = 4\pi j^\alpha, \label{100}\ee
the system of field plus dust has the divergence-free stress-energy tensor
\begin{eqnarray*}
T^{\alpha\beta} &=& T^{\alpha\beta}_{\text{dust}} + T^{\alpha\beta}_{\text{e-m}}\\
&=& \rho u^\alpha u^\beta + \frac{1}{4\pi} \left[ F^{\alpha\gamma}
F^\beta{}_\gamma -\frac{1}{4} \eta^{\alpha\beta} F^{\gamma\delta}
F_{\gamma\delta}\right]:
\end{eqnarray*}
\begin{eqnarray}
u_\alpha \nabla_\beta T^{\alpha\beta} &=& -\nabla_\beta (\rho u^\beta
)-F^{\alpha\gamma}u_\alpha j_\gamma\nonumber\\
&=& -\nabla_\beta (\rho u^\beta )-
\rho_e{\underbrace{F^{\alpha\gamma}u_\alpha u_\gamma}_0}\nonumber\\
&=& -\nabla_\beta (\rho u^\beta ) 
\label{101}
\end{eqnarray}
and
\[ q^\alpha{}_\gamma\nabla_\beta T^{\beta\gamma} = \rho u^\beta\nabla_\beta
u^\alpha - F^{\alpha\beta}j_\beta \hspace{4mm} {\mbox{(using Ex.\ 16 and
Eq.\ (\ref{101})):}}\]
that is, given Maxwell's equations, the equation of motion for the dust
together with conservation of mass follows from $\nabla_\beta
T^{\alpha\beta} =0$:
\begin{eqnarray}
u_\alpha\nabla_\beta T^{\alpha\beta} &=& 0 \Rightarrow \nabla_\beta (\rho
u^\beta )=0\nonumber\\
q^\alpha{}_\gamma \nabla_\beta T^{\beta\gamma} &=& 0 \Rightarrow \rho
u^\beta\nabla_\beta u^\alpha = F^{\alpha\beta} j_\beta .\label{102}
\end{eqnarray}

\subsection{Perfect Fluids}\index{fluid!perfect fluid}\index{perfect fluid}
\label{s:perfect_fluid}

	We will obtain the relativistic equations of motion and energy conservation for a fluid from Eq.~\eqref{e:divT}, $\nabla_\beta T^{\alpha\beta}=0$, and will then verify that 
they agree with the Newtonian equations in
the limit $v\ll c, P\ll \rho c^2$. We again assume a fluid-element description, boxes large 
enough that pressure, density, and fluid velocity are defined, and small enough to 
neglect their change across the box.  Saying that the fluid is shear-free is then 
equivalent to saying that a comoving observer sees the fluid as isotropic:  In a box 
small enough to ignore changes in density and pressure there is no preferred spatial direction.  
That is, the components of the fluid's stress-energy tensor in the observer's basis must have no 
preferred direction:   
$T^{\mu\nu}$ must be invariant under rotations.  We can see this as follows:  
The components $T^{0i}$ $(i=1-3)$ transform as a 3-vector under rotations and so can be
invariant only if
\be T^{0i} = 0 \label{112}\ee
(for a comoving observer, the fluid element's 3-momentum vanishes).  
The spatial part of the tensor can be invariant under rotations only if it 
is a multiple of $\delta^{ij}$:  Pick a basis in which the 
symmetric matrix $T^{ij}$ is diagonal. Unless all three eigenvalues are the 
same, the largest or smallest eigenvalue picks out a direction -- along its eigenvector.
It follows that the only nonzero components of $T^{\mu\nu}$ are the
rotational scalars $T^{00}$ and $\delta^{ij}T^k{}_k/3$:
\footnote{Equivalently, we can decompose the spatial tensor $T^{ab}$ into its symmetric tracefree part, the shear tensor, $S^{ab} = T^{ab} - \frac13 \d^a_b T_c^c$, and its trace:   $T^{ab} =S^{ab} + \frac13\d^a_b T_c^c$. A symmetric tracefree tensor transforms 
as an irreducible $j=2$ representation of the rotation group and  
is therefore rotationally invariant only if it vanishes (as our quick eigenvalue 
argument shows).}
\be \| T^{\mu\nu}\| 
	= \left\|\begin{matrix}\rho &&&\cr &P&&\cr &&P&\cr &&&P\cr
	\end{matrix}\right\| , 
\label{114a}\ee
where we have defined $\rho$ and $P$ by
\begin{eqnarray*}
\rho &=& T^{00}\\
P &=& \frac{1}{3} T^k{}_k .
\end{eqnarray*}
In the comoving basis, the fluid's four velocity has components $(u^\mu) = (1,
0, 0, 0)$, and $\rho$ is the energy density measured by a comoving observer.
\index{energy density!perfect fluid}\index{energy density!comoving}
The projection operator orthogonal to $u^\a$, defined in Eq.~\eqref{e:qab}, has the contravariant form 
\[
	q^{\alpha\beta} = \eta^{\alpha\beta} + u^\alpha u^\beta,
\]
with components
\be \| q^{\mu\nu}\| 
	= \left\|\begin{matrix} 0 &&&\cr &1&&\cr &&1&\cr &&&1\cr
	\end{matrix}\right\| . 
\label{114b}\ee
Then 
\[ T^{\mu\nu} = \rho u^\mu u^\nu + Pq^{\mu\nu},\]
and we have
\be\crv T^{\alpha\beta} = \rho u^\alpha u^\beta + Pq^{\alpha\beta} 
		    = (\rho +P)u^\alpha u^\beta + P\eta^{\alpha\beta} .
\label{e:Tfluid}\ee
\index{density!of fluid}\index{fluid!perfect fluid|textbf}
\index{perfect fluid!stress-energy tensor}
\index{stress-energy tensor!perfect fluid}
\index{energy-momentum tensor!perfect fluid}

\noindent {\sl Conservation of energy.}\index{conservation laws!energy} \hspace{20mm}
\index{energy!conservation for perfect fluid}
$u_\alpha\nabla_\beta T^{\alpha\beta} = 0$
\begin{align}
0 &= u_\alpha\nabla_\beta T^{\alpha\beta} = u_\alpha\nabla_\beta [\rho
u^\alpha u^\beta + Pq^{\alpha\beta}]\nonumber\\
&= -\nabla_\beta (\rho u^\beta ) +Pu_\alpha\nabla_\beta (\eta^{\alpha\beta} +
u^\alpha u^\beta )\nonumber\\
&= -\nabla_\beta (\rho u^\beta ) -P\na_\b u^\b \nonumber\\
 \crv\nabla_\beta (\rho u^\beta ) &=\crv -P\na_\b u^\b 
\label{e:econsderive}\end{align}
The equation means that the mass of a fluid element decreases by the work,
\be
 P\; dV = PV\; \nabla\cdot \bm u \, d\tau,
 \label{e:dVdivu} \ee
it does in proper time $d\tau$.  To see this, note that in proper time $d\tau$, the fluid element is 
displaced by $\xi^\alpha = u^\alpha d\tau$. Eq.~\eqref{e:gauss00} implies, for a volume $V$ 
orthogonal to $u^\a$, the quantity $dV/V$ is the 
spatial divergence of $\xi^\alpha$, the divergence in the subspace orthogonal to $u^\a$: 
\[
  \frac{dV}V = q^{\a\b}\na_\a \xi_\b = q^{\a\b}\na_\a u_\b d\tau = \na_\b u^\b d\tau.
\]
\vspace{-3mm}
This is Eq.~\eqref{e:dVdivu}.\\

\noindent 
{\sl Equation of motion} (Relativistic Euler equation): \hspace{3mm} $q^\alpha{}_\gamma \nabla_\beta
T^{\beta\gamma} = 0$
\begin{eqnarray*}
0 &=& q^\alpha{}_\gamma\nabla_\beta [\rho u^\beta u^\gamma +
P q^{\beta\gamma} ]\\
&=& q^\alpha{}_\gamma\rho u^\beta\nabla_\beta u^\gamma +
q^{\alpha\beta}\nabla_\beta P + q^\alpha{}_\gamma P\nabla_\beta (u^\beta
u^\gamma )\\
&=&\rho u^\beta\nabla_\beta u^\alpha + q^{\alpha\beta}\nabla_\beta P +
Pu^\beta\nabla_\beta u^\alpha
\end{eqnarray*}
\be\crv (\rho +P)u^\beta\nabla_\beta u^\alpha =-q^{\alpha\beta}\nabla_\beta P
.\label{e:pcons}\ee
\index{equation of motion!relativistic Euler equation}

\noindent
{\sl Newtonian limit}. Let $e$ be a small parameter of order $v/c$ or $v_{\rm sound}/c$, whichever is larger. 
\begin{align*}
u^\mu &= (1,v^i) + O(e^2), \qquad  P /\rho = O(e^2)\\
 \rho &= {\mbox{ rest mass density }} + O(e^2).
\end{align*}
Then Eq.~\eqref{e:econsderive} implies 
\begin{align*}
\partial_t(\rho u^t) + \partial_i (\rho u^i) &= -P(\partial_t u^t+\partial_iu^i)\\
 \partial_t\rho + \partial_i(\rho v^i) &= 0 + O(e^2),
\end{align*}
in agreement with the conservation of mass equation \eqref{e:masscons}.
\index{conservation laws!energy}\index{conservation laws!mass}
Similarly, Eq.~(\ref{e:pcons}) implies
\begin{align*}
\rho u^\mu\nabla_\mu u^i &= -\nabla^i P \\
 \rho (\partial_t+v^j\nabla_j)v_i&=-\nabla_i P ,
\end{align*}
in agreement with Euler's equation (\ref{110}).

Relativistic energy conservation, Eq.~(\ref{e:econsderive}), also implies the 
Bernoulli equation, expressing energy conservation in a Newtonian 
flow.  We have only looked at its lowest-order form, obtaining 
conservation of mass at order $e^0$; to extract Newtonian energy 
conservation, one must keep terms at the next nonvanishing order, 
order $e^2$. 

To summarize what we have just shown:\\
\noindent The fact that a perfect fluid is locally isotropic implies 
the form of its stress-energy tensor.  The fact that the stress-energy tensor is divergencefree (true for any system) implies all of the equations 
governing a perfect fluid --  the equations that we first obtained by 
following in detail the motion of a fluid element. \\
\newpage

\benr 
\index{entropy!conservation for perfect fluid}
\item Show that the entropy of a fluid element is conserved for a
perfect fluid: The 2nd law of thermodynamics can be stated in the
form $ dE=TdS-PdV$, where\\
$dV=$ the change in a volume $V\perp u^\alpha\ \ $ 
(i.e., $V$ is a comoving volume)\\
$dS=$ the change in entropy of the fluid element\\
$dE=$ the change in energy of the fluid element as measured by a
comoving observer.
\begin{enumerate}\vspace{-3mm}
\item[a.]   Introduce a conserved baryon number density $n$:\\
Conservation of baryons is $\nabla_\alpha(nu^\alpha)=0$.
\index{conservation laws!baryons}
Show that if the 
number of baryons in a fluid element is constant, 
its volume satisfies $\dis u^\alpha\nabla_\alpha V =  V\nabla_\alpha u^\alpha$.
\item[b.]  Show that $u_\beta\nabla_\alpha T^{\alpha\beta}=0$ implies
\[ u\cdot\nabla E=-P u\,\cdot\nabla V.\]
\item[c.] Conclude from the 2nd law that the entropy of a fluid
element is constant along its world line:
\[u\cdot\nabla S\equiv\frac{dS}{d\tau}=0. \]
\end{enumerate}

\item {\sl Doppler Shift.}  \label{ex:doppler}\label{redshift!Doppler shift}
\index{energy!photon}
\index{Doppler shift}
Photons move along null geodesics and  
have null four-momenta: $ p^\alpha p_\alpha = 0$.  An observer 
with velocity $t^\alpha$ decomposes $p^\alpha$ into energy and 
3-momentum in the usual way: $p^\alpha = E t^\alpha + p_\perp^\alpha$,
where $t_\alpha p_\perp^\alpha = 0$.  Because $p^\alpha$ is null, 
the decomposition has the form
\[ 
p^\alpha = E (t^\alpha + n^\alpha),
\]
where $n^\alpha$ is a unit vector along $p_\perp^\alpha$. 
  We can identify the spatial 4-vector $p_\perp^\alpha$ with a 3-vector 
$p^a$ that lies in the 3-space orthogonal to $t^\alpha$: 
In an orthonormal basis with timelike vector $t^\alpha$, 
\[ 
 p^\mu = (E,p^i) = E(1,n^i) \qquad p_\perp^\mu = (0,p^i)\qquad n^\mu = (0,n^i).
\]  
\index{momentum!photon}\index{photon!momentum}\index{photon!frequency}
\index{photon!energy}\index{frequency!photon}
Photon energy and 3-momentum are related to frequency by 
$E = \hbar \omega,\ \ p^a = \hbar k^a$.  If we define a   
wave 4-vector $k^\alpha$ by $\hbar k^\alpha = p^\alpha$, then 
the observer $u^\alpha$ sees frequency $\omega = - k_\alpha t^\alpha$
and wave 3-vector $\hbar k_\perp^\alpha = p_\perp^\alpha$, or 
$k^a = \omega n^a$.   
 
Let $u^\alpha$ be the velocity of a second observer, 
moving at 3-velocity $\bf v$ relative to the first.  
By computing the dot product $ u^\alpha k_\alpha$, 
show that this observer sees a frequency $\omega^\prime$ given by  
\[
 \omega^\prime 	= \omega\gamma (1-{\bf v}\cdot {\bf n}) 
 		= \omega\frac{1-v\cos\alpha}{\sqrt{1-v^2}, } 
\]
with $\alpha$ the angle between the velocity of the second observer 
and the direction of the photon, as seen by the first observer.  

\item\label{beaming} (Similar to Hartle 5.17) {\em Relativistic beaming}.
\index{beaming, relativistic}\index{relativistic beaming} 
This is a challenging problem, and hints follow this set of exercises. Hartle's answer book gets part of it wrong.\\
 A body emits radiation of 
frequency $\omega$, with $j$ photons per unit time emitted isotropically
in the source's rest frame.  An observer sees the body move toward her with speed 
$v$ along her $x'$ axis.     
\benalph
\item Let $\alpha'$ be the angle between a photon path and the source's 
velocity, measured in the observer's basis, and let $\alpha$ be the 
corresponding angle (between the photon path and the source's x-axis), 
measured in the source basis.  Using the 4-vector $k^\alpha$, show
\be 
	\mu' = \frac{\mu+v}{1+v\mu},  \mbox{ where } \mu = \cos\alpha.\tag{$*$}
\label{1}\ee
\item  Find the number flux of photons $j'(\alpha')$ seen by the 
observer, at a fixed distance $R$ from the source (as measured by 
the observer).  Note that, if $\alpha'$ and $\alpha$ are related by 
($*$), the photons emitted by the source at angles 
between $\alpha$ and $\alpha + d\alpha$ are seen by the observer 
at angles between $\alpha'$ and $\alpha'+d\alpha'$.  

\item Find the intensity $f'(\alpha')$, the energy flux measured by 
the observer.   

\item 
 Discuss the beaming of number flux and energy flux as $v\rightarrow 1$:\\  
For $\gamma >> 1$, show that $v= 1-\frac1{2\gamma^2} + O(\gamma^{-4})$.  
Using this and $\cos\alpha'= 1-\frac12\alpha'^2 + O(\alpha'^4)$, show that 
$\dis j' = \frac{8\gamma^3}{(1+\gamma^2{\alpha'}^2)^3} j $, ignoring terms of higher 
than quadratic order in $\alpha', 1/\gamma$.  \vspace{2mm}

For what angles is $j'/j\gg 1$?  For what angles is $j'/j\ll 1$?  
\een

\item {\sl GZK cutoff} (Also adapted from Hartle).
\index{GZK cutoff}\index{Greisen}\index{Zatsepin}\index{Kuz'min}
The highest energy cosmic rays have
energies of about $3\times 10^{20}$ eV.  A celebrated argument by
Greisen,\cite{greisen66} and by Zatsepin, and Kuz'min\cite{zk66} 
shows that they cannot have traveled
farther than about $10^7$ ly, not much farther than our Local Group of
about 30 galaxies (Our local group extends slightly past Andromeda).  
GZK note that particles with high enough energy interact
with photons of the cosmic microwave background (CMB)\index{cosmic microwave background} to produce
pions.  A typical cosmic ray is a proton, and the dominant reactions
are
\[ 
  p + \gamma \rightarrow \ n + \pi^+ \qquad p+\gamma \rightarrow p + \pi^0 . 
\]
The reactions can proceed only if, in the center of mass frame, the energy
is high enough to allow the final particles to emerge with zero speed.
\[ 
p_n^\alpha = m_n u_{CM}^\alpha, \quad p_\pi^\alpha = m_\pi u_{CM}^\alpha.
\]
(At higher energy, they emerge with nonzero relative velocity.)  
Find this threshold energy as follows:
\benalph
\item Write conservation of 4-momentum and square both sides of the equation 
to deduce
\[ 
2 p_{(\gamma)}^\alpha\ p_{{\rm p}\,\alpha} - m_p^2 = - (m_n+m_\pi)^2. 
\]
\item What fractional error does one make in $m_n+m_\pi$ if one 
replaces $m_n$ by $m_p$?
\item Suppose that in a frame for which the CMB looks isotropic (e.g., the 
Earth's rest frame, to about 0.1\% accuracy) the proton and photon 
are moving toward each other. 
Use the 3+1 decomposition of $p_{(\gamma)}^\alpha$ and $p_p^\alpha$ 
in this frame to show (with $m_p = m_n$ and $v=1$) 
\[ 
	E_p^{CMB} = \frac{m_p m_\pi}{2E_\gamma^{CMB}} (1+\frac{m_\pi}{2m_p})
		\approx 3\times 10^{20} \mbox{eV}.
\]

\item \index{photon!mean free path}
The mean free photon path is $\dis\ell = \frac1{n_\gamma\sigma}$, with a 
cross section $\sigma \sim 2\times 10^{-28}$ cm$^2$ and $n_\gamma$ the 
number density of photons in $2.7^\circ$ blackbody radiation, about 
400/cm$^3$.  Check that $\ell$ is the $10^7$ ly mentioned above.  
Protons with lower energy, however,
will also interact if they encounter photons with higher energy in the 
thermal blackbody distribution.  As a result, the interaction will 
also cut off protons of lower energy that come from larger distances.
Use your knowledge of the energy distribution of particles (photons) 
at temperature $T$ to estimate the minimum energy that a proton 
must have to have mean free path $\ell$ less than the radius of the 
visible universe. \\

Hint: First find the number density of photons for which 
$\ell$ is the size of the visible universe; next estimate the energy 
of photons with this number density, in a thermal distribution at 
about $3^\circ $K; finally, for a photon of this energy find the 
threshold energy of the proton, from (c). \\
\een

\een
\vspace{4cm} 
Hints for \ref{beaming}
\begin{figure}[H]
	\begin{center}
		\includegraphics[width=.9\textwidth,angle=180]{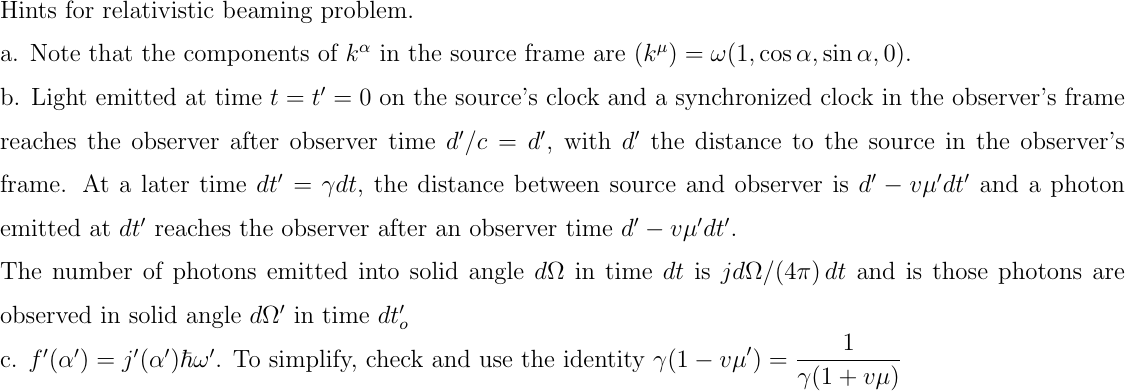}
	\end{center}
\end{figure}

\newpage

\section{Curvilinear Coordinates}\index{curvilinear coordinates|textbf}\index{coordinates!curvilinear|textbf}

	A coordinate system ({\sl chart}, in the mathematics literature) on an $n$-dimensional space $M$ 
is a smooth map assigning to each 
point $P$ of $M$ $n$ numbers $(x^1(P),\dots , x^n(P)\,)$, called the coordinates of
$P$.  That is, a coordinate system is a map $x:  M \rightarrow {\mathbb R}^n$ 
with $x$ smooth and invertible.  
\begin{figure}[h]
\begin{center}
\includegraphics[width=.6\textwidth]{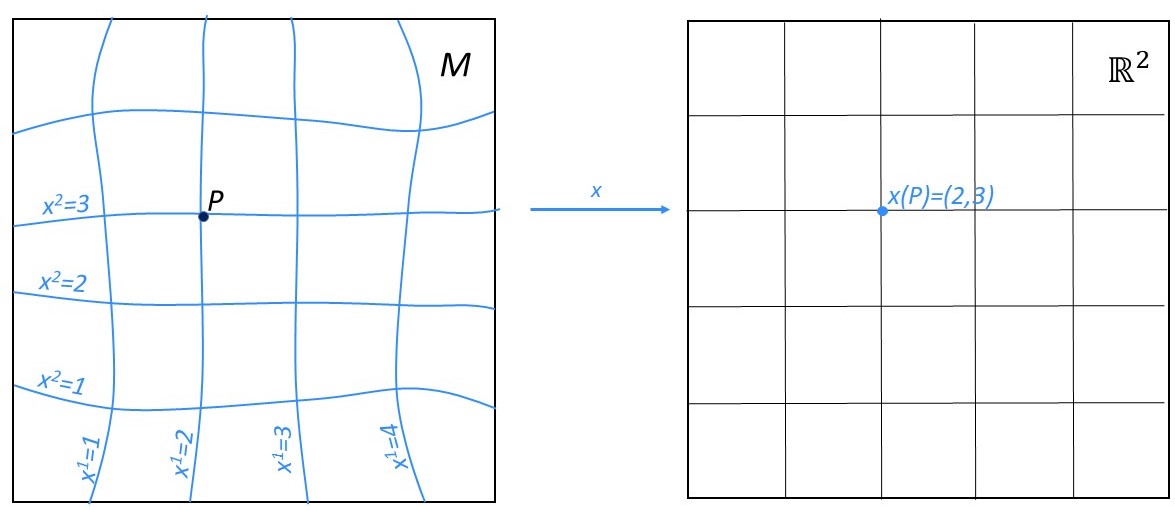}
\end{center}
\caption{A coordinate system $x:M\rightarrow {\mathbb R}^2$ in two dimensions.}
\label{fig:coord1}
\end{figure} 
The natural coordinates $(t, x, y, z)$ on Minkowski space provide an obvious example.   
Another is spherical coordinates 
$(t, r, \theta , \phi )$, although strictly speaking these are coordinates 
on $M$ with the points $x=y=0$ (the history of the symmetry axis) removed:   
$P\ \rightarrow\  (t(P),\, r(P),\, \theta(P))$ is not smooth
at $r=0$, and  $\phi(P)$ is not defined on the z-axis. 

The two-dimensional figure below shows coordinate systems $x:M\rightarrow {\mathbb R}^2$ 
and $y:M\rightarrow {\mathbb R}^2$.  
\begin{figure}[h]
\begin{center}
\includegraphics[width=.6\textwidth]{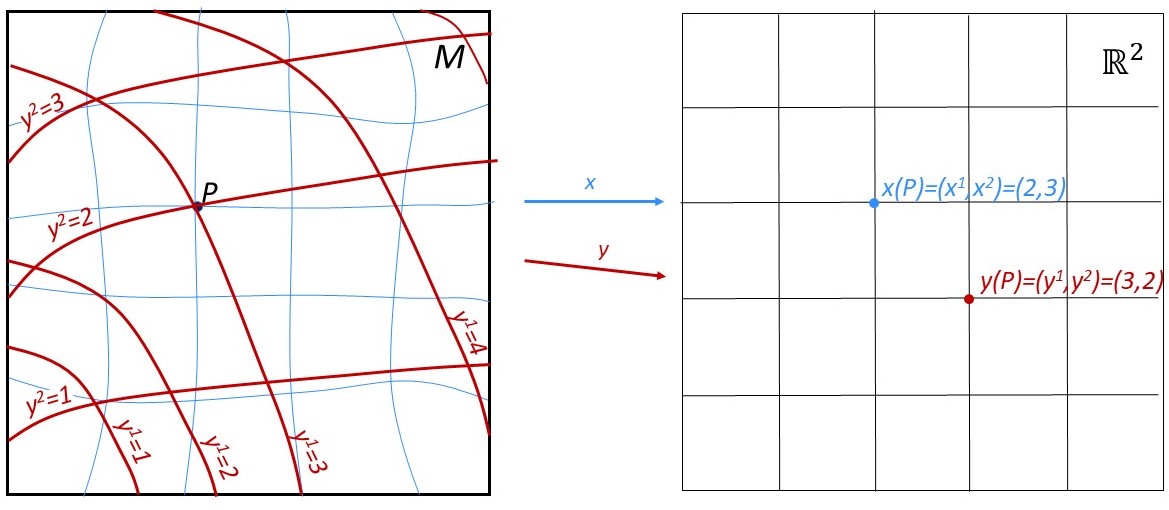}
\end{center}
\label{fig:coord2}
\end{figure} 
The change in coordinates $(y^1,y^2) \mapsto (x^1,x^2)$ is the map 
$x\circ y^{-1}: {\mathbb R}^2 \rightarrow {\mathbb R}^2$, 
smooth and invertible if and only if the Jacobian matrix 
\index{coordinates!change of coordinates}
\[
  \left\| \frac{\partial x^i}{\partial y^j } \right\|
 \]
of $x\circ y^{-1}$ is smooth and has nonzero determinant. In physicists' 
notation, the coordinate transformation is written $x(y)$ instead of 
$x\circ y^{-1}$.  We will see that the Jacobian takes components of 
vectors and tensors from one coordinate system to another.  

\subsection{Components of Vectors and Tensors}
\index{component}
\label{s:components}\index{basis!coordinate basis}\index{vector!component}\index{tensor!component}

A coordinate system on $M$ provides a basis at each point $P$ for the vectors at $P$.  
The radial basis vector $\bm e_1 = \widehat{\bm r}$ at a point $P$ in spherical coordinates, for example, is 
tangent to the path through $P$ with constant $\theta$ and $\phi$ coordinates, the 
path $c(\lambda)$ with coordinates $(\,x^i(\lambda)\,)= (r+\lambda,\theta,\phi)$.  
In general, for a coordinate system $x$, the path along which only $x^1$ changes is the path 
with coordinates
\be 
   \lambda\rightarrow (\, x^1(P)+\lambda, x^2(P), \ldots, x^n(P)\ )
\label{e:path1}\ee
and can be regarded as the $x^1$-axis through $P$. The tangent to this path is the 
first basis vector ${\bm e}_1$ at $P$.
\begin{figure}[ht]
\begin{center}
\includegraphics[width=.3\textwidth]{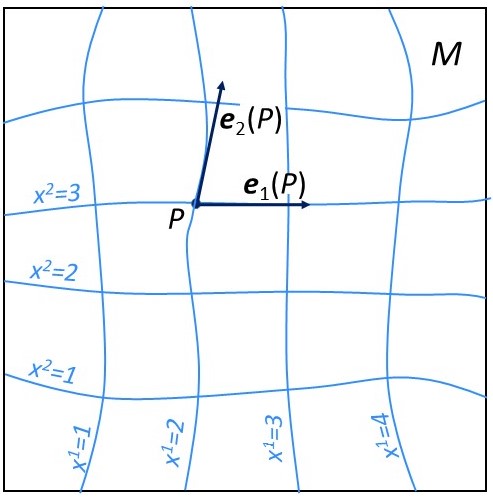}
\end{center}
\label{fig:coord3}
\end{figure} 
\vspace{-5mm}
 
\noindent
Similarly, the tangent to the $x^2$-axis through $P$, to 
the path with  coordinates 
\[ 
	\lambda \rightarrow (\ x^1(P),\, x^2(P) + \lambda ,\, x^3(P),\ldots,\,  x^n(P)\ ),
\]
is the second basis vector ${\bm e}_2$.  

 If $f$ is any function on $M$, the directional derivative of $f$ in the direction 
${\bm e}_1$ is the derivative of $f$ along the path \eqref{e:path1} tangent to ${\bm e}_1$, 
\[ 
  \frac{\partial f}{\partial x^1}(P)= \left.\frac {d} {d \lambda }  f\left(x^1(P) +
			\lambda , x^2(P), \ldots, x^n(P)\right) \right|_{\lambda = 0} 
		= {\bm e}_1\cdot\nabla f.
\]
In abstract index notation, 
\be
 	{\bm e}_1\cdot\nabla f = e_1{}^a \nabla_a  f = \frac{\partial f}{\partial x^1}\ .
\label{ef}\ee 
For this reason, as mentioned earlier, the vectors ${\bm e}_i$ of a coordinate
basis are often written ${\bm\partial}_i$.  Regarded in this way as a directional derivative, the vector $\bm e_i$ is the map, 
\[ 
   \bm e_i(f) = \bm e_i\cdot\nabla f = \partial_i f\,,
\] 
of Eqs.~\eqref{e:vf} and \eqref{e:tf}.

    Return now in notation from a generic dimension to 4-dimensional Minkowski space.  
The basis dual to \{${\bm e}_\mu $\} is just the set of dual vectors
\be 
\bm\omega^0 = {\bm\nabla} x^0,\ \bm\omega^1 = {\bm\nabla} x^1,\ 
\bm\omega^2 = {\bm\nabla} x^2,\ \bm\omega^3 = {\bm\nabla} x^3, 
\ee
because
\be \bm\omega^\mu({\bm e}_\nu ) = \omega^\mu{}_\alpha   e_\nu{}^\alpha  
= e_\nu{}^\alpha \nabla_\alpha  x^\mu 
\tikzmark{First}{=} \frac{\partial x^\mu}{\partial x^\nu} = \delta^\mu_\nu .
\ee

	\begin{tikzpicture}[overlay, remember picture,node distance =.5cm]
		\node[,text width=3cm] (FirstObjectdescr) [below left=of First ]{use (\ref{ef})};
		\draw[,->,thick] (FirstObjectdescr) to [in=-80,out=30] (First);
 	\end{tikzpicture}
\vspace{3mm} 

	The components $v^\mu $ of a vector ${\bm v} $ with respect to the
coordinate basis \{${\bm e}_\mu $\} are called the components of ${\bm v} $
in the coordinate system $x$.  If $x'$ is another coordinate system, with
\{${\bm e}_{\mu'}$\} the associated basis, ${\bm e}_{\mu'}$ can be
related to ${\bm e}_\mu$ in this way:

\be 
{\bm e}_{\mu'}(f) = 
\frac{\partial f}{\partial x^{\mu'}} 
= \frac{\partial f}{\partial x^{\mu}}\frac{\partial x^\mu }{\partial x^{\mu'}}  
= \frac{\partial x^\mu}{\partial x^{\mu'}} {\bm e}_\mu (f).
\ee
That is, 
\be 
 {\bm e}_{\mu'} = \frac{\partial x^\mu}{\partial x^{\mu'}} {\bm e}_\mu  
\label{transf3}\ee
(where $\dis\left\|\frac{\partial x^\mu}{\partial
x^{\mu'}}\right\|$ is the Jacobian matrix of the map $x\circ x'^{-1}$).  
Then
\[ {\bf v}  = v^{\mu'}{\bm e}_{\mu'} = v^\mu{\bm e}_\mu  =
v^\mu \frac{\partial x^{\mu'}}{\partial x^\mu}{\bm e}_{\mu'}\ , \]
\be \Longrightarrow v^{\mu'} 
 = \frac{\partial x^{\mu'}}{\partial x^\mu} v^\mu,
\ee
where we used (\ref{transf3}) with primed $\leftrightarrow$ unprimed.

	Similarly, the dual basis vectors $\bm\omega^{\mu'}$ 
are related to $\bm\omega^\mu $ by 
\be 
\bm\omega^{\mu'} = {\bm\nabla} x^{\mu'} 
		=  \frac{\partial x^{\mu'}}{\partial x^\mu} {\bm\nabla} x^\mu 
		= \frac{\partial x^{\mu'}}{\partial x^\mu}\bm\omega^\mu,
\label{omega'}\ee
whence the components $\sigma_\mu $ of a dual vector transform in the manner
\[ 
\sigma_{\mu'}\bm\omega^{\mu'} = \sigma_\mu \bm\omega^\mu
	= \sigma_\mu \frac{\partial x^\mu}{\partial x^{\mu'}}\bm\omega^{\mu'} 
\]
\be \Rightarrow 
\sigma_{\mu'} = \frac{\partial x^\mu }{\partial x^{\mu'}}\sigma_\mu\ \ . 
\ee
(This also follows from (\ref{omega'}) together with $\sigma_\mu v^\mu =
\sigma_{\mu'} \, v^{\mu'}$.)\\

	For arbitrary tensors $T^{\alpha \cdots \beta}{}_{\gamma\cdots\delta}$,
we have
\be 
  T^{\mu'\cdots\nu'}{}_{\sigma' \cdots \tau'} 
  = \frac{\partial x^{\mu'}}{\partial x^\mu } \cdots 
    \frac{\partial x^{\nu'}}{\partial x^\nu} 
    \frac{\partial x^\sigma}{\partial x^{\sigma'}} \cdots 
    \frac{\partial x^\tau}{\partial x^{\tau'}} 
    T^{\mu \cdots \nu}{}_{\sigma \cdots \tau}. 
\label{t'}\ee
\index{passive transformation!coordinate transformation}
\noindent This is just a special case of the transformation law (\ref{transf2}),
where, in the case of coordinate bases, the $a^{\mu'}{}_\mu$ and
$a^\mu {}_{\mu'}$ matrices are the Jacobian matrices of the
transformations $x'\circ x^{-1}$ and $x\circ x'^{-1}$:

\be a^{\mu'}{}_\mu  =  \frac{\partial x^{\mu'}}{\partial
x^\mu },\qquad a^\mu {}_{\mu'} = \frac{\partial x^\mu }{\partial
x^{\mu'}} .\ee
Because $P$ has coordinates $x'(P)$ in the primed system and $x(P)$
in the unprimed system, Eq. (\ref{t'}) is sometimes written with explicit arguments $x'$ and $x$: 

\be 
T^{\mu' \cdots}{}_{ \cdots \tau'}(x') 
= \frac{\partial x^{\mu'}}{\partial x^\mu} 
	\cdots \frac{\partial x^\tau}{\partial x^{\tau'}} 
	T^{\mu \cdots }{}_{\cdots \tau}(x). 
\ee

\subsection{Covariant Derivatives}
\index{covariant derivative!flat spacetime}
\index{derivative!covariant derivative}

	We want to find the coordinate components of 
$\nabla_\alpha  T^{\beta\cdots \gamma}{}_{\delta\cdots\epsilon}$ 
in terms of the partial derivatives
\[ 
\frac{\partial }{\partial x^\lambda} 
		T^{\mu\cdots\nu}{}_{\sigma\cdots\tau} .
\]
In abstract-index notation, the covariant derivative of a vector $v^\alpha$ 
is unambiguously written $\nabla_\alpha v^\beta$.  An ambiguity in 
notation arises when we write the components of $\nabla_\alpha v^\beta$
in a basis: Because the $\nu$th component 
$v^\nu \equiv  \bm \omega^\nu(\bm v)$ of a vector field 
$v^\alpha$ is a scalar field, the $\mu$th component of its gradient 
is $\partial_\mu v^\nu$. Since we write the gradient of a scalar  
in the form $\nabla_\alpha f$, the $\mu$th component of $v^\nu$ could 
be written $\nabla_\mu v^\nu$, but this is {\sl not} the way the 
symbol $\nabla_\mu v^\nu$ is used by physicists. Instead, it is natural to 
write the $\mu$-$\nu$th component of the tensor $\nabla_\alpha v^\beta$ 
as $\nabla_\mu v^\nu$.  When we use curvilinear coordinates (or a basis that 
is not a coordinate basis, or when we work in a curved
spacetime) the two possibilities are not equivalent, because the basis vectors 
are not constant vector fields: The gradient of 
the scalar $v^\nu$ has as its $\mu$th component  
\[ 
\nabla_\mu\, (v^\nu) = \nabla_\mu (v^\alpha \omega^\nu_{\ \alpha}) 
		    = \partial_\mu (v^\alpha \omega^\nu_\alpha) 
		    = \partial_\mu v^\nu; 
\]
while the $\mu$-$\nu$th component of the tensor $\nabla_\alpha v^\beta$ is 
\be
(\nabla_\mu\, v^\alpha) \omega^\nu_{\ \alpha} 
	= \nabla_\mu (v^\alpha\omega^\nu_{\ \alpha}) 
		- v^\alpha \nabla_\mu\omega^\nu_{\ \alpha}
	= \partial_\mu v^\nu - v^\alpha \nabla_\mu\omega^\nu_{\ \alpha}.
\label{delv}\ee
The second expression involves the covariant directional derivative 
of the basis vector $\bm\omega^\nu$ in the direction of ${\bm e}_\mu$.
For the gradient of the scalar $v^\nu$ (rarely needed), these notes will use the 
expression $\nabla_\alpha(v^\nu)$, with $\mu$th component 
\be
\nabla_\mu (v^\nu):= \partial_\mu v^\nu\ ;
\ee
the parentheses means that the component is found {\em before} the 
derivative is evaluated.  We could similarly write the $\mu$-$\nu$th 
component of $\nabla_\alpha v^\beta$
in the form $(\nabla_\mu v)^\nu$ to make clear that the component was 
computed {\em after} the covariant derivative was evaluated, but we will 
follow the standard physics notation, omitting the parentheses, and writing 
\be
 \nabla_\mu\, v^\nu := (\nabla_\mu v)^\nu 
		     = (\nabla_\mu v^\alpha) \omega^\nu_{\ \alpha}
		     = (e_\mu^{\ \beta} \nabla_\beta v^\alpha) \omega^\nu_{\ \alpha}\ .
\ee

The expression analogous to Eq. (\ref{delv}) for the covariant derivative
of a dual vector $\sigma_\alpha$ is then 
\begin{eqnarray*} 
\nabla_\mu \sigma_\nu 
& = & e_\mu{}^\alpha  e_\nu{}^\beta \nabla_\alpha \sigma_\beta  \nonumber \\
&=& 	  \nabla_\mu(e_\nu {}^\beta \sigma_\beta)
	- (\nabla_\mu e_\nu{}^\beta)\sigma_\beta. 
\end{eqnarray*}
Again, for a coordinate basis, $\nabla_\mu f = \partial_\mu f$, and we have 
\be 
\nabla_\mu \sigma_\nu = \partial_\mu \sigma_\nu 
			-(\nabla_\mu e_\nu{}^\beta)\sigma_\beta. 
\label{delsigma}\ee

The vector $\nabla_\mu \bm e_\nu$ is the covariant directional derivative of 
$\bm e_\nu$ in the direction of $\bm e_\mu$.  Because it is a vector, 
it can be written as a linear combination of basis vectors, namely 
\be \crv
 \nabla_\mu {\bm e}_\nu = \Gamma^\lambda{}_{\nu \mu }\bm e_\lambda \cb
\label{christ1}\ee
The coefficients $\Gamma^\lambda{}_{\nu \mu}$ are called 
Christoffel symbols (or, for an orthonormal basis, Ricci rotation coefficients).
\index{Christoffel symbol|textbf}\index{Ricci rotation coefficients}
We have
\be
\Gamma^\lambda{}_{\nu \mu} 
 = e_\mu{}^\alpha\nabla_\alpha (e_\nu{}^\beta)\ \omega^\lambda_\beta . 
\label{christ2}\ee

Using the defining equation, (\ref{christ1}), Eq. (\ref{delsigma}) for the 
components of $\nabla_\alpha\sigma_\beta$ becomes
\be 
\cblue\nabla_\mu \sigma_\nu \cb f
		= \partial_\mu \sigma_\nu 
		  - \Gamma^\lambda{}_{\nu\mu}\,e_\lambda^{\ \beta}\,\sigma_\beta
		= \cblue \partial_\mu \sigma_\nu
			- \Gamma^\lambda{}_{\nu\mu}\,\sigma_\lambda .\cb
\label{delsigma2}\ee
Observing that
\begin{eqnarray*}  
\nabla_\mu (\omega^\nu{}_\beta)e^\beta {}_\lambda
&=& \nabla_\mu (\omega^\nu{}_\beta  e^\beta{}_\lambda) - 
(\nabla_\mu e^\beta{}_\lambda )\omega^\nu{}_\beta   \\
&=& \partial_\mu \delta^\nu{}_\lambda - \Gamma^\nu{}_{\lambda\mu }\\
&=& -\Gamma^\nu{}_{\lambda\mu } ,
\end{eqnarray*} 
we obtain from (\ref{delv}) the components of $\nabla_\beta v^\alpha$
\be \cblue
	\nabla_\mu v^\nu 
		= \partial_\mu v^\nu + \Gamma^\nu{}_{\lambda \mu }v^\lambda. \cb
\label{delv2}\ee
Equation (\ref{delv2}) could also have been obtained from (\ref{delsigma2}),
using
$\nabla_\mu (v^\nu \sigma_\nu ) = \partial_\mu (v^\nu \sigma_\nu )$ 
and the Leibnitz rule, 
$\nabla_\mu (v^\nu \sigma_\nu) = (\nabla_\mu v^\nu)\sigma_\nu 
				+ v^\nu \nabla_\mu \sigma_\nu$.

With components of the covariant derivative of vectors and 
dual vectors computed, the components of 
$\nabla_\epsilon T^{\alpha\cdots\beta}{}_{\gamma\cdots\delta}$ follow 
quickly. One acquires one $\Gamma$ with a + sign for each up index and 
one $\Gamma$ with a $-$ sign for each down index: 
\bea
\nabla_\lambda T^{\mu \cdots \nu}{}_{\sigma \cdots \tau } 
&=&	\omega^\mu{}_\alpha \cdots \omega^\nu{}_\beta  
	e_\sigma{}^\gamma \cdots e_\tau{}^\delta 
	\nabla_\lambda T^{\alpha\cdots\beta}{}_{\gamma\cdots\delta}
\nonumber\\
&=& \nabla_\lambda (T^{\alpha\cdots\beta}{}_{\gamma\cdots\delta} 
 		\omega^\mu{}_\alpha \cdots \omega^\nu{}_\beta  
		e_\sigma{}^\gamma \cdots e_\tau{}^\delta)
\nonumber\\
&& - T^{\alpha\cdots\beta}{}_{\gamma\cdots\delta}
 \nabla_\lambda (\omega^\mu {}_\alpha) \cdots \omega^\nu{}_\beta  
 e_\sigma{}^\gamma \cdots e_\tau{}^\delta
 \ - \cdots
   - \ T^{\alpha\cdots\beta}{}_{\gamma\cdots\delta}
  	\omega^\mu {}_\alpha \cdots (\nabla_\lambda \omega^\nu{}_\beta )
	e_\sigma{}^\gamma \cdots e_\tau{}^\delta 
\nonumber\\
&& - T^{\alpha\cdots\beta}{}_{\gamma\cdots\delta}
	\omega^\mu {}_\alpha   \cdots \omega^\nu{}_\beta   
	(\nabla_\lambda e_\sigma^{\ \gamma})\cdots e_\tau^{\ \delta} 
 \ - \cdots
   -  T^{\alpha\cdots\beta}{}_{\gamma\cdots\delta}
 	\omega^\mu{}_\alpha\cdots \omega^\nu{}_\beta 
 	e_\sigma^{\ \gamma}\cdots (\nabla_\lambda  e_\tau^{\ \delta})
\nonumber\\
\nabla_\lambda T^{\mu \cdots \nu}{}_{\sigma \cdots \tau } 
 &=& \partial_\lambda T^{\mu \cdots \nu}{}_{\sigma \cdots \tau } +
\Gamma^\mu{}_{\iota\lambda}T^{\iota \cdots \nu}{}_{\sigma \cdots \tau }
 +\cdots 
 + \Gamma^\nu{}_{\iota\lambda} T^{\mu \cdots \iota}{}_{\sigma \cdots \tau }
 \nonumber\\
&&\phantom{xxxxxxxx}   
   - \Gamma^\iota{}_{\sigma\lambda} T^{\mu \cdots \nu}{}_{\iota \cdots \tau}
   -\cdots
   - \Gamma^\iota{}_{\tau\lambda}T^{\mu \cdots \nu}{}_{\sigma \cdots \iota} . 
\label{delt}\eea
\noindent 

Eq. (\ref{delt}) could also have been obtained from Eqs. (\ref{delsigma2}) and 
(\ref{delv2}), using the Leibnitz rule and the fact that any 
$T^{\alpha \cdots \beta}{}_{\gamma\cdots \delta}$ is the sum of
outer products of vectors and dual vectors.  The derivation above
is a little more direct.  In this derivation, the only place where we 
used a the fact that $\{e_\mu \}$ is a coordinate basis is in replacing
the directional derivative $\nabla_\mu f$ by $\partial_\mu f$. 
In a general basis, Eq. (\ref{delt}) has the form
\bea
\nabla_\lambda T^{\mu \cdots \nu}{}_{\sigma \cdots \tau }
 &=& \nabla_\lambda ( T^{\mu \cdots \nu}{}_{\sigma \cdots \tau } )
 +\Gamma^\mu{}_{\iota\lambda}T^{\iota \cdots \nu}{}_{\sigma \cdots \tau }
 +\cdots
 + \Gamma^\nu{}_{\iota\lambda} T^{\mu \cdots \iota}{}_{\sigma \cdots \tau }
  \nonumber\\
 &&\phantom{xxxxxxxx}
 - \Gamma^\iota{}_{\sigma\lambda} T^{\mu \cdots \nu}{}_{\iota \cdots \tau}
 -\cdots
 - \Gamma^\iota{}_{\sigma\lambda}T^{\mu \cdots \nu}{}_{\sigma \cdots \iota} . 
\label{delt2}
\eea

We next evaluate $\Gamma^\lambda{}_{\nu \mu }$ when
$\{e_\mu \}$ \textit{is} a coordinate basis.  

\par First note that in a preferred coordinate system 
\[ 
\nabla_\mu \nabla_\nu f = \frac{\partial^2f}{\partial x^\mu \partial x^\nu}; 
\]
because partial derivatives commute for smooth scalars $f$,
\[ \nabla_{[\mu }\nabla_{\nu ]}f = 0. 
\]
(In fact, because all components of $\nabla_{[\alpha}\nabla_{\beta]}f$ 
vanish in some basis, they must vanish in any basis:
$ \nabla_{[\alpha}\nabla_{\beta]}f = 0 $.)

Now in a coordinate basis, we have $\nabla_\mu f = \partial_\mu f$,
and our equation for the covariant derivative of a dual vector, 
(\ref{delsigma2}), implies 
\[ 
 \nabla_\mu \nabla_\nu f
  = \partial_\mu \partial_\nu f - \Gamma^\lambda{}_{\nu \mu} \partial_\lambda f,
\]
\[
\nabla_{[\mu}\nabla_{\nu]}f = 0 
	= \partial_{[\mu}\partial_{\nu]}f 
	- \Gamma^\lambda{}_{[\nu\mu]} \partial_\lambda f
	= - \Gamma^\lambda{}_{[\nu \mu]} \partial_\lambda f
\]
That is, {\cblue for a coordinate basis}, $\Gamma^\lambda{}_{[\mu \nu]} = 0$: 
$\Gamma^\lambda{}_{\mu \nu}$ is symmetric in $\mu$ and $\nu$:
\be \cblue
\Gamma^\lambda {}_{\mu \nu } = \Gamma^\lambda{}_{\nu \mu }  \cb\ .
\ee
Once the symmetry of $\Gamma^\lambda{}_{\mu \nu }$ is established, its value
can be found in terms of the components $\eta_{\mu \nu}$ of the metric.
To do this one uses the equation
\be 
	\nabla_\alpha  \eta_{\beta\gamma} = 0 .
\label{deleta}\ee
The components of (\ref{deleta}) are
\[ \nabla_\mu \eta_{\nu\lambda} 
 = \partial_\mu \eta_{\nu\lambda} - \Gamma^\sigma {}_{\nu\mu}\eta_{\sigma\lambda } 
	- \Gamma^\sigma{}_{\lambda\mu} \eta_{\nu \sigma } = 0 .\]
Defining
\[ \Gamma_{\lambda\mu\nu} := \Gamma^\sigma{}_{\mu\nu}\eta_{\sigma\lambda}, \]
we have
\[ \partial_\mu \eta_{\nu\lambda} = \Gamma_{\lambda \nu \mu }
+ \Gamma_{\nu \lambda \mu }.\]
To find $\Gamma{}_{\lambda \nu \mu }$ one cyclically permutes $\lambda, \mu,
\nu $:
\[ \partial_\nu \eta_{\lambda \mu } = \Gamma_{\mu \lambda \nu }
+ \Gamma_{\lambda\mu \nu  } \]
\[ \partial_\lambda\eta_{\mu\nu} = \Gamma_{\nu \mu \lambda }
+ \Gamma_{\mu \nu \lambda } \]
Add the first two equations and subtract the last:
\[ 
\partial_\mu \eta_{\nu\lambda } + \partial_\nu \eta_{\lambda\mu } 
	- \partial_\lambda \eta_{\mu \nu } 
= \Gamma_{\lambda\nu\mu} + \cancel{\Gamma_{\nu\lambda\mu}}{-10}
 + \cancell{\Gamma_{\mu \lambda \nu}}{-10} + \Gamma_{\lambda \mu \nu} 
 - \cancel{\Gamma_{\nu\mu\lambda}}{-10} - \cancell{\Gamma_{\mu\nu\lambda}}{-10}
\]
\[\partial_\mu \eta_{\nu \lambda } + \partial_\nu \eta_{\mu
\lambda } - \partial_\lambda \eta_{\mu \nu } = 2\Gamma_{\lambda \mu
\nu }. \]
(The symmetry of $\Gamma$ was used to combine or cancel each pair of terms.)  Then  
\be
 \Gamma_{\lambda \mu \nu } = \frac{1}{2}[\partial_\mu \eta
_{\nu \lambda} + \partial_\nu \eta_{\mu \lambda } -\partial_\lambda
\eta_{\mu \nu }]
\label{gammadown}\ee
and 
\index{Christoffel symbol!coordinate basis}
\be \cblue
\Gamma^\lambda_{\ \mu \nu} = \frac{1}{2}\eta^{\lambda\iota}
		[\partial_\mu \eta_{\nu \iota} + \partial_\nu \eta_{\mu\iota} 
		- \partial_\iota \eta_{\mu \nu }].\cb
\label{gamma up}\ee
As soon as we go to a curved spacetime, the same steps will give the same result 
in terms of an arbitrary metric $g_{\alpha\beta}$.

\subsection{Examples}
\index{Christoffel symbol!geometric example}

\noindent{\sl Example 1}: Consider Minkowski space with two spatial dimensions 
in the coordinate system $t, r, \phi$:
\be x = r\;\cos\phi, \hspace{.5in} y = r\;\sin\phi,\hspace{.5in} \phi =
\tan^{-1 }\frac{y}{x},\hspace{.5in}r = (x^2+y^2)^{1/2}\ee 
The covariant basis vectors are 
\be 
\omega^t{}_\alpha = \nabla{}_\alpha t, \  \ 
\omega^\phi{}_\alpha = \nabla_\alpha \phi \ \ \mbox{ and } \ \ 
\omega^r{}_\alpha   = \nabla_\alpha r \ .
\ee
with
\begin{eqnarray}
\nabla_\alpha  x &=& \frac{\partial x}{\partial \phi } \nabla_\alpha  \phi +
\frac{\partial x}{\partial r} \nabla_\alpha  r   \nonumber\\
&=&\cos\phi \;\nabla_\alpha  r-r\;\sin\phi \;\nabla_\alpha   \phi \label{delx}\\
\nabla_\alpha  y &=& \frac{\partial y}{\partial r } \nabla_\alpha  r 
+ \frac{\partial y}{\partial \phi} \nabla_\alpha   \phi  \nonumber\\
&=&\sin\phi \;\nabla_\alpha  r + r\; \cos\phi \;\nabla_\alpha  \phi. 
\label{dely}\end{eqnarray}
Then
\begin{eqnarray} 
\eta_{\alpha\beta}  &=& -\nabla_\alpha  t \nabla_\beta  t 
	+ \nabla_\alpha  x \nabla_\beta  x
	+ \nabla_\alpha  y \nabla_\beta  y \label{e:eta}\\
&=& -\nabla_\alpha  t \nabla_\beta  t 
	+ (\cos\phi \nabla_\alpha  r - r\sin\phi \nabla_\alpha\phi)
	  (\cos\phi \nabla_\beta r - r\sin\phi \nabla_\beta  \phi )\nonumber \\
&& + (\sin\phi \nabla_\alpha  r + r\cos\phi \nabla_\alpha\phi)
	(\sin\phi \nabla_\beta  r + r\cos\phi \nabla_\beta  \phi)\nonumber \\
&=& -\nabla_\alpha  t \nabla_\beta  t + (\cos^2\phi + \sin^2\phi)\nabla_\alpha r\nabla_\beta r 
	+ \underbrace{(-r\sin\phi \cos\phi + \sin\phi r\cos\phi )}_0
	   \nabla_\alpha  r\nabla_\beta \phi 	\nonumber \\
&&+ r^2(\sin^2\phi + \cos^2\phi )\nabla_\alpha  \phi \nabla_\beta  \phi 
   + \underbrace{(-r\sin\phi \cos\phi + r\sin\phi \cos\phi )}_0
           \nabla_\alpha  \phi \nabla_\beta  r,  
\label{e:etasph}\end{eqnarray}
giving
\be 
	\eta_{\alpha\beta}  = -\nabla_\alpha  t \nabla_\beta  t 
				+ \nabla_\alpha  r \nabla_\beta  r 
				+ r^2\nabla_\alpha  \phi \nabla_\beta   \phi .
\label{eta'}\ee 
Thus
\[ 
\eta_{tt} = -1, \hspace{.5in} \eta_{rr} = 1, \hspace{.5in} \eta_{\phi
\phi} = r^2, \hspace{.5in} \text{and} \ \  \eta_{\mu \nu } = 0,
\ \ \mu \neq \nu.
\]
One could also have found $\eta_{\mu \nu }$ by, e.g.,
\begin{eqnarray*}
	\eta_{\phi \phi} &=& \frac{\partial t}{\partial \phi}\frac{\partial t}{\partial \phi}~\eta_{tt} 
			    + \frac{\partial t}{\partial\phi} \frac{\partial x}{\partial \phi}\eta_{tx} 
			    + \cdots + \frac{\partial y}{\partial \phi} \frac{\partial y}{\partial \phi}\eta_{yy}\\
			  &=& \left(\frac {\partial x}{\partial \phi}\right)^2 
				+ \left(\frac {\partial y}{\partial\phi}\right)^2 = r^2\sin^2\phi + r^2\cos^2\phi 
			   = r^2 ,
\end{eqnarray*}
but the calculation from \eqref{e:eta} to \eqref{e:etasph} is more efficient.  

The following notation makes coordinate calculations look simpler and more natural: Write
\be
 \mbox{Write }\quad dx^\mu \hspace{.3in} \mbox{for} \hspace{.3in}
			\nabla_\alpha  x^\mu = \omega^\mu{}_\alpha  .
\ee
Then, writing $\partial_\mu$ for $e^\alpha_\mu$, we have 
$dx^\mu (\partial_\nu)=\delta^\mu_\nu$; 
the metric tensor 
$\eta_{\alpha\beta} = \eta_{\mu \nu }\nabla_\alpha x^\mu \nabla _\beta  x^\nu$ 
is $\eta_{\mu \nu}dx^\mu dx^\nu $ and is abbreviated $ds^2$:
\be
 ds^2 = \eta_{\mu \nu }dx^\mu dx^\nu.  
\ee
The change of basis resulting from $x^\mu \rightarrow x^{\mu'}$ is
written
\[
 dx^{\mu'} = \frac{\partial x^{\mu'}}{\partial x^\mu }dx^\mu \qquad
 \partial_{\mu'} = \frac{\partial x^\mu }{\partial x^{\mu'}}\partial_\mu, 
\] 
and the change in metric components is then
\beaa
ds^2 &=& \eta_{\mu'\nu'}dx^{\mu'}dx^{\nu'} 
	= \eta_{\mu \nu }dx^\mu dx^\nu \\ 
&=& \left(\eta_{\mu \nu }\frac{\partial x^\mu}{\partial x^{\mu'}}
	\frac{\partial x^\nu}{\partial x^{\nu'}}\right)dx^{\mu'}dx^{\nu'}.
\end{eqnarray*}
Eq. (\ref{e:eta}) takes the form 
\be 
	ds^2 = - dt^2 + dx^2 + dy^2 , 
\label{dsq}\ee
Eqs. (\ref{delx}) and (\ref{dely}) become
\begin{eqnarray} 
	dx &=& \cos\phi \;dr - r\sin\phi \;d\phi \nonumber \\
	dy &=& \sin\phi \;dr + r\cos\phi \;d\phi ,
\label{delxy'}\end{eqnarray}
and, by substituting (\ref{delxy'}) into (\ref{dsq}), we obtain
\be 
	ds^2 = -dt^2 + dr^2 + r^2d\phi^2 .
\label{dsq'}\ee
Thus, by identifying the ``infinitesimals" $dx^i$ that had no precise
meaning (by themselves) with the coordinate basis vectors $\nabla_\alpha  x^i$ --
the gradients of the scalar fields $x^i$ -- the intuitively clear but
mathematically ill-defined equations (\ref{dsq})-(\ref{dsq'}) are given 
a precise meaning:  they are the tensor equations (\ref{delx})-(\ref{eta'}).\\
An occasionally encountered notation for the contravariant 
metric is 
\[ \partial^2_s = -\partial^2_t + \partial^2_r +
\frac{1}{r^2}\;\partial^2_\phi .\]
(This does {\sl not} denote and is not the Laplacian in polar coordinates.)
 
Christoffel symbols:  Because the only non-constant metric component in 
(\ref{dsq'}) is \\
$\eta_{\phi \phi } = r^2$, the only nonzero components 
 $\Gamma_{\lambda \nu \mu}$ are
\begin{align} 
  \Gamma_{r\phi \phi } &= - \frac{1}{2}\partial_r\eta_{\phi \phi} 
	                = -r \ ,\nonumber \\ 
  \Gamma_{\phi r\phi } &= \Gamma_{\phi \phi r} 
		        = \frac{1}{2}\;\partial_r\eta_{\phi \phi} 
		        = r, 
\end{align}
and the only nonzero $\Gamma^\lambda {}_{\nu \mu }$'s are
\be 
\Gamma^r{}_{\phi \phi} = -r \hspace{.5in} \Gamma^\phi{}_{r\phi } = \
\Gamma^\phi{}_{\phi r} = \eta^{\phi \phi }\Gamma_{\phi r \phi } = \frac{1}{r}. 
\label{gammar}
\ee
To see what (\ref{gammar}) means, look at the basis vectors $\bm\partial_r$ and
$\bm\partial_\phi $ in an $r$-$\phi$ plane (a $t =$ constant plane):\\
\be 
\bm\partial_r  \mbox{ is a unit vector: } \qquad\qquad\sqrt{\eta(\bm\partial_r, \bm\partial_r)} = \sqrt{\eta_{rr}} = 1
		\hspace{5cm}
\ee
\be 
  \bm\partial_\phi  \mbox{ has length $r$: } \ \ \qquad\qquad\sqrt{\eta(\bm\partial_\phi, \bm\partial_\phi)} = \sqrt{\eta_{\phi \phi }} = r\hspace{5.2cm}
\ee
That is, if $\widehat{\bm\phi}$ is a unit vector in the $\bm\partial_\phi $ direction, 
$\dis\widehat{\bm\phi}(f) =\widehat{\bm\phi}\cdot \nabla f = \frac{1}{r}\partial_\phi f$, whereas 
$\bm\partial_\phi(f) = \partial_\phi f$, so
$\dis\bm\partial_\phi = r\widehat{\bm\phi}$.
\begin{figure}[ht]
\begin{center}
\includegraphics[width=.4\textwidth]{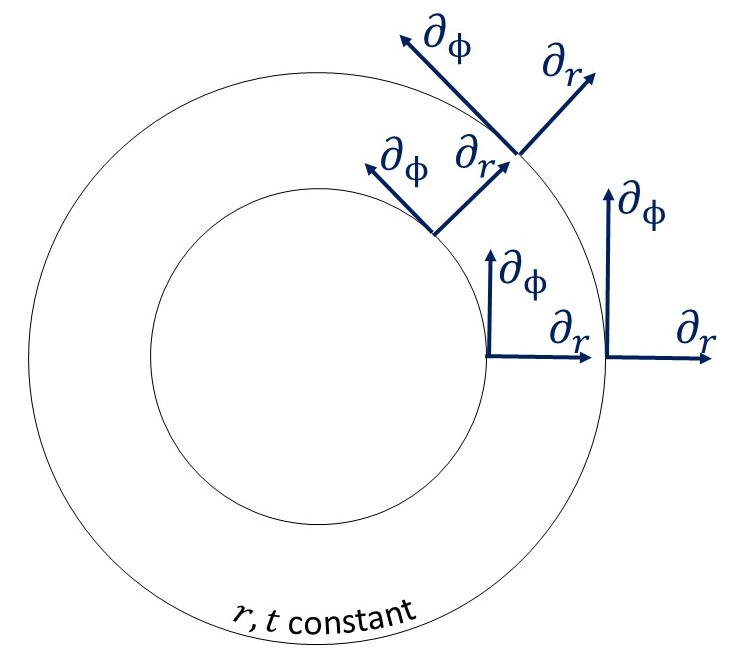}
\end{center}
\caption{The basis vectors $\bm\partial_r$ and $\bm\partial_\phi$}
\label{fig:partial}
\end{figure}

\newpage
  Along a curve of constant $r$ and $\phi$, when only $t$ changes,
$\bm\partial_\phi$ and $\bm\partial_r$ don't change, so\\
 $\nabla_t\bm\partial_\phi = \nabla_t\bm\partial_r = 0$,
implying $\Gamma^\lambda{}_{\phi t} = 0,\ \ \Gamma^\lambda {}_{rt} = 0$.  \\
The $r$-$\phi $ plane is more interesting:
Along a curve of constant $\phi $ and $t$, $\bm\partial_r$ is constant but
$\bm\partial_\phi $ changes its length: Since the length of $\bm\partial_\phi $
is $r$, and along a $\phi $ = constant curve the direction of 
$\bm\partial_\phi $ doesn't change,\\
$\dis\left.\bm\partial_\phi\, \right|_{r+\Delta r} 
	= \frac{r+\Delta r}{r} \left.\bm\partial_\phi\, \right|_r$ \\
$\dis\nabla_r\bm\partial_\phi 
 = \lim_{\Delta r \rightarrow 0}
 \left(\bm\partial_\phi\, |_{r+\Delta r}-{\bm\partial }_\phi\, |_r\right) 
 \frac{1}{\Delta r} = \frac{1}{r}\bm\partial_\phi $ .\\
Then
\[ 
  \Gamma^\phi{}_{\phi r} = \frac{1}{r} \hspace{.5in} \text{and}\qquad 
  \Gamma^r{}_{\phi r} = \Gamma^t{}_{\phi r} = 0.
\]

 We can similarly find $\nabla_\phi \bm\partial_r$ and 
$\nabla_\phi \bm\partial_\phi $: Along an $r, t$ =
constant curve, the lengths of $\bm\partial_\phi $ and $\bm\partial_r$ are
constant but their directions change -- see Fig.~\ref{fig:partial}.\\

$\dis\bm\partial_\phi |_{\phi + \Delta \phi} 
 = \cos\Delta\phi \; \bm\partial_\phi|_\phi 
 		- r\sin\Delta \phi \; \bm\partial_r|_\phi$\\

$\dis\bm\partial_r|_{\phi + \Delta \phi} 
	= \cos\Delta \phi \; \bm\partial_r|_\phi
	  +  \frac1r\sin\Delta \phi \; \bm\partial_\phi|_\phi$\\  

$\dis\nabla_\phi \bm\partial_\phi 
 = \lim_{\Delta \phi\rightarrow 0} \frac{1}{\Delta \phi }\;
 	[\bm\partial_\phi |_{\phi + \Delta\phi}	- \bm\partial_\phi |_\phi ] 
 	= -r\; \bm\partial_r$ \\

$\dis\nabla_\phi \bm\partial_r 
  = \lim_{\Delta \phi\rightarrow 0} \frac{1}{\Delta \phi}\;
  	[\bm\partial_r|_{\phi + \Delta \phi}
	- \bm\partial_r|_\phi ] = \frac{1}{r}\; \bm\partial_\phi $ \\
\beaa \Rightarrow \;\Gamma^r{}_{\phi \phi } &=&  -r \qquad
		    \Gamma^t{}_{\phi \phi} =\Gamma^\phi{}_{\phi \phi} = 0\\
		   \Gamma^\phi {}_{r \phi} &=& \frac{1}{r}\qquad\ 
		   \Gamma^t{}_{r \phi} = \Gamma^r{}_{rr} = \Gamma^r{}_{r \phi} = 0.
\eeaa
\newpage

\index{coordinates!null coordinates}
\noindent {\sl Example 2}:  A spherical null coordinate system
for Minkowski space:  The coordinates are $u, r, \theta , \phi $\ .
\begin{alignat*}{2}
	u &= t - (x^2+y^2+z^2)^{1/2} \hspace{2in} && t = u+r \\
	r &= (x^2+y^2+z^2)^{1/2}     && x = r\sin\theta \cos \phi \\
   \theta &= \tan^{-1}\left[\frac{(x^2+y^2)^{1/2}}{z}\right] 
			     && y = r\sin\theta\;\sin\phi \\
   \phi  & = \tan^{-1} \frac{y}{x} && z = r\cos\theta,\\
\end{alignat*}

\vspace{-8mm}

\noindent where, of course,  $t, x, y, z$ is a natural coordinate system.

Here's a diagram of a $\theta =$ constant hypersurface with some 
constant-$u$ light cones.
\begin{figure}[h!]
                \centering 
                \includegraphics[width=5cm]{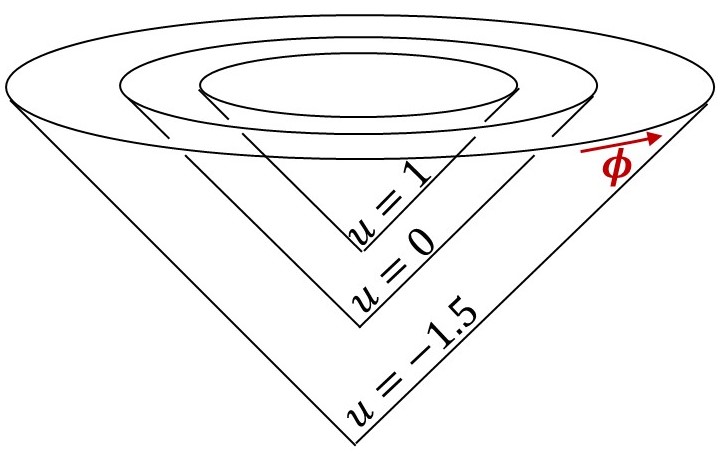} 
\label{fig:nullcoord}
\end{figure}
\vspace{-5mm}

Notice that the $u=$ constant surfaces are not orthogonal to the 
$r=$ constant surfaces.  This implies that the basis vectors $\bm\partial_u$ 
and $\bm\partial_r$ are not orthogonal, because 
$\bm\partial_u$ is tangent to a line of constant $r, \theta, \phi $ and 
{$\bm\partial_r$ is a null vector tangent to a line of constant 
$u, \theta, \phi$.   
\begin{figure}[ht]
\begin{center}
\includegraphics[width=.5\textwidth]{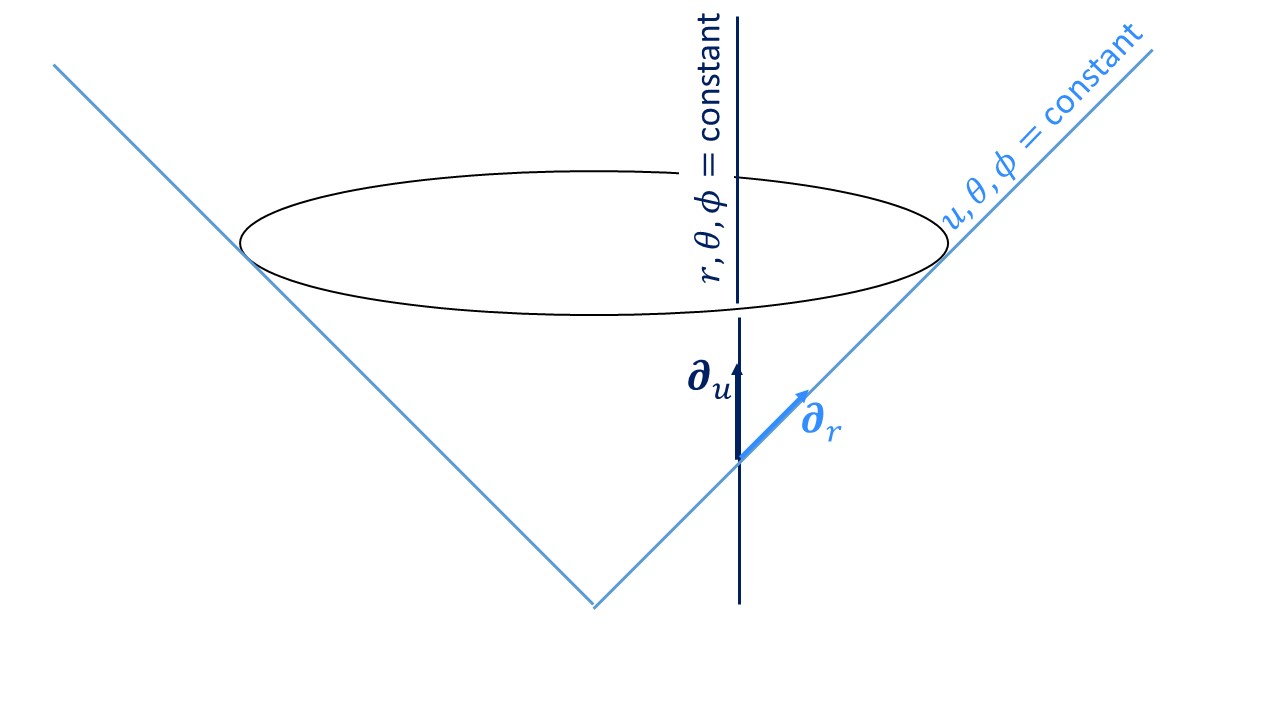}
\end{center}
\vspace{-12mm}
\caption{The basis vectors $\bm\partial_u$ and $\bm\partial_r$}
\label{fig:partial_ur}
\end{figure} 

\noindent
On the other hand, $\bm\partial_\theta$ and $\bm\partial_\phi$ are still orthogonal, 
because the $\theta=$ constant and $\phi=$ constant surfaces are
orthogonal. 
\begin{figure}[h!]
\begin{center}
\includegraphics[width=.3\textwidth]{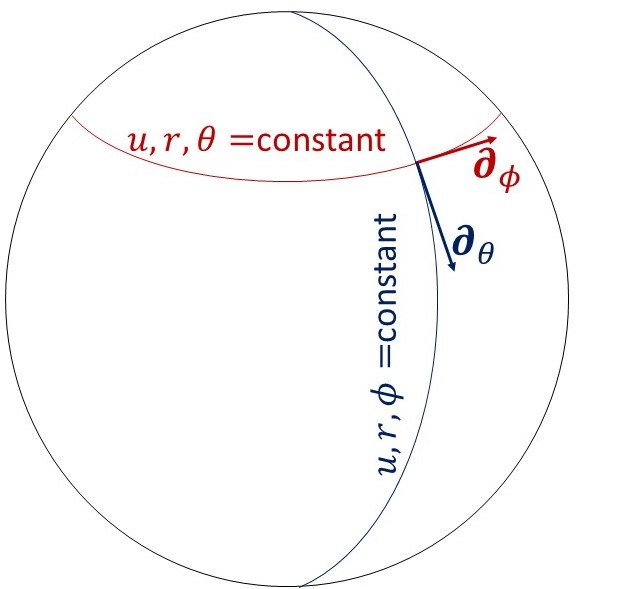}
\end{center}
\vspace{-6mm}
\caption{The basis vectors $\bm\partial_\theta$ and $\bm\partial_\phi$}
\label{fig:partial_thetaphi}
\end{figure} 

Because $\bm\partial_u$ and $\bm\partial_r$ are not orthogonal, the 
$tr$ component of the metric is nonzero:  
\begin{align*}
 ds^2&= -dt^2 + dx^2 + dy^2 + dz^2\\
     &= -(du+dr)^2 + (dr\sin\theta \cos\phi + r\cos\theta \cos\phi d\theta
	-r\sin\theta \sin\phi d\phi)^2 \\
     &\ \ \ \ + (dr\sin\theta \sin\phi + r\cos\theta \sin\phi d\theta 
	  + r\sin\theta \cos \phi d\phi)^2 + (dr\cos\theta - r\sin\theta d\theta)^2;
\end{align*} 
collecting terms and using $\cos^2+\sin^2=1$ a few times gives 
\be
 	ds^2 = -du^2 - 2dudr+r^2d\theta^2 + r^2\sin^2\theta d\phi^2,
\label{e:dsnull}\ee
or
\be 
 \| \eta_{\mu \nu}\| = \left| \left|
\begin{array}{rrrr}
-1 & -1& \\
-1 & 0 &\\
&&r^2& \\
&&&r^2\sin^2\theta 
\end{array}\right| \right|\ .
\ee 
\noindent
Note that the off-diagonal components $\dis\eta_{ur}$ and $\eta_{ru}$ are $-1$, not $-2$:  Eq.~(\ref{e:dsnull}) is a shorthand for 
\be 
\eta_{\alpha\beta}  = 
 - \nabla_\alpha u \nabla_\beta u - \nabla_\alpha u \nabla_\beta r 
 -\nabla_\alpha  r\nabla_\beta u + r^2\nabla_\alpha\theta \nabla_\beta\theta 
 + r^2\sin^2\theta \nabla_\alpha \phi \nabla_\beta \phi; 
\ee
that is, $dudr$ means the symmetric tensor $\frac{1}{2}
(\nabla_\alpha  u\nabla_\beta  r + \nabla_\alpha  r \nabla_\beta  u)$.
The contravariant metric components are
\be 
\| \eta^{\mu \nu}\| = \left\| \begin{array}{lcrr}
					\ \ 0 & -1& \\
					-1 &\ \ 1 &\\
					&&r^{-2}& \\
					&&&r^{-2}\sin^{-2}\theta 
				\end{array}\right\| .
\label{e:dsup}\ee

\noindent The basis vectors $\bm\partial_u, \bm\partial_r, \bm\partial_\theta $
and $\bm\partial_\phi $ have lengths 
$\dis 
	\left|\bm\partial_u\right|^2 = \left|\eta(\bm\partial_u, \bm\partial_u)\right| 
		= \left| \eta_{uu}\right| = 1$ or
\begin{align*}
 \left| \bm\partial_u\right| &= \sqrt{\left|\eta_{uu}\right|} = 1, \\
 \left| \bm\partial_r\right| &= \sqrt{\left|\eta_{rr}\right|} = 0,\\
 \left| \bm\partial_\theta\right| & = \sqrt{\left|\eta_{\theta \theta}\right|} = r,\\
 \left| \bm\partial_\phi\right| & = \sqrt{|\eta_{\phi\phi}|} = r\sin \theta.
\end{align*}

The covariant basis vectors have lengths\\

\indent $\dis\left| du \right| = \sqrt{\left|\eta(du,du)\right|} 
= \sqrt{\left|\eta^{\alpha\beta} \nabla_\alpha u \nabla_\beta u\right|} 
= \sqrt{\left|\eta^{uu}\right|} = 0$\\

\indent $\dis\left| dr \right| = \sqrt{\eta^{rr}} = 1$\\

\indent $\dis\left| d\theta \right| = \sqrt{\eta^{\theta \theta}} = \frac{1}{r}$\\

\indent $\dis\left| d\phi \right| = \sqrt{\eta^{\phi \phi}}=
\frac{1}{r\sin\theta}$\\
\newpage

\noindent Gradients of functions, of which covariant basis vectors are an example, 
can be visualized in either of two ways: 
The first is the way countour lines show the changing height of a hill or valley.  
\begin{enumerate}\item[1.]
 To picture $dx^\mu $, draw a set of $x^\mu $ = constant surfaces a given
coordinate distance $\Delta x^\mu $ apart.  When $dx^\mu $ is large,
the surfaces will be close together, and when $dx^\mu $ is small, far
apart.  If $v^\alpha $ is a vector tangent to some curve $c(\lambda)$,
then $v^\mu=v^\alpha \nabla_\alpha  x^\mu = \frac{d}{d\lambda } x^\mu (c(\lambda ))$ 
is the rate at which a coordinate distance $x^\mu $  is crossed in parameter 
length $\lambda $ along $c$.  So if you draw $v^\alpha $ as an arrow with 
length proportional to the components $v^\mu $, then its length will be 
proportional to the number of surfaces the arrow crosses.  \vspace{-6mm}

\begin{figure}[ht]
\begin{center}
\includegraphics[width=.3\textwidth]{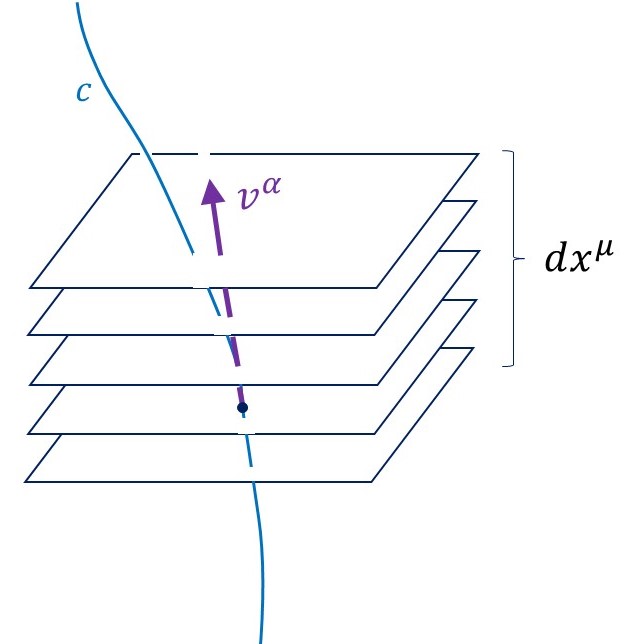}
\end{center}
\vspace{-7mm}
\label{fig:dxmu}
\end{figure} 
The natural 
way to visualize the gradient $df = \nabla f$ of a function is then as a 
set of planes tangent to $f=$ constant surfaces, and it is presented that way 
in Schutz pp.61-2 and in MTW 2.5 and 2.6. 
\index{gradient!visualized as set of planes}

\item[2.]  You can raise the index of the naturally covariant vector $\nabla_\alpha x^\mu$ and then draw the resulting contravariant vector $\nabla^\alpha  x^\mu $ as an arrow.
With an orthogonal basis and Euclidean metric, the result is an arrow that 
looks perpendicular to the surface in a diagram.  Here, without some 
practice, the contravariant vector often looks peculiar, dramatizing the 
fact that, for any function $f$, $\ \nabla_\alpha f$ is naturally a covariant vector.
\end{enumerate}   \vspace{-4mm} 
\begin{figure}[ht]
\begin{center}
\includegraphics[width=.6\textwidth]{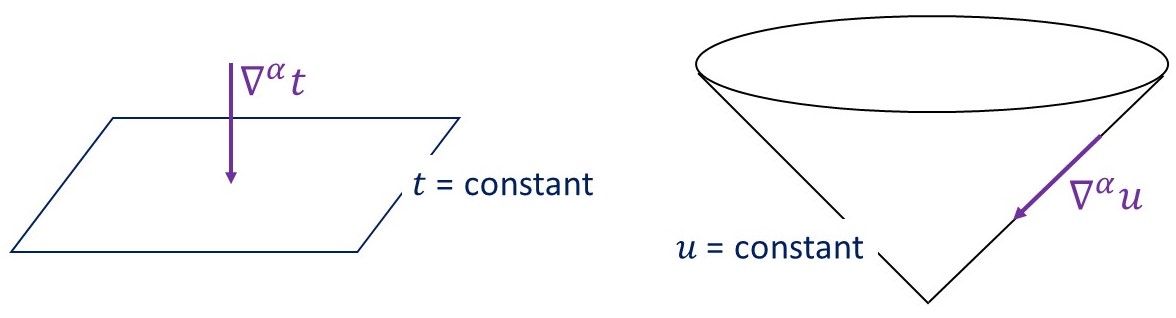}
\end{center}
\vspace{-6mm}\hspace{-1cm}
\caption{The vectors $\nabla^\alpha t$ and $\nabla^\alpha u$ acquired by raising the naturally covariant vectors $\nabla_\alpha t$ and $\nabla_\alpha u$.}
\label{fig:raiseindex}
\end{figure} 
\vspace{-4mm} 

\noindent
In the first example, despite the fact that $t$ increases to the future, 
$\nabla^\alpha t$ is a past-pointing vector.  This is due to the $-+++$ signature
of the Minkowski metric:  The $t,x,y,z$ components of $\nabla^\alpha t$ are \\ 
$(\nabla^\mu t) = (-1,0,0,0)$. 

\index{null hypersurface}\index{hypersurface!null}
\label{p:nullsurface}
In the second example, $\nabla_\alpha u$ is null: $\eta^{\alpha\beta}\nabla_\alpha u \nabla_\beta u = 0$, and we can see as follows that the raised normal to a 
null surface, $\nabla^\alpha u$, is a vector that lies in the surface!  Start with 
the fact that, for any scalar $f$, if $\bm v$ is a vector orthogonal to 
$\bm \nabla f$ (i.e., $v^\alpha\nabla_\alpha f=0$), then $f$ is constant along $\bm v$: 
That is, $v^\alpha$ lies in a surface of constant $f$. Here, with $f=u$ and $v^\alpha = \nabla^\alpha u$, the relation $\nabla^\alpha u\nabla_\alpha u = 0$ implies $\nabla^\alpha u$ is a vector lying in the $u=$ constant surface.\\
{\sl Check}:  
\[
  \nabla^\mu u = \eta^{\mu \nu }\nabla_\nu u = \eta^{\mu\, u} = -\delta^\mu_r\,,
\]
where we have used Eq.~\eqref{e:dsup} for the contravariant components $\eta^{\mu\nu}$ in $u,r,\theta,\phi$ coordinates.\\
So, in these coordinates, $\nabla^\alpha u$ is $-\bm\partial_r$.

\newpage
\index{coordinates!Rindler coordinates}\index{Rindler space}
\benr
\item Consider 2-dimensional Minkowski space, with natural coordinates $t$ and $x$.  A family of 
uniformly accelerated observers follow the trajectories 
\index{accelerated observer}
\[
  t(T)=X\sinh T, \qquad x(T) = X \cosh T,
\]
with $X$ fixed for a given observer.
\ben
\item[a.] Show that, at each point, the curve with $T$ fixed and $X$ changing is orthogonal to the accelerated trajectory, the curve with $X$ fixed and $T$ changing.  
\item[b.] 
Show that the coordinates $T,X$ cover only half of 2-dimensional Minkowski space.  Show that the 
coordinates are not defined on the lines $x=\pm t$, so they really cover two disjoint quadrants.  
The right-hand 
quadrant in these coordinates is call {\sl Rindler space}.\index{Rindler space}\index{black hole!Rindler space model}  Show that the metric in these coordinates, 
is $ds^2 = -X^2 dT^2 + dX^2$.  
\item[c.] The boundary line $x=t$ is analogous to a black hole horizon for uniformly accelerated observers: 
Show that signals from the far side of that line never cross it and therefore never reach observers who 
remain in Rindler space.  (Observers at rest relative to a Schwarzschild black hole must uniformly 
accelerate to remain at rest; they are in this way analogous to our Rindler observers. The 
analogy is part of an equivalence-principle relation between Rindler space and a black hole spacetime.)  
\een 
\index{horizon!Rindler space}

\item 
\benalph
\item Show that $\dis \epsilon_{ijk}A^i B^j C^i = \bm A\times\bm B\cdot \bm C = 
\left|\begin{array}{ccc} A^1 & A^2 & A^3\\
		    B^1& B^2 & B^3\\
		    C^1& C^2 & C^3\\
       \end{array} \right| $.
\item Using part (a), write the determinant of a $3\times3$ matrix $a_{ij}$ in terms of $\epsilon_{ijk}$.  
\item Show for any dual vector $A_a$ that $\nabla_a A_b-\nabla_b A_a$ has 
components $\partial_i A_j - \partial_j A_i$ in any chart. 
\item Similarly show, for any antisymmetric tensor $A_{a\ldots b}$, that the 
components of $\nabla_{[a} A_{b\ldots c]}$ are simply  $\partial_{[i} A_{j \ldots k]}$. 
\een

\item 
Let $g_{ab}$ be the metric of flat 3-space, $ds^2=dx^2+dy^2+dz^2$.  
Let $\{\bm e_{i'}\}$ be any basis, and write the relation between $\{\bm e_{i'}\}$ 
and the natural coordinate basis $\{\bm e_{i}\}$ in the usual way, 
$\bm e_{i'}=\bm e_{i}a^i{}_{i'}$.
\benalph
\item Show that the determinant $g$ of the metric has the value $g = [\det(a)]^2$ in the primed basis.
\index{determinant!of the metric}
\item Show that $\epsilon_{i'j'k'} = \det(a)\epsilon_{ijk}$, and conclude that 
$\epsilon_{1'2'3'} = \sqrt g$.
In other words, dropping the primes, in an arbitrary basis $\{\bm e_{i}\}$, 
$\epsilon_{123} =  \sqrt g$. 
\item What is $\epsilon_{ijk}$ in an arbitrary basis, in terms of $g$ ?

\een

\index{coordinates!spherical}
\item The metric of 3-dimensional Euclidean space in Cartesian
coordinates is $\dis\| g_{ij}\|=\|\delta_{ij}\|$ or\\ $ds^2 = dx^2+dy^2+dz^2$.
\benalph
\item  Find the metric in spherical coordinates
$r,\theta,\phi$.
\item Find the components of the contravariant metric $g^{ab}$ in
spherical coordinates.
\item What is the value of 
$\epsilon_{r\theta\phi}$?  Of $ \epsilon^{r\theta\phi}$?
\item Find all nonzero Christoffel symbols $\Gamma^i{}_{jk}$
\item Find the volume of a sphere by writing
$V=\int\sqrt{g}\;drd\theta d\phi$
\end{enumerate}

\newpage

\item Divergence and Laplacian (still in 3-d).
\label{p:div}
\benalph
\item Show that the divergence $\nabla_a A^a$ of a vector can be written in the form
\[
  \frac12 \epsilon^{abc}\nabla_a(\epsilon_{bcd}A^d).
\]
\item Using the fact that this expression involves the derivative $\nabla_{[a}T_{bc]}$ of the antisymmetric tensor
$T_{bc} = \epsilon_{bcd} A^d$, show that the divergence can be written in any chart in the form
\[
	\frac1{\sqrt g} \partial_i(\sqrt g\, A^i). 
\] 
\item Conclude that the Laplacian $\nabla_\alpha \nabla^\alpha f$ can be written in any chart in the form
\[
	\frac1{\sqrt g} \partial_i(\sqrt g\, g^{ij}\partial_j f). 
\]
\item Write the Laplacian in spherical and cylindrical coordinates. Cylindrical coordinates are often written $\varpi,z,\phi$ in astrophysics. “$\varpi$” is script $\pi$, used instead of $\rho$ to avoid confusion with density. The relation to Cartesian coordinates is  $x = \varpi\cos\phi$, $y=\varpi\sin\phi$.
\een

\item Short answer questions:
\benalph
\item What are the components of a flat 3-dimensional metric in the orthonormal basis \\
$\bm e_r\equiv \hat{\bm r},\ \ \bm e_\theta\equiv \hat{\bm \theta},\ \ \bm e_\phi\equiv \hat{\bm \phi}$.
\item What are the components $\dis \epsilon_{\hat i \hat j \hat k}$ in that basis?
\item Show that the redshift from a source of light moving away from you at speed $v$ has the nonrelativistic limit $\dis z \equiv \frac{\Delta\lambda}\lambda = v$.
\index{redshift}%
\item Show that the sourcefree Maxwell equation $\nabla_{[\alpha}F_{\beta\gamma]}$ is equivalent to 
\[  \nabla_{\alpha}F_{\beta\gamma}+ \nabla_\beta F_{\gamma\alpha} + \nabla_\gamma F_{\alpha\beta} = 0. \]
\een

\een
\index{flat spacetime, Chap.~\ref{c:fst}|)}
\index{spacetime!flat, Chap.~\ref{c:fst}|)}
\index{special relativity|)}\index{relativity!special relativity|)}
\newpage

\chapter{Curved spacetime}\label{c:cst}
\index{spacetime!curved, Chap.~\ref{c:cst}|(}
\index{curved spacetime, Chap.~\ref{c:cst}|(}

\section{Manifolds, Smooth Functions (Scalar Fields), Tensor Fields}

A smooth $n$-dimensional manifold can be thought of as a smooth $n$-dimensional
subset of some ${\mathbb R}^m$, $m\geq n$.  For example a curve in ${\mathbb R}^m$ is a 1-dimensional
manifold; any $S^n$ is a smooth $n$-dimensional subset of ${\mathbb R}^{n+1}$; on the 
other hand, a self intersecting surface is not a manifold, and a curve with 
a cusp is a manifold, but not a smooth manifold. The embedding space ${\mathbb R}^m$, however, plays no real role here, and $n$-dimensional manifolds 
are defined intrinsically by piecing together open blocks of ${\mathbb R}^n$.

 For example, a closed strip can be made from two pieces of ${\mathbb R}^2$, each one the piece $\begin{array}{c} 0<x<4\\  0<y<1\end{array}\ $, by gluing the squares of the same color with two different identifications: 

\begin{figure}[ht]
\begin{center}
\includegraphics[width=.7\textwidth]{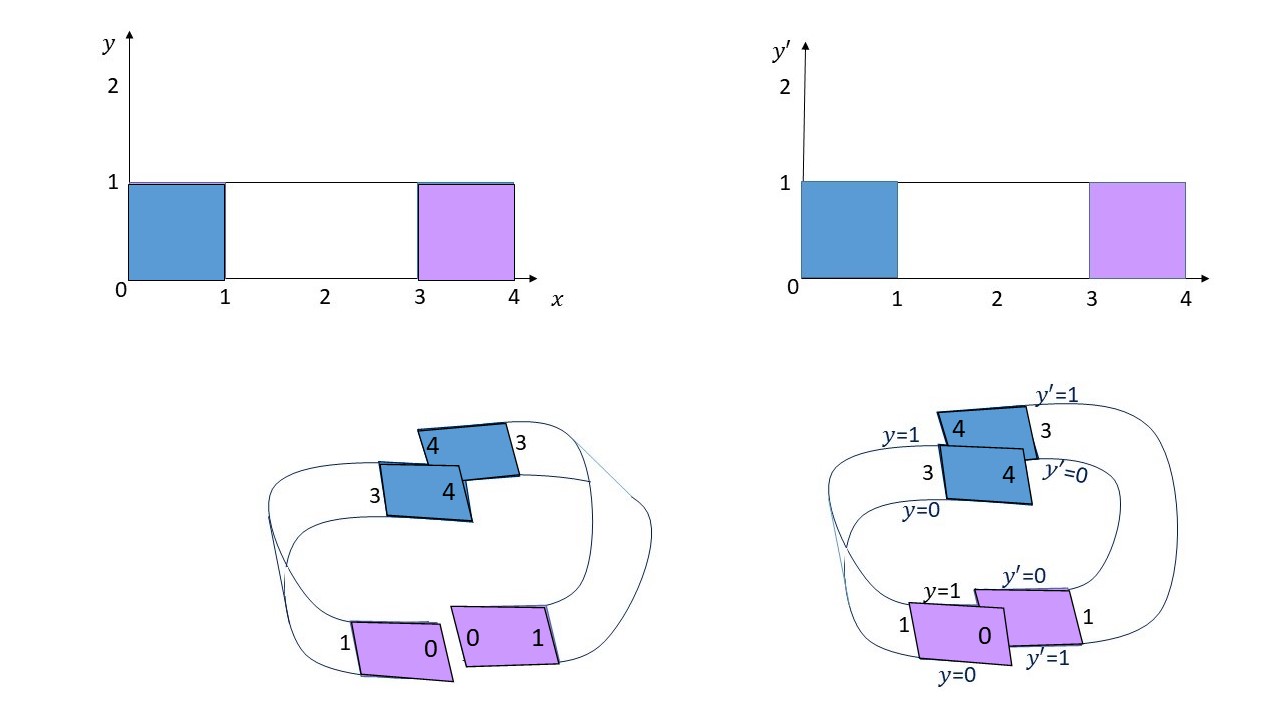}
\end{center}
\caption{Two different gluings of the two pieces of $\mathbb R^2$ give an orientable strip and a M\"obius strip.}
\label{mobius}\end{figure}

For the ordinary (orientable) strip, glue the point $(x,y)$ to the point $(x',y')$ where 
\[
 \begin{array}{l} (x',y')= (1-x,y), \qquad  0<x<1,\ 0<y<1\\
		          (x',y') = (7-x,y), \qquad 3<x<4,\ 0<y<1\ \ .
	\end{array}
\]
Similarly, a M\"obius strip is made by the overlap maps (gluing instructions)
\[
 \begin{array}{l} (x',y') = (1-x,1-y), \qquad  0<x<1,\ 0<y<1\\
		          (x',y') = (7-x,y), \qquad 3<x<4,\ 0<y<1\ \ .
	\end{array}
\]

\index{chart}\index{coordinates!on manifold|textbf}\index{coordinates|textbf}
\index{manifold|textbf}
More formally, one defines a {\sl manifold} $M$ by requiring that each point $P$
of $M$ be in the domain $U$ of some coordinate patch $x: U\rightarrow {\mathbb R}^n$
whose range $x(U)$ is an open set of ${\mathbb R}^n$, and that in the regions of
${\mathbb R}^n$ corresponding to the intersection $U\cap V$ of two coordinate patches
$x: U\rightarrow {\mathbb R}^n$, $y: V\rightarrow {\mathbb R}^n$, the maps $x\circ y^{-1}$ and
$y\circ x^{-1}$ are differentiable inverses of each other.
\begin{figure}[h!]
   \begin{center}
    \includegraphics[width=.7\textwidth]{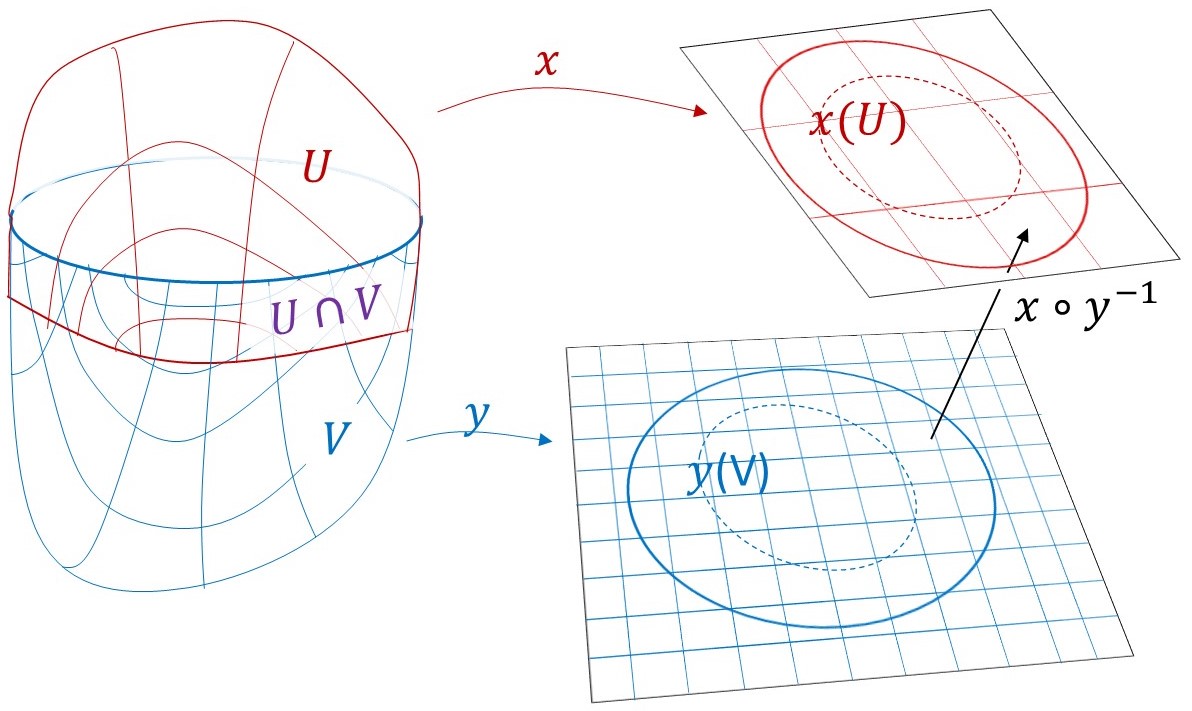}
   \end{center}
\label{charts}
\end{figure}

A smooth function (scalar field) on $M$ is a map from $M$ into $\mathbb R$
smooth in each coordinate system.  That is, $f: M\rightarrow \mathbb R$ is smooth
at $P$ if $f\circ x^{-1}: {\mathbb R}^n\rightarrow \mathbb R$ is smooth at $x(P)$.

  Mathematicians use the word ``chart'' instead of ``coordinate system''
because it has one instead of six syllables and because of the ordinary
language/technical language statement that a chart is a map.  ``Chart''
will often be used in the notes because it's shorter. \\

\begin{wrapfigure}[13]{r}{6cm}
		\begin{center}\vspace{-12mm} 
                \includegraphics[width=8cm]{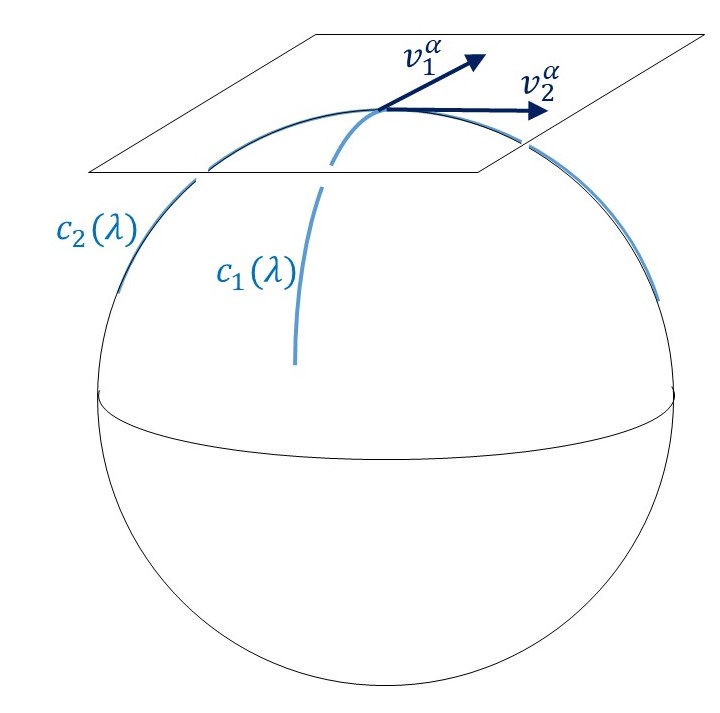}
		\end{center}
                 \label{vectors_embed}
\end{wrapfigure}

\noindent{\sl Vectors}\\
\color{white}.

\cb

 In Minkowski space a vector $v^\a$ at a point $P$ could
simply be identified with the directed line from $P$ to the point $Q$ with
$Q^\a-P^\a=v^\a$.  
In a curved space, if
one embeds the manifold in $\mathbb R^m$ (see picture at right), the tangent to a
curve in $M$, regarded as a curve in $\mathbb R^m$, is an arrow that sits not in
$M$, but in the (hyper)plane tangent to $M$ at $P$.\\

The figure shows a sphere embedded in ${\mathbb R}^3$.  The tangent vectors $v^a_1$ and $v_2^a$ to the
curves $c_1$ and $c_2$ (regarded as curves in ${\mathbb R}^3$) lie in the tangent
plane $T_PS$ to $S$ at $P$.
\newpage

\noindent If we wanted to carry
the embedding space along, we could identify the tangent vector with this
arrow in $\mathbb R^m$; but the macroscopic universe in which we live is 4-dimensional, so we 
are handed $M$, not the embedding, and keeping $\mathbb R^m$ needlessly encumbers the 
mathematics.\\ 

Instead, as in Eq.~\eqref{e:vf}, one identifies the tangent to a curve $c(\lambda)$ at $P$ with
the linear map from $\{$smooth functions on $M\}$ to $\mathbb R$ given by
\be 
	\bm v: f\mapsto \bm v(f) 
		= \left.\frac{d}{d\lambda} f[c(\lambda)]\right|_{\lambda =0}\  . 
\ee
\index{vector!tangent vector to a curve}\index{tangent vector}
Once we have defined the gradient $\na_a f$, we will write this as 
$\bm v(f)= v^a\nabla_a f $.
As in flat space, the components of $v^a$ in a chart $\{x^i\}$, are just 
\index{component!of a vector|textbf}
$\dis v^i(P) = \left.\frac{d}{d\lambda}x^i(\lambda)\right|_{\lambda=0}$.\\
Then $\bm v(f) = v^i \partial_i f$, and the $i^{\rm th}$ component of $v^a$ is
\be
   v^i = \bm v(x^i), \ \mbox{because}\ \bm v(x^i) = v^j\partial_j x^i = v^j\delta^i_j = v^i.
\ee
 
A vector is a linear map from functions to $\mathbb R$, but the definition has 
one more piece to it, because if $f$ and $g$ are two scalar fields we have the Leibnitz rule
\be
  v^a\nabla_a(fg) = v^a\nabla_a f\ g(P) + f(P)\ v^a\nabla_a g .
\ee
With that included, we can define the space of vectors at a point of a smooth manifold $M$.  \\
\index{vector\textbf}
{\bf Definition}. \crv A vector $V$ at a point $P$ in $M$ is a linear map from 
\{smooth functions on $M$\} to $\mathbb R$ satisfying\\
\phantom{xxxxxxxxx} $V(fg)=V(f)g(P)+f(P)V(g)\quad$ \cb (the Leibnitz rule).\\ 

We had now better check that, with this definition, every vector can be written as the 
tangent to a curve.  

\noindent{\sl Theorem}.  If $V$ is any linear map from \{smooth functions on $M$\} to $\mathbb R$
satisfying $V(fg)=V(f)g(P)+f(P)V(g)$, then there is a
curve $c(\lambda )$ through $P$ whose tangent $v^a$ is the map $V$.\\

\noindent {\sl Proof}.  First note that $V(k) =0$ for any constant $k$ (regarded as
a constant function on $M$).  This is quick: Given any $f$,\vspace{-5mm}
\begin{align*}
	V(kf) &= kV(f)\hspace{2mm} {\mbox{by linearity \hspace{2mm} and}}\\
	V(kf) &= kV(f) + f(P)V(k) \hspace{2mm} {\mbox{by Leibnitz}}\\
	\Rightarrow f(P)V(k) &= 0, \hspace{2mm} \mbox{all } f \Rightarrow V(k) = 0.
\end{align*}
Next, let $x^i$ be a chart about $P$ with $x^i(P)=0$.  Define $n$ numbers $v^i $ by
\be v^i = V(x^i) .\ee

We need first to write $V(f) = v^i \left. \frac{\partial f}{\partial
x^i }\right|_P$, and then to show that the $v^i$ are components of the tangent to some curve $c$. 
For the first part,  we expand $f$ in a Taylor series about $P$:
\be 
	f=f(P) + \left.\frac{\partial f}{\partial x^i }\right|_P x^i  +O(|x|^2) ,
\ee
where $O(|x|^2)$ means some function for which $ |O(|x|^2)|\leq K|x|^2$, some constant $K$. (This is the formal way of stating the intuitive idea that $O(|x|^2)$ goes to zero as fast as $|x|^2$.)   Because $f$ is smooth, we can write $O(|x|^2)$ as ${\mathpzc f}_{ij}x^ix^j$, with each ${\mathpzc f}_{ij}(x)$ a smooth function.  Now the 
value $f(P)$ of $f$ at the given point $P$ is a constant, so
\be 
	V(f) = \left. \frac{\partial f}{\partial x^i }\right|_P V(x^i) + V(O(|x|^2)).
\label{e:Vf}\ee
The first term on the right is the term we want, because $V(x^i) = v^i$.  We need to show that the 
other term vanishes at $P$, where $x^i=0$.  We have 
\[
 V(O(|x|^2)) = V[{\mathpzc f}_{ij}x^ix^j] = V({\mathpzc f}_{ij})x^i(P) x^j(P)  + {\mathpzc f}_{ij}(P) V(x^i) x^j(P)+ {\mathpzc f}_{ij}(P) x^i(P)V(x^j) 
	     = 0 \,,\ 	{\mbox{ because  }} x^i (P) = 0.
\]
\noindent Thus \eqref{e:Vf} becomes\\
\centerline{$\dis V(f) = \left. \frac{\partial f}{\partial x^i}\right|_P v^i$.}
Finally, we need a curve $c$ whose tangent $v^a$ has components $v^i$, and this is easy: 
Let $c$ be the path with coordinates
\be x^i(\lambda ) = \lambda v^i .\ee
Then 
\be
 \frac{d}{d\lambda} f(c(\lambda))|_{\lambda =0} 
	= \left. \frac{\partial f}{\partial x^i }\right|_P 
	   \left.\frac{dx^i}{d\lambda}\right|_{\lambda =0} 
	= \left.\frac{\partial f}{\partial x^i}\right|_P v^i  ,
\ee
and we have proved that the tangent $v^a$ to the path $c$ is the map $v^a\nabla_af = V(f)$.
\begin{figure}[h!]
		\begin{center}
                \includegraphics[width=11cm]{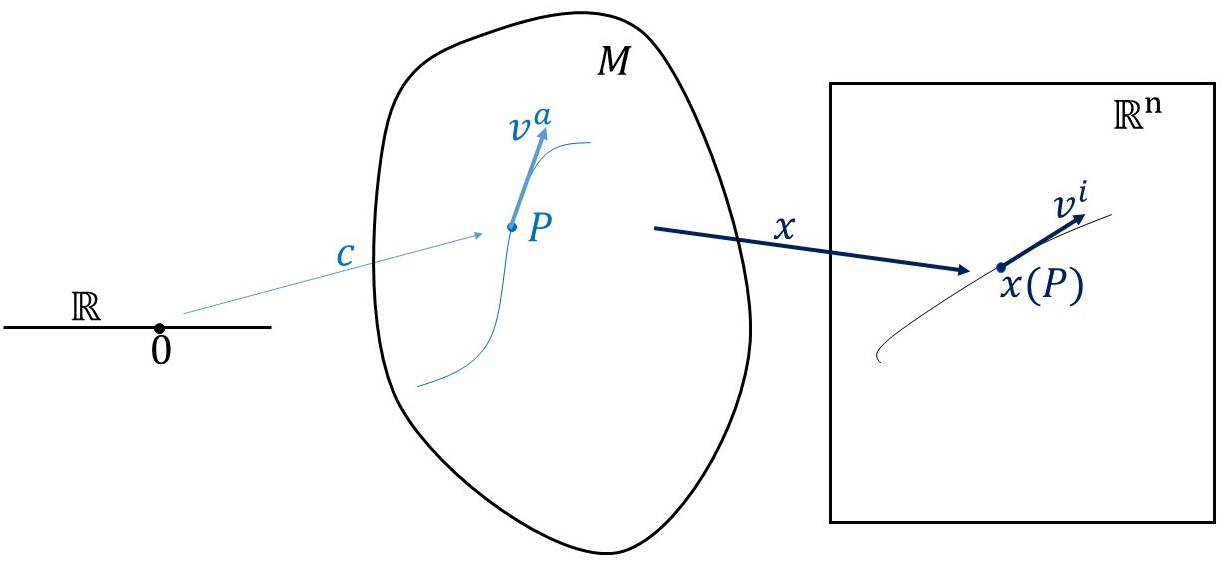}
		\end{center}
                 \label{vector_tangent}
 \end{figure}
\vspace{-7mm}

\noindent\hrulefill\\
\vspace{-5mm}

\noindent {\sl Technical Note}\\
As is endemic to physicists, we have sluffed over the distinction between
the function $f: M\rightarrow \mathbb R$ and ``$f(x)$'', the function $f\circ
x^{-1}: {\mathbb R}^4\rightarrow {\mathbb R}$.  To be more precise one should have said that
$f\circ x^{-1}$ -- call it $\bar f$ -- can be Taylor expanded about 0 in
$\mathbb R^n$.  Denoting a point in ${\mathbb R}^n$ by $r^i$, we could write
\centerline{$\dis 
	\bar f(r) = \bar f(0) + \partial_i \bar f|_0 r^i  
		   +  O(|r|^2) $ .}
Then $f=\bar f\circ x$ so if $Q$ is a point in $M$ near $P$,
\[ 
	f(Q) = \bar f [x^i (Q)] + \partial_i \bar f|_0 x^i (Q)
			+ O(|x(Q)|^2).
\]
But $\bar f[x^i (Q)] = f(Q)$ and $\partial_i \bar f|_0$ is
$e_i {}^a\nabla_af|_P$, where $e_i {}^a$ is tangent to the curve with coordinates 
$\lambda\rightarrow (x^1(P),\ldots ,x^i (P)+\lambda ,\ldots, x^n(P))$.
 So denoting $e_i {}^a\nabla_af|_P$ by the shorthand $\partial_i
f|_P$ we recover $\frac{d}{d\lambda} f\circ c(\lambda )|_{\lambda =0} =
v^i \partial_ i f|_P$.  In the same formal spirit, note that
$x^i (\lambda)=x^i \circ c(\lambda )$ and $v^i  = \frac{d}{d\lambda} x^i (\lambda )$ 
is the tangent in $\mathbb R^n$ to the curve $x^i (\lambda )$ in $\mathbb R^n$, as 
shown in the figure above.    
(Normally, we will use the convenient abuses of notation.)\\
\noindent\hrulefill\\

We have shown that the set of tangent vectors to curves through a point $P$
can be identified with the set of {\em all} linear maps at $P$ satisfying
Leibnitz.  This space of tangent vectors at $P$ with addition and scalar multiplication
defined by
\[(U+V)(f) = U(f) + V(f),~~ (k\, U)(f) = k\,U(f)\]
is the {\sl\crv tangent space} to $M$ at $P$.  
\index{tangent space}
The tangent space is a vector space:  
We need only show closure under addition and scalar multiplication:
\begin{eqnarray*}
(U+V)(fg) &=& U(fg) + V(fg)\\
&=& U(f)g(P) + U(g)f(P) + V(f)g(P) + V(g)f(P) \\
&=& (U+V)(f)g(P)+(U+V)(g)f(P)\\
(k\, V)(fg) &=& k(V(fg)) = k(V(f)g(P) + V(g)f(P)) = (k\,V)(f)g(P) +
(k\,V)(g)f(P).
\end{eqnarray*}

\noindent{\sl Notation}. \label{p:TM} The tangent space at $P$ is denoted by $V_P$ in Wald and by $T_P M$ in much of the mathematical literature.  
dual space is then $V_P^*$ or $T_P^*M$.  The set of all 
vectors at all points of the manifold is written as $TM$ and is called the tangent bundle. \index{tangent space|textbf}  These notes will 
use the notation $T_PM$.\\

	As in Minkowski space, a chart $x: U\rightarrow \mathbb R^n$ provides a natural
basis for the tangent space at each point $P$ of $U$, namely the tangent
vectors $e_i{}^a$ to the curves $\lambda\rightarrow (x^1,\ldots ,x^i+ \lambda ,\ldots, x^n)$; 
again
\be 
	e_i{}^a \nabla_\a f=\frac{d}{d\lambda} f(x^1,\ldots ,x^i+\lambda ,\ldots ,x^n) 
			   = \partial_i f|_P   ,
\label{e:basis}\ee
and a coordinate basis vector $\bm e_i$ is again written $\bm\partial_i $.
\index{coordinates!coordinate basis}\index{basis!coordinate basis}

	Because the tangent space at each point $P$ is a vector space, 
the definitions of dual vectors and tensors introduction of dual vectors and tensors at $P$ are those given in Sect.~\ref{ss:tensors}:
\index{vector!dual vector, covector|textbf}
\index{covector|textbf}\index{dual vector|textbf}
\index{tensor}
{\crv Dual vectors (covectors) at $P$ are linear maps from vectors at $P$ to numbers:
$\sigma_a: v^a  \rightarrow \sigma_av^a$. Tensors are multilinear maps
from covectors and vectors to numbers}
\be \crv T^{a\cdots b}{}_{c\cdots d}:  (\sigma_a,\ldots ,\tau_b, u^c, \ldots ,
v^d) \rightarrow T^{a\cdots b}{}_{c\cdots d}\sigma_a\cdots\tau_bu^c\cdots
v^d .\cb
\ee 
A {\sl tensor field} assigns to each point $P$ of $M$ a tensor $T^{a\cdots
b}{}_{c\cdots d}(P)$ at $P$.
\index{tensor!tensor field}

	$\delta^a{}_b$ is now the tensor field that takes
\be (\sigma_a, u^b) \hspace{2mm} {\text{to}} \hspace{2mm}
\delta^a{}_b\sigma_au^b = \sigma_au^a ,\ee
again, it has components $\delta_j^i$ in any basis for the tangent
space at any point $P$.  A basis at each $P$ of $M$ (a field of bases) will
be called a basis on $M$.\\

Just as the tangent to a curve is naturally a vector, 
\cblue  the gradient of a scalar is naturally a dual vector.\\ \cb
{\bf Definition}. The {\cblue gradient} of a scalar $f$ at a point $P$ is the linear map $\nabla f: \bm v \mapsto \bm v(f)$, where $\bm v$ is any vector at $P$.   
\index{gradient!of a scalar|textbf}\\
  Just as any vector $v^a$ at 
a point $P$ can be written as the tangent vector to some curve with $c(0)=P$, any dual vector 
$\sigma_a$ at $P$ is the gradient of some function $f$ defined near $P$:  
$\sigma_a(P)= \nabla_a f(P)$.  
\\

\noindent {\sl Smooth}

	A vector field is smooth $(C^\infty )$ if $v^a\nabla_af$ is smooth for all
smooth $f$.  A tensor field is smooth if $T^{a\cdots b}{}_{c\cdots
d}\sigma_a\cdots\tau_b u^c\cdots v^d$ is smooth for all smooth
$\sigma_a,\ldots ,\tau_b, u^c, \ldots , v^d$, where, in particular,
$\sigma_a$ is smooth if $\sigma_av^a$ (i.e. the map
$P\rightarrow\sigma_a(P)v^a(P))$ is smooth for all smooth $v^a$.

	As we noted, a function $f$ provides a dual vector field
$\nabla_af$ given by
\begin{eqnarray} \left.\nabla_af\right|_P v^a(P) &=&
\frac{d}{d\lambda} \left.f(c(\lambda )) \right|_{\lambda =0}\nonumber\\
&=& \left. v^a\nabla_af\right|_P \hspace{2mm} {\mbox{where $c(\lambda )$
has tangent $v^a$ at $P$.}} \end{eqnarray}

Given a basis $\{\bm e_i\}$ for the space of vectors at $P$, we again denote by 
$\bm \omega^i$ the dual basis,\index{basis!dual basis|textbf} the basis for the space of dual vectors for which 
$\bm \omega^i(\bm e_j) = \delta^i_j$.  The components $\sigma_i$ of a dual vector 
$\bm \sigma$ are then given by 
\be
   \sigma_i = \bm\sigma(\bm e_i), \mbox{ and } \bm \sigma = \sigma_i \bm \omega^i.
\ee
Equivalently, $\sigma_i = e_i{}^a \sigma_a$ and $\sigma_a = \sigma_i\, \omega^i{}_a$.

A chart $x$ again provides the basis
\[ 
	\omega^i{}_a = \nabla_ax^i 
\]
dual to the coordinate basis ${\bm e}_i  = \bm\partial_i$, of \eqref{e:basis}, 
and the discussion of coordinate components
given for Minkowski space in Sect.~\ref{s:components} goes through as written.

	Finally, there is an equivalent and sometimes more useful way to describe
a tensor field: \\
\index{tensor!tensor field|textbf}
\cblue A tensor field $T^{a\cdots b}{}_{c\cdots d}$ is a map from vector fields
and dual vector fields to functions
\[ T^{a\cdots b}{}_{c\cdots d}: (\sigma_a,\cdots , \tau_b, u^c,\cdots , v^d)
\mapsto T^{a\cdots b}{}_{c\cdots d}
\sigma_a\cdots\tau_bu^c\cdots v^d \]
that is linear in each argument under addition and under multiplication by
{\sl scalar fields}. \cb  
\label{p:tensorfield} 
That is, if $T^{a\cdots b}{}_{c\cdots d}$ is a tensor
field, obviously
\be 
   T^{a\cdots b}{}_{c\cdots d}(f\sigma_a+g\widehat\sigma_a)\cdots\tau_bu^c\cdots v^d 
		= fT^{a\cdots b}{}_{c\cdots d}\sigma_a\cdots\tau_bu^c\cdots v^d 
		 + gT^{a\cdots b}{}_{c\cdots d}\widehat\sigma_a\cdots\tau_bu^c\cdots v^d . 
\ee
And if $T$ is any multilinear map (in this strong sense of scalar field
multiplication), then $T$ is a tensor field: e.g.\ let $T$ be a map from
vector fields to functions with 
$T(f\bm u  +g{\bf v} ) = fT(\bm u  ) + gT(\bm v)$.  Then the dual vector field 
$T_a$ to be identified with $T$ has the value
$T_a(P)$ at $P$ given by $T_a(P)v^a(P)=T({\bf v} )(P)$.  The subtlety is to
show that $T_a(P)$ is well defined:  If $v^a$ and $\widehat v^a$ are any vector
fields which agree at $P$, $v^a(P) = \widehat v^a(P)$, then $T_a(P)v^a(P) =
T_a(P)\widehat v^a(P)$.  Thus the definition will work only if 
$T(v)(P) =
T(\widehat v)(P)$ whenever $v^a(P) = \widehat v^a(P)$.  And this
follows from the strong linearity:  write $v^a-\widehat v^a$ in terms of a basis\\
\centerline{
	$\bm v-\widehat{\bm v} = \underbrace{(v^i -\widehat v^i)}_{\mbox{scalar field}}\bm e_i$.
}
Then
\begin{eqnarray*}
T(\bm v-\hat{\bm v})(P) &=& [(v^i-\widehat v^i)T(\bm e_i)](P) , {\mbox{  by
strong linearity}} ,\\
&=& [v^i (P)-\widehat v^i(P)]T(\bm e_i)(P)\\
&=& 0 ,
\end{eqnarray*}
and since $T(\bm v-\hat{\bm v} ) = T(\bm v)-T(\hat{\bm v})$, $T(\bm v)(P) = T(\hat{\bm v})(P)$.\\

 An example of a linear $T$ that is not a tensor is
$T(\bm v)(P) = \int\limits_M \sigma_av^a d V$, where $\sigma$ is a dual vector
field on $M$. Here the value $T({\bm v} )$ is a constant scalar field that 
depends on what ${\bm v} $ looks like everywhere.  Consequently, unless 
$f$ is constant, there is some $\bm v$ for which 
$T(f\bm v)\neq fT({\bm v})$, and $T$ cannot be represented
by a tensor field. \\ 

We will ordinarily use the term {\sl basis} to refer to a field $\{\bm e_i\}$ 
of basis vectors, as well as to a basis at one point, when the meaning is clear from the context.  \\ 

\noindent {\sl Commutators (Lie brackets)}\\
\index{commutator!of vector fields|textbf}\index{Lie bracket}
	If $u^a$ and $v^a$ are vector fields, one can define the linear map
$f\rightarrow u^a\nabla_a[v^b\nabla_bf]$:  that is, $v^b\nabla_bf$ is a
function on $M$ when $f$ is a function, so we can take its directional
derivative along $u^a$, $u^a\nabla_a(v^b\nabla_bf)$.  But there is no
vector $w^a$ for which $w^a\nabla_af = u^a\nabla_a(v^b\nabla_bf)$ because
$f\mapsto u^a\nabla_a(v^b\nabla_bf)$ does not satisfy the Leibnitz rule:
\begin{align*}
	fg \mapsto u^a\nabla_a[v^b\nabla_b(fg)] 
		& = u^a\nabla_a[(v^b\nabla_bf)g+fv^b\nabla_bg]\\
		&= u^a\nabla_a(v^b\nabla_bf)g+v^b\nabla_bf u^a\nabla_ag
		  + u^a\nabla_a(v^b\nabla_bg)f + u^b\nabla_bf v^b\nabla_bg\\
		&\neq u^a\nabla_a(v^b\nabla_bf)g + u^a\nabla_a(v^b\nabla_bg)f ;
\end{align*}
the cross terms $v\cdot\nabla f ~u\cdot\nabla g + v\cdot\nabla g~
u\cdot\nabla f$ are not zero.

	If, however, one takes the antisymmetric combination of $u$ and $v$,
\be\crv
 [u,v]f := u\cdot\nabla (v\cdot\nabla f)-v\cdot\nabla (u\cdot\nabla f) \cb
\ee the symmetric cross terms will drop out:
\begin{align}
 u\cdot\nabla [v\cdot\nabla (fg)]-v\cdot\nabla [u\cdot\nabla (fg)] 
	= &[u\cdot\nabla (v\cdot\nabla f) -v\cdot\nabla(u\cdot\nabla f)]g \nonumber\\
	 &  + [u\cdot\nabla (v\cdot\nabla g) -v\cdot\nabla(u\cdot\nabla g)]f \nonumber\\
	=&([u,v] f) g + ([u,v] g)f.
 \end{align}
Thus the map $f\mapsto [u,v]f$ is linear and satisfies Leibnitz, and so there is a vector field $[u,v]^a$
with
\[ \nabla_a f\equiv [u,v]^a\nabla_af = [u,v]f .\]

	In a chart, $[u,v]^a$ has components $[u,v]^i $ given by
\begin{align*}
   [u,v]^i \partial_i f 
	&= u^j \partial_j(v^i\partial_i f) -v^j \partial_j (u^i \partial_if)\\
	&= (u^j \partial_j v^i )\partial_i f-(v^j\partial_j u^i )\partial_i f 
	  + u^j v^i \underbrace{ (\partial_i \partial_j f-\partial_j\partial_i f)}_0 \\
	&= (u^j \partial_j v^i -v^j \partial_j u^i )\partial_i f ,
\end{align*}
whence \vspace{-3mm}
\be\cblue 
	[u,v]^i  = u^j \partial_j v^i -v^j \partial_j u^i \,.
\label{e:commutator}\ee

Notice that if $u^a$ and $v^a$ are coordinate basis vectors, $\bm\partial_1$ and $\bm\partial_2$, say, then $[u,v]=0$: 
\be
 [u,v]^a\nabla_a f = \partial_1\partial_2 f - \partial_2\partial_1 f  = 0.
\label{e:commute}\ee
Or note that the components are constants, $u^i=\delta^i_1$, $v^i= \delta^i_2$, so 
$[u,v]=0$ by Eq.~\eqref{e:commutator}.

The converse is that, if two nonvanishing vector fields commute, then one can choose 
a coordinate system for which the two vector fields are coordinate basis vectors $\bm\partial_1$ and 
$\bm\partial_2$; that is, the vector fields are tangent to a family of surfaces on which only 
$x$ and $y$ change.  This is true as long as $u$ and $v$ are linearly independent.  A  
cleaner,  statement is that, if $u$ and $v$ are linearly independent and $[u,v]=0$, 
then the vector fields are {\sl surface forming}, tangent to families of 2-surfaces.  

We'll return later to the commutator as the Lie derivative of $v$ along $u$ (or of $u$ along $v$ 
after a sign change), giving it a geometric meaning and extending it to the Lie derivative of tensors.  
 
\section{Metric and Covariant Derivative}
\label{s:g_nabla}

\index{metric|textbf}
A {\em metric} on a manifold $M$ is a symmetric covariant tensor $g_{ab}$.
A metric $g_{ab}$ is nondegenerate (we may never look at any other kind) if
$g_{ab}u^b=0\Rightarrow u^a=0$. If nondegenerate, a metric defines an isomorphism
from contravariant to covariant vectors $u^a\rightarrow g_{ab}u^b=u_a$; the
inverse isomorphism is $g^{ab}u_b=u^a$, where $g^{ab}$ is the contravariant
metric given by
\be g^{ab}g_{bc} = \delta^a_{\,c}, \hspace{4mm} g^{ab} {\mbox{ symmetric}}.\ee

A  metric $g_{ab}$ is positive definite if $g_{ab}u^au^b >0$, all $u^a\not= 0$.\\
{\bf Definition}.   $g_{\a\b}$ is a {\crv \sl Lorentz metric} on a 4-dimensional manifold $M$ if the matrix
$\|g_{\mu\nu}\|$ of the components of $g_{\a\b}$ in a basis has 1 negative and 3 
positive eigenvalues at each point of $M$.\\
Then $g_{\a\b}(P)$ is a Minkowski metric for the tangent space at 
$P$.  Equivalently, with with this metric, the tangent space 
$T_PM$ is a copy of Minkowski space.  \\

\noindent 
{\sl \bf Definition} A {\em\crv spacetime} is a 4-dimensional manifold with a Lorentz metric.  
\index{spacetime|textbf}

Note the distinction here between a Minkowkski metric and a Lorentz metric.  A spacetime with a Minkowski metric is a flat spacetime: 
There are natural coordinates in which the metric has the form 
diag$[-1,1,1,1]$ {\sl at every point}. That is, the coordinate basis is orthonormal at every point.  A spacetime with a Lorentz metric 
is, in general, curved; there is no coordinate system for which 
the coordinate basis is orthonormal everywhere.  As we will see, 
the best one can do is to choose coordinates for which the 
metric components are diag$[-1,1,1,1]$ at a point $P$ and their 
first derivatives vanish at $P$.  The curvature of the space 
is then measured by their second derivatives.\\  

\noindent {\sl Example 0}.  Minkowski space

\noindent{\sl Example 1}. 
Take a slice of Minkowski space between $t=0$
and $t=1$ ($t$ a natural time coordinate) and identify all points on the
$t=0$ surface with corresponding points on $t=1$:
$(0,x,y,z)\leftrightarrow (1,x,y,z)$
\begin{figure}[h!]
		\begin{center}
                \includegraphics[width=13cm]{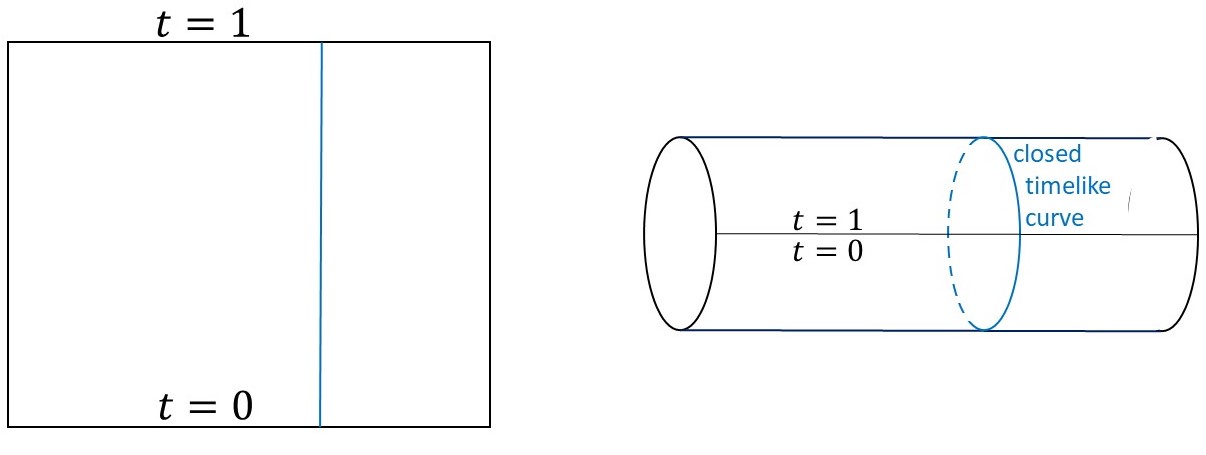}
		\end{center}
                 \label{mfld_ctc}
 \end{figure}
\noindent The resulting 4-dimensional cylinder $S^1\times {\mathbb R}^3$ 
is a smooth (flat!) spacetime with same the flat Minkowski metric of the space before identifying points.  It has closed timelike curves---e.g.\ the
blue $x,y,z = $ constant curve running between the identified points at $t=0$ and $t=1$.\\
\noindent{\sl Example 2}.  1-dimensional, 2-dimensional and 3-dimensional spheres 
of radius $a$ are are manifolds with the positive definite metrics 
\bsube\begin{align}
   ds^2 & = a^2 d\phi^2 ,\\ 
   ds^2 & =  a^2(d\theta^2 + \sin^2\theta d\phi^2), \\
   ds^2 & = a^2[d\chi^2 + \sin^2\chi(d\theta^2 + \sin^2\theta d\phi^2)].
\end{align}\esube
The metric is smooth, but the coordinates are singular at the poles.  \\
\noindent{\sl Example 3}.  An expanding universe is a spacetime that has at each time the geometry of a 3-sphere of radius $a(t)$:
\[
 ds^2  = - dt^2 + a^2(t)[d\chi^2 + \sin^2\chi(d\theta^2 + \sin^2\theta d\phi^2)].
\]

	Given a metric, one wants to define a covariant derivative 
$\nabla_a$.  In Minkowski space, we used the natural constant vector 
fields and required that $\nabla_a v^b =0$ for any constant vector field $v^b$.
$\nabla_a$ was then extended to arbitrary tensors by
$\nabla_a(f)=\nabla_af$ on functions $f$ and by Leibnitz
\be 
	\nabla_\cdot (S^{\cdots}{}_{\cdots}T^{\cdots}{}_{\cdots})
		= (\nabla_\cdot S^{\cdots}{}_{\cdots}) T^{\cdots}{}_{\cdots} 
		  + S^{\cdots}{}_{\cdots}\nabla_\cdot T^{\cdots}{}_{\cdots} 
\ee
and linearity.  We can still require $\nabla_a(f)=\nabla_a f$, the gradient
of a scalar $f$, and also linearity and Leibnitz, but in curved manifolds 
there are no natural parallel vector fields; and, as we shall see, 
in a curved space one can't define (in general) vector fields for 
which $\nabla_a v^b=0$ everywhere.  
The idea is that a vector $v^a$ can be parallel transported
around a curve, but the result of parallel transport from a point $A$ to a
point $B$ depends on the curve.\\

\noindent {\sl Example}.
\vspace{-10mm}

\begin{figure}[H]
		\begin{center}
                \includegraphics[width=13cm]{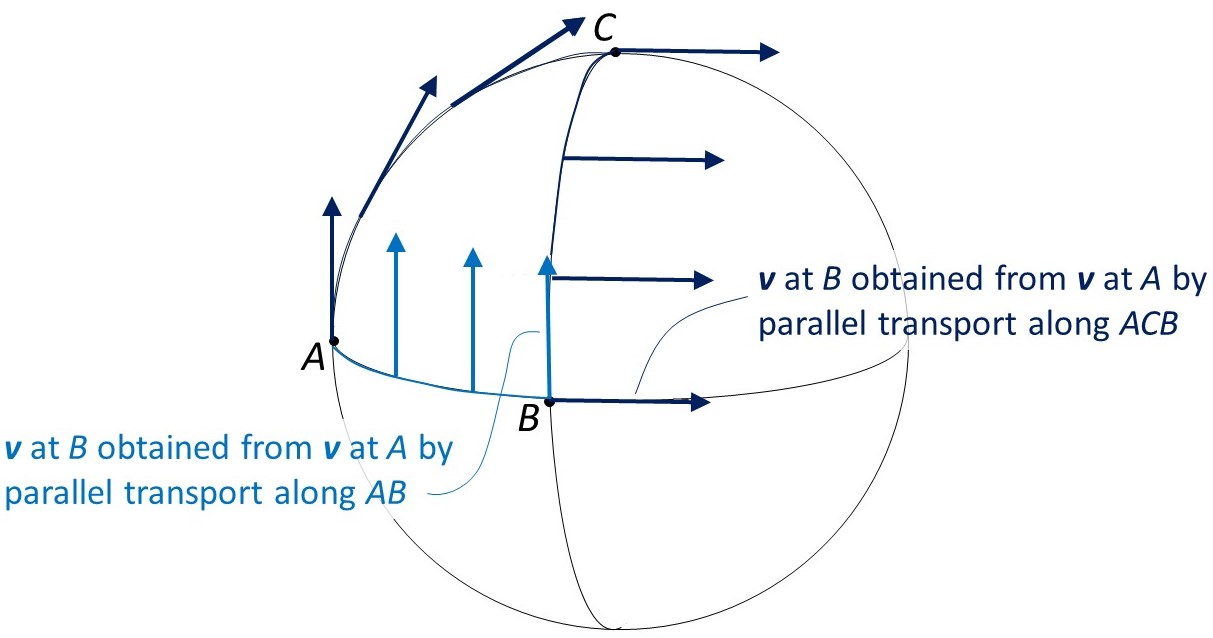}
		\end{center}
                 \label{transport}
 \end{figure}

It is intuitively clear how to parallel transport a vector $\bm v$ around each
of the great quarter circles $AC$, $CB$ and $AB$ of the sphere shown here.
The resulting vector at $B$ depends on whether the path is $ACB$ or $AB$.
So given a vector $\bm v$ at $A$, there can be no vector field on $S$
everywhere parallel to $\bm v$.  It would have to be both vertical and
horizontal at $B$.

	$\nabla_a$ is to be defined so that if $u^a$ is tangent to the curve
$c(\lambda )$, then a vector $v^a$ will be parallel transported along $c$
when $u^b\nabla_b v^a=0$.  Parallel transport should preserve length and
angles, so if $v^a$ and $w^a$ are parallel transported along $c$, the dot
products $v^av_a$, $v^aw_a$ and $w^aw_a$ should be constant along $c$:
that is, if
\be 
	u^b\nabla_b v^a = 0\hspace{3mm} {\text{and}} \hspace{3mm} u^b\nabla_b w^a=0 \,,
\label{udvw}\ee
then
\[ 
	u^a\nabla_a(g_{bc}v^b w^c)=0 .
\]
By Leibnitz
\begin{eqnarray*}
u^a\nabla_a(g_{bc}v^bw^c) &=& u^a(\nabla_ag_{bc})v^bw^c +
g_{bc}(u^a\nabla_av^b)w^c + g_{bc}v^bu^a\nabla_aw^c\\
&=& u^a(\nabla_ag_{bc})v^bw^c
\end{eqnarray*}
by \eqref{udvw}.  So $\nabla_ag_{bc}\ u^av^bw^c =0$ for all $u^a, v^a,
w^a$ satisfying \eqref{udvw}, and it follows%
\footnote{It follows because you can
pick $u^a, v^a, w^a$ arbitrarily at a point $P$ and then parallel transport $v$
and $w$ along $c$ (with tangent $u^a$) to get $u$, $v$ and $w$ satisfying
\eqref{udvw}.  Then $\left. \nabla_ag_{bc}u^av^bw^c\right|_P =0$ all vectors $u, v,
w$ at $P$, implying $\nabla_ag_{bc}(P)=0$.}
that
\be 
	\nabla_ag_{bc}=0 .
\label{e:deriv}\ee

We can now define our covariant derivative $\nabla_a$ by requiring, 
for all smooth functions $f$ and tensor fields $S$, $T$, 
\index{covariant derivative|textbf}\index{derivative!covariant derivative|textbf}
\begin{enumerate}
\item[(i)] $\nabla_a$ linear
\item[(ii)] $\nabla_\cdot (S^{\cdots}{}_{\cdots}T^{\cdots}{}_{\cdots}) =
S^{\cdots}{}_{\cdots}\nabla_\cdot T^{\cdots}{}_{\cdots} + (\nabla_\cdot
S^{\cdots}{}_{\cdots})T^{\cdots}{}_{\cdots}$\quad (Leibnitz) 
\item[(iii)] Commutes with contraction $\nabla_a(T^b{}_b) = \nabla_a T^b{}_b$ 
\item[(iv)] $\nabla_a f$ on scalars is the gradient 
\item[(v)] $\nabla_{[a}\nabla_{b]}f =0$
\item[(vi)] $\nabla_ag_{bc}=0$\,.
\end{enumerate}
Note that (ii) is the Leibnitz rule for the outer product of two tensors: No index of $S$ is 
contracted with an index of $T$.  

We saw that condition (v) holds in Minkowski space because, in a natural
coordinate system,
\be 
 \nabla_{[\mu}\nabla_{\nu]}f = \frac{1}{2} \left(\frac{\partial^2f}{\partial x^\mu \partial x^\nu} 
				-\frac{\partial^2f}{\partial x^\nu \partial x^\mu}\right) = 0 .
\ee
Here we must put it in by hand.  In the context of spacetime, the
condition will turn out to be equivalent to demanding that the spacetime 
be {\em locally Minkowskian}, that
there be inertial coordinates in the neighborhood of any point 
$P$,  coordinates in which the equations of nongravitational fields take the
same form at $P$ they have in special relativity, with $\nabla_\mu
v^\nu (P) = \partial_\mu v^\nu (P)$.
\index{coordinates!inertial coordinates}\index{inertial coordinates}
This requirement, that there be coordinates in 
which all metric derivatives vanish at a point, is the way Schutz and Hartle impose 
this condition.  

	We will show that (i)--(vi) uniquely determine $\nabla_a$, by explicitly
constructing the components of $\nabla_aT^{b\cdots c}{}_{d\cdots e}$ in an
arbitrary basis $e_i{}^a$.  One could do this more elegantly, but we'll
need the components in any event.  We'll begin by looking at the components
of the covariant derivative of a dual vector $\sigma_a$.  As in flat space, 
we define $\Gamma^i{}_{jk}$ by
\be 
	\nabla_k {\bm e}_j = {\bm e}_k{}\cdot \nabla\, {\bm e}_j
			   = \Gamma^i{}_{jk}\ {\bm e}_i. 
\label{e:Gamma_up}\ee
{\sl Terminology}:\index{Christoffel symbol!terminology}\index{Ricci rotation coefficients}
For a coordinate basis, the coefficients $\Gamma^i{}_{jk}$
are historically called Christoffel symbols; for an orthonormal basis, 
they are called Ricci rotation coefficients, because an orthonormal basis 
parallel-transported from point $P$ to point $Q$ differs from the 
value of the original orthonormal basis at $Q$ by a rotation (or, for 
a Lorentzian signature metric, by a Lorentz transformation).  In general, 
MTW calls the $\Gamma$'s {\em connection coefficients}.\index{connection coefficient}

Then $\Gamma^i_{\ jk}$ will be expressed in terms of the
partial derivatives of the metric components,\\  
$\begin{array}[t]{l}
\hspace{52mm} {e_i^{\ a}}\underbrace{\nabla_a (g_{jk})},\\ 
\phantom{.}{\hspace{45mm}\mbox{\small gradient of the scalar field $g_{jk}$}}\end{array}$\\
and of the commutators (or {\sl commutation coefficients}) $c^i_{\ jk}$ defined by
\index{commutator}\index{commutator!commutation coefficients}
\be \cblue
	[{\bm e}_j ,{\bm e}_k] = c^i_{\ jk} {\bm e}_i .\cb
\ee
Note the antisymmetry $c^i_{\ jk} = c^i_{\ [jk]}$.  For a coordinate basis, 
$[{\bm e}_j ,{\bm e}_k]=[\bm\partial_j,\bm\partial_k]=0$. 

We have been through nearly the same calculation in Minkowski space.  The
components of $\nabla_a\sigma_b$ are
\begin{eqnarray}
 \nabla_i \sigma_j  := e_i {}^a e_j{}^b\nabla_a\sigma_b 
 &=& e_i{}^a\nabla_a(e_j{}^b\sigma_b)-e_i {}^a(\nabla_ae_j{}^b)\sigma_b 
 				\hspace{4mm} {\mbox{by (iii)}}\nonumber\\
 \nabla_i \sigma_j  &=& \bm e_i\!\cdot\!\partial\, \sigma_j -\Gamma^k{}_{ji}\sigma_k 
 				\hspace{30mm} {\mbox{by (\ref{e:Gamma_up})}}
\label{e:delsigma}\end{eqnarray}
\index{covariant derivative!of dual vector}
To find $\Gamma^k{}_{j  i  }$ we'll write the components of
$\nabla_ag_{bc}=0$, as we did in flat space.  This time, however,
we'll do the computation for an arbitrary basis, not restricting 
ourselves to a coordinate basis.  We could have done the same 
thing for flat space, but there the object was to give a simpler 
computation for our first encounter with Christoffel symbols.
Using the notation $e_i f \equiv e_i(f) = e_i^a\nabla_a f$, we'll 
denote the directional derivative of the scalars $g_{jk}$ in the 
direction ${\bm e}_i$ by $e_i g_{jk}$. Then  
\begin{align*}
 0 &= \nabla_i g_{jk} = e_i{}^a e_j {}^b e_k{}^c\nabla_ag_{bc} \hspace{40mm}{\mbox{by (iv)}}\\
   &= e_i[e_j {}^be_k{}^cg_{bc}] - (\nabla_i e_j{}^b)g_{bc}e_k{}^c 
	- (\nabla_i e_k{}^c)e_j{}^bg_{bc} \hspace{4mm} {\mbox{by (iii)}}\\
   &= e_i g_{jk} -\Gamma^l{}_{ji} g_{l k} - \Gamma^l{}_{ki}g_{jl}
				\hspace{40mm} {\mbox{by (\ref{e:Gamma_up})}}
\end{align*}
Defining
\be 
	\Gamma_{kji} = g_{kl}\Gamma^l{}_{ji}  , 
\ee
we have
\be 
	0 = e_i g_{jk} -\Gamma_{kji}-\Gamma_{jki} .
\label{egjk}\ee
For a coordinate basis,  $\Gamma_{ijk}$ is symmetric in $j$ and $k$,  
but orthonormal bases are often more useful, and for a general basis
$\Gamma$ has no index symmetry.  Instead, we'll write  $\Gamma_{ijk}$ 
as a sum of its symmetric and antisymmetric parts,
 $\Gamma_{ijk} = \Gamma_{i(jk)} + \Gamma_{i[j k]}$.  The antisymmetric 
part is present when the basis vectors do not commute and is related 
to their commutator in a simple way: 
\be 
 \Gamma_{i[jk]} =-\frac{1}{2} c_{ijk}
 		\equiv -\frac{1}{2} g_{il}c^l_{\ jk}. 
\label{gasym}\ee

To see (\ref{gasym}), one notes that 
$[u,v]^a = u^b\nabla_b v^a - v^b\nabla_b u^a$ :
\begin{align}
 [u,v]^a\nabla_af  
 &= u^b\nabla_b(v^a\nabla_af)-v^b\nabla_b(u^a\nabla_af) \nonumber\\   
 &= (u^b\nabla_bv^a)\nabla_af - (v^b\nabla_bu^a)\nabla_af 
+ \smash{\underbrace{u^bv^a\nabla_b\nabla_af - v^bu^a\nabla_b\nabla_af}_0} \nonumber\\
	  &= (u^b\nabla_bv^a-v^b\nabla_bu^a)\nabla_a f  \quad \Longrightarrow   \nonumber\\
 \cblue [u,v]^a  &\cblue = u^b\nabla_b v^a-v^b\nabla_b u^a,
\label{e:commutator1}\end{align}
where we used $(\nabla_b\nabla_a-\nabla_a\nabla_b)f=0$ in the second line.
In particular,
\begin{align*}
 [{\bm e}_j , {\bm e}_k ] 
  &= {\bm e}_j\cdot\nabla {\bm e}_k - {\bm e}_k\cdot\nabla {\bm e}_j  \nonumber\\
 c^i_{\ jk} {\bm e}_i &= \Gamma^i{}_{kj}{\bm e}_i-\Gamma^i{}_{jk}{\bm e}_i \nonumber\\
c^i_{\ jk}{}&= \Gamma^i{}_{kj} - \Gamma^i{}_{jk} ,
\end{align*}
identical to (\ref{gasym}).  Finally we'll write (\ref{egjk}) three times and 
add:
\begin{eqnarray}
   -e_i g_{jk} &=& -\underline{\Gamma_{k (ji)}} - \cancel{\Gamma_{j(ki)}}{-12} 
		   - \Gamma_{k[ji]} -\Gamma_{j[ki]}\\
 \begin{tikzcd}[column sep=small]
 & i\arrow[dl,leftarrow]\arrow[dr] & \\
   k && j\arrow{ll}
\end{tikzcd} \hspace{15mm} 
e_j g_{k i  } &=& \Gamma_{i(kj)} + \underline{\Gamma_{k (ij)}} +
	\cancell{\Gamma_{i[kj]}}{-12} + \Gamma_{k[ij]}\\
\begin{tikzcd}[column sep=small]
 & i\arrow[dl,leftarrow]\arrow[dr] & \\
   k && j\arrow{ll}
\end{tikzcd}
\hspace{15mm} e_k g_{ij} &=&
\cancel{\Gamma_{j(ik)}}{-12} + \Gamma_{i(jk)} + \Gamma_{j[ik]} 
				+ \cancell{\Gamma_{i[jk]}}{-12}
\end{eqnarray}

\begin{align*}
e_j g_{ki} + e_k g_{ij} - e_i g_{jk} 
 	&= 2\Gamma_{i(jk)} +2\Gamma_{j[ik]} + 2\Gamma_{k[ij]}\\
 	&= 2\Gamma_{ijk }-2\Gamma_{i[jk]} +2\Gamma_{j[ik]} + 2\Gamma_{k[ij]}\\
2\Gamma_{ijk} &= e_j g_{ik} + e_k g_{ij} - e_i g_{jk}
		+ 2\Gamma_{i[jk]} -2\Gamma_{j[ik]}-2\Gamma_{k[ij]}
\end{align*}

\index{Christoffel symbol!for a generic basis|textbf}\index{Christoffel symbol}
\be\crv
  \Gamma_{ijk} = \frac{1}{2} [e_j g_{ik} + e_k g_{ij} - e _i g_{jk} 
 			+ c_{kij} - c_{jki} - c_{ijk}] \cb 
\label{gamijk}\ee 

The covariant derivative of any tensor then follows from Leibnitz (iii), 
exactly as in Eq. (1.152).  We'll keep the $e_i T^j_k$ notation
instead of  $\nabla_i (T^j_k)$, but they mean the same thing:
\index{covariant derivative!of tensor}
\be
\nabla_i T^{j\cdots k}{}_{m \cdots n} =
e_i T^{j  \cdots k}{}_{m \cdots n} +
\Gamma^j{}_{li}T^{l \cdots k}{}_{m\cdots n}  + \cdots 
	+ \Gamma^k{}_{li}T^{j \cdots l}{}_{m\cdots n}\nonumber\\ 
-\Gamma^l{}_{mi}T^{j \cdots k}{}_{l\cdots n} - \cdots   
- \Gamma^l{}_{ni}T^{j \cdots k}{}_{m\cdots l} 
\ee

	Two special cases of Eq. (\ref{gamijk}) are all that one 
ordinarily encounters.  With a coordinate basis, 
\[
	[{\bm e}_i, {\bm e}_j]=[\partial_i,\partial_j] = 0,
\]
implying $c^k_{\ i  j } = 0$, and one recovers the usual 
Christoffel symbols;  in greater detail,   	
\be
\bm e_i  = \frac{\partial}{\partial x^i} \hspace{3mm} {\rm and}
\hspace{3mm} 
	[\bm e_i ,\bm e_j]f 
		= \frac{\partial^2f}{\partial x^i \partial x^j} 
		  - \frac{\partial^2f}{\partial x^j\partial x^i} = 0.
\ee
Then 
\index{Christoffel symbol!for a coordinate basis}
\be \cblue
\Gamma_{ijk} 
   = \frac{1}{2} (\,\partial_k g_{ij} + \partial_j g_{ik} - \partial_i g_{jk}\,) 
\cb\label{gamg}\ee

With an orthonormal basis, the metric components are constant, with values $1,-1$ or $0$,
so $e_i g_{jk} = 0$, and only the commutators contribute to $\Gamma$:
For example, for a spacetime,
\be \| g_{\mu\nu}\| 
 = \left\| e_\mu {}^\alpha e_\nu{}^\beta g_{\alpha\beta}\right\|
 = \left\|\begin{array}{cccc} -1 &&&\\ &1&&\\ &&1&\\ &&&1\end{array}\right\|.
\ee
Thus, for an orthonormal basis or any basis with respect to which the metric has constant components,
\index{Christoffel symbol!for an orthonormal basis}\index{Ricci rotation coefficients} 
\be \cblue
\Gamma_{ijk} = \frac{1}{2} (\, c_{kij} - c_{jki} - c_{ijk}\,).
\label{gamc}\ee
\vspace{3mm}

Cartan calculus (Appendix \ref{s:cartan}), an efficient way to compute the Riemann tensor, 
and the Newman-Penrose formalism (Chap. \ref{c:np}) each use bases of this kind.  \\

Having used the intuitive properties of parallel transport to define the covariant 
derivative operator of a metric, we now have a formal definition:\\
\noindent{\bf Definition}. A vector $v^a$, defined on a curve with tangent $w^a$, is {\sl parallel transported} along the curve if
\be 
    w^b\na_b v^a = 0.  
\ee   
\index{parallel transport|textbf}
Because the covariant directional derivative $w^b\na_b v^a$ involves only the values of $v^a$ on the curve, it is well-defined even if $v^a$ is defined only on the curve.   \\ 

\noindent
{\sl The n-form $\epsilon_{a_1\ldots a_n}$ and its covariant derivative.}

In Sect.~\ref{s:eta_delta_ep}, we introduced the totally antisymmetric tensor 
$\epsilon_{\a\b\c\d}$ whose one independent component was $\epsilon_{0123} = 1$ in an orthonormal basis.  In flat space, it is covariantly constant, $\na_\ep\epsilon_{\a\b\c\d}=0$, because 
it has constant components in a natural chart.  In curved space, we can again 
define $\epsilon_{\a\b\c\d}$ as the unique totally antisymmetric tensor (4-form) 
whose one independent component in an orthonormal basis is $\epsilon_{0123} = 1$.
Again its covariant derivative vanishes:  Start with the relation, 
$\epsilon_{\a\b\c\d} \epsilon^{\a\b\c\d} = -4!$; this follows from the 4! permutations of 4 indices and the value $\epsilon^{0123} = -1$ in an orthonormal basis at a point.  Take the covariant derivative of this relation, 
\[
 0 = \na_\iota(\epsilon_{\a\b\c\d} \epsilon^{\a\b\c\d}) 
 = \na_\iota (\epsilon_{\a\b\c\d}\epsilon_{\ep\zeta\eta\theta})g^{\a\ep}g^{\b\zeta} g^{\c\eta} g^{\d\theta}  
	= 2\epsilon^{\a\b\c\d}\na_\iota \epsilon_{\a\b\c\d}\ .
\] 
The $\mu^{th}$ component of the last equality is 
$0=4!\ep^{0123} \na_\mu\epsilon_{0123}$, so $\na_\mu\epsilon_{0123}=0$. 
Total antisymmetry of $\epsilon_{\a\b\c\d}$ implies $\na_\ep\epsilon_{\a\b\c\d}=0$. \\ 
We can similarly define the unique totally antisymmetric tensor 
$\epsilon_{a_1\cdots\, a_n}$ on an $n$-dimensional space with metric $g_{ab}$,
 and the same argument shows that $\na_b\epsilon_{a_1\cdots a_n} = 0$. \\
 
\noindent{\sl Local flatness:  Locally inertial or ``normal'' coordinates at a point $P$} \\
\index{coordinates!intertial coordinates}\index{coordinates!normal coordinates}\index{inertial coordinates}

Spacetime near any point looks like flat Minkowski space to linear order in the distance 
from a point.  In particular, one can always choose a chart about $P$ in which the metric has the form, 
\[ 
	\| g_{\mu\nu}(P)\| = \| \eta_{\mu\nu}\| 
			\equiv \left\|\begin{array}{cccc} -1 &&&\\ &1&&\\
					&&1&\\ &&&1\end{array}\right\|, 
\]
of the Minkowski metric in natural coordinates, and the first derivatives of the metric all vanish. In such a ``locally inertial'' chart, 
\[
  \Gamma_{\lambda\mu\nu}(P) = 0 = \Gamma^\lambda{\ \mu\nu}(P).
\]

\benr\item Show that  $\Gamma_{\lambda\mu\nu}(P) = 0$ implies that all first derivatives of 
the metric components vanish, $\partial_\lambda g_{\mu\nu}(P)=0$. (Hint: Add Eq.~\eqref{gamg} to 
Eq.~\eqref{gamg} with $i\leftrightarrow j$). 
\een  

More generally, in an $n$-dimensional space with a Lorentzian (signature $- + \cdots +$) or 
Riemannian (signature $+ + \cdots +$) metric, one can always 
choose a chart for which the coordinate basis at a point 
$P$ is orthonormal and for which all the derivatives of the metric components 
vanish at $P$.  We'll explicitly construct a chart of this form at the end of the next 
section.   

\newpage
\subsection{Geodesics} 
\index{geodesic|(}

\noindent {\sl Geodesic Equation}:
\index{geodesic!geodesic equation}\index{equation of motion!geodesic equation}

   Geodesics are paths of extremal length.  Consider a curve $c(\lambda)$ with 
$c(\lambda_1)=P$ and $c(\lambda_2) = Q$.  We show that the demand that the curve be of extremal length --- that
\[
	s:=\int\limits_{\lambda_1}^{\lambda_2} \left| g_{ab}v^av^b\right|^{1/2} d\lambda,
\]
 where
$v^a$ is tangent to $c(\lambda )$, be an extremum among all nearby curves
from $P$ to $Q$ --- is equivalent to demanding that the unit tangent vector
$u^a$ be parallel transported along $c$:
\be 
 u^a\nabla_au^b=0 \Leftrightarrow 
  \delta\int ds\equiv \delta \int \left| g_{ab}v^av^b\right|^{1/2} d\lambda = 0. 
\ee

Consider another curve $\bar c(\lambda )$ with the same endpoints $P$ and $Q$ as $c$.  
We again follow Wald in writing the coordinates of the curves $c$ and $\bar c$ in a chart  
as $x^i(\lambda) := x^i\circ c(\lambda)$, $\bar x^i(\lambda) = x^i\circ \bar c(\lambda)$. 
Then $\bar x^i(\lambda) = x^i(\lambda) + \delta x^i(\lambda )$ where 
$\delta x^i(\lambda_1) = \delta x^i(\lambda_2) = 0.$
\begin{align*}
s=\int ds &= \int_{\lambda_1}^{\lambda_2}\, \left| g_{ij }[x(\lambda)] \,
 	        \dot x^i (\lambda )\dot x^j(\lambda )\right|^{1/2}d\lambda , 
	 {\mbox{where ($\dot{\phantom{x}}$) means }} \frac{d}{d\lambda}\ \ .\\
\delta\int ds 
	&= \int_{\lambda_1}^{\lambda_2} \left\{\left| g_{ij}(x+\delta x)
	   (\dot x^i +\delta\dot x^i )(\dot x^j+\delta\dot x^j)\right|^{1/2} 
	    - \left| g_{ij}(x)\dot x^i \dot x^j\right|^{1/2}\right\}d\lambda
\end{align*}
Now $g_{ij}(x+\delta x) = g_{ij}(x) + \partial_k g_{ij}(x)\delta x^k 
	+ O(\delta x)^2.$
Then, to $O(\delta x)$,
\begin{eqnarray*}
 \delta\int ds 
 =\int_{\lambda_1}^{\lambda_2} \frac{1}{2} \left| g_{i j }\dot x^i\dot x^j\right|^{-1/2} 
 	\left[\partial_k g_{i j }(x(\lambda ))\delta x^k\dot x^i\dot x^j 
 		+ 2g_{ij}\dot x^i\delta \dot x^j\right] d\lambda 
\end{eqnarray*}
The last half of the calculation is easier if we take $\lambda$ to be 
proper distance along the unperturbed curve $c$ --- so that the tangent 
vector $v^a$ is the unit tangent $u^a$.  We then have $\dot x^i = u^i$, 
$\left| g_{i j }\dot x^i \dot x^j \right| = 1$,    
and
\[
	\delta\int ds = \int_{\lambda_1}^{\lambda_2}  \frac {1}{2} [\partial _k g_{i j } u^i u^j\delta x^k 
			+  2g_{i j } u^i \delta \dot x^j ] d\lambda . 
\]
Last term:
\begin{align*}
\int_{\lambda_1}^{\lambda_2} g_{ij }(x(\lambda ))u^i(x(\lambda ))
\frac{d}{d\lambda} \delta x^j(\lambda )d\lambda
&= -\int_{\lambda_1}^{\lambda_2} \frac{d}{d\lambda} \left[ g_{i j }(x(\lambda
))u^i(x(\lambda ))\right]\, \delta x^j d\lambda\\
&= - \int_{\lambda_1}^{\lambda_2} [ g_{i j }\dot u^i 
	+ \partial_k g_{i j }\dot x^k u^i]\,\delta x^j d\lambda
= -\int_{\lambda_1}^{\lambda_2} [ g_{ik}\dot u^i + \partial_j g_{ik}u^i u^j]\,\delta x^k d\lambda\\
&=-\int_{\lambda_1}^{\lambda_2} [ g_{ik}\dot u^i + \frac12\partial_j g_{ik}u^i u^j
			+ \frac12\partial_i g_{jk}u^i u^j]\,\delta x^k d\lambda. 
\end{align*}
Thus
\begin{align*}
\delta\int ds &= \int_{\lambda_1}^{\lambda_2} [-g_{i  k }\dot u^i - \frac12\partial_j g_{ik}u^i u^j
 	- \frac12\partial_i g_{jk}u^i u^j+ \frac{1}{2} \partial_k g_{ij}u^i u^j ]
 	\,\delta x^k  \\
&= -\int_{\lambda_1}^{\lambda_2} g_{kl} (\dot u^k  + \Gamma^k{}_{ij} u^i u^j )\,\delta x^l \ . 
\end{align*}
Then $\dis  \delta\int ds = 0 \ \ \mbox{ for all } \delta c\ \Rightarrow $\
\bsube
\begin{align} 
	 \dot u^k  + \Gamma^k {}_{ij} u^i u^j &= 0, \quad {\mbox{or }} 
\label{e:geodesica}\\
	 \frac{d^2 x^k}{d\tau^2} + \Gamma^k {}_{ij} \frac{dx^i}{d\tau} \frac{dx^j}{d\tau}&=0 \quad {\mbox{or }} \label{e:geodesicb}\\
\crv 	 u^b \nabla_b u^a  &\crv = 0.   \cb
\label{e:geodesicc}\end{align}\esube
\index{geodesic!geodesic equation|textbf}

For spacelike separated points, the shortest path between them is a geodesic, but there can be 
more than one geodesic joining two points; for example, an infinite set of longitude 
lines join the north and south poles of a sphere.  Each spacelike geodesic 
is a {\sl local} minimum of length.  Geodesics between timelike separated 
points, however, are local {\sl maxima} of length (of proper time).  This is just the usual behavior of time dilation: Clocks moving along accelerated paths run slow, registering shorter proper times. As in the figure on p. \pageref{hartle4.8},
the twin traveling on a roundtrip to a star returns home younger than the twin left behind.\\

Finally, light travels along geodesics $c(\lambda)$ whose tangent vector $k^a$ is everywhere null, 
$k^a k_a = 0$, but one can choose the parameter $\lambda$ so that the geodesic equation again 
has the form  
\be
	k^b\nabla_b k^a = 0, 
\ee
or 
\be
  \frac{d^2 x^k}{d\lambda^2} + \Gamma^k {}_{ij} \frac{dx^i}{d\lambda} \frac{dx^j}{d\lambda}=0.
\label{geodesic_affine}\ee
The parameter $\lambda$ is now not distance, but is called an {\sl affine} parameter.\index{affine parameter|textbf}  
It is not unique:  If we replace $\lambda$ by $\bar \lambda = C\lambda$ the geodesic equation 
keeps the same form.  For timelike or spacelike geodesics, proper time $\tau$ and 
spatial distance $s$ are affine parameters, because the geodesic equation has the form 
\eqref{e:geodesicc}, but so are $Cs$ and $C\tau$.  We'll use an affine parameter in the 
next section, as part of a construction of coordinates that exhibit the local flatness of spacetime. \\ 

The geodesic equation in the form \eqref{e:geodesicb} is a coupled set of 
ordinary differential equations (ODEs) for the functions $x^i(\tau)$.  
The equations are second-order (involving only first and second derivatives), 
and second order ODEs have unique solutions for any given initial values 
$x^i(0)$ and $\dot x^i(0)$.  That is, \crv given an initial starting point $c(0)=P$ and 
an initial tangent vector $T^a$ at $P$, there is a unique geodesic $c(\lambda)$  through $P$. \cb
(This is the fundamental existence theorem for solutions of ODEs. For a proof, see, for example, 
Coddington \& Levinson, {\sl Theory of Ordinary Differential Equations}\cite{coddington55}. The 
statement of the theorem is in Wikipedia's ODE article.) \\
\newpage

\benr
\item Because a geodesic is a path of extremal length between two points, one can obtain the 
geodesic equation by requiring \vspace{-4mm}
\index{Lagrangian!for geodesic equation} 
\[
  0 = \delta \int_{\tau_1}^{\tau_2} d\tau =  \delta \int_{\tau_1}^{\tau_2} L(\dot x, x)  d\tau , 
\quad \mbox{where} \quad L = \sqrt{-g_{\mu\nu}(x)\dot x^\mu\dot x^\nu}
\]
and where $\tau$ is proper time along the geodesic $x^i(\tau)$ (but not along the perturbed path 
$x^\mu(\tau)+\delta x^\mu(\tau)$).  
\benalph \vspace{-3mm}
\item 
By using $L =1$ for a geodesic, 
show that the geodesic equation also follows for  \vspace{-3mm}
\[
   0 = \delta  \int_{\tau_1}^{\tau_2} F[L(\dot x, x)] d\tau ,\vspace{-3mm} 
\] 
if $dF/dL\neq 0$ at the geodesic, where $F(L)$ is some smooth function of $L$. 
You will only need the fact that $L=1$ along the geodesic (implying $F'(L)= F'(1)$), 
so use an arbitrary function $L(\dot x, x)$ rather than replacing $L$ 
by its explicit form in terms of $g$. 
\item Derive the geodesic equation as the Euler-Lagrange equations of the Lagrangian \\
$\wt L(\dot x,x) = -L^2(\dot x, x) = g_{\mu\nu}(x)\dot x^\mu \dot x^\nu$.   
\een \vspace{-3mm}
\item
\benalph
\item Find the four components of the geodesic equations for flat space with polar coordinates 
as the Euler-Lagrange equations for the Lagrangian 
$-\dot t^2+\dot r^2 +r^2\dot\theta^2 +r^2\sin^2\theta \dot\phi^2$.
\item This is an efficient way to compute the Christoffel symbols.  Using your result 
for the last part, and the geodesic equation in the form 
$   \ddot x^\mu + \Gamma^\mu_{\nu\lambda} \dot x^\nu\dot x^\lambda$, \\
find the nonzero Christoffel symbols of the Minkowski metric in polar coordinates. 
\een
\item \phantom{xx}  
\benalph\vspace{-2mm}
\item Write the components of the equation ${\bf v\cdot \nabla v} = 0$ 
for a geodesic in Euclidean 3-space in polar coordinates, where  
$\dis v^i = \dot x^i$, with $\dis (^.) := \frac d{ds}$.  Replace $v^j\partial_j v^i$ by $\dot v^i$ 
or $\ddot x^i$. (Use the coordinate basis, as in the previous problem, not an orthonormal basis.)
\item Specialize now to the $\theta=\pi/2$ plane, and use 
$ds^2=dr^2+r^2 d\phi^2$ to deduce $1=\dot r^2+ r^2\dot\phi^2$. Check that the derivative of this 
equation, with respect to $s$ yields the $r$-component of the 
geodesic equation.
\item Check that the $\phi$-component of the geodesic equation has 
the first integral $r^2\dot\phi = l$, $l=$ constant.  
\item What equation for central force motion corresponds to the equation $r^2\dot\phi = l$?
Why does this agree for motion in 
a straight line, when $t$ for central-force motion is replaced here by $s$?   
\een
\een

\noindent{\sl The exponential map and geodesic coordinate systems}
\index{exponential map}\index{geodesic coordinates} \\

As mentioned above, spacetime (and, in general, a manifold with a smooth metric) is locally flat.  If you are a small bug on a surface with radius of curvature much larger than the bug, the surface will look flat to you. Our goal here is to find a coordinate system in which the metric components are as close as possible to $\eta_{\mu\nu}$, coordinates that are as close as possible to Cartesian.     

In Minkowski space, the set of all geodesics--straight lines--through any point 
$P$ exactly fills the entire space.  Regarding a vector $X^\a$ as a connecting vector from $P$ 
gives a one-one map from $X^\a$ to a point $Q$: $X^\a \mapsto Q^\a = P^\a + X^\a$. 
The components $X^\mu$ are then the coordinates of $Q$ in a chart with origin at $P$.  
We now restate this construction in a way that can be carried over 
to a curved spacetime.  The idea is again to identify each tangent vector $X^a$ at $P$ with the point $Q$ a distance $|X|$ from $P$ along the geodesic from $P$ to $Q$ with tangent $X^a$, but with 
a description that also works for light-like vectors.   

Given a vector $X^\a$, there is a unique geodesic through $P$ with tangent $X^\a$. The coordinates 
$x^\mu(\lambda)$ of the geodesic satisfy (still in Minkowski space) 
\be
   \frac{dx^\mu}{d\lambda} = X^\mu, \qquad \frac{d^2x^\mu}{d\lambda^2} = 0.
\ee
That is, the geodesic has the form 
\be
  x^\mu(\lambda) = \lambda X^\mu,   
\ee 
and $x^\mu=X^\mu$ at $\lambda=1$.

  For spacelike or timelike geodesics parametrized 
by proper distance or proper time, the tangent vectors are unit vectors, 
but here the tangent vector has length equal to the distance to $Q$.  The parameter 
$\lambda$ is  $s/|X|$ or $\tau/|X|$, differing from proper distance (or proper time) by 
the constant factor $1/|X|$. By definition, it is an affine parameter,\index{affine parameter} and the map 
is defined for null vectors $X^\a$ where the proper distance vanishes.  
Like the original connecting-vector description, our construction of the map 
$X^\alpha\mapsto Q$ doesn't mention $|X|$ and doesn't care whether 
$X^\a$ is spacelike, timelike or null.  
 
The map from vectors at a point $P$ to points near $P$ is called the {\sl exponential map}. 
We can now say nearly the same words to define the exponential map in a curved space $M$.    
Let's start with a sphere and consider the tangent space at a point $P$, which we can call 
the north pole, as in Fig.~\ref{exp_map}.  
\begin{figure}[h!] 
\centerline{\includegraphics[width=.6\textwidth]{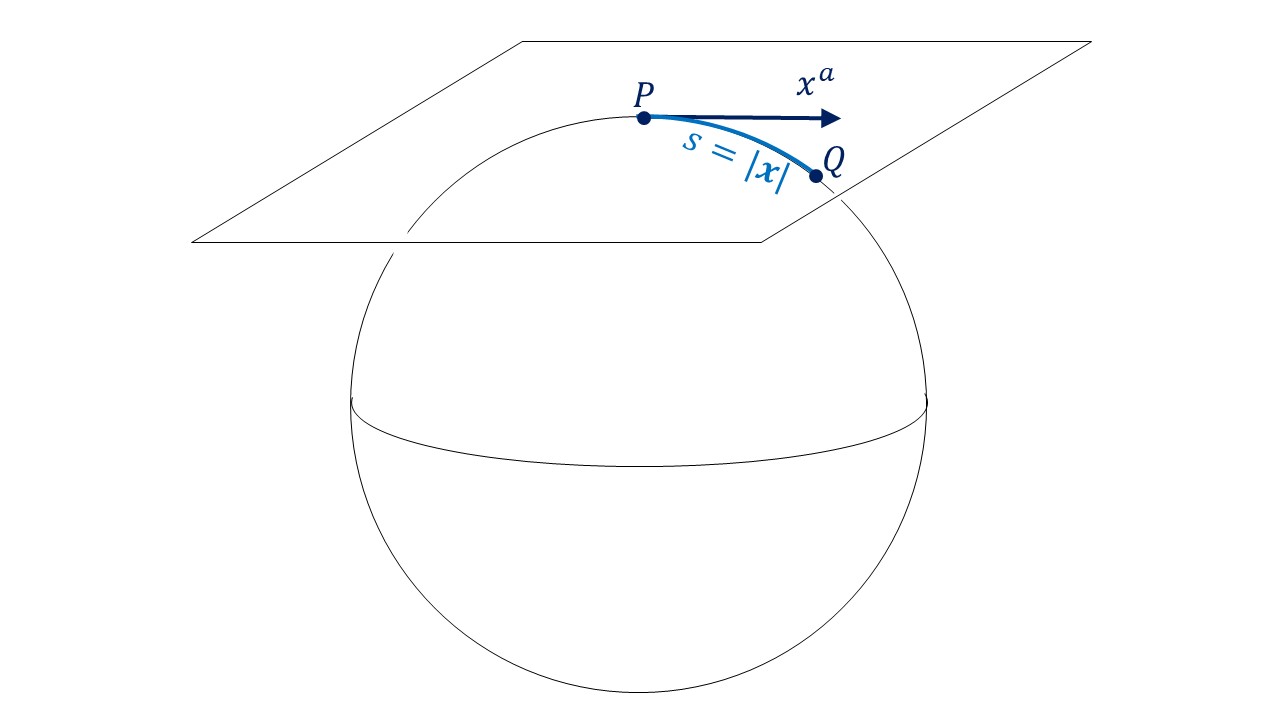} } 
\caption{The exponential map takes a vector $X^a$ at $P$ to the point $Q$ a parameter distance $\lambda = s/|X|=1$ from $P$ along the geodesic through $P$ with tangent $X^a$.}  
\label{exp_map} 
\end{figure} 
Given a vector $X^a$ at $P$, 
there is a unique geodesic (longitude line) through $P$ with tangent $X^a$. We map $X^a$ 
to the point $Q$ of the sphere at parameter distance $\lambda=1$ along the geodesic (longitude line) through $P$.  This gives a 1-1 map as long as the distance $|\bm X|$ is less than half the circumference of the sphere.  At the south pole, of course, all geodesics through $P$ meet.

Here is the statement for spacetime (or any smooth manifold with a metric): 
Every point $P$ has a finite neighborhood in which geodesics from $P$ do not cross each other. 
Let $T_PM$ be the space of vectors at $P$ (the tangent space introduced on p. \pageref{p:TM}).       
{\crv The {\sl exponential map} is the map $T_PM\rightarrow M$ that takes each vector 
$X^a$ at $P$ to the point $Q$ a parameter distance $\lambda=1$ along the geodesic through 
$P$ with tangent $X^a$.} 
\newpage    

\noindent {\sl Riemann normal coordinates}.\\
\index{coordinates!Riemann normal|textbf}\index{Riemann normal coordinates|textbf}

We can now use the exponential map for a curved space $M$ to construct a chart with 
with origin at $P$ in $M$, analogous to the chart $\{ x^i\}$ of $\mathbb R^n$ with origin at a point $P$ of $\mathbb R^n$.   Let $\{\bm e_i\}$ be an orthonormal basis at $P\in M$.  Any vector $x^a$ 
at $P$ is a sum of the basis vectors, $\bm x = x^i \bm e_i$.  The components $x^i$ are the coordinates of the point $Q$ associated with $x^a$. That is:  \\ 
\crv Each point $Q$ in a neighborhood of $P$ is joined to $P$ by a unique geodesic $c(\lambda)$, and the coordinates of $Q$ are $(x^1, \ldots, x^n)$, where $\bm x=x^i\bm e_i$ is the tangent to $c$ at $P$.   \cb 
\\

We now show that, in this chart for spacetime, the metric has components $g_{\mu\nu}(P) =\eta_{\mu\nu}$ and that its first derivatives vanish at $P$: $\partial_\lambda g_{\mu\nu} = 0$. To do this, we 
show that (1) the coordinates of geodesics starting from $P$ with tangent $X^\mu$ have the exact flat space form \eqref{exp_map}, $x^\mu(\lambda) = \lambda X^\mu$; and (2)  the  
geodesic equation \eqref{geodesic_affine} implies $\Gamma^\lambda_{\,\mu\nu}(P)=0$ in this chart. \\
(1) Suppose $R=c(\lambda_R)$ is any point on the geodesic from $P$ to $Q$ with tangent $X^\a$.
$R$ is the same direction as $Q$ but is at $\lambda=\lambda_R$ instead of at $\lambda=1$. 
The geodesic that meets $R$ at $\lambda=1$ is then $\bar c(\lambda) := c(\lambda_R\,\lambda)$ with 
coordinates $\bar x^\mu(\lambda)= x^\mu(\lambda_R\, \lambda)$.  The 
tangent vector to that geodesic is $\bar X^\alpha = \lambda_R X^\alpha$:  That is, 
by the chain rule,
\be
  \left.\frac{d\bar x^\mu(\lambda)}{d\lambda}\right|_{\lambda=0} 
	= \lambda_R \left.\frac{d x^\mu(\lambda)}{d\lambda}\right|_{\lambda=0}=\lambda_R X^\mu.
\ee
By our definition of the coordinates, $R$ has coordinates $x^\mu = \lambda_R X^\mu$.  
Since this is true for each point $R$ on the geodesic, $x^\mu(\lambda) = \lambda X^\mu$ as claimed.\\
(2) For $x^\mu(\lambda) = \lambda X^\mu$, we have $\ddot x^\mu=0$, where $({}^\cdot) := d/d\lambda$.   The geodesic equation 
\eqref{geodesic_affine} then implies $\Gamma^\lambda_{\mu\nu}\dot x^\mu\dot x^\nu=\Gamma^\lambda_{\mu\nu}X^\mu X^\nu=0$.  But this is true for all geodesics through $P$, implying that, at $P$ 
it is true for all vectors $X^\alpha$.  Then $\Gamma^\lambda_{\mu\nu}(P)=0$ as claimed, implying 
$\partial_\lambda g_{\mu\nu}(P)=0$.  

Finally, because the basis is orthonormal, the metric at $P$ has components 
$g_{\mu\nu} = \bm e_\mu\cdot \bm e_\nu = \eta_{\mu\nu}$.  $\Box$

\noindent{\sl Terminology.} {\sl Inertial coordinates} in a neighborhood of a point $P$ 
are coordinates for which the metric at $P$ has the form diag$[-1,1,1,1]$ and 
all derivatives of the metric components vanish at $P$. Riemann normal coordinates 
are a subclass of inertial coordinates.  Riemann normal coordinates are sometimes 
called {\sl geodesic coordinates}. But in the mathematical literature,
geodesic coordinate systems include the more general class of coordinates obtained by 
the exponential map from a general basis at $P$, instead of the orthonormal basis that gives Riemann 
normal coordinates.\index{coordinates!geodesic}\index{coordinates!inertial|textbf}\index{geodesic coordinates|textbf}\index{inertial coordinates|textbf} 
\index{geodesic|)}

\subsection{Curvature Tensor}
\index{Riemann tensor|textbf}\index{curvature tensor|textbf}

In a flat spacetime, covariant derivatives commute.  In curved spacetime,
however,
\[ 
	\nabla_{[a}\nabla_{b]} v_c \neq  0 
\]
\noindent and as we will see, the fact is closely related to the statement
that parallel transport from $A$ to $B$ depends on the path from $A$ to $B$.
Write
\[
\frac{1}{2} R_{abc}({\bm v}) \equiv \nabla_{[a}\nabla_{b]} v_c;
\]
so far, this just gives a new name to the tensor $\nabla_{[a}\nabla_{b]} v_c$.
As a function of $\bm v$, $R_{abc}$ is linear under addition and 
multiplication by scalar {\em fields}:
\begin{eqnarray*}
\nabla_{[a} \nabla_{b]}(fv_c) &=& \nabla_{[a}(\nabla_{b]}^{}f v_c +
f\nabla_{b]} v_c)\\
&=& \underbrace{ \nabla_{[a} \nabla_{b]} f}_{0} v_c +
 \cancel{ \nabla_{[b} f\nabla_{a]} }{-7} v_c +
 \cancel{ \nabla_{[a} f\nabla_{b]} }{-7} v_c + f\nabla_{[a}\nabla_{b]}v_c\\
&=& f\nabla_{[a}\nabla_{b]} v_c
\end{eqnarray*}
Therefore $R_{abc}(v) = R_{abc}{}^d v_d$.  The tensor $R_{abc}{}^d$ is called 
the {\em Riemann tensor} or the {\em curvature tensor}.
\index{curvature!curvature tensor|textbf}
\be 
	\nabla_{[a}\nabla_{b]}v_c = \frac{1}{2} R_{abc}{}^d v_d\ , 
\ee
or
\be\crv
	[\nabla_a, \nabla_b]v_c = R_{abc}{}^d v_d\ .\cb
\label{rabcd}\ee
It has the {\crv index symmetries}
\begin{eqnarray}
\left.\begin{array}{ll}
&{\crv R_{abcd} = -R_{bacd}} \hspace{8mm} {\mbox{from its definition (\ref{rabcd}))}}\vspace{2mm}\\
&\crv R_{abcd} =-R_{abdc} \vspace{2mm}\\
&\crv R_{abcd}=R_{cdab}\vspace{2mm}\\
{\text{and}} \hspace{20mm} &R_{a[bcd]} =0=R_{abcd}+R_{adbc}+R_{acdb}\ ; 
\end{array}\right\}
\label{e:Rsymm}
\end{eqnarray}
these are not difficult to show, are part of the problems at the end of Sec.\ref{s:bianchi}, 
and also follow from the form \eqref{e:RatP} of its components in a locally inertial chart.   

 Acting on a vector $v^a$, the commutator of two covariant derivatives  
is again given in terms of the curvature tensor:  Using 
$\nabla_ag^{bc} = 0$, we have
\index{commutator!of two covariant derivatives|textbf}
\[ 
	[\nabla_a, \nabla_b] v^c =  R_{ab}{}^c{}_d v^d ;\]
and if $T^{a\cdots b}{}_{c\cdots d}$ is any tensor,
\begin{eqnarray}
[
 \nabla_e, \nabla_f] T^{a\cdots b}{}_{c\cdots d} 
  &=& R_{ef}{}^a{}_g T^{g\cdots b}{}_{c\cdots d} +\cdots + R_{ef}{}^b{}_g T^{a\cdots g}{}_{c\cdots d} \nonumber\\
&& 	+ R_{efc}{}^gT^{a\cdots b}{}_{g\cdots d} +\cdots 
	+ R_{efd}{}^g T^{a\cdots b}{}_{c\cdots g},
\end{eqnarray}
which follows by Leibnitz, writing $T^{a\cdots b}{}_{c\cdots d}$ as a
linear combination of outer products of vector and dual vector fields:
\begin{eqnarray*}
{\text{e.g.}}\hspace{3mm} 
\nabla_{[a}\nabla_{b]}(\sigma_c\tau_d) 
	&=& (\nabla_{[a}\nabla_{b]}\sigma_c)\tau_d 
	  +\cancel{\nabla_{[a}\sigma_{|c|}\nabla_{b]}\tau_d }{-10}
	  +\cancel{\nabla_{[b}\sigma_{|c|}\nabla_{a]}\tau_d }{-10}
	  +\sigma_c\nabla_{[a}\nabla_{b]}\tau_d \\
	&=& \frac{1}{2} R_{abc}{}^e\sigma_e\tau_d 
	  +\frac{1}{2} R_{abd}{}^e\sigma_c\tau_e .
\end{eqnarray*}

	We will write the coordinate components of $R_{abcd}$ in terms of the
Christoffel symbols $\Gamma^i_{jk}$ and show that in a locally inertial chart 
about $P$, $R_{ijkl}$ has a simple form in which the symmetries (\ref{e:Rsymm}) are 
manifest. First the components in an arbitrary chart:
\begin{eqnarray*}
R_{i j  k}{}^l v_l &=& \nabla_i \nabla_j v_k - \nabla_j\nabla_i v_k\\
&=& \partial_i \nabla_j v_k - \cancel{\Gamma^l_{\ ji}\nabla_l v_k}{-8} 
	- \Gamma^l_{\ ki}\nabla_j v_l\\ 
&&   -\partial_j \nabla_i v_k +
\cancel{\Gamma^l_{\ ij}\nabla_l v_k}{-8} + \Gamma^l_{\ kj}\nabla_i v_l,
\end{eqnarray*}
where the symmetry $\Gamma^l_{\ ji} = \Gamma^l_{\ ij}$ for a coordinate 
basis was used in the cancellation of the second and fifth terms.

Then
\begin{eqnarray*}
R_{ijk}{}^l v_l  
&=& \partial_i ({\underline {\partial_j v_k }}-\Gamma^l_{\ kj}v_l) 
	  -\Gamma^l_{\ ki}(\partial_j v_l - \Gamma^m_{\ lj}v_m)\\
&& -\partial_j( {\underline{\partial_i v_k }} - \Gamma^l_{\ ki}v_l ) 
	  + \Gamma^l_{\ kj}(\partial_i v_l - \Gamma^m_{\ li}v_m)\\
&=& -\partial_i  \Gamma^l_{\ kj}v_l  
	- \cancel{\Gamma^l_{\ kj}\partial _i v_l}{-6} 
	- \cancell{\Gamma^l_{\ ki}\partial_j v_l }{-7} 
	+ \Gamma^l_{\ ki}\Gamma^m_{\ lj}v_m \\ 
&&  	+ \partial_j\Gamma^l_{\ ki}v_l  
	+ \cancell{\Gamma^l_{\ ki}\partial_j v_l }{-7} 
	+ \cancel{\Gamma^l_{\ kj}\partial_i v_l }{-6} 
	- \Gamma^l {}_{k j } \Gamma^m_{\ li}v_m  \end{eqnarray*}
\[
 R_{ijk}{}^l 
	= \partial_j\Gamma^l{}_{ki} - \partial_i \Gamma^l{}_{kj} 
  	  +\Gamma^m_{\ ki}\Gamma^l_{\ mj} - \Gamma^m {}_{kj} \Gamma^l_{\ mi}
\]
Using the index symmetries to interchange $k$ and $l$ and exchange the 
pairs $i,j$ and $k,l$,
we can order the indices in a way that allows comparison with, e.g., MTW's inside cover:
\be\cblue
R^k_{\ lij} 
	= \partial _i \Gamma^k{}_{lj}-\partial_j\Gamma^k{}_{li}  
  	  +\Gamma^k_{\ mi}\Gamma^m_{\ lj} - \Gamma^k {}_{mj} \Gamma^m_{\ li}\cb.
\label{e:rijklup}\ee
\index{curvature tensor!components in general chart}
\index{Riemann tensor!components in general chart}
Components of the Riemann tensor in a general basis are given in Eq.~\eqref{rijkl1} of 
the Appendix, where they are used for the efficient Cartan-calculus calculation of 
Riemann tensor components in an orthonormal basis.   \\

In a locally inertial chart about $a$ point $P$

\[
 \Gamma^i {}_{j  k}|_P = 0 ,\qquad  \partial _i g_{jk}| _P   = 0 .
\]
Then, at $P$, only derivatives of $\Gamma$ and second derivatives of $g_{ij}$
survive, and we have  
\be
\left. R_{ijk}{}^l\right|_P 
 	= \left. \partial_j\Gamma^l{}_{ki}\right|_P 
 	 -\left. \partial _i \Gamma^l{}_{kj}\right|_P, \qquad
\left. R_{ijkl}\right|_P 
 	= \left. \partial_j\Gamma_{lki}\right|_P 
 	 -\left. \partial _i \Gamma_{lkj}\right|_P; 
\label{rinert}\ee
\begin{eqnarray}
\partial_ j  \Gamma^l_{\ ki} |_P 
&=& \frac{1}{2} \partial_j [ g^{lm}(\partial_k g_{mi} 
	+ \partial_i g_{mk} -\partial_m g_{ik}) ] |_P
= \frac{1}{2} g^{lm} \left(\partial_j \partial_k g_{im} 
	+ \partial_i \partial_j g_{km} 
	- \partial_m \partial_j  g_{ik}\right)|_P, \nonumber\\
\partial_ j  \Gamma_{lki} |_P  
&=& \frac12\left(\partial_j \partial_k g_{il} 
	+ \partial_i \partial_j g_{kl} 
	- \partial_l \partial_j  g_{ik}\right)|_P.
\end{eqnarray} 
Substituting this last expression and its (i$\leftrightarrow$j) counterpart 
into (\ref{rinert}), we obtain (only at the single point $P$) 
\index{Riemann tensor!components in inertial chart}\index{curvature tensor!components in inertial chart}
\be 
R_{ijkl}(P) = \frac{1}{2} [\partial _j\partial _k g _{i  l } 
		  + \partial _i \partial_l g_{jk} - \partial _i \partial _k g _{jl } 
		  - \left.\partial _j \partial_l g _{ik}]\right|_P ,
\label{e:RatP}\ee

\noindent from which $R_{ijkl} = R_{[ij][kl]} = R_{klij}$ is evident, and
$R_{ijkl} + R_{kijl} + R_{jkil} = 0$ follows as soon as it's written.  If
all the components of a tensor, e.g.,\ the tensor
$R_{abcd}-R_{[ab][cd]}$,\  vanish in any basis, the tensor vanishes, implying 
the index symmetries \eqref{e:Rsymm}.

From the Riemann tensor $R_{acb}{}^d$ one constructs the {\sl Ricci} tensor by
contracting: \index{curvature tensor!Ricci tensor, Ricci scalar|textbf}
\index{Riemann tensor!Ricci tensor, Ricci scalar|textbf}
\index{Ricci tensor, Ricci scalar|textbf}
\be 
	{\crv R_{ab} := R_{acb}{}^c }= R_b{}^c{}_{ac} = R_{bca}{}^c = R_{ba} 
\ee
[the other contractions give the same tensor up to sign $(-R_{ad}=R_{ab}{}^b{}_d)$ 
or zero $(R_a{}^a{}_{cd}=0)$].  A further contraction produces the {\sl Ricci scalar}:
\be {\crv R = R^a{}_a }= R^{ab}{}_{ab}. \ee

\subsection{Bianchi Identities}\index{Bianchi identities|textbf}
\index{Riemann tensor!Bianchi identities}\index{curvature tensor!Bianchi identities}
\label{s:bianchi}
From the coordinate components (\ref{e:rijklup}) of $R_{abc}{}^d$, 
it is easy to show the Bianchi identity
\be \crv
	\nabla_{[a}R_{bc]de} = 0 ,
\label{bianchi}\ee 
and (\ref{bianchi}) may also be proven directly---without reference 
to components (Exercise below).  In Riemann normal coordinates at $P$, 
the covariant derivative $\nabla_aR_{bcd}{}^e$ has components 
\begin{eqnarray*}
\left.\nabla_i R_{jkl}{}^m \right|_P 
&=& \left.\partial_i R_{jkl}{}^m  \right|_P\\
&=& 	  \left.\partial_i \partial_k \Gamma^m_{\ jl}\right|_P 
	- \left.\partial_i\partial_j\Gamma^m _{\ kl}\right|_P,
\end{eqnarray*}
where we have used  
\be
\left.\partial_i(\Gamma^\cdot_{\cdot\cdot}\Gamma^\cdot_{\cdot\cdot})\right|_P 
= \left(\left.\partial_i\Gamma^\cdot_{\cdot\cdot}\right|_P\right)
	\underbrace{\left.\Gamma^\cdot_{\cdot\cdot}\right|_P}_0 +
	\underbrace{\left.\Gamma^\cdot_{\cdot\cdot}\right|_P}_0
	\left.\partial_i \Gamma^\cdot_{\cdot\cdot}\right|_P\ .\\
\ee
Thus, at $P$, 
\begin{eqnarray*}
\nabla_i  R_{jkl}{}^m  + \nabla_j R_{kil}{}^m + \nabla_k R_{ijl}{}^m 
&=& \underline{\partial_i \partial_k\Gamma^m{}_{jl}} 
- \cancel{\partial_i\partial_j\Gamma^m{}_{kl}}{-6}\\ 
&&  + \cancel{\partial_j\partial_i  \Gamma^m{}_{kl}}{-6} 
- \cancell{\partial_j\partial_k\Gamma^m {}_{il}}{-6} \\ 
&& \mbox{} + \cancell{\partial_k \partial _j \Gamma^m{}_{il}}{-6} 
- {\underline {\partial _k \partial_i \Gamma^m{}_{jl}}}\\ 
&=& 0 \end{eqnarray*} 

By contracting $\nabla_aR_{bcde} + \nabla_bR_{cade} + \nabla_cR_{bade} = 0$
on $c$ and $e$ we obtain
\[
 \nabla_aR_{bd} - \nabla_bR_{ad} + \nabla_cR_{abd}{}^c = 0 
\]
Contracting on a different pair of indices produces either the same
equation (e.g.\ $a$ and $e$) or $0=0$ (e.g.,  $a$ and $b$).

\noindent Contracting on a further pair of indices $b$ and $d$ gives
\begin{eqnarray}
0 &=& \nabla_aR-\nabla_bR_a{}^b -\nabla_cR_a{}^c\nonumber\\
&=& \nabla_aR - 2\nabla_bR_a{}^b . \end{eqnarray} 
Then, defining the Einstein tensor $G_{ab}$ by
\be\crv
 G_{ab}=R_{ab}-\frac{1}{2} g_{ab}R ,\cb
\ee
one can write this last contracted Bianchi identity in the form
\be \crv
	\nabla_b G^{ab}=0 .
\label{cbianchi}\ee 
The contracted identity (\ref{cbianchi}) plays a major role in relativity---the field
equation is $G_{\alpha\beta}=8\pi G T_{\alpha\beta}$ and so the equation of motion
$\nabla_\beta T^{\alpha\beta}=0$ is a consequence of the field equation and the Bianchi
identity (\ref{cbianchi}).\index{Einstein field equation}\index{field equation}\\ 

\benr\item  Use the definition of the
curvature tensor,
\[ \nabla_{[a}\nabla_{b]}\sigma_c = \frac{1}{2} R_{abc}{}^d\sigma_d ,\]
the relation
\[ \nabla_ag_{bc} =0,\] 
and the Leibnitz rule to show that for any tensor $T^{ab}{}_c$~,
\[ 
\nabla_{[a}\nabla_{b]}T^{cd}{}_e 
=   \frac{1}{2} R_{ab}{}^c{}_f T^{fd}{}_e + \frac{1}{2} R_{ab}{}^d{}_f T^{cf}{}_e 
  + \frac{1}{2}R_{abe}{}^fT^{cd}{}_f
\]

\item  
\index{curvature tensor!index symmetries}\index{Riemann tensor!index symmetries}
By writing
$\nabla_{[a}\nabla_{b]}g_{cd}=0$, show that $R_{ab(cd)} =0$.  Show that
$R_{[abc]d} =0$, by looking at $\nabla_{[a}\nabla_b\nabla_{c]}f$.\\ 

\item  Show that the symmetries
$R_{abcd}=R_{[ab][cd]}$ and $R_{a[bcd]}=0$ imply $R_{abcd} = R_{cdab}$.\\

\item  Prove the Bianchi identity
$\nabla_{[a}R_{bc]de} =0$ this way:  The two equations, 
\begin{eqnarray*}
\nabla_a\nabla_b\nabla_cv_d-\nabla_a\nabla_c\nabla_bv_d &=&
\nabla_aR_{bcd}{}^ev_e + R_{bcd}{}^e\nabla_av_e\\
\nabla_a\nabla_b\nabla_cv_d-\nabla_b\nabla_a\nabla_cv_d &=&
R_{abc}{}^e\nabla_ev_d + R_{abd}{}^e\nabla_cv_e ,
\end{eqnarray*}
allow you to write $\nabla_{[a}\nabla_b\nabla_{c]}v_d$ in two ways.  Do so,
and equate the resulting expressions to obtain $\nabla_{[a}R_{bc]de} =0$.

\item\ben\item[a.] Eq.~\eqref{e:commutator1}, $[u,v]^a =u^b\nabla_b v^a-v^b\nabla_b u^a$, is proved by applying $[u,v]^a$ to an arbitrary scalar $f$.  
Prove this relation in a different (shorter) way, by showing that the components of 
$u^b\nabla_b v^a-v^b\nabla_b u^a$ in a coordinate basis agree with Eq.\eqref{e:commutator}, 
$[u,v]^i  = u^j \partial_j v^i -v^j \partial_j u^i $.   
\item[b.] Fill in the missing steps in the equations leading from Eqs.~\eqref{rinert} to \eqref{e:RatP} in the 
notes, to obtain the components of the Riemann tensor in locally inertial coordinates.
\een

\item Calculate the Riemann tensor for a cylindrical surface, using cylindrical coordinates.  You should find that it vanishes.  In constructing a cylinder from a flat sheet, you don't stretch the sheet:  All angles and intrinsic lengths (measured along the cylinder, not in the 3-dimensional space in which it's embedded) are unchanged.

\een
\subsection{Meaning of \texorpdfstring{$\bm{R^a{}_{bcd}}$}. }

	We now carry through two calculations that show the geometrical and 
physical meaning of the Riemann curvature tensor $R^a{}_{bcd}$.  
The first relates the Riemann tensor to the path dependence of parallel 
transport:  When the curvature is nonzero, a vector $v^a$ parallel 
transported around a closed path returns not to its original value, but to
a new value $v^a+\delta v^a$ related to $v^a$ by a Lorentz transformation. 
For small loops, $\delta v^a$ is proportional to the Riemann tensor 
and to the area of the loop.  The second calculation shows that the 
relative acceleration of nearby free particles (geodesics) 
is proportional to $R^a{}_{bcd}$, which can thus be regarded as the 
gravitational force that can't be transformed away by making
measurements in a freely falling frame. \\

\index{commutator!of two covariant derivatives} 
\noindent{\sl First calculation: parallel transport around a loop} \\ 
\index{curvature tensor!parallel transport around a loop}
\index{Riemann tensor!parallel transport around a loop}
\index{parallel transport!around a loop}

We look at a loop in a two-dimensional surface.  Following Wald's notation, we 
choose coordinates $t,s$ (you could just as well call them $x,y$ or $t,x$) and look at 
a loop with sides $\Delta t$ and $\Delta s$ as shown in Fig.~\ref{loop}.  
We denote by $T^a$ and $S^a$ the coordinate basis vectors $\bm\partial_t$ and 
$\bm\partial_s$, so that $T^a\nabla_a f(t,s) = \partial_t f(t,s)$ 
for a function $f$ on the surface. 
We are expecting to get a commutator $[\bm T\cdot\nabla, \bm S\cdot \nabla]$ of 
derivatives along $\bm S$ and $\bm T$ from going around the path, so look for 
antisymmetry under $S\leftrightarrow T$ as $s$ and $t$ first increase and then 
decrease along the path.  
\begin{figure}[h] 
\centerline{\includegraphics[width=.7\textwidth]{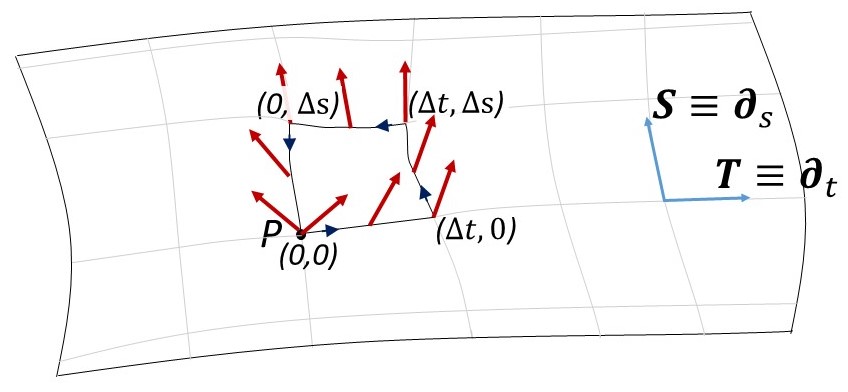} } 
\caption{The red vector at $P$ pointing rightward is parallel transported counterclockwise around the loop, coming back to $P$ as the vector pointing leftward. Gray lines of constant coordinates $t$ and 
$s$ are shown with their coordinate basis vectors $T^a$ and $S^a$ 
($\bm \partial_t$ and $\bm \partial_s$).}  
\label{loop} 
\end{figure} 

We start with a vector $v^a$ at any point $P$ and take the origin of the coordinates to be at $P$.  
The goal is to parallel-transport $v^a$ counterclockwise around the loop, 
from $(0,0)$ to $(\Delta t,0)$ to $(\Delta t,\Delta s)$ to $(0,\Delta s)$ and back to 
the origin. We could find the equation for the components $v^i = v^a \omega^i_a$ along a basis
dual vector but will follow Wald's more concise treatment, picking an arbitrary dual vector field $w_a$ 
and computing $v^a w_a$. 
For the first leg of the loop, $s=0$, and $t$ changes from $0$ to $\Delta t$. We have%
\footnote{Wald evaluates the first derivative on the right side of \eqref{e:vw1} at $(\Delta t/2,0)$ to get rid of the 
order $(\Delta t)^2$ term.  This is clever, but he pays for it later by having to 
imagine a second construction, parallel transporting $v^a$ along another segment at 
$\Delta t/2$.} 
\be
    (v^a w_a)(\Delta t,0) 
	   = v^a w_a (P) + \left.{\cblue \partial_t(v^a w_a)}\right|_{(0,\, 0)}\Delta t 
		  +\frac12\left. \partial_t^2 (v^a w_a)\right|_{(0,\, 0)}(\Delta t)^2 				  + O(\Delta t^3).
\label{e:vw1}\ee 
Now 
\[
  {\cblue \partial_t(v^a w_a)}  = T^b\nabla_b(v^a w_a)
	  			  =  v^a T^b\nabla_b w_a 
\] 
where the last equality is the requirement that $v^a$ be parallel translated: 
$T^b\nabla_b v^a=0$.  The change $\delta_1$ in $v^aw_a$ along the first part of the loop is then 
\[
  \delta_1 =  v^a \left.T^b\nabla_b w_a \right|_{(0,\, 0)} \Delta t 
		  +\frac12\left. \partial_t^2 (v^a w_a)\right|_{(0,\, 0)}(\Delta t)^2 
		  + O((\Delta t)^3).
\]
    
Along the second leg, $t=\Delta t$, and $s$ changes from $0$ to $\Delta s$.  The change in 
$v^aw_a$ is then
\[
   \delta_2  =  v^a \left.S^b\nabla_b w_a \right|_{(\Delta t,\, 0)} \Delta s 
  		  +\frac12\left. \partial_s^2 (v^a w_a)\right|_{(\Delta t,\, 0)}(\Delta s)^2 
		  + O((\Delta s)^3).
\] 
Similarly along the third leg, where $s=\Delta s$ and $t$ {\rm decreases} from $\Delta t$ to $0$, 
we expand around the final point of that leg at $(0,\Delta s)\ $ (go backwards from $t=0$ to $t=\Delta t$ and change sign): 
\[
   \delta_3 = - v^a \left.T^b\nabla_b w_a \right|_{(0,\,\Delta s)} \Delta t 
		  -\frac12\left. \partial_t^2 (v^a w_a)\right|_{(0,\,\Delta s)}(\Delta t)^2 
		  + O((\Delta t)^3).
\] 
Along the fourth leg, where $t=0$ and $s$ {\rm decreases} from $\Delta s$ to $0$, we 
expand about the final point at $(0,0)$: 
\[
   \delta_4 = - v^a \left.S^b\nabla_b w_a \right|_{(0,\, 0)} \Delta s 
  		 -\frac12\left. \partial_s^2 (v^a w_a)\right|_{(0,\, 0)}(\Delta s)^2 
		  + O((\Delta s)^3).
\]
In the sum $\delta_1 + \delta_3$, the $O(\Delta t^2)$ terms cancel at second order (their sum is order 
$\Delta t^2\Delta s$). The order $\Delta t$ terms give, to quadratic order, 
\[
  \delta_1 + \delta_3 = -\left[v^a \left.T^b\nabla_b w_a \right|_{(0,\,\Delta s)} 
			     -v^a \left.T^b\nabla_b w_a \right|_{(0,\, 0)}\right] \Delta t 
			= -v^a S^c\nabla_c \left.\left(T^b\nabla_b w_a\right) \right|_{(0,\, 0)}\Delta t\Delta s, 
\]  
where we have used $S^c\nabla_c v^a = 0$ along the segment from $(0,0)$ 
to $(0,\Delta s)$.
Similarly,  
\[
 \delta_2 + \delta_4 = \left[ v^a\left.S^b\nabla_b w_a \right|_{(\Delta t,\,0)} 
			     - v^a\left.S^b\nabla_b w_a \right|_{(0,\, 0)}\right] \Delta s 
			= v^a T^c\nabla_c \left.\left(S^b\nabla_b w_a\right) \right|_{(0,\, 0)}\Delta t\Delta s.
\]
Adding the last two results, gives us a total change that is antisymmetric under $S\leftrightarrow T$: 
\be
   \delta (v^aw_a) \equiv \delta_1 + \delta_2+ \delta_3 + \delta_4
	= v^a \left[T^c\nabla_c \left(S^b\nabla_b w_a\right) 
		-S^c\nabla_c \left(T^b\nabla_b w_a\right)\right] \Delta t\Delta s.   
\ee
Because $T^a$ and $S^a$ are the coordinate basis vectors, they commute: 
$T^c\nabla_c S^b - S^c\nabla_c T^b = 0$-- this is Eq.~\eqref{e:commute}.  Then
\[
  \delta (v^aw_a) 
		= v^a T^c S^b (\nabla_c\nabla_b-\nabla_b\nabla_c) w_a \Delta t\Delta s,
\] 
and we have obtained the desired commutator of two covariant derivatives that defines the Riemann tensor 
in Eq.~\eqref{rabcd}: 
\[ 
 \delta (v^aw_a) 
		= v^a T^c S^b R_{cba}{}^d w_d \Delta t\Delta s.
\]
Changing the names of dummy indices on the right and using on the left the fact that 
$\delta w_a=0$ (because $w^a$ is continuous), we can write 
this equation as   
$w_a\delta v^a 
		= w_a v^d T^c S^b R_{cbd}{}^a \Delta t \Delta s$.
Finally, 
because this holds for all $w_a$, we have 
\be\crv
     \delta v^a = v^d T^c S^b R_{cbd}{}^a \Delta t \Delta s. \cb 
\ee

The area of the parallelogram spanned by two vectors $\bm A$ and $\bm B$ is 
the magnitude of their cross product, proportional to the magnitude of 
$A_{[a} B_{b]}$: Area = $\sqrt{\frac12(A^a B^b - A^b B^a)(A_aB_b-A_b B_a)}$. Here, because $\bm S$ and $\bm T$ 
occur in this antisymmetric combination, $R_{bcd}{}^a (\Delta t T^b) (\Delta s S^c)$ 
is proportional to the area of the parallelogram spanned by $ \Delta t\,\bm T$ and 
$\Delta s\, \bm S$ -- proportional to the area of the loop, at this order, as 
claimed.  

Because parallel transport preserves lengths and angles, 
the change in $v^a$ after parallel transport around a loop is a rotation for a 
positive-definite metric and a Lorentz transformation for a Lorentzian metric.    
Then $\delta v^a$ is the change in $v^a$ under an infinitesimal rotation 
or Lorentz transformation, $\delta \Lambda^a{}_d$.    
\be
   \delta v^a = \delta \Lambda^a{}_d v^d, \ \mbox{where} \ \delta \Lambda^a{}_d = R_{bcd}{}^a (\Delta t T^b) (\Delta s S^c).
\ee
But  $\delta \Lambda_{ad}$ (lowered by the metric) is an infinitesimal rotation or Lorentz transformation 
if and only if it is antisymmetric.  \cblue That's why the second pair of indices of the Riemann tensor is antisymmetric.\cb
\index{curvature tensor!index symmetries}\index{Riemann tensor!index symmetries}\index{Lorentz transformation!parallel transport around loop}\\

\newpage

\noindent {\sl Second calculation: geodesic deviation}. 
\index{geodesic deviation}\index{curvature tensor!geodesic deviation}\index{Riemann tensor!geodesic deviation}\index{deviation!geodesic}

In flat space free particles separate with constant speed.  Because you can't tell the difference between flat space and a spacetime with a uniform gravitational field (constant gravitational acceleration $\bm g$ in a Newtonian context), the relative acceleration of free particles depends not on $\bm g$ but on the derivative of $\bm g$ --  on the tidal force. In GR that means the relative acceleration of nearby geodesics is a measure of the curvature, proportional to the Riemann tensor. \index{tidal force}\index{force!tidal} 

Formally, let $T^\alpha$ be the 
unit tangent to a first geodesic.  In flat space, let $S^\alpha(t)$ be a family of 
connecting vectors perpendicular to this first geodesic at proper time $t$, and meeting 
a second geodesic.  The length of the connecting vector is its magnitude, $s=|S|$, and its acceleration vanishes  
\be
  \ddot S^\alpha := \frac{d^2}{dt^2}S^\alpha 
		  = T^\gamma\nabla_\gamma (T^\beta\nabla_\beta S^\alpha) = 0.  
\ee  

In curved space nearby geodesics accelerate.      
\begin{figure}[h] 
\centerline{\includegraphics[width=0.7\textwidth]{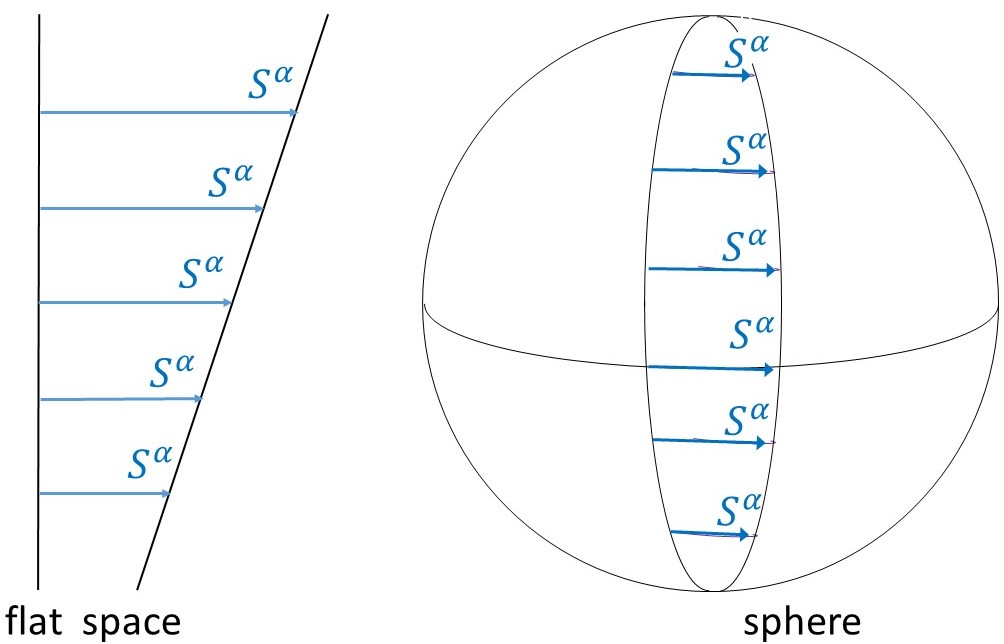}} 
\caption{At left, geodesics in flat space have no relative acceleration: A connecting 
vector joining them increases its length at a constant rate.  At right, on the 
surface of a sphere, the relative acceleration between geodesics is nonzero.  }  
\label{geodesic_dev} 
\end{figure} 

 We want to define and calculate a relative acceleration between neighboring 
free particles in a curved spacetime. As in the first calculation, we'll use a two-dimensional surface with coordinates $t,s$.  
This time the curves $t\rightarrow c(t,s)$ for fixed $s$ will be timelike geodesics.  
To construct the surface, start with a smooth spacelike curve $s\rightarrow c(0,s)$ 
and at each point of that curve, choose a future directed timelike geodesic perpendicular 
to the curve, each geodesic, \mbox{$t\rightarrow  c(t,s)$}, parameterized by proper time.  

Then, regarding each geodesic as the trajectory of a free particle, the set of 
particles at proper time $t$ comprise the curve $s\rightarrow c(t,s)$.   
The vector $S^\alpha$ tangent to $s\rightarrow c(t,s)$ 
is called a connecting vector between neighboring particles.%
\footnote{Wald changes his notation 
here and calls the connecting vector $X$ instead of $S$.  These notes keep $S$ for 
consistency.}
Nearby particles a parameter distance $\Delta s$ apart remain at the same parameter distance, 
and their proper distance is then $|\bm S|\Delta s$ at linear order in $\Delta s$. 
 We will see that $\bm S$ remains orthogonal to $\bm T$, so 
$|\bm S|\Delta s$ is the proper distance between nearby particles 
measured orthogonal to either trajectory (to $O(\Delta s)$ it doesn't matter 
which).\\

As in Eq.~\eqref{e:commute}, because the vectors $\bm S$ and 
$\bm T$ are the coordinate basis vectors $\bm \partial_s$ and 
$\bm \partial_t$, they commute 
\be 
	T^\beta \nabla_\beta S^\alpha  = S^\beta \nabla_\beta  T^\alpha,   
\label{lie1}\ee
and now  $T^\alpha$ is the velocity $u^\alpha$ of 
the particles.  Were we to choose spacelike geodesics,  
$T^\a$ would be a unit spacelike vector.

From (\ref{lie1}), the statement that $S^\alpha $ 
remains orthogonal to $T^\alpha $ along the particle trajectories follows immediately from the 
geodesic equation for $T^\alpha$ and the fact that it is a unit vector: 
\begin{eqnarray*}
T^\beta \nabla_\beta (T^\alpha S_\alpha ) &=& \underbrace{ T^\beta \nabla_\beta T^\alpha }_0 S_\alpha 
			+ T^\beta T^\alpha \nabla_\beta S_\alpha \\
&=& T^\alpha S^\beta \nabla_\beta  T_\alpha  \hspace{5mm}   \mbox{by (\ref{lie1})}\\
&=& 0\,, \hspace{4mm} \mbox{because $T^\a$ is a unit vector.}
\end{eqnarray*} 
The separation vector $S^\alpha $ changes at the rate $\dot S^\alpha $ with
$
	\dot S^\alpha  = T^\beta \nabla_\beta S^\alpha 
$,
and the relative acceleration of neighboring geodesics is then 
$\displaystyle\ddot S^\alpha\,\Delta s $.  We have
\begin{align}
\ddot S^\alpha  &=T^\beta \nabla_\beta (T^\gamma \nabla_\gamma  S^\alpha )\nonumber\\
&=T^\beta \nabla_\beta (S^\gamma \nabla_\gamma T^\alpha ) \hspace{4mm} \mbox{by (\ref{lie1})}\nonumber\\
&=(T^\beta \nabla_\beta S^\gamma )\nabla_\gamma T^\alpha  + T^\beta S^\gamma \nabla_\beta \nabla_\gamma T^\alpha \nonumber\\
&=(S^\beta \nabla_\beta T^\gamma )\nabla_\gamma T^\alpha  + T^\beta S^\gamma (\nabla_\gamma \nabla_\beta T^\alpha  +
[\nabla_\beta ,\nabla_\gamma ]T^\alpha )\nonumber\\
&=(S^\g\nabla_\g T^\b )\nabla_\b T^\alpha  
   + T^\beta S^\gamma \nabla_\gamma \nabla_\beta T^\alpha  
   + T^\beta S^\gamma R_{\beta\gamma}{}^\alpha{}_\delta  T^\delta \nonumber\\
&= S^\g\nabla_\g \underbrace{(T^\beta\nabla_\b T^\alpha)}_0  
	+T^\beta S^\gamma  R_{\beta\gamma}{}^\alpha {}_\delta T^\delta \nonumber\\
\crv \ddot S^\alpha  &=\crv R^\alpha {}_{\beta\gamma\delta}T^\beta T^\gamma S^\delta  
\label{e:geodev}\end{align}
This is the {\sl\crv equation of geodesic deviation}, first obtained in 1926 by Levi-Civita \cite{levicivita26} and by \href{https://www.jstor.org/stable/1968486?seq=2#metadata_info_tab_contents}{Synge}.\cite{synge26}    

An observer moving along a geodesic (a freely falling observer) 
does not detect a uniform gravitational field --- e.g. a man in a freely falling elevator in
the Earth's field sees objects behaving almost as if there were no gravity.  
But because the Earth's field is not really uniform, he will see their relative 
acceleration, the relative acceleration of two particles falling toward the center of the Earth.  

\begin{wrapfigure}[10]{r}{6cm}
		\vspace{-7mm}
                \includegraphics[width=6cm]{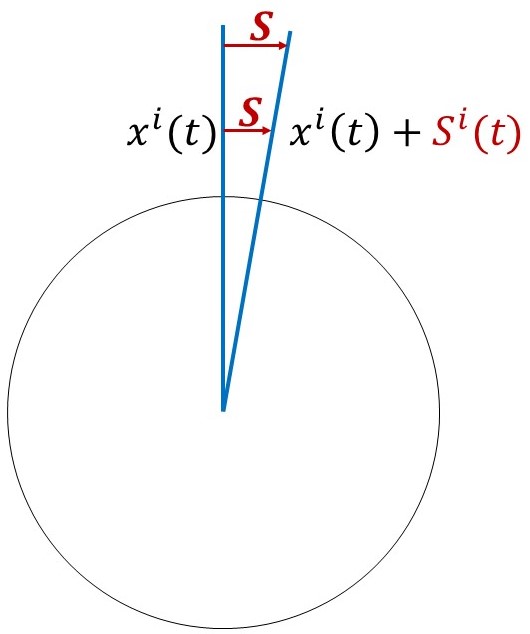}
                \label{deviation}
\end{wrapfigure}

Newtonian approximation:  For two nearby particles with separation vector $\bm S$ 
(here, for brevity, $\bm S$ is written instead of $\Delta s \bm S$), 
\begin{align}
	\ddot S^i  &= \ddot x^i  + \ddot S^i  - \ddot x^i \nonumber\\
		   &= -\nabla^i \phi (\bm x+\bm S) + \nabla^i \phi (\bm x)\nonumber\\
		   &= -\nabla^i \nabla_j\phi (\bm x)S^j .
\label{e:tide}\end{align}
The magnitude of this relative acceleration is of order $\frac{|S|}{R} \nabla\phi$, 
smaller than the ``gravitational acceleration'' $\nabla\phi$ by the ratio of the
particle-particle distance to the radius of the Earth.  It is this tidal
force that alone reveals the presence of a gravitational field to a local
observer.\\
\vspace{11mm}

	In III these remarks will be made again in the context of a discussion of
the equivalence principle and will motivate the form of the field equations
via the correspondence between equations (\ref{e:geodev}) and (\ref{e:tide}).\\

A note on interpretation:  The calculation required only that the curves 
$s\rightarrow c(t,s)$ be smooth and spacelike.  To interpret $s$ as a 
distance between nearby particles, $s$ should be proper distance along each curve, 
and each of these curves should be a spacelike geodesic orthogonal to the 
particle trajectory $t\rightarrow c(t,0)$.  We can always choose $s$ to 
be proper distance along the curve $s\rightarrow (t,s)$, 
and we can choose the initial curve $s\rightarrow (0,s)$ to 
be a spacelike geodesic.  But once we choose a set of timelike geodesics 
through each point of $c(0,s)$, the remaining curves $s\rightarrow c(t,s)$ are 
fixed: They are the position of the particles at proper time $t$; equivalently, 
they are the result of Lie dragging the initial curve $s\rightarrow c(0,s)$ a parameter distance 
$t$ along $T^\a$. It is only at zeroth order in $s$ that these remaining 
curves are geodesics and only at zeroth order are they orthogonal to a 
particle trajectory.\\

A final comment on the Riemann tensor:  If one chooses Riemann normal coordinates 
in a neighborhood 
of a point $P$, the components of the metric near $P$ can be written in terms 
of the Riemann tensor and its derivatives at $P$.  That is, the Taylor series 
of $g_{\mu\nu}(x)$ about $P$ has the form \index{coordinates!Riemann normal}\index{Riemann normal coordinates}
\be
  g_{\mu\nu}(x) = \eta_{\mu\nu} 
  		- \frac13 R_{\mu\sigma\nu\tau} x^\sigma x^\tau + O(|x|^3),   
\ee
where the higher order terms involve derivatives of $R_{\mu\sigma\nu\tau}$.  
(See, for example, MTW Sect. 11.6, or Leo Brewin's notes \href{https://users.monash.edu.au/~leo/research/papers/files/lcb96-01.pdf}{brewin96}.) 
The expansion is used in quantum field 
theory on curved spacetimes and in work on extreme mass-ratio inspiral (EMRI). 
In EMRI work, one uses point-particles moving in a background spacetime to 
approximate stellar size black holes spiraling into the vastly more massive 
black holes at the centers of galaxies.

\benr
\index{Christoffel symbol!as a component of a tensor}
\item Wald's way of defining Christoffel symbols.  Given a chart $\{x^i\}$,
one can write down quantities like $\partial_i v^j$ and the Christoffel symbols 
that are not naturally the components of tensors. 
One simply {\sl defines} $\partial_a v^b$ to be the tensor whose components in 
the chart  $\{x^i\}$, are $\partial_i v^j$.
Then, in another chart $x'$, the components will {\sl not} be $\partial_{i'} v^{\j'}$.  
Check that defining $\partial_a v^b$ in this way is equivalent to defining $\partial_a$ 
to be the covariant derivative of the artificial flat metric 
$(dx^1)^2+\cdots + (dx^n)^2$ or $-(dx^0)^2+\cdots + (dx^3)^2$. (This should be 
a one- or two-line argument.)  
\label{ex:christoffel}   

\item Show that the difference $\nabla_a-\partial_a$ between the two derivative operators 
acts as a tensor, that it is linear under multiplication by scalar fields:  
$(\nabla_a-\partial_a)(fv^b) = f\ (\nabla_a-\partial_a)v^b$, for any 
smooth scalar field $f$.  Then the action of $\nabla_a-\partial_a$ has the form 
\[
  (\nabla_a-\partial_a)v^b =\Gamma^b{}_{ac} v^c, 
\]  
where $\Gamma^b{}_{ac}$ is a tensor field.
\label{ex:nabla-partial}
\item 
Show that the components of $\Gamma^b{}_{ac}$ in the coordinates $\{x^i\}$ are 
the usual Christoffel symbols $\Gamma^j{}_{ik}$.  (This should be quick.  
You do not have to compute the form of $\Gamma$ in terms of metric derivatives.) 

\item The notation has a problem if one wants to write components of $\partial_a v^b$ in 
in a new chart $x'$.  In polar coordinates $\theta, \phi$ for the unit sphere, 
let $\partial_a$ be the covariant 
derivative associated with the flat metric $d\theta^2+d\phi^2$.   
Show that the components of $\partial_a v^b$ in a chart $x,z$, where $z=a\cos\theta$, 
$x=a\sin\theta\cos\phi$ are not the partial derivatives:  To avoid confusion, 
write $T^a{}_b\equiv \partial_a v^b$.  Find $T^x{}_x$ and $T^x{}_z$, and 
compare to $\dis \frac{\partial v^x}{\partial x}, \frac{\partial v^x}{\partial z}$.    

\een

\benr
\item \label{e:riemann1} The next two problems are simple Riemann tensor examples. If you have already done 
similar calculations, you would learn more by finding components in 
an orthonormal basis: See the starred version of the problems below. 
\ben\item[a.]  Find the independent components of the curvature tensor for a sphere 
of radius $a$, with metric $ds^2 = a^2(d\theta^2+\sin^2\theta d\phi^2)$. 
No need to recompute Christoffel symbols that you already know.   
\item[b.] Find all components of the Ricci tensor and Ricci scalar.  
\een 

\item \label{e:riemann2}(Essentially Problem 7 of Wald, Chap. 3).  
An arbitrary Lorentzian 
metric on a 2-manifold can always be written locally in the form 
$ds^2 = \Omega^2(-dt^2+dx^2)$. Find the Riemann tensor of this 
metric.  

\item Show that the Riemann tensor in $n$ dimensions has $\dis\frac{n^2(n^2-1)}{12}$ 
independent components.  First show that the symmetry $R_{abcd} = R_{cdab}$ is 
implied by the other index symmetries.  Because of that, it is not used in the 
counting.  

\een

Variants of \ref{e:riemann1} and \ref{e:riemann2}:  

\ben
\item[]  Do \ref{e:riemann1} using the orthonormal basis 
$\dis\bm e_1 = \frac1a \bm\partial_\theta, \quad \bm e_2=\frac1{a\sin\theta}\bm\partial_\phi$ or the dual basis \\
$\omega^1 = a d\theta$, $\omega^2= a\sin\theta d\phi$.  You can either use commutators or Cartan calculus to find the Christoffel symbols.   

\item[] Do \ref{e:riemann2} using the orthonormal basis $\dis\bm e_0 = \frac1\Omega \bm\partial_t, \quad \bm e_1=\frac1\Omega\bm\partial_x$ or the dual basis \\
$\omega^0 = \Omega dt$, $\omega^1= \Omega dx$.   

\een

\newpage
\subsection{Electromagnetism and perfect fluids in curved spacetime}
\label{s:em_fluid_cst} 
\index{electromagnetism|(}\index{electromagnetism!curved spacetime|(}

The equations governing the electromagnetic field and perfect fluids 
must be consistent with their form in flat space.  In each case, 
there is an obvious candidate for the curved-space equations: Simply replace the flat space derivative operator -- the covariant derivative operator of the flat metric $\eta_{\a\b}$ -- by the covariant derivative of the curved metric $g_{\a\b}$.  Keep the flat space form, with the covariant derivative operator of the flat metric 
now regarded as the covariant derivative operator of the curved metric.    
In each case, to within current experimental accuracy, this is the 
right prescription.  \\

\noindent{\sl\crv Electromagnetism} 

Then Maxwell's equations in a curved spacetime again have the form 
\index{electromagnetism!Maxwell's equations in terms of $F_{\a\b}$}
\index{Maxwell's equations!in terms of $F_{\a\b}$}
\eqref{e:Maxwell}, namely
\bsube\begin{align} 
	\nabla_{[\alpha}F_{\beta\gamma]} & =0 
\label{e:maxwell1}\\
\nabla_\beta   F^{\alpha\beta} &=4\pi j^\alpha .
\end{align}\label{e:maxwell2}\esube

The first of these equations {\sl does not depend on the metric}: 
In any chart, its components are \mbox{$\pa_{[\lambda}F_{\mu\nu]}=0$},
because $\Gamma^\sigma_{\ [\mu\nu]}= 0$:
\[
  \nabla_{\lambda}F_{\mu\nu} = \pa_{\lambda}F_{\mu\nu} 
  	- \Gamma^\sigma_{\ \mu\lambda}F_{\sigma\nu} 
	- \Gamma^\sigma_{\ \nu\lambda}F_{\mu\sigma} \ \Longrightarrow\ \ 
\nabla_{[\lambda}F_{\mu\nu]}= \pa_{[\lambda}F_{\mu\nu]}. 
\]

\noindent{\sl Conservation of charge}\index{charge!charge conservation}\index{conservation laws!charge} \index{electromagnetism!charge conservation}
\benr
\item Check that, as in flat space, $\nabla_\b F^{\a\b} = 4\pi j^\a$  implies charge conservation, 
\be
\na_\a j^\a = 0.
\ee  
\een 

In any chart $\na_\b F^{\a\b}$ has the form that we saw in \ref{p:div}\hspace{-3pt},
for the divergence of a vector, namely  
\be
   \na_\nu F^{\mu\nu} = \frac1{\sqrt{|g|}}\pa_\nu (\sqrt{|g|} F^{\mu\nu} ).
\label{e:divF1}\ee

\benr\item 
\benalph\item 
Show again that $\nabla_\mu A^\mu = \frac1{\sqrt{|g|}}\partial_\mu(A^\mu\sqrt{|g|})$ 
this time by showing the relation \\
$\dis
   \Gamma^\nu{}_{\mu\nu} = \pa_\mu \log\sqrt{|g|}\,$.

Note that, in the determinant $g$, $g_{\mu\nu}$ multiplies 
its cofactor $\Delta^{\mu\nu}$, so $\dis\frac{\pa g}{\pa g_{\mu\nu}} = \Delta^{\mu\nu}$.  Because $g^{\mu\nu} = \Delta^{\nu\mu}/g$, we have 
$\dis 
   \frac{\pa g}{\pa g_{\mu\nu}} = g g^{\mu\nu}$. 
\item Show that Eq.~\eqref{e:divF1} holds for any antisymmetric tensor $F^{\a\b}$. 
\een

\een

Because the sourcefree Maxwell equation never heard of the metric, it continues to 
imply the existence of a vector potential $A_\a$ with
\index{vector potential|textbf}\index{electromagnetism!vector potential|textbf}
\be
   F_{\a\b} = \na_\a A_\b - \na_\b A_\a.   
\ee
$A_\a$ is determined only up to the addition of a gradient:  $A_\a+\na_\a\psi$ 
gives the same field $F_{\a\b}$, since $\na_{[\a}\na_{\b]}\psi = 0$.  A choice 
of $A_\a$ is a choice of gauge, and, as in flat space, one can choose a Lorenz 
gauge in which 
\be
   \na_\a A^\a = 0.  
\ee  
That is, given a vector potential $A^{\rm old}_\a$ one can find a new 
vector potential $A_\a = A_\a^{\rm old} + \na_\a\psi$, satisfying the 
Lorenz gauge condition, by finding a solution $\psi$ to the wave equation 
\[
  \na^\a \na_\a \psi = - \na^\a A^{\rm old}_\a.
\]

Now, however, $A_\a$ does not satisfy the flat space form of the wave equation.  Instead, 
we have 
\begin{align}
  4\pi j^\a &= \na_\b F^{\a\b} = \na_\b(\na^\a A^\b - \na^\b A^\a) 
   		= -\na_\b \na^\b A^\a + [\na_\b, \na^\a] A^\b 
   		  +\color{gray}\na^\a \na_\b A^\b\\
	&= -\na_\b \na^\b A^\a + R^\a_{\ \b} A^\b, \\ 
    \na_\b \na^\b A^\a &= R^\a_{\ \b} A^\b +4\pi j^\a .
\end{align}
In contrast to the form \eqref{e:maxwell2} of Maxwell's equations in terms of the 
Faraday tensor, the presence of two covariant derivatives has led to a Ricci-tensor term.   
This departure from the flat-space form of the wave equation exemplifies the ambiguity inherent in generalizing the flat space form of an equation to a curved spacetime. 
\footnote{In particular, the agreement of Maxwell's equations in the form \eqref{e:maxwell2} with observation does not rule out corrections that involve the 
small ratio of the Planck length $\ell_P$ to the radius of curvature.  For example, 
quantum corrections to the electromagnetic (or Yang-Mills) Lagrangian can include terms such as $\ell_P^2 R F_{\a\b}F^{\a\b}\sqrt{|g|}$ and $\ell_P^2 R_{\a\b\c\d}F^{\a\b}F^{\c\d}\sqrt{|g|}$ that explicitly involve the curvature tensor.  The factor $\ell_P^2$ here maintains the dimension $L^{-2}$ of a Lagrangian density in geometrical units. }
\vspace{3mm}  

\noindent{\sl 3+1 decomposition}\\
\index{electromagnetism!3+1 decomposition}
An inertial observer in flat space has a natural way to decompose the field $F_{\a\b}$ into electric and magnetic parts, along and orthogonal to her 4-velocity -- or, equivalently, orthogonal and tangent to her surfaces of simultaneity.  But for an observer in a curved spacetime (or an accelerated observer in flat space -- see Fig.~\ref{accelerated_observer}), there is in general no preferred choice of 
time, no preferred way to decompose the spacetime into space and time.  A 3+1 
decomposition of the electromagnetic field is still useful in many contexts:  for an description of the local field seen by an observer, for the initial 
value problem, discussed in Chap.~\ref{c:iv}, and for a static spacetime, where 
there is again a preferred choice of time.  

In a generic spacetime, one must choose a time coordinate, a scalar $t$ whose $t=$ constant surfaces are spacelike, with $\na_\a t\na^\a t < 0$ everywhere.  Each choice of a coordinate $t$ decomposes spacetime into a set of spacelike {\sl slices}, and it gives a corresponding $3+1$ decomposition of the electromagnetic field:  Use the unit normal $n^\a = -\na^\a t/\sqrt{-\na_\a t\na^\a t }$ to write 
the decomposition corresponding to Eq.~\eqref{e:FEB} as 
\be
 F_{\alpha\beta}  = n_\alpha  E_\beta   -n_\beta  E_\alpha   
		  + \epsilon_{\alpha\beta\gamma} B^\gamma ,
\ee 
\index{magnetic field!curved spacetime}\index{electric field!curved spacetime}
where now $\epsilon_{\a\b\c} := \ep_{\d\a\b\c}n^d$.  Although this form 
is identical to the flat-space decomposition, with $n^\a$ replacing $t^\a$, 
the 3+1 Maxwell equations are more cumbersome, because the derivative  
$\na_\a n_\b$ is now nonzero.
\vspace{3mm}   

\noindent{\sl Notation and definitions}.  You are likely to encounter the following definitions in the literature when dealing with the antisymmetric derivative of an antisymmetric tensor.
  
\noindent{\bf Definition}. A {\sl p-form} $\sigma_{a\cdots b}$ is an antisymmetric, covariant tensor  with $p$ indices. \\
\index{forms, differential}\index{differential forms}
\noindent{\bf Definition}. The exterior derivative $d\sigma$ of a $p$-form $\sigma$
is the $p+1$ form 
\be 
 (d\sigma)_{ab\cdots c} = (p+1)\nabla_{[a}\sigma_{b\dots c]}.
\label{e:exteriord0}\ee
\index{exterior derivative}
\noindent The factor $p+1$ is the number of independent ways of distributing the $p+1$ indices 
between $\nabla$ and $\sigma$.  The antisymmetry implies that $d\sigma$ is independent 
of the derivative operator; in any chart it has components
\be
(d\sigma)_{ij\cdots k} = (p+1)\partial_{[i}\sigma_{j\dots k]}.
\ee 

In particular, a scalar $f$ is a 0-form, a covariant vector $A_a$ is a 1-form, and an antisymmetric 2-index tensor $F_{ab}$ is a 2-form. The sourcefree Maxwell equation is then $dF=0$, and the implied vector potential satisfies $F = dA$. 
Appendix \ref{appendix} gives a detailed presentation, with relations among  
forms, densities, Lie derivatives, exterior derivatives, and Stokes's theorem.    
\vspace{3mm}
\index{electromagnetism|)}\index{electromagnetism!curved spacetime|)}

\noindent{\sl\crv Perfect fluids}
\index{perfect fluid!curved spacetime}\index{perfect fluid}
\label{p:perfect fluid}

The stress-energy tensor of a perfect fluid was obtained in Sect.~\ref{s:perfect_fluid} by demanding that the matter be shear-free. 
Again this is the requirement that $T^{\a\b}$ at any point $P$ be invariant under rotations in the space orthogonal to the fluid 4-velocity $u^\a$, 
and again it implies the form \eqref{e:Tfluid},
\be
  T^{\a\b} = \rho u^\a u^\b + P q^{\a\b}, 
\label{e:Tab_fluid}\ee
\index{perfect fluid!stress-energy tensor}\index{stress-energy tensor!perfect fluid}
with 
\be
   q^\a_\b = \delta^\a_\b + u^\a u_\b
\ee
the projection orthogonal to $u^\a$.  
Equivalently, an inertial observer, instantaneously comoving with the fluid, sees the fluid as isotropic, if looking only in a box small enough to ignore changes in density  and pressure, and now small compared to the radius of curvature.     
     
Replacing the covariant derivative operator of a flat metric by that of 
a generic metric $g_{\a\b}$ automatically extends Eq.~\eqref{e:divT} 
to a curved spacetime:  
\be
 \na_\b T^{\a\b} = 0.  
\label{e:divT1}\ee
It is now easy to check that the projections of this equation along and 
orthogonal to $u^\a$ reproduce the energy conservation equation \eqref{e:econsderive}, 
\be
\nabla_\beta (\rho u^\beta ) = -P\nabla_\b u^\b, 
\ee
and the relativistic Euler equation \eqref{e:pcons}, 
\be
(\rho +P)u^\beta\nabla_\beta u^\alpha =-q^{\alpha\beta}\nabla_\beta P.
\ee
\benr\item Show that the derivations preceding Eqs.~\eqref{e:econsderive} 
and \eqref{e:pcons} go through as written, with $\eta_{\a\b}$ replaced by 
$g_{\a\b}$.  

\een

What is lost in a curved spacetime is the relation between Eq.~\eqref{e:divT1} 
and the integral energy and momentum conservation laws given by Gauss's 
theorem in Sect.~\ref{s:conservation_laws}.  We will see in Sect.~\ref{s:gauss0} that Gauss's theorem holds for the divergence of a vector field $A^\a$, but not for the divergence of 
a tensor field $T^{\a\b}$.  Only if the spacetime has a symmetry is there 
a corresponding conservation law.  If, for example, the metric is  symmetric under time-translation or under rotations, the energy or angular momentum 
of a field propagating on that spacetime will be conserved.  The description 
of these symmetries relies on the formalism of the next section.  

\newpage

\section{Lie Derivatives}
\label{sec:lie}
\index{Lie derivative|(}
This section and the sections on forms and integration that follow are largely taken from 
an appendix of the book {\sl Rotating Relativistic Stars}, by JF and Nikolaos Stergioulas.\cite{fsbook}\\

Lie derivatives arise naturally in the context of fluid flow. They are needed to understand symmetries of spacetime and of matter in curved spacetimes, and they simplify calculations 
and aid one's understanding of relativistic fluids.  

We have just seen that the connecting vector $\bm S$ between nearby particles with 
velocity $\bm T$ satisfies $[\bm T,\bm S]=0$, and this will be the statement that 
the Lie derivative of $\bm S$ by the vector field $\bm T$ vanishes. 
We begin, for simplicity, in a Newtonian context, with a stationary fluid flow with 3-velocity $\bf v(\bf r)$.  A function $f$ is said to be {\em dragged along} 
by the fluid flow, or {\em Lie-derived} by the vector field $\bf v$ that 
generates the flow, if the value of $f$ is constant on a fluid element, that is, constant 
along a fluid trajectory $\bm r(t)$: 
\be 
\frac d{dt} f[\bm r(t)]={\bf v}\cdot \nabla f = 0.
\ee
The {\em Lie derivative} of a function $f$, defined by 
\be 
 {\cal L}_{\bf v} f={\bf v}\cdot \nabla f,
\label{lief}\ee
is the directional derivative of $f$ along $\bf v$, the rate of change of 
$f$ measured by a comoving observer.   If $f$ is density or temperature, say, $f$ is 
dragged along by the flow if the density or temperature of a fluid element remains 
constant as the fluid element moves with the flow.  
  
For a vector, imagine an arrow painted on a river of sliding jello, the arrow dragged along as the jello moves and oscillates.  That is, consider a vector that joins two nearby fluid elements, 
two points $\bm r(t)$ and ${\bf\bar r}(t)$ that move with the fluid: 
Call the connecting vector $\lambda \bf w$, so that for small $\lambda$
the fluid elements are nearby:   
$\lambda{\bf w} =  {\bf\bar r}(t)-\bm r(t)$. Then $\lambda\bf w$ is said to be 
{\em dragged along} by the fluid flow, as shown in Fig.~(\ref{fig:lie_drag}).
In the figure, the endpoints of $\bm r(t_i)$ and $\bar{\bm r}(t_i)$ are labeled $\rm r_i$ 
and $\rm\bar r_i$.  
\begin{figure}[h!]
\begin{center}
\includegraphics[width=10cm]{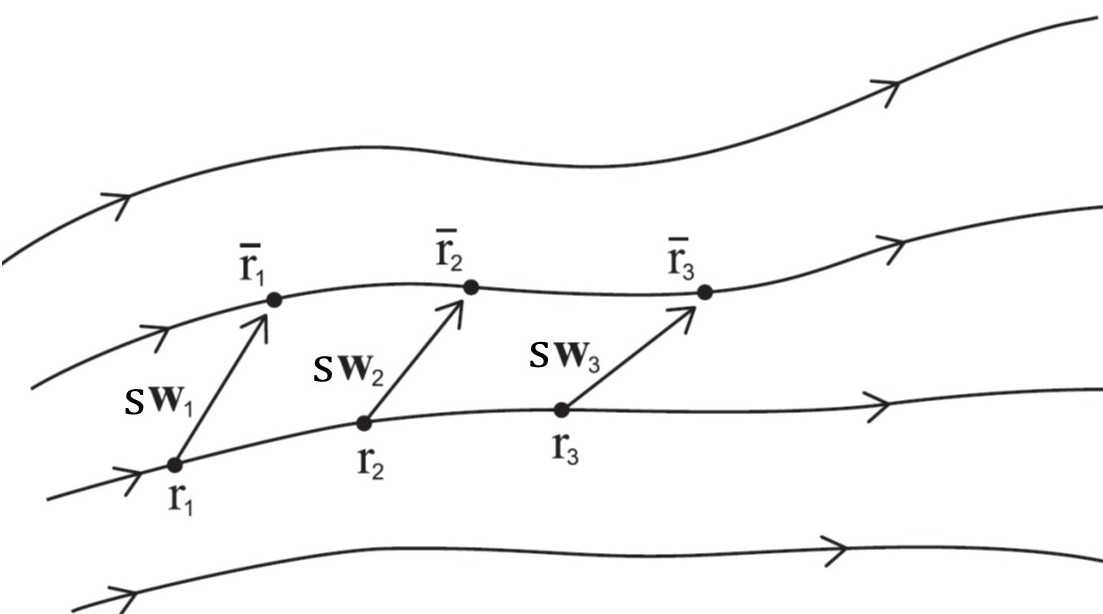}
\end{center}
\vspace{-5mm}
\caption{Two nearby fluid elements move along the flow lines, their successive 
positions labeled $\rm r_i$ and $\rm\bar r_i$.  A vector field $s \bf w$ is said to 
be dragged along by the flow 
when, as shown here, it connects successive positions of two nearby fluid elements.}
\label{fig:lie_drag}
\end{figure}

A vector field $\bf w$ is {\em Lie-derived} by $\bf v$ if, for small $s$, 
$s {\bf w}$ is dragged along by the fluid flow.  To make this precise, 
we are requiring that the equation
\be
  \bm r(t)+s {\bf w}(\bm r(t)) = {\bf \bar r}(t)
\ee 
be satisfied to $O(s)$.  Taking the derivative of both sides 
of the equation with respect to $t$ at $t=0$, we have 
\bea
  {\bf v}(\bm r)+s {\bf v}\cdot\nabla {\bf w}(\bm r) 
&=& {\bf v}({\bf \bar r})= {\bf v}[\bm r+s {\bf w}(\bm r)]
\nonumber\\
&=& {\bf v}(\bm r)+s {\bf w}\cdot\nabla 
{\bf v}(\bm r)+O(s^2), 
\eea   
which holds if and only if 
\be 
 {\bf [v,w]} \equiv {\bf v}\cdot\nabla {\bf w} - {\bf w}\cdot\nabla {\bf v}= 0.
\ee 
The commutator $[\bf {v,w}]$ is the {\em Lie derivative} of $\bf w$ with 
respect to $\bf v$, written
\index{commutator!and Lie derivative} \index{Lie derivative!of vector|textbf}
\be 
      \bf \Lie_{\bf v} w = [v,w].
\label{liev}\ee
Then $\bf w$ is Lie-derived by $\bf v$ when $\bf \Lie_{\bf v} w = 0$.  
The Lie derivative $\bf \Lie_{\bf v} w$ compares the change in the 
vector field $\bf w$ in the direction of $\bf v$ to the change that 
would occur if $\bf w$ were dragged along by the flow generated 
by $\bf v$.  

In a curved spacetime the Lie derivative of a function $f$ is again its 
directional derivative, \index{Lie derivative!of scalar|textbf}
\be
\Lie_{\bf u}f = u^\alpha\nabla_\alpha f.
\ee
If $u^\alpha$ is the velocity of a fluid, generating the fluid trajectories 
in spacetime, $\Lie_{\bf u} f$ is commonly termed the convective derivative of 
$f$.  The Newtonian limit of $u^\alpha$ is the 4-vector ${\bm\partial}_t+{\bf v}$, 
and $\Lie_{\bf u} f$ has as its limit the Newtonian convective derivative 
$(\partial_t+{\bf v}\cdot\nabla)f$, again the rate 
of change of $f$ measured by a comoving observer. (Now the flow is 
arbitrary, not the stationary flow of our earlier discussion.)  

A connecting vector is naturally the tangent to 
a curve joining nearby points in a flow, and in a curved spacetime, the Lie derivative of a contravariant vector field is again defined 
by Eq.~(\ref{liev}),
\be\cblue
 	\Lie_{\bf u} w^\alpha = u^\beta\nabla_\beta w^\alpha - w^\beta\nabla_\beta u^\alpha.\cb
\ee
We have used a fluid flow generated by a velocity $u^\alpha$ to motivate 
a definition of Lie derivative; the definition, of course, is the same 
in any dimension and for any vector fields:\index{commutator!in terms of covariant dervatives}
\be \crv
  \Lie_{\bf v} w^a = v^b\nabla_b w^a - w^b\nabla_b v^a. \cb
\label{liev2}\ee
Although the covariant derivative operator $\nabla$ appears in the 
above expression, we have already seen that the commutator is in fact independent 
of the choice of derivative operator.  This was implied by our definition of 
the commutator (which did not involve covariant derivatives); it also follows from the 
symmetry $\Gamma^i_{jk}= \Gamma^i_{(jk)}$, which 
implies that the components have in any chart the form 
\be
 \Lie_{\bf v} w^i = v^j\partial_j w^i - w^j\partial_j v^i.
\ee

We now extend the definition of Lie derivative to arbitrary 
tensors.  This just involves using the definitions for vectors and scalars and 
the Leibnitz rule.  For any dual vector $\sigma_a$ and vector $w^a$, we have  
\begin{align} 
\Lie_{\bf v}(\sigma_a w^a) &= (\Lie_{\bf v}\sigma_a)w^a + \sigma_a\Lie_{\bf v} w^a 
\label{e:Liecov}\\
v^b\nabla_b \sigma_a w^a + \cancel{\sigma_a v^b\nabla_b w^a}{-25} 
	& = (\Lie_{\bf v}\sigma_a)w^a + \cancel{\sigma_a v^b\nabla_b w^a}{-25} - \sigma_a w^b\nabla_b v^a \ \Longrightarrow \nn\\
w^a \Lie_{\bf v}\sigma_a &= w^a(v^b\nabla_b \sigma_a  + \sigma_b \nabla_a v^b).\nn
\end{align}
Because this holds for all $w^a$, we have
\be
\crv\Lie_{\bf v}\sigma_a = v^b\nabla_b \sigma_a  + \sigma_b \nabla_a v^b.
\label{liecov2}\ee 
Again the definition (\ref{liecov2}) is independent of the choice 
of derivative operator, because the Lie derivatives of the scalar $\sigma_a v^a$ 
and the vector $v^a$ in Eq.~\eqref{e:Liecov} do not depend on $\nabla$.  The 
components in any chart are given by
\be
   \cblue \Lie_{\bf v} \sigma_i = v^j\partial_j \sigma_i 
			+\sigma_j \partial_i v^j. \cb
\label{liecov3}\ee 
Finally, the Lie derivative of an arbitrary tensor 
$T^{a\cdots b}{}_{c\cdots d}$ again follows from
the Leibnitz rule (writing a tensor as a sum of products $u^a\cdots v^b\sigma_a\cdots \tau_d$ 
as usual).  You can quickly see from Eqs.~\eqref{liev2} and \eqref{liecov2} that the result is 
\crv\bea 
\Lie_{\bf v} T^{b\cdots c}{}_{d\cdots e} 
&=& v^a\nabla_a T^{b\cdots c}{}_{d\cdots e} \nonumber \\
&&- T^{a\cdots c}{}_{d\cdots e}\nabla_a v^b
-\cdots - T^{b\cdots a}{}_{d\cdots e}\nabla_a v^c
\nonumber\\
&& 
+T^{b\cdots c}{}_{a\cdots e}\nabla_d v^a
+\cdots 
+ T^{b\cdots c}{}_{d\cdots a}\nabla_e v^a,\cb
\label{eq:liedef1}\eea\cb
independent of the derivative operator, and with components in a chart 
given by 
\bea 
\Lie_{\bf v} T^{j\cdots k}{}_{m\cdots n} &=&
v^i\partial_i T^{j\cdots k}{}_{m\cdots n} \nonumber \\
&&
- T^{i\cdots k}{}_{m\cdots n}\partial_i v^{j}
-\cdots - T^{j\cdots i}{}_{m\cdots n}\partial_i v^{k}
\nonumber\\
&&
+T^{j\cdots k}{}_{i\cdots n}\partial_{m}v^i
+\cdots 
+ T^{j\cdots k}{}_{m\cdots i}\partial_{n} v^i.
\label{eq:liedef2}\eea

\vskip 0.4cm
We began by using the flow generated by a velocity field $\bf v$ to motivate the definition \eqref{liev} of Lie derivative of a vector field $\bf w$. 
We then extended the definition algebraically to tensors, by using the 
Leibnitz rule. We now return to the geometric picture, generalizing 
the flow of a fluid to a family of smooth invertible maps generated by an arbitrary vector field.  We can then give a geometric definition of Lie derivative by extending the action of that flow from vectors to tensors. \\  
{\bf Definition}. \crv A smooth 1-1 map of a manifold onto itself or onto another manifold is called a {\em diffeo} (or diffeomorphism) if its inverse is also smooth.} \index{diffeomorphism|textbf}\\
Note first that the trajectory of a fluid element is an {\em integral curve} of 
the vector field $u^\alpha$, where: \\
{\bf Definition}. An integral curve $c(\lambda)$ of a vector field $\dis v^a$ 
is a curve whose tangent vector at each point $P=c(\lambda)$ is $\dis v^a(P)$.\\
In a chart $\{x^i\}$, the tangent $\dis v^i$ to a curve $c(\lambda)$ has 
components $\dis \frac d{d\lambda}x^i(\lambda)$; and the statement that 
$c(\lambda)$ is an integral curve has the form   
$\dis \frac{d}{d\lambda}x^i(\lambda)= \dis v^i[c(\lambda)]$.\\

\noindent
{\bf Proposition}. Any smooth vector field $\dis v^a$ in an n-dimensional manifold
$M$ has a family of {\em integral curves}, one through each point of $M$.
\footnote{This result is equivalent to the existence theorem for solutions 
to ordinary differential equations, proved, for example, in Coddington and Levinson, {\em Theory of Ordinary Differential Equations}, McGraw-Hill, 1995.} \\

\noindent {\bf Example 1}:  As noted, the velocity field $u^\alpha$ of a 
fluid has as its integral curves the fluid trajectories parameterized by proper time. The 3-dimensional 
vector field $\bf v$ of a stationary Newtonian flow has as its integral curves the 
flow lines, parameterized by Newtonian time. \\

\noindent {\bf Example 2}:  The vector field ${\bm\partial}_\phi =
x{\bm\partial}_y-y{\bm\partial}_x$ has as integral curves the lines of constant $t,
r, \theta$, 
\[ \lambda\rightarrow (t, r, \theta , \phi + \lambda )\]
Note that when a vector field vanishes at $P$ (e.g.,\ ${\bm\partial}_\phi$
vanishes on the symmetry axis $x=y=0$), the integral curve simply stays at
$P$:  $c(\lambda)=P$.\\

We can view the 4-dimensional flow of a fluid as a family of smooth maps of 
the fluid to itself in the following way:  In a given proper time  $\tau$ 
each point $P$ in the fluid moves along the fluid trajectory through $P$ from $c(0)=P$ 
to the point $c(\tau)$.   
As in the case of the 4-velocity, we    
can use the integral curves of any vector field to define a family $\psi_\lambda$ of diffeos 
of a manifold to itself (for a star, the fluid has a boundary, and the map $\psi_\tau$ is from the support of the fluid to itself): \\
For each point $P$, let $c(\lambda )$ be the integral curve of $v^a$ for which 
$P=c(0)$.  For a fixed value $\lambda$, define the map $\psi_\lambda$ by 
\be 
   \psi_\lambda(P) = c(\lambda).
\ee
That is, $\psi_\lambda$ maps each point $P$ to the point a parameter distance 
$\lambda$ from $P$ along the integral curve through $P$.  The vector field 
$v^a$ is said to {\em generate} the family $\psi_\lambda$ of diffeos.
In a chart $\{x^i\}$, we have 
\be
\dis v^i(x) = \left.\frac{d}{d\lambda}\psi_\lambda^i(x)\right|_{\lambda=0}.
\label{xipsi}\ee

\noindent {\bf Example}:  The vector field $\bm\phi:={\bm \partial}_\phi$ generates the
family of diffeos\\
 \centerline{$(t, r, \theta , \phi ) \rightarrow (t, r,\theta, \phi+\lambda)$,}
rotations by $\lambda$ in the $x$-$y$ plane about the axis where
${\bm\partial}_\phi$ vanishes.  A spacetime is axisymmetric if its metric is 
invariant under rotations, if $\Lie_{\bm \phi} g_{\alpha\beta}=0$ or, 
equivalently, $\psi_\lambda^*g_{\alpha\beta} = 0$.

We can now repeat for manifolds the relation with which we began this section,  
between the flow -- the diffeos -- generated by a vector field and the Lie derivative. 
We need the action of a smooth map $\psi$ on a tensor, beginning with its action on 
a curve and on a function, and from that finding its action on a vector as a tangent 
to the curve, and on a dual vector as the gradient of the function. For Lie derivatives, 
we could look only at diffeos from a manifold to itself, but the action of $\psi$ on 
vectors and dual vectors needs only a smooth map, and you'll use the more general 
description when you encounter the 3+1 formalism and the initial value problem 
in the GR literature or later in the notes. \\

\label{p:diffeo}
Let $\psi:M\rightarrow N$ be a smooth map from one manifold to another.  
A curve on $M$ is mapped by $\psi$ to a curve on $N$.  One says that the curve 
$c:{\mathbb R}\rightarrow M$ is {\sl dragged} or {\sl carried along} or {\sl pushed forward} 
by $\psi$ to the curve $\psi\circ c:  {\mathbb R}\rightarrow N$. 
\begin{figure}[H]
\begin{center}
\includegraphics[width=10cm]{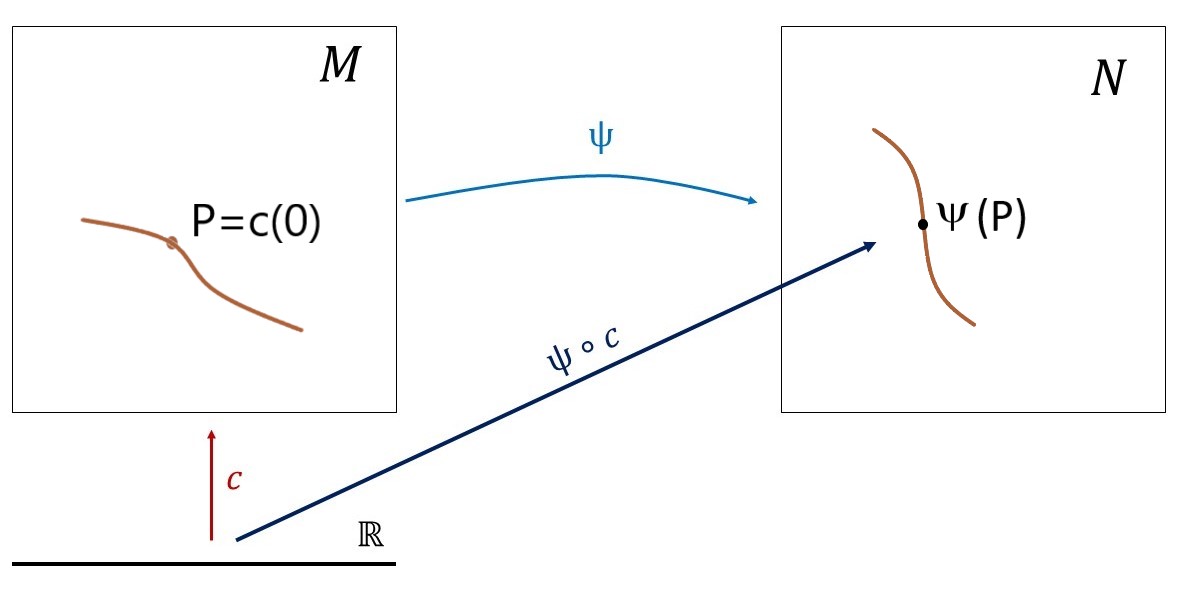}
\end{center}
\vspace{-5mm}
\caption{}
\label{fig:psic}
\end{figure}
A function $f: N\rightarrow {\mathbb R}$ on $N$ is {\sl pulled back} to the function 
$f\circ\psi: M\rightarrow \mathbb R$ on $M$.  

Then the tangent $v$ to the curve $c$ at 
a point $P=c(0)$ is dragged (or carried along) to the tangent $\psi^*v$ to the 
curve $\psi\circ c$ at the point $\psi(P)$.  And the gradient $\nabla f$ at a 
point $\psi(P)$ in $N$ is pulled back to the gradient $\psi_*\nabla f:=\nabla(f\circ\psi)$ of 
$f\circ\psi$ at $P$.  \footnote{In general, $\psi$ need not be 1-1, and this pullback 
gives a gradient at all points whose image is $\psi(P)$.} 
\begin{figure}[h!]
\begin{center}
\includegraphics[width=10cm]{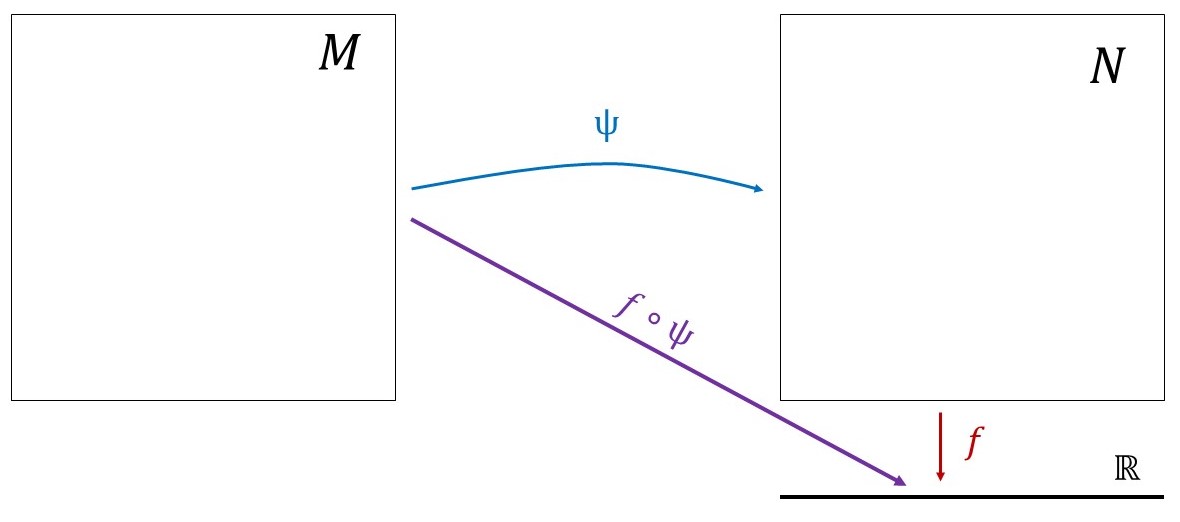}
\end{center}
\vspace{-5mm}
\caption{}
\label{fig:psi*}
\end{figure}

Because $\psi^*v$ is tangent to $\psi\circ c$, the vector $\psi^*v$ is the map $\{f\}\rightarrow \mathbb R$ given by 
\be\crv
   \psi^*v (f) = \frac d{d\lambda}f\circ \psi\circ c(\lambda)]|_{\lambda=0}. 
\label{e:psi*v}\ee
Because $\psi_*\nabla f$ is the gradient of $f\circ \psi$, 
the right side of Eq.~\eqref{e:psi*v} is also $\psi_*\nabla f\ (v)$. That is, 
the action of the dual vector $\psi_*\nabla f$ on $v$ is 
\be
   \psi_*\nabla f\ (v) = \psi^*v (f) = \frac d{d\lambda} f\circ \psi\circ c(\lambda)]|_{\lambda=0}. 
\label{e:psi*df}\ee 

In charts $x$ and $y$ on $M$ and $N$, the curve $c$ has coordinates $x^i(\lambda)$,
the dragged curve $\psi\circ c$ has coordinates $y^\mu[x(\lambda)]$, and we have
\[
  (\psi^*v)^\mu = \frac d{d\lambda}y^\mu[x(\lambda)] 
	=  \frac{\partial y^\mu}{\partial x^i} \frac{dx^i}{d\lambda}
	=\frac{\partial y^\mu}{\partial x^i} v^i, 
\] 
or 
\be
   (\psi^*v)^\mu|_{\psi(P)} =  \frac{\partial y^\mu}{\partial x^i} v^i|_P .
\label{e:psi*component}\ee
The different indices here are used because $M$ and $N$ need not have the same dimension.  

Similarly, the pulled-back gradient $\psi_*\nabla f =\nabla (f\circ \psi)$ has components 
given by 
\be
\nabla_i(f\circ\psi)=  \frac{\partial}{\partial x^i} f[y(x)] 
		=\frac{\partial f}{\partial y^\mu} \frac{\partial y^\mu}{\partial x^i} 
		=  \nabla_\mu f \frac{\partial y^\mu}{\partial x^i}.  
\ee
Because any dual vector $\sigma$ at a point $\psi(P)$ of $N$ can be represented by a gradient on $N$, 
its components are 
\be\cblue
  (\psi_*\sigma)_i = \sigma_\mu\frac{\partial y^\mu}{\partial x^i}.
\ee
The action of $\psi_*\sigma$ on the vector $v$ is then  
\be
  \psi_*\sigma (v) = \sigma(\psi^* v), 
\label{e:psi^*}\ee
implied by Eq.~\eqref{e:psi*df} or by multiplying both sides of Eq.~\eqref{e:psi*component} by 
$\sigma_\mu$.  (Wald takes \eqref{e:psi^*} as the definition of $\psi_*$.) 

Summary:  \\
 \cblue 
\noindent A map $\psi: M\rightarrow N$ drags a curve $c$ and its tangent vector $v$ on $M$ to 
the curve $\psi\circ c$ with tangent $\psi^* v$ on $N$.\cb\\
It pulls back a function $f$ and its gradient $\nabla f$ on $N$ to a function $f\circ\psi$ 
and its gradient $\psi_*\nabla f$ on $M$. Hence\\
\cblue $\psi$ pulls back a dual vector $\sigma$ on $N$ to a dual vector $\psi_*\sigma$ on $M$\cb.   

Any contravariant tensor $T^{a\cdots b}$ is dragged by $\psi$ from $M$ to 
$(\psi^*T)^{\alpha\cdots\beta}$ on $N$ because $T^{a\cdots b}$  can be written as a 
sum of products $u^a\cdots v^b$, each of which is dragged by $\psi$.  Similarly, 
any covariant tensor $T_{\alpha\cdots\beta}$ is pulled back by $\psi$ from $N$ to 
$(\psi_*T)_{a\cdots b}$ on $M$ because $T_{\alpha\cdots\beta}$  can be written as a 
sum of products $\sigma_\alpha\cdots \tau_\beta$, each of which is pulled back by $\psi$.  
Equivalently, acting on a covariant tensor $T_{\alpha\cdots \beta}$, 
\be
\psi_* T(u, \ldots ,v) = T(\psi^* u, \ldots, \psi^* v), 
\ee
and on a contravariant tensor $T^{a\cdots b}$, 
\be
  \psi^* T(\sigma, \ldots ,\tau) = T(\psi_*\sigma, \ldots, \psi_*\tau). 
\ee

\noindent{\bf Notation.} We have followed Wald in writing $\psi^* v$ for the dragged vector, 
and $\psi_*\sigma$ for the pulled back dual vector.  This is opposite to the convention 
in most of the mathematics and mathematical physics literature.\\  
 
\noindent
{\sl Exercise}.  Find the components $(\psi_* T)_{i\cdots j}$ and $(\psi^*T)^{\mu\cdots\nu}$.\\

But $\psi$ does not know how to drag a dual vector or pull back a vector.  To talk about 
the action of a map on a general tensor, the map $\psi$ must have an inverse -- it 
must be a diffeo, and we now consider only diffeos.  
By using $\psi^{-1}:N\rightarrow M$ to drag a dual vector from $N$ to $M$, we can extend 
$\psi^*$ to arbitrary tensors.  Given a tensor $T^{a\cdots b}{}_{c\cdots d}$ on $M$, 
we define the dragged tensor $\psi^* T^{a\cdots b}{}_{c\dots d}$ on $N$ by 
\be
 \psi^* T(\sigma, \ldots, \tau, u,\ldots, v) = T(\psi^{-1}_*\sigma,\dots,\psi^{-1}_*\tau,\psi^* u,\ldots,\psi^* v).
\ee
Dragging a tensor from $M$ to $N$ by $\psi^*$ is then equivalent to pulling it back from 
$M$ to $N$ by $\psi^{-1}_*$. \\

\noindent
{\sl Exercise}. Take $M=N$. Because there is only one coordinate system, denote by 
$\psi^i(P)$ the coordinates of $\psi(P)$ and write $\psi^i(x)$ for $\psi$ regarded 
as a function of the coordinates: 
$\psi^i$ maps $(x^1, \ldots, x^n)$ to $(\psi^1(x), \ldots, \psi^n(x))$.   
Show for a tensor $T^a_b$ that the components of $\psi_* T$ at $\psi(P)$ are
given in terms of the components of $T$ at $P$ by  
\be
  (\psi_*T)^i_j = \frac{\partial\psi^i}{\partial x^k}\frac{\partial\psi^{-1\, l}}{\partial x^j} T^k_l.  
\ee

\noindent
{\sl Lie derivatives from a geometric standpoint.} \\ 

The geometric definition of the Lie derivative $\Lie_{\bm u}$ of functions and 
vectors just restates the way we introduced Lie derivatives for a fluid flow.
For a fluid with velocity $u^\alpha$, the Lie derivative $\Lie_{\bm u}f$ of 
a function on the fluid is the directional derivative of $f$ along $u^\alpha$, 
\be
   \Lie_{\bm u} f = \frac d{d\tau} f\circ \psi_\tau (P)|_{\tau=0} =u^\alpha\nabla_\alpha f.
\ee
Now $f\circ\psi_\tau$ is the pullback of $f$ by $\psi_\tau$ or, equivalently. the 
result of dragging $f$ by $\psi_\tau^{-1}=\psi_{-\tau}$. The Lie derivative subtracts the 
the value of $f$ at $P$ from its value at $\psi_\tau(P)$, and we have stated this difference 
in a way that makes sense for vectors in curved space, where one knows only how to 
add and subtract vectors at the same point:    
That is, to define the Lie derivative $\Lie_u w^\alpha$ of a vector,  
we again subtract the value of $w^\alpha$ at $P$ from the result of dragging $w^\alpha$ from 
$\psi_\tau(P)$ back to $P$ by $\psi_{-\tau}$:  
\be 
 \Lie_{\bf u} w^\alpha = \frac d{d\tau} \psi_{-\tau}^* w^\alpha.
\ee 
And in general, for any for any smooth scalar $f$ and two smooth vector fields $v^a$ and $w^a$ on a manifold, 
\be
\Lie_{\bm v} f = \frac d{d\lambda} f\circ \psi_\lambda (P)|_{\lambda=0},
\label{e:lief}\ee
\be 
 \Lie_{\bf v} w^a = \frac d{d\lambda} \psi_{-\lambda}^* w^a,
\qquad  
\label{e:liegoemv}\ee 
where $\psi_\lambda$ is the family of diffeos generated by $v^a$.  

A vector field $w^a$ is Lie-derived by a vector field $v^a$ if the 
family of diffeos $\psi_\lambda$ generated by $v^a$ leave $w^a$ unchanged, 
if $\psi_\lambda w^a = w^a$.

To check that our geometric definition \eqref{e:liegoemv} of the Lie derivative of a vector 
agrees with the commutator $[v,w]$, we compute the right side of Eq.~\eqref{e:liegoemv}, writing 
\be
\psi_{-\lambda}^* w^i= \frac{\partial\psi^i_{-\lambda}}{\partial x^j}
		w^j\circ\psi_\lambda(P).
\ee
When we take the derivative $\dis \frac d{d\lambda}$ of this expression, we get two terms, with the 
derivative hitting each $\psi_\lambda$ while the other $\psi_\lambda$ is evaluated at $\lambda=0$ where 
it is the identity:
\be
\frac{d}{d\lambda}\left[\frac{\partial\psi^i_{-\lambda}}{\partial x^j}
		w^j\circ\psi_\lambda(P)\right]_{\lambda =0} 
= \frac{d}{d\lambda}\left(\frac{\partial\psi^i_{-\lambda}}{\partial x^j }\right) w^j(P) 
	+\left. \frac{\partial\psi^i_{-\lambda}}{\partial x^j}\right|_{\lambda=0}
	  \frac{d}{d\lambda} w^j \left(\psi_\lambda(P)\right) 
\ee  
Now $\dis \left.\frac d{d\lambda} w^j\circ\psi_\lambda(P)\right|_{\lambda =0}$ is just the Lie derivative 
\eqref{e:lief} of the scalar $w^j$:
\[  
    \left.\frac d{d\lambda} w^j\circ\psi_\lambda(P)\right|_{\lambda =0} =v^k\partial_k w^j;
\]
the first term is what distinguishes the Lie derivative of a vector from that of a scalar:  
\[
   \frac{d}{d\lambda}\left[\frac{\partial\psi^i_{-\lambda}}{\partial x^j}\right]_{\lambda =0} 
	=\frac{\partial}{\partial x^j }
	\left(\frac{d\psi^i_{-\lambda}}{d\lambda}\right)_{\lambda =0}=-\partial_j v^i.
\]
So we have 
\begin{align*}
\frac{d}{d\lambda} \psi_{-\lambda}^* w^i 
	&= \frac{d}{d\lambda}\left(\frac{\partial\psi^i_{-\lambda}}{\partial x^j }\right) w^j(P) 
	+\left. \frac{\partial\psi^i_{-\lambda}}{\partial x^j}\right|_{\lambda=0}
	  \frac{d}{d\lambda} w^j \left(\psi_\lambda(P)\right) \\
  & =- \partial_j v^i w^j + v^k \partial_k w^i , 
\end{align*}
where we have used $\psi^i_0(x)=x^i$ to write 
$\dis
  \left. \frac{\partial\psi^i_{-\lambda}}{\partial x^j}\right|_{\lambda =0} 
	= \frac{\partial x^i}{\partial x^j}=\delta^i _j
$
(the Jacobian of the identity map is the identity map).  We have shown 
\[ 
  \frac{d}{d\lambda} \psi_{-\lambda}^* w = [v,w], 
\]
as claimed.  
 
A nearly identical calculation checks the components of the Lie derivative $\Lie_{\bm v}\sigma_a$ of a dual vector field:  Suppressing ``$\lambda=0$,''  we have
\begin{align*}
\frac{d}{d\lambda} \psi_{-\lambda}^* \sigma_i = \frac{d}{d\lambda} \psi_{\lambda*} \sigma_i 
	&= \frac{d}{d\lambda}\left(\frac{\partial\psi^j_{\lambda}}{\partial x^i }\right) \sigma_j(P) 
	+ \frac{\partial\psi^i_{\lambda}}{\partial x^j}
	  \frac{d}{d\lambda} w^j \left(\psi_\lambda(P)\right) \\
  & = \partial_i v^j w_j + v^j \partial_j w^i , 
\end{align*}
in agreement with Eq.~\eqref{liecov3}.

Finally, the Lie derivative of any tensor is given by 
\be\crv
  \Lie_{\bm v} T^{a\cdots b}{}_{c\cdots d} = \frac{d}{d\lambda} \psi_{-\lambda}^*  T^{a\cdots b}{}_{c\cdots d},
\label{eq:liepsi}\ee
and a calculation essentially identical to those we have just done for vectors and dual vectors verifies 
that the definition yields Eq.~(\ref{eq:liedef2}).
\vspace{2mm}

\noindent{\em Killing Vectors}
\index{Killing vector|textbf}\index{isometry|textbf}
\index{symmetry!Killing vectors and isometries|textbf}

A common use of Lie derivatives in relativity arises from vector fields
that generate {\sl isometries}, symmetries of the spacetime.\\      
{\bf Definition}: An {\sl\crv isometry} $\psi$ is a diffeo that preserves the 
metric: $\psi^*~g_{\alpha\beta}~=~g_{\alpha\beta}$.

If a vector field $\xi^\alpha $ generates the family of diffeos
$\psi_\lambda$, then 
\[ 
 \psi_\lambda^*g_{\alpha\beta} = g_{\alpha\beta} \Rightarrow 
	\Lie_{\bm\xi} g_{\alpha\beta}=0.
\]
By (\ref{eq:liedef1}),
\begin{eqnarray*}
\Lie_{\bm\xi} g_{\alpha\beta} 
&=& \xi^\gamma\nabla_\gamma g_{\alpha\beta} 
	+ g_{\alpha\gamma}\nabla_\beta\xi^\gamma+
	+ g_{\gamma\beta}\nabla_\alpha\xi^\gamma.\\
&=& \nabla_\alpha \xi_\beta  + \nabla_\beta \xi_\alpha .
\end{eqnarray*}
{\bf Definition}: \crv $\xi^\alpha $ is a Killing vector field if
\be	
	\crv\Lie_{\bm\xi} g_{\alpha\beta} = 0\cb,
\label{e:kvdef}\ee\cb
or, equivalently,
\be 
	\nabla_{(\alpha}\xi_{\beta)} = 0 .
\label{e:lieg}\ee
 
	Now for any vector $\xi^\alpha $, one can choose coordinates 
(where $\xi^\a\neq 0$) in the following way: \\
Take a hypersurface $S$ cutting the integral curves of $\xi^\alpha $.  Let $x^0$ be 
a scalar field with $x^0=0$ on $S$ and $x^0(P)$ the parameter distance
from $S$ to $P$ along the integral curve through $P$. Arbitrarily choose the 
remaining three coordinates  $x^1$, $x^2$ and $x^3$ on $S$, and extend them
 to $M$ by requiring them to be constant along the integral curves.

Then $\dis \Lie_{\bm\xi} x^0 = \frac{dx^0(\lambda )}{d\lambda} =
\frac{d\lambda}{d\lambda} = 1, \quad 
\Lie_{\bm\xi} x^1 = \Lie_{\bm\xi} x^2 = \Lie_{\bm\xi} x^3 = 0$, \ \ and 
\[ (\Lie_{\bm\xi} g)_{\mu\nu} = \xi^\lambda\partial_\lambda g_{\mu\nu}
+ g_{\lambda\nu}\partial_\mu \xi^\lambda  +
g_{\mu\lambda}\partial_\nu\xi^\lambda. \]

\noindent But
\[\xi^\mu   = \xi (x^\mu   ) = \Lie_{\bm\xi} x^\mu   = \delta^\mu  _0\]
\[ \Longrightarrow (\Lie_{\bm\xi} g)_{\mu\nu} = \partial_0 g_{\mu\nu} .\]

We have chosen coordinates for which $\bm \xi = \bm\pa_0$: $\xi^\mu  $ is just the tangent to
$\dis \lambda\rightarrow (x^0+\lambda, x^1, x^2, x^3)$.  The symmetry  
$\dis (\Lie_{\bm\xi} g)_{\mu\nu} = 0$ 
implies that $\partial_0 g_{\mu\nu} = 0$, that the components
$g_{\mu\nu}$ are independent of the coordinate $x^0$.  Conversely, if
there is a chart in which the components $g_{\mu\nu}$ of the metric
are independent of a coordinate $x^0$, then the vector with components
$\xi^\mu = \delta_0^\mu$ in that chart is a Killing vector of the spacetime:
\begin{eqnarray*}
(\Lie_{\bm\xi} g)_{\mu\nu} &=& \xi^\lambda \partial_\lambda g_{\mu\nu} 
+ g_{\lambda\nu}\partial_\mu \xi^\lambda+g_{\mu\lambda}\partial_\nu\xi^\lambda    
= \partial_0 g_{\mu\nu} = 0 \\
\Rightarrow (\Lie_{\bm\xi} g)_{\mu\nu} &=& 0.
\end{eqnarray*}
\noindent {\bf Example 1}:  Minkowski space has 10 independent
Killing vector fields, 
\index{Minkowski space!Killing vectors}\index{Killing vectors!of flat space}
\begin{eqnarray*}
\begin{array}{ll} \bm\pa_t, \; \bm\pa_x,\; \bm\pa_y,\; \bm\pa_z &
{\mbox{  4 translational}}\\
x\bm\pa_y-y\bm\pa_x, \; z\bm\pa_x-x\bm\pa_z, \; y\bm\pa_z
-z\bm\pa_y &{\mbox{  3 rotational}}\\
x\bm\pa_t+t\bm\pa_x, \; y\bm\pa_t + t\bm\pa_y, \;
z\bm\pa_t+t\bm\pa_z &{\mbox{  3 boosts ,}} \end{array}
\end{eqnarray*}
and they generate 10 one-parameter groups of diffeos that leave $\eta_{\mu\nu}$
invariant:  4 translations, 3 rotations, 3 boosts, which together close
under composition to form the Poincar\'e group. 
\index{Poincar\'e group}\index{isometry!of flat space} 
Here's the check that 
$\eta_{\alpha\beta}$ is invariant under boosts:
\begin{eqnarray*}
\xi^\mu &=&  (x,t,0,0) = x^1\delta_0^\mu + x^0\delta_1^\mu\\
(\Lie_{\bm\xi}\eta)_{\mu\nu} 
 &=& \underbrace{\xi^\lambda \partial_\lambda\eta_{\mu\nu} }_0 
    + \partial_\mu \xi^\lambda \eta_{\lambda\nu} 
    + \partial_\nu\xi^\lambda \eta_{\mu\lambda} \\
&=&  \partial_\mu  \left(x^1 \delta_0^\lambda 
    + x^0 \delta_1^\lambda \right)\eta_{\lambda\nu}  
 + \partial_\nu\left(x^1\delta_0^\lambda+x^0\delta_1^\lambda\right)\eta_{\mu\lambda}
 \\
&=& \underline{\delta_\mu^1\eta_{0\nu} }
  + \cancel{\delta_\mu^0\eta_{1\nu}}{-6}
  + \cancel{\delta_\nu^1\eta_{\mu 0}}{-6} 
  + \underline{\delta_\nu^0\eta_{\mu 1}}\\
  &=& 0.
\end{eqnarray*}
That $\eta_{\alpha\beta}$ is invariant under translations is immediate:
$\partial_t\eta_{\mu\nu} = 0$.

\noindent That $\eta_{\alpha\beta}$ is invariant under rotations is 
immediate in spherical coordinates with the axis of rotation taken 
to be the symmetry axis:
\[ x\partial_y - y\partial_x = \partial_\phi , \; \xi^\mu   =
\delta^\mu  _\phi,\; (\Lie_{\bm\xi}\eta)_{\mu\nu} = \partial_\phi
\eta_{\mu\nu} = 0.\]

\noindent {\bf Example 2}:  A rotating black hole\index{black hole! Kerr spacetime} 
has the (Kerr) metric
\[ ds^2 =-\frac{\Delta}{\rho^2} (dt-a\sin^2\theta\, d\phi)^2 
	+ \frac{\sin^2\theta}{\rho^2} [(r^2+a^2)d\phi -adt]^2 
	+ \frac{\rho^2}{\Delta} dr^2 + \rho^2d\theta^2,
\]
where
\[ 
\Delta =r^2-2Mr+a^2 \qquad \rho^2 = r^2+a^2\cos^2\theta, 
\]
with $a := $ angular momentum/mass. \index{angular momentum} 
Since $\| g_{\mu\nu}\|$ is independent of the coordinates $t$ and
$\phi$, $\partial_t$ and $\partial_\phi$ are Killing vectors and the
spacetime is invariant under time translations and rotations about the
$\theta =0$ axis.

\section{Introduction to integration on manifolds}
\label{sec:integration}
\index{integration!on manifolds|(}

In flat space, the area of a parallelogram spanned by the 
vectors $\bf A, B$ is $|{\bf A}\times {\bf B}| = |\epsilon_{ab}A^aB^b|$;
and the volume spanned the vectors $\bf A, B, C$ is \\
$|{\bf A}\times {\bf B}\cdot {\bf C}| = |\epsilon_{abc}A^aB^bC^c|$.
Similarly, in Minkowski space, requiring that the volume spanned by four orthonormal 
vectors $t^\alpha $, $x^\alpha $, $y^\alpha $, $z^\alpha $ be 1 
implies that a parallelepiped $\Omega$ spanned by any four vectors 
$A^\alpha $, $B^\alpha $, $C^\alpha $, $D^\alpha $ is
\[ 
|\Omega| = |\epsilon_{\alpha\beta\gamma\delta} A^\alpha B^\beta C^\gamma D^\delta|.
\]
The set of four vectors is positively oriented if 
$\epsilon_{\alpha\beta\gamma\delta} A^\alpha B^\beta C^\gamma D^\delta > 0$.

	The volume of an arbitrary region $\Omega$ is obtained by adding volumes of
infinitesimal parallelepipeds spanned by vectors along the coordinate axes,
$e^\alpha_0$, $e^\alpha_1$, $e^\alpha_2$, $e^\alpha_3$ with lengths 
$\Delta x^0$, $\Delta x^1$, $\Delta x^2$, $\Delta x^3$
\begin{eqnarray*}
\Delta^4V &=& \epsilon_{0123} \Delta x^0\Delta x^1\Delta x^2\Delta x^3\\
&=& \frac{1}{4!} \epsilon_{\mu\nu\sigma\tau} \Delta x^\mu \Delta
x^\nu  \Delta x^\sigma \Delta x^\tau (-1)^\pi,
\end{eqnarray*}
where $(-1)^\pi = 1$ when $\mu,\nu\,\sigma,\tau$ is an even permutation $\pi$  
of $0,1,2,3$ and $(-1)^\pi = -1$ for an odd permutation.
Because $\epsilon_{\alpha\beta\gamma\delta}A^\alpha B^\beta C^\gamma D^\delta$ is 
a scalar, the volume of a region of flat space is given in any oriented chart by 
\[ |\Omega | = \int_\Omega \epsilon_{0123}\, dx^0dx^1dx^2dx^3 \equiv\int_\Omega\,
d{}^4V.\]
The Jacobian, $\left|\frac{\partial x}{\partial x'}\right|$, that relates 
the volume element in two different coordinate systems arises from the 
coordinate transformation of the totally antisymmetric tensor 
$\epsilon_{\alpha\beta\gamma\delta}$:\\
$\dis \epsilon_{0'1'2'3'}= \frac{\partial x^\mu}{\partial x'^0}\frac{\partial x^\nu}{\partial x'^1}\frac{\partial x^\sigma}{\partial x'^2}\frac{\partial x^\tau}{\partial x'^3}\epsilon_{\mu\nu\sigma\tau} = \left|\frac{\partial x}{\partial x'}\right|\epsilon_{0123}$.
In the mathematics literature the volume integral is written as
\be 
   \int_\Omega \bm \epsilon.
\ee

A curved space is locally flat in the sense that, in a locally inertial coordinate 
system, the metric components are flat up to quadratic order in the coordinates.  
By demanding that the volumes of small regions, to first order in the length of a side, are those measured by a locally inertial observer using her local Minkowski metric,  
one uniquely picks out the volume element
\be 
	d{}^4V = \epsilon_{0123}\, dx^0dx^1dx^2dx^3.  
\ee
Because the totally antisymmetric tensor $\epsilon_{0123}$ has the value $\sqrt{|g|}$, 
the volume element can be written in the equivalent form 
\be 
	d{}^4V = {\sqrt{|g|}}\, dx^0dx^1dx^2dx^3,  
\ee
and in n-dimensions
\be 
   d{}^nV = \epsilon_{1\cdots n} dx^1\cdots dx^n = {\sqrt{|g|}}\,dx^1\cdots dx^n.
\label{dnv}\ee
In index notation, one writes
\[ 
 d{}^4V = \epsilon_{a\cdots b} dS^{a\cdots b}, 
\]
and thinks of $dS^{a\cdots b} $ having components
\[
  {\mbox{`` }}\ dS^{\mu\nu\sigma\tau}  
  	= \pm\frac{1}{4!} dx^\mu dx^\nu dx^\sigma dx^\tau\ .{\mbox{ ''}}
\]
The integral, 
\be
 \int_\Omega f d{}^nV = \int_\Omega f\, \epsilon_{1\cdots n} \,dx^1\cdots dx^n 
	= \int_\Omega f\,{\sqrt{|g|}}\,dx^1\cdots dx^n,
\ee 
over a region $\Omega$ is well-defined (that is, its value is independent of 
the choice of coordinates), because, under a change of coordinates, 
the integrand on the right side is multiplied by the Jacobian 
$\left|\frac{\partial x}{\partial x'}\right|$.

\subsection{Gauss's theorem}\index{Gauss's theorem|textbf}\index{conservation laws!Gauss's theorem|textbf}\index{divergence theorem}\index{integration!Gauss's theorem}
\label{s:gauss0}

We'll need Gauss's theorem, whose form and proof for curved spacetime is nearly the same as that 
for flat space.  The more extensive Appendix \ref{appendix}, including forms, densities, and Stokes's theorem, is useful but not needed for a one-semester GR course. 
 \\

Gauss's theorem relates the volume integral of a divergence of a vector to the surface integral of 
its normal component.  We have already seen in a flat-space context its meaning as a conservation 
law, relating the change in charge or mass in a volume to the flux leaving the volume.  

	In ${\mathbb R}^n$, in Cartesian coordinates, the integral of a
divergence over an $n$-cube can be expressed as a surface integral
after an integration by parts:
\begin{align*}
\int_V\partial_iA^i & d^nx = \int_V\partial_1A^1dx^1dx^2\cdots dx^n +\cdots
	+\int_V \partial_nA^ndx^ndx^1\cdots dx^{n-1}\\
&= \ \int_{\partial_{1+} V}A^1dx^2\cdots dx^n 
	-\int_{\partial_{1-} V}A^1dx^2\cdots dx^n
	+\cdots 	+  \int_{\partial_{n+}V} A^n dx^1\cdots dx^{n-1}
	-  \int_{\partial_{n-}V} A^n dx^1\cdots dx^{n-1}\\
&= \int_{\partial V} A^idS_i  \hspace{5mm} (\partial V {\mbox{ means
the boundary of $V$)}}\\
& {\mbox{where }} dS_i  = \pm \epsilon_{ij\cdots k} dx^j  \cdots dx^k
\frac{1}{(n-1)!}\ , \mbox{with}\\
& dS_1 = +dx^2\cdots dx^n\,\mbox{ for $x^1$ increasing outward,}\\
&	dS_1 = -dx^2\cdots dx^n\ ,   \mbox{ for $x^1$ increasing inward.}
\end{align*}
 
 More generally the integral over any volume $V$ in ${\mathbb R}^n$ of a
divergence is related to a surface integral by
\[ \int_V\partial_iA^i\,d^nV  = \int_{\partial V} A^idS_i  =
\int_{\partial V} A^i  n_idS\]
where $n_i$ is the unit outward normal to $S$ and $dS$ the area element of
$S$.  \\

In curved space the analogous result is easily obtained:
\begin{eqnarray*}
\int_\Omega \nabla_a A^a d{}^n V  
&=& \int_\Omega \frac{1}{\sqrt{|g|}} 
	\partial_i \left( {\sqrt{|g|}} A^i \right) {\sqrt{|g|}} d^4x\\
&=& \int_\Omega \partial_i  \left( {\sqrt{|g|}} A^i \right) d^nx
\end{eqnarray*}

When $\Omega$ is a coordinate box,
\begin{align*}
   \int_\Omega \partial_i  \left(  A^i\sqrt{|g|} \right) d^n x 
	&= \int_{\partial_{1+} V} A^1\sqrt{|g|} dx^2\cdots dx^n 
	   - \int_{\partial_{1-} V}A^1\sqrt{|g|} dx^2\cdots dx^n \\
	&\phantom{xx}   +\cdots 	+  \int_{\partial_{n+}V} A^n\sqrt{|g|} dx^1\cdots dx^{n-1}
	-  \int_{\partial_{n-}V} A^n\sqrt{|g|} dx^1\cdots dx^{n-1}\\
&= \int_{\partial V} A^idS_i , 
\end{align*}
where again each term on the right side has the form 
$\dis\epsilon_{ij\cdots k} dx^j \cdots dx^k \frac{1}{(n-1)!}$.  We can then write
\be
\int_\Omega \nabla_a A^a d{}^n V = \int_{\partial\Omega} A^adS_a  
\label{e:gauss0}\ee
where  $\dis dS_i = \pm\epsilon_{ij\cdots k} dx^j \cdots dx^k \frac{1}{(n-1)!}$. 

Equivalently, the integrand of each term on the right has the form 
$\pm A^k\nabla_k x^i \sqrt{|g|}$ for the part of the boundary 
that is a surface of constant $x^i$.  
To get the sign right, we need to choose the outward normal.  For example consider 
a 4-dimensional slab $\Omega$ whose boundary consists of two $t$=constant slices, 
$t=t_1$ and $t=t_2$,  and a part of the boundary at spatial infinity where $A^\alpha$ vanishes. On the future boundary, 
\[
  \int_{t=t_2} A^\alpha dS_\alpha = \int_{t=t_2} A^t \sqrt{|g|} d^3x 
	= \int_{t=t_2} A^\mu \nabla_\mu t \sqrt{|g|} d^3x, \quad t \mbox{ increasing outward}, 
\]
so $dS_\mu = \nabla_\mu t \sqrt{|g|} d^3x$. On the past boundary, 
\[
  \int_{t=t_1} A^\alpha dS_\alpha= -\int_{t=t_1} A^t \sqrt{|g|} d^3x 
	=  \int_{t=t_1} A^\mu \nabla_\mu(- t)\sqrt{|g|} d^3x,
		\quad -t \mbox{ increasing outward}.
\]
In general, choose a scalar $f$ that increases outward and is constant  
on a part of the surface.  Then take $f$ to be 
a coordinate $f=x^1$ and add coordinates on the surface, with  
$x^1\equiv f, x^2, \ldots, x^n$ a right handed chart, so that 
$\epsilon_{12\cdots n} = \sqrt{|g|}$.  We then have 
\be
 dS_a = \nabla_a f\sqrt{|g|} dx^2 dx^3\cdots dx^n.
\ee

If, as usual, we decompose the volume of integration into a set of 
cubes and take the limit as the size of each cube shrinks to zero, 
we find
\be 
\int_\Omega \nabla_a  A^a  d{}^nV  
	= \int_{\partial\Omega} A^a  dS_a.  
\label{e:gauss}\ee
As usual, surface terms from cubes that share a surface cancel,
because the outward normal to one cube is the inward normal to the
adjacent cube.

This form is correct for any (smooth, orientable) volume in a space 
with a metric, independent of the signature of the metric.  
When $\partial\Omega$ has a unit normal (when it is not a null surface), 
one can write $dS_\alpha$ in a more
familiar form. With outward unit normal $n_a$ (along the gradient of  
a scalar that increases outward) $dS_a  = n_a  dS$, with $dS$ the 
$n-1$ dimensional volume element, given by 
\be
    dS = \sqrt{|{}^{(n-1)}g|}\ d^{n-1}x,
\label{e:dS1}\ee
where  
${}^{(n-1)}g_{ab}$ is the metric on the $n-1$-dimensional surface.   
We'll pick $n=4$ and 
look at a $t=$ constant surface.  You can always change $t$ to $f$ and $4$ to another dimension. The 
unit normal (for $t$ increasing outward) is related to $\nabla_\alpha t$ by 
\be
    n_\alpha = \nabla_\alpha t \frac1{\sqrt{|\nabla_\beta t \nabla^\beta t|}} 
		=\nabla_\alpha t\frac1{\sqrt{|g^{tt}|}}.
\ee
Now we need the metric $^3g_{ab}$ on a $t=$constant surface.  First, a simple-minded version using components.
At the end, we'll circle back and define $^3g_{ab}$ as the pullback of $g_{\alpha\beta}$ to $S$. 
If $v^\alpha$ is any vector lying in a $t=$constant surface, $v^t=v^\alpha\nabla_\alpha t=0$.
The metric in the surface is, by definition, the map from a pair
of vectors $v^\alpha$ and $w^\alpha$ in the surface to $\mathbb R$,  and that is already given 
by the original metric on $M$,
\be
  g_{\alpha\beta} v^\alpha w^\beta = g_{ij}v^i w^j.
\ee
So the components of the 3-metric are the spatial components of the 4-metric. 
\[
^3g_{ij} = g_{ij}.
\]
We can relate the determinant $^3g$ to $g$ and the unit normal $n_\alpha$ as follows:  The inverse metric 
has component $g^{tt}$ given by
$\dis g^{tt} = |\mbox{cofactor of } g_{tt}|/g$, 
and the cofactor of $g_{tt}$ is the determinant of the spatial components $g_{ij}={}^3g_{ij}$ of the metric:
\[
    g^{tt} = \frac{^3g}{g}.
\]
Then 
\begin{eqnarray*} 
dS_\mu &=& \nabla_\mu t \sqrt{|g|}~dxdydz = \sqrt{|g^{tt}|} n_\mu  \sqrt{|g|}~dxdydz 
= n_\mu \sqrt{^3g} ~dxdydz\\
&=& n_\mu dS ,
\end{eqnarray*}
with $dS=\sqrt{^3g} ~d^3x$ in agreement with Eq.~\eqref{e:dS1}.\\

Finally, we show that $^3g_{ab}$ is the pullback of $g_{\alpha\beta}$ from the spacetime $M$ to 
the submanifold $S$: That is,  $^3g_{ab}=(\psi_*g)_{ab}$ where $\psi:S\rightarrow M$ takes 
each point in $S$ to itself, the identity diffeo, but with its image regarded as a point of $M$. 
Then $(\psi^\mu(x)) = (0,x^i)$, and $\dis \frac{\partial\psi^\mu}{\partial x^i} = \delta^\mu_i$, giving 
\be
   ^3g_{ij}=\frac{\partial\psi^\mu}{\partial x^i}\frac{\partial\psi^\nu}{\partial x^j}g_{\mu\nu} 
	   =\delta^\mu_i \delta^\nu_j g_{\mu\nu} = g_{ij}.
\ee

\noindent {\bf Example}:  The integral form of mass
conservation $\nabla_\alpha  (\rho u^\alpha  )=0$ for dust is \index{conservation laws!mass}
\begin{eqnarray*}
0 &=& \int_\Omega \nabla_\alpha  (\rho u^\alpha  )d{}^4V  = \int_{V_2} \rho u^\alpha  dS_\alpha  -
\int_{V_1} \rho u^\alpha  dS_\alpha \\
&& {\mbox{or   }} \int_V \rho u^\alpha  dS_\alpha  = {\mbox{ constant}}
\end{eqnarray*}
Similarly, the number of baryons in a fluid is\index{conservation laws!baryons}
\[ N = \int_V nu^\alpha  dS_\alpha  = \int_V nu^\alpha  n_\alpha  dS\]
(and $N = \int_V ndS$ only if $u^\alpha  \perp V$ so that $u^\alpha  n_\alpha 
=1$---i.e.\  if the fluid is at rest relative to $V$.)

\noindent Note that in a chart with $V$ a $t =$ constant surface,
\[ 
	\int_V A^\alpha  dS_\alpha  = \int_VA^t {\sqrt{-g}}~d^3x ,
\]
so there is no need to introduce $n_\alpha $ and ${\sqrt{^3g}}$ to evaluate the
integral.  This fact is {\em essential} if one is evaluating an 
integral $\int j^\alpha dS_\alpha$ over a null surface, where there is no unit 
normal.  \\
\index{null hypersurface|textbf}\index{hypersurface!null|textbf}
\noindent{\bf Definition}: {\sl Null surface}.  A hypersurface 
is {\sl null} if, at each point of the surface, the direction normal to the surface is null. \\
{\sl Exercise}.  Show that every vector tangent to a null surface and 
not in the direction of the normal is spacelike. (Proof on next page.)\\   

We will see that the event horizon of a black hole is a null surface, 
and the flux of energy or of baryons across the horizon can be computed in coordinates 
that are regular there.  
Consider coordinates $v,r,\theta,\phi$ with the surface of constant $r$ a null surface 
(If you are familiar with the Schwarzschild geometry, think of the $r=2M$ 
surface.)  Then \index{energy!energy flux across surface}\index{flux!across a surface}
\[
 \int j^\alpha dS_\alpha = \int j^r \sqrt{-g}\ dv d\theta d\phi.
\]
There is no unit normal, but the flux across the surface doesn't care.  

\newpage
\noindent
{\sl Solution to Exercise}. Let $\ell^\a$ be normal to a null surface. We are to show that every vector $v^\a$ tangent to the surface is either spacelike or is a multiple of $\ell^\a$.  \\

The normal $\ell^a$ can be written as a sum $\hat t^\a+\hat x^\a$ of 
orthonormal timelike and spacelike vectors.  A vector $w^\a$ is tangent to 
the surface if $w^\a \ell_\a = 0$.  Any vector $w^\a$ is a sum 
$A \wh t^\a + B\wh x^\a + v^\a$, with $v^\a$ orthogonal to $\wh t^\a$ and 
$\wh x^\a$. Because $v^\a$ is orthogonal to $\wh t^\a$, $v^\a$ is spacelike.  
We have
\[
  0 =w^\a \ell_\a = -A + B \Rightarrow A=B \Rightarrow w^a = A\ell^\a + v^\a.
\]
Then $w^\a w_\a = v^\a v_\a$, implying either $w^\a = A \ell^\a$ or 
$v^\a\neq 0$ and $w^\a w_\a >0.\ \Box $ 

Notice that the normal $\ell^\a$ to a null surface is also tangent to the surface.
We already encountered (on p.~\pageref{p:nullsurface}) this peculiarity of a metric with Lorentz signature in our discussion of the normal $\na_\a u$ to the null $u=$~constant surface in Minkowski space.  In that example, 
the null surface is a light cone, the surface generated by the future-directed null rays from a point of flat space. 
\index{integration!on manifolds|)}

\section{Symmetries and conserved currents}
\index{symmetry!relation to conserved current}\index{conservation laws!relation to Killing vector}
We saw in Sect.~\ref{s:conservation_laws} that, in Minkowski space, 
conservation of momentum followed from the vanishing divergence 
$\na_\a T_\a^{\ \b}=0$.  For each direction, $\wh t^\a,\wh x^\a, \wh y^\a, \wh z^\a$, the corresponding momentum current $j^\a$ is conserved: 
e.g., with $ j^\a = T^\a_{\ \b} \wh t^\b$, 
\[
   \na_\a j^\a = \na_\a(T^\a_{\ \b}) \wh t^\a+ T^\a_{\ \b}\na_\a \wh t^\b = 0. 
\]
Corresponding to each conserved current is a conserved component of the 
total momentum, the integral of the current over a spacelike hypersurface $V$: 
e.g., the energy, 
\[
   E = \int_V T_\a^{\ \b} \wh t^\a dS_\b,  
\]
is unchanged from one spacelike hypersurface to the next if there is 
no flux of energy at infinity (or, for finite $V$, across the boundary of $V$).  

We will now see that the same steps relate each continuous 
symmetry of a curved spacetime 
to a conserved current and thus to a corresponding conserved quantity, the 
integral of the current over a spacelike hypersurface. The relation 
follows from the form \eqref{e:lieg} of $\Lie_\xi g_{\a\b}=0$, 
the antisymmetry of $\na_\a\xi_\b$,  
\[
    \na_{(\a}\xi_{\b)} = 0.  
\]
Then $\na_\a T^{\a\b}=0$ implies $j^\a = T^\a_{\ \b}\xi^\b$ is conserved,
\be
\crv \na_\a j^\a = \na_\a (T^\a_{\ \b}\xi^\b) =0:
\label{e:Txi}\ee  
\[
  \na_\a j^\a = \na_\a (T^{\a\b}\xi_\b) 
   	=\na_\a (T^{\a\b})\xi_\b + T^{\a\b}\na_\a \xi_\b 
   	= 0. 
\]
The term $T^{\a\b}\na_\a \xi_\b $ vanishes because $T^{\a\b}$ is symmetric and 
$\na_\a \xi_\b$ is antisymmetric:\\
 {\sl Claim}. Let $S_{ab}$ and $A^{ab}$ be symmetric and antisymmetric tensors.
 Then 
 \be 
 	S_{ab}A^{ab} =0.
\label{e:SA}\ee
Check:   
\begin{align*}
  S_{ab}A^{ab}  &= - S_{ba}A^{ba} \mbox{\ \ (exchanging indices in each tensor)} \\
  	       &= -S_{ab}A^{ab} \mbox{\ \ (changing the names of the dummy indices)}\\
 \Longrightarrow\ \  S_{ab}A^{ab}  & = 0.
\end{align*} 

\noindent{\bf Example}:  The conserved quantities associated with the rotational Killing vectors of Minkowski space are the components of angular momentum:
For $\bm \xi = x\bm\pa_y - y\bm\pa_x$ and $V$ a $t=$ constant surface, 
we have 
\be
  \int_V T^\a_{\ \b}\xi^\b dS_\a = \int_V (xT^t_{\ y} - yT^t_{\ x})\ dV= L_z.
\ee
\noindent{\sl Symmetry and conserved momentum of a particle} \\

The same index-symmetry argument shows that, for each Killing vector 
$\xi^\a$, the momentum of a free  particle along $\xi^\a$ is conserved.
Here $p_\a = m u_\a$, $m$ the particle's rest mass, and $p_\a \xi^\a$ 
is conserved along the particle's trajectory: 
The geodesic equation, $u^\b\na_\b u^\a = 0$, implies   
\be
  \crv u^\b\na_\b (p_\a \xi^\a) = 0.
\label{e:pxicons}\ee
\benr\item  Check this equation and show that the conserved quantities 
associated with translational and rotational Killing vectors are 
components of momentum and angular momentum.  What are the conserved 
quantities associated with the boost Killing vectors of Minkowski space? 
\een 
\index{spacetime!curved, Chap.~\ref{c:cst}|)}\index{curved spacetime, Chap.~\ref{c:cst}|)}
\newpage

\chapter{\texorpdfstring{$\bm{G_{\alpha\beta} = 8\pi T_{\alpha\beta}}$}.}

\index{general relativity|(}\index{relativity!general relativity|(}
\section{``Derivation" of the Equation}
\index{Einstein field equation!derivation}

 To conform to the state of knowledge in 1910, a theory of gravity should satisfy:

\noindent I.\indent With no gravity, special relativity is valid.\\

\noindent II. \indent Principle of equivalence:  ``Locally" one can't
distinguish uniform acceleration from a uniform gravitational field --
or free fall from no gravitational field.\\

\noindent III. \indent Newtonian gravity is valid in the Newtonian regime, i.e. when
the sources are nonrelativistic in character.  

The observed solar-system departure from Newtonian gravity was the anomalous $43''$ precession of Mercury's orbit each century, a correction of about one part in $10^7$. (Mercury's year is 88 days, 
so 100/88 orbits per century for a total of about $(100/88)3600\times 360\approx 5\times 10^8$ arcseconds.)
\\

\index{accelerated observer}
I and II partly cease to make sense once the theory is accepted because
there will be no gravitational ``field" -- just curved spacetime -- and no
distinction between a {\sl uniform field observed by a stationary observer} and
{\sl no field observed by a uniformly accelerated observer}.  In place of I and
II will be the statement that free particles move on geodesics of the
spacetime.  
\index{equivalence principle|textbf}\index{principle of equivalence|textbf}
\index{equivalence principle!Eotvos@E\"otv\"os experiments}
One can, however, continue to test the equivalence principle prediction 
that the gravitational acceleration of small objects in a time independent gravitational 
field is independent of the objects' internal structure and composition (\href{https://en.wikipedia.org/wiki/E%C3%B6tv%C3%B6s_experiment}{E\"otv\"os experiments}), \index{Eotvos@E\"otv\"os experiments} 
and that local Lorentz invariance holds. (Here one is measuring acceleration 
relative to a stationary observer in a time-independent gravitational field.) 

	Prior to writing the theory, I and II together are taken to imply that a
freely falling, nonrotating observer, measuring components of tensors
along her orthonormal basis finds that matter with stress-energy tensor
$T^{\mu \nu }$ obeys $\partial _\nu T^{\mu \nu } = 0$ + terms
of order $\ell^2\mid \partial _i \partial _j\Phi \mid$, with $\Phi $ the
gravitational potential, $\ell $ a length characteristic of the size of her
laboratory. She will similarly find that electromagnetic fields satisfy
\[\partial_\nu F^{\mu \nu } = 4\pi j^\mu + \;\;\text{terms\;\;
smaller\;\; by\;\; a\;\; factor} \sim \ell ^2\mid \partial _i\partial
_j\Phi \mid \] 
\[ \partial_{[\mu}F_{\nu \sigma ]} + \;\;\text{terms\;\; smaller\;\;
by}\;\; \ell ^2\mid \partial _i\partial _j\Phi \mid = 0\;\;.\]\\  
Finally, III means that when source velocities are small compared to 
the speed of light, when  
\begin{eqnarray*}
\begin{array}[t]{r|l}
v \ll 1 &\ \  \displaystyle{\frac{v}{c} \ll 1} \\
&\\
\mbox{pressure} ~~ P \ll \rho &\ \  \displaystyle{ P \ll \rho c^2}  \\
\\
\mbox{and} ~~ \displaystyle{\Phi \sim \frac{M}{R} \ll 1}, &\ \  \displaystyle{\frac{\Phi}{c^2}\sim \frac{GM}{Rc^2} \ll 1} \\
{}& \displaystyle{} \end{array} \end{eqnarray*}
then  $\nabla ^2\Phi = 4\pi G\rho$, and ${\bf g} = -\nabla \Phi $ is the acceleration due to gravity.

These are the constraints on the theory.\\

Why does gravity involve a metric?  We have already mentioned Geroch's argument:
Anything you use to decide what a straight line is–- or, more generally to determine the geometry of spacetime–-is curved by gravity.
so it is perhaps not surprising that gravity is tied to geometry.

	The more specific, equivalence-principle argument arises from the following 
experiment, an experiment that uses the first two assumptions above.  
Consider the following two situations.  In the 
first, two identical clocks A and B are placed on the wall of a rocket that
accelerates uniformly with acceleration $g$.  In the second situation, the clocks are
at rest, one above the other in the nearly uniform field of the Earth.

\begin{figure}[h] 
\centerline{\includegraphics[width=7cm]{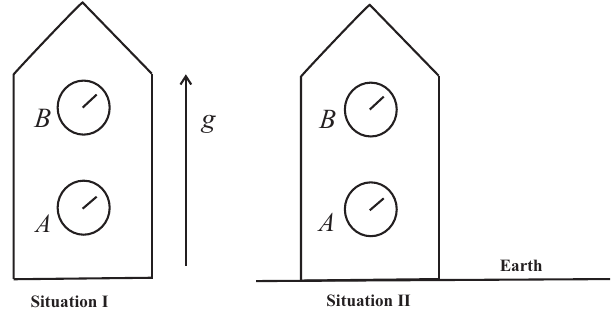}} 
\label{w118all} 
\end{figure} 

	Technically, if each clock in situation I feels a uniform acceleration
$g$, the spatial distance between them will not remain constant:  two
particles satisfying (I.66) have a time dependent separation defined as the length of a 
connecting line orthogonal to one of their trajectories.  But the difference will be of order $\dis\frac{v^2}{c^2}
\sim \frac{(g\tau )^2}{c^2}$ and will be neglected:  In this calculation we
will work to $\dis O\left(\frac vc\right)$.

	In situation I, let $h$ be the initial distance between clocks measured
orthogonal to their trajectories.  Situation II will then be ``equivalent" if
$h$ is the distance measured by a meter stick at rest with respect to the
clocks and if each clock feels constant acceleration $g$. For example, a pair of 
clocks at rest relative to the Earth are equivalent to a pair of uniformly accelerating 
clocks at fixed distance from each other.  We will show that light
signals sent a proper time $\Delta \tau _A$ apart from A to B are received
at a different interval $\Delta \tau _B$ measured on B's clock.  In the
``equivalent" situation, where the clocks are at rest on Earth, the
implication is that again $\Delta \tau _A \neq  \Delta \tau _B$ - that the
clocks move at different rates.\\

\benr \item Use the Newtonian trajectories of two uniformly accelerated clocks 
separated by a distance $h$ and each at rest at $t=0$.  Show that photons 
emitted by clock $A$ at $t=0$ and at $t=\Delta t_A$ reach $B$ at times
differing by $\Delta t_B = \Delta t_A(1+gh)$, neglecting 
terms of order $g(\Delta t)^2$ and order $\Delta t (gh)^2$.  
\een
\newpage

Here is the same result using the relativistic trajectories and proper time measured by $A$ and $B$.\\   

\begin{figure}[h] 
\centerline{
\includegraphics[width=6cm]{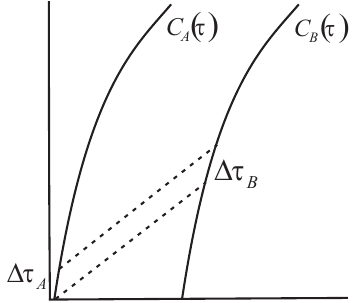}} 
\label{w119} 
\end{figure}

 In units with $c=1$, the time for light to go from $A$ to $B$ is roughly $h$.
In that time, $A$ and $B$ gain velocity $v$ of order $gh$  (i.e. $v\sim gh/c$), and we assume 
that this speed $v$ is small: $gh\ll 1$  (i.e. $gh/c^2\ll 1$).  

The trajectory of clock $B$ is 
\bea c_B^\alpha(\tau) &=& \frac{1}{g}~[\sinh(g\tau)~t^\alpha +\cosh(g\tau) x^\alpha]
		+ \left(h - \frac{1}{g}\right)x^\alpha 
\nonumber\\
&=&  \tau~t^\alpha + \left(h + \frac{1}{2} g\tau^2\right)\,x^\alpha + O[(g\tau)^2\tau]
\eea
 
1st photon has path $\lambda (t^\alpha + x^\alpha)$ hitting $c_B(\tau )$ at the time $\tau_1$ on clock $B$ for 
which 
\bea
\lambda (t^\alpha + x^\alpha)&=& c_B^\alpha(\tau_1) \ \Longrightarrow \ 
\lambda = \tau_1\ \mbox{and } \lambda = h + \frac{1}{2} g\tau ^2_1 + O[(g\tau)^2\tau], \nonumber\\
\tau _1 &=& h + \frac{1}{2} g\tau ^2_1 	+ O[(g\tau_1)^2\tau_1]\nonumber\\
        &=& h + \frac{1}{2} gh^2 +O[(gh)^2h]\,.
\eea

2nd photon has path $\lambda(t^\alpha+x^\alpha) + \Delta \tau _A\, t^\alpha$ hitting
$c_B(\tau )$ at
\bea
\tau _2 &=& \Delta \tau _A + h + \frac{1}{2} g\tau^2_2 + O[(gh)^2 h],
O[(\Delta \tau )^2] \nonumber \\
&=& \Delta \tau _A + \tau _1 + \frac{1}{2} g(\tau ^2_2- \tau^2_1) \nonumber \\
&=& \Delta \tau _A + \tau _1 + \frac{1}{2} g(\tau _2-\tau _1)(2\tau _1)\nonumber\\
&=&  \Delta \tau _A + \tau _1 +   g\tau_1 \,(\tau _2-\tau _1)
\eea
\bea
\Delta \tau _B &=&  \tau _2 - \tau _1 = \Delta \tau_A + g\tau_1\, \Delta \tau _B\nonumber \\
&=& \Delta \tau _A(1 + gh) + O[(gh)^2\Delta \tau _A].
\end{eqnarray}

	Then, for clocks on Earth, the equivalence principle implies
\begin{equation} \Delta \tau (h) 
			= \left(1 + \frac{\Phi}{c^2}\right) \Delta \tau(h = 0)\;\;, 
\label{equiv} \end{equation}
where $\Phi = gh$ is the gravitational potential.\\

	From this result one infers that the metric of spacetime depends on $\Phi$, because 
the proper time read on a clock is proportional to $\sqrt{g_{tt}}$.  Here is the argument.  Assume 
that the spacetime geometry of the two clocks and the Earth is time independent.  Then there is a scalar $t$ for which the metric components $g_{\mu \nu}$ are independent of $t$ in a coordinate system $\{t,x^i\}$ for which the spatial coordinates of the clocks and the Earth are independent of $t$.
 By saying ``there is a metric," one means, in particular, that the proper time elapsed on a clock
with trajectory $c(\lambda )$ is $\tau = \int^\lambda \mid g_{\alpha\beta}\xi^\alpha
\xi^\beta \mid ^{1/2}d\lambda $ (and, in general, the spacetime distance between two events $P$
and $Q$ is the geodesic distance with respect to $g_{\alpha\beta}$). 

Consider now two photons sent from clock $A$ at two times a coordinate interval $\Delta t$ apart.  Because the geometry is independent of $t$, the trajectory of the second 
photon is identical to that of the first photon, translated in time by $\Delta t$: 
\be
	x^\mu_2(\lambda ) = x^\mu _1(\lambda) + \delta^\mu _t \,\Delta t.
\ee 
Therefore signals sent a coordinate time $\Delta t$
apart are received at $B$ a coordinate time $\Delta t$ apart.

But \hspace{.3in}$\Delta \tau _A = \;\mid g_{tt}(A)\mid ^{1/2} \Delta t, 
	   \qquad \Delta \tau _B = \;\mid g_{tt}(B)\mid ^{1/2} \Delta t$.\\
If the equivalence principle is to hold, these two time intervals must be related as they are 
in the uniformly accelerating spacecraft: They must satisfy Eq. (\ref{equiv}).  
Thus 
\[  
 \mid g_{tt}(h) \mid ^{1/2} = \left(1 + \frac{\Phi}{c^2}\right)\mid  g_{tt}(0)\mid ^{1/2}.
\]
\noindent Normalizing $t$ to make $\Delta \tau (h=0) = \Delta t$ gives\\
\be g_{tt} = -\left(1 + \frac{2\Phi }{c^2}\right).
\label{gtt}\ee 

	In the more sophisticated language of the last section, one assumes that
the metric has a timelike Killing vector, $t^\alpha$, and that the Earth and
clocks are simply time translated by its family of diffeos.  Then, assuming
each photon trajectory is similarly time translated, the parameter
intervals $\Delta t$ between transmitted photons are equal to the parameter
intervals $\Delta t$ between received photons, and (\ref{equiv}) implies
\be g_{\alpha\beta}t^\alpha t^\beta(h) = \left(1 + \frac{2\Phi }{c^2}\right)g_{\alpha\beta}t^\alpha t^\beta(0)\;\;.\ee
  
   The equivalence principle argument has led us to a metric having in the Newtonian limit a
component $g_{tt}$ of the form (\ref{gtt}), 
\[
	g_{tt} = -\left(1 + \frac{2\Phi }{c^2}\right) ,
\] 
where $\Phi $ is the Newtonian potential
\be 
	\nabla ^2\Phi = 4\pi G\rho. 
\ee 
Now comes the kind of spectacular coincidence that happens when one is on the right track: 
If we suppose that 
\be
 \frac{1}{c}\partial _t g_{\mu \nu} \sim \frac{v}{c} \partial _i g_{\mu \nu} \ll \partial _ig_{\mu \nu},
\qquad  i = 1-3,
\ee
as befits a field produced by slowly moving sources, then the
geodesic equation
\[
	 u^\nu\nabla _\nu u^\mu = 0 \;\;, \hspace{.5in} \mbox{for} \hspace{.5in}  
	 u^\mu = (1,v^i) + O(v^2),
\]
is 
\[ 
 \frac{dv^i}{dt} + \Gamma ^i {}_{\mu \nu} u^\mu u^\nu = 0 	\hspace{.5in}
 \left(\frac{dv^i}{d\tau } = \frac{dv^i}{dt} +O(v^2)\,\right),
\]
\[ 
 \frac{dv^i}{dt} + \frac{1}{2}(-\partial _ig_{tt} +2\partial _tg_{ti}) = 0 
\]
or
\be 
	\frac{dv^i}{dt} = -\partial _i\Phi , 
\label{e:geod_Newtonian}\ee
the Newtonian equation of free particle motion!  More generally, the relativistic equation of motion
\[
0 = \nabla_\nu T^{i\nu } = \partial _\nu T^{i\nu } + \Gamma ^i{}_{\sigma \nu }T^{\sigma
\nu } + \Gamma ^\nu {}_{\sigma \nu }T^{i\sigma}
\]
has Newtonian limit 
\[
\partial _\nu T^{i\nu } + \Gamma^i{}_{tt}T^{tt}, \]
using   $T^{it},T^{ij} \ll T^{tt}$ for nearly Newtonian matter.
Then 
\be \partial _\nu T^{i\nu } = -\rho \partial ^i\Phi \;\;, \;\;(T^{tt}
\simeq \rho )\ee
so a $g_{tt}$ component of the metric of the form (\ref{gtt}), $-(1+ 2\Phi/c^2)$, gives rise to the gravitational
force, in the Newtonian regime.\\

Consequently, if one assumes that spacetime has a Lorentz metric $g_{\alpha\beta}$,
that matter with stress-energy tensor $T^{\alpha\beta}$ satisfies $\dis\nabla_\beta T^{\alpha\beta}=0$, 
and that $g_{\alpha\beta}$ is related to $T_{\alpha\beta}$ by an equation whose
Newtonian limit is
\[ g_{tt} = -\left(1 + \frac{2\Phi }{c^2}\right),\]
with
\be
	 \nabla ^2\Phi = 4\pi G\rho,
\label{dsqphi}\ee
then virtually all relevant experimental knowledge (I-III) will be
accounted for.

     What remains is to find the relation between $g_{\alpha\beta}$ and $T^{\alpha\beta}$.  One
way to get it is to say one wants a scalar Lagrangian whose variation gives
equations with no higher than 2nd derivatives of $g_{\alpha\beta}$ -- because
$\nabla ^2\Phi = \nabla ^2g_{tt}$ has 2nd derivatives of $g_{\a\b}$.  The only
scalar that doesn't involve higher derivatives is $R$ (really $R+$ constant, 
written as $R-2\Lambda$, where $\Lambda$ is the {\sl cosmological constant}).  
We'll discuss this in Chap.~\ref{s:action} -- it's Hilbert's
method.  A different approach is as follows (due to Geroch).

	From our discussion of the equivalence principle, it is clear that the
form of $g_{tt}$ obtained for a uniform gravitational field is the same
$g_{tt}$ that would be quoted by a system of uniformly accelerating
observers (with no gravitational field) -- that was how we found the
``gravitational" $g_{tt}$.  A real gravitational field, however, is
nonuniform, and only by its nonuniformity can it be recognized.  Thus to
seek a relation between $g_{\alpha\beta}$ and matter, it is natural to look at the
$\it{relative}$ acceleration of free particles because it is that which
depends on the fact that $\nabla \Phi $ is not constant -- i.e. relative
acceleration is given in the Newtonian regime by Eq.~\eqref{e:tide},
\be 
	\ddot S^i = -\nabla ^i\nabla_j\Phi \, S^j \ ,
\ee
\index{force!tidal}\index{tidal force}
relating the separation vector $\ddot S^i$ to the tidal forces, the second derivatives of
$\Phi$.  Because $S^t = 0$ and $\dis\frac{1}{c}\partial _t\Phi \ll \partial _i \Phi$, this can be written
\be 
	\ddot S^\alpha = -\nabla^\alpha \nabla_\beta \Phi\,  S^\beta \ . 
\ee
In relativity the same experiment is described by the deviation of
neighboring geodesics (2.66),
\be 
  \ddot S^\alpha = R^\alpha_{\ \beta\gamma\delta}u^\beta u^\gamma S^\delta
\ee
\index{geodesic deviation}
so that
\[ 
    R^\alpha{}_{\gamma\delta\beta}u^\gamma u^\delta 
	\mbox{ must correspond in the Newtonian approximation to }\ -\nabla^\alpha\nabla_\beta \Phi\ . 
\]
Because $\Phi$ satisfies $\nabla ^2\Phi = 4\pi G\rho$, the curvature  
tensor must satisfy, in the Newtonian approximation,
\be 
	R^\gamma_{\ \alpha\gamma\beta}u^\alpha u^\beta 
		= R_{\alpha\beta}u^\alpha u^\beta = 4\pi G\rho. 
\ee
	There is, however some ambiguity in what to choose as the scalar that
becomes $\rho $: From $T_{\alpha\beta}$ and $u^\alpha$ one can construct two scalars
\be 
	T_{\alpha\beta}u^\alpha u^\beta \qquad {\rm and }\qquad -T\equiv -T^\alpha{}_\alpha, 
\ee
each of which has the mass density $\rho$ as its Newtonian limit.  Thus one
could generalize (\ref{dsqphi}) to any equation of the form
\be R_{\alpha\beta}u^\alpha u^\beta = 4\pi G[rT_{\alpha\beta} + (1-r)T\;\;g_{\alpha\beta}]u^\alpha u^\beta\;\;,\ee
and in order that it hold for any timelike $u^\alpha$, one has
\be 
	R_{\alpha\beta} = 4\pi G[rT_{\alpha\beta} + (1-r)T\;\;g_{\alpha\beta}] 
\label{rab}\ee
or
\be 
	4\pi GrT_{\alpha\beta} = R_{\alpha\beta} - \frac{r-1}{3r-4} g_{\alpha\beta}\;\;R\;, 
\label{tab}\ee
where (\ref{rab}) was contracted to give $R = 4\pi G(4-3r)T$.  But (\ref{tab}) must be
consistent with 
\[ \nabla _\beta T^{\alpha\beta} = 0 \;.\]
Recalling that the contracted Bianchi identity is 
\[ \nabla_\beta G^{\alpha\beta} \equiv \nabla_\beta (R^{\alpha\beta} - \frac{1}{2}g^{\alpha\beta}R) = 0 \]
one has a unique choice of r which makes these equations consequences of
each other, namely $r=2$.  Thus we are led to Einstein's equation
\be G^{\alpha\beta}\equiv R^{\alpha\beta} - \frac{1}{2} g^{\alpha\beta}R = 8\pi GT^{\alpha\beta}\;.\ee
\index{Einstein field equation|textbf}\index{field equation|textbf}

\noindent{\sl Hilbert's derivation}.\\
Hilbert obtained the field equations immediately before Einstein, by adopting  
for a Lagrangian density the simplest nontrivial scalar density, $R\sqrt{-g}$, that 
can be constructed from the metric (see Sect.~\ref{s:action} for the variation of 
the Hilbert action.) Here is a quote from Kip Thorne,\cite{thorne94}
\begin{verse} 

\hspace{6mm}{\sl 
In autumn 1915, even as
Einstein was struggling toward the right law, making mathematical mistake 
after mistake, Hilbert was mulling over the things he had learned from 
Einstein’s summer visit to Göttingen. While he was on an autumn vacation 
on the island of Rugen in the Baltic the key idea came to him, and within 
a few weeks he had the right law–derived not by the arduous trial-and-error 
path of Einstein, but by an elegant, succinct mathematical route.
Hilbert presented his derivation and the resulting law at a meeting of the Royal Academy of Sciences in Göttingen on 20 November 1915, just five days before Einstein’s presentation of the same law at the Prussian Academy meeting in Berlin.}
\end{verse}

There is some debate over whether Hilbert presented the explicit field
equation before Einstein (see, e.g., \cite{renn_stachel07,winterberg04} 
and references therein). The proof sheets from Hilbert's paper predate 
by a few days Einstein's talk presenting the equation, but part of a 
page has been cut off.  The truncated proof sheets have the field equation 
implicitly, as the variation $\delta I/\delta g^{\a\b}$ of the action 
with Lagrangian density $R\sqrt{-g}$.   The explicit form of the 
variational derivative, $R_{\a\b}-\frac12 g_{\a\b}$, appears in 
Hilbert's printed paper, and it is plausibly also in the missing part 
of the proof sheets.  But whether or not it was in Hilbert's talk or 
in the 1915 proof sheets, both Hilbert and Einstein knew the explicit 
form of that variational derivative.  
\index{action!Hilbert(Einstein-Hilbert) action} 
\index{Lagrangian!gravitational}\index{Hilbert action}

\vspace{5mm} 
\noindent
{\sl Gravitational units}.\\
  Beginning now, units are chosen to make $c=G=1$.  
These are called {\sl gravitational units} or 
{\sl geometrized units}.  With $c=1$, length and time have the 
same dimension; if time is measured 
in seconds, length is measured in light-seconds,  where 
1 light-second = $3\times 10^{10}$ cm, the distance light travels in 1 
second.  MTW gives distance in cm, and the 
unit of time is then $1/(3\times 10^{10})$ s.  With $G = 1$, mass, 
length and time all have the same dimension, which MTW takes to be length 
with unit 1 cm.  Then the unit of mass is 
(1 cm)$\times c^2/G = 1.3468\times 10^{28}$ g.  Here is the MTW 
table:  
\centerline{\includegraphics[width=\textwidth]{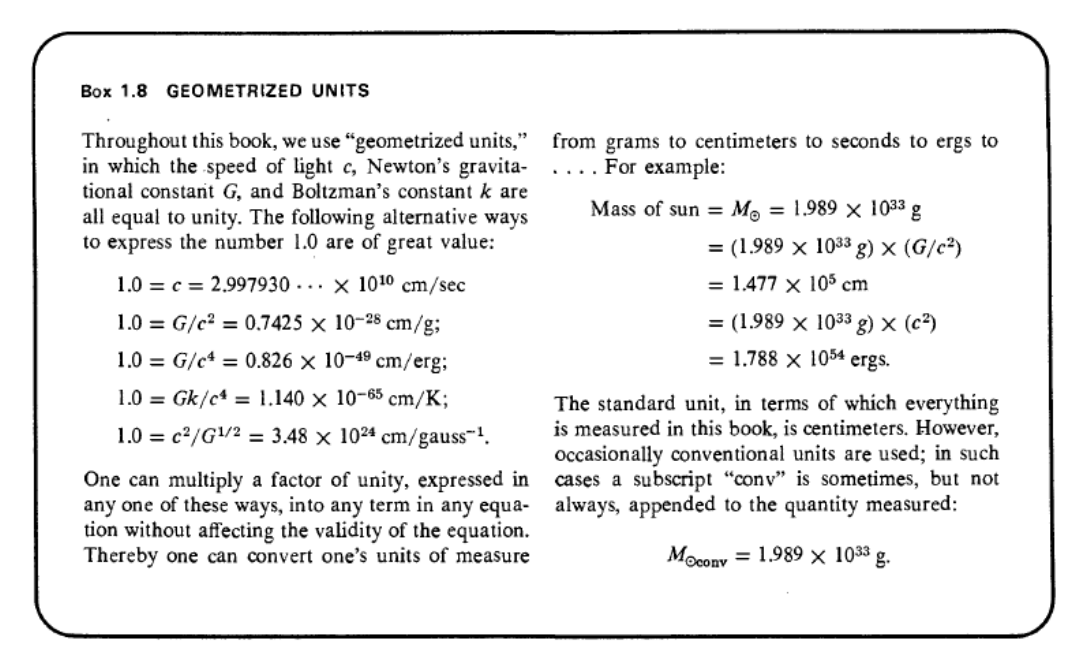}} 
Denoting by $M,L,T$ dimensions of mass, length and time and denoting by $[Q]$ the dimension of 
a quantity $Q$, we have $[G]=M^{-1}L^3/T^2$, $\ [c]=L/T$.  In units with $c=1$, $L=T$; in gravitational 
units $M$ also has the same dimension:  $G = 1$ implies $M=L^3/T^2 = L =T$. 

In particular, the gravitational potential $\Phi$ has conventional units of energy/mass because, by its definition, 
the gravitational potential energy of a small mass $m$ is $m\Phi$.  Then  
$[\Phi]_{\rm conventional} = L^2/T^2$.  With $c=1$, length and time have the same dimension, and $\Phi$ is dimensionless. To convert back to conventional units, one must multiply by the factors of $G$ and $c$ that 
restore the conventional dimension, so in this case 
$[\Phi]_{\rm conventional} = c^2 [\Phi]_{\rm gravitational}$.  The gravitational potential outside a 
spherical object of mass $M$ is $\Phi=-GM/R$ in conventional units and $\Phi= -M/R$ in gravitational units.
In conventional units, the dimensionless quantity equal to $M/R$ is $\dis \frac{GM}{Rc^2}$.\\

\benr\item  {\sl Converting conventional to gravitational units.}\\   
Consider a quantity $Q$ with conventional dimension $M^mL^\ell T^t$ and hence dimension 
$L^{m+\ell+t}$  in gravitational units.  Show that there is a unique product $G^i c^j$ for which 
$[G^i c^j] M^mL^\ell T^t =L^{m+\ell+t} $ and find $i$ and $j$ in terms of $m,\ell$ and $t$.  
(Using $M$ or $T$ as the single dimension for gravitational units, one could similarly find 
the associated unique powers of $G$ and $c$ that convert from conventional units
to these gravitational units.)
\een 

In gravitational units, the field equation is
\[
   G_{\a\b} = 8\pi T_{\a\b}.
\]
With a cosmological constant $\Lambda$, the equation is  
\be
  G_{\a\b} + \Lambda g_{\a\b} = 8\pi T_{\a\b}.
\label{e:Lambda}\ee   
Because $\Lambda$ {\sl is} a constant, the Bianchi identity again implies $\nabla_\b T^{\a\b}=0$.
\index{Einstein field equation!with cosmological constant}\index{field equation!with cosmological constant}\index{cosmology!cosmological constant}      
\newpage

\section{Newtonian Regime}
The way we arrived at $G_{\alpha\beta} = 8\pi GT_{\alpha\beta}$ does not imply 
that this differential equation for $g_{\alpha\beta}$ will really lead in the
Newtonian regime to a metric with $g_{tt} = -(1+2\Phi)$.  That it does is the final part of the  
miracle.  

\noindent{\sl Newtonian limit.}  In a system for which the Newtonian approximation is valid, there is 
a basis in which the system's constituents move at speeds $v$ slow compared to the speed of light.  
We define a small parameter $e$ to be of order $v$ (i.e., $e\sim v/c$, in conventional units).
If the system's gravity does not produce larger speeds, the escape velocity from parts of the system 
must be of order $e$, implying the potential energy $m\Phi$ of particles of mass $m$ must 
be of order $mv^2$, or $\Phi\sim e^2$.  
\index{energy!gravitational potential energy (Newtonian)} 
In gravitational units, the potential outside a spherical 
mass $M$ is $\Phi = M/r$, so $\dis \frac Mr\sim e^2$  (i.e. $GM/(rc^2) \sim e^2$). 
\index{Newtonian limit}

\begin{center}
\def\arraystretch{1.8}\arraycolsep=1.4pt
$\begin{array}{rl!{\quad\vline\quad}rl}
\MCA{gravitational units} & \MCB{conventional units}\\
v &\sim e\ll 1 & \dis \frac vc  &\sim e \ll 1\\ 
\Phi &\sim e^2 & \dis\frac{\Phi }{c^2} &\sim e^2 \\
\rho \sim n m_{\rm molecule},\ \ P \sim \frac{1}{2} &n m_{\rm molecule} v^2
\dis \Rightarrow  \frac{P}{\rho }\sim e^2 & \dis
\frac{P}{\rho c^2} &\dis \sim \frac{v^2}{c^2}  \sim e^2\\
\text{If}~$Q$ \mbox{ is a quantity that depends on the} &  \mbox{ matter, } & & \\
\dis \frac{\partial_t Q}{\partial_x Q} &\sim e 
& \ \  \dis \frac{\frac{1}{c} \partial_t Q}{\partial_x Q} &\sim \dis \frac{1}{c} \frac{\frac{Q}{t}}{\frac{Q}{x}} \sim
\frac{v}{c} \sim e \\
\end{array}$
\end{center}

\noindent The departure of $g_{\alpha\beta}$ from $\eta_{\alpha\beta}$ 
will be of $O(\Phi)$, that is $O(e^2)$, so
$\Gamma = O(e^2)$ and $\Gamma ^2=O(e^4)$.\\

\noindent The dominant contribution to $T^{\mu \nu }$ is $T^{tt} =
\rho$.  The field equation $R^{\alpha\beta} - \frac{1}{2} g^{\alpha\beta}R 
= 8\pi T^{\alpha\beta}$ is equivalent to  
\[
  R^{\alpha\beta} = 8\pi (T^{\alpha\beta} - \frac{1}{2}g^{\alpha\beta}T).
\] 
\index{Einstein field equation}\index{field equation}

We will see that the $R_{tt}$ equation is, to lowest order in $e$, 
an equation for $g_{tt}$.  First evaluate the leading contribution to 
the right side, using 
\[
P\sim \rho e^2 \Rightarrow T= \rho u^\alpha u_\alpha (1+O(e^2))
	= -\rho\ (1+O(e^2)).
\]    
\begin{eqnarray}R_{tt} &=& 8\pi (T_{tt}-\frac{1}{2}g_{tt}T)\nonumber \\
&=& 8\pi (\rho - \frac{1}{2}(-1)(-\rho))\nonumber \\ 
&=& 4\pi \rho.  \end{eqnarray}

Next evaluate the left side, \( R_{tt} = R_{it}{}^i{}_t\):
\begin{eqnarray}R_{itj}{}^t 
&=& \partial _t\Gamma^t{}_{ij} - \partial_i\Gamma^t{}_{jt} + O(\Gamma ^2) 
								\nonumber \\
&=& \partial_i\Gamma_{tjt} + O(e^3) \hspace{.5in} (\partial _t \Gamma
\sim v\partial _x\Gamma \sim e^3) \nonumber \\
&=& \partial _i[\frac{1}{2}(\partial_j g_{tt} + \partial_tg_{jt} -
\partial _tg_{tj})] \nonumber \\
&=& \frac{1}{2}\partial_i \partial_j g_{tt} + O(e^3). 
\label{ritjt}\end{eqnarray}
Thus
\be 
	R_{tt} =- \frac{1}{2} \nabla^2 g_{tt},
\label{rtt}\ee
and the field equation becomes
\be
  \nabla^2 g_{tt}  = -8\pi \rho.
\ee
\index{Einstein field equation!Newtonian limit}\index{field equation!Newtonian limit}
\index{Newtonian limit!field equation}

Then $\dis g_{tt}\xrightarrow{r\rightarrow\infty} -1$ implies the 
final requirement for the theory to agree with special relativity and 
with Newtonian gravity in the Newtonian limit,  
\[
   g_{tt} = -1 - 2\Phi .
\]

Our derivation implies that, with this $g_{tt}$, the equation
of geodesic deviation has the right Newtonian form:  That was our starting point.
But here's a direct check:

\begin{eqnarray}
\ddot S^i &=& R^i{}_{ttj}\, S^j \hspace{.75in} 
\mbox{using $u^\mu = \delta^\mu_t + O(e)$}\nonumber \\
&=& \frac{1}{2} \partial ^i\partial_j g_{tt}\ S^j \hspace{.5in}\text{from\;\;
(\ref{ritjt})} \nonumber \\
&=& -\partial^i\partial_j \Phi\ S^j.
\end{eqnarray}

\section{Pound-Rebka Experiment}
\index{Pound-Rebka experiment}\index{energy!conservation in time-independent field}
\index{conservation laws!energy conservation in time-independent field}
 A discussion of experimental verification of the theory beyond that
discussed here will be deferred to discussions of precession in Sect. \ref{s:geodesics}, 
of frame dragging in Sect. \ref{s:kerr}, and of gravitational waves from binary inspiral 
in Sect. \ref{s:inspiral}.  Meanwhile, we can get cheaply
the prediction that light traveling in a time-independent gravitational
field has different frequencies as seen by observers ``at rest" (moving
along the  Killing vector) at different spatial positions.  This is essentially 
equivalent to our calculation of the time between two photons sent from $A$ to $B$, 
with that time interval replaced by the time between two crests.  In the new version, the result  
will come in the guise of a conservation law-- energy conservation -- associated with 
time-translation invariance.  \\

Let $t^\alpha$ be the timelike Killing vector.
Observers A and B are ``at rest" if they move in spacetime along the timelike Killing vector.  Their 
velocities are then 
\be 
	u^\alpha_A = Ct^\alpha \hspace{1in} u^\alpha_B = C't^\alpha, 
\ee
and 
$u^\alpha u_\alpha  = -1 \Rightarrow $ \hspace{.3in} $C = (-g_{\alpha\beta}t^\alpha t^\beta)^{-1/2}$.
Then 
\be
  u^\alpha_A = (-g_{\beta\gamma}t^\beta t^\gamma)^{-1/2}(A)\ t^\alpha \qquad 
  u^\alpha_B = (-g_{\beta\gamma}t^\beta t^\gamma)^{-1/2}(B)\ t^\alpha.
\label{e:ut}\ee

\noindent A light ray sent from A to B has frequency
\beaa \omega _A = -k_\alpha u_A^\alpha  \ \ &&\mbox{ seen by $A$, and }\\
      \omega _B = -k_\alpha u_B^\a  \ \ &&\mbox{ seen by $B$}. 
\eeaa
$k^\alpha$, the photon's wavevector, satisfies the geodesic equation
\[ 
	k^\beta\nabla_\beta k^\alpha = 0 .
\]
Because $t^\alpha$ is a Killing vector, it satisfies Eq.~\eqref{e:lieg}, 
$\nabla_{(\alpha}t_{\beta)} = 0$, and, as in Eq.~\eqref{e:pxicons}, we have
\bea 
 (k^\beta\nabla_\beta k^\alpha)t_\alpha  
  &=& k^\beta\nabla_\beta (k^\alpha t_\alpha)
   -\underbrace{k^\beta k^\alpha\nabla_\beta t_\alpha }_0 \nonumber\\
  &=& k^\beta\nabla_\beta (k^\alpha t_\alpha ) 
\eea
Thus $k^\beta\nabla_\beta (k_\alpha t^\alpha ) = 0$ and \crv $k_\alpha t^\alpha$ is
conserved along the trajectory\cb.  At $\infty $, $t_\alpha $ is the velocity 
of an inertial observer $(t^\alpha t_\alpha  = -1)$, so we write 
\index{conservation laws!relation to Killing vector}
\be
	-k_\alpha t^\alpha  = \omega_\infty,
\label{e:k_t}\ee
the frequency of the photon seen by an observer at $\infty $.  Because $k_\a t^\a$ is 
constant, the frequency seen 
by a stationary observer at a point $A$ is, by Eq.~\eqref{e:ut}, 
\be
   \crv \omega_A = (-t_\b t^\b)^{-1/2}(A)\, \omega_\infty \cb
	    = |g_{tt}(A)|^{-1/2}\, \omega_\infty .
\label{e:redshift}\ee
\index{redshift!gravitational}
Then 
\be 
  \omega _A-\omega _B 
	= \omega _{\infty}[\,|g_{tt}(A)|^{-1/2}-|g_{tt}(B)|^{-1/2}\,]. 
\ee

Nearly Newtonian:  $g_{\alpha\beta}t^\alpha t^\beta = g_{tt} = -1 -2\Phi $
\be
\mid g_{tt}\mid^{-1/2}(A) - \mid g_{tt}\mid^{-1/2}(B) = \Phi_B - \Phi_A
		= \mid \Phi _A \mid - \mid \Phi _B \mid 
\ee
\begin{eqnarray} \frac{\Delta \omega }{\omega } &=& \Delta \Phi \\
\mbox{or} \hspace{1in} \frac{\Delta \omega }{\omega } &=& -g h\;\;\;
\mbox{ on Earth.} \quad \mbox{Reinstating $c$, we have } 
\frac{\Delta \omega }{\omega } = - \frac{gh}{c^2}.\nonumber  
\end{eqnarray} 
\noindent A photon sent from $A$ up to $B$ a height $h$ above $A$ loses
energy in climbing up and is redshifted by $|\Delta \omega |/\omega = g h $.  
Because a photon's energy is $\hbar\omega$, conservation of $\hbar k_\a t^\a$
is called energy conservation, with\index{conservation laws!relation to Killing vector}\index{conservation laws!energy}
\[
\mbox{${\cal E}:= -\hbar k_\a t^\a = \hbar\omega_\infty$} 
\]
the conserved energy. 
Conservation of $\cal E$ gives the energy
$-\hbar k_\a u^\a$ measured by a local observer with velocity $u^\a$.

The redshift was first experimentally measured by Pound and Rebka \url{https://journals.aps.org/prl/abstract/10.1103/PhysRevLett.3.439}\cite{pr59}.  To repeat: 
The experiment really tests only the equivalence principle.  Two crests sent a
proper time $\Delta \tau _A $ apart at $A$ are received with proper time interval 
$\dis\Delta \tau _B,$ where $\dis\frac{\Delta \tau _B-\Delta \tau _A}{\Delta \tau} =
 g h$.\\
\index{equivalence principle!Pound-Rebka experiment}
 \index{general relativity|)}\index{relativity!general relativity|)}


\chapter{Spherical Relativistic Stars}

\section{Spherically symmetric spacetimes}
\label{s:spherical_symmetry}
\index{spherically symmetric spacetimes}\index{spacetime!spherically symmetric}

	A spacetime is spherically symmetric if the rotation group acts on it as a
group of smooth maps from $M$ to itself that preserve the metric:  for each
rotation $R$ in $O_3$ there is a diffeo 
$\psi(R)$:  $M \rightarrow M$
that leaves the metric invariant,
\be \psi(R) g_{\alpha\beta} = g_{\alpha\beta}, \ee
such that $R \rightarrow \psi(R)$ is an action of $O_3$:
\be \psi(\rm id) = id \ee
\be \psi(R_1R_2) = \psi(R_1)\psi(R_2), \ee
and such that almost every point of $M$ sweeps out a sphere under the action
of $O_3$ on ${\mathbb R}^3$.  This last requirement eliminates the trivial action
$\psi (R) = $id, all $R$. Picture a symmetric spacelike surface of $M$
this way:

\begin{figure}[h!]
\includegraphics[width=.5\textwidth]{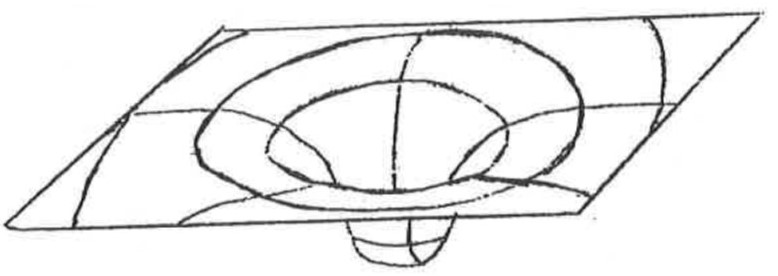}
\end{figure}
\noindent The circles represent
symmetry spheres of $M$, orbits of points of $M$ under the action of
$O_3$. 

One acquires coordinates $\theta $ and $\phi $ on a
symmetric sphere $S$ in $M$ this way:  Denote by ${\cal R}(\psi{\bf n})$ 
the rotation of $\mathbb R^3$ by angle $\psi$ about the axis $\bf n$. 
Acting on a symmetry sphere of $M$, the rotation ${\cal R}(\psi \bf k)$ 
leaves two points fixed.  
One can call one of the fixed points $\theta = 0$ and give
coordinates $(\theta,0)$ to the image of $\theta = 0$ under 
$R(\theta{\bf j})$, just as one would do if the sphere were in 
flat space.  Finally, the image of $(\theta,0)$ under 
$R(\phi{\bf k})$ has coordinates $(\theta,\phi)$.  

Let $t$ and $r$ be coordinates that label the symmetry spheres -
i.e., $t$ and $r$ are scalars constant on each sphere.  Then under a
rotation, $\theta $ and $\phi $ change as they would in ${\mathbb R}^3$
\be \theta \rightarrow \theta '(\theta , \phi ) \hspace{.5in} \phi
\rightarrow \phi '(\theta , \phi ) \ee
or\hspace{2in}$\theta = \theta (\theta ', \phi ')$ \hspace{.5in} $\phi =
\phi(\theta ', \phi ') $,\\
 
\noindent but $t$ and $r$ are constant \hspace{.7in} $t' = t$,
\hspace{.5in} $r' = r$.

\noindent Thus
\begin{eqnarray} g_{t't'} &=& 
\frac{\partial x^\mu}{\partial t'}\frac{\partial x^\nu}{\partial t'} g_{\mu \nu}
= \delta^\mu_t \delta^\nu_t g_{\mu\nu}= g_{tt}, \nonumber\\
g_{t'r'} &=& g_{tr} \nonumber\\
g_{r'r'} &=& g_{rr}. 
\end{eqnarray}
	In other words, $g_{tt}, g_{tr}$ and $g_{rr}$ are rotational scalars,
while $g_{t\theta}$ and  $g_{t\phi}$ transform as the $\theta $ and $\phi
$ components of a vector in ${\mathbb R}^3$ under rotations, and ($g_{r\theta },
g_{r\phi })$ are again components of a rotational vector. Finally, the metric components
\noindent
\[\left| \left|  \begin{array}{ll}
g_{\theta \theta} & g_{\theta \phi }\\
g_{\phi \theta } & g_{\theta \theta }
\end{array}\right| \right|\]\\
transform under rotation as the $\theta \theta $, $\theta \phi $, and $\phi
\phi $ components of a symmetric tensor on ${\mathbb R}^3$.  

Because the only scalars on
${\mathbb R}^3$ invariant under rotations are functions independent of $\theta $ and
$\phi $, we have
\be
	g_{tr} = g_{tr}(t,r)\qquad 
	g_{tt} = g_{tt}(t,r)\qquad
	g_{rr} = g_{rr}(t,r).
\ee
Because the only vector field on ${\mathbb R}^3$ independent of rotations is a 
radial vector field (whose $r$-component is independent of $\theta $ and $\phi $ )
only the r-components of $g_{ti}$ and $g_{ri}$ are nonzero: \\
\be 
	g_{t\theta } = g_{t\phi } = 0,\ \ \mbox{ and }\ \  g_{r\theta } =
g_{r\phi } = 0.  
\ee
Finally any symmetric tensor 
$\left| \left|  \begin{array}{ll}
g_{\theta \theta} & g_{\theta \phi }\\
g_{\phi \theta } & g_{\phi \phi }
\end{array}\right| \right|$ invariant under rotations is proportional to 
the metric on the 2-sphere:
\be
 f \left| \left| \begin{array}{ll}
1 & \\
& \sin^2\theta \end{array}\right| \right|,
\ee
where $f$ is independent of
$\theta $ and $\phi $. %
\footnote{Here's a proof: Let $f_{ab}$ be a symmetric tensor on the 2-sphere.  
Because it is symmetric, 
it has two distinct eigenvectors, where an eigenvector $v^a$ satisfies $f_{ab} v^b = \lambda v_b$. (Equivalently, there is an orthonormal basis in which $f_{ij}$ is diagonal.) The two eigenvalues must be the same - otherwise we could distinguish a direction, an eigenvector with smallest eigenvalue, violating spherical symmetry. But any vector on the sphere is a linear combination of the two eigenvectors, implying $f_{ab} v^b = \lambda v_b$, for all vectors $v^b$.  Thus $f_{ab} = \lambda\ ^2\!g_{ab}$, with $^2\!g_{ab}$ the metric on the 2-sphere.}   

Thus\\ 
\centerline{
$\dis \left| \left|g_{\mu\nu }\right| \right|$ = 
$\left| \left|  \begin{array}{llll}
g_{tt}(t,r) & g_{tr}(t,r)  \\
g_{tr}(t,r) & g_{rr}(t,r)  \\
&&f(t,r)\\ 
&&&f(t,r)\sin^2\theta \end{array}\right| \right|$.}\\

Now pick a new radial coordinate $r' = \sqrt{f}$ .  Calling the new
radial coordinate $r$ again, we have for $\left| \left| g_{\mu \nu }\right|
\right|$ the form 

\[\left| \left|  \begin{array}{llll}
g_{tt} & g_{tr}  \\
g_{tr} & g_{rr}  \\
&& r^2 \\ 
&&&r^2\sin^2 \theta \end{array}\right| \right|\].\\

\noindent (Here $g_{tt}, g_{tr}$, and $g_{rr}$ are the new components 
that result from transformation to the new radial coordinate.) With this new radial
coordinate, $4\pi r^2$ is the area of a symmetry sphere.  

Next, by a coordinate transformation $t \rightarrow T(r,t)$ we can
make $g^{Tr} = 0$ \hspace{1mm} (and so $g_{Tr} = 0$).  This can 
be seen as follows.  We seek a function $T$ for which

\[ 0 = g^{Tr} = \frac{\partial T}{\partial t} g^{tr} + \frac{\partial
T}{\partial r} g^{rr}.  \]
Equivalently, 
\[
 {\bm\xi} \cdot \nabla T = 0,\;\;\mbox{where } (\xi^\mu) = (g^{tr}, g^{rr},0,0).
\]
\noindent Let $c(\lambda )$ be the integral curve of the vector field
${\bm\xi}$ through a point $(t_0,r_0,\theta_0,\phi_0)$.  If we 
set $T=$ constant along $c(\lambda)$, then ${\bm\xi} \cdot \nabla T = 0$
along $c(\lambda)$.  Extend 
this $T=$ constant curve to the 3-surface of revolution consisting of 
the rotated images of $c(\lambda)$ that pass through the points 
$(t_0,r_0,\theta,\phi)$, all values of $\theta,\phi$. 
Then ${\bm\xi} \cdot \nabla T = 0$  on this surface, implying  
$g^{Tr} = 0$, as claimed.

 Hence, changing the name ``$T$'' back to ``$t$'', $\|g_{\mu\nu }\|$ can always be
cast in the form 
\[\left| \left|g_{\mu\nu }\right| \right| =
\left| \left|  \begin{array}{llll}
g_{tt} &  \\
& g_{rr}  \\
&&r^2\\ 
&&&r^2\sin^2 \theta \end{array}\right| \right|\].\\
We will write $g_{tt} \equiv -e^{2\Phi},\ \ g_{rr} \equiv e^{2\lambda},$ and
$d\Omega^2 \equiv d\theta^2 + \sin^2\theta d\phi ^2:$

\be 
	ds^2 = -e^{2\Phi }dt^2 + e^{2\lambda } dr^2 + r^2d\Omega ^2 .
\label{e:gss}
\ee
For a collapsing star, $\Phi$ and $\lambda$ depend on time, but we will 
be concerned with static equilibria until we look at a collapsing ball of 
dust and the expanding universe in Chap.~\ref{c:cosmology}. \\ 

\noindent
{\em Static, spherically symmetric spacetimes}\\

\indent A static spacetime has a timelike Killing vector $t^\alpha$ that is 
orthogonal to a family of spacelike hypersurfaces.  There is then a natural time 
coordinate $t$ that can be defined as follows.  Let $S$ be one of the hypersurfaces orthogonal 
to $t^\a$.  Set $t=0$ on $S$, and define $t$ elsewhere by $t^\a \na_\a t = 1$.  
Then $t$ is the parameter distance from $t = 0$ along the integral curves of $t^\alpha$.  
For example, the $t = 1$ surface is the result of dragging the $t = 0$ surface a parameter 
length $1$ along the Killing vector.  Any two constant $t$ surfaces $S_t$  
are identical-- they have the same geometry.  Because $t^\alpha$ is perpendicular to $S$, it is perpendicular to $S_t$.  If we extend $t$ to a chart $t,\{x^i\}$, then orthogonality is 
equivalent to the vanishing of the time-space components of the metric:
\[ 
	g_{ti} = 0,
\] 
in a chart with $g_{\mu\nu}$ independent of $t$.

Note that a rotating star has a timelike Killing vector $t^\a$, but $t^\a$ is not 
hypersurface orthogonal.  In particular, natural spacelike hypersurfaces are axisymmetric, 
with the rotational Killing vector $\phi^\alpha$ tangent to each hypersurface; but 
$t^\alpha$ is not perpendicular to $\phi^\a$: 
$g_{t\phi} = t^\alpha \phi_\alpha \neq 0$.
\index{hypersurface!hypersurface orthogonal}
  
	Our next objective is to write and solve the Einstein field equation for a
static spacetime with the metric (\ref{e:gss}).  First we need to compute $R_{\mu
\nu }$.  We'll sketch the derivation in a coordinate basis here, leaving 
as an exercise the somewhat quicker computation using Cartan calculus of the 
components along an orthonormal basis (see Appendix \ref{s:cartan}).

	We first find the nonzero components $\Gamma^\mu{}_{\nu \lambda }$, 
reading off $\Gamma^\mu $'s from the Euler-Lagrange equation  
\[
\frac{d}{ds} \left(\frac{\partial L}{\partial \dot {x}^\mu}\right)-\frac{\partial L}{\partial
x^\mu} = 0, 
\]
with $L = \frac{1}{2} g_{\alpha\beta} \dot{x}^\alpha \dot{x}^\beta. $

For example, writing $\Phi':= \partial_r\Phi$, we have
\begin{eqnarray*}
	\frac{\partial L}{\partial \dot{t}}= -e^{2\Phi}\dot{t} \ \qquad\
	\frac{d}{ds} \frac{\partial L}{\partial \dot{t}} &=&
	-e^{2\Phi}\ddot{t} - 2e^{2\Phi}\Phi '\dot t\dot r \ \qquad\ 
			\frac{\partial L}{\partial t} = 0\\
\\
	\ddot{t} + 2\Phi'~\dot t\dot r &=& 0.
\end{eqnarray*}
Then \\
\indent$\dis\ \ 
	\Gamma^t_{tr} =	\Phi' .
$
\hspace{10mm} Here are the remaining nonzero $\Gamma$'s:
\begin{align}
 \Gamma^r_{\ tt}&=\Phi'e^{2\Phi-2\lambda} \quad  &\Gamma^r_{\ rr}&=\lambda' 
\quad  &\Gamma^r_{\ \theta\theta}=-re^{-2\lambda}\quad 
 &\Gamma^r_{\ \phi\phi}=-r\sin^2\theta e^{-2\lambda}\nonumber\\
 \Gamma^\theta_{\ r\theta}&=\frac1r\quad 
 &\Gamma^\theta_{\ \phi\phi}&=-\sin\theta\cos\theta  \nonumber\\
 \Gamma^\phi_{r\phi} &= \frac1r  &\Gamma^\phi_{\theta\phi} &= \cot\theta.
\label{e:schgamma}\end{align}
	Next, the Ricci tensor components $R_{\mu \nu }$ are computed in
terms of the $\Gamma$'s using the simplification
\be
\Gamma^\sigma{}_{\mu\sigma} = \frac12 g^{\sigma\tau}\pa_\mu g_{\sigma\tau} 
	=  \partial_\mu \log\sqrt{-g}, 
\label{e:Gammajij}\ee 
This relation is implied by \ref{p:div}, but here's an alternative derivation 
that starts by showing that the derivative of the determinant $g$ has the form 
\be
 \frac{\pa g}{\pa g_{\mu\nu}} = g g^{\mu\nu}. 
\label{e:dgdgij}\ee
To obtain Eq.~\eqref{e:dgdgij}, write the determinant as the sum 
of the matrix elements $g_{\mu\lambda}$ in the $\mu^{\rm th}$ row times their cofactors $\Delta^{\mu\lambda}$:
\[
   g = \sum_\lambda g_{\mu\lambda}\Delta^{\mu\lambda}\   \mbox{( no sum over $\mu$)}.
\]  
Because none of these cofactors involve $g_{\mu\nu}$, we have 
\[
  \frac{\pa g}{\pa g_{\mu\nu}}  = \Delta^{\mu\nu}.
\] 
Finally, the inverse of a matrix is given in terms of its cofactors by 
\[
  g^{\mu\nu} = \frac{\Delta^{\nu\mu}}g, 
\]
whence, for the symmetric matrix $g^{\mu\nu}$,  $\Delta^{\nu\mu} = gg^{\mu\nu}\qquad\Box$.\\
Using Eq.~\eqref{e:dgdgij}, we have
\[
  \pa_\mu \log\sqrt{-g} = \frac12 g^{-1}\pa_\mu g 
  = \frac12 g^{-1}\frac{\pa g}{\pa g_{\sigma\tau}}\pa_\mu g_{\sigma\tau} 
  = \frac12 g^{\sigma\tau} \pa_\mu g_{\sigma\tau} 
  = \Gamma^\sigma{}_{\mu\sigma}.  
\]

From Eq.~\eqref{e:rijklup},
\bsube\begin{align} 
  {\cblue R_{\mu \nu }} &= \partial_\lambda \Gamma^\lambda{}_{\mu \nu }
			- \partial_\mu \Gamma^\lambda{}_{\lambda \nu } 
			+ \Gamma^\lambda{}_{\mu \nu }\Gamma^\sigma {}_{\lambda\sigma} 
			- \Gamma^\lambda{}_{\sigma \nu }\Gamma^\sigma {}_{\lambda \mu}\\
               &\cblue = \partial_\lambda \Gamma ^\lambda {}_{\mu \nu }
			- \partial_\mu \partial _\nu \log \sqrt{-g} 
			+ \Gamma ^\lambda{}_{\mu \nu }\partial _\lambda \log \sqrt{-g} 
			- \Gamma^\lambda{}_{\sigma \nu }\Gamma^\sigma {}_{\lambda \mu} .
\label{e:Ricci_compute}\end{align}\esube
\benr\item  Find $G_{\mu\nu}$ for the static,
spherically symmetric metric (\ref{e:gss}); start by computing the  
components $\Gamma^\lambda{}_{\mu\nu}$ of Eq.~\eqref{e:schgamma} with $\lambda=r,\theta,\phi$.
\een
The computation gives
\bsube\begin{eqnarray}
G^t{}_t &=& e^{-2\lambda}\left(\frac{1}{r^2}- \frac{2}{r}\lambda'\right) - \frac{1}{r^2}\\
	&=& -\frac{1}{r^2} \frac{d}{dr} \left[r(1-e^{-2\lambda })\right]\nonumber
\label{e:Gtt}\\
G^r{}_r &=& e^{-2\lambda}\left(\frac{1}{r^2}+ \frac{2}{r}\Phi '\right) - \frac{1}{r^2}
\label{e:Grr}\\
G^\theta{}_\theta 
	&=& G^\phi{}_\phi 
	 = e^{-2\lambda} \left[\Phi''+(\Phi')^2+\frac{1}{r}(\Phi'-\lambda') 
	 		 - \Phi'\lambda'\right] \,.
\label{e:Gthetatheta}\end{eqnarray}\esube
All other $G_{\mu \nu }$'s except $G_{tr}$ vanish (by the argument showing that off-diagonal $g_{\mu \nu }$'s are zero).  For a static spacetime, symmetry  under time reversal implies $G_{tr}=0$:   \\
For the coordinate transformation $t'=-t$, 
\begin{align*}
  G_{t'r}(t',r,\theta,\phi) &= \frac{\pa t}{\pa t'} G_{tr}(t,r,\theta,\phi)
  			   = -G_{tr}(t,r,\theta,\phi) \\
 \mbox{Independent of $t$ } \Longrightarrow G_{t'r}(r,\theta,\phi) &= -G_{tr}(r,\theta,\phi).
\end{align*}
Invariant under the transformation $\Longrightarrow$ 
\[
 G_{t'r}(r,\theta,\phi) = G_{tr}(r,\theta,\phi)\ \Longrightarrow 
 G_{tr} =0.
\]
For a time-dependent, spherically symmetric spacetime, one can choose coordinates $t$ and $r$ to make $G_{tr}=0$.

\subsection{Schwarzschild solution}
\index{black hole!Schwarzschild spacetime|textbf}\index{Schwarzschild spacetime}

The Schwarzschild solution is the geometry outside a spherically symmetric star and 
the geometry of a black hole: 
It is the geometry of any asymptotically flat spherically symmetric vacuum spacetime.
\index{asymptotically flat spacetime}  

In a vacuum, the field equations are
\[ G^\mu_{\ \nu} = 0. \]
The $G^t_t$-equation \eqref{e:Gtt},
\[
	G^t{}_t = -\frac{1}{r^2} \frac{d}{dr}[r(1-e^{-2\lambda})] = 0, 
\]
has the first integral
\be
r(1-e^{-2\lambda}) = 2M, \mbox{ for some constant } M, 
\ee
implying
\be
	e^{2\lambda} = \left(1 - \frac{2M}{r}\right)^{-1}.
\ee
From the combination,
\be
	 G^r{}_r - G^t{}_t = \frac{2}{r}~e^{-2\lambda}(\Phi'+\lambda') = 0,
\ee
we have
\be
	\Phi' = - \lambda '
\ee
or
\[ 
 \Phi = -\lambda + k \ \qquad \ e^{2\Phi} = k\left(1 - \frac{2M}{r}\right).	
\]
\indent Reparametrizing the time by writing $\tilde t = \frac{1}{\sqrt{k}}t$, and changing
the name of $\tilde t$ back to $t$, gives \index{Schwarzschild spacetime!Schwarzschild coordinates}\index{coordinates!Schwarzschild coordinates}
\be\crv
	ds^2 = -\left(1 - \frac{2M}{r}\right)dt^2 
		+ \left(1 - \frac{2M}{r}\right)^{-1}dr^2  + r^2d\Omega ^2\cb,
\label{e:schwarzschild}
\ee
the {\sl exterior Schwarzschild metric}. The geometry is asymptotically flat: 
\index{asymptotically flat spacetime}
For large $r$, the metric is the Minkowski metric,
\be 
   ds^2 = -dt^2 + dr^2 + r^2d\Omega^2. 
\ee
For $r = 2M$ the components $g_{\mu \nu }$ are singular, so the
form (\ref{e:schwarzschild}) provides a metric for a spacetime with a hole in it:
$\dis\infty > r > 2M, \ \ -\infty < t< \infty$.  When we discuss black holes, 
we'll see that this is a coordinate singularity, like the poles in spherical 
coordinates.  Changing to coordinates that are smooth at $r=2M$ reveals what 
you know: The surface $r=2M$ is an event horizon.  
For large $r$, the metric (\ref{e:schwarzschild}) takes the post-Newtonian form
\be 
	ds^2 = -\left(1 - \frac{2M}{r}\right)dt^2 
		+ \left(1 + \frac{2M}{r}\right)dr^2 + r^2d\Omega^2. 
\ee
For nearly Newtonian stars, $g_{tt}$ determines the effect of the gravitational field 
on matter, and $-\dis\frac{M}{r}$ is the Newtonian potential $\Phi $.
Because the trajectories of particles at large $r$ are those of Newtonian 
particles about a mass $M$, one calls $M$ the mass of the spacetime.

	The Bianchi identity \ref{cbianchi} implies that the remaining equations, $G^\theta{}_\theta = 0,\ \ G^\phi{}_\phi = 0$, are 
automatically satisfied once $G^t{}_t = 0$ and $G^r{}_r = 0$ 
(not in general, just in this spherically symmetric case).\index{Bianchi identities!and independent field equations}\\

There is a different choice r of radial coordinate for which the exterior Schwarzschild 
metric has the form\index{black hole!isotropic coordinates}\index{metric!Schwarzschild in isotropic coordinates}\index{Schwarzschild spacetime!isotropic coordinates}\index{coordinates!isotropic coordinates}
\be\cblue
  ds^2 = -\frac{\left(1-\frac M{2\textrm r}\right)^2}{\left(1+\frac M{2\textrm r}\right)^2}dt^2 
	+ \left(1+\frac M{2\textrm r}\right)^4 (d\textrm r^2 + \textrm r^2 d\Omega^2).
\label{e:giso}\ee
The coordinates $t,\mbox{r},\theta,\phi$ (with $t,\theta,\phi$ unchanged) are called {\sl isotropic} coordinates.   The next problem asks you to find the coordinate transformation that gives this
form.   A metric of the form $g_{ab} = F^2\ f_{ab}$, with $f_{ab}$ flat and
$F\neq 0$ is said to be conformally
flat, and $F^2$ is called the conformal factor.\index{conformal factor}  That the spatial part of the Schwarzschild metric
can be written in isotropic form means that it is conformally flat,\index{conformal flatness} here with conformal
factor $\dis\left(1+\frac M{2\textrm r}\right)^4$.\\

\newpage

\benr\item (This is problem 1 of Wald, Chapter 6).
\benalph\item Show that any spherically symmetric Riemannian 3-metric can be written in the
form
\[
  ds^2 = [F(\bar r)]^2(d\bar r^2 + \bar r^2 d\Omega^2).
\]
From our earlier discussion, we already know that a metric of this kind can be written as
the spatial part of \eqref{e:gss}, namely $\dis ds^2 = e^{2\lambda} dr^2 + r^2 d\Omega^2$.
So the problem is to find functions $\bar r(r)$ and $F(r)$ for which
$\dis [F(\bar r)]^2(d\bar r^2 + \bar r^2 d\Omega^2) = e^{2\lambda(r)} dr^2 + r^2 d\Omega^2$.
\item  Specialize to the exterior Schwarzschild solution to recover Eq.~\eqref{e:giso}
with $\textrm r\equiv \bar r$.

\een

\een

\noindent{\sl Gravitational mass $M$} \index{mass!gravitational mass}
\index{gravitational mass}

The mass $M$ that appears in the Schwarzschild metric is called the 
{\sl gravitational mass} of the spacetime.  The terminology is natural, 
because a particle in orbit at large $r$ follows a Keplerian orbit 
of the Newtonian gravitational field of a spherical mass $M$. 
Checking this claim involves only the asymptotic form of the 
Schwarzschild metric, and that $O(1/r)$ form is the same for 
{\sl any stationary source}.  To describe the asymptotic 
form more carefully, let $\{x^i\}$ be the standard Cartesian 
coordinates associated with 
the isotropic spherical coordinates of Eq.~\eqref{e:giso}. 
The Schwarzschild metric then has the asymptotic form 
\be 
  ds^2 = -\left(1-\frac{2M}r\right) dt^2 
	+ \left(1+\frac{2M}r\right) \delta_{jk}dx^j dx^k +O(r^{-2}),   
\ee
and the metric of any stationary, asymptotically flat gravitational field has 
this form in some chart.
    
The check of Keplerian motion is quick: 
For a particle at large $r$ with speed $v\ll 1$ relative 
to a stationary observer, the derivation of Eq.~\eqref{e:geod_Newtonian} holds, 
giving 
\[
   \frac{dv^i}{dt} = -\pa_i\Phi +O(r^{-3})= -\frac{M}{r^2}\wh r^i +O(r^{-3}).   
\]
(The earlier derivation required $\pa_t g_{\mu\nu}$ to be higher order in $1/c$, 
and here $\pa_t g_{\mu\nu} = 0$.)  \\



\noindent
{\sl Stellar Interior} 

 Equilibrium configurations of stars are accurately modeled as 
perfect fluids.  
For static, spherical stars, $\rho$ and $P$ depend 
only on $r$, while the velocity $u^\alpha$ is along the Killing vector 
$t^\alpha:$
\[ u^\alpha = kt^\alpha .\]
But $t ^\alpha t_\alpha  = g_{tt} = -e^{+2\Phi}$, so $u^\alpha u_\alpha  = -1$ implies
\[  u^\alpha = e^{-\Phi }t^\alpha. \]
Then, from Eq.~\eqref{e:Tfluid},
\[
 T^{\alpha\beta} =  \rho u^\alpha u^\beta + Pq^{\alpha\beta},
\]
we have 
\[
	 T^t{}_t = -\rho  \qquad  T^r{}_r = P .
\]

The field equation components are\\
$\dis 
G^t{}_t = 8\pi T^t{}_t 
$:
\begin{align}
 \frac{1}{r^2}~\frac{d}{dr}[r(1-e^{-2\lambda })]
    			&= 8\pi \rho 		\nonumber \\
    r(1-e^{-2\lambda }) &= 2\int^r _0 4\pi \rho r^2dr=:2m(r) \label{e:m}\\
							\nonumber \\
  \crv	   e^{2\lambda} &\crv= \left(1 - \frac{2m(r)}{r}\right)^{-1}. 
\label{lambda}\end{align}
Here $m(r)$ is a kind of mass within a radius $r$. For $r\geq R$, $\ m(r) = M$, the mass measured at infinity. \\
$G^r{}_r = 8\pi T^r{}_r:$
\begin{align}
e^{-2\lambda }\left(\frac{2\Phi'}{r} + \frac{1}{r^2}\right) - \frac{1}{r^2} &= 8\pi P
\nonumber \\
(1 - \frac{2m}{r})(\frac{2\Phi'}{r} + \frac{1}{r^2}) 
		&= 8\pi P + \frac{1}{r^2} = \frac{8\pi Pr^2+1}{r^2} \nonumber \\
\frac{2\Phi'}{r} + \frac{1}{r^2} &= \frac{8\pi Pr^2+1}{r(r-2m)} \nonumber \\
2\Phi' &= \frac{8\pi Pr^2+1}{r-2m} - \frac{1}{r} 
	= \frac{8\pi Pr^3+2m}{r(r-2m)} \nonumber \\
\crv  \Phi '& \crv= \frac{m+4\pi Pr^3}{r(r-2m)} 
\label{phip}\end{align}

Note that the Newtonian limit $(P \ll \rho ,\ \  m \ll r)$ of (\ref{phip}) is
\[  \Phi ' = \frac{m}{r^2}~,\]
so that $\Phi$ is again the Newtonian potential.

The remaining field equation components, $G^\theta {}_\theta = 8\pi
T^\theta {}_\theta , \ G^\phi {}_\phi =  8\pi T^\phi {}_\phi$,  are
identical to one another and are implied by (\ref{lambda}),
(\ref{phip}) and the equation of hydrostatic equilibrium 
$q^\alpha{}_\gamma\nabla_\beta T^{\gamma\beta}$, which we now obtain:
Recall that  $q^\alpha{}_\gamma\nabla_\beta T^{\gamma\beta} = 0$ has the form
\be 
 u^\beta\nabla_\beta u_\alpha  = - \frac{1}{\rho+P}~q_\alpha^\beta\nabla_\beta P .
\ee
On the RHS, $\ \dis q_\alpha ^\beta\nabla_\beta P = (\delta _\alpha {}^\beta+u_\alpha u^\beta)\nabla_\beta P$.  The second term vanishes because $\dis \partial _t P = 0$.
Then
\[  
q_\alpha{}^\beta\nabla_\beta P = \nabla _\alpha P.
\]

Next, we have 
\beaa
u^\beta\nabla_\beta\ u_\alpha  &=& e^{-\Phi }t^\beta\nabla _\beta(e^{-\Phi }t_\alpha ) \nonumber \\ 
&=& e^{-2\Phi }t^\beta\nabla _\beta t_\alpha  ~(\text{using}~~t^\beta\nabla _\beta\Phi = 0)\nonumber
\\ 
&=& -e^{-2\Phi }t^\beta\nabla _\alpha t_\beta~~~(\nabla _\alpha t_\beta + \nabla _\beta t_\alpha  = 0 \rm~{~-~
Killing~~vector~~eq.})\nonumber \\ 
&=& -\frac{1}{2} e^{-2\Phi }\nabla _\alpha (t^\beta t_\beta) \nonumber \\ 
&=& \frac{1}{2} e^{-2\Phi }\nabla _\alpha (e^{2\Phi})\nonumber \\
&=& \nabla _\alpha \Phi \nonumber \\
\nabla _\alpha \Phi = &-& \frac{1}{\rho+P}~ \nabla _\alpha P \nonumber 
\eeaa
or
\be \cblue
\Phi' = -~\frac{1}{\rho +P}~P'\ \ .
\label{pp}\ee
Eqs. (\ref{phip}) and (\ref{pp}) imply the equation of hydrostatic equilibrium - the TOV  
(Tolman-Oppenheimer-Volkoff) equation:
\be \crv
	 P' = -(\rho +P) \frac{m+4\pi r^3 P}{r(r-2m)}~~.
\label{tov}\ee
\index{hydrostatic equilibrium!TOV equation}

A spherical relativistic star is a solution to equations (\ref{lambda}), 
(\ref{phip}), and (\ref{tov}) together with an equation of state; 
the numerical models that have been constructed usually involve an equation of state of the simplest form
\be 
	P = P(\rho),
\label{eos}
\ee\index{equation of state}\index{neutron star}
\index{numerical model of star!relativistic}
accurate for neutron stars . A general equation of 
state has the form $P = P(\rho, s, z_1,\dots, z_n)$, with $s$ the entropy per
baryon and $z_i$ the concentration of the ith particle species. Neutron stars and 
white dwarfs are cold enough (KT $\ll$ Fermi energy of electrons in white dwarfs and 
of nucleons in neutron stars) that they
are nearly isentropic ($s = $ constant), and their nuclear reactions have proceeded 
to completion, so each $z_i$ is itself a function of $\rho$. 
\index{entropy!per baryon} 
\index{energy!Fermi energy}\index{Fermi energy} That's why 
$P = P(\rho)$ is a good approximation. At absolute zero of course, $s$ 
is constant.

	One obtains a star by integrating Eqs. (\ref{tov}) and (\ref{eos}) 
together with the defining equation \ref{lambda} for $m$.  That is, one integrates 
the system
\be \crv
	m(r) = \int^r_0 \rho \ 4\pi r^2dr, \quad 
	\frac{dP}{dr} = -(\rho  + P)\frac{m+4\pi r^3 P}{r(r-2m)}~,\quad
	P = P(\rho )~. 
\label{e:relstar0}\ee
\index{hydrostatic equilibrium}
One begins with a central density $\rho_c$ and integrates up to the radius
$R$ at which $P$ drops to zero ($P$ is a decreasing function of $r$).  This
is the boundary of the star.  The metric inside the star is then given by

\be 
	e^{2\lambda} = \left(1 - \frac{2m}{r}\right)^{-1},
\label{e:e2lambda}
\ee
\be
  \Phi = \Phi(R) - \int_r^R dr\frac{m+4\pi Pr^3}{r(r-2m)},
\ee
and outside by
\be 
	e^{2\Phi} = e^{-2\lambda } = \left(1 - \frac{2M}{r}\right)~.
\label{e:philambda}\ee

To restore $G$ and $c$ to these equations, one can 
multiply by the (unique) factors of $G$ and $c$ that allow each quantity 
to have the desired conventional units.  For example, in Eq.~\eqref{e:e2lambda},
both sides are dimensionless, and the unique factor built from $G$ and $c$ that makes 
$m/r$ dimensionless is $G/c^2$. That is, $\dis\frac G{c^2} \frac mr$ is 
dimensionless, so the right side of \eqref{e:e2lambda} becomes 
$\dis \left(1-\frac{2Gm}{c^2 r}\right)^{-1}$.

The Newtonian limit of the structure equations (\ref{tov}) of a spherical star are
of course the Newtonian equations
\be 
	m = \int^r_0 \rho 4\pi r^2dr,~~\frac{dP}{dr} 
	  = -\frac{Gm\rho}{r^2}~,~~~~P = P(\rho)~.
\label{e:newt_star}\ee
The Newtonian potential $\Phi$ satisfies 
\be
    \na^2\Phi = 4\pi \rho, \quad \lim_{r\rightarrow\infty} \Phi = 0.  
\ee
Gauss's law then implies $\Phi' = m/r^2$, for a spherical star. 


\subsection{Stellar models}\index{stellar models|(}

The simplest stellar model is a uniform-density Newtonian star. 

\benr\item Find the solutions $P(r)$ and $\Phi(r)$ to the Newtonian equations 
governing hydrostatic equilibrium for a uniform-density star, \eqref{e:newt_star} and $\Phi' = m/r^2$, 
and show that they can be written in the form \eqref{e:newt_star2} below.
\index{hydrostatic equilibrium!Newtonian}

A star of radius $R$ and constant density $\rho$ has $m(r)$, $P(r)$ and $\Phi(r)$ given by  
\bsube\begin{align}
  m(r) &= \frac43\pi r^3\rho = M\frac{r^3}{R^3}, \\
  P(r) & = P_c\left(1-\frac{r^2}{R^2}\right), \quad \mbox{where } 
	   P_c = \frac{\rho M}R,\\
 \Phi(r)&= \frac12 \frac{m}{r} - \frac32\frac MR. 
\end{align}\label{e:newt_star2}\esube
Here $P_c$ is the pressure at the center of the star.  
Outside the star, 
\be
   \Phi = - \frac Mr.  
\label{e:Phiexterior}\ee
Inside the star, $\Phi'$ has the simple form $\dis\frac m{r^2}$ given by Gauss's law, 
and a common error is to think of Gauss's theorem as meaning $\Phi$ itself is given 
by $-\dis\frac mr$.  But $\Phi$ involves the integral $\dis\int_r^\infty \Phi' dr$.
Inside a uniform density spherical shell, for example, $\Phi$ is constant, like 
the electric potential inside a charged spherical shell.  

\newpage
\item Relativistic uniform-density star. Find the pressure $P$ inside and the
 metric potentials $\Phi$ and $\lambda$ inside a relativistic star with uniform energy density $\rho$. Use Eq.~\eqref{e:m} to find $m(r)$, the TOV equation \eqref{tov} to find $P(r)$, and Eq.~\eqref{pp} to find $\Phi$.  Then show that 
they can be written in the form \eqref{e:relstar} below.
\label{ex:relstar}\een

We have already found the solution \ref{e:philambda} outside the star. 
Inside the star, it is given by 
\bsube\begin{align}
  m(r) &= \frac43\pi r^3\rho, \\
  P(r) &= \rho\ \frac{\sqrt{1-2m/r}-\sqrt{1-2M/R}}{3\sqrt{1-2M/R}-\sqrt{1-2m/r}},\\ \nn\\
  e^{\Phi(r)} &= \frac32\sqrt{1-2M/R} - \frac12\sqrt{1-2m/r},\\ \nn\\
  e^{\lambda(r)} &= \frac1{\sqrt{1-2m/r}}.
\end{align}\label{e:relstar}\esube
If you get stuck, a hint is given after \ref{ex:buchdahl}

\benr \item Write the TOV equation \eqref{tov} and the solution \ref{e:relstar} for a relativistic star in conventional units, restoring $G$ and $c$ to the expressions. 

\item Show that the Newtonian limit of the relativistic solution is the Newtonian solution 
for $P$ and $\Phi$.  Either use the form in gravitational units, noting that $m/r\lesssim 1, M/R\lesssim 1$,
and discarding terms of quadratic order in these quantities; or use the form  
with $G$ and $c$ restored, discarding terms with $c^2$ in the denominator. If you use conventional 
units, what is the small dimensionless quantity whose quadratic and higher orders you are discarding.  

\item \label{ex:buchdahl} The smallest possible value of $\dis\frac{2M}{R}$ at the surface
of a static (spherical) star is attained when the star is perfectly
incompressible -- when it is one of the uniform density models. 
(This is Buchdahl's Theorem\cite{buchdahl59}, with details given in 
Sect. \ref{s:buchdahl} below.   
\benalph\item  Show for this sequence of stars that the central density becomes 
infinite when $\dis\frac{2M}{R}=\frac{8}{9}$.
\item Deduce from (a) the maximum redshift $z$,
\index{redshift!maximum gravitational redshift}
\[
	1+z := \frac{\lambda_\infty}{\lambda_{\rm star}} = \frac{\omega_{\rm star}}{\omega_\infty},
\]
from the surface of a spherical star.  (Recall the redshift relation~\eqref{e:redshift}.) 
\een 
\een
{\sl Hint for} \ref{ex:relstar}\hspace{-1mm}: From the TOV equation, show 
$\dis\frac{dP}{dr} = \frac43\pi \frac{(\rho+P)(\rho+3P)}{1-\frac83\pi\rho r^2}$,  
divide by \mbox{$(\rho+P)(\rho+3P)$}, and integrate from $P$ to $0$ on the left side, and 
$r$ to $R$ on the right to obtain $P(r)$.  Then show $\Phi' = [\ln(\rho + P)]'$ and integrate 
to find $\Phi(r)$, using the value of $\Phi(R)$ from 
the form \eqref{e:Phiexterior} of $\Phi$ outside the star.    
\index{stellar models|)}

\section{White dwarfs and neutron stars} 
\subsection{Estimates}

At the endpoint of stellar evolution, when nuclear reactions cease, stars
with mass less than about 10 $M_\odot$ die as white dwarfs.
\footnote{The critical mass of the main-sequence progenitor star, above which 
it ends its life in a core-collapse supernova, depends on the 
initial {\sl metallicity} of the star -- the fraction of elements 
heavier than helium 
(see, e.g., \href{https://arxiv.org/pdf/1301.5783.pdf}{Ibeling and Heger 2023}).} 
In white dwarfs the major pressure supporting the
star arises from the Fermi degeneracy pressure -- the Pauli exclusion
principle requiring the fermions to have nonzero momentum.  
\index{degeneracy pressure} Here is first
an intuitive, then a more careful discussion of the inability of a degenerate
Fermi gas to support more than about   1.4 $M_\odot$ against its own
gravity.  (It starts with rocks for practice.) 

\noindent{\em Rocks}

 Intuitive:  For small cold masses (rocks), gravity is unimportant and the
degeneracy pressure balances the electrostatic attraction of electrons to nuclei: 
We use this as follows to obtain a rough estimate of the ground-state size of 
atoms and hence of the density of rock.  Let $n$ be the number of
electrons per unit volume, $m_e$ the mass of an electron.  Pauli exclusion
restricts each fermion to its own volume $\dis {\mathfrak v} = \frac{1}{n}$ 
and the uncertainty principle then implies a momentum, 
\index{Fermi momentum}
\be 
	p \sim \frac{\hbar}{{\mathfrak v}^{1/3}} = \hbar n^{1/3}, 
\ee
and a corresponding pressure (momentum crossing unit area/unit time)
\be
	\dis P \sim p v n.  
\ee 
For the nonrelativistic electrons of rock, $v=p/m_e$ and the pressure is
\index{degeneracy pressure!non-relativistic}
\be 
	P \sim \frac{p^2}{m_e}~n\sim \frac{\hbar ^2n^{5/3}}{m_e} .
\label{e:P}\ee
This must balance the electrostatic energy density:
\index{energy!electrostatic energy density}\index{energy density!electrostatic}
\be  \frac{e^2}{{\mathfrak v}^{1/3}}\  n = e^2n^{4/3} ,
\label{es}\ee
or 
\[ 
	\frac{\hbar ^2n^{5/3}}{m_e}\sim e^2n^{4/3}.
\]
The corresponding number density of electrons is then
\[
   n \sim \frac{m_e^3e^6}{\hbar ^6} 
\]
With $m_p$ the mass of a proton or neutron, we have
\be 
\rho \sim m_p n \sim
\frac{e^6m_e^3m_p}{\hbar ^6} \sim 10\ {\rm g/cm}^3~. 
\ee

 Gravity is dominant in objects massive enough that the gravitational 
binding energy density is greater than the electrostatic energy density.
\index{binding energy!gravitational compared to electrostatic}
Writing $M \sim \rho R^3$, the gravitational energy density is 
\index{energy!gravitational energy density (Newtonian)}
\[
  \frac{GM}R\rho \sim G\rho^{4/3}M^{2/3}.
\]
Substituting $\rho = m_p n$ in this expression, and equating it to the 
RHS of Eq. (\ref{es}) for the electrostatic energy density, we have 
\[
	e^2n^{4/3} \sim n^{4/3}m_p^{4/3}GM^{2/3},
\]
when the electrostatic and gravitational energy densities are equal.
Solving this equation for $M$, we find the mass 
\be
  M \sim \frac{e^3}{G^{3/2} m_p^2} \sim 2\times 10^{30}{\rm g}\sim M_{\text{Jupiter}}.
\label{mcrit}\ee  

The critical size above which an asteroid or moon must be spherical is much smaller than 
this, because the molecular bonds that keep rock rigid have much lower energy than 
the binding energy (ionization energy) of their atoms.  The next problem estimates 
that critical size, using the melting energy of quartz.  This is an upper limit, 
ignoring weakness from, for example, grain boundaries and faults.  

\benr \item The maximum height of mountains can be estimated by requiring that the
energy needed to melt the rock in a layer of thickness $h$ below the mountain is equal 
to the gravitational energy lost when the mountain sinks by the distance $h$. 
\benalph\item Using an energy of about 0.1 eV/molecule to melt quartz, estimate 
the maximum height of a mountain on the Earth and on Mars.  
\item Asteroids and rocky moons with large enough radius are spherical (if nonrotating).  
Estimate the radius above which a rocky moon is spherical (i.e., with bumps smaller than 
the moon's radius).  
\een
\een   

\subsection{White Dwarfs}\index{white dwarf|(}

 The cores of objects of mass greater than $M_{\text{Jupiter}}$  are compressed 
beyond planetary densities and for masses $>50~M_{\text{Jupiter}}$ balls of 
hydrogen begin nuclear reactions and are -- by definition -- stars. As noted above, 
a white dwarf is the final state of a star whose {\em initial} mass is
less than about 10 $M_\odot$.  By the end of its evolution, the star
has ejected its outer envelope of hydrogen, and the core that remains
contracts and eventually cools to a dead ball of He, C and O, most 
commonly dominated by a C-O interior.  (Although most stars have 
masses too small to burn He to C and O, the evolution time of stars with 
masses below about 0.8 $M_\odot$ is longer than the age of the universe; 
and He will fuse in cores of stars with initial masses above about 0.5 $M_\odot$.)  
The most massive stars that end as white dwarfs include heavier elements: Ne and Mg.   

\subsubsection{Structure: Mass-Radius Relation}
\index{white dwarf!mass radius relation}

Because its nuclear reactions have turned off, a dead star is held
apart by its degeneracy pressure.  The size of such a star turns out to
{\em decrease} as its mass increases:  adding baryons increases the
gravitational attraction enough that more baryons are packed in a
smaller total volume.  This relation between mass and radius can be
found from our equations of hydrostatic equilibrium and the equation of
state of a degenerate gas. Here we'll again obtain an estimate based 
on the averaged form of the equation of hydrostatic equilibrium with 
$dP/dr$ approximated by $-P/R$ and $m(r)$ by $\rho r^3$, with $M$ 
the total mass. Then the Newtonian equation of hydrostatic equilibrium, 
$\dis \frac{dP}{dr} = -~\frac{Gm(r)\rho }{r^2}$, 
is roughly approximated by  
\index{hydrostatic equilibrium!Newtonian}
\be 
\frac{P}{R} \sim \frac{GM\rho }{R^2} 
\label{pr}\ee
Similarly dropping numerical factors, we have $\dis \rho \sim \frac{M}{R^3}$, 
giving 
\be 
  P \sim \frac{GM^2}{R^4}.   
\label{e:heq}\ee
\index{degeneracy pressure|textbf} 
Using the degeneracy pressure $\dis P\approx \frac{\hbar^2}{m_e} n^{5/3}$ of \eqref{e:P} and    
\[
\rho = m_p n \sim \frac{M}{R^3} \Rightarrow 
	n \sim \frac{M}{m_pR^3} 
\]
gives 
\begin{align}
  \frac{GM^2}{R^4} 
	 &\sim P\sim \frac{\hbar^2}{m_e} \left(\frac{M}{m_pR^3}\right)^{5/3} 
\nonumber\\
	&\sim \frac{\hbar^2}{m_e}
			\frac{M^{5/3}}{m_p^{5/3}R^5}\nonumber\\
R&\sim\frac{\hbar^2}{Gm_em_p^{5/3}} \cdot \frac{1}{M^{1/3}} .
\label{mr}\end{align}
With the right numerical factors for a Helium dwarf,
\[ R = 1.4 \frac{\hbar^2}{Gm_em_p^{5/3}} M^{-1/3}\]
\[ {\text{or,}} \hspace{7mm} \frac{R}{R_\odot} = 0.014 \left(
\frac{M_\odot}{M}\right)^{1/3} .\]
(When $M=M_\odot$, we have $R=.014 R_\odot$).  So we obtain the relation
$R\propto M^{-1/3}$, valid when the star is
dense and cold enough for degeneracy and not so massive that it collapses.
\index{collapse, gravitational}\index{gravitational collapse}

\subsubsection{Upper mass limit}
\index{Chandrasekhar limit}\index{black hole!and upper mass limit}\index{Chandrasekhar limit}
\index{white dwarf!Chandrasekhar limit (upper mass limit)}\index{upper mass limit!white dwarf}\index{maximum mass!white dwarf}

The speed limit set by causality-- the speed of light-- also sets an upper limit on the 
mass of white dwarfs, on the dead iron cores of massive stars, and 
on the mass of neutron stars (this one takes more work).  

\begin{verse}
 {\sl\hspace{3mm} 
The star has to go on radiating and radiating and 
contracting and contracting until, I suppose, it gets down
to a few km.~radius, when gravity becomes strong enough 
to hold in the radiation, and the star can at last 
find peace. Dr. Chandrasekhar had got this result 
before, but he has rubbed it in in his latest 
paper; and, when discussing it with him, I felt
driven to the conclusion that this was almost 
a {\em reductio ad absurdum} of the relativistic 
degeneracy formula.}  
\hspace{3mm} A.~S.~Eddington  (1935), published version of comments that 
followed a talk by Chandra on the upper mass limit. \cite{ch_edd35} 
\end{verse}\index{Eddington}

Notice that Chandrasekhar and Eddington recognized the possibility of collapse to 
a black hole resulting from the upper limit on a mass supported by 
degeneracy pressure, but they failed to make the connection between collapse  
and supernovae. Baade and Zwicky\cite{bz34} had proposed in the previous year that 
a supernova was the result of collapse to a neutron star, 
but didn't relate collapse to the upper mass limit.  
The connection between the limiting mass of a degenerate stellar core (or a white
dwarf) and the collapse to a neutron star did not appear in print until 1939
articles by Gamow\cite{gamow39} and by Chandrasekhar\cite{chandra39}.
\index{collapse, gravitational}\index{gravitational collapse} 
\index{degenerate electron gas}

Initially the gas will be nonrelativistic, but as the mass increases and the radius decreases, the density rises quadratically in $M$:  Writing hydrostatic equilibrium in the form \eqref{e:heq} with $R$ replaced by 
$\dis \left(\frac M\rho\right)^{1/3} = \left(\frac M{m_p n}\right)^{1/3}$ gives
\be
  P\sim GM^{2/3}(m_p n)^{4/3} , 
\label{e:he1}\ee
implying
\be
M \sim \frac{\hbar^3}{G^{3/2}}~\frac{n^{1/2}}{m_p^2m_e^{3/2}} 
\quad \mbox{(nonrelativistic)}.
\ee
As the density rises, the Fermi momentum becomes relativistic when $p = \hbar n^{1/3}$ 
is of order $m_ec$.  
Because $v$ is limited by $c$, it no longer grows as $n^{1/3}$, and that limit on $v$ 
means the pressure is given by
\index{degeneracy pressure!relativistic}
\begin{align*}
 P &\sim  pvn < \hbar n^{1/3}cn \hspace{.5in} (v \sim
c~\text{now, or})\\ 
\cblue P &\cblue <  \hbar cn^{4/3} .
\end{align*}
With $P$ growing no faster than $n^{4/3}$ instead of its nonrelativistic $n^{5/3}$ behavior,  
hydrostatic equilibrium in the form \eqref{e:he1} now implies
\[
 GM^{2/3}m_p^{4/3}n^{4/3} \sim P  < \hbar cn^{4/3}  \ ;
\]  
the $n^{4/3}$ cancels (!) and one finds a maximum mass
\be 
  \cblue M \lesssim \left(\frac{\hbar c}{G}\right)^{3/2}~~\frac{1}{m_p^2}\cb = 1.8~M_\odot
\hspace{.3in} (3.7\times10^{33}g). 
\ee
For typical dwarfs (e.g., helium or helium-carbon dwarfs) the actual 
calculation gives 
\be \crv
  M_{\rm max} = 0.76 \left(\frac{\hbar c}{G}\right)^{3/2}~~\frac{1}{m_p^2} = 1.4 M_\odot. 
\ee
\noindent This is the Chandrasekhar limit -- matter supported by the 
pressure of degenerate electrons and with masses greater than this must 
collapse.

To summarize:  As long as the gas is nonrelativistic, $P \propto
n^{5/3}$, and as a star contracts, its pressure therefore rises faster than
its gravitational energy density $\frac{GM\rho }{R}$, which is proportional
to $n^{4/3}$.  There will then always be a density, $n$, high enough
that the pressure gradient, $\frac{P}{R}$, balances the gravitational
force, $\frac{GM\rho }{R^2}~$.  At that density the star will cease to
contract.  If, however, the gas becomes relativistic before equilibrium is
reached, then the pressure will be bounded by $kn^{4/3}$ because $v$ is
bounded by $c$; and it will never catch up to gravity if the coefficient of
$n^{4/3}$ in the pressure is smaller than the coefficient of $n^{4/3}$ in
the gravitational energy density -- if
\[  M > \left(\frac{\hbar c}{G}\right)^{3/2} \frac{1}{m_p^2}~.\]

Some additional history:  
In 1931 shortly before Chadwick's discovery of the neutron and shortly after the 
first paper by Chandrasekhar \cite{chandra31} (following approximate computations by Anderson \cite{anderson29} and Stoner \cite{stoner30}) on an upper mass limit of white dwarfs, 
Landau \cite{landau32} submitted a paper that independently argued that there was an upper limit on the mass of a collection of degenerate fermions and speculated on the existence of stars with cores of nuclear density now called Thorne-\.Zytkow objects \cite{tz75}.
L\'eon Rosenfeld \cite{rosenfeld} gives a widely 
repeated but false description: ``when the news on the neutron's discovery reached Copenhagen, 
we had a lively discussion on the same evening about the prospects opened by this discovery. 
In the course of it Landau improvised the conception of neutron stars – ``unheimliche 
Sterne,'' weird stars, which would be invisible and unknown to us unless by colliding 
with visible stars they would originate explosions, which might be supernovae.'' 
A careful historical study by Yakovlev {\it et al.} \cite{yhbp}, however, finds that 
Landau was in Copenhagen in 1931 {\em before} Chadwick's experiments -- and just 
after Landau had submitted his paper on nuclear-density cores.

\benr
\item Estimate the radius of a neutron star, assuming the degeneracy pressure 
of free nonrelativistic neutrons supplies the pressure that supports it.  

\item Suppose that a species of neutrino has mass
$m_{\nu}=10$ eV. 
\benalph\item Find the upper mass limit on a degenerate 
neutrino star. \index{upper mass limit!degenerate neutrinos}
\item Consider a neutrino star with mass of $1/10$ this upper limit.
Estimate the radius of the star, as in the previous problem for for neutron stars.
\item Use this radius to estimate the temperature above which  
the neutrinos in the neutrino star in part (b) would cease being degenerate. 
\een

\een
\index{white dwarf|)}

\subsection{Neutron Stars}\label{s:neutron stars}
\index{neutron star|(}
A very short introduction is given here.  Standard references are Shapiro-Teukolsky\cite{st86} 
and Camenzind\cite{camenzind07}.  

	The only stars in whose structure general relativity plays a 
major role are neutron stars.  The first observations were made in two ways: 
First, in 1962, a team led by Giacconi, using an Aerobee rocket with an x-ray detector, 
found the first x-ray source outside the Solar System, Sco X-1 (the label means 
the brightest x-ray source in the direction of the constellation Scorpius).   
Shklovsky \cite{shklovsky65} suggested 
that Sco X-1 might be a neutron star and that synchrotron radiation from 
relativistic electrons might power the Crab nebula \cite{shklovsky66}).
The ``binary hypothesis'' emerged from a discussion at the Noordwijk Symposium 
in August 1966 that included G. and M. Burbidge, Ginzburg, Shklovsky, Savedoff, 
Woltjer, Prendergast, and Herbig. 
After Rossi presented observations suggesting that Sco X-1 was a binary, Burbidge writes, ``several of us then realized that \ldots the very powerful x-ray source might naturally 
arise through mass exchange between the secondary and primary \ldots''  
Articles with accretion from a companion to a neutron star, beginning with \href{https://articles.adsabs.harvard.edu/pdf/1967ApJ...148L...1S}{Shklovsky}\cite{shklovsky67}, 
quickly followed.  Here's what is happening: 
Old neutron stars in binary systems\index{binary star}\index{neutron star!binary} can accrete mass from their companion after the companion evolves off the main sequence to become a red giant. (The star vastly expands because of greatly increased energy from a contracting core that gets hotter as it contracts and from the reactions in a shell around the core.  The reaction rate is much faster than for the main-sequence star because the core is hotter.)  Matter falling onto the neutron star is moving fast enough to emit x-rays.   
\index{x-ray binary} 
 
The second observation came within two years of Shklovsky's first paper.  
In 1967, Jocelyn Bell, 
using as a radio telescope a field of wires designed 
by her advisor, Anthony Hewish, \href{https://articles.adsabs.harvard.edu/full/1967ApJ...148L...1S}{discovered}\cite{bell68} pulsing light signals (radio frequency) 
arriving $\sim$ every second, with fantastic regularity .\\
The first one seen had period $P = 1.3373011$ s, another $P = .253065 $s.
The regularity
\[ 
	\frac{1}{P}~\frac{dP}{d\tau } \approx \frac{1}{\text{2000 ~years}} 
	\]
is much too great to be an oscillating star.  They were nevertheless 
called {\sl pulsars} to describe the observation of periodic pulses of 
radio waves.   

 A rotating star would fly apart unless its density were great enough that\\ 
(rotational kinetic energy) $< \frac{1}{2}$ (gravitational binding energy). 
\index{binding energy!gravitational} \\
A rough estimate is enough to rule out rotating white dwarfs:  
\begin{align*}
 \frac{1}{2} I\Omega^2 &< \frac12 \mbox{ binding energy}\\
	MR^2\Omega^2 &< \frac{M^2}{R}
		\hspace{.5in}\mbox{(ignoring numerical factors)}\\
\Omega^2 &< \rho \hspace{.5in} (\Omega^2 < G\rho) \\
\Omega = 60\pi\ {\rm s}^{-1} ~\Longrightarrow ~ \rho &> 10^{11} {\rm g/cm}^3
\end{align*}
Because $\rho_{\rm dwarf} < 10^8 {\rm g/cm}^3$, only neutron stars are dense enough. 

Nuclear density is 
\be
 \rho_n = m_pn = 2.8\times 10^{14}{\mbox{g cm}}^{-3} , 
\ee 
corresponding to a separation between nucleons of about 
$ \ell = 1.8\times 10^{-13} {\rm{cm}} $.  The average density of neutron stars is a few times 
greater than this, and their corresponding maximum spin is larger than 1000 Hz. 
As of 2023 the spin rate of the fastest observed pulsar is 716 Hz.

Additional strong evidence that pulsars are neutron stars was not long in coming.  
Just before the pulsar discovery, Pacini \cite{pacini67} proposed the magnetic 
field of a rotating neutron star as the energy source of the Crab and other 
supernova remnants; and, of course, searches for pulsars in the Crab and some 
of the other remnants were successful.  The pulses can be reasonably ascribed to synchrotron
radiation from particles spiraling in the magnetic field near the poles, with 
a contribution from radiation produced by a spinning magnetic field.  The 
beam sweeps past you each rotation, and often one observes double
pulses, as one would expect for a beam from each pole, if they are 
not close to the spin axis. \index{magnetic field!pulsar} \index{pulsar}

Pulsars spin down slowly; $\dot P/P\sim \frac{1}{2000~\text{yrs}}$ for Crab,  and the \\
\centerline{loss of rotational energy = observed luminosity of the Crab nebula,}
 if one 
assumes that the Crab pulsar is a rotating neutron star, with 
\[ 
	dE/dt \sim MR^2 \omega \dot\omega.
\]
This check, due to Tommy Gold\cite{gold68}), together with the pulsar's location 
in a supernova remnant was the most compelling evidence that pulsars are 
neutron stars. 
\benr\item Do Gold's calculation:  Estimate the moment of inertia of a neutron star and 
the rotational energy of the Crab pulsar. 
Data:\href{https://www.atnf.csiro.au/people/pulsar/psrcat/}{ATNF Catalogue}: 
$P=33$ ms, $\dot P = 4.2\times 10^{-13}$ Hz/s. 
You should roughly find the observed luminosity of the Crab Nebula, namely \mbox{$L\sim 75,000 L_\odot = 3\times 10^{38}$ erg/s}\href{https://en.wikipedia.org/wiki/Crab_Nebula}{Wikipedia}.  
\een 

Here's a check that the gravitational field of a neutron star is strong enough to accelerate
infalling electrons to x-ray energies, about $m_e/100$, beginning with an estimate of 
the star's radius.  \\
A star of nuclear density and mass $1.4 M_\odot$ is roughly 
\[
   R\sim \left(\frac3{4\pi}\frac{1.4M_\odot}{\rho_n}\right)^{1/3} = 13\ \rm km, 
\] 
suprisingly close to the best current measurements of neutron-star radii.  
For example, the radius of a 1.4 $M_\odot$ neutron star inferred from observations 
by the 
\href{https://iopscience.iop.org/article/10.3847/2041-8213/ac089b/pdf}{NICER and XMM-Newton satellites} is\cite{nicer21}
\be
  	R = 12.45\pm0.65\mbox{ km},  
\ee 
and the LIGO/Virgo collaboration's estimate from the gravitational-wave observation of coalescing $1.4 M_\odot$ neutron stars\cite{gw170817} is $11.9\pm 1.4$ km. 

An electron falling to the surface of a neutron star can emit x-rays if  
\[ 
\frac{m_eM}{R} > m_e/100 \Rightarrow \frac{M}{R} > 1/100, 
\]
consistent with 
\[ 
\left(\frac{M}{R}\right)_{\text{neutron~star}} \approx 1.4M_\odot/12 \mbox{ km} = 
\frac{1.4\times 1.5\mbox{ km}}{12\mbox{ km}} = 0.18. 
\]
In contrast, for a white dwarf, 
\[   
\left(\frac{M}{R}\right)_{\text{dwarf}} \sim \frac{M_\odot}{10^4{\rm km}} 
	\sim 10^{-4}.
\]
So matter in an accretion disk around a neutron star can emit an intense 
beam of x-rays from radii up to more than 10 times the radius of the star, 
while the x-ray intensity from accreting white dwarfs is low.
\vspace{2mm}

\noindent{\sl Energy of a supernova}
\index{supernova}

As mentioned above, Baade and Zwicky, armed with Landau's suggestion and 
with the observations of the previous ten years showing that observed supernova 
occurred in other galaxies, wrote in 1934  
``With all reserve we advance the view that a super-nova represents 
the transition of an ordinary star into a neutron star, consisting mainly 
of neutrons.'' Their suggestion was due in part to the following
computation of the energy that would be produced in such a collapse,
although they were unaware that Chandrasekhar's upper limit on white dwarfs
would imply that a collapse of this kind was in fact possible.
\index{collapse, gravitational}\index{gravitational collapse}

Supernovae can outshine the entire galaxy in which they
occur, with brightness as great as $10^{43}$ erg/s ($> 10^9 L_\odot$) and
total energy emitted as large as $10^{53}$ erg.  (Energy in light is 
$10^{51}-10^{52}$ erg, in neutrinos about $10^{53}$ erg).  If one could convert the
entire mass of the Sun to light, the energy emitted would be $M_\odot c^2 =
(2\times 10^{33}$g)($3\times 10^{10}$cm s$^{-1}$)$^2 = 2\times 10^{54}$
erg, so the total energy emitted by a supernova is nearly $\frac{1}{10}
M_\odot c^2$. 

The implicit calculation underlying the Baade-Zwicky paper is simple:  
What is the gravitational energy 
released in the collapse of a star of mass $M_\odot$ to the size of a
neutron star?  
\be
\Delta E = E_{\rm{initial}} - E_{\rm{final}} =
-\frac{M_\odot^2}{R_{\rm{initial}}} - \left( -\frac{M^2_\odot}{12\mbox{km}}\right)
= \frac{M_\odot^2}{12 \mbox{ km}} .
\ee
In the middle equalities, the first term is negligible compared to the second, because the initial radius, $R_{\rm{initial}}$, is 1000 times the final 
10 km radius for collapse of a stellar core (at white dwarf size) or of 
a white dwarf to a neutron star.  Let's write $\Delta E$ as a fraction 
of $M_\odot$:
\be
\Delta E = \frac{M_\odot}{12 \mbox{ km }} M_\odot 
 	  =\frac{1.5 \mbox{ km}}{12 \mbox{ km}} M_\odot = 0.1 M_\odot.
\ee 
Thus the collapse of a stellar core of iron to a neutron star produces total 
energy equal to that observed in supernovae.  As mentioned above, 
most of that is carried off by neutrinos.   

Iron cores collapse to neutron stars instead of simply fusing to make 
heavier elements because iron is the stablest element - the element 
with the greatest binding energy per nucleon.\index{binding energy!nuclear}  
White dwarfs, on the 
other hand, are the endpoints of evolution of stars too light to have 
formed iron.  If they accrete enough matter from a companion to reach 
their upper mass limit (or if two degenerate stellar cores of elements 
lighter than iron merge) their initial collapse can lead to explosive 
nucleosynthesis, carbon fusion, well before neutron-star density is 
reached. As a result, much less energy is emitted: The total emitted 
energy is about $10^{51}$ erg $<10^{-2} M_\odot$.  
Instead of collapsing to a neutron star, the stapdfsr 
blows itself apart, and the event is identified with a type Ia supernova. 
Heavier C-O and O-Ne-Mg dwarfs, however, may again collapse to neutron stars (see, 
e.g., \href{https://articles.adsabs.harvard.edu/pdf/1991ApJ...367L..19N}{Nomoto and Kondo 1991}
and citations to this paper for studies of initial conditions leading to 
accretion-induced collapse). \\

\noindent{\sl Maximum mass}\index{upper mass limit!neutron star} \index{maximum mass!neutron star}\index{neutron star!maximum mass}

The upper limit on the mass of neutron stars depends on an equation of state that is 
imperfectly understood.  
By ignoring nuclear interactions and treating the neutrons in a neutron star as 
gas of free fermions, \href{https://journals.aps.org/pr/abstract/10.1103/PhysRev.55.374}{Oppenheimer and Volkoff}, in 1939, found an upper limit of $0.7M_\odot$ on the mass of 
a neutron star. Remarkably, they also considered the possibility of repulsive interactions 
at high density, but they restricted a stiffer EOS to $\rho >10^{15}$ g/cm$^3$
and thereby enforced an unrealistically low maximum mass of order $M_\odot$, below the Chandrasekhar limit
on white dwarfs.  

It is easy to show (Hartle and Sabbadini\cite{hs77}) that for \textit{any}
equation of state there is a maximum mass for a spherical star of average density $\geq$ nuclear density, $\rho_n = 2.7\times 10^{14}\rm g/cm^3$.  Here is the quick 
calculation:  

With no horizon, $R<2M$. Using Eq.~\eqref{e:relstar0} for $m(r)$, we have
\begin{eqnarray*}\rho > \rho _n \ \Longrightarrow\ M &>& \int_0^R \rho
_n4\pi r^2dr > \frac{4}{3}\pi r^3\rho _n > \frac{4}{3}\pi \rho _n(2M)^3 \\
\Longrightarrow\ M &<& \left(\frac{3}{32\pi \rho _n}\right)^{1/2} \approx 8M_{\odot}.  
 \end{eqnarray*}
Matching to a known equation of state below a density $\rho_{\rm match}$ and 
{\cblue imposing causality in the form \mbox{(speed of sound) $< c$}}, Rhoades and Ruffini and, in the form given here, 
Hartle and Chitre find 
\be\cblue
   M < 4.1\left(\frac{\rho_n}{\rho_{\rm match}}\right)^{1/2}.  \cb
\ee
Candidate equations of state have maximum masses below 3 $M_\odot$.

The four largest accurately measured neutron star masses have values near and slightly above 2.0$M_\odot$.%
\footnote{The stars, J1614-2230\cite{demorest10}, J0348+0432\cite{antoniadis13}, 
J0740+6620\cite{Cromartie2020}, and J09520607\cite{romani22} have, respectively, measured masses $1.97\pm 0.04M_\odot$, $2.01\pm 0.04$, $2.14\pm 0.1 M_\odot$, and $2.35\pm 0.17 M_\odot$ .} 
The first observed inspiral and merger of a binary neutron star system, GW1701817 ended in a  
collapse to a black hole of matter with mass not larger than about 2.7 $M_\odot$.\index{binary system}\index{neutron star!binary}   
Because the matter was hot and supported by differential rotation, the corresponding 
upper mass limit on cold nonrotating neutron stars is significantly smaller than this, 
with several authors giving estimates below $2.3 M_\odot$.\cite{Margalit2017a,rezzolla18,kyoto17,
ruiz18,Shunke2020,Margalit2019}.  
\index{neutron star|)}

\section{Buchdahl's theorem}\index{Buchdahl's theorem}
\label{s:buchdahl} 

The upper mass limit on neutron stars is larger than that of white dwarfs 
because of repulsive nuclear interactions when 
nucleons are compressed above nuclear density.  The equation of state 
that gives the most massive star with density above 
nuclear density is that of a completely incompressible star, 
with $\rho=$ constant, and \ref{p:buchdahl} asks you to find $M_{\rm max}$ 
for that configuration.  The incompressible model is also maximally compact: Compactness is defined as the ratio $M/R$, 
and for this model, you show in this exercise that incompressible 
models can have $M/R$ as large as $4/9$ and find the corresponding maximum redshift.  

\href{https://journals.aps.org/pr/abstract/10.1103/PhysRev.116.1027}{Buchdahl's theorem}\cite{buchdahl59} states that any other equation of state gives a less compact star.  
Note that the speed of sound in a fluid is $v_{\rm sound} = \sqrt{dP/d\rho}$  (true in both a Newtonian context and still true in relativity, with $\rho$ the energy density). For $\rho =$ constant, $d\rho/dP = 0$, implying an infinite speed of sound, so the equation of state is unphysically stiff.  Causality, $v_{\rm sound}<1$ is weaker than the unphysical constraint of uniform density, but it is enough to force $R$ to be at least twice as large as the Buchdahl limit for the same mass, giving a greatest possible compactness about twice the value from Buchdahl's theorem.\\  
+
\noindent{\sl Proof of the theorem}.  The proof relies on the fact that the density $\rho$ of a spherical star decreases outward and so is maximum at the center.  This implies that the average density inside a radius $r$ also decreases outward, with minimum value at $r=R$. Because $m(r)=\int_0^r dr 4\pi r^2 \rho$, the average density is 
$\dis \frac{m(r)}{\frac43 \pi r^3} \ $, implying
\be
    \frac{m(r)}{ r^3} \geq \frac M{R^3} >0. 
\label{e:mr3}\ee

We use the $G^\theta_\theta$ and $G^r_r$ equations, \eqref{e:Gthetatheta} and \eqref{e:Grr}.  Because each has source $8\pi P$,  

 \[
 e^{2\lambda} G^\theta_\theta = e^{2\lambda}G^r_r \Longrightarrow 
      \Phi''-\frac1r(\lambda'-\Phi') +\Phi'(\Phi'-\lambda')  = 
       \frac2r\Phi' - e^{2\lambda}\frac1{r^2}\left(1-e^{-2\lambda}\right)\ .
\]
Grouping all terms involving $\Phi$ on the left, we have 
\begin{align*}  \Phi'' +\Phi'\left(-\frac1r+\Phi'-\lambda'\right)  & = \frac1r\lambda' - e^{2\lambda}\frac1{r^2}\left(1-e^{-2\lambda}\right) \quad\mbox{or} \\
r e^{\lambda-\Phi}\left(\frac1r \Phi' e^{\Phi-\lambda}\right)'&= \frac r2 e^{2\lambda} \left[\frac1{r^2}(1-e^{-2\lambda})\right]'\ .
\end{align*}  
On the right side, $\dis \frac1{r^2}(1-e^{-2\lambda})= \frac{2m}{r^3}$, proportional 
to the average density inside radius $r$. Because the average density decreases outward,
the right side is negative (or zero in the limiting case of uniform density). 
Then 
\[
    \left(\frac1r \Phi' e^{\Phi-\lambda}\right)'\leq 0.
\]
Integrate from $r$ to $R$:  
\[
\int_r^R \left(\frac1r \Phi' e^{\Phi-\lambda}\right)' dr
	= \frac1R\left. \Phi'e^{\Phi-\lambda}\right|_R - \frac1r \Phi' e^{\Phi-\lambda} \leq 0.  
\]
At the surface $\Phi$ and $\Phi'$ are continuous as long as the density and pressure are continuous, so we can evaluate $\Phi'(R)$ by taking the derivative outside the star and using its value $r=R$.  Outside, because $\lambda=-\Phi$, we have $\dis\Phi'e^{\Phi-\lambda} = \frac12(e^{2\Phi})'= \frac12(1-2M/r)'=\frac{M}{r^2}$.  Our inequality now becomes 
\[
   \frac{M}{R^3} \leq \frac1r\Phi' e^{\Phi-\lambda}\quad\mbox{  or }\quad \Phi'e^\Phi \geq r e^\lambda\frac M{R^3}.
\]  
Using $e^\lambda = (1-2m(r)/r)^{-1/2}$, we have 
\[
   (e^\Phi)'\geq \frac M{R^3}r (1-2m/r)^{-1/2}. 
\] 
We again use the fact that the average density decreases outward, this time in the form \eqref{e:mr3}:  Then 
$m/r \geq Mr^2/R^3$, so
\[
   (e^\Phi)' \geq \frac M{R^3}r(1-2Mr^2/R^3)^{-1/2}.
\]
We integrate once more, using 
\[
  \int_0^R r(1-2Mr^2/R^3)^{-1/2} dr = - \frac{R^3}{2M} [(1-2M/R)^{1/2}-1] 
\]
to get
\[
   e^{\Phi(R)} - e^\Phi|_0 \geq -\frac12\left[\left(1-\frac{2M}R\right)^{1/2}-1\right].
\]
Finally, $e^{\Phi(R)}=(1-2M/R)^{1/2}$ and $e^{\Phi(0)}\geq 0$, giving 
\[
  \sqrt{1-2M/R} \geq  \frac13, \qquad \crv\frac MR \leq \frac49,  
\]
as claimed.

\section{Particle and photon orbits}\index{Schwarzschild spacetime!photon orbits (null geodesics)}\index{black hole!photon orbits}\index{Schwarzschild spacetime!particle orbits}\index{black hole!particle orbits}
\index{photon!orbits in Schwarzschild spacetime}
\label{s:geodesics}

As in the Newtonian central force problem, one reduces the geodesic equation 
to a first-order ordinary differential equation for $r(\tau)$ by using 
conservation of energy and conservation of momentum, and by picking 
coordinates for which the orbit is in the $\theta = \pi/2$ plane.
The Newtonian 
equation $\dis E_{\rm newt} = T +V $ comes from the relativistic equation
$p^\alpha p_\alpha = -m^2$, or, equivalently, $u^\alpha u_\alpha = -1$.

Writing $\dis(^\cdot)$ for $\dis \frac{d}{d\tau}$, we have   
$u^\alpha = \dot{t}t^\alpha  + \dot{\phi} \phi^\alpha  + \dot{r}r^\alpha$,
or $u^\mu = \dot x^\mu$.  Associated with the timelike Killing vector 
$t^\alpha$ is the conserved energy $-p_\alpha t^\alpha$; we'll use 
energy per unit rest-mass, 
\be 
\crv E := - u_\alpha t^\alpha \cb = -u_t = \left(1-\frac{2M}{r}\right)\dot t.
\label{energy}\ee 
Associated with the rotational Killing vector $\phi^\alpha$\index{Killing vector!rotational}
is the conserved angular momentum $p_\alpha \phi^\alpha$;
\index{angular momentum|textbf}\index{conservation laws!angular momentum} 
again we'll use angular momentum/rest-mass
\be
 \crv L :=u_\alpha \phi^\alpha \cb= u_\phi = r^2\dot\phi.
\label{angmom}\ee 
($\sin\theta=1$, for an orbit in the $\theta = \pi/2$ plane). As in the Newtonian description of central force motion, conservation of angular momentum implies a planar orbit.
\index{conservation laws!angular momentum}  

The Newtonian equation 
$E = \frac12 mv^2 + V(r) = \frac12 m\dot r^2 + V_{\rm eff}(r)$ is 
the Newtonian limit of the relativistic equation $p_\alpha p^\alpha = - m^2$ 
or $E_{\rm relativistic}^2 = p_\perp^2 + m^2$.  We again eliminate the 
rest mass, dividing by $m^2$ and using the equation in the form 
$u_\alpha u^\alpha = -1$, for a timelike geodesic.
\index{Newtonian limit!of orbit in Schwarzschild spacetime}
\begin{align}
   -1 &= g_{\alpha\beta}u^\alpha u^\beta = g^{\alpha\beta}u_\alpha u_\beta 
	= g^{tt} (u_t)^2 + g^{\phi\phi}(u_\phi)^2 + g_{rr} (u^r)^2
\nonumber\\
	& = -\left(1-\frac{2M}r\right)^{-1} E^2 + \frac{L^2}{r^2} 
		+\left(1-\frac{2M}r\right)^{-1}\dot r^2 \nonumber\\
  \frac12\dot{r}^2 &= \frac12( E^2-1)- V_{\rm eff}
\label{rdot1}\end{align} 
where $\dot r = dr/d\tau$ and 
\beq\cblue
	V_{\rm eff}= - \frac{M}{r}+ \frac{L^2}{2r^2} -\frac {ML^2}{r^3}.\cb
\label{rdot2}\eeq
That is, the orbit is described as motion in an effective potential.   
\index{effective potential!Schwarzschild spacetime, particle orbits|textbf}
\index{Schwarzschild spacetime!effective potential}

For $r>>2M$, $V_{\rm eff}$ is a Newtonian effective potential, the sum 
of a $1/r$ attractive potential and an $ L^2/r^2$ centrifugal barrier.
(The term $ML^2/r^3$ is of order $M/r\times L^2/r^2 \sim M v^2/r$, smaller 
than $M/r$ by order $v^2 = v^2/c^2$.)
At small $r$, however, relativistic gravity overcomes the centrifugal 
barrier. By $r=2M$, $V_{\rm eff}$ has fallen to zero, and for $r<2M$ it 
becomes increasingly negative.  As a result, in addition to bound orbits
and the analog of hyperbolic unbound orbits, the Schwarzschild geometry 
has a class of orbits with no Newtonian analog.

Write $e_N := \frac12( E^2-1)$; asymptotically,  
\[
 E = \frac{dt}{d\tau}=\gamma_\infty=\sqrt{\frac1{1-v_\infty^2}}\ \  \Longrightarrow  \ \ 
			e_N = \frac12 v_\infty^2\ \ \frac1{1-v_\infty^2},
\]
agreeing with the Newtonian kinetic energy/mass for small $v$.  A particle
moving in the effective potential $V_{\rm eff}$ is allowed to be in the
region $e_N > V_{\rm eff}$, so three types of orbits are possible from
the figure below. 
\begin{figure}[bh!]
\centering\includegraphics[width=.8\textwidth]{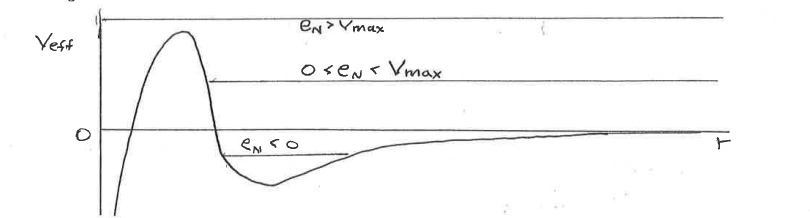}
\end{figure}

For $e_N < 0$, the orbits are bound -- the particle moves
between an inner and outer radius.  For $e_N \geq 0$ but $e_N < V_{\rm
max}$, the particle is unbound and moves from $\infty $ to a minimum
radius and out again. These orbits are similar to the Newtonian
hyperbolic orbits.
But for $e_N > V_{\rm max}$,
the particle never turns around, and we will see that it reaches $r = 2M$
{\em in finite proper time}.  In an extended geometry that includes
the black hole interior (see Sect.\ref{s:ef} below) , the particle falls to $r = 0$, where it hits an infinite
curvature singularity.  By making $e_N $ larger for fixed
$L $, you increase $\dot r$ and thus aim the particle more directly inward: 
You decrease the impact parameter, which we now define.  

In flat space, for a particle moving in a central potential, the impact parameter $b$ is defined in terms of the straight line tangent to the particle's asymptotic trajectory.  That is, $b$ is the minimum distance  between that free-particle trajectory and the 
center of the potential.  A particle with asymptotic momentum $p =m\gamma v$ has angular momentum $\texttt L = bp$ and energy $\texttt E=m\gamma$,\index{angular momentum!asymptotic}  so its impact parameter is $b=\texttt L/\texttt E=L/(Ev)$.  Using 
\be
   b := \frac L {Ev_\infty}
\ee 
as the definition of $b$ is appropriate because the spacetime is asymptotically flat, and $L$ and $E$ are the 
energy and angular momentum per unit mass measured by a stationary observer at infinity. 
\newpage

Here's the effective potential for different values of $\frac LM$:
\begin{figure}[bh!]
\centering\includegraphics[width=.8\textwidth]{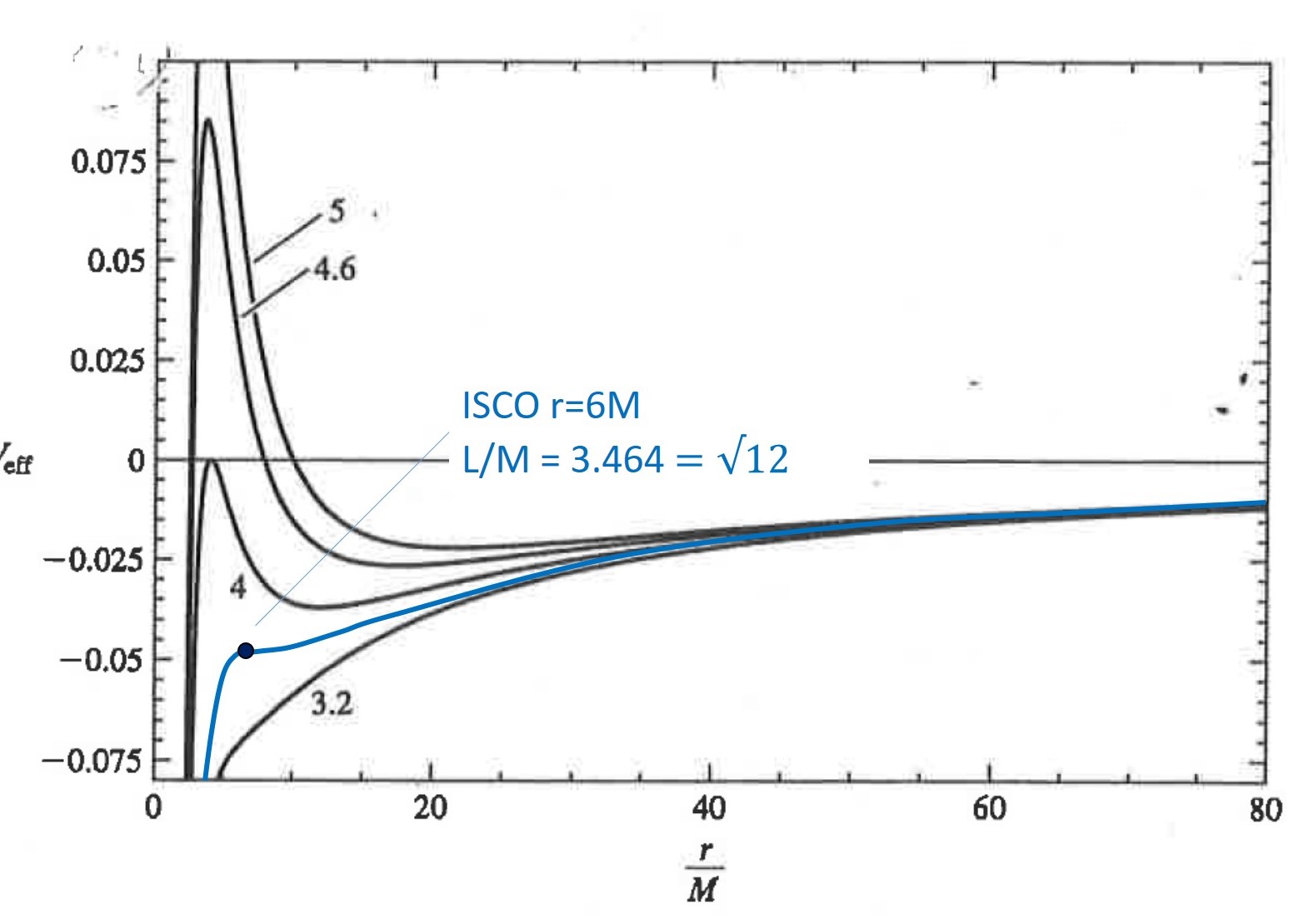}
\caption{Figure adapted from Hartle, with each curve labeled by $\frac LM$.}
\label{f:hartlepot}\end{figure}

For $L^2 < 12 M^2, V_{\rm eff}$ has no maxima or minima and all orbits
eventually hit $r = 0$.  For each value of $L/M > \sqrt{12}$, there is a stable 
circular orbit whose energy $e_N$ is the minimum of the potential.  It clear from the figure that the family of stable circular orbits has 
a minimum radius, represented by the black dot on the blue curve in the figure.  And that minimum radius is well outside the 
event horizon. Called the radius of the {\sl innermost stable circular orbit} (ISCO), it is at $r=6M$.  

\vspace{2mm} 

\noindent{\sl Circular orbits}

For a circular orbit, $\dot r = 0$ implies $e_N\equiv\frac12( E^2-1) = V_{\rm eff}$, and 
$\ddot r=0$ implies $V_{\rm eff}' = 0$. With $r$ the radius of the 
circular orbit, the second condition is 
\begin{align}
   \frac{M}{r^2} -\frac{L^2}{r^3} + 3\frac{ML^2}{r^4}&=0 \Longrightarrow
\nonumber\\
 L^2 = \frac{Mr^2}{r-3M}.
\label{e:scirc_l}\end{align}
Then Eq.~\eqref{rdot2}, $e_N= V_{\rm eff}$ implies
\beq
    E^2 = \frac{(r-2M)^2}{r(r-3M)}, 
\label{e:scirc_e}\eeq
and the Kepler relation between $\Omega$ and $r$ is  
\begin{align}
  \crv \Omega &= \frac{\dot\phi}{\dot t} =g^{\phi\phi}L\,\frac1{-g^{tt}E}
		= \frac1{r^2} \sqrt{\frac{Mr^2}{r-3M}}\ 
		  (1-2M/r)\frac{\sqrt{r(r-3M)}}{r-2M} 
\nonumber\\
  &\crv = \sqrt{\frac{M}{r^3}},
\end{align}
identical, with this choice of radial coordinate, to the Newtonian relation. 
For later reference, we will use $\omega_\phi:=\dot \phi$, the orbital frequency measured in terms of the proper time $\tau$ of an observer in circular orbit: 
\be
   \omega_\phi = \frac1{r^2}L = \sqrt{\frac M{r^2(r-3M)}}.
\label{omega_phi}\ee 

Eqs.~\eqref{e:scirc_l} and \eqref{e:scirc_e} allow circular orbits for 
every $r>3M$.  For $r<6M$, however, the orbits are unstable: $V_{\rm eff}''<0$.
They correspond to the maxima of the potentials in Fig.~\ref{f:hartlepot}.    
Using $V_{\rm eff}' = 0$ for a circular orbit, we have 
\begin{align}
  r^4 V_{\rm eff}''&= \frac d{dr}\left(r^4\frac{dV_{\rm eff}}{dr}\right) 
	= \frac d{dr}\left(Mr^2 -L^2 r + 3ML^2 \right) = 2Mr-L^2 
	 = 2Mr - \frac{Mr^2}{r-3M} \nonumber\\
 	& = \frac{Mr(r-6M)}{r-3M},    
\label{Veff''}\vspace{-3mm}\end{align} 
positive for circular orbits only for $r>6M$. As claimed, the innermost 
stable circular orbit is at \vspace{-3mm}
\index{innermost stable circular orbit (ISCO)!Schwarzschild spacetime}
\index{Schwarzschild spacetime!innermost stable circular orbit (ISCO)}

\be 
\crv r_{\rm ISCO} = 6M.
\ee 
For a 1.4 $M_\odot$ neutron star, this is at \mbox{$r=6(1.4)(1.477\,\rm km) = 12.4$ km}, either just outside or 
just inside the surface of the star: We haven't yet measured neutron star 
radii accurately enough to know. 

Particles do not move along the unstable orbits with $r<6M$; they plunge into the black hole. 
Formally, however, along the family of unstable orbits, as the orbital radius decreases, the 
orbital speed increases, and the smallest circular orbit, at $r=3M$, corresponds to the 
largest speed: As we will see, it is the unstable circular orbit of a photon. 
From Eq.~\eqref{Veff''}, the unstable orbits are then in the range $3M \leq r\leq 6M$. \\

A key goal of the future space-based gravitational-wave observatory LISA is to observe 
gravitational waves from stellar-size black holes ($<100 M_\odot$) spiraling 
in to supermassive black holes in the centers of galaxies, as the binary systems lose energy to gravitational waves.\index{LISA}\index{binary system}\index{black hole!binary}\index{inspiral}\index{gravitational waves!extreme mass ratio inspiral}\index{extreme mass-ratio inspiral}
An {\sl extreme mass-ratio inspiral} (EMRI) is accurately modeled as a point particle 
orbiting the large black hole.  For an eccentric orbit about a spherical black hole, more  gravitational radiation is emitted when the particle is closest to the black hole, and that leads to 
a nearly circular orbit long before the particle reaches $r=6M$. When it reaches $r=6M$, 
it plunges into the black hole.  As a result, a good approximation to the total energy 
emitted in gravitational waves during the inspiral is the decrease in the energy of a 
circular orbit between $E=1$ at infinity and $E$ at $r=6M$. This is the binding energy 
$E_B$ of the orbit at $6M$:\index{binding energy!orbital}   From Eq.\eqref{e:scirc_e}, 
\begin{align}
  E_B &= 1-E = 1 - \frac{r-2M}{r(r-3M)} = 1 - \frac{r-2M}{\sqrt{r(r-3M)}}
	= 1 - \frac{4M}{\sqrt{6M(6M-3M)}} \nonumber\\
	&=  1-\sqrt{\frac89} \approx 0.06.
\label{e:gwemri}\end{align}
That is, about 6\% of the particle's rest mass is radiated in gravitational waves.  \\

\noindent{\sl General bound orbits}  \\

At large radii, it's clear from the effective potential for radial motion that a 
particle in bound orbit moves periodically between a minimum radius (periastron)  
and maximum radius (apastron) in its orbit about a star or black hole. 
\index{periastron}\index{apastron}
(For orbits about the Sun, the more familiar terms perihelion and aphelion are from Helios, god of the Sun).  At large distances from neutron stars or black holes, 
the orbits are nearly the Keplerian elliptical orbits, differing only by a slow 
precession, with $\phi$ increasing by an angle $\Delta\phi$ slightly larger than 
$2\pi$ between each periastron.  \\

\noindent{\sl Zoom-whirl orbits} \\
The unstable circular orbits at the maximum of the effective potential turn out to have an important implication
for future observations of LISA and perhaps of ground-based observatories.  
The figure on the left below shows the effective potential with a value of $E$  corresponding to 
an unstable orbit at $r_c \approx 5M$.  From the figure you can see that 
another orbit has this critical energy: A particle with energy $E$ can move 
from a maximum value of $r$ near $30 M$ to a minimum value at $r_c$. 
Like any ball with exactly the energy needed to reach the top of a hill,  
the particle takes an infinite amount of time to reach $r_c$.  But, as $r\rightarrow r_c$, 
$\ \Omega = \frac{d\phi}{dt} \rightarrow \sqrt{M/r_c^3}$, a constant, finite value.
So the particle spirals around the black hole an infinite number of times as 
it approaches $r_c$. 
 
\begin{figure}[h!]
\centering\includegraphics[width=\textwidth]{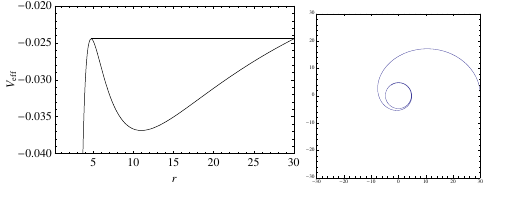}
\caption{From Healy et al.\cite{hls09}}
\end{figure}

More generally, for noncircular orbits, when $e_N$ is near the peak of the potential, 
$\dot r$ is near zero for $r$ near $r_{\mbox{\scriptsize min}}$, while $\dot\phi$ has its largest value (because $r$ has its smallest value for the orbit).  So the particle again circles the black hole several times before 
going back out.   The figure below shows the resulting {\sl zoom-whirl} orbit for a single whirl between each zoom.
\begin{figure}[H]
\centering\includegraphics[width=.45\textwidth]{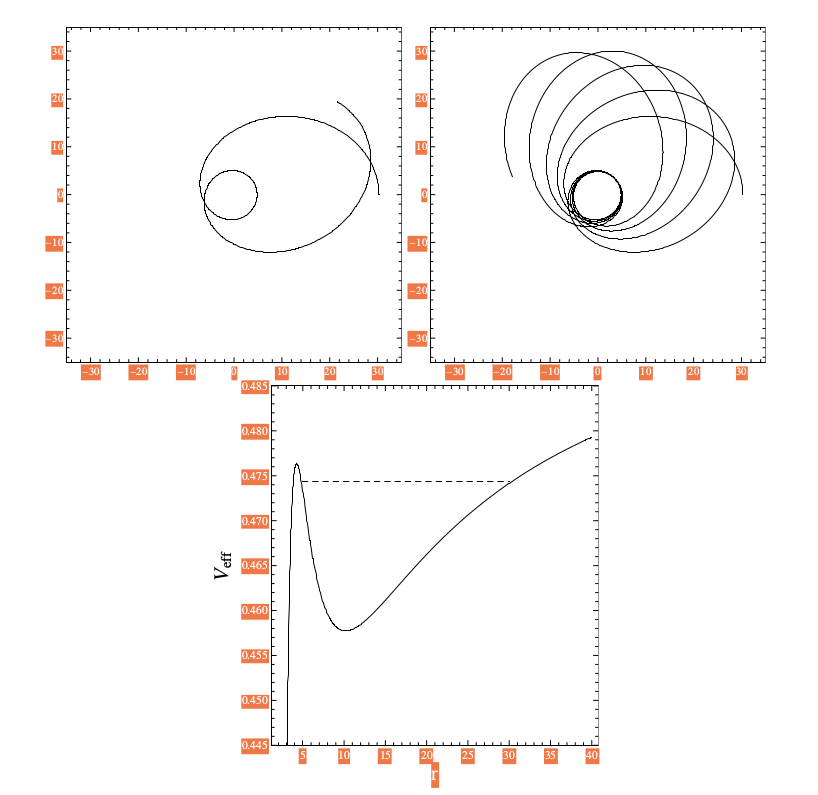}
\caption{Source of figure not known to JF}
\end{figure}  
Zoom-whirl orbits, then, are particle orbits that graze the black hole, 
orbits with an impact parameter just large enough to avoid being swallowed.  
Typical comets in the 
solar system are the result of chance encounters of ice balls (in the Kuiper belt 
and Oort cloud) with passing stars that nudge them in a direction that removes 
enough of their orbital energy that they fall toward the Sun.  
Because orbital speed at that distance is already small, they start their 
infall in nearly parabolic orbits with the small impact parameter 
associated with their small initial speed.  
Zoom-whirl orbits are most likely to arise from a similar scenario:  
Stars initially not too 
near a galactic black hole that are nudged into a trajectory that grazes  
the black hole have the best chance of being in 
the zoom-whirl parameter range.  \\

\noindent{\sl Precession Angle}\\

\index{periastron!precession}\index{precession!of periastron}
We now look at the precession angle $\Delta\phi_P = \Delta\phi-2\pi$ for nearly circular orbits, with $\Delta\phi$ the angle through which the particle moves 
between successive periastrons.  The radial component of the geodesic equation
has the form 
\be 
   \ddot r = - V_{\rm eff}' .
\ee 
If this is not obvious, note that, as in the Newtonian case, differentiating the expression for the conserved energy, Eq.~\eqref{rdot1},  
gives the equation of motion: 
\[
 \dot r\ddot r = \frac d{d\tau}(\frac12\dot r^2) 
	      =  -\frac d{d\tau}V_{\rm eff}(r(\tau)) 
	      =  - V_{\rm eff}' \dot r\ \Longrightarrow \ \ddot r = - V_{\rm eff}'.
\]

The radial period is the coordinate time 
$t=T$ between periastrons;  recall that $t$ is Killing time and is the time measured 
by an observer at infinity whose velocity coincides with $t^\alpha$.  Let 
$r$ be the radius of a circular orbit and $r+\delta r(\tau)$ the radius 
of a particle in a nearby orbit with the same angular momentum $L$.  For a Newtonian system, $r+\delta r(t)$ is 
the radial coordinate of a nearby elliptical orbit. Using $V_{\rm eff}'(r)=0$, we 
have  
\begin{align*}
  \delta\ddot r = -V_{\rm eff}'(r+\delta r) &=-V_{\rm eff}''(r)\delta r +O(\delta r^2), \ \ 
\\
 \mbox{or} \qquad \delta\ddot r + \omega_r^2\,\delta r &=0,  
\end{align*}
with 
\be
   \omega_r := \sqrt{V_{\rm eff}''(r)} = \sqrt{\frac{M(r-6M)}{r^3(r-3M)}}.
\label{omega_r}\ee  
Then $\delta r$ oscillates with frequency $\omega_r$ defined in terms 
of proper time $\tau$ along the circular geodesic.  The corresponding frequency 
$\Omega_r$ in terms of Killing time is $\Omega_r = \omega_r/\dot t$ and 
the radial period is $T=2\pi/\Omega_r$.  In one radial period, the change in 
angle is 
\begin{align*}
  \Delta \phi &= \Omega T = 2\pi \frac{\Omega}{\Omega_r} 
		= 2\pi\frac{\omega_\phi}{\omega_r} 
		= 2\pi \sqrt{\frac{M}{r^2(r-3M)}\, \frac{r^3(r-3M)}{M(r-6M)}}\\
		&=2\pi\frac1{\sqrt{1-6M/r}}
\end{align*}
where we have used \eqref{omega_phi} and \eqref{omega_r} for $\omega_\phi$ and 
$\omega_r$. The angle of precession is then 
\be\crv
   \Delta \phi_P = 2\pi\left(1-\frac1{\sqrt{1-6M/r}}\right)\cb.   
\ee
For $M/r$ small, it becomes 
\be
  \crv \Delta \phi_P = 6 \pi \frac{M}{r}\cb.
\ee

In particular, Mercury's orbit is nearly circular with $r = 5.55\times 10^7$ km.  
The mass in the formula is that of the Sun, $ M = M_\odot = 1.477$ km, giving
\begin{align} 
\Delta \phi_P &= 6 \pi \frac{1.477\,\rm km}{5.55\times10^7\,\rm km} = 4.99\times 10^{-7} {\rm radians/orbit},\nonumber\\ \nonumber\\ 
 \Delta \phi_P 
 	&= \frac{4.99 \times 10^{-7}}{.24{\rm yr} } \times \left(\frac{180}\pi\times 3600''\right)
\nonumber\\
 	& = 0.43''/{\rm yr}  = 43''\text{/century.}
\end{align}

At the other extreme, for $r$ near $6M$, the precession angle is large, and the particle follows a zoom-whirl orbit, circling the black hole many times in each radial period. In this case there's not much of a zoom, because of our $\delta r \ll r$ assumption.     
\newpage

\noindent{\em Photon Orbits: Null Geodesics}\\ 
\index{effective potential!Schwarzschild spacetime, photon orbits (null geodesics)|textbf}
\index{photon!orbits in Schwarzschild spacetime}\index{Schwarzschild spacetime!photon orbits (null geodesics)|textbf}

The effective potential for null geodesics is simpler: Writing $(^\cdot)$ for $d/d\lambda$ with $\lambda$ an 
affine parameter,\index{affine parameter} we have   
\bea
 0 = k^\alpha k_\alpha  &=&  -\frac1{\displaystyle 1-\frac{2M}r}\ E^2 
   	     	+\frac{ L^2}{r^2} + \frac1{\displaystyle 1-\frac{2M}r}\ \dot{r}^2\nonumber\\
     	\nonumber\\
\frac12\dot{r}^2 &=& \frac12 E^2 - V_{\rm eff}, \quad \mbox{with}\quad 
	\cblue V_{\rm eff} = \left(1 - \frac{2M}{r}\right) \frac{ L^2}{2r^2}\cb.
\label{ngeod}\eea
\index{effective potential!Schwarzschild spacetime, null geodesics|textbf}

\noindent{\sl The Photon Sphere}\\
\index{photon sphere}
Like massive particles, photons moving inward with large $E$ are trapped by 
the black hole, even if they have nonzero angular momentum.  And there 
is a surprise: The potential has a maximum at $r=3M$, implying an unstable 
circular orbit for photons at this radius. \\
Check:  
\beq
0=\frac1{L^2}\ \frac{d}{dr}V_{\rm eff} = \frac{d}{dr}\left(\frac{1}{2r^2} - \frac{M}{r^3}\right) 
	= -\frac1{r^3} + \frac{3M}{r^4}  \Longrightarrow r = 3M~.
\eeq
One calls $r = 3M$ the {\em photon sphere}.

Photons coming in from infinity hit the black hole when $\dot r >0$ at the maximum 
of $V_{\rm eff}$, and miss it when $\dot r = 0$ before they reach the top of the 
potential.  At its $r=3M$ peak, 
\[
  V_{\rm eff, max} = \frac{L^2}{54 M^2},
\] 
and photons miss the black hole if and only if 
\beq 
   V_{\rm eff, max}  > \frac12E^2\ \Longleftrightarrow \ \frac{L^2}{54 M^2} >\ \frac12 E^2 \Longleftrightarrow
	b := \frac LE > 3\sqrt{3}\ M.   
\eeq 
That is, the critical impact parameter for which photons just reach the top 
of the potential -- spiraling an infinite number of times about the $r=3M$ 
circular orbit -- is $b =3\sqrt{3}\ M$.  The corresponding absorption cross section  is then $\sigma = \pi b^2 = 27 \pi M^2$.  

In the images of black holes recently obtained by the \href{https://eventhorizontelescope.org/}{EHT} (Event Horizon Telescope) most of the light passes the black hole near the photon sphere.

\begin{figure}[H]
\begin{center}
\includegraphics[width=0.6\textwidth]{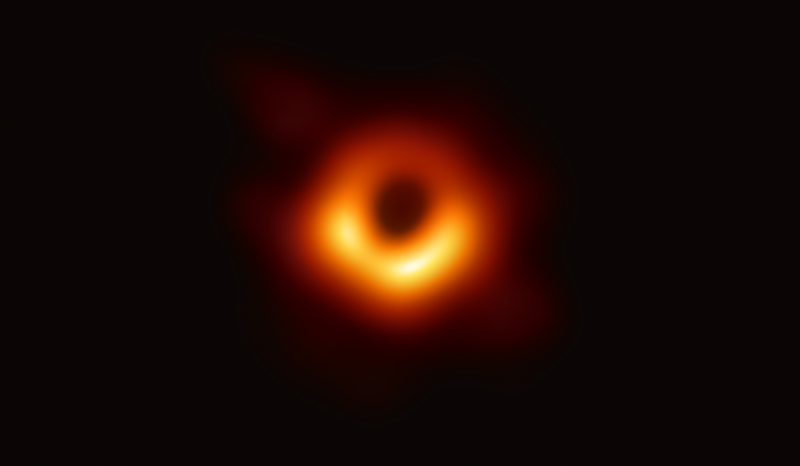}
\caption{EHT image of the black hole at the center of the giant elliptical galaxy M87}
\end{center}
\end{figure}
\index{Event Horizon Telescope}
\index{horizon!EHT image}

\noindent{\sl Deflection of Light}\\
\index{deflection of light}\index{Schwarzschild spacetime!deflection of light}

Null geodesics that do not hit the black hole or star follow a symmetric trajectory 
that looks similar to a hyperbolic orbit, with the angle between the ingoing and outgoing asymptotes called the deflection angle $\Delta\phi$.  
\begin{figure}[H]
\includegraphics[width=.7\textwidth]{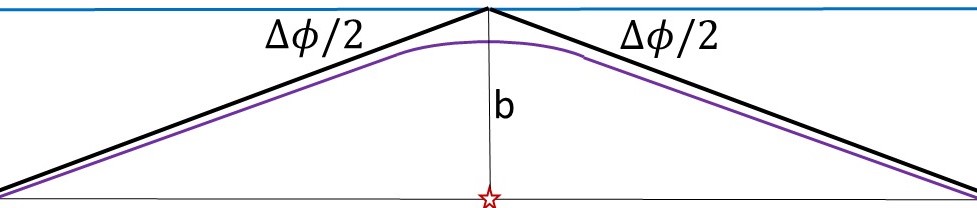}
\caption{Flat-space trajectory in blue, actual trajectory in purple.} 
\end{figure}
In the figure, the path chosen to make the diagram symmetric about the point of closest approach.    

\noindent {\em Mathematical Prerequisite}

\noindent
$\dis \hspace{.8in} y'' + y = \sin^2 x \qquad \mbox{has solution}\qquad 
	y = \frac13(2-\sin^2x ) $. \\  

\noindent{\em Spatial Path}\\
The spatial path, $r=r(\phi)$ can, as usual be found by writing  
\[ 
r'\equiv \frac{dr}{d\phi} 
	= \frac{\dot r}{\dot\phi} 
	= \frac{\dot{r}r^2}{ L} \Rightarrow \dot{r} = \frac{ L}{r^2}~r'.
\]
It is convenient to use 
\[ 
	u = \frac{1}{r}, \hspace{.5in} u' = -\frac{r'}{r^2} = -\frac{\dot{r}}L.
\]
In terms of $u$, the geodesic equation \eqref{ngeod} is
\be
 	u'^2 = \frac{E^2}{L^2} -u^2 + 2Mu^3\\
\label{u'}\ee
Differentiate: 
\begin{align*}
 	2u'u'' &= - 2u'u + 2u'3Mu^2\\
	u'' +u & = 3Mu^2 
\end{align*}

We'll solve this to order $M/b$, with $b= L/E$ the impact parameter.  We first find the zeroth-order 
(flat space) solution:

At zeroth order, with $M=0$, we are in flat space, and we choose as the geodesic a straight line at constant $y=r\sin\phi$ : 
\[
u'' + u = 0
\]
has a solution 
\[
u = \frac1b\sin\phi, \quad \mbox{or}\quad  y=r\sin\phi = b, 
\]
with $r$ ranging from $\infty$ to $-\infty$ as $\phi$ runs from $0$ to $\pi$, 
with minimum value $b$ as claimed. Far from the star, space is flat, and we have just checked that 
the flat-space trajectory with $b=L/E$ has $b$ as the distance of closest approach.\\

\noindent{\sl Perturb: Order $M/b$ correction}.\\
  Let $u+\delta u$ be the perturbed solution, with $\delta u$ of order $M/b$.  Then to linear order in $M/b$ we have 
\begin{align*}
\delta u'' + \delta u &= \frac{3M}{b^2}\sin^2\phi  \\
\delta u &= \frac M{b{}^2}(2- \sin^2\phi )\qquad \mbox{from the prerequisite above}\\
u + \delta u &= \frac1b\sin\phi + \frac{2M}{b^2} -
\frac M{b^2} \sin^2\phi\,.
\end{align*}
The perturbed orbit starts $\phi = -\Delta\phi/2$ with the particle at $r+\delta r =\infty$, $u+\delta u = 0$.  As the photon moves from right to left, $\phi$ increases to $\pi/2$ at maximum $u+\delta u$ and then increases from $\pi/2$ to $\pi+\Delta\phi$ as the photon flies out to infinity and $u+\delta u$ symmetrically decreases to $0$.   
To find $\Delta\phi$ we just write 
\begin{align}
	0 = u + \delta u &= \frac{1}{b}\sin(-\Delta\phi/2)  
	      + \frac M{b^2}\left[2-\sin^2(-\Delta\phi/2\right] \nonumber\\
	 &= -\frac{\Delta\phi}{2b} +\frac {2M}{b^2} + O((\Delta\phi)^2) \nonumber\\
 \Longrightarrow\  \crv\Delta\phi & \crv= 4~\frac M{b}.
\end{align}

For light grazing the Sun,  $M=M_\odot = 1.477$ km, $b = R_\odot = 6.96 \times 10^5$ km, implying
\bea
  \Delta\phi &=& 4 \, \frac{1.477}{6.96\times 10^5}\left(\frac{180}{\pi}\times 3600\right) 
								\nonumber\\  
	 &=& 1.75''. \nonumber
\eea
This, of course, is the value first measured by the 1919 eclipse expedition led by Frank Dyson and Arthur Eddington. An interesting detailed history by Jos\'e Lemos et al.:   \href{https://arxiv.org/pdf/1912.08354.pdf}{arXiv:1912.08354}.

     It is not too hard to find the exact deflection as an integral.  
Following Wald, we solve Eq.~\eqref{u'} for $u'$ to write
\be
  \Delta\phi = 2\int_0^{1/R_0} \frac{du}{\sqrt{b^{-2} -u^2+2Mu^3}},
\ee 
where $R_0$ is the minimum value of $r$ along the exact trajectory, where $u'=0$.  
To find $R_0$, we have to solve the cubic equation 
$2Mu^3-u^2+b^{-2} =0$ or $r^3-b^2 r+2M b^2 =0$ to obtain
\be
	\dis R_0 = \frac{2b}{\sqrt 3} \cos\left[\frac13\cos^{-1}\left(-\frac{3^{3/2}M}b\right)\right].
\label{e:minimumr}\ee

\index{cubic equations}
If you have never solved a cubic, or never seen the solution in this form, 
here is how to do it:   First change variables to get rid of any quadratic term -- in our case it's already gone. The next step isn't necessary, but it gives a
canonical form with a solution that's easy to understand:  Rescale $r$ to 
make the coefficient of the linear term -3:  $\dis y = \sqrt 3 \frac rb$, giving 
 \\
\centerline{$y^3 - 3y + q =0$, with $\dis q= \frac{3^{3/2}M}b$. }\\
Now, the key step: Write $y= a+b$, $y^3 = 3ab(a+b) + a^3+b^3$.  
Our cubic then becomes\\
 \mbox{$3ab\, y + a^3+b^3 -3\,y +2q =0$}, satisfied 
when $ab=1$, $a^3+b^3 +2q=0$. Then $b=1/a$, and $a^3+b^3 +2q=0$ becomes   
 $(a^3)^2 + 2q a^3 + 1 = 0$, a simple quadratic equation for $a^3$:
Here $q<1$, and the roots are 
$\dis   a^3 = -q\pm i\sqrt{1-q^2}, \ \ \mbox{with}\quad |q\pm i\sqrt{1-q^2}| = 1$.
Because the roots have unit norm, we can write 
$a=e^{i\psi}$, $b=e^{-i\psi}$. Then $a^3+b^3 +2q=0 \Longrightarrow 2\cos3\psi + 2q = 0$, 
and we have
\[
 y=a+b=2\cos\psi = 2\cos\left[\frac13\cos^{-1}(-q)\right], \quad \mbox{giving the claimed expression for $R_0$}.
\]

By the end of the last century, light deflection in the guise of gravitational lensing began to play an important role in astronomy, in particular for measurements of the mass of dark matter halos that dominate the mass of galaxies and clusters of galaxies.  DES (the Dark Energy Survey) uses lensing in a survey of 26 million galaxies to improve the precision of cosmological parameters (see, for example, 
\href{https://room.eu.com/news/most-accurate-measurement-of-universes-dark-matter-revealed-by-des-survey}{DES}, \href{https://www.darkenergysurvey.org/des-year-3-cosmology-results-papers/}{DES papers}).  \\

\noindent{\sl Shapiro time delay}
\index{time delay, Shapiro}\index{Shapiro time delay}
    
\begin{figure}[h!]
\centerline{\includegraphics[width=.7\textwidth]{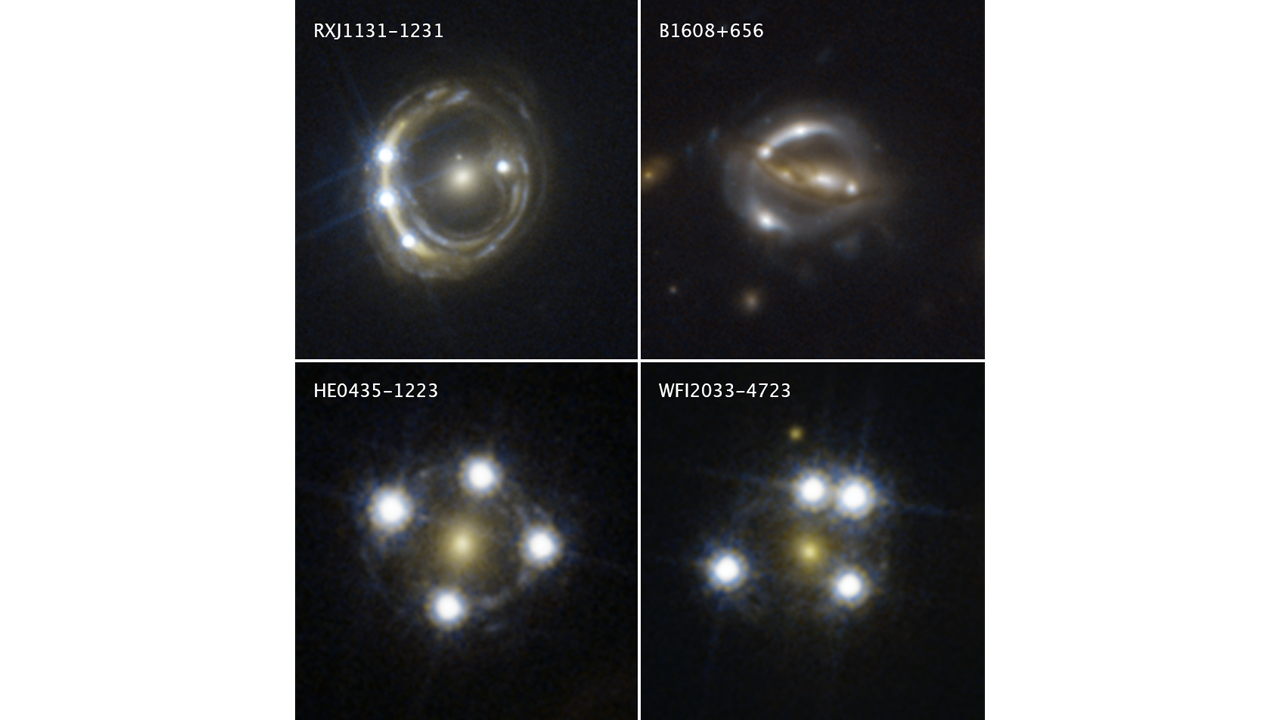} }
\caption{Each photo in the figure shows multiple \href{https://hubblesite.org/contents/media/images/2020/04/4606-Image}{Hubble images} of a single quasar. }
\end{figure}

Many quasars are seen as more than one image, their light bent around intervening 
galaxies or clusters of galaxies.  
In the images, light from a quasar arrives at different times, depending on the path.  Quasar intensity changes in less than a day, implying an emitting region less than a light-day across; that fact made accretion onto a black hole the leading candidate for the engine.  The same bright flares can be identified in different images, arriving months or years apart:  For example the same flare has been identified in 3 of the 6 images of SDSSJ1029+2623 (SDSS for a Sloan Digital Sky Survey object), with a 48-day and a 722-day delay between the brightest and two of the dimmer images.\href{https://arxiv.org/pdf/1505.06187.pdf}{Dahle et al. 15}  

First computed and measured by Irwin Shapiro  
\href{https://journals.aps.org/prl/abstract/10.1103/PhysRevLett.13.789}{PRL}\cite{shapiro64}, the delay has provided an accurate test of GR, first in the solar system and more recently in double pulsars.  It is also used in measurements of neutron star masses.  The presentation here uses {\sl isotropic coordinates} for Schwarzschild. It just involves changing to a new radial coordinate (still written here as $r$) for which the metric has the form 
\be 
\crv ds^2 = - \left(\frac{1-M/2r}{1+M/2r}\right)^2 dt^2 + (1+M/2r)^4 (dx^2+dy^2+dz^2)\cb, 
\label{e:sch_isotropic}\ee
with $r^2=x^2+y^2+z^2$. 
To get to this form starting from the standard radial coordinate $r_S$, 
write 
\[
   \left(1-2M/r_S\right)^{-1} dr_S^2 + r_S^2 d\Omega^2 = F^2 (dr^2 + r^2 d\Omega^2);  
\]
then equate the coefficients of $d\Omega^2$ and require 
$ \left(1-2M/r\right)^{-1} dr_S^2 = F^2 dr^2 $.  (Filling in the steps is 
\ref{p:sch_isotropic} below.)  

We again look at the same geodesic with impact parameter $b$ and calculate the 
correction to the travel time between the emitter (which could be a pulsar 
or a planet) at initial large radius $r_p$ and the Earth at large radius $r_e$.  
The fractional correction to the travel time will be of order $M/(r_e+r_p)$ and can 
be found from the flat-space path:  The curved path deviates from the 
flat trajectory by an angle of order $O(b/M)$, its spatial length changing by 
$O\left((r_e+r_p)b^2/M^2\right)$: Think of the length of the hypotenuse of a right triangle of $r_p$ and angle $M/b$. The corrected time delay from the change in path length is then of order $(M/b)^2(r_e+r_p)$, and can be ignored.    

We take as the spatial trajectory the path $y=b$, with $x$ running from $x=-r_p$ 
to $x = r_e$, with $x=0$ the point of closest approach.  The null path has 
$ds=0$, with 
\be
  0=ds^2 = - \left(\frac{1-M/2r}{1+M/2r}\right)^2 dt^2 + (1+M/2r)^4 dx^2.
\ee
Integrating to find the time from the source to the point of closest approach and the time from closest approach to Earth and adding them gives 

\be
     t = r_p+r_e 
    + 2M\log \frac{\left(r_p+\sqrt{r_p^2+b^2}\right)\left(r_e+\sqrt{r_e^2+b^2}\right)}
		  {b^2}.
\label{e:shapiro1}\ee
For $b<<r_p, r_e$, the Shapiro delay (correction to $r_p+r_e$) is then
\be\crv
   \Delta t = 2M\log \frac{4r_e r_p} {b^2}.\cb
\label{e:shapiro2}\ee
There is also a $b$-independent delay from the redshift of a photon leaving a pulsar and 
its binary system and a blueshift as it arrives at Earth, but only the $b$-dependent part is 
measured.  
\begin{figure}[h!]
\centerline{\includegraphics[width=.55\textwidth]{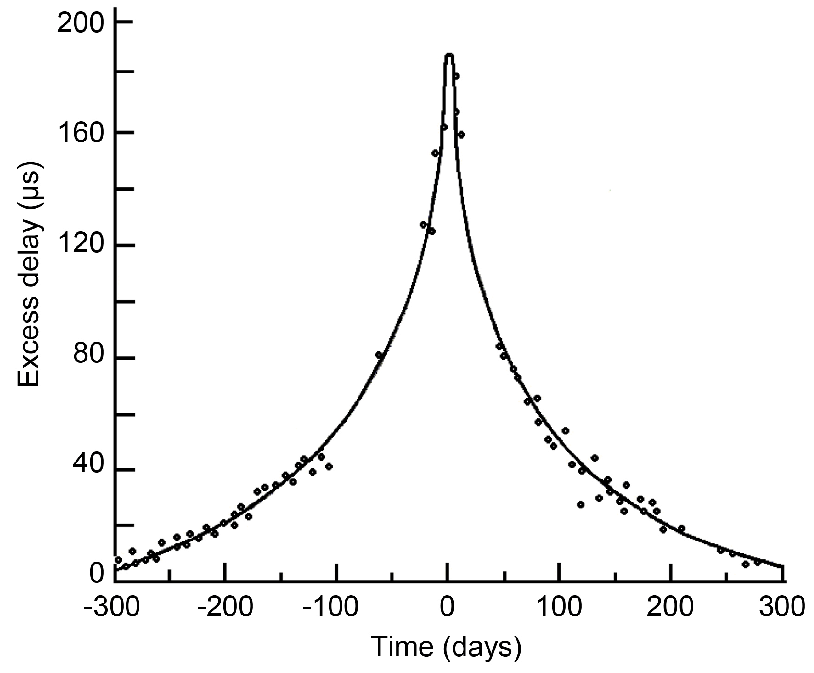} }
\caption{The measured delay of light traveling from Earth to Venus to Earth: Radar bounced off Venus as Venus passes the Sun. From Jacob Schaf, \href{https://www.scirp.org/pdf/JMP_2017022715463111.pdf}{Journal of Modern Physics, 2017}\cite{schaf17}}
\end{figure}  

A final photon orbit, that of a radially infalling photon, begins the black-hole section.
\newpage

\benr
\item \label{p:sch_isotropic} (Schwarzschild in isotropic coordinates).  Complete the calculation above to 
obtain the form \eqref{e:sch_isotropic} of the Schwarzschild metric in isotropic 
coordinates. 

\item Complete the computation of Eqs.~\eqref{e:shapiro1} and \eqref{e:shapiro2} to show that 
the Shapiro time delay of a photon traveling from a pulsar to the earth and 
passing another star of mass $M$ along the way has the form \index{neutron star!pulsar}\index{pulsar!Shapiro time delay}
\[
   \Delta t = 2M\log \frac{4r_e r_p} {b^2}  + \mbox{terms smaller by factors}\ \ 
	\frac b{r_e}, \frac b{r_p}, \frac bM.  
\]
Ignore the effect of the pulsar's and the Earth's gravity.  $M$ is the mass of 
the intervening star, and $r_p$ and $r_e$ are the distances from the pulsar to 
that star and from the Earth to that star.  \\
An optional variant: Wald Chapter 6 Problem 5, finding the exact form of $\Delta t$.     

\item\label{p:buchdahl}  The smallest possible value of $\dis 2M/R$ at the surface
of a spherical star is attained when the star is perfectly
incompressible -- when it is one of the uniform density models you constructed. 
\ben\item[a.]  Show for this sequence of stars that the central density becomes 
infinite when $\dis\frac{2M}{R}=\frac{8}{9}$.
\item[b.] Deduce from (a) the maximum redshift $z$ from the surface of a spherical star.\\ $\dis 1+z := \frac{\lambda_\infty}{\lambda_{\rm star}} = \frac{\omega_{\rm star}}{\omega_\infty}$.
\een 
\index{redshift!maximum gravitational redshift}

\item  For spherical (i.e. nonrotating) stars, there is an exact solution to the Newtonian 
equations of hydrostatic equilibrium
$\dis
 \nabla P =-\rho\nabla \Phi, \qquad \nabla^2\Phi = 4\pi\rho
$,
with the equation of state $P = K\rho^2$.  Show that 
$\dis\rho = \rho_c\frac{\sin(\alpha r)}{\alpha r}$ and the corresponding $\Phi$ satisfy the system of equations, find $\alpha$, and find the radius of the star 
for $K=150\ \rm km$.  \\
{\sl Hint}: Write $\nabla^2\Phi$ in the form $\dis \frac1r\frac{d^2}{dr^2}(r\Phi) = 4\pi\rho$, and solve for $r\Phi$, with $\Phi=0$ at infinity.

\item   {\sl Numerical model}.  
\index{numerical model of star!Newtonian}
A model of a neutron star more realistic than the 
uniform-density model is an equation of state of 
the form $P=K\rho_0^2$, where $\rho_0$ is the baryon mass density, called a polytropic equation of state.
To make things simpler, use $P=K\rho^2$ as in the last problem, but this time for a general relativistic 
model.  In gravitational units with km the only unit, take 
$K=150\ \rm km^2$. 
\benalph
\item A typical central density of a neutron star is about $10^{-3}\rm km^{-2}$.  What is that 
in $\rm g/cm^3$?  
\item Starting with this central density, integrate the TOV equation from the center of the 
star to the surface, determined by the requirement that $P$ vanish.  Find the radius of the 
star and plot $P$ and $\rho$ as functions of $r$.  The equation of hydrostatic equilibrium looks like 
$0 = 0/0$ at $r=0$, so start near $r=0$ and integrate outward, with $\rho=\rho_c$, $P = P(\rho_c)$ 
and $m=\frac43 \pi r_0^3 \rho_c$ as initial values.  What is $M/M_\odot$?  

Because $P\propto \rho^2$, it never becomes negative.   If you replace $dP/dr$  by   $d \rho/dr$   and integrate to compute $\rho(r)$,  you can find the surface more easily: $\rho$ becomes negative when $r$ is larger than the radius of the star.  
\een

\item Upper limit on the mass of a star with $\rho > \rho_n$, where $\rho_n$ is nuclear density, 
about $2.7\times 10^{14}$ g/cm$^3$ (Hartle-Sabbadini).   Use the inequality 
$\rho > \rho_n$ and the equation for the mass of a relativistic star in terms of $\rho$ 
to show that $M> K \rho_n R^3$, and find $K$.  Combine this inequality with the requirement that 
the star is not inside a black hole, $R> 2M$ to find an upper limit on $M$ in terms of $\rho_n$.  
Evaluate $M_{\rm max}/M_\odot$ for $\rho_n = 2.7\times 10^{14}$ g/cm$^3$.

\een
\newpage


\chapter{Black Holes} \index{black hole}
\section {Schwarzschild Black Holes}
\index{black hole!Schwarzschild spacetime}
The Schwarzschild metric \eqref{e:schwarzschild},\index{metric!Schwarzschild vacuum}

\beq
ds^2 = -\left(1 - \frac{2M}r\right)~dt^2 + \left(1 - \frac{2M}r\right)^{-1}~dt^2 +r^2d\Omega ^2, 
\label{ds2}\eeq
provides a smooth spacetime in the region

\noindent I: \indent $-\infty < t < \infty $ \hspace{.5in} $2M < r < \infty
$\\
and another (disconnected) spacetime corresponding to the
coordinate values\\
\noindent II: \indent $-\infty < t < \infty $ \hspace{.5in} $0 < r < 2M $.\\
\indent In region I, $t^\alpha =\bm\partial_t$ is a timelike Killing vector, 
but in region II, $g_{tt} = t^\alpha t_\alpha > 0$, so that $t^\alpha$, 
although still a Killing vector, is no longer timelike.  Similarly,
$\partial_r$ is spacelike in I and timelike in II.  In obtaining the form
(\ref{ds2}), we assumed the existence of a timelike Killing vector orthogonal to a
family of spacelike ($t$ = constant) hypersurfaces, but in the region $r <
2M$ the field is so strong tht even light is constrained to move inward --
no matter can remain at rest (i.e. at constant $r$), and everything follows 
trajectories that end at $r = 0$.\\

 Remaining at rest means moving along $t^\alpha$, so the fact that
$t^\alpha$ is spacelike is exactly the statement that physical particles
(which must move along timelike trajectories) cannot remain at rest.  One
might hope to join regions I and II by relaxing the condition that
there be coordinates $\{t,x^i\}$, with $\partial _t$ a Killing vector and $t$ =
constant a spacelike surface.  Before trying this, it is helpful to learn
about the geometry outside $r = 2M$, and one way to do it is to look at the
geodesics (another way is to embed the geometry in some ${\mathbb R}^n$ - 
we'll do that later).\\

\noindent{\sl  Radial photons}
\index{photon!orbits in Schwarzschild spacetime}

	We now show that radially ingoing light rays reach $r = 2M$ in finite
affine parameter length, although the coordinate $t$ becomes infinite along
their trajectories.  Similarly, radially ingoing massive particles reach $r = 2M$,
in finite proper time.  When $L  = 0$, Eq.~\eqref{ngeod} implies  $dr = Ed\lambda
$, so $r$ is itself an affine parameter.\index{affine parameter}  Thus photons travel from
8M to 2M in parameter length $\Delta r = 8M$, or $\dis\Delta \lambda =
\frac{8M}{E}$.  The coordinate $t$, however, becomes infinite as $r\rightarrow 2M$:
\[
	\left(1-\frac{2M}{r}\right)dt^2 = \left(1 - \frac{2M}{r}\right)^{-1}dr^2 .
\]
\beq
	dt = - \frac{dr}{\displaystyle 1 - \frac{2M}{r}} 
\eeq
\beq
	t = v_0 - r - 2M \log \left| \frac{r}{2M} - 1 \right|, 
\eeq
with $v_0$ the constant of integration. It is useful to introduce 
a coordinate $r_*$ (called the ``tortoise'' coordinate), writing
\beq
	t = v_0 - r_*, 
\label{tin}\eeq
\beq
 \crv r_* := \int^r \frac{dr}{1 - \frac{2M}{r}} =
r + 2M \ln\left| \frac{r}{2M} - 1 \right|\cb.
\label{rstar}\eeq
\noindent As $r \rightarrow 2M$, $r_* \rightarrow - \infty$, whence $t
\rightarrow \infty$.\\

	The same calculation shows that light emitted from $r = 2M(1 +\epsilon )$
reaches a distant observer after a time $\Delta t$ that blows up like
$|\log \epsilon |$ as $\epsilon \rightarrow 0$. From (\ref{tin}) with a + sign for
outgoing photons,
\begin{align} 
\Delta t &= r_*(8M) - r_*[2M(1+\epsilon )] \nonumber \\
	&= (8M + 2M \log 3) - [2M(1+\epsilon ) + 2M \log \epsilon ] \nonumber \\
	&=  6M -2M \log (\epsilon/3) + O(M\epsilon).
\end{align}
Thus an observer who remains a finite distance outside $r = 2M$ never sees
an infalling particle reach $r = 2M$. Moreover, the observed redshift
becomes infinite:  A photon's frequency as measured by an
observer moving along $t^\alpha$ (an observer at fixed $r, \theta, \phi $) is
$-k_\alpha u^\alpha$, with 
\index{photon!frequency}\index{frequency!photon}
\[ 
u^\alpha = \frac{t^\alpha}{\sqrt{-g_{tt}}} = \frac{t^\alpha}{\sqrt{1-2M/r}}.
\] 
\beq \omega = \frac{E}{\sqrt{1 - 2M/r}} 
	= \frac{\omega_\infty}{\sqrt{1 - 2M/r}}  
\eeq
Thus 
\beq
	\frac{\omega (r = 2M(1+\epsilon))}{\omega (r = R)} =
\left[\frac{(1+\epsilon)(1 - \frac{2M}{R})}{\epsilon }\right]^{1/2}
\rightarrow ~ \infty \mbox{ as } \epsilon\  \rightarrow\ 0.
\eeq

Although the formal expression shows light from the infalling star reaching an outside 
observer at all future times, the light comes from less than a Planck length ($10^{-33}$ cm)
of $r=2M$ surface after about a millisecond for a star of about 10 $M_\odot$ (see \ref{p:bhtime}).

The $r=2M$ surface in the Schwarzschild geometry is an {\sl event horizon}.
As we will see in the next section, no signal from events inside 
the horizon can reach the spacetime outside. 
\index{horizon!Schwarzschild} \index{Schwarzschild spacetime!horizon}

\section{Eddington-Finkelstein Coordinates}\index{Schwarzschild spacetime!Eddington-Finkelstein coordinates}
\index{Eddington-Finkelstein coordinates}\index{Finkelstein}\index{coordinates!Eddington-Finkelstein}
\label{s:ef}
	The fact that ingoing photons reach $r = 2M$ in finite value of their
affine parameter\index{affine parameter} and that ingoing finite rest mass particles reach $r = 2M$
in finite proper time suggests that one might be able smoothly to join
regions I $(r > 2M)$ and II $(r < 2M)$ with coordinates tied to infalling
particles.  This is easier to do with photons, because they have simpler
trajectories. Because the method leads to null coordinates, we'll first revisit null coordinates in Minkowski space, to better understand their meaning.

\noindent{\em Null coordinates in flat space}:\\
\index{coordinates!null coordinates}\index{null coordinates}
\indent Future null cones are $u$ = constant surfaces, where $u = t-r$ 
\begin{figure}[H]
\centering\includegraphics[width=.2\textwidth]{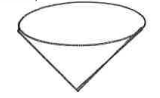}
\end{figure}

Past null cones are $v$ = constant surfaces, with $v = t+r$ 
\begin{figure}[H]
\centering\includegraphics[width=.2\textwidth]{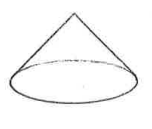}
\end{figure}

Then
\[ t = \frac{u+v}{2}~,~~~r = \frac{v-u}{2} \]
Outgoing null coodinates: $u, r, \theta , \phi $
\begin{eqnarray}
 ds^2 &=& -dt^2  + dr^2 + r^2d\Omega^2 
 	= -(du+dr)^2 + dr^2 + r^2d\Omega^2 \nonumber \\
 &=& \cblue -du^2 - 2dudr + r^2d\Omega ^2 
\end{eqnarray}
Ingoing null coordinates: $v, r, \theta , \phi $
\beq
	ds^2 = -dv^2 + 2dvdr + r^2d\Omega ^2.
\label{gin}\eeq 
Double null coordinates:  $u, v,  \theta , \phi $
\begin{eqnarray}
ds^2 &=& -[\frac{1}{2}(du+dv)]^2 + [\frac{1}{2}(dv-du)]^2 + r^2(u,v)d\Omega^2 \nonumber \\
ds^2 &=& -dudv + r^2d\Omega^2. 
\label{gnull}\end{eqnarray}
In this chart, $r$ is a function of $u$ and $v$: $\ r^2 = \frac{1}{4}(u-v)^2$.

In the ingoing null chart $v, r, \theta , \phi$, the path $v, \theta , \phi
$ = constant  is the trajectory of a radially
ingoing photon, and $r$ is an affine parameter along the path.\\

\noindent{\em Ingoing and outgoing null coordinates in Schwarzschild}\\
To mimic
this in the Schwarzschild geometry, one must take 
\be
	v = t + r_* ,
\label{e:ef0}\ee
because by (\ref{tin}) the trajectory of an ingoing photon is
\[ t + r_* = \text{constant,} ~~~ \theta , \phi ~\text{constant.} \]
Again, $r$ is an affine parameter along the geodesic so we try a ($v, r,
\theta , \phi $) chart:
\begin{eqnarray*}
ds^2 &=& -\left(1 - \frac{2M}r\right)dt^2 + \left(1 - \frac{2M}r\right)^{-1}dr^2 + r^2d\Omega^2
								\nonumber \\
&=& -\left(1 - \frac{2M}r\right)(dv- \frac{dr_*}{dr}dr)^2 
 +\left(1 - \frac{2M}r\right)^{-1}dr^2 + r^2d\Omega^2 
								\nonumber \\
&=& -\left(1 - \frac{2M}r\right)\left[dv - \left(1 - \frac{2M}r\right)^{-1}dr\right]^2 
	+ \left(1 - \frac{2M}r\right)^{\!-1}dr^2 + r^2d\Omega ^2~, 
\end{eqnarray*}
where we used Eq.~(\ref{rstar}) to write 
$\dis \frac{dr_\ast}{dr} = \left( 1- \frac{2M}{r}\right)^{-1}$.  
The form of the metric is surprisingly simple:

\beq
\crv ds^2=-\left( 1-\frac{2M}{r}\right) dv^2 + 2dvdr + r^2d\Omega^2\cb .
\label{e:e-f-in}\eeq
The metric components are well behaved for $ 0< r < \infty$.  For $r\gg 2M$, 
(\ref{e:e-f-in}) looks like the flat space form (\ref{gin}) for a single null coordinate.  
For $r=2M$ it is
\beq
 ds^2 = 2dv dr+r^2d\Omega^2,
\label{e:nullflat}\eeq
equivalent after a rescaling to the form (\ref{gnull}) for two null coordinates; 
and for $r < 2M$, the coefficient of $dv^2$ is 
positive.

The sign change in $g_{vv}$ shows that that the ``timelike'' Killing vector, 
$t^\alpha = \bm\partial_v$, becomes null at $r=2M$ and spacelike for $r < 
2M$.  Note that, at $r=2M$, the completely null form $(g_{vv}=g_{rr}=0)$ reflects 
the fact that both $\bm\partial_v$ and $\bm\partial_r$ are null there.

How do regions I and II in the old $(t, r, \theta ,\phi )$ charts fit into 
the new extended geometry in the $(v, r, \theta ,\phi )$ chart?  It's 
easier to visualize the $v, r, \theta , \phi$ description with $v$ drawn so 
that the $v$ = constant $(\theta = \frac{\pi}{2}$, $\phi = 0)$ null 
geodesics are 45$^\circ$ lines (representing ingoing photons).

\begin{figure}[bh!]
\centering\includegraphics[width=.8\textwidth]{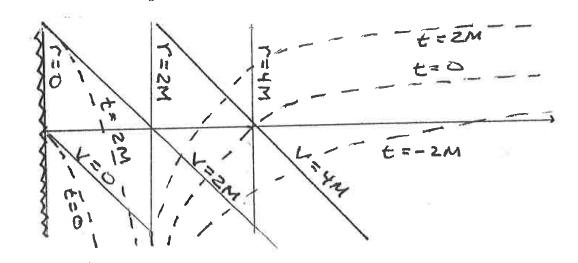}
\end{figure}
\noindent The images of some $t$ = constant lines are shown in the $v, r$ diagram as dotted lines.

Note that the $r = 2M$ line in $t, r, \theta , \phi $ coordinates has 
been pushed to $v = -\infty $.\\

\noindent {\sl Light cones}:
\begin{figure}[H]
\centering\includegraphics[width=.5\textwidth]{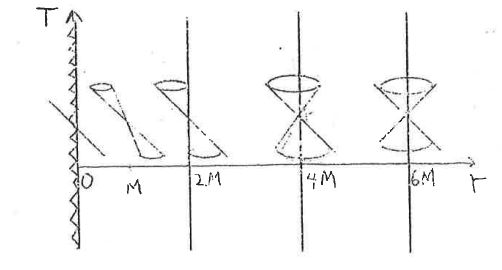}
\caption{Light cones in an Eddington-Finkelstein diagram, with $T=v-r$.}
\label{f:cone}\end{figure}
\index{light cone!of black hole}

Ingoing radial photons move along $v$ = constant lines. Outgoing photons 
move along $u = t-r_* = $constant lines. Then $u = v - 2r_*$ = constant, 
and along the $u =$ constant lines (i.e., along the outward radial null geodesics) 
$dv  = 2dr_* = 2(1 - 2M/r)^{-1}~dr.$ Or, with $T = v-r $, ingoing rays 
are given by $dT = - dr$, outgoing rays by
\[
	dT = \frac{r+2M}{r- 2M}dr. 
\]

\textit{Stellar collapse} can be described using
Eddington-Finkelstein coordinates: Inside the star the geometry is regular
initially, and outside the geometry is vacuum Schwarzschild.  The surface of
the star follows a timelike trajectory and so once inside $r = 2M$, it is
constrained to go to $r = 0$ in finite proper time.
\index{collapse, gravitational}\index{gravitational collapse}
\begin{figure}[H]
\centering\includegraphics[width=.8\textwidth]{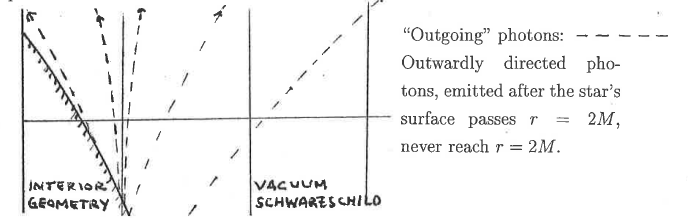}
\end{figure}

	Finkelstein\cite{finkelstein58}\href{https://link.aps.org/doi/10.1103/PhysRev.110.965}{PRD 1958}, after reinventing Eddington's coordinates, realized the 
the $r = 2M$ surface was an event horizon. Observers outside it are
unable to receive signals from within.  A spacelike slice of the region
inside an event horizon is a black hole.
\index{horizon|textbf}

	In a general asymptotically flat spacetime, event horizons and black holes
are characterized this way:  The set of all future directed null geodesics
that make it out to infinity (that eventually leave any compact region) is
called the past of future null infinity (in Schwarzschild this in the
region outside $r = 2M$).  The boundary of this region, {\sl the boundary of
the past of future null infinity}, is called the event horizon; and a
spacelike section of the spacetime inside the event horizon is called a
black hole.\index{black hole|textbf}\index{event horizon|textbf}
\index{horizon!Schwarzschild}\index{Schwarzschild spacetime!black hole} 

\begin{figure}[H]
\centering\includegraphics[width=.8\textwidth]{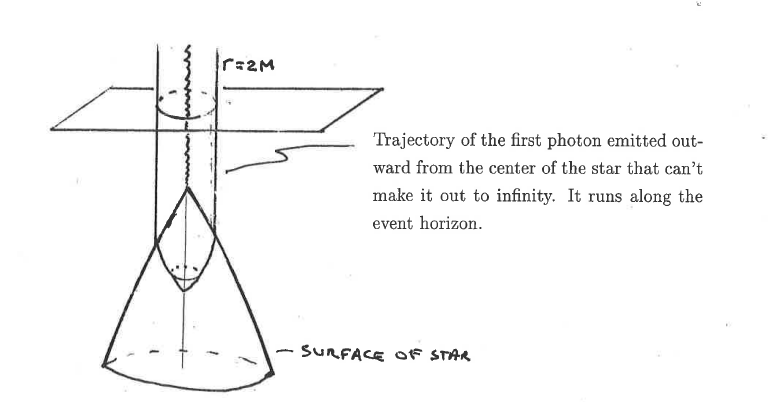}
\end{figure}

\newpage
\noindent {\sl Trapped Surfaces}:
\index{trapped surface}\\
  From the light cone diagram, Fig.~\ref{f:cone}, 
it is apparent that any photon emitted within $r = 2M$ eventually hits $r
= 0$.  Because the radial coordinate $r$ has the meaning that $4\pi r^2$ is
the area of a $v,r$ = constant surface, the following peculiar phenomenon
occurs.  Consider the set of all outgoing photons emitted from an $r$ =
constant sphere.\\

\begin{figure}[H]
  \centering\includegraphics[width=0.25\textwidth]{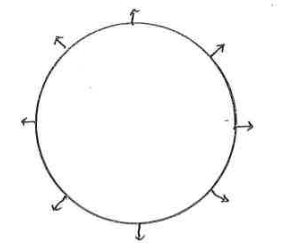}
\end{figure}

For large $r$, the emitted
photons comprise a sphere of increasing area -- they are diverging. 
For $r < 2M$ however, even the ``outgoing'' photons are moving to smaller values
of $r$, so the sphere of outgoing photons is shrinking in area  
-- the photons are converging, trapped.
\vspace{3mm}

Note, however, that an outgoing photon, emitted by a flashlight
pointing radially outward, moves away from the flashlight in the direction 
that it is emitted at the speed of light, according to an observer at 
rest relative to the flashlight.  This is consistent, because the flashlight 
itself is falling inward faster that the emitted photons. In general, a 
{\em trapped surface} $S$ is one
whose outward directed null geodesics ($\perp $ to $S$) are all converging.
Thus in Schwarzschild, all $r, v$ constant surfaces with $r < 2M$ are
trapped.\\

\index{collapse, gravitational!singularity theorem}\index{gravitational collapse!singularity theorem}
\index{singularity theorems}
It used to be thought likely that the singularity at $r = 0$ in
Schwarzschild represented the end of spherical collapse only -- that
nonspherical collapsing stars would never encounter singularities.  In
1965, however, Penrose proved that if energy dominates pressure in the sense
that 
\beq (T_{\alpha\beta} - \frac{1}{2} g_{\alpha\beta}T)u^\alpha u^\beta  \geq 0 
\label{strong}\eeq
all timelike $u^\alpha$, and if there is a trapped surface, then the
spacetime is singular.   For a fluid this is satisfied unless 
$P < -\frac{1}{3} \rho$ (attractive pressure exceeding the relativistic 
limit); for an electromagnetic field it's always true; for a massive scalar 
field the condition can fail in regions on the order of the Compton wavelength, 
$\sim 10^{-16}$ cm for the Higgs boson.  Eq.  (\ref{strong}) is called the 
{\em strong energy condition}.  The
singularity may not be as violent as the $r=0$ singularity of Schwarzschild
where $R_{\alpha\beta\gamma\delta}$ becomes singular ($R_{\alpha\beta\gamma\delta}R^{\alpha\beta\gamma\delta}$, for example, blows
up).  The theorem only says that at least one null geodesic is incomplete -
the geodesic leaves the spacetime in finite affine parameter length.\index{affine parameter}  The coordinate
singularity at $r = 2M$ had that character, but that was the result of 
picking coordinates that cover only a piece of the spacetime -- 
in effect, by choosing Schwarzschild coordinates, one is  
cutting a hole out of the spacetime at $r=2M$.  In that case, one can smoothly extend 
the truncated geometry.  Penrose's theorem guarantees that there is 
no way to avoid the singularity by, in particular, extending the spacetime.
As we have already seen, null geodesics hit $r = 0$ in finite parameter 
length, and this is the singularity implied by Penrose's theorem. 
Penrose's theorem is proved in Chapter 9 of Wald, where it is Theorem 9.5.3; 
a somewhat stronger version due to Hawking and Penrose is 9.5.4.
 
The theorems only guarantee geodesic incompleteness.   
Timelike geodesics within the horizon of the analytically extended Kerr 
geometry do not generically encounter infinite curvature; whether they 
do so in the actual geometries of rotating black holes is not yet known.   

\section{Embedding Diagrams}
\index{embedding diagram}

If you are handed the components of a metric in some coordinates, it is
often difficult to understand what the metric means physically or
geometrically.  Commonly one draws light cones and constructs the
geodesics, as we have done for the Schwarzschild geometry.  Another,
more powerful, technique, when it can be used, is to construct an
embedding diagram.  The idea is to regard a section of the spacetime
(e.g., a $t$ = constant surface) as a curved surface in flat space
(some ${\mathbb R}^n$).  For example, a sphere can be characterized
intrinsically by the metric
\[
	ds^2 = r^2(d\theta^2+\sin^2\theta d\phi^2)
\]
or pictured as an $r$ = constant subset of ${\mathbb R}^3$.  In other
words, one looks for a submanifold of ${\mathbb R}^n$ whose metric is
that of the surface you want to understand.\\

	Let us first embed a $t$ = constant surface of a spherical star
\beq ds^2 = \left(1 - \frac{2m(r)}{r}\right)^{-1} dr^2 + r^2d\Omega^2\,. \eeq
To visualize, we can't look at more than a 2-dimensional surface embedded in 
${\mathbb R}^3$, so we'll take a $\theta = \frac{\pi}{2}$ plane:
\beq
	ds^2 = (1 - \frac{2m}{r})^{-1} dr^2 + r^2d\phi^2. 
\label{embed}\eeq
The object is to find a surface $z = z(r, \phi )$ in ${\mathbb R}^3$ with this
metric.  The flat metric of ${\mathbb R}^3$ has $z, r, \phi $ components
\beq
	ds^2 = dz^2 + dr^2 + r^2d\phi^2 
\eeq
The metric of a $z = z(r, \phi )$ surface is thus
\beq
	ds^2 = \left(\frac{\partial z}{\partial r} dr 
		+ \frac{\partial z}{\partial\phi }d\phi\right)^2 + dr^2 + r^2d\phi^2. 
\eeq
Axisymmetry $\Rightarrow  z = z(r)$,
\[ 
	ds^2 = (z'^{\,2}+1)dr^2 + r^2d\phi^2. 
\]
To agree with (\ref{embed}) we need
\[ 	
	z'^{\,2} + 1 = \left(1 - \frac{2m}{r}\right)^{-1} \,.
\]
\beq
 z = \mbox{\Large$\displaystyle \int_0^r$}\frac{dr}{\left[\mbox{$\displaystyle\frac r{2m(r)}$} 
		- 1\right]^{1/2}}, 
		~~~\mbox{ when there is no black hole -- when }~2m(r) 
		\mbox{ is everywhere greater than } r. 
\eeq

Near the center of the star
\[
	m = \frac{4}{3}\pi~\rho_c ~r^3 + O(r^4) \ \Longrightarrow
\]
\begin{align}
z &= \int_0^r \left(\frac{8\pi\rho_c}{3}\right)^{1/2}r~dr + O(r^3)\nonumber \\
&= \left(\frac{2\pi\rho_c}{3}\right)^{1/2}r^2 + O(r^3)\,.
\end{align}
Outside the star 
\beq
	z = z_0 + [8M(r - 2M)]^{1/2} 
		~~\text{or}~~(r-2M) = \frac{1}{8M} (z-z_0)^2\,.
\label{zr}\eeq
Check:
\[
	\frac{dz}{dr} 	= (8M)^{1/2} \frac{1}{2}(r-2M)^{-1/2} 
			= \frac1{\left(\mbox{$\displaystyle\frac r{2M}$ }-1\right)^{1/2} }\,.
\]
\index{collapse, gravitational}\index{gravitational collapse}
\begin{figure}[h!]
\centering\includegraphics[width=.8\textwidth]{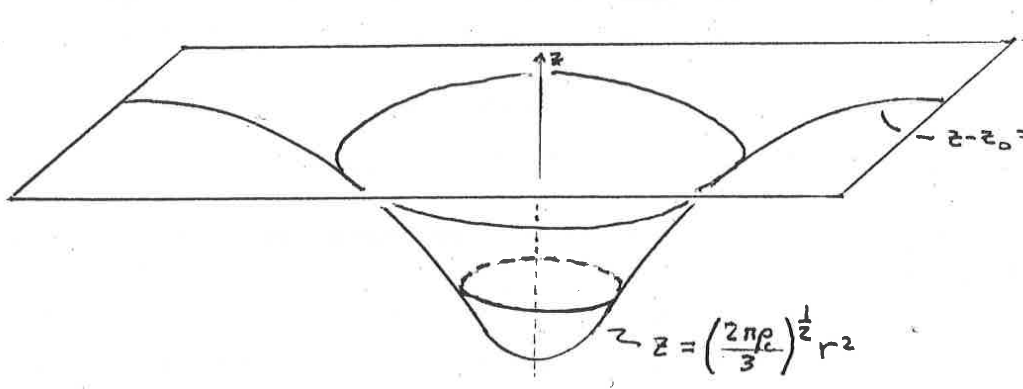}
\end{figure}
\noindent 
Note that the surface is coordinate independent -- specified by
demanding that its metric inherited from the flat metric on 
${\mathbb R}^3$ be the same as the metric of
an equatorial plane in the Schwarzschild geometry.\\

After the star collapses one can embed a $T$ = constant surface:
\begin{figure}[h!]
\centering\includegraphics[width=.5\textwidth]{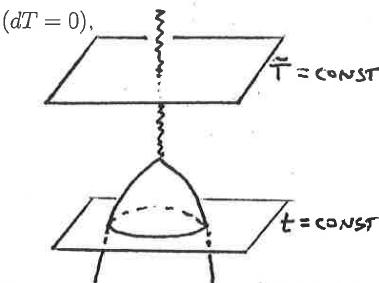}
\end{figure} 
From (\ref{e:e-f-in}), with $dv = dr$ \,\,\,($dT = 0)$,

\[ ds^2 = \left[-(1- \frac{2M}{r}) + 2\right]dr^2 + r^2 d\Omega ^2 
	  = \left(1 + \frac{2M}{r}\right)dr^2 + r^2d\Omega ^2 
\]
Embed:
\begin{align}
 (z')^2 + 1 &=  1 + \frac{2M}{r} \nonumber\\
z' &= \sqrt {\frac{2M}{r}}\nonumber\\
z & = (8Mr)^{1/2} \quad \mbox{ or }\quad r = \frac{1}{8M}z^2.
\end{align} 
After the collapse the surface is no longer smooth at $r=0$; 
there is a cusp, implying infinite curvature. 
(Old figure has $\tilde T$ instead of $T$.)
\begin{figure}[h!]
\centering\includegraphics[width=\textwidth,angle=-2]{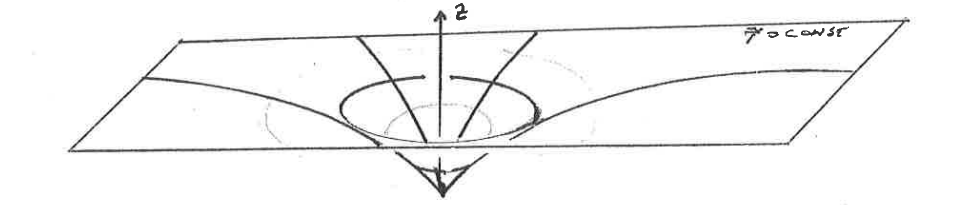}
\end{figure}

	Although Eddington-Finkelstein ingoing coordinates are fine for
describing collapse, and provide a completion for the future-directed
null and timelike geodesics that are cut off in the $\{t, r, \theta, \phi\}$
chart, the $\{v, r, \theta, \phi\}$ coordinates are not good for
describing past-directed geodesics that move inward (equivalently, the past 
of future-directed outward geodesics).  These run off the manifold in
finite length as before.
   When the past was a star, the problem disappears, but if one wants to
look formally at the vacuum Schwarzschild geometry a further completion
is needed.  One way to see what to expect is to look at the embedding
of a $t$ = constant surface of vacuum Schwarzschild.  From (\ref{zr}), the
embedding looks like $r - 2M = \frac{1}{8M}~(z-z_0)^2$, and it ends
abruptly at $r = 2M,~z = 0$ (where, without loss of generality, we have 
taken $z_0=0$).\index{black hole!embeddding diagram}

\begin{figure}[th!]
\centering\includegraphics[width=.5\textwidth]{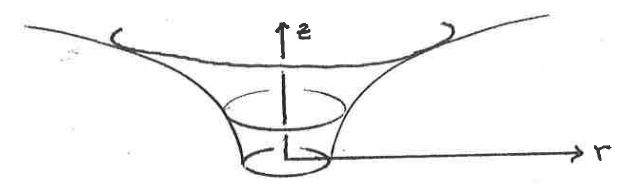}
\end{figure}

\noindent But the bottom half of the paraboloid is identical to the top
half and is also described by the metric
\beq
	ds^2 = \left(1 - \frac{2M}{r}\right)^{-1}dr^2 + r^2d\Omega ^2 
\label{ssch}\eeq
of a $t$ = constant surface.
 
\noindent Thus the full paraboloid is a smooth extension of the 
$t$ = constant surface, and it has the spatial Schwarzschild metric 
(\ref{ssch}) everywhere.  
\begin{figure}[h!]
\centering\includegraphics[width=.5\textwidth]{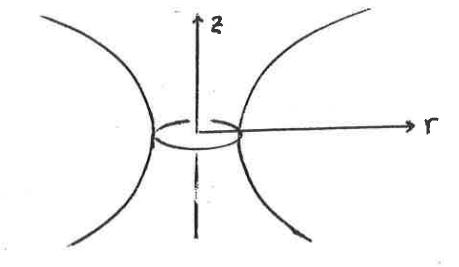}
\label{f:schwarzschild_embed}\end{figure}

This suggests that one can extend the spacetime 
in such a way as to get two identical asymptotically flat regions.
One way is to patch together different charts.  But it turns
out to be possible to find a single chart that covers the complete
extension of the geometry.

\section{Kruskal-Szekeres Coordinates} \index{Kruskal-Szekeres coordinates}\index{Schwarzschild spacetime!Kruskal-Szekeres coordinates}\index{coordinates!Kruskal-Szekeres coordinates}
The papers: \href{https://journals.aps.org/pr/abstract/10.1103/PhysRev.119.1743}{Kruskal}\cite{kruskal60}, \href{https://link.springer.com/article/10.1023/A:1020744914721}{Szekeres}\cite{szekeres60,szekeres60a}.  

	Part of the problem with Eddington-Finkelstein is that,
by choosing ingoing coordinates, one imposes a preferred time direction
(future) on a geometry that doesn't know future from past.  As in flat
space, one could have picked outgoing null coordinates $u = t-r_*, r,
\theta, \phi $ and found a metric $\displaystyle ds^2 = -\left(1 -
\frac{2M}r\right)du^2 - 2dudr + r^2d\Omega ^2$ for a spacetime with
complete past directed geodesics, but incomplete future geodesics.

	So to regain past-future symmetry one could try null coordinates $u, v,
\theta , \phi $.  Then, using
\beq
	t = \frac{u+v}{2},\ r_* = \frac{v-u}{2},
		\ {\rm and}\ \frac{dr}{dr_*} = \left(1-\frac{2M}{r}\right), 
\label{trstar}\eeq
we have
\begin{eqnarray}
ds^2 &=& -\left(1 - \frac{2M}r\right)\frac{1}{4} (du+dv)^2 + \left(1 - \frac{2M}r\right)^{-1}
\frac{1}{4} (dv-du)^2 \left(1 - \frac{2M}{r}\right)^2 + r^2d\Omega ^2 \nonumber\\
&=& -\left(1 - \frac{2M}{r}\right)dudv + r^2d\Omega ^2 
\end{eqnarray}
Unfortunately, at $r = 2M$ the 4-dimensional metric is degenerate (not invertible):  
$ds^2 = r^2d\Omega^2$, implying \mbox{$\sqrt {-g} = 0$}.
This, however, is the only remaining obstacle, and it is easily overcome 
by a change of scale:
\beq
	U = -e^{-u/4M} \qquad
	V = e^{v/4M} 
\label{tildeuv}\eeq
\[
	dU = \frac{1}{4M} e^{-u/4M} du \qquad dV 
		= \frac{1}{4M} e^{v/4M} dv 
\]
\bea 
 du\, dv &=& 16M^2 e^{(-v+u)/4M} dU dV
 	  = 16M^2 ~e^{-r_*/2M}dU dV 
								\nonumber\\	
 &=& 16M^2 \exp\left[-\frac{r}{2M} - \log \left| \frac{r}{2M}
 			     -1 \right|\right]dU dV
								\nonumber\\
 &=& 16M^2 e^{-r/2M}\frac{1}{\displaystyle\frac{r}{2M}-1} dU dV,
 \ \ r>2M.
\eea
Finally, 
\bea 
	ds^2 &=& -16M^2~e^{-r/2M}\, \frac{2M}{r} dU dV
		 + r^2(U, V)d \Omega ^2, \nonumber\\
 \crv ds^2 &\crv =& \crv-\frac{32M^3}{r}~e^{-r/2M}\, dU dV 
 	  + r^2 d\Omega ^2. 
\label{kruskal}\eea
\noindent This looks good  for $r > 0$, but we still need to find out when
$r(U, V)$ is well behaved.  From (\ref{rstar}), (\ref{trstar}) and 
(\ref{tildeuv}), $r$ is given implicitly for $r>2M$ by
\beq
	\cblue \left(\frac{r}{2M} - 1\right)e^{r/2M} = -UV \equiv f(r),
\label{f}\eeq
and we {\em define} $r(U,V)$ by this relation for all $r>0$.
$f(r)$ is invertible to $r > 0$ when $-U V$ is in the 
range of $f$:
\beq U V < 1. \eeq

\begin{figure}[h!]
\centering\includegraphics[width=.5\textwidth]{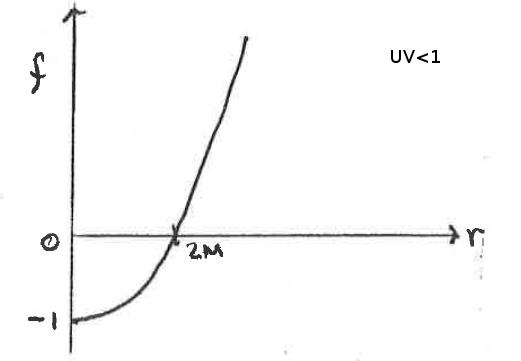}
\end{figure}
\noindent One commonly introduces the spacelike and timelike 
coordinates  
\beq 	
	X = \frac{1}{2}(V-U)\qquad 
	T = \frac{1}{2}(V+U).
\eeq
Then 
\vspace{-3mm}
\beq
	f(r) = X^2-T^2 
\label{e:xt}\eeq
and the metric has components
\[ 
ds^2 =   \frac{32M^3}{r}~e^{-\frac{r}{2M}}(-dT^2+dX^2) + r^2d\Omega^2. 
\]
Eq. \eqref{e:xt} is invertible to a unique $r > 0$ when $-X^2+T^2 < 1$, or 
$T^2 < 1 + X^2$, a region between two hyperbolae; it has four parts, I, II, III, IV.\\

\begin{figure}[h!]
\centering\includegraphics[width=.45\textwidth]{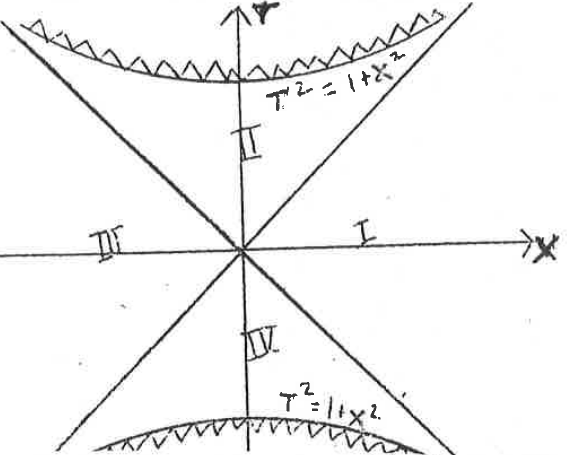}
\end{figure}
In I, one obtains a $t, r $ patch with $r > 2M$ by setting
\begin{align}
 {\rm I}\ (r > 2M):\hspace{1in}
	X &= \left(\frac{r}{2M} - 1\right)^{1/2}~e^{r/4M}\cosh \frac{t}{4M}\nonumber\\
	T &= \left(\frac{r}{2M} - 1\right)^{1/2}~e^{r/4M}\sinh \frac{t}{4M}\nonumber
\end{align}
Region II provides a $t, r$ patch with $r<2M$
\begin{align} 
 {\rm II}\ (r < 2M):\hspace{1in}
	X &= \left(1-\frac{r}{2M}\right)^{1/2}~~e^{r/4M}~\sinh \frac{t}{4M} \nonumber\\
 	T &= \left(1-\frac{r}{2M}\right)^{1/2}~~e^{r/4M}~\cosh \frac{t}{4M}
\nonumber\end{align}
Similarly regions III and IV give $t, r$ patches isometric to 
I and II, respectively, with
\begin{eqnarray*}
{\rm III}\ (r > 2M): \hspace{1in}X = -\left(\frac{r}{2M} - 1\right)^{1/2}~e^{r/4M} \cosh
\frac{t}{4M}\\
T = -\left(\frac{r}{2M} - 1\right)^{1/2}~e^{r/4M} \sinh \frac{t}{4M}\\
\\
{\rm IV}\ (r < 2M): \hspace{1in}X = -\left(1-\frac{r}{2M}\right)^{1/2}~e^{r/4M} \sinh
\frac{t}{4M}\\
\\
T = -\left(1-\frac{r}{2M}\right)^{1/2}~e^{r/4M} \cosh \frac{t}{4M}\end{eqnarray*}

As we assumed to begin with in (\ref{f}) and (\ref{e:xt}), $r$ is given
everywhere by \mbox{$\dis X^2-T^2 = \left(\frac{r}{2M}-1\right)e^{r/2M}$}, 
and the $r =$ constant curves are then hyperbolae:\vspace{-2cm}

\begin{figure}[H]
\centering\includegraphics[width=.55\textwidth]{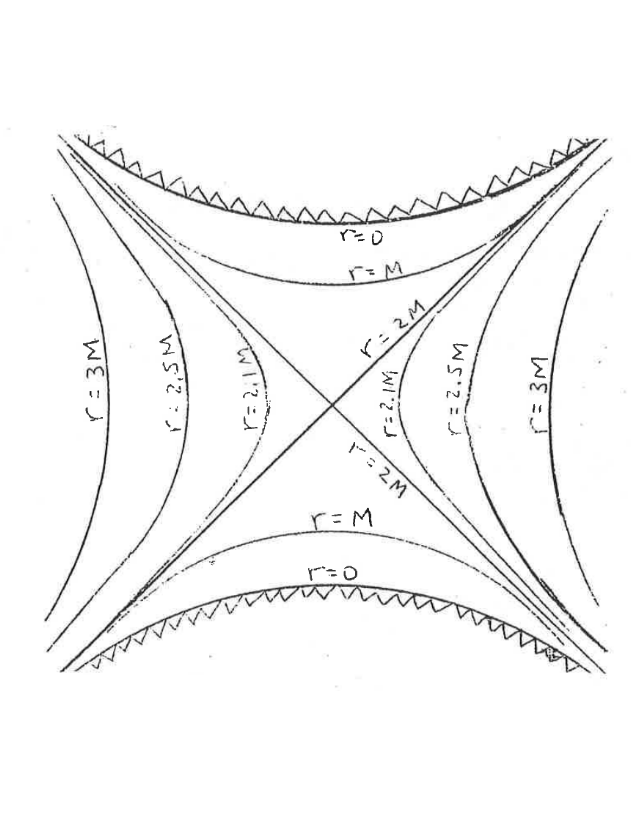}
\end{figure}

\newpage

The $t=$ constant curves ares straight lines:\\
\\
\centerline{In regions I and III, $\dis\tanh \frac{t}{4M} = \frac{T}{X}.\qquad$
In regions II and IV, $\dis \tanh \frac{t}{4M} = \frac{X}{T}$.}\\

\begin{figure}[H]
\centering\includegraphics[width=.4\textwidth]{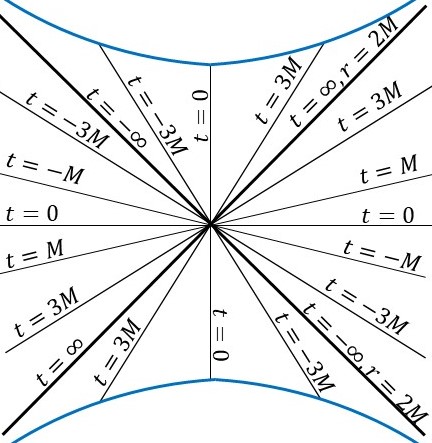}
\end{figure}

Features of the complete Schwarzschild geometry in Kruskal
coordinates: All values of $t$ for $r = 2M$ are mapped to a single point, 
the center point of the figure above.  With the angular dimensions included, 
this is a sphere of radius $r = 2M$ in spacetime.  
This problem with the coordinate $t$ is analogous to the coordinate singularity in polar coordinates where all
values of $\phi$ at $\theta = 0$ are mapped to a single point (a line, the
z-axis, when one includes $r$). \\

{\crv In a Kruskal diagram, the radial null geodesics from any point are 45$^\circ$
lines.    }  \index{Kruskal diagram}\index{Schwarzschild spacetime!Kruskal diagram}\\
They are the radial lines 
\[
\crv T = \pm X + \mbox{constant};  
\] equivalently, they are the the lines $U,\theta,\phi = $ constant 
and the lines $V,\theta,\phi=$ constant.  This means that the 
2-dimensional light cones in a Kruskal diagram look like the light 
cones of 2-dimensional Minkowkski space. Each point in the diagram 
however, represents a sphere in the 4-dimensional geometry, as

\begin{figure}[H]
\centering\includegraphics[width=.3\textwidth]{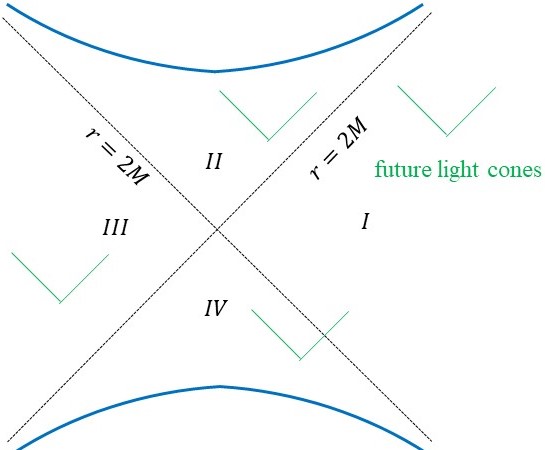}
\caption{The full Schwarzschild vacuum spacetime has two asymmptotic regions, I and III, a black hole II, bounded by a future horizon, and a white hole IV, bounded by a past horizon.}  
\label{f:Kruskal}\end{figure}

\noindent There are two asymptotically flat regions. Spacelike
surfaces, for example the $t$ = constant surfaces of regions I and III,
extend from one asymptotic region to another, via a wormhole (initially
called an Einstein-Rosen bridge).  It is immediately clear from the
Kruskal diagram that no particle can go through through the wormhole
and reach the other asymptotically flat region:  Only spacelike curves
do that -- future directed timelike curves all hit the $r = 0$ singularity 
and don't make make it through from I to III or III to I.  

This is a generic property of asymptotically flat spacetimes with 
non-Euclidean topology (there are a countably infinite number of such 
topologies).  No null geodesic that starts and ends at infinity can thread  
a topological structure:  Positive energy implies that any topological structure 
collapses too quickly for observers to find out that it's there (JF, Schleich, Witt)
\href{https://arxiv.org/abs/gr-qc/9305017}{topological censorship}.\cite{fsw93} 
(See also \href{https://arxiv.org/pdf/1906.02151}{Chru\'sciel \& Galloway}\cite{cg19} and references therein.)   
\index{topological censorship}  
A corollary is a strengthened version of a theorem by Hawking that any spacelike 
slice of the event horizon of a stationary black hole has spherical topology 
\href{https://arxiv.org/pdf/gr-qc/9410004.pdf}{(Chrusciel, Wald)}\cite{cw94,galloway95,jv95}.
\index{black hole!horizon topology}   
 
  It is also apparent from the diagram that no particle in II, in the 
black-hole interior, can escape from the black hole to regions 
I or III.   \\

\noindent{\sl Future and past horizon; black hole and white hole}\\
\index{white hole|textbf}
\index{Schwarzschild spacetime!black hole}\index{Schwarzschild spacetime!white hole}\index{black hole!Schwarzschild}  

The full Schwarzschild geometry, as is clear from Fig.~\ref{f:Kruskal} 
is symmetric under time-reversal, a reflection in a horizontal plane 
through the center of the diagram.  The reflection maps a point with 
Kruskal coordinates $(T,X)$ to $(-T,X)$. It takes the black 
hole, region I, to region III.  Region III is a {\sl white hole}, 
the time reverse of a black hole.  No light ray or physical object 
outside the white hole can enter it:  That is, no future directed 
light cone with vertex outside III intersects III.  

The future horizon is the boundary of the black hole; it is then, 
the boundary of the exterior of the black hole.  
That is, as defined previously, 
the future horizon is the boundary of the region from which 
some future-directed causal curve reaches arbitrarily large values of $r$, 
leaving any compact region.\\  
The {\sl past horizon} is the boundary of the white hole and so 
the boundary of the exterior of the white hole.  It is the 
boundary of the region from which some {\sl past-directed} timelike 
or null curve reaches arbitrarily large values of $r$.  \\

The spacetime of a collapsing star has only one asymptotic region, and 
it does not include the left half of the Kruskal diagram or the white 
hole. But if one time-reverses the history of the universe, black holes 
turn into white holes.  The absence of white holes in the visible 
universe is then tied to the arrow of time, set by the direction in which entropy increases-- the thermodynamic arrow of time.   
\index{collapse, gravitational}\index{gravitational collapse}
\begin{figure}[h!]
\centering\includegraphics[width=.3\textwidth]{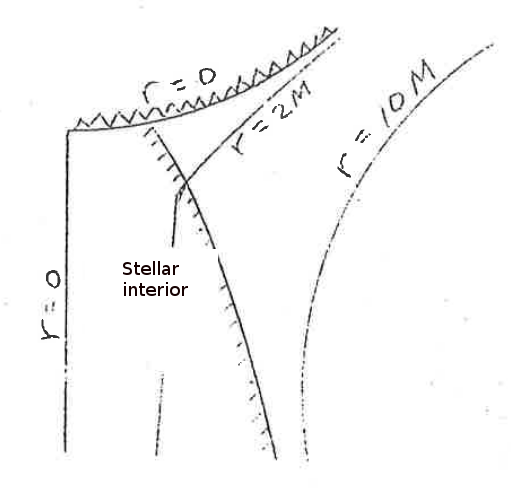}
\caption{A collapsing star in a Kruskal diagram 
	looks like this figure.  The left half of the Kruskal 
	spacetime doesn't exist here.  Because the Eddington-Finkelstein 
	ingoing chart is mapped to regions I and II, an 
	Eddington-Finkelstein chart also covers the full collapse spacetime, 
	if one continues the coordinates $v,r$ to the star's interior, as null and radial coordinates inside the star.}  
\end{figure}

\newpage	

The Schwarzschild geometry is time dependent:  The Killing vector that
is timelike outside $r = 2M$ becomes spacelike inside.  Here are two
ways to slice the history (the spacetime) given as in MTW as sequences of 
embedding diagrams.\index{black hole!embedding diagram}.  
\\

\begin{wrapfigure}[12]{l}{0.45\textwidth}
\begin{center}
    \vspace{-12mm}
\includegraphics[width=0.43\textwidth]{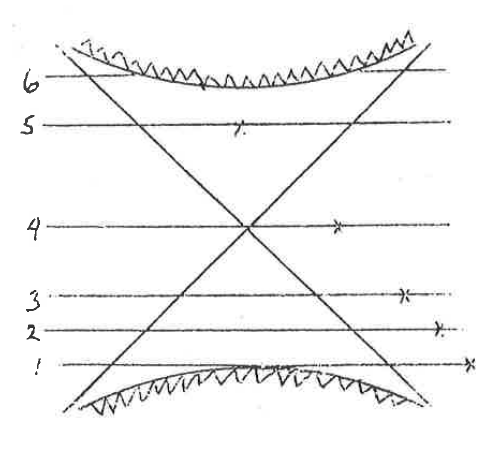}
  \end{center}
\end{wrapfigure}
\noindent In figuring out what the embedding of each slice should look like qualitatively, 
the key point is 
that distances on the embedded surface are proper distances in the spacetime.  
In particular, a circle with radial coordinate $r$ has proper circumference 
$2\pi r$ and thus is always at a distance $r$ from the z-axis in the embedding
diagram.  A curve in the embedding diagram corresponding to a radial curve 
in spacetime (($t,\theta,\phi$ constant)) is perpendicular to the circles of 
constant $r$ and has proper distance along the curve equal to the proper 
spacetime distance.  \\ 

 \noindent
Slice 4:  $r$ begins on the right at $\infty$, then goes 
to $r=2M$ (the central point in the diagram), and then out to $\infty$ again:
\vspace{-2mm}

\begin{figure}[h!]
\centering\includegraphics[width=\textwidth]{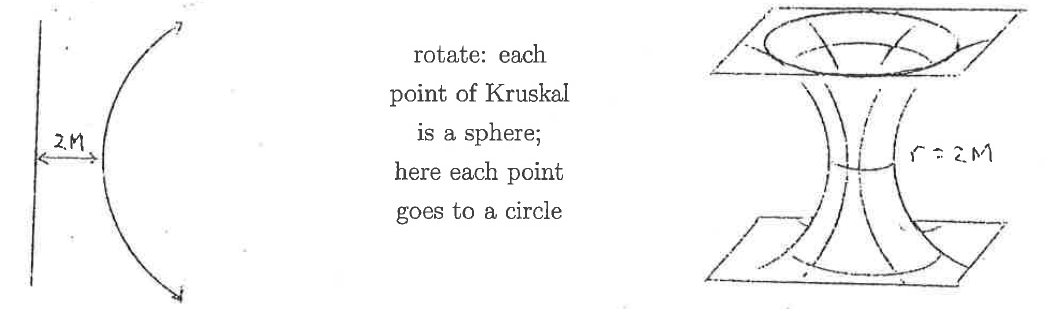}
\end{figure}

\noindent
Slice 5: $r$ begins at $\infty$, goes to $r=1.5$, then out to $\infty$:
 \vspace{-3mm}

\begin{figure}[h!]
\centering\includegraphics[width=.7\textwidth]{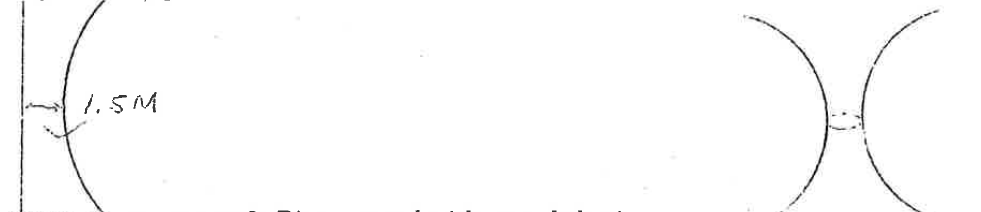}
\end{figure}
Slice 6: $r$ begins at $\infty$, goes to $r=0$; a disconnected piece starts at $r=0$ and goes to $\infty$:
\vspace{-3mm}
 
\begin{figure}[h!]
\centering\includegraphics[width=.7\textwidth]{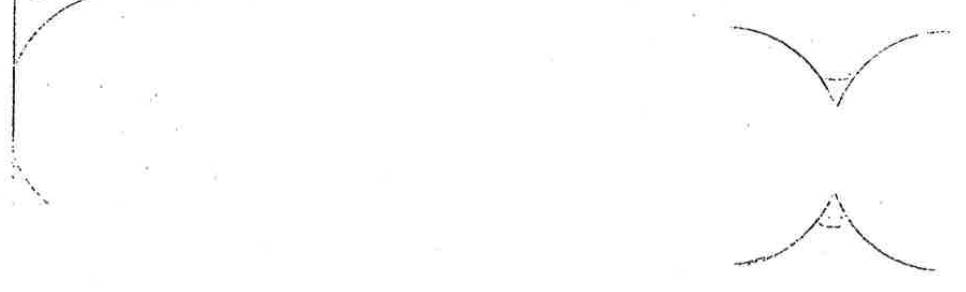}
\end{figure}

\newpage

So the slicing 1-6 of the history looks like this.\\
\begin{figure}[h!]
\centering\includegraphics[width=.28\textwidth]{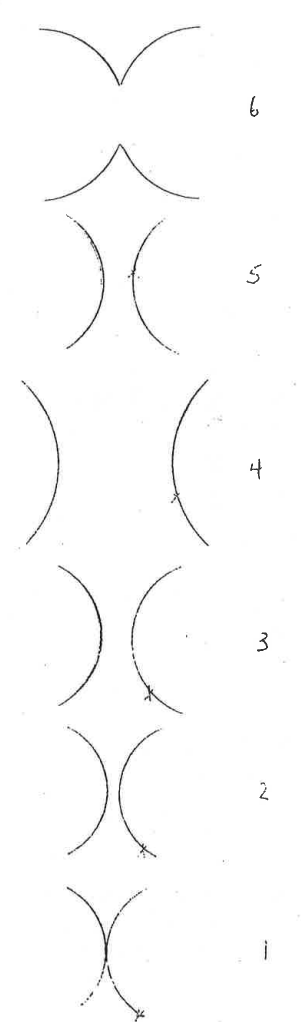}
\end{figure}

The crosses mark the trajectory of an ingoing particle that fails to make it 
through the wormhole before the asymptotic regions pinch off.
\newpage
\begin{figure}[h!]
\centering\includegraphics[width=\textwidth]{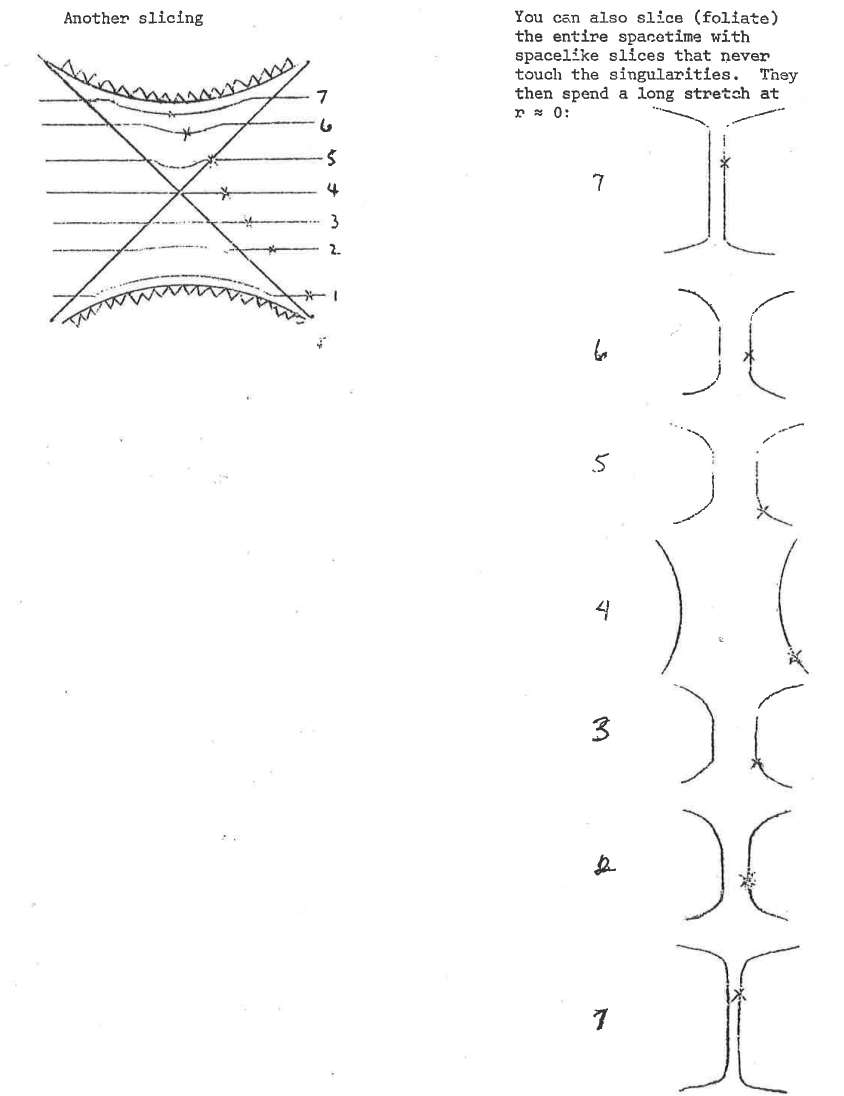}
\end{figure}
\newpage

\benr

\item Embedding diagram.\index{embedding diagram}
\benalph\item Sketch a surface in flat $\mathbb R^3$ whose metric is the wormhole metric 
\mbox{$ds^2 = dx^2 + (x^2+a^2)d\phi^2$}, $-\infty<x<\infty, \quad 0\leq\phi<2\pi$.  The 
surface should have two asymptotically flat regions.  
\item With $z,\varpi,\phi$ polar coordinates for $\mathbb R^3$ ($\varpi$ the distance from the 
$z$-axis) find the embedded surface $z=z(\varpi)$ whose metric is the wormhole metric.  

\een 
\item Calculate the cross section for capture of particles by a Schwarzschild 
black hole for \mbox{$v\ll 1$}.  Hint: First show that $V_{\rm eff}\ll 1$ at the 
largest impact parameter for which the particle can avoid capture. 
\index{black hole!capture cross section} 

\item (Hartle 12-12) Working in Eddington-Finkelstein coordinates, check that 
the normal vector to the horizon 3-surface of a Schwarzschild black hole is a null vector.

\item\label{p:bhtime}  The time for light to reach an observer outside a black hole horizon from radius \mbox{$r=2M(1+\epsilon)$} becomes infinite as $\epsilon\rightarrow 0$, 
but the divergence is only logarithmic in $\epsilon$.  
\benalph\item Consider a point $P$ at proper distance $d=2M\delta$ to 
the horizon along a $t=$ constant surface. Find $\delta$ to lowest nontrivial order in in terms of $\epsilon$.   
\item Find the time for light from $P$ to reach an observer at $r=8M$ along a radial geodesic.    
\item Find that time for $d$ the Planck length and $M=10 M_\odot$. 
(You should find a time of order a millisecond).  What is 
the contribution to that time from the term involving $\epsilon$?  
\end{enumerate}

\index{flux!baryon}\index{flux!scalar field}
\item Recall that conservation of baryons has the form $\nabla_\alpha j^\alpha = 0$, 
where $j^\alpha = nu^\alpha$, with\\ $n=$ baryon density of a fluid with velocity $u^\alpha$.
\index{conservation laws!baryons}  
\benalph
\item Using ingoing Eddington-Finkelstein coordinates, find the number of baryons per unit time entering a Schwarzschild black hole.  That is, find $dN/dv$, with $dN$ the number of 
baryons that cross the horizon between successive slices at $v=$ constant and 
$v+dv =$ constant.  
\item The Killing energy flux of a scalar field into the Schwarzschild horizon
is positive: Show that 
\[
   \int j^\alpha dS_\alpha \geq 0, 
\]
where
\[
  j^\alpha = -T^\alpha_{\ \beta} t^\beta, \qquad 
  T_{\alpha\beta} = \nabla_\alpha\Phi \nabla_\beta\Phi 
  		- \frac12 g_{\alpha\beta}\nabla_\gamma\Phi\nabla^\gamma\Phi.
\]
Note that the Killing vector $t^\alpha$ is $\bm \partial_v$ in this chart.  See p. \ref{p:bcons} for both parts of the problem.   
\een

\item An observer falls radially into a spherical black hole of mass $M$.  She 
starts from rest relative to a stationary observer at $r=8M$.  What time does her watch
read just before she reaches the singularity at $r=0$?

\item Og is at rest outside a spherical neutron star, at radius $r=8M$.
\benalph
\item 
Exhibiting strength characteristic of neutron-star denizens, 
Og throws a particle radially upward with speed $v$ as measured in Og's orthonormal basis, large enough to reach infinity with zero speed: That is, $v=v_{\rm escape}$.  What is $v$?
\item
Og throws a second particle horizontally 
(tangent to the $r=8M$ sphere), with speed \mbox{$v=v_{\rm escape}$}.  What is the maximum 
value of $r$ that the particle reaches?  

\een

\item (Adapted from Hartle) An observer falls feet first into a Schwarzschild black hole 
looking down at her feet.  Is there a radius at which she cannot see her feet? In particular, 
\benalph
\item When her eyes are at the horizon, can she see her feet? If so, at what radius does she see them?
\item Does she see her feet hit the singularity, assuming she remains sentient until her head 
hits the singularity?    
\item {\em  Is it dark inside a black hole?} An observer outside sees a star become dark as it collapses 
to a black hole.  But would it be dark inside a black hole, assuming a collapsing star continues 
to radiate at a steady rate, measured by an observer on its surface? 
\item Finally, if the observer looks up from within a black hole in the disk of the Milky Way, would she 
see stars above her?
\een
\vspace{-4mm}
\phantom{xx}This exercise is most naturally done using a 
Kruskal diagram.

\een
\newpage

\section{Kerr Black Holes}\index{black hole!Kerr black hole|textbf}\index{Kerr spacetime|textbf}\index{rotating black holes|see {Kerr spacetime}}
\label{s:kerr} 

\noindent{\sl Rotating stars and black holes}\\
\index{rotating stars}

The metric of a rotating star or a black hole has a timelike Killing vector $t^\alpha$ 
and a rotational Killing vector $\phi^\alpha$.  We have seen that a system with 
stress-energy tensor $T^{\alpha\beta}$ has a conserved current associated with each Killing 
vector.  In the case of a rotational Killing vector, the current $j^\alpha = T^\alpha_\beta\phi^\beta$ is associated with angular momentum:  $j^t=T^t_\phi$ is the angular momentum density,\index{angular momentum!angular momentum density} \index{conservation laws!angular momentum}
and the integral $J = \int j^\alpha dS_\alpha$ is the total angular momentum of the spacetime.\index{angular momentum!total angular momentum of spacetime|textbf}   
In a nearly Newtonian star, this has the form \\
\centerline{ $J=\int \rho v_i\phi^i dV= \int\rho(x v_y - y v_x)$.  }
For uniform rotation, $\bm v= \Omega \bm \phi$, with  
\be
  \bm \phi = \widehat{\bm z}\times\bm r = x\widehat{\bm y} - y\widehat{\bm x}, 
\quad\mbox{and}\quad J = \int \rho v_i\phi^i = \int\rho (x^2+y^2)\Omega dV.
\label{e:vdipole}\ee

The Coulomb potential of a ball of charge of electric charge density $\rho_e$ is, up to a constant, identical in form to the Newtonian potential $\Phi$ of a mass with density $\rho$, each satisfying $\nabla^2\Phi = C \rho$ for some constant $C$. Similarly the magnetic field of 
a stationary current density $\bm j_e=\rho_e \bm v$ is identical up to a constant to components of the metric associated with the mass current $\bm j = \rho \bm v$ of a nearly Newtonian fluid. 

Let's start with the magnetic field.  In a Lorenz gauge, a time-independent vector potential 
satisfies $\nabla\cdot \bm A=0$, and the magnetic field of a stationary current is given by
\[
  \nabla \times \bm B = 4\pi \bm j_e,  \ \bm B = \nabla\times\bm A, \quad\Longrightarrow \ \nabla^2 A_i = -4\pi j_{e\,i}, 
\]
with solution for the Cartesian components
\be
   A_i = \int\frac{\rho_e v_i}{|\bm r - \bm r'|} dV'.    
\ee

The gravitational analog of the magnetic field produced by an electric current 
is the gravitational field produced by a mass current, called a {\sl gravitomagnetic field}.  
\index{gravitomagnetic field}\index{magnetic field!gravitomagnetic field} 
Consider the stress tensor of nearly Newtonian fluid with velocity field $\bm v$. It  
has the form 
\[
   (T^{\mu\nu})=\begin{pmatrix} \rho & \rho \bm v\\	
				\rho \bm v & 0 
		\end{pmatrix} + O(e^2), 
\]
with $e$ as before a small parameter of order the largest of $v$ and $v_{\rm sound}$.
(Recall that $P=O(\rho v_{\rm sound}^2$).)   For a nearly Newtonian source, the metric has the form 
$g_{\alpha\beta} = \eta_{\alpha\beta} + h_{\alpha\beta}$, and we can ignore terms 
quadratic in the small correction $h_{\alpha\beta}$ to the flat metric. 
The Ricci and Einstein tensors are then linear in $h_{\alpha\beta}$;  
because the $\Gamma\Gamma$ terms in the Riemann tensor are quadratic in $h_{\alpha\beta}$, we can use Eq.~\eqref{e:RatP}, valid when terms quadratic in $\Gamma$ are discarded.  Then, 
using $\partial_\sigma g_{\mu\nu}= \partial_\sigma h_{\mu\nu}$, we have  
\be 
R_{\mu\nu\sigma\tau} = \frac{1}{2} [\partial_\nu\partial _\sigma h_{\mu\tau} 
		  + \partial_\mu \partial_\tau h_{\nu\sigma} 
		  - \partial_\mu \partial_\sigma h_{\nu\tau} 
		  - \partial_\nu \partial_\tau h_{\mu\sigma}]. 
\label{e:RP1}\ee 
In the nearly-Newtonian regime, the diagonal components of $h_{\mu\nu}$ are given by the diagonal components of $T^{\mu\nu}$, as in Eq.~\eqref{ritjt}, and we can quickly see, as follows, that the off-diagonal components of the Einstein equation are satisfied when $h_{ti}$ satisfies 
\be
   \nabla^2 h_{ti} = 16\pi \rho v_i\, 
\label{e:gti}\ee
with conservation of mass for a stationary flow, $\partial_i (\rho v^i)=0$,
\index{conservation laws!mass} 
implying $\partial^i h_{ti}=0$.
For $\mu\neq \nu$, $G_{\mu\nu} = R_{\mu\nu} + O(h^2)$, and contracting \eqref{e:RP1} on $\nu$ and $\tau$ gives  
\[
   R_{ti} = \frac{1}{2} [\partial^\nu\partial_i h_{t\nu} 
	 + \partial_t \partial^\nu h_{\nu i} - \partial_t \partial_i h_\nu^\nu 
		  - \partial^\nu \partial_\nu h _{ti}].
\]

Because the star is stationary, the time derivatives vanish:   
\[
  R_{ti} = -\frac12 \nabla^2 h_{ti} = 8\pi T_{ti}, 
\]
implying \eqref{e:gti} -- notice the sign: $T_{ti}= - \rho v_i$.  The off-diagonal spatial components $R_{ij} = 8\pi T_{ij} = 0$, 
are automatically satisfied by $h_{ij} = 0$, $i\neq j$.  Then  
\be
   h_{ti} = -4 \int\frac{\rho v_i(x')}{|\bm r - \bm r'|} dV'
\ee  

For a rotating star, $v=\Omega\bm\phi$, and the asymptotic form of $h_{ti}$ is proportional to the total angular momentum $J$.\index{angular momentum!total angular momentum of spacetime} This follows from the expansion 
\be 
   \frac1{|\bm r - \bm r'|} = \frac1{|r^2-2\bm r\cdot\bm r'+r'^2|^{1/2}} 
	= \frac 1r + \frac{\bm r\cdot\bm r'}{r^3} + O(r^{-3}).  
\ee
Because the total linear momentum vanishes, $\int \rho {v_i} dV = 0$, and the 
leading asymptotic term is the dipole contribution, 
\be 
   h_{ti} = -4\frac{1}{r^3}\bm r\cdot\int \bm r' \rho v_i dV' +O(r^{-3}).
\label{e:asymp}\ee 
Now $g_{ti} = h_{ti} + O(h^2)$. For a rotating star, 
$v^i = \Omega\phi^i$, and Eq.~\eqref{e:asymp} implies
\be
  g_{ti} = h_{ti} = 2\epsilon_{ijk} \frac{x^j J^k}{r^3} +O(r^{-3}).
\label{e:asymp2}\ee

\benr 
\item Asymptotic behavior of a rotating star.
\benalph\item  Show Eq.~\eqref{e:asymp} implies \eqref{e:asymp2}. 
\item The equations assume Cartesian coordinates.  Convert (5.41) to spherical 
coordinates, showing that the asymptotic $g_{ti}$ has the single nonzero component 
\be
   g_{t\phi} = -\frac{2J}r \sin^2\theta.
\label{e:asymp3}\ee 
\een\een

\noindent{\sl The Kerr spacetime}

Had anyone taken seriously the idea that stars might really collapse to
within their Schwarzschild radius, the Kerr solution might have been found
in the 1920s by asking what would happen if a rotating star collapsed.
Instead, Roy Kerr found it forty years later by looking for exact vacuum
solutions whose Riemann tensor had a particularly simple form.\footnote{Ezra Newman nearly 
found it -- he, Unti and Tambourino used the Newman-Penrose formalism to classify all solutions of that kind but made a error that Kerr caught; Newman was not happy that Kerr didn't share credit with him. In Kerr's version of the history, he and Alan Thompson had written an equivalent set of equations and that allowed Kerr to immediately spot the error \url{https://arxiv.org/abs/0706.1109}. Newman and Janis found a simple 
way to find the Kerr metric \url{https://aip.scitation.org/doi/10.1063/1.1704350} from 
a complex coordinate transformation of Schwarzschild, and Newman generalized that to obtain the
charged, rotating black hole solutions that generalize the charged Schwarzchild solutions 
of Riessner and Nordstrom.}  

 Kerr black holes are the unique stationary asymptotically flat vacuum
spacetimes with event horizons.  For each mass M, there is a 1-parameter
family  of Kerr black holes with that mass; the parameter is the angular
momentum $J$ about the symmetry axis, and at $J = 0$ the family begins with 
Schwarzschild.  There is a maximum angular momentum for each mass $M$:\index{angular momentum!maximum angular momentum of black hole}\index{Kerr spacetime!maximum angular momentum}\index{black hole!maximum angular momentum}
No star can contract past the point where its rotational speed exceeds the speed of 
light. This limits the angular momentum of a black hole
that can be physically generated.  The physical limitation is mirrored in
the solution set:  For $J > M^2$ (think of $ J_{\rm max} = Mv_{\rm max}R 
= M c M = M^2$), Kerr vacuum metrics exist, but they have no horizons.
Instead a naked singularity deprives these spacetimes of physical meaning.\\  
 
Several features of a rotating black hole could have been predicted at the
outset: Its two Killing vectors $ \phi^{\alpha}$ and $t^{\alpha}$ and 
its asymptotic form \eqref{e:asymp2}. 
Note that the $t$-$\phi$ part of the metric just involves the Killing vectors: 
\[
 g_{t\phi } = g_{\alpha \beta } t^{\alpha }\phi ^{\beta }= t^{\alpha }\phi _{\alpha},\qquad   g_{tt} = t^{\alpha }t_{\alpha },\qquad
 g_{\phi \phi } = \phi^{\alpha }\phi _{\alpha}.
\]
Furthermore, the geometry should be symmetric under a
simultaneous reversal of time and the direction of rotation, under
\be
	 t \rightarrow~-t , \qquad \phi \rightarrow~-\phi\,.
\ee
This condition eliminates the $tr$, $t\theta$, $\phi r$, and $\phi\theta$ components 
of $g_{\alpha\beta}$ and $G_{\alpha\beta}$; and one can choose coordinates to eliminate  
$g_{r\theta}$.     

	The Kerr metric, in a chart that becomes the ($t, r, \theta , \phi$)
chart of Schwarzschild when  $ J = 0$, is  
\index{Kerr spacetime!Boyer-Lindquist coordinates}\index{Boyer-Lindquist coordinates}\index{coordinates!Boyer-Lindquist}\index{metric!Kerr}
\beq
 ds^2 = - \frac{\Delta-a^2\sin^2\theta}{\rho^2}~dt^2~
	 -~2~\frac{2Mar\sin^2\theta}{\rho^2} dt~d\phi 	
	+ \frac{(r^2+a^2)^2 -a^2\Delta \sin^2 \theta}{\rho ^2}~\sin^2\theta d\phi ^2 
	+ ~\frac{\rho^2}{\Delta } dr^2~+~ \rho ^2d\theta ^2~,  
\label{e:kerr1}
\eeq 
where  
\beq
 a =J/M, ~~\rho^2 = r^2+a^2\cos^2\theta~=\Sigma_{Wald}, ~~\Delta = r^2-2Mr+a^2~.
\label{e:notation}\eeq
Notation:  The notes follow MTW and Chandrasekhar's {\sl Mathematical Theory of Black Holes}\cite{chandra83} 
in using $\rho^2$, while Wald and Shapiro-Teukolsky \cite{st86} use $\Sigma$.   

 Here $(t,r,\theta ,\phi)$ are called Boyer-Lindquist coordinates,
after their discoverers.  The metric (\ref{e:kerr1}) can also be written 
in the factored form
\beq\crv
 ds^2 = - \frac{\Delta }{\rho^2}~(dt-a\sin^2\theta d\phi)^2 
 	+ \frac{\sin^2 \theta }{\rho^2} [(r^2+a^2)d\phi - adt]^2
	+ \frac{\rho ^2}{\Delta } dr^2 + \rho ^2 d\theta ^2
\label{e:kerrfactored}\eeq
Its determinant is 
 \beq
	\sqrt{-g} = \rho^2\sin\theta,
\eeq
and the determinant of the  $t$-$\phi$ part of
the metric is 
\beq
   ^2g = \bm{ t \cdot t ~\phi \cdot \phi} -(\bm {t \cdot \phi})^2 
	 =  g_{tt}g_{\phi\phi}-(g_{t\phi})^2 \\
	= \Delta \sin^2\theta.
\label{dettp}\eeq 
The inverse metric is then
\beq
(g^{\mu\nu}) 
= \begin{pmatrix}
\dis-\frac{(r^2+a^2)^2  -a^2\Delta\sin^2\theta}{\rho^2\Delta}&0&0
				&\dis-2\frac{aMr}{\rho^2\Delta}\\
				0 & \dis\frac{\rho^2}\Delta &   0	& 0 \\
				0 &      0 		    & \rho^2	& 0 \\
\dis-2\frac{aMr}{\rho^2\Delta}& 0&& \dis \frac{\Delta-a^2\sin^2\theta}{\rho^2\Delta\sin^2\theta} 
\end{pmatrix}
\eeq

When $a = 0$ and $M\neq 0$, $\rho^2 = r^2, \ \Delta = r^2(1-\frac{2M}{r})$, and
the metric is 
\[
	ds^2 = -\left(1 - \frac{2M}r\right)dt^2 +
r^2\sin^2\theta d\phi^2 + \left(1 - \frac{2M}r\right)^{-1}dr^2 + r^2d\theta ^2,
\]
the Schwarzschild metric in Schwarzschild coordinates.

The most important difference between the geometry of a rotating star 
or black hole and the geometry of a static object is the {\sl dragging 
of inertial frames} (or just {\sl frame dragging}),
\index{Kerr spacetime!dragging of inertial frames}\index{dragging of inertial frames}
a nonzero $t$-$\phi$ 
metric component, whose meaning is that time-translation is not orthogonal to 
rotation, $t^\alpha \phi_\alpha \neq 0$. The angular velocity 
\beq\crv
  \omega := -\ \frac{t^\alpha \phi_\alpha}{\phi^\beta  \phi_\beta} ,
\eeq
measures the frame dragging in the sense that particles with zero 
angular momentum move along trajectories whose angular velocity 
relative to infinity is $d\phi/dt =\omega$. To see this, note that the angular 
momentum of a particle is $p_\alpha \phi^\alpha = p_\phi = m u_\phi$; 
we'll again use the angular momentum per unit rest mass, $L = u_\a \phi^\a$.  
For a freely falling particle (and hence for an inertial observer), $L$ 
is conserved.  But the angular velocity of an observer, measured from 
infinity is $\Omega = d\phi/dt = u^\phi/u^t$.  Now 
\beq
   L = u_\phi = g_{\phi t} u^t + g_{\phi\phi} u^\phi +  
	= u^t\bm{\phi\cdot t} + u^\phi \bm{\phi\cdot\phi}, 
\eeq
so a particle with zero angular momentum\index{angular momentum!zero angular momentum particle} has angular velocity
\beq
   \Omega \equiv \frac{u^\phi}{u^t} 
		= - \frac{\bm{t\cdot\phi}}{\bm{\phi\cdot\phi}} = \omega.  
\eeq 
A zero-angular momentum observer (ZAMO)\index{angular momentum!zero angular momentum observer}\index{zero angular momentum observer (ZAMO)} at constant radius $r$ has velocity 
$u^\alpha$ normal to a \mbox{$t=$ constant} surface:  That is, the vectors $\bm \partial_\phi = \bm\phi$, $\bm\partial_r$ and $\bm\partial_\theta$ span the subspace of vectors tangent to a 
constant $t$ surface;  $u^\alpha$ is proportional 
to $t^\alpha+\omega\phi^\alpha$, which is obviously orthogonal to the vectors 
in the $r$ and $\theta$ directions, and it is orthogonal to $\phi^\alpha$ because 
\beq
  (t^\alpha + \omega\phi^\alpha)\phi_\alpha 
	= t^\alpha\phi_\alpha - t^\alpha\phi_\alpha = 0.
\eeq 
For Kerr, from its definition and the form \eqref{e:kerr1} of the metric, $\omega$ is 
\beq
    \omega = \frac{2Mar}{(r^2+a^2)^2-a^2\Delta\sin^2\theta}.  
\label{e:omega}\eeq
Frame dragging was measured to about 20\% accuracy by gravity probe B. 
\url{https://ui.adsabs.harvard.edu/abs/2011PhRvL.106v1101E/abstract}  
More on frame dragging when we get to particle orbits.  \\

\noindent{\sl The horizon}\label{s:horizon} \index{Kerr spacetime!horizon location}\index{horizon!of Kerr spacetime}\\
Because the spacetime is stationary and axisymmetric, rotations and time-translations map
the horizon to itself. This implies that $t^\alpha$ and $\phi^\alpha$ are tangent to the 
horizon $H$.  (This is because $\phi^\alpha$ is tangent to the circular 
trajectories of rotated points, and $t^\alpha$ is tangent to the trajectories of time-translated points.)
We now show that the horizon of the Kerr geometry is the set of points 
where $\Delta = 0$.  We begin with an outline and then fill in the argument for each step. \newpage

\noindent Here's the outline: 

\noindent
1.  The event horizon is a null surface; that is, it is a hypersurface whose 
tangent space includes one null vector and no timelike vectors. 
\index{hypersurface!null} \\
2.  The null vector field of the Kerr horizon (and of any stationary axisymmetric 
horizon) is a linear combination, $t^\alpha + f\phi^\alpha$, of the two Killing vectors. \\
3.  The metric on a null surface has vanishing determinant. \\ 
4. Because the space of vectors spanned by $t^\alpha$ and $\phi^\alpha$ is 
a plane and that plane is null on the horizon, the determinant $^2g= \Delta\sin^2\theta = 0$ 
on the horizon.  \\

\noindent Here are the details:  \\
1.  The fact that the horizon $H$ is a null surface follows from 
its definition as boundary of the set of points from which timelike 
curves can reach null infinity. From $H$ 
and from points inside $H$, no null or timelike curves reach null infinity.
From each point outside $H$, outwardly directed null rays do reach null 
infinity.  Outwardly directed null rays beginning at points of the horizon, 
however, remain on the horizon.  Thus a null geodesic passes through each 
point of $H$.  Because one can regard $H$ as formed by these 
rays, the horizon is said to be {\em generated} by its null geodesics 
or by the null vector field $\ell^\alpha$ tangent to them. A horizon then 
is a null hypersurface, a hypersurface whose tangent space includes one 
null direction and no timelike vectors. 
\index{event horizon!as null surface}\index{horizon!as null surface}
\index{hypersurface!null|textbf}

Why can $H$ have no timelike vectors?
If $u^\alpha$ is a timelike vector on $H$ and $v^\alpha$ a spacelike 
vector pointing out from $H$, then for small enough $\lambda$, 
$u^\alpha+\lambda v^\alpha$ is still timelike, and it points out of $H$. 
Then a timelike curve tangent to $u^\alpha+\lambda v^\alpha$ reaches a point 
outside $H$.  Because all points outside $H$ are joined by timelike curves 
to null infinity, $H$ itself would be joined by a timelike 
curve to null infinity, contradicting its definition.
\\

\noindent 2.  We next show that the generators $\ell^\alpha$ of the horizon 
are linear combinations of $t^\alpha$ and $\phi^\alpha$.  Because $t^\alpha + k\phi^\alpha$ 
is a Killing vector for every constant $k$, it must be tangent to the horizon.  
(Combinations of rotations and time translations map the horizon to itself.) Because $H$ is 
a null surface, $t^\alpha + k\phi^\alpha$ is either spacelike or null.  
Suppose that the null generator at a point of $H$ has a part along $\theta$ or $r$:  
$\ell^\alpha = (t^\alpha + f \phi^\alpha) + v^\alpha$,
with $v^\alpha$ orthogonal to the $t$-$\phi$ plane.  Because the $r$ and $\theta$ directions 
are spacelike, $v_\alpha v^\alpha\geq 0$, vanishing only if $v^\alpha=0$. But     
\[
  0 = \ell_\alpha \ell^\alpha = (t_\alpha+f\phi_\alpha)(t^\alpha + f \phi^\alpha) + v_\alpha v^\alpha.
\]
Because each term on the right is non-negative, $v^\alpha = 0$ and 
$\ell^\alpha = t^\alpha + f \phi^\alpha$ as claimed.\\  

\noindent 3. Consider an element of area in a null plane with one segment along the null direction, 
and one in a perpendicular direction (call them the 1 and 2 directions). Because the 
null segment has proper length $d\ell_1 = 0$, the area is \mbox{$dA = d\ell_1 d\ell_2 = 0$}.   
Thus a null plane has $^2g = 0$.  \\

\noindent 4. At each point of the horizon of Kerr, the $t$-$\phi$ 
plane is null, whence Eq.~(\ref{dettp}) implies $\Delta = 0$. $\Box$\\

The argument holds for distorted black holes, black holes with stationary, axisymmetric matter 
(e.g., an accretion disk) outside the horizon: $H$ is again composed of $t$-$\phi$
surfaces for which 
\[
	^2g = \bm{t \cdot t ~\phi \cdot \phi} - (\bm{t \cdot \phi})^2 = 0.
\]

For a Schwarzschild
black hole, the asymptotically timelike Killing vector $t^\alpha$ becomes 
null on the horizon and is therefore its null generator.  For Kerr, the vector field 
$t^\alpha + \omega \phi ^\alpha $ is null on the horizon, because 
\begin{eqnarray}
 (t^\alpha + \omega \phi ^\alpha )(t_\alpha +\omega \phi _\alpha ) 
 	&=& t^\alpha t_\alpha + 2\omega t^\alpha\phi _\alpha 
 		+ \omega^2\phi^\alpha\phi_\alpha \nn\\
	&=&(\phi^\beta \phi_\beta)^{-1}[(t^\alpha t_\alpha )(\phi_\beta\phi_\beta)
	   - 2(t^\alpha\phi_\alpha )^2 + (t^\alpha\phi_\alpha )^2]\nn\\
	&=&(\phi^\beta \phi_\beta)^{-1}[(t^\alpha t_\alpha )(\phi^\beta \phi_\beta) 
		- (t^\alpha\phi_\alpha )^2] \nn\\
	&=&0 ~~ \text{at} ~~\Delta = 0.
\label{e:omega1}\end{eqnarray}

We have shown that the horizon is at the radius $r$ where $\Delta = 0$ and that it is 
generated by the null vector field $t^\alpha + \omega_H \phi^\alpha$, 
where $\omega_H$ is the value of the frame-dragging angular velocity 
$\omega$ on $H$. Now $\Delta\equiv r^2-2Mr+a^2 = 0$ has the roots 
\be
   r_\pm = M \pm (M^2-a^2)^{1/2}.  
\label{e:rpm}\ee
The solution $r=r_+$ is the horizon.\footnote{The second solution $r=r_-$ is also a null surface, 
lying inside the horizon and is called the inner horizon. It is not part of the spacetime of 
a collapsing star.}   
At the horizon, we then have 
\be
  \omega_H = - 	\left.{\frac{g_{t\phi}}{g_{\phi\phi}} }\right|_{r{_+}}  
  	=  \frac{a}{2Mr_+}.
\label{e:omegaH}\ee
Because the generators of the horizon rotate relative to infinity with angular
velocity $\omega_H$, one regards the horizon itself as an object rotating
with the angular velocity $\omega_H$, constant on $H$.        
 
Within the horizon (for $r < r_+$), there are no nonspacelike constant 
$r$ paths, and $g_{rr} < 0 \Rightarrow \bm\partial _r$ is a timelike
vector; future directed timelike curves are then constrained to have
$u^r < 0 $. (One can analytically extend Kerr across $\Delta = 0$, 
as in the Kruskal or Eddington-Finkelstein
extensions of Schwarzschild.)

\indent Asymptotically, the metric (\ref{e:kerr1}) has the form 
\[
ds^2 = - \left(1 - \frac{2M}r\right)dt^2 
	- \frac{4Ma}{r} \sin^2\theta dtd\phi  
       + \left(1 + \frac{2M}{r}\right)dr^2 + r^2(d\theta ^2 
	+ \sin^2\theta d\phi^2).
\]
This is also the asymptotic form of the metric of a rotating star, with 
$g_{t\phi}$ given by Eq.~\eqref{e:asymp2} and 
the $O(r^{-1})$ part agreeing with the Schwarzschild metric, 
the general form of a stationary, asymptotically flat geometry.  In 
the case of a rotating star, the star's angular momentum is $J=Ma$, and 
that is defined to be the angular momentum of a black hole as well.\index{angular momentum!total angular momentum of spacetime}%
\footnote{One can also define a symmetry group of rotations and translations at 
spatial infinity and use the symmetries to define the associated 
asymptotic conserved quantities.}
Note however, that the rest of the geometry of Kerr -- involving the quadrupole and higher 
multipole moments of the metric -- does not agree with the geometry of any rotating star. 
And rotating stars with different equations of state have different shapes and 
different gravitational fields (in both GR and in the Newtonian approximation).

  The asymptotically timelike Killing vector becomes null on a 
surface that lies outside the horizon (except at $\theta = 0, \pi$ where 
it touches the horizon).  $t^\alpha$ is  spacelike   
$(t^\alpha t_\alpha > 0)$ inside the $t^\alpha t_\alpha = 0$ surface:
\begin{align}
0 = t^\alpha t_\alpha &= g_{tt} = -\frac{\Delta-a^2\sin^2\theta}{\rho^2} 
\ \Longrightarrow \ r^2 - 2Mr + a^2 - a^2\sin^2\theta = 0 \nonumber \\
	r^2 &- 2Mr + a^2\cos^2\theta = 0. 
\end{align}
The region where $t^\alpha$ is spacelike is called the {\em ergosphere}
and is discussed in some detail below.

\noindent The rotational Killing vector is spacelike everywhere (at least for $r
> 0$):
\[
  \phi^\alpha\phi _\alpha = g_{\phi\phi} 
 	= \frac{(r^2 + a^2)^2 - a^2\Delta \sin^2\theta}{\rho^2}\sin^2\theta.
\]
Then
\[\rho^2 > 0,~~r^2+a^2>a^2, ~~r^2+a^2>r^2-2Mr+a^2 = \Delta\ \ \mbox{and }\ 
	1 > \sin^2\theta
\]
$\Rightarrow$
\[
	\frac{(r^2+a^2)^2}{\rho^2} >
\frac{a^2\Delta\sin^2\theta}{\rho^2} \Rightarrow \phi^\alpha\phi_\alpha > 0~.
\]
\noindent
[If $\phi^\alpha\phi _\alpha < 0$ as happens if we allow $r<0$, there are closed
timelike curves -- namely the curves of constant $t, r, \theta$, the 
circular trajectories along the vector field $\phi^\alpha $.]

\index{Kerr spacetime!$a>m$}
When $a > M, (M^2 - a^2)^{1/2}$ is imaginary and thus $\Delta$ is
nowhere zero.  The components of the Kerr metric (\ref{e:kerr1}) are
singular at $r=0$, and this singularity cannot be avoided by extending
Kerr via a coordinate change:  The scalar
$R^{\alpha\beta\gamma\delta}R_{\alpha\beta\gamma\delta}$, for example,
blows up as $r\rightarrow 0, (\theta=\pi/2)$.  In this case, unlike
Schwarzschild, or Kerr with $a<M$, null geodesics from all points
outside $r=0$ escape to infinity, so the
singularity is not shrouded by a horizon and is said to be naked.
As mentioned above, however, it appears that naked singularities
cannot, in fact, form from nonsingular initial data of physical fields,
and, in particular, attempts to find processes that take a black hole
with $a<M$ and add enough angular momentum to make $a>M$ have failed
(see e.g.  Wald, Annals of Physics, {\bf 82}, 548 (1974)).\index{angular momentum!maximum angular momentum of black hole}\index{black hole!maximum angular momentum}\index{Kerr spacetime!maximum angular momentum}  
\label{p:censor}
One reason they fail is this:  A typical attempt is to drop a gyroscope
with $a_{\rm gyro} \equiv \frac{J_{\rm gyro}}{M_{\rm gyro}} \gg M_{\rm gyro}$ into the
$a<M$ black hole.  But there is a spin-spin repulsion, and in order to push the
gyroscope into the black hole, you have to do so much work that 
$(W+ M_{\rm gyro} + M )^2>J_{\rm gyro}+J$, so the new mass of the hole, 
$W+M+M_{\rm gyro}$ exceeds its new angular momentum per unit mass.  The
statement that naked singularities cannot evolve from nonsingular
initial data of physical fields is called the cosmic censorship
hypothesis and its proof is a fundamental unsolved problem of
classical relativity.  \index{cosmic censorship}
We henceforth assume $a<m$.\\

\noindent{\sl Particle orbits: General features}
\index{Kerr spacetime!particle orbits}\index{black hole!particle orbits}

     The appearance of accreting black holes depends in part on the trajectories 
of photons and orbits of massive particles.  We'll restrict ourselves here 
to orbits in the equatorial plane.  
If, at $\infty$, $u^\phi = 0$, then $L= u_\phi = 0$ as well, because
the metric is asymptotically Minkowski; but, as we have seen, 
zero angular momentum particles have nonzero angular velocity
\beq
	\Omega = \omega.
\eeq
Because $\omega \sim r^{-3}$ as $r \rightarrow \infty$ this
dragging is hard to measure unless one observes objects near
a dense, rapidly rotating object -- in this case, a Kerr black hole.

Consider an arbitrary timelike or null trajectory:
\[
	u^\alpha = u^t(t^\alpha + \Omega \phi ^\alpha + v^\alpha)
\]
where $v^\alpha \perp t^\alpha, \ \phi^\alpha$.  In Minkowski space, 
a particle can have a timelike or null trajectory only if 
$r\sin\theta\Omega < 1$, implying $-1/r < \Omega < 1/r$.
Here the angular velocity $\Omega$ is limited to a range of values that
is not symmetric as seen by an observer at infinity. 
(The range is symmetric as seen locally by a observer with zero angular momentum).
\index{angular momentum!zero angular momentum observer}  
From $u_\alpha u _\alpha = -1 <0$, we have 
\[ 
	t^\alpha t_\alpha + 2\Omega t^\alpha\phi_\alpha 
	+ \Omega^2\phi^\alpha\phi_\alpha \leq 0.
\]
Then $g_{rr}>0,\ g_{\theta \theta} >0 \Longrightarrow v^\alpha v_\alpha > 0$, 
$\Rightarrow \Omega _- \leq \Omega \leq \Omega _+$, where
\[
	\Omega _{\pm} = - \bm { \frac{t\cdot \phi }{\bm \phi\cdot\phi } } 
			  \pm \left[\left(\bm{ \frac{ t\cdot\phi }{\phi\cdot \phi} }\right)^2 
			-\bm{\frac{t\cdot t }{\phi\cdot\phi}} \right]^{1/2}
\]
or
\beq
  \Omega _{\pm} 
	= \omega \pm \left(\omega^2 - \bm{ \frac{ t\cdot t }{\phi\cdot\phi}}\right)^{1/2}. 
\eeq
These extrema are reached when $ v^\alpha = 0$, that is, for motion 
in the equatorial plane at constant $r$.\\

For large $r$, 
\[
  \omega \sim  \frac{2J}{r^3} \qquad   t^\alpha t_\alpha = g_{tt} \sim -1
 \qquad \phi^\alpha\phi_\alpha = g_{\phi\phi} \sim r^2,
\]
and $\Omega_\pm \approx  \pm\dis \frac1r$, as in Minkowski space.  
As long as $t^\alpha t_\alpha < 0, \Omega
_{-} < 0$ and $\Omega_+ > 0$ and particles can rotate with or opposite to
the black hole rotation.  But when $t^\alpha t_\alpha > 0$, both $\Omega_+$ and 
$\Omega _-$ are greater than zero: 
{\em All timelike or null trajectories rotate with the geometry} -- no
physical particle can remain at rest as seen by an observer at infinity.
In other words, when $t^\alpha$ is spacelike, no physical particle can move
along this Killing vector.  In contrast to the Schwarzschild case, however,
particles within the region where $t^\alpha$ is spacelike can escape to 
infinity because for $\Omega_- < \Omega <\Omega _+,\ \  u^r$ can be positive. \\

 The boundary of the ergosphere, where $t^\alpha t_\alpha = 0$, is called the static limit by MTW (stationary limit by most other people).  
Observers inside the static limit can remain at  constant distance 
from the black hole (constant $r$) as long
as they rotate in the positive direction with angular velocity $\Omega$  in
the range $\Omega_{-} < \Omega <\Omega _+.$  But as 
$\Delta \rightarrow 0$, $\Omega _-\rightarrow \Omega _+$:
\bea
    \Omega _+ - \Omega_{-} 
	&=&\left(\omega^2 - \frac{t^\alpha t_\alpha }{\phi^\beta \phi_\beta}\right)^{1/2} 										\nonumber \\
&=& (\phi^\alpha\phi_\alpha )^{-1}
	\left[(t^\beta \phi_\beta)^2 - (t^\beta t_\beta)(\phi^\gamma \phi_\gamma)\right]^{1/2}\,.
\eea
From eq. (\ref{dettp}), namely\,  $(t^\beta \phi{}_\beta)^2 
- (t^\beta \phi{}_\beta)(\phi^\gamma\phi _\gamma) = \Delta\sin^2\theta$, 
\[\Omega _+ - \Omega _{-} \rightarrow 0 ~~\text{as}~~ \Delta \rightarrow 0~.\]
So the allowed range of rotation shrinks to zero as $\Delta \rightarrow
0$, and for $\Delta = 0$ the only nonspacelike constant $r$ path is the
null trajectory along the generator of the horizon, 
with $\Omega = \Omega _+ = \Omega _{-} = \omega_H$.  

\section{Ergospheres: The Penrose process, superradiant scattering, and ergosphere instabilities}
\index{ergosphere|textbf}\index{ergosphere!instability}\index{ergosphere!Penrose process}\index{Kerr spacetime!Penrose process}\index{Kerr spacetime!ergosphere|textbf}\index{Penrose process}

\noindent{\bf Definition} An {\sl ergophere} is a region in which an asymptotically timelike Killing vector $t^\alpha $ is spacelike.  \\
Ergospheres are present not only in rotating 
black holes, but also in models of dense, rapidly rotating stars
(perfect fluids) which have been constructed numerically by \href{https://articles.adsabs.harvard.edu/full/1975ApJ...200L.103B}{Ipser \&
Butterworth}\cite{bi75} (earlier, for dust disks,
by Bardeen \& Wagoner\cite{bw71}). 
If a star is compact enough, and it spins fast 
enough, its rotation of a star can force all physical particles 
within a region to rotate with respect to a distant observer.  For later 
references, see \href{https://journals.aps.org/prd/pdf/10.1103/PhysRevD.101.064069}{Tsokaros et al.} \cite{trs20}, who look at the rotation laws and equations of state that give rise to these {\sl ergostars}.
The term is similar to {\em atmosphere}; the ergosphere of a dense, rapidly 
rotating stellar model is a solid torus, not a topological sphere.)  

\newpage
\begin{figure}[h!]
\centering\includegraphics[width=\textwidth]{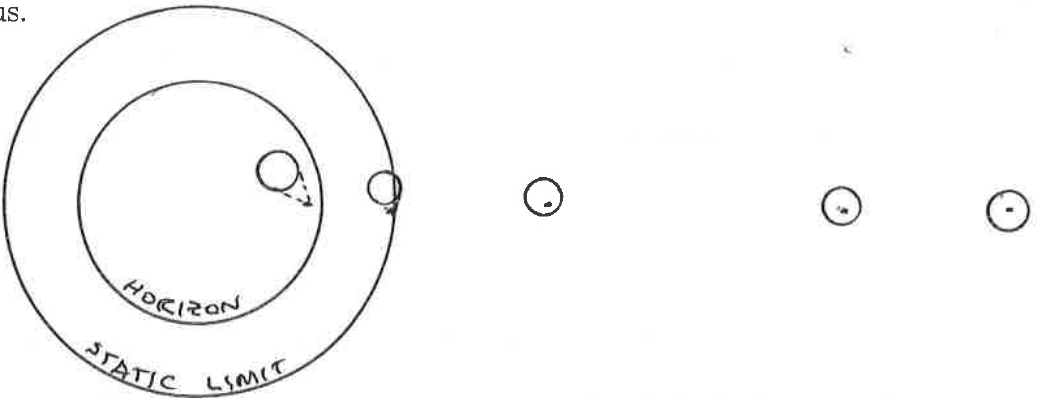}
\caption{A $t$ = const, $\theta = \frac{\pi}{2}$ plane
showing the horizon, static limit, and ergosphere.  Light cones are
represented by dots for their vertices and circles for the cone - a circle
represents the position of flash of light a short time after it 
was emitted from the dot.  Picture the light cone by imagining time out of the paper, so the
circles sit just above the paper.}
\end{figure}

	It is possible to extract rotational energy from a geometry with an
ergosphere in the following way.  A particle (1) sent from infinity to the
ergosphere can decay into particles (2) and (3), leaving (2) in the 
ergosphere and sending (3) back out to $\infty$ with greater energy 
than (1) had to begin with.  Let particle (1) begin with momentum 
$p_1{}^\a$. Along its trajectory, $-p_{1\,\a} t^\alpha = E_1$ is constant; $E_1$ is its energy measured by an observer at infinity who sends 
the particle in (at infinity, $t^\alpha $ is a unit timelike vector 
and can be chosen as the velocity of our observer).

\begin{figure}[H]
\centering\includegraphics[width=.6\textwidth]{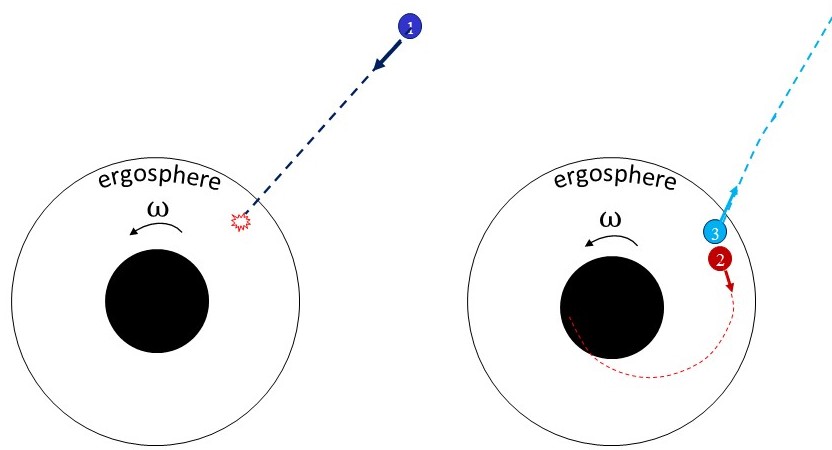}
\caption{Particle 1 falls into the ergosphere, where $t^\a$ is spacelike. 
It decays into particles 2 and 3.  Particle 2 has $E= - p_t >0$, and particle 
3 emerges with energy greater than the initial energy of particle 1.  
The diagram shows trajectories 
relative to zero-angular-momentum observers (ZAMOs); a ZAMO trajectory 
here is a straight line directed inward. Were they drawn as seen from infinity, all 
trajectories inside the ergosphere would proceed in a positive $\phi$ direction. }
\end{figure}

\noindent Inside the ergosphere, $t^\alpha $ is spacelike, so 
$-p_{1\a} t^\a$ is not the energy measured by any observer.  In
particular, consider a ZAMO, an observer with velocity $u^\a =
u^t(t^\a+\omega \phi^\a)$.  Let (1) decay into (2) + (3).  The
observer $u^\a$ must see $u^\a p_{2\a} < 0$ and 
$u^\alpha p_{3\,\alpha} < 0$ because the locally measured energy is positive:  
\centerline{
$0 > u^\alpha p_{2\alpha} 
	= u^t(t^\alpha p_{2\alpha} + \omega \phi^\alpha p_{2\alpha})
	= u^t(t^\alpha p_{2\alpha} + \omega L)$,}
$L$ the angular momentum of (2). 

Since $t^\alpha$ is spacelike, $t^\alpha p_{2\alpha}$ can be positive. 
If we choose $L < 0$, the momentum $\ t^\alpha p_{2\alpha}$ 
need only satisfy $t^\alpha p_{2\alpha} < |\omega L|$; because all 
decays that satisfy conservation of 4-momentum can occur, we can 
choose a decay for which $t^\alpha p_{2\alpha} >0$.\\
Momentum conservation in the decay is
\[
	p_{1\a} = p_{2\a} + p_{3\a} .
\]
Its component along $t^\alpha$,
\[
p_{1\a} t^\a = p_{2\a} t^\a + p_{3\a} t^\a,
\]
implies
\[
E_3 = -p_{3\a} t^\a = -p_{1\a} t^\a +  p_{2\a} t^\a >E_1,
\]
and particle (3) reaches infinity with energy greater than
that particle (1) had.  The energy comes from the rotational energy of 
the black hole:  Particle 3 carries off some of the black hole's angular 
momentum.  

Note that particle (2) cannot escape to
infinity, or even to a region where $t^\alpha $ is timelike, because
then the physical energy of (2) measured by an observer with velocity
along $t^\alpha $ would be negative.  In the case of an ergosphere
about a Kerr black hole, particle (2) spirals into the black
hole.  When a geometry has an ergosphere and no horizon, however, a particle 
with $L < 0$ (or a field in the ergosphere with negative ``energy'') 
will radiate and its ``energy,'' $-p_\alpha t^\a $, will
grow increasingly negative until ultimately enough angular momentum
is radiated to $\infty$ that no ergosphere remains. 
This growth of a particle's negative energy, defined as 
$-p_\a t^\a$, is a simple example of the instability of an ergostar, 
described below. 

There is no similar mechanism to extract energy from a nonrotating black hole.  
In his celebrated 1971 paper,\cite{hawking71} Hawking proved that the area of a 
black hole cannot decrease (the proof is given in Sect.~\ref{laws} below).  
Because the area of a Schwarzschild black hole is $4\pi(2M)^2 = 16\pi M^2$, 
its mass can only increase.  A rotating black hole, however, has area 
$A = 8\pi M r_+ = 8\pi M (M+\sqrt{M^2 - a^2})$, so $A$ can increase in a 
process that carries away both energy and angular momentum.  \\

\noindent{\sl Superradiant scattering and ergosphere instabilities.}  

The counterpart of the Penrose process for a classical field is called superradiant scattering.  A scalar, electromagnetic or linearized gravitational field can radiate negative energy across the horizon, here meaning that the current $J^\a = -T^\a{}_\b t^\b$ has negative flux across the horizon, that $\int_H J^\a dS_\a < 0$.  For a massless scalar field, an electromagnetic field or a gravitational wave, this implies that the outgoing field 
carries more energy to future null infinity than the energy of the initial incoming field. 

In the case of an ergostar (with no horizon), the conserved energy of radiation 
fields on a background with negative values in the
ergosphere leads to instability \href{https://projecteuclid.org/journals/communications-in-mathematical-physics/volume-63/issue-3/Ergosphere-instability/cmp/1103904565.full}{Friedman '78}\cite{f78}:  Because the negative energy part of the field is again
trapped within the ergosphere, and because a time-dependent,
nonaxisymmetric field radiates positive energy to infinity, fields of
this kind are unstable in spacetimes with an ergosphere and no horizon. 
As linear fields on a background spacetime, they grow without bound; 
in the full theory, they radiate away the angular momentum of the 
ergostar until no ergosphere remains.    
The associated growth time, however, is comparable to the age of the
universe \cite{cs78}.

There is a related instability of rotating black holes, associated in this case 
with a {\sl massive} scalar field (or any massive boson field) 
outside the ergosphere. Because a massive scalar field 
falls off exponentially at spatial and null infinity, scalar radiation flows only into 
the horizon, not to infinity.  If the field is superradiant, radiating negative 
energy across the horizon, the field outside the horizon again grows without 
bound on a fixed background spacetime (Damour '76\cite{damour76},\href{http://users.physics.uoc.gr/~tzouros/References/Zouros_AoP118_139_1979.pdf}{Zouros \& Eardley 79}\cite{ze79},
\href{https://journals.aps.org/prd/pdf/10.1103/PhysRevD.22.2323}{Detweiler '80}\cite{detweiler80}).

  With spacetime dynamics included, the black hole spins down and is surrounded by a condensate of the 
massive boson field.  For very small scalar masses, whose Compton wavelength 
is larger than the black hole's radius, the growth time is small compared to the age of the universe. As pointed out by \href{https://journals.aps.org/prd/pdf/10.1103/PhysRevD.83.044026}{Arvanitaki and Dubovsky}\cite{ad11},
this means that any observation of the mass and spin of a black hole can set an upper bound on the mass of light bosons.  
Because ultralight bosons (axions, for example) are a candidate for dark matter, a ballooning literature has followed the Arvanitaki-Dubuvsky paper. \\

\benr
\item An observer is stationed at fixed values of $r$ and $\phi$ in the equatorial plane of a rotating black hole 
of mass $M$ and angular momentum $Ma$.   A proton is moving in the $\phi$-direction as it traverses the laboratory 
of the observer, who measures its energy $E$ and momentum $p_\perp$.  
\benalph
\item What is the relation between $E$ and $p_\perp$? 
\item What are the components of the proton's 4-momentum in the Kerr coordinate basis  
in terms of $E$ and $p_\perp$? 
\item What is the proton's speed $v$ as measured by the observer? (Give $v$ in terms of $E$, $p_\perp$, and 
rest mass $m$?  
\item What are the components of the proton's 4-velocity along an orthonormal basis aligned with the coordinate 
directions in terms of $v$?
\item What are the values of the two conserved quantities $E = - u_\alpha t^\alpha$ and 
$L = u_\alpha\phi^\alpha$?
\een

\een

\newpage

\section{Orbits of Kerr black holes}  
This section partly follows corresponding discussions in Chandrasekhar \cite{chandra83} 
and in Shapiro-Teukolsky\cite{st86} 12.7.\\

\subsection{Photon orbits} \index{Kerr spacetime!photon orbits (null geodesics)}\index{black hole!photon obits (null geodesics)}

We will look at orbits in the equatorial plane.  Massive particles have an orbit 
through any point for each value of the particle's velocity.  
Photon orbits have only one speed and so only one orbit for a given spatial 
direction.  Where massive particles can adjust their speed to have circular orbits 
for all radii greater than some minimum value, photons have only 
a single circular orbit outside the horizon.  

As in Schwarzschild, we write $k^\alpha k_\alpha=0$ in terms of 
$\dot r$ and the  conserved energy/$\hbar$ and angular momentum/$\hbar$,  
\beq
   E = -k_\alpha t^\alpha = -k_t, \qquad L  = k_\alpha\phi^\alpha = k_\phi. 
\eeq
At $r=\infty$, a stationary observer has velocity $t^\alpha$, so 
$-\hbar k_t$ and $\hbar k_\phi$ are the 
energy and angular momentum of a photon, if the affine parameter is 
chosen to equal $t$ at infinity.  The orbit of a photon moving in 
the equatorial plane is determined by its impact parameter%
\footnote{As in the case of a massive particle, the definition is natural because 
the energy and angular momentum measured by a stationary observer at infinity are 
$\texttt L = \hbar L$ and $\texttt E = \hbar E$, $\texttt L/\texttt E=L/E$ is the impact parameter in flat space, and the spacetime is asymptotically flat. 
Note that changing $E$ and $L$ to $kE$ and $k L$ in \eqref{e:dotr} 
changes 
$\dot r = dr/d\lambda$ to $k\dot r$:  This corresponds to a change 
$\lambda\rightarrow \lambda/k$ in the affine parameter, and it leaves 
$dr/dt = \dot r/\dot t$ unchanged. Adopting our convention that $\dot t=1$ at 
infinity fixes $\dot r$. } 
\[
   b:= \frac L E.
\]  

The equatorial plane is the $\theta=\pi/2$ surface, with
\[
   \sin\theta = 1, \quad \rho^2 = r^2.  
\]
We have 
\begin{align}
0 &=  g^{tt} E^2 - 2 g^{t\phi} E L + g^{\phi\phi}L^2 + g_{rr}\dot r^2
\nonumber\\
     &= -\frac{(r^2+a^2)^2 - a^2\Delta}{r^2\Delta}  E^2 +4\frac{aM}{\Delta r}  EL 
       + \frac{\Delta-a^2}{r^2\Delta} L^2 + \frac{r^2 \dot r^2}\Delta
 \nonumber\\     
  r^2\dot r^2 &= \left(r^2+a^2 + \frac{2M}r a^2\right)  E^2 -4\frac{aM}r  EL 
	  	  -\left(1-\frac{2M}r\right) L^2.         
\label{e:dotr}\end{align} 
Dividing by $r^2$ and grouping terms with the same power of $r$ gives
\beq
    \dot r^2 =  E^2 +\frac{2M}{r^3}(L-a E)^2 - \frac{L^2-a^2 E^2}{r^2},
\label{e:rdot}\eeq
the equation of a particle moving in an effective potential 
$\dis  U = - \frac{2M}{r^3}(L-a E)^2 + \frac{L^2-a^2 E^2}{r^2}$.\\
\index{effective potential!Kerr spacetime, equatorial photon orbits|textbf}

\newpage\noindent{\sl Ingoing and outgoing photons: Principal null geodesics}\\

Eq.~\eqref{e:rdot} has its simplest form when $L=aE$: 
\be
   \dot r \equiv \frac{dr}{d\lambda} = \pm E.
\label{e:rdot1}\ee 
This means $r$ is an affine parameter, as it was for Schwarzschild. 
Without loss of generality, we can take $\lambda = \pm r$, $E=1$.   

To complete the solution, we need  
$\dot\phi = k^\phi$ and $\dot t = k^t$ as functions of $r$: 
\begin{align}
\dot\phi = k^\phi &= g^{\phi\phi}k_\phi+g^{\phi t}k_t 
	= \frac{\Delta-a^2\sin^2\theta}{\rho^2\Delta\sin^2\theta} L + 2\frac{aMr}{\rho^2\Delta}\sin^2\theta  E
\label{e:phidot}\nonumber\\
	&= \frac1\Delta\left[ \left(1-\frac{2M}r\right)L +a\frac{2M}r  E\right],\quad \mbox{ at } \theta=\pi/2,\\
\dot t = k^t &= \frac1\Delta\left[ \left(r^2+a^2+2\frac{a^2M}r\right) E - a\frac{2M}r L\right],
\label{e:tdot}\end{align}
Here, with $L=aE=a$, the relations are simply 
\be
 \dot \phi = \frac a\Delta, \qquad \dot t = \frac{r^2+a^2}\Delta.
\label{e:kgeod}\ee
 
Eqs.~\eqref{e:kgeod} now give the coordinates $t$ and $\phi$ of outgoing and 
ingoing geodesics as functions of $r$:  Up to a choice of initial values, we have
\bsube\begin{align}
  \phi &= \pm\int \frac{a}\Delta dr = \pm \frac{a}{r_+-r_-} \ln\left|\frac{r-r_+}{r-r_-}\right| \\
   t   &= \pm\int \frac{r^2+a^2}\Delta dr = \pm r_*, \quad \mbox{with}\ \  r_* = r+\frac{Mr_+}{r_+-r_-} \ln\left|\frac{r-r_+}{r_+}\right| 
   	  				- \frac{Mr_-}{r_+-r_-} \ln\left|\frac{r-r_-}{r_-}\right|. 
\end{align}\label{e:kgeod1}\esube
The corresponding null vectors tangent to the outgoing and ingoing geodesics then have 
$t,r,\theta,\phi$ components
\bsube\begin{align}
      (\ell^\mu) &=  \left(\frac{r^2+a^2}\Delta,\,1,\,\frac a\Delta,\, 0\right), 
\label{e:laffine}\\
         (n^\mu) &=  \left(\frac{r^2+a^2}\Delta,\,- 1,\,\frac a\Delta,\, 0\right).
\label{e:naffine}\end{align}\esube
 
These vectors will reappear in Sect.~\ref{s:teukolsky} (with $\bm n$ rescaled to make $\bm\ell\cdot \bm n = 1$) as two of the Newman-Penrose tetrad vectors for the Kerr geometry; and the ingoing null geodesics 
will, as in Schwarzschild, be used in Sect.~\ref{s:kcoord} to construct a chart regular on the future horizon.  

Although we have found these geodesics by assuming $\theta = \pi/2$, the equations hold for 
any value of $\theta$: The ingoing and outgoing solutions describe null geodesics with $\dot\theta=0$.\\
\newpage
  
\noindent{\sl Circular orbits}\\

For a circular orbit, $\dot r=0$, implying $ E^2 = U$, and $dU/dr=0$: 
\begin{align}
 E^2 +\frac{2M}{r^3}(L-a E)^2 - \frac{L^2-a^2 E^2}{r^2} &= 0, 
\label{e:c1}\\
 - \frac{6M}{r^4}(L-a E)^2 + 2\frac{L^2-a^2 E^2}{r^3} & =0.
\label{e:c2}
\end{align}
These are two equations relating $r, E$ and $L$; notice, however, that they 
are homogeneous in $( E,L)$, that every term is proportional to $ E^2$, $L^2$ or $ EL$. 
Dividing each equation by $L^2$ then gives two equations for the two 
variables $r$ and $b=L/E$ 
and so determines the values of each, giving $r$ and the angular 
velocity of the orbit.  
The second equation immediately gives 
\beq
   r = 3M\frac{L-a E}{L+a E} = 3M\frac{b-a}{b+a}. 
\label{e:rc}\eeq
Substituting this value of $r$ in the first equation \eqref{e:c1}
yields for $b$ the cubic equation 
\beq
  1 = U/ E^2 =  \frac1{27M^2}\frac{(b+a)^3}{b-a} 
\eeq 
To solve, we simplify as usual by appropriately picking dimensionless 
variables 
\beq
   y:= \frac{b+a}{3M}, \quad \chi = \frac aM.  
\label{e:y}\eeq
Then $\dis\frac{b-a}M = 3y-2\chi$, and the  equation is surprisingly simple:  
\beq
   y^3 - 3y + 2\chi = 0, 
\eeq
with a real solution 
that depends on whether the orbit is corotating or counter-rotating 
with the black hole.  (See the discussion of the solution to a cubic after 
Eq.~\eqref{e:minimumr}).  Assuming $\chi=a/M>0$, the corotating orbit is 
the solution given by    
\[
   y = 2\cos\left[\frac13\cos^{-1}(-\chi)\right] =: 2\cos\psi.    
\]
and, from \eqref{e:y} and \eqref{e:rc}, respectively, 
\beq
  b = 6M \cos\psi - a, \quad r = 3M (1-2\chi/y) = M\left(3-\frac\chi{\cos\psi}\right).  
\eeq
As $\chi$ increases from $0$ to $1$ (as $a$ increases from $0$ to $M$), 
$\psi$ increases from from $\pi/6$ to $\pi/3$ and \\ $r$ decreases from 
\beq
   r=3M\ \mbox{ at } \chi = 0,\quad \mbox{ to } \quad r=M \ \mbox{ at } \chi = 1.
\eeq 
(When $\chi =1$, the horizon itself has shrunk to $r=M$.)  

Counter-rotating orbits have $\Omega$ and $a$ with opposite signs.  We can 
just switch the sign of $a$ (of $\chi$) to write 
\beq
    r = M\left(3+\frac{|\chi|}{\cos\psi}\right), \quad \cos3\psi = |\chi|\quad \mbox{ (counter-rotating)}, 
\eeq
with $r$ increasing from $3M$ to $4M$ as $|\chi|$ increases from $0$ to $1$.

Eqs.~\eqref{e:phidot} and \eqref{e:tdot} again give $\Omega$ in terms of $b$ and $r$, 
\beq
   \Omega = \frac{\left(1-2M/r\right)b +2aM/r}
		{\left(r^2+a^2+2a^2M/r\right) - (2aM/r) b}.
\label{e:Omega}\eeq

\subsection{Particle orbits}\index{Kerr spacetime!particle orbits}\index{black hole!particle orbits}

Finding the circular orbits in the equatorial plane is done in essentially the 
same way as for photons, and the basic equations are easy to write down, 
just changing $k^\alpha k_\alpha = 0$ to $u^\alpha u_\alpha = -1$:
Again multiplying by $\Delta/r^2$ gives Eq.~\eqref{e:rdot} with the 
additional term $-\Delta/r^2$ on the right side: 
\beq
 \dot r^2 =  E^2 -U, \quad 
U = - \frac{2M}{r^3}(L-a E)^2 
		+ \frac{L^2-a^2 E^2}{r^2}+\frac\Delta{r^2}.
\label{e:U}\eeq
\index{effective potential!Kerr spacetime, equatorial particle orbits|textbf}

The conditions for a circular orbit are again $\dot r=0$, $\ddot r=0$, or  
$  E^2 -U = 0$, $dU/dr = 0$.  Again grouping powers of $r$, we write the 
first relation as
\beq
   E^2 -1 + \frac{2M}{r^3}(L-a E)^2 
		- \frac{L^2-a^2 (E^2-1)}{r^2}+\frac{2M}r = 0.
\eeq
Because $ E^2 -U=0$, we can write the second relation, $U'=0$, in the form 
$\dis\frac1{2r}[r^2( E^2 -U)]'=0$:  
\beq
   E^2 -1 -\frac M{r^3}(L-a E)^2 + \frac Mr = 0.
\label{e:esq} \eeq 
Our goal is to use the two equations to find the generalized Kepler 
relation $\Omega(r)$ and to find the innermost stable circular orbit. 

We have two equations for the three variables $r$, $E$ and $L$.
The strategy is to solve them for $E$ and $L$ in terms of $r$, 
to write $\Omega$ in terms of  $E$ and $L$, and so to obtain 
$\Omega(r)$.  First the surprisingly simple result, then the calculation.
The Newtonian Keplerian angular velocity is $\dis\Omega_N = \sqrt{\frac M{r^3}}$, 
and we have seen the form of $\Omega$ is identical for Schwarzschild.  
For Kerr it has the form   
\begin{align}
  \cblue\Omega & \cblue= \frac{\Omega_N}{1+a\Omega_N}, \quad\ \mbox{ \cb corotating}, \nonumber\\
  \cblue\Omega & \cblue= -\frac{\Omega_N}{1-a\Omega_N}, \quad\mbox{\cb counter-rotating} . \cb  
\end{align} \cb 
At order $a$, the change is due to frame dragging because the order 
$a$ change in the metric is the nonzero value of $g_{t\phi}$. From Eq.~\eqref{e:omega}, $\omega= 2a\Omega_N^2 + O(a^2)$, implying 
$\Omega = \pm(\Omega_N\mp\omega/2) +O(a^2)$. 
\footnote{The term linear in $a$ in the effective potential raises its height for a corotating orbit.   
In heuristic language, a larger value of $L$ is then needed for the centrifugal barrier to balance 
the gravitational $M/r$ potential. There is an opposite effect from the relation between 
$L$ and $\Omega$, the relation $L = r^2(\Omega-\omega)\dot t$ requiring 
a larger $\Omega$ for the same $L$, but the change in the effective potential wins.}    \\

\noindent The calculation:  \\
This time use the variable%
\footnote{If you're going to try 
the calculation, you might want to set $M=1$ or divide variables by $M$ to 
make everything dimensionless. On the other hand, keeping $M$ has the advantage 
that you can spot errors by checking that every term has the right dimension as 
a power of $\texttt L=[M]$.}
$x=L-a E$
and write the equations in terms of $x, E$ and $r$, using 
$L^2-a^2 E^2 = x^2+2a Ex$ in the first equation. 
The equations are now   
\begin{align}
 E^2 &= -\frac{2M}{r^3} x^2 + \frac{x^2+2a Ex}{r^2} +\frac\Delta{r^2} 
\nonumber\\
 E^2 &= \frac{Mx^2}{r^3} - \frac Mr+1.  
\label{e:e2}\end{align}
We now eliminate $E$ to get an equation for $x$ and $r$ alone 
and then solve for $x$ in terms of $r$.  First  get rid of $E^2$ by subtracting the first equation from the second. 
This lets us find $E$ in terms of $x$ and $r$ without having 
to solve a quadratic equation: 
\beq
   2ax E = - \left(1-\frac{3M}r\right)x^2 +Mr-a^2
\label{e:E}\eeq
 We now have independent expressions \eqref{e:E} for 
$ E$ and \eqref{e:e2} for $ E^2$.  So we get an 
equation for $x$ and $r$ alone by\\
 $[$right side of \eqref{e:E}$]^2 =$ [right side of \eqref{e:e2}]$\times(2ax)^2$:
Notice that, although it's quartic in $x$, it's quadratic in $x^2$! 
So we're almost done. The equation has the form   
\beq
  \alpha x^4 -2 \beta x^2 + \gamma = 0, 
\eeq   
with 
\begin{align}
  \alpha &= \left(1-\frac{3M}r\right)^2-4a^2\frac M{r^3}
  	  =(1-3M/r-2a\Omega_N)(1-3M/r+2a\Omega_N)  ,\nonumber\\
  \beta & = \left(1-\frac{3M}r\right)\left(Mr-a^2\right) 
		+2a^2\left(1-\frac Mr\right),\nonumber\\
 \gamma & = (Mr-a^2)^2, 
\end{align}    
and the next happy fact is that, although $\alpha$ and $\beta$ are somewhat 
long, the discriminant is short
\beq
   \beta^2 - \alpha\gamma = 4a^2 \Delta^2 M/r^3 = (2a\Delta\,\Omega_N)^2\,.
\eeq
(This is one line in Mathematica, using \texttt{Factor[]}. By hand 
it took a page.) 
The solution is then 
\footnote{Intermediate steps are $x^2 = (\b - 2a\Delta\,\Omega_N)/\a,\quad \beta -2a\Delta\,\Omega_N = (1-3M/r-2a\Omega_N)(r^2\Omega_N - a)^2$.}
\beq
  x = \frac{r^2\Omega_N-a}{\sqrt{1-3M/r+2a\Omega_N}}\,.
\eeq 
Everything has this same denominator, and it just gets 
carried along, so we'll call it $Q$: 
\beq
   Q = \sqrt{1-3M/r+2a\Omega_N}\,.
\eeq
Eq.~\eqref{e:e2} gives 
\beq
   E = \frac{1-2M/r+a\Omega_N}Q, 
\eeq
and we have
\beq
 L = x+a E 
	= \frac{(r^2+a^2)\Omega_N-2aM/r}Q.
\eeq

Finally, we find $\Omega(r)$ from Eq.~\eqref{e:phidot} and 
\eqref{e:tdot} for $\dot\phi$ and $\dot t$. This is quick:  
\begin{align}
   \dot\phi &= \frac{\Omega_N}Q\ , \nonumber\\
   \dot t &= \frac{1+a\Omega_N}Q\ , 
\end{align}  
whence 
\beq
\Omega = \frac{\Omega_N}{1+a\Omega_N}\,,
\eeq
as claimed.  This is the solution for a corotating orbit, with $a>0$. 
For a counter-rotating (retrograde) orbit, change $a$ to $-a$ with 
the convention that $\Omega > 0$, or replace $a$ by $-a$ and 
change the sign of $\Omega$ for the standard convention that $a>0$ 
and $\Omega < 0$ for counter-rotating orbits, 
$\Omega = -\Omega_N/(1-a\Omega_N)<0$.     

\index{Kerr spacetime!ISCO (innermost stable circular orbit)}
\index{innermost stable circular orbit (ISCO)!Kerr spacetime}
The condition for stability of the circular orbit, $U''>0$ can 
be written as $[r^3( E^2 -U)]'' = 0$, because 
$ E^2 -U = 0$ and $( E^2 -U)' = -U' =0$. 
From Eq.~\eqref{e:U}, we have at the ISCO (innermost stable circular orbit), 
\beq
    E^2 = 1-\frac23\frac Mr.
\eeq 
The radius of the ISCO decreases from $6M$ to $M$ as 
$a$ increases from 0 to $M$ for corotating orbits, and it increases from 
$3M$ to $9M$ for counter-rotating orbits.  Its explicit form is given by 
\href{https://articles.adsabs.harvard.edu/pdf/1972ApJ...178..347B}
{Bardeen, Press and Teukolsky 1972}:\cite{bpt72}
\begin{align*}
r_{\rm ISCO} &= M\left\{ 3+Z_2\mp \left[(3-Z_1)(3+Z_1+2Z_2) \right]^{1/2}\right\},\\
Z_1 &= 1+\left(1- \frac{a^2}{M^2}\right)^{1/2}\left[\left(1+\frac aM\right)^{1/3}
		+  \left(1-\frac aM\right)^{1/3} \right],\\
Z_2&= \left(3\frac{a^2}{M^2} +Z_1^2\right)^{1/2}.
\end{align*} 

\section{Kerr coordinates: Coordinates regular on the future (or past) horizon}  
\label{s:kcoord}\index{Kerr spacetime!Kerr coordinates|textbf}\index{coordinates!Kerr coordinates|textbf}

Kerr coordinates are the analog for rotating black holes of the Eddington-Finkelstein coordinates for spherical black holes (Sect.~\ref{s:ef}). 
As in Eq.~\eqref{e:ef0}, we introduce a null coordinate $v$ for which the 
\mbox{$v=$ constant} surfaces are null surfaces generated by ingoing null geodesics.   From Eq.~\eqref{e:naffine}, we see that along the principal ingoing null geodesics of Kerr, $\dis d\phi + \frac a\Delta dr= 0$ and $\dis dt + \frac{r^2+a^2}\Delta dr = 0$. 
Then, if we define $v$ and a new angular coordinate $\widetilde\phi$ by 
\bsube\begin{align}
   d\widetilde\phi & = d\phi + \frac a\Delta dr, \\
   d v &= dt + dr_* = dt + \frac{r^2+a^2}\Delta dr, 
\label{e:vkerr}\end{align} \esube
$\wt \phi$ and $v$ will be constant along the geodesics.  Replacing $d\phi$ and $dt$ in 
the factored form \eqref{e:kerrfactored} of the metric 
by $\dis d\widetilde\phi - \frac a\Delta dr$ and $\dis dv-\frac{r^2+a^2}\Delta dr$ gives 
\begin{align}
  \cblue ds^2 &= - \frac\Delta{\rho^2}~
			[(dv-\frac{\rho^2}\Delta dr-a\sin^2\theta d\widetilde\phi]^2 
 	 	+ \frac{\sin^2 \theta }{\rho^2} [(r^2+a^2)d\wt\phi -adv]^2
	 	+ \frac{\rho ^2}{\Delta } dr^2 + \rho ^2 d\theta ^2 \nonumber\\
\cblue	&\cblue= - \frac{\Delta }{\rho^2}~(dv-a\sin^2\theta d\widetilde\phi)^2 + 2 dv\,dr 
	   + \frac{\sin^2 \theta }{\rho^2} [(r^2+a^2)d\wt\phi-adv]^2-2 a\sin^2\theta dr d\widetilde\phi 
	   +\rho^2 d\theta^2\cb,
\label{e:kmetric}\end{align}
smooth at the horizon where $\Delta = 0$.  With $a$ set to zero, the metric in Kerr coordinates 
is the Eddington-Finkelstein form \eqref{e:e-f-in}.  By construction, the surfaces of constant 
$v$ are null, and $v$ and $\widetilde\phi$ are constant along the ingoing null geodesics.  
In the $(v,r,\tilde\phi)$ chart, the ingoing null vector $n^\alpha$ is 
then simply 
\be
    \bm n = -\bm\pa_r.  
\ee

\noindent Direct check:\\
For the change of coordinates $(v,r,\tilde\phi) \rightarrow (t,r,\phi)$
\[
   \frac{\pa t}{\pa r} = -\frac{r^2+a^2}\Delta, \ \frac{\pa r}{\pa r} = 1, \ 
   \frac{\pa \phi}{\pa r} = -\frac a\Delta, 
\] 
implying 
\be
-\partial_r \mbox{(with $v,\widetilde\phi$ fixed)} 
	=  \frac{r^2+a^2}\Delta\partial_t -\partial_r\mbox{(with $t,\phi$ fixed)} 
	   +\frac a\Delta\partial_\phi\ .
\label{e:partialr}\ee 
\vspace{1cm}

For treatments of the analytically extended Kerr solution, analogous to but more complex 
than the full Schwarzschild solution in a Kruskal-Szekeres chart, see, for example, 
Hawking and Ellis\cite{he75}, Wald\cite{waldbook}, or Poisson \cite{poisson07}.  

\chapter{Linearized gravity and gravitational waves}
\index{gravitational radiation|see {gravitational waves}}\index{gravitational waves|(}
  Some recent treatments of gravitational waves:\\
  Jolien Creighton \& Warren Anderson, Gravitational-Wave Physics and Astronomy,\cite{creightonanderson}\\
Michele Maggiore, Gravitational Waves \cite{maggiore08}\\
  Thorne\&Blandford \cite{thorneblandford}\href{http://www.pmaweb.caltech.edu/Courses/ph136/yr2012/1227.1.K.pdf}{Chap. 27}\cite{thorneblandford}\\
 Schutz \cite{schutzbook}, Chap. 9\\

\section{Linearized field equations} 
\index{linearized field equations|(}\index{field equation!linearized|textbf}
\index{Einstein field equation!linearized|textbf}
\subsection{Preview} 

Before doing any calculation, we can quickly estimate the maximum amplitude 
of a gravitational wave from the inspiral of two black holes:  When  
black holes of mass $M$ are about to merge, the fractional change in 
length near the black holes is of order $1$: That is, at a distance of order $M$ 
from the black holes, a length of order $M$ changes by order $M$.  Because 
the amplitude of a gravitational wave is proportional to $1/d$, with $d$ 
the distance to the source, and that amplitude is the magnitude of the 
metric perturbation, the maximum fractional change in length at the observer 
is of order $M/d$.  Using $M_\odot =1.5$ km, $d = 100\text{ Mpc} = 3\times 10^{21}$ km, 
we have 
\be
   \frac Md \approx \frac12 \times 10^{-21}\ \frac M{M_\odot}\frac{100\mbox{ Mpc}}d .
\label{e:gw_est}\ee 
Black hole masses in detected inspirals (\href{https://media.ligo.northwestern.edu/gallery/mass-plot}{LIGO graveyard}) are larger than this and the distances to them have been greater than 100 Mpc.  
A detailed estimate of the amplitude of gravitational 
waves from GW170817, the closest observed NS-NS event as of 2022, at 30 Mpc, is \ref{ex:170817}.  

With the exception of spacetime in or near black holes and neutron stars, the geometry 
of the universe is nearly flat.  We can write the metric in the form 
\be
   g_{\alpha\beta} = \eta_{\alpha\beta} + \frac12b
\ee
where the components $h_{\mu\nu}$ are small compared to 1 in a chart for which $(\eta_{\mu\nu})$ 
is diag(
-1,1,1,1).  It is the magnitude of this metric perturbation at the observer 
that we just estimated in Eq.~\eqref{e:gw_est} and thus 
the magnitude of the corresponding change in distance between the test masses 
of a gravitational-wave detector (see Eq.~\eqref{e:gw}).

Terms quadratic in $h_{\alpha\beta}$ can be neglected compared to 
linear terms, and we similarly assume that terms quadratic in $\partial_\lambda h_{\mu\nu}$
are small compared to $\partial_\sigma\partial_\tau h_{\mu\nu}$. In the Einstein equation, we will then ignore terms of quadratic and higher orders in $h_{\alpha\beta}$ and its derivatives. The inverse metric to linear order in $\frac12b$ is 
\be
  g^{\alpha\beta} = \eta^{\alpha\beta} -h^{\alpha\beta}
\ee 
where indices are raised using $\eta^{\alpha\beta}$.  We will denote by $\partial_\alpha$ the 
flat covariant derivative operator, the covariant derivative associated with $\eta_{\alpha\beta}$, and we denote by $R^{(1)}_{\alpha\beta\gamma\delta}$ the 
Riemann tensor at linear order in $\frac12b$.         

The linearized Riemann tensor $R^{(1)}_{\alpha\beta\gamma\delta}$ has the form we found in 
Eq.~\eqref{e:RP1} by discarding terms quadratic in derivatives of $\frac12b$,
 
\be
 R^{(1)}_{\alpha\gamma\beta\delta} 
	    =  \frac{1}{2} (\pa_\beta \pa_\gamma h_{\alpha\delta} 
		+ \pa_\alpha \pa_\delta h_{\beta\gamma} - \pa_\alpha \pa_\beta h_{\gamma\delta} 
		- \pa_\gamma \pa_\delta h_{\alpha\beta}), 	
\label{e:delta_riemann}\ee
The corresponding linearized Ricci tensor is then
\be 
  R^{(1)}_{\alpha\beta} = R^{(1)}_{\alpha\gamma\beta}{}^\gamma 
	= \frac{1}{2} (-\pa_\gamma  \pa^\gamma  h_{\alpha\beta} 
					+ \pa_\beta \pa^\gamma h_{\alpha\gamma} 
					+ \pa_\alpha  \pa^\gamma h_{\beta\gamma} 
					- \pa_\alpha  \pa_\beta h),
\quad \mbox{where}\ \  h:= h_\alpha{}^\alpha,
\label{e:dRab}
\ee
and the resulting $G^{(1)}_{\a\b}$ looks complicated. It can be greatly simplified by 
a choice of \textit{gauge}. Preview: 
An infinitesimal diffeo (or an infinitesimal coordinate transformation) 
 gives a physically 
equivalent $h_{\alpha\beta}$ satisfying an analog of the Lorenz gauge for electromagnetism:
Outside the matter, we can choose $\partial^\gamma h_{\a\gamma} = 0$, $h=0$. 
Then $R^{(1)}=0$, and we immediately get the wave equation, 
\begin{align} 
  G^{(1)}_{\a\b} = R^{(1)}_{\alpha\beta} 
		&= - \frac12\, \raisebox{-1mm}{\text{\Large$\Box$}}\, h_{\alpha\beta}.
\label{e:Gh}\end{align}
Now a diffeo has only four components $\psi^\mu$, and when matter is present, we can 
only satisfy four of the five conditions $\partial^\nu h_{\mu\nu} = 0$, $h=0$.  Inside the 
matter, one instead imposes the 4-component condition  $\partial^\b \bar h_{\a\b} = 0$ with 
$\bar h_{\a\b} = \frac12b -\frac12 \eta_{\a\b} h$. Again one obtains a wave equation, this time for $\bar h_{\a\b}$:
\[
   \raisebox{-1mm}{\text{\Large$\Box$}} \bar h_{\alpha\beta} = -16\pi T_{\a\b}.
\]

Here are the details.

\subsection{Gauge transformations} 
\label{s:gauge}
\index{gauge transformation}\index{infinitesimal diffeomorphism}\index{infinitesimal coordinate transformation}\index{coordinates!infinitesimal coordinate transformation}

A less abstract treatment using coordinate transformations is given on p. \pageref{p:gtransf}. 

We describe the universe and its inhabitants by a set of tensors (and spinors, which we'll ignore).
The geometry and physics do not depend on the choice of coordinates we choose, and this statement 
is equivalent to saying that the geometry and physics are unchanged by a diffeomorphism $\psi$ 
that acts on all the tensors.  Let's start more familiarly with rotations of ordinary 3-space, with 
each point $P$ a vector from the origin: Rotating the space by a rotation $R$ and 
simultaneously rotating the coordinates leaves the components of the vector unchanged. 
That is, if $R:M\rightarrow M$ is a rotation and $x:M\rightarrow \mathbb R^3$ is a coordinate system, the coordinates $x(P)$ are the same as the coordinates $\bar x$ of the point $R(P)$ if $\bar x[R(P)] = x(P)$ or 
$\bar x=x\circ R^{-1}$.   

Here is a more formal way of saying the same thing using our notation for dragging a vector by a map, 
in this case by the coordinate map $x:M\rightarrow \mathbb R^3$. The components of a vector $v$ 
are just $x^*v= (v^1,v^2,v^3)$.   The rotated vector is $R^*v$ and its components in the chart 
$\bar x$ are $\bar x^*(R^*v) = (x\circ R^{-1})^*R^* v = (x\circ R^{-1}\circ R)^* v = x^*v$.   

Written in this way, the statement can be immediately generalized to any tensor field and any 
diffeo $\psi$: The coordinates $x(P)$ of a point $P$ are the same as the coordinates $\bar x$ of 
the point $\psi(P)$ if $\bar x = x\circ\psi^{-1}$.   Then the components of a tensor field $T$ 
in the chart $x:M\rightarrow \mathbb R^n$ are the same as the 
components of the dragged along tensor $\psi^* T$ in the dragged along chart $x\circ\psi^{-1}$: 
\be
 \bar x^*(\psi^*T) = (x\circ \psi^{-1})^*\psi^* T = (x\circ\psi^{-1}\circ \psi)^* T = x^*T
\ee

  What does this mean for a perturbed metric?  Write a small perturbation 
of a metric as $ g_{\a\b} + \lambda h_{\a\b} $ with $\lambda$ small.  That is, 
we look at a family of metrics
\be
  g_{\a\b}(\lambda) =  g_{\a\b} + \lambda h_{\a\b}, 
\ee
with 
\be
h_{\a\b} = \left.\frac d{d\lambda}g_{\alpha\beta}(\lambda)\right|_{\lambda=0}. 
\ee
When no other fields are present, the metric $g_{\alpha\beta}$ is physically 
equivalent to $\psi^* g_{\alpha\beta}$.  
A gauge transformation is defined by considering a smooth family of diffeos $\psi_\lambda$,
with $\psi_0$ the identity.  
Now to linear order in the vector field $\xi^a$ generating a family of diffeos, the change in a tensor $T$ is its Lie derivative: Let $\psi_\lambda$ be generated by $\xi^a$.  
From the geometric definition Eq.~\eqref{eq:liepsi}, of Lie derivative, we have 
\be
  \psi_\lambda^* T^{\cdots}_{\cdots} 
    = \psi_0 T^{\cdots}_{\cdots} + \lambda\frac d{d\lambda} \psi_\lambda^* T^{\cdots}_{\cdots}|_{\lambda=0}
		+O(\lambda^2) 
    = T^{\cdots}_{\cdots} -\lambda \Lie_{\bm\xi} T^{\cdots}_{\cdots} +O(\lambda^2) .  
\ee

Let's change notation from $\lambda \xi^a$ to $-\xi^a$; then $\xi^a$ itself 
is small, and we get a plus sign for the Lie derivative. 
The family of metrics $g_{\alpha\beta}(\lambda)$ 
is then physically equivalent to the family $\psi^*_\lambda g_{\alpha\beta}(\lambda)$, 
and the metric perturbation {\cblue $h_{\alpha\beta}$ is physically equivalent to the 
gauge-related metric perturbation }
\be
 \left.\frac d{d\lambda}\psi^*_\lambda g_{\alpha\beta}(\lambda)\right|_{\lambda=0} = \left.\frac d{d\lambda} g_{\alpha\beta}(\lambda)\right|_{\lambda=0}+\left.\frac d{d\lambda}\psi^*_\lambda g_{\alpha\beta}(0)\right|_{\lambda=0}
 = {\cblue h_{\alpha\beta} +\Lie_{\bm\xi}g_{\alpha\beta}}.
\ee

And if $(h_{\alpha\beta}$, $T^{(1)}_{\alpha\beta})$ is a solution to the linearized field equations
\[ 
	G^{(1)}_{\alpha\beta} = 8\pi T^{(1)}_{\alpha\beta}, 
\]
then a physically equivalent solution is 
\beq h_{\alpha\beta} + \Lie_{\bm\xi} g_{\alpha\beta} \quad  
 T^{(1)}_{\alpha\beta} + \Lie_{\bm\xi} T_{\alpha\beta},
\label{5.25}\eeq
where $\xi^\alpha$ is any vector field. In our case, the unperturbed metric is a flat vacuum 
spacetime. Because the background (i.e., unperturbed) $T^{\alpha\beta}$ is zero for us, 
$T^{(1)}_{\alpha\beta}$ is {\sl gauge invariant}.  Similarly, because the background 
Riemann tensor vanishes, $R^{(1)}_{\a\b\gamma\delta}$ is gauge invariant.  

This ends our formal discussion of gauge transformations, and we return to 
our earlier notation, replacing $\lambda h_{\a\b}$ by $h_{\a\b}$ and writing 
the perturbed metric as $g_{\a\b} + h_{\a\b}$.  \\

We can now choose a gauge to simplify the field equation.  It will be
analogous to the Lorenz gauge for electromagnetism, $\pa_\beta  A^\beta  = 0.$
Define 
\[ \color{blue}
	\overline h_{\alpha\beta} := h_{\alpha\beta} - \frac{1}{2} g_{\alpha\beta} h.
\]
{\sl Claim}:  One can always pick a gauge in which
\beq \color{blue}
	 \pa_\beta  \overline h^{\alpha\beta} = 0. 
\label{e:dedonder}\eeq
This is called the deDonder (or transverse or harmonic or Lorenz) gauge.
\index{gravitational waves!transverse gauge}\index{gauge!deDonder,harmonic,transverse}\index{transverse gauge}\index{harmonic gauge}\index{deDonder gauge}

\noindent {\sl Proof}:  Under a gauge transformation generated by $\xi^\a$,
\begin{align}
\overline h_{\alpha\beta} = h_{\alpha\beta} 
		- \frac{1}{2} g_{\alpha\beta} h 
\ \ \rightarrow \ \ & \overline h_{\alpha\beta} + \pa_\alpha \xi_\beta  
	+ \pa_\beta \xi_\alpha - \frac12 g_{\alpha\beta}2\pa_\gamma \xi^\gamma  \nonumber\\
	&= \overline h_{\alpha\beta} + \pa_\alpha \xi_\beta  + \pa_\beta \xi_\alpha  
	   - g_{\alpha\beta}\pa_\gamma \xi^\gamma  
\label{e:gauge3}\\ 
\pa_\beta \overline h_\alpha {}^\beta  
  \ \ \rightarrow\ \  & \pa_\beta \overline h_\alpha {}^\beta  
		+ \pa_\alpha \pa_\beta \xi^\beta  
		+ \pa_\beta \pa^\beta \xi - \pa_\alpha \pa_\gamma \xi^\gamma  
\nonumber\\
	&=\pa_\beta \overline h_\alpha {}^\beta  + \pa_\beta \pa^\beta \xi_\alpha
\end{align}

 Thus by choosing as $\xi_\alpha $ a solution to the wave equation with source
$-\pa_\beta  \overline h_\alpha {}^\beta $, 
\beq 
	\raisebox{-1mm}{\text{\Large$\Box$}}\, \xi_\alpha 
	  \equiv \pa_\beta \pa^\beta \xi_\alpha  
	     = -\pa_\beta \overline h_\alpha {}^\beta,
\label{e:boxxi}\eeq
one obtains a new gauge in which
\[ 
	\pa_\beta \overline h_\alpha {}^\beta  = 0. 
\]
In this deDonder gauge, $ G^{(1)}_{\alpha\beta}$ is simple: From Eq.~\eqref{e:dRab},
with $\pa^\gamma h_{\a\gamma} = \frac12\pa_\a h$, we have 
\bsube\begin{align}
R^{(1)}_{\alpha\beta} & =-\frac12 \pa_\gamma  \pa^\gamma h_{\alpha\beta} \\
R^{(1)} & = -\frac12 \pa_\gamma \pa^\gamma h, \\
G^{(1)}_{\a\b}&= -\frac{1}{2} \pa_\gamma  \pa^\gamma \bar h_{\alpha\beta} .
\label{e:G1}\end{align}\esube
As claimed, the linearized field equation is then
\beq \crv
	\raisebox{-1mm}{\text{\Large$\Box$}}\, \overline h_{\alpha \beta}= - 16\pi T_{\alpha\beta} \cb,
\label{e:boxgamma}\eeq
and the Cartesian components of $\bar h_{\a\b}$ are
\beq \crv
\overline h_{\mu\nu}(x)  = 4 \int \frac{T_{\mu\nu}|_{ret}}{|\bm r -\bm r'|}~dV' 
	= 4 \int \frac{T_{\mu\nu}(t-|\bm r -\bm r'|,\bm r')}{|\bm r -\bm r'|}~dV'\cb , 
\label{e:hret}\eeq

In observing gravitational waves, one sees only the $1/r$ term in the asymptotic expansion. 
We will call the size of the source ${\cal R}$, taking $T^{\mu\nu}(t,\bm r ')$ to be nonzero only when $\bm r ' < \cal R$. Then 
\be
  \overline h_{\mu\nu}(x)  = 4 \int \frac{T_{\mu\nu}|_{ret}}{|\bm r -\bm r'|}~dV' 
	= 4 \int \frac{T_{\mu\nu}(t-|\bm r -\bm r'|,\bm r')}{r}~dV' \ [1+O({\cal R}/r)].
\label{e:barh_asympt}\ee
This $1/r$ term also includes the time-independent $M/r$ part of the asymptotic metric, which 
always agrees with Schwarzschild at this order. 

\index{charge!dipole}
We now turn to a slow-motion approximation, in which successively higher multipoles (values of 
$\ell$) are smaller by factors of $\omega{\cal R}$.  A scalar field has all multipoles, 
starting with monopole radiation, $\ell=0$, corresponding to loss of scalar charge.  For an electromagnetic field, charge is conserved, and the radiation starts with a dipole ($\ell=1$) field.  For gravity, there is again no monopole radiation and now there is no dipole radiation
(this is exact, not just for slow motion). The first nonzero multipole is quadrupole, $\ell =2$.  
We will see that this follows from 
asymptotic conservation of momentum as well as mass at linear order in $h_{\alpha\beta}$.
In a Newtonian context, the center of charge of a rotating dipole has the acceleration 
of the circular motion of its opposite charges, with $\ddot {\bm d} = \frac{d^2}{dt^2} \int \rho_e \bm x\, dV\neq 0$; but in a binary system, the masses have the same sign, and the center of mass does not accelerate, $\frac{d^2}{dt^2} \int \rho \bm x\, dV = 0$.   
\index{linearized field equations|)}

\subsection{Slow-motion source}
\index{gravitational waves!slow-motion source|(}\index{slow-motion source}
  A source with an oscillation frequency or an orbital frequency $\omega$ has velocity of order 
$v = {\cal R}\omega$. When the source moves slowly, $v/c = {\cal R} \omega \ll 1$.
Then \vspace{-2mm}
\[ 
	{\cal R}\,\partial _t T_{\mu\nu} \ll T_{\mu\nu} .\vspace{-2mm}
\]
The multipole expansion of the radiation field involves an expansion of 
$\dis T_{\mu\nu}(t-|\bm r-\bm r'|, \bm r')$ in powers of ${\cal R}\omega$, i.e., in powers of the 
post-Newtonian small parameter $e \sim v/c$.  We start with
\be
  |\bm r-\bm r'| =  [r^2 - 2\bm r\cdot\bm r'+r'^2]^{1/2} = r - \hat{\bm r} \cdot \bm r' + O(\frac{\cal R}{r}).
\ee
Then \vspace{-3mm}
\begin{align} 
   T_{\mu\nu}(t-|\bm r-\bm r'|,\bm r') 
	&= T_{\mu\nu}(t-r,\bm r')+\hat{\bm r}\cdot\bm r'\partial_t T_{\mu\nu} \
		+ \frac12 (\hat{\bm r}\cdot\bm r')^2 \partial_t^2 T_{\mu\nu}-\cdots 
\label{e:multipole}\\
	&=  T_{\mu\nu}(t-r,\bm r') [1 +\ \  O({\cal R}\omega)\ \ + \ \ O(\ ({\cal R}\omega)^2\ )\quad
		-\cdots]. \nonumber
\end{align}

Now for a nearly Newtonian source, $T^{00} = \rho$, $T^{0i} = \rho v^i$, $T^{ij} = \rho v^i v^j$, 
so $T^{ij}$ is already smaller than $T^{00}$ by order $v^2$.  But the leading terms from 
$T^{00}$ and $T^{0i}$ are just the components of the 4-momentum, and 
the equation of motion
\[ \partial_0T^{0\mu} = - \partial_jT^{j\mu},
\]
implies that they are conserved: 
\beq
   \partial_0 \int T^{0\mu}(t-r,\bm r') dV' = -\int \frac{\partial}{\partial x'^j} T^{j\mu} dV' = 0. 
\label{e:paT}\eeq 
The next-order (dipole) term in $T^{00}$ from Eq.~\eqref{e:multipole} is again time independent, because the center of mass $\dis \int \rho \bm r' dV'$ is fixed:
\[
   \int\left[\hat{\bm r}\cdot\bm r'\partial_t T^{00}\right] dV' 
	= \hat{\bm r} \cdot \partial_t \int  \rho \bm r' dV'.
\]  
Ordinarily, one picks a frame in which the center of mass is at the origin.  It stays there 
because of conservation of momentum.  \index{conservation laws!momentum}

With the monopole and dipole parts of $T^{00}$ time-independent, we are left for its 
radiative part with the quadrupole contribution 
\be
   \int\left[\frac12 (\hat{\bm r}\cdot\bm r')^2\partial_t^2 T^{00}\right] dV' 
	= \frac12 \hat x^i \hat x^j \ddot I_{ij}(t-r), \mbox{ where } 
	\cblue I_{ij}(t):=\int  T^{00} x_i x_j dV\cb.
\label{e:Iij}\ee
Now the leading term $T^{ij}$ is already quadrupole and as we now show, its value is related 
to the radiative part of $T^{00}$. (We'll soon see that the trace $I_i^i$ does not contribute 
to the radiation field, and the quadrupole contribution is the tracefree part of $I_{ij}$, 
namely $I_{ij} -\frac13\delta_{ij}I^k_k$.) 

Again using the equation of motion $\partial _\nu T^{\mu\nu}=0$, we have
\begin{align*} 
   \pa_0 \int (T^{i0}x^j + T^{j0}x^i)dV &=
		-\int(x^j\partial_kT^{ik} + x^i\partial_kT^{jk})dV 
	    = 2 \int T^{ij}dV   \quad\mbox{after integration by parts}, \nonumber\\
  \pa_0 \int T^{00}x^ix^j dV 
	    &= - \int x^ix^j\pa_k T^{0k} dV =  \int(x^iT^{0j} + x^jT^{0i})dV\  
			\quad \mbox{after integration by parts},
\end{align*}
whence   
\be
 \int T^{ij}(t-r,\bm r')dV' = \frac{1}{2}\partial^2_0 \int T^{00}(t-r,\bm r')x'^i x'^jdV' 
	    =: \frac{1}{2} \ddot I^{ij}(t-r).
\label{e:ddotI}\ee
Then at leading order in $1/r$, 
\be
   \bar h_{ij} = \frac4r\left(\frac12 \ddot I_{ij}\right) 
		= \frac{2}{r} \ddot I_{ij}(t-r). 
\label{e:hslow}\ee

\noindent Eliminating a cheat:  \\
This derivation is fine for a spinning rod, where we can ignore the Newtonian gravitational field, but writing $\pa_\b T^{\a\b}=0$ for a Newtonian star 
means ignoring gravity!  That is, $\nabla_\mu T^{i\mu}=0$ includes the term $\rho\pa^i\phi$.  We'll see that the solution we just found for $\bar h_{\a\b}$ is still valid for a binary system.  But in the derivation we should include Newtonian gravity in an effective stress tensor $t^{\a\b}$. In Newtonian physics, the gravitational stress tensor is 
\be
  t^{ab} = \frac1{4\pi}\left(\na^a\Phi \na^b\Phi -\frac12g^{ab}\na_c\Phi \na^c\Phi\right) . 
\label{e:tab_newt}\ee
\index{stress tensor!gravitational, Newtonian} 
with $\na^2\Phi = 4\pi \rho$, and the Newtonian stress tensor including gravity 
is $T^{ab} + t^{ab} = \rho v^a v^b + P g^{ab} + t^{ab}$.
There is no Newtonian correction to the time components of $T^{\a\b}$: At Newtonian order,
$t^{t\mu} = 0$.\\
Then $\pa_\nu (T^{\mu\nu} + t^{\mu\nu} ) =0$ reproduces the Newtonian equations:    
\begin{align*}
T^{tt} &= \rho\quad T^{ti} = \rho v^i \\ 
\pa_\mu T^{t\mu} &= \pa_t\rho + \pa_j(\rho v^j) = 0\\
\pa_\mu (T^\mu{}_i +t^\mu{}_i)& = \pa_t(\rho v_i) + \pa_j(\rho v_i v^j) +\pa_i P +\rho\pa_i\Phi =0\\
\Longrightarrow 
\rho(\pa_t +v^j\pa_j) v_i &= -\pa_i P - \rho\pa_i\Phi.
\end{align*}

\noindent
{\sl Claim}: For a self-gravitating system in the Newtonian approximation, 
one can just replace $T^{\a\b}$ by $T^{\a\b}+t^{\a\b}$ as the source for 
$h_{\a\b}$, writing   
\[
   \raisebox{-.5mm}{\text{\Large$\Box$}}\, \bar h_{\alpha \beta} 
	= - 16\pi (T_{\alpha\beta} +t_{\a\b})\cb.
\] 

Here's why. Write the exact metric as 
$g_{\a\b} = \eta_{\a\b} + h_{\a\b}$, and divide $h_{\a\b}$ into 
its Newtonian part and the rest, which we'll call $h_{\a\b}^{\rm\, rad}$ 
for the radiative part -- the non-Coulomb part:  
 \begin{align*}
g_{\a\b} &= \eta_{\a\b} + h_{\a\b}\\
  &= \eta_{\a\b}+ h_{\a\b}^{\rm rad} + h^{\rm Newt}_{\a\b}, \\
\na^2 \bar h^{\rm Newt}_{\a\b}  &= -16\pi \rho\ \na_\a t\na_\b t 
\\
\bar h^{\rm Newt}_{\a\b} &= -4\Phi\ \na_\a t\na_\b t ,
\end{align*}
where we require $\pa^\b \bar h_{\a\b} = 0, \quad 
\pa^\b \bar h_{\a\b}^{\rm Newt}= 0$. Note that $\pa_t\Phi/|\bm\na\Phi| = O(e)$. 

 Then \vspace{-4mm}
\bsube\begin{align}
G_{\a\b} = &-\frac12\Box \left(\bar h^{\rm rad}_{\a\b} 
	+ \bar h^{\rm Newt}_{\a\b}\right) + G^{\rm nonlinear}_{\a\b} 
\label{e:Gnonlinear}\\
\Box \left(\bar h^{\rm rad}_{\a\b}+ \bar h^{\rm Newt}_{\a\b}\right) 
        =& -2 (8\pi T_{\a\b} - G^{\rm nonlinear}_{\a\b} ) 
        =: -16\pi(T_{\a\b}+t_{\a\b}),
\end{align} \label{e:Gnonlinear1}\esube
where
\begin{align*}
  t_{\a\b}: =& -\frac1{8\pi} G^{\rm nonlinear}_{\a\b}, \\
  G^{\rm nonlinear}_{\a\b} =& 
	  {\cblue\mbox{ terms in $G_{\a\b}$ quadratic in } \Phi }  
	 + \mbox{terms in $G^{\a\b}$ involving } \Phi h..   
            \mbox{ and } h..h..\\
        & +  \mbox{higher order terms} 
\end{align*}
At Newtonian order, $\bar h^{\rm rad}_{\a\b} = 0$, and \\
(terms quadratic in $\Phi) = 8\pi$(Newtonian gravitational stress tensor) 
\[
\mbox{{\cblue terms quadratic in $\Phi$} in }\cblue G_{ab}
 = -8\pi\ \frac1{4\pi} \left(\na_a\Phi \na_b\Phi 
 	-\frac12g_{ab}\na_c\Phi \na^c\Phi \right).
\]

Because $\pa_\b(T^{\a\b} + t^{\a\b})=0$ and $t^{\a\b}$ vanishes at spatial infinity, 
the steps leading to Eq.\eqref{e:ddotI} hold with $T^{\a\b}$ replaced by 
$T^{\a\b} + t^{\a\b}$.  
The relation \eqref{e:hslow}, $\bar h_{ij} = \frac2r \ddot I_{ij}(t-r)$, is then correct as written, for a nearly Newtonian self-gravitating system. \\

Because the derivative $\pa_\a$ appearing in 
$G^{\rm nonlinear}_{\a\b}$ is the covariant derivative operator of the the flat 
background metric, $G^{\rm nonlinear}_{\a\b}$ and $t_{\a\b}$ are tensors.  
Without this definition of $\pa_\a$, one simply writes the components 
$t^{\mu\nu}$ as expressions involving partial derivatives of $h_{\mu\nu}$; 
then as in MTW, the set of components $t_{\mu\nu}$ is called the ``stress-energy pseudotensor for the gravitational field.'' (p. 996, Sect. 36.9).  
With $\pa_\lambda h_{\mu\nu}$ the partial derivative, the components do not transform as tensor.  With our convention, $\pa_\g h_{\a\b}$ is a tensor, 
but its definition depends on the choice of a flat background metric.    \\ 
index{gravitational waves!slow-motion source|)}

We now show that a choice of gauge reduces the asymptotic form of the 
radiation field $h_{\mu\nu}$ to two free functions.\\

\subsection{Transverse tracefree gauge}
\label{s:TTgauge}
\index{gravitational waves!transverse tracefree gauge}
\index{gauge!transverse tracefree}\index{transverse tracefree gauge}
Requiring that the gauge vector be transverse -- that it satisfy Eq.~\eqref{e:boxxi}, does not completely specify it:  We can 
add to $\xi^\a$ any vector $\zeta^\a$ satisfying $\raisebox{-1mm}{\text{\Large$\Box$}}\zeta^\a=0$.
Outside the source, this allows us to further restrict the gauge to make $h_{\a\b}$ tracefree and to require 
$h_{t\mu}=0$.    

We follow a simple standard derivation using plane waves, waves of the form 
$h_{\alpha\beta} = A_{\alpha\beta}e^{ik_\mu x^\mu}$.  Because any solution to $\Box \Phi = 0$ 
is a superposition of planes waves, it follows that  
any vacuum $h_{\alpha\beta}$ can be written in the $TT$ gauge.  Wald avoids using 
plane waves in favor of an initial value argument; for readers following Wald, his argument,  
with details filled in, is given below.\pageref{p:waldTT} \\ 

In the transverse gauge \eqref{e:dedonder},  $h_{\mu\nu}$ is given by the retarded solution to the wave equation.  To obtain that gauge, we required the gauge vector $\xi^\alpha$
to satisfy the wave equation \eqref{e:boxxi}, and the new gauge will satisfy that and the additional conditions  Then, in the new gauge, 
$\overline h^{NEW}_{\mu\nu} = h^{NEW}_{\mu\nu}$.  The new  $h_{\mu\nu}$ is 
is denoted by $h^{TT}_{\mu\nu}$, ``TT'' abbreviating ``transverse tracefree.'' For a wave of 
the form $h_{\mu\nu}\propto e^{ik_\mu x^\mu}$, with $k_\mu$ along the null direction $\hat t^\mu + \hat z^\mu$, the transverse condition is 
\[
   0 = k^\mu h_{\mu\nu} \Longrightarrow 0= h_{t\mu} +h_{z\mu} = h_{z\mu}.
\]
The nonzero components of $h^{TT}_{\mu\nu}$ are then
\beq
  h_+:= h^{TT}_{xx} = -h^{TT}_{yy},\qquad  h_\times:=h^{TT}_{xy} \ .
\label{e:htt1}\eeq 
The gauge conditions have thus reduced the wave to the two degrees of freedom, 
$h_+$ and  $h_\times$. \\ 

In spherical coordinates $t,r,\theta,\phi$, the nonzero components of $h^{TT}_{\mu\nu}$ are 
\beq
  h^{TT}_{\hat\theta\hat\theta} = -h^{TT}_{\hat\phi\hat\phi},\qquad  h^{TT}_{\hat\theta\hat\phi} \ , 
\eeq 
with $k^\a$ along $\widehat r^\a$.  The wave looks to a distant observer at 
$r_0,\theta_0,\phi_0$ like a plane wave of the form \eqref{e:htt1} 
with $\hat x, \hat y$ and $\hat z$ taken along 
$\hat\theta,\hat\phi$, and $\hat r$ at the observer.\\ 
 
\noindent{\sl Constructing the gauge}:\\

Let $\zeta^\a$ be the gauge vector from our transverse $\overline h_{\a\b}$ 
to $h^{NEW}_{\a\b} = h^{TT}_{\a\b}$: 
\beq
  h^{TT}_{\mu\nu} = h_{\mu\nu} + \pa_\mu \zeta_\nu  + \pa_\nu \zeta_\mu.  
\label{e:htt}\eeq
We are to find $\zeta^\alpha = B^\alpha e^{ik_\mu x^\mu}$ to make 
\[ 
  h^{TT} = 0, \qquad h^{TT}_{t\mu} = 0.
\]
Impose 4 conditions to get the 4 components $\zeta^\mu$:
\begin{align*}
(1)\quad 0 = h^{TT} &=  h +2 \pa_\mu \zeta^\mu \Rightarrow\ 
	ik_\mu B^\mu = -\frac12 A^\mu_\mu \\ 
 i\omega(B_t+B_z) & = -\frac12 A^\mu_\mu\\
(2-4)\quad 0 = h^{TT}_{ti}&=  h_{ti}+ \pa_t \zeta_i + \pa_i \zeta_t 
  \Rightarrow \\
 i\omega B_x &  = A_{tx}, \\
 i\omega B_y & = A_{ty} \\
  i\omega (B_z - B_t) & = A_{tz}
\end{align*}\cb
Then $ k^\nu h^{TT}_{\mu\nu} = 0$ implies $ h^{TT}_{tt}=0 , \ \ h^{TT}_{z \mu} =  0$. 

In the TT gauge, the only nonvanishing components of the perturbed metric 
are $h^{TT}_{xx}, h^{TT}_{xy}$, and 
$h^{TT}_{yy}$.  That is, $h^{TT}_{t\mu}=0$ and $\partial^\nu h_{\mu\nu}=0$ imply 
$0 = -\partial_t h^{TT}_{t\mu} + \partial_z h^{TT}_{z\mu} = \partial_z h^{TT}_{z\mu} \Longrightarrow h_{z\mu}=0$. \\ {\cblue These nonvanishing components constitute the tracefree part of} $\dis\begin{pmatrix}
				\bar h_{xx} & \bar h_{xy}\\
				\bar h_{xy} &\bar h_{yy}
			\end{pmatrix} $ {\cblue in our retarded solution}:   

To see this, use the gauge vector to write the two independent components 
$h_+$ and $h_-$ of \eqref{e:htt1} of $h^{TT}_{\mu\nu}$ in terms of 
the retarded solution $\bar h_{\mu\nu}$ to the wave equation \eqref{e:hret} 
as follows. \\ Because $h^{TT}_{\mu\nu} =h_{\mu\nu}  + \pa_\mu\zeta_\nu + \pa_\nu\zeta_\mu$ and 
$\zeta_\mu = \zeta_\mu(t-z)$, we have 
\be
	h^{TT}_{xx} = h_{xx} = \bar h_{xx} - \frac12 h.   
\ee
Then 
\bsube\begin{align}
   h_+ &= \frac12(h^{TT}_{xx} - h^{TT}_{yy}) = \frac12(h_{xx} - h_{yy}) 
   	=\frac12(\overline h_{xx} - \overline h_{yy}), \\
   h_\times &= h^{TT}_{xy} = \overline h_{xy}.
\end{align}\esube
Finally, returning the names of components along $x$ and $y$ back to 
$\hat\theta$ and $\hat\phi$,  
\beq
h^{TT}_{\hat\theta\hat\theta} = -h^{TT}_{\hat\phi\hat\phi} =  h_+ ,\qquad 
	h^{TT}_{\hat\theta\hat\phi} =  h_\times. 
\label{e:htt+x}\eeq

In particular, in the slow-motion approximation, the gravitational waves described 
by $h^{TT}_{\mu\nu}$ have as their source the {\sl quadrupole tensor}, the 
tracefree part of the tensor $I_{ij}$,  
\index{quadrupole tensor}
\be
	\Ibar_{ij} = I_{ij} - \frac{1}{3} \delta_{ij}I^k{}_k~. 
\ee
From Eq.\eqref{e:hslow}, at leading order in $1/r$, 
\be\cblue
  h_+ = \frac1r (\ddot \Ibar_{\hat\theta\hat\theta} - \ddot \Ibar_{\hat\phi\hat\phi}), \quad 
  h_\times = \frac2r \ddot \Ibar_{\hat\theta\hat\phi}.
\label{e:h+hx}\ee

Far from a periodic source, the wave looks like a plane wave.  That is, 
our wave has the asymptotic behavior 
$h_{\mu\nu} = \frac1{r_0} A_{\mu\nu}(t-r_0 - z) +O(r^{-2}) = h_{\mu\nu}(t-z) + O(r^{-2})$, 
and $\partial_r h_{\mu\nu} = \partial_z h_{\mu\nu} +O(r^{-2})$.  Treating the wave as a plane wave $h_{\mu\nu}(t-z)$ is then equivalent to ignoring terms of order $r^{-2}$, terms 
smaller than $h_{\mu\nu}$ itself by a factor of order ${\cal R}/r$ .  This is a spectacularly good approximation:  
For a source of size $10^3$ km at a distance of 30 Mpc $\sim 10^{15}$ km, the error is one part in $10^{12}$. 

The waveform of an inspiraling binary,\index{binary system!inspiral}\index{inspiral}\index{gravitational waves!binary inspiral} however, has a time-dependent frequency; in late inspiral and merger, it is far from periodic.  To avoid introducing plane waves 
and making the argument than any vacuum solution can be written 
as a superposition,\footnote{
The superposition argument has a subtlety because the spacetime is not everywhere vacuum.  Outside any finite neighborhood of the source, one can duplicate the actual solution for a finite time by a sourcefree solution that includes incoming waves from past null infinity. To obtain the sourcefree solution, one uses initial data between slices of future null infinity and evolves the data backwards in time.  
The incoming and outgoing waves overlap only in the neighborhood of what was the source.  The simplicity of the plane-wave derivation justifies ignoring the subtlety.}  
Wald avoids plane waves in favor of an initial-value argument.  If you are not 
following Wald, you can go ahead to the next section, \ref{s:quadrupole}.     \\

\subsubsection{\sl Wald's derivation}
\label{p:waldTT}

Given the values of a scalar $\Phi$ and its time derivative $\partial_t\Phi$ on an initial surface, 
say at $t=0$, the wave equation $\raisebox{-1mm}{\text{\Large$\Box$}}\Phi=0$ uniquely determines 
$\Phi(t,x)$ everywhere.  Our job is to find a gauge vector for which $h^{TT}_{\mu\nu} = \partial_t h^{TT}_{\mu\nu} = 0$ and $h^{TT}_{0\mu} = \partial_t h^{TT}_{0\mu}=0$, at $t=0$.  The wave equation,
 $\raisebox{-1mm}{\text{\Large$\Box$}}h^{TT}_{\mu\nu}=0$, 
will then force each component to be zero at all times.  

To obtain a transverse gauge, we required the gauge vector $\xi^\alpha$  
to satisfy the wave equation \eqref{e:boxxi}. This does not quite exhaust our 
gauge freedom, because we can choose initial conditions -- the values 
of $\xi^\alpha$ and $\partial_t\xi^\alpha$ on an initial spacelike surface--
at, say, $t=0$.  This allows us to make $h_{\alpha\beta}$ tracefree, $h=h_\alpha{}^\alpha=0$,
and to choose $h_{0\mu}=0$ at $t=0$.  Inside the source, the evolution of $h$ 
is given by the wave equation 
with source $-16\pi T_\alpha{}^\alpha$, so $h$ does not remain tracefree, nor does 
$h_{0\mu}$ remain zero.   
Outside the source, however, because $h_{\alpha\beta}$ satisfies the homogeneous wave equation,
\[
   \raisebox{-1mm}{\text{\Large$\Box$}} \,h_{\alpha\beta} = 0, 
\]
they remain zero.

Because the old  $\overline h_{\a\b}$ is already transverse, $\zeta^\a$ keeps 
it transverse when $\zeta^\mu$ satisfies $\raisebox{-1mm}{\text{\Large$\Box$}}\,\zeta^\a = 0$. 
The tracefree 
condition means $\overline h_{\a\b}^{TT} = h^{TT}_{\a\b}$, so the gauge will have 
$\pa^\b h^{TT}_{\a\b}=0$.  Our gauge conditions are then
\bsube\begin{align}
  0 & = \raisebox{-1mm}{\text{\Large$\Box$}}\,\zeta^\mu \hspace{4.6cm}\mbox{transverse},\\
  0 &= h^{TT\,\mu}_\mu =  h +2 \pa_\mu \zeta^\mu, 
\qquad\quad\ \ \ \mbox{tracefree}
\label{e:cb}\\
  0 &= h^{TT}_{t\mu} =  h_{t\mu}+ \pa_t \zeta_\mu + \pa_\mu \zeta_t  
			\quad\mbox{no $t$ component}.
\label{e:cc}\end{align}\esube
Any gauge vector of the form $\zeta^\mu = \zeta^\mu(t-z)$ satisfies the first condition, 
$\raisebox{-1mm}{\text{\Large$\Box$}}\,\zeta^\mu = (-\pa_t^2 + \pa_z^2)\zeta^\mu = 0$. 

The tracefree condition gives 
 \[
  0 = h +2(\pa_t \zeta^t+\pa_z \zeta^z) \Longrightarrow \dot \zeta^t - \dot \zeta^z = -\frac12 h.  
\]

The spatial components of the last condition determine  $\partial_t\zeta^i$ and $\dot\zeta^t$ 
\begin{align}
   -h_{tz} & = \partial_t\zeta_z +\partial_z\zeta_t = \dot \zeta^z + \dot\zeta^t 
	\Longrightarrow \nonumber\\
\dot\zeta^t &= -\frac12 h_{tz} -\frac14 h, \quad \dot\zeta^z = -\frac12 h_{tz} + \frac14 h.
\label{e:zeta1}\end{align}
The $tx$ and $ty$ components of the no-$t$-component condition are immediate: 
\beq 
   \dot\zeta_x = -h_{tx}, \qquad \dot\zeta_y = - h_{ty}. 
\label{e:zeta2}\eeq 

The $tt$ component gives no additional 
condition on $\dot\zeta^t$, because $\partial^\mu \overline h_{t\mu}=0$ implies \\
$\partial_t (h_{tt}+\frac12 h +h_{tz})=0 = -\partial_z(h_{tt}+\frac12 h +h_{tz})$, whence
\[  
 h_{tt}+\frac12 h +h_{tz} =0.
\] 
The $h^{TT}_{tt} =0$ condition is then automatically satisfied:
\[
  \frac12 h_{tt} - 2\dot\zeta^t = h_{tt} +\frac12 h + h_{tz} = 0.     
\] 

The gauge conditions \eqref{e:zeta1} and \eqref{e:zeta2} are now satisfied, and
we have completely determined $\dot\zeta^\mu|_{t=0}$. We now need to use our freedom to choose 
$\zeta^\mu$ itself at $t=0$ to enforce $\dot h^{TT} =0$ and $\dot h^{TT}_{0\mu}=0$.
To write these conditions, replace $\zeta^\mu$ 
by $\dot\zeta^\mu$ and $h_{\mu\nu}$ by $\dot h_{\mu\nu}$ in Eqs.~\eqref{e:zeta1} and \eqref{e:zeta2}:
\be
         \ddot\zeta^t = -\frac12 \dot h_{tz} -\frac14 \dot h, \quad \ddot\zeta^z = -\frac12 \dot h_{tz} + \frac14 \dot h,\qquad \ddot\zeta_x = -\dot h_{tx}, \qquad \ddot\zeta_y = - \dot h_{ty}
\label{e:hdot}\ee
But these are conditions on $\ddot\zeta$ at $t=0$, not on $\zeta$ itself.  
What are we to do?  We use the wave equation for $\zeta^\mu$, and simply replace 
$\ddot\zeta^\mu|_{t=0}$ by $\nabla^2\zeta^\mu$, writing
\be
         \nabla^2\zeta^t = -\frac12 \dot h_{tz} -\frac14 \dot h, \quad \nabla^2\zeta^z = -\frac12 \dot h_{tz} + \frac14 \dot h,\qquad \nabla^2\zeta_x = -\dot h_{tx}, \qquad \nabla^2\zeta_y = - \dot h_{ty}.
\ee
Each Poisson equation has a unique solution (vanishing at spatial infinity) for $\zeta^\mu$, 
and with this initial data, the wave equation, 
$\ddot\zeta^\mu = \nabla^2\zeta^\mu$, for $\zeta^\mu$  enforces the initial conditions \eqref{e:hdot} for $\dot h^{TT}_{\mu\nu}$.\\

\section{The quadrupole formula and gravitational-wave energy}
\label{s:quadrupole}
\index{gravitational waves!quadrupole formula|(}

\noindent{\sl Effective stress-energy tensor for gravitational waves}\\
The Coulomb and radiative parts of $h_{\a\b}$ play opposite roles 
near and far from the source.  \\
\index{stress-energy tensor!gravitational waves}\index{effective stress-energy tensor, gravitational waves}\index{gravitational waves!effective stress-energy tensor|(}

\noindent 
{\sl Near zone}:  We have already written, in Eq.~\eqref{e:tab_newt}, the Newtonian limit of the effective gravitational stress-energy tensor $t^{\a\b}$. Close to 
a nearly Newtonian source, $h_{\a\b}^{\rm rad}$ is negligible 
compared to $\Phi$, because the Coulomb part $\Phi$ of the field is of 
is of order $M/r$, while the radiative part of the field is of order 
\[
  h_{..} \sim \frac{\ddot I_{..}}r \sim \frac {MR^2\omega^2}r\sim \Phi v^2.  
\]  

\noindent
{\sl Far zone}:  The situation is reversed at large $r$.  Because $\pa_i\Phi\pa_j\Phi$ falls off as $r^{-4}$, the contribution of the Coulomb field to $t^{\a\b}$ 
is negligible at large $r$.  With $\pa_t h_{ij}^{\rm rad}$ and $\pa_r h_{ij}^{\rm rad}$ of order $r^{-1}$, $h_{\a\b}^{\rm rad}$ dominates the effective stress-energy tensor $t^{\a\b}$. The resulting $\frac1{r^2}$ behavior of the energy 
flux gives a finite power radiated at $\infty$. \\

  We now calculate this power radiated by a gravitational wave.  Like the stress-energy tensor of the electromagnetic field, which is quadratic in first derivatives of the vector potential $A_\a$, the effective stress-energy of gravitational waves is quadratic in first derivatives of $h_{\a\b}$.  The asymptotic radiated flux of the electromagnetic field in a Lorenz gauge is proportional to $\dot{\bm A}^2$, involving only components orthogonal to the direction  of propagation.  Similarly, in our analogous 
TT gauge for  $h_{\a\b}$, the radiated energy is quadratic in $\dot h_{ij}$, 
involving only components orthogonal to the direction of propagation.

Given a stress-energy $T^{\a\b}+t^{\a\b}$ tensor satisfying 
$\partial_\b (T^{\a\b}+t^{\a\b}) = 0$ on a flat background, we have the 
conservation law   
\beq
   \frac{dE_{GW}}{dt} = \partial_t\int_V (T^{tt}+t^{tt}) d^3x 
		= - \int_V \partial_i (T^{ti}+t^{ti}) d^3x
		= - \int_{r=\rm constant} t^{tr} r^2 d\Omega,
\eeq 
\index{conservation laws!energy of gravitational waves}
where $V$ is bounded by an $r=$ constant sphere outside the source (so 
$T^{\a\b}$ vanishes on the sphere).  

Now $t^{\a\b}$ is defined in Eq.~\eqref{e:Gnonlinear} as the nonlinear 
part of $G^{\a\b}$.  We could directly compute this to quadratic order in 
$h_{\a\b}$; but it is easier and perhaps more instructive to use the fact 
that the field equation $G^{\a\b} = 8\pi T^{\a\b}$ can be obtained by varying 
an action $I=I_G+I_M$, where $I_G$ is the gravitational action and $I_M$ 
the action for the matter fields.  In fact, as we will see in 
Sect.\ref{s:action}, the gravitational action is given by 
\[
I_G = (16\pi)^{-1} \int R \sqrt{-g} d^4x,  
\]
or, in a form involving only first derivatives of the metric, by Eq.~\eqref{e:lag1b}.  
But everything follows just from the existence of an action, from the fact that varying an action with respect to the metric 
$g_{\alpha\beta}$ gives  
\beq
   \delta  I = \delta  I_G + \delta  I_M 
	= C \int \left(\frac1{8\pi}G^{\a\b} 
		- T^{\a\b}\right)\delta g_{\alpha\beta} \sqrt{-g}d^4x. 
\label{e:dI}\eeq

Here $I_G$ depends only on $g_{\a\b}$, while $I_M$ is a function 
of both the matter fields and the metric, and Eq.~\eqref{e:dI} implies 
$\dis    T^{\a\b} = -\frac1C\frac{\delta  I_M}{\delta  g_{\a\b}}$. 
The form of $t^{\a\b}$ will follow from this equation; 
all that is missing is the factor $C=-1/2$, and this follows from the explicit form of any matter action.  Let's use a free particle (the sign of the action is determined by  
the sign of the Newtonian kinetic energy, $mv^2/2)$)  
\begin{align}
  I_M &= -\int d\lambda\ m\  
	\sqrt{-g_{\mu\nu}\dot x^\mu\dot x^\nu}, 
\nonumber\\
\frac{\delta  I_M}{\delta  g_{\mu\nu}} & = \frac12 m\ u^\mu u^\nu 
		= \frac12 T^{\mu\nu} \Longrightarrow C=-\frac12.
\end{align} 
With $C$ known, we have for any matter action,   
\beq
  \crv T^{\a\b} = 2\frac{\delta I_M}{\delta g_{\a\b}}\cb. 
\eeq 
\index{stress-energy tensor!from variation of an action}   

 Then from Eq.\ref{e:dI}, 
\beq
  \crv \delta I_G = -\int\frac1{16\pi}G^{\a\b} \delta g_{\a\b}\sqrt{-g}d^4x\cb.
\eeq
Now $G^{\a\b}$ vanishes for the background spacetime.  We have seen that, in
a TT gauge,  it has the form \eqref{e:Gh}, 
$G^{(1)}_{\a\b} = - \frac12\, \pa_\g  \pa^\g \,  h_{\alpha\beta},$
at first order in $h_{\a\b}$. 
We can then write the action for the perturbed 
field equations as
\beq
I^{(2)}_G 
    = \frac1{64\pi} \int \pa_\g  h_{\a\b}\pa^\g h^{\a\b}\sqrt{-g}d^4x.
\label{e:I2a}\eeq
noting that 
\[
\delta I^{(2)}_G = 2 \frac1{64\pi} \int \pa_\g  \delta h_{\a\b}\pa^\g h^{\a\b}\sqrt{-g}d^4x 
       = -\frac1{16\pi}\int \delta h_{\a\b}\left[-\frac12 \pa_\g  \pa^\g h^{\a\b}\right]\sqrt{-g}d^4x = -\frac1{16\pi}\int G^{(1)\a\b}\delta h_{\a\b}\sqrt{-g}d^4x.
\]

We need to find the part of $G^{\rm nonlinear}_{\a\b}= t_{\a\b}$ that is quadratic
in $h_{\a\b}$, and this is easy: Write the action \eqref{e:I2a} as 
\beq
I^{(2)}_G(h,g) = \frac1{64\pi} 
	    \int \pa_\c \bar h_{\a\b}\, \pa_\d h_{\ep\zeta}\ 
		g^{\c\d}g^{\a\ep}g^{\b\zeta} \sqrt{-g}d^4x,
\label{e:I2}\eeq
with $g_{\a\b} = \eta_{\a\b} + h_{\a\b}$.  If one ignores the dependence of 
$g^{\a\b}$ on $h_{\a\b}$, the variation of $\pa_\c h_{\a\b} \pa^\g h^{\a\b}$
immediately yields the vacuum field equation $\Box h_{\a\b} = 0$ 
for the first-order perturbation in a $TT$ gauge.  Including the variation of
the factor $g^{\c\d}g^{\a\ep}g^{\b\zeta} \sqrt{-g}$ gives the correction to
the field equation quadratic in $h_{\a\b}$.%
\footnote{It is helpful to think of the action  $I^{(2)}_G$ as a function of
independent variables $g_{\a\b}$ and $h_{\a\b}$, as it would be if $h_{\a\b}$ were a matter 
field.  
Because $\dis\frac{\delta g_{\a\b}}{\delta h_{\gamma\delta}} = \delta^\gamma_{(\alpha}\delta^\delta_{\beta)}$, the total variation of a function $I(h) = I(h,g(h))$ with respect to $h$ is the sum
$\dis \frac{\delta I(h,g)}{\d h_{\a\b}} +\dis \frac{\delta I(h,g)}{\d g_{\a\b}} 
	= -\frac12 \Box\bar h^{\a\b} - 8\pi t^{\a\b}$, 
giving the first- and second-order equations, $\ \dis \Box h^{(1)}_{\a\b} = 0, \ \ 
\Box \bar h^{(2)\ \a\b} = -16\pi t^{\a\b}$.\\

\noindent
Note that the exact action is given by Eq. \eqref{e:lag1b}. It is only in the far
zone that the action has the form \eqref{e:I2}, and only if the perturbed metric
$h_{\a\b} = g_{\a\b}-\eta_{\a\b}$ is transverse up to second order in the sense 
$\eta^{\b\g}\pa_\g h_{\a\b} = 0$.  
}  

We need this quadratic correction, the effective stress-energy tensor 
$t^{\a\b} = 2\delta I^{(2)}/\delta g_{\alpha\beta}$, only at large $r$, 
where it is simplest to evaluate.
We once again use the relation  
$\partial_t h^{TT}_{\mu\nu} = - \partial_r h^{TT}_{\mu\nu} +O(1/r^2)$ 
for the components in a Cartesian chart:
Dropping the $TT$ and using dots to represent arbitrary indices, we have  
\beq
  \partial_\gamma h_{..} \partial^\gamma h_{..} 
    =  -\partial_t  h_{..} \partial_t h_{..} + 
 	\partial_r h_{..} \partial_r h_{..}
	= O(1/r^3).
\eeq    
Now unless $\dis\frac\delta{\delta g_{\alpha\beta}}$ hits $g^{\kappa\lambda}$ 
on the right side of \eqref{e:I2}, 
the resulting term in the stress-energy tensor will be proportional to 
$\partial_\gamma h_{. .} \partial^\gamma h_{. .}$ and so will not 
contribute to the radiated energy.
The only surviving term is then 
\beq
  {\cblue t^{\alpha\beta} }= 2\frac{\delta I^{(2)}}{\delta g_{\b\c}}
      {\cblue = \frac1{32\pi} \pa^\alpha h^{TT}_{\c\d} \pa^\beta h^{TT\ \c\d}}.  
\eeq 

\index{flux!gravitational wave}\index{gravitational waves!flux}
\index{gravitational waves!effective stress-energy tensor|)}\index{gravitational waves!effective stress-energy tensor|textbf}
The radiated energy flux is $t^{tr}$ and the asymptotic form of the radiated 
energy is (using Eqs.~{e:htt+x} and \eqref{e:h+hx})
\bsube
\begin{align}
 \frac{dE}{dtd\Omega} & = r^2 t^{tr}
       = \frac{r^2}{32\pi}
	   \left[(\dot h^{TT}_{\hat\theta\hat\theta})^2 
	  + (\dot h^{TT}_{\hat\phi\hat\phi})^2  
	   + 2(\dot h^{TT}_{\hat\theta\hat\phi})^2 \right]  + O\left(\frac{1}{r}\right)  
\nonumber\\
	&=\frac{r^2}{16\pi}(\dot h_+^2 + \dot h_\times^2) 
\label{e:dotEdoth}\\ 
   & = \frac{1}{4\pi}\left[\left(\frac{\dddot I_{22}- \dddot I_{33}}{2}\right)^2 
		         + (\dddot I_{23})^2 \right], 
\label{5.51}\end{align}\esube
where, for ease of notation in the next part, we write the indices as $\hat\theta = 2$ 
and $\hat\phi=3$.   We can write either $I_{ij}$ or $\Ibar_{ij}$ here 
because $I_{22}-I_{33}$ and $I_{23}$ involve only the tracefree part of $I_{ij}$.

Then \eqref{5.51} implies 
\[ 
\frac{dE}{dtd\Omega} 
    = \frac{1}{4\pi} \left[\left(\frac{\dddot \Ibar_{22} - \dddot \Ibar_{33}}{2}\right)^2 
			   + (\dddot \Ibar_{23})^2 \right] 
\]
Denote by $P_{ij}$ the projection operator orthogonal to the direction of
$\hat{\bm r}$: 
$\ P_{ij} = \delta_{ij} - \hat{r}_i \hat{r}_j$. Then
\begin{eqnarray}
	4\pi \frac{dE}{dt d\Omega} &=& \frac{1}{2}
(\dddot \Ibar_{22} \dddot \Ibar_{22} +  2\dddot \Ibar_{23}
\dddot \Ibar_{23} + \dddot \Ibar_{33} \dddot \Ibar_{33}) 
	- \frac{1}{4} (\dddot \Ibar_{22} + \dddot \Ibar_{33})^2 \nonumber \\
&=&  \frac{1}{2} P^{i\ell}P^{jm} \dddot \Ibar_{ij}\stackrel{...}
\Ibar_{\ell m} - \frac{1}{4}(P^{ij} \dddot \Ibar_{ij})^2 \nonumber \\
&=& \frac{1}{2}[\dddot \Ibar_{ij} \dddot \Ibar^{ij} - 2\hat{r}^j
{\dddot \Ibar}_{ij} \dddot \Ibar^i {}_k \hat{r}^k +
\frac{1}{2}(\hat{r}^i\stackrel{...}\Ibar_{ij} \hat{r}^j)^2] 
\label{5.52}\end{eqnarray}
To find the total power radiated we need to evaluate
\[ 
 \int \hat{r}_i\hat{r}_jd\Omega \hspace{.5in}\mbox{and} \hspace{.5in} 
 \int \hat{r}_i\hat{r}_j\hat{r}_k\hat{r}_\ell d\Omega.
\]
Because the integral over angles sums over all directions of $\bm r$,  
$\dis\langle\hat{r}_i \hat{r}_j\rangle := \frac{1}{4\pi} \int \hat{r}_i \hat{r}_j
d\Omega$ is a symmetric tensor that knows no direction. So it must be
constructed from $\delta_{ij}$:
\[ \langle\hat{r}_i\hat{r}_j\rangle  = k\delta_{ij} \]
Take its trace:
\[
 \delta^{ij}\langle\hat{r}_i\hat{r}_j\rangle = \langle1\rangle = 1 = 3k \Longrightarrow 
	k = \frac{1}{3} ,
\]
\beq 
\langle\hat{r}_i\hat{r}_j\rangle = \frac{1}{3} \delta_{ij} .
\label{5.53}\eeq
Similarly 
\[ \langle\hat{r}_i\hat{r}_j \hat{r}_k\hat{r}_\ell\rangle =~k' ~\delta_{(ij}\delta_{k
\ell)} = k''(\delta_{ij}\delta_{k\ell} + \delta_{ik}\delta_{j\ell} +
\delta_{i\ell}\delta_{jk}) \]
\[ \delta^{ij}\delta^{k\ell}\langle\hat{r}_i\hat{r}_j \hat{r}_k\hat{r}_\ell\rangle = 1
= k'' (9+3+3) \]
\[ k'' = \frac{1}{15} \] 
\beq 
\langle\hat{r}_i\hat{r}_j \hat{r}_k\hat{r}_\ell\rangle 
   = \frac{1}{15}(\delta_{ij}\delta_{k\ell} + \delta_{ik}\delta_{j\ell} 
	+ \delta_{i\ell}\delta_{jk}) . 
\label{5.54}\eeq
Finally Eqs.~\eqref{5.52}-\eqref{5.54} $\Longrightarrow $ 
\[
\frac{dE}{dt} 
= \frac12 \left[\dddot \Ibar_{ij} \dddot \Ibar^{ij} 
		- 2(\frac{1}{3}\delta^{k\ell})\dddot \Ibar_{ik}\dddot \Ibar^i_\ell 
		+ \frac{1}{2} (\frac{1}{15})(\delta^{ij}\delta^{k\ell} 
		+ \delta^{ik}\delta^{j\ell} + \delta^{i\ell}\delta^{jk})\dddot \Ibar_{ij} \dddot \Ibar_{k\ell}
	   \right], 
\]
giving
\beq
 \crv \frac{dE}{dt} = \frac15 \dddot \Ibar_{ij} \dddot \Ibar^{ij}\cb. 
\label{e:dedt}\eeq
Restore $G$ and $c$ to the right side to recover the conventional dimension $[E/T] = ML^2T^{-3}$:  
Because $\dddot I_{ij}\dddot I^{ij}$ has conventional dimension $M^2 L^4T^{-6}$, we multiply 
by the combination of $c$ and $G$ that has dimension $M^{-1} L^{-2} T^3$, namely $[G/c^5]$: 
\beq
\cblue \frac{dE}{dt} = \frac G{c^5} \frac15 \dddot \Ibar_{ij} \dddot \Ibar^{ij}\cb. 
\label{dedt2}\eeq
\index{quadrupole formula|textbf}\index{gravitational waves!quadrupole formula|)}
\index{gravitational waves!quadrupole formula|textbf}

\section{Binary Inspiral}\index{binary system!inspiral}\index{inspiral}\index{gravitational waves!binary inspiral|(}
\label{s:inspiral}
Consider now a binary system with masses $m_A$ and $m_B$, in circular orbits about their center of mass, 
and with separation $a$ between the masses.  The distance from each mass to the center of mass is \\
$r_A = a\,m_B/M,\ r_B = a\,m_A/M$.  For an orbit in the $x$-$y$ plane, 
we have 
\begin{align*}
  x_A &= r_A \cos\Omega t, \quad x_B = -r_B \cos\Omega t,\\
  y_A &= r_A \sin\Omega t, \quad y_B = -r_B \sin\Omega t, 
\end{align*}
\bsube\begin{align}
  I_{xx} &= m_A x_A^2 + m_B x_B^2 
   = \frac12 \mu a^2 \cos 2\Omega t + \mbox{constant} ,\\
I_{yy} &=-\frac12 \mu a^2 \cos 2\Omega t+ \mbox{constant},\\
I_{xy} &= \frac12\mu a^2 \sin 2\Omega t, 
\end{align}\label{e:Ixy}\esube
implying that the wave has frequency $\omega = 2\Omega$.
\index{gravitational wave!frequency}\index{frequency!gravitational wave}

From Eq.~\eqref{e:h+hx}, the gauge-invariant parts of the perturbed asymptotic metric 
are given by
\be
 h_+ = -\frac1r \mu a^2 \omega^2 \cos\omega t, \qquad h_\times = -\frac1r\mu a^2\omega^2\sin\omega t.
\label{e:hbinary}\ee 

We can now use Eq.~\eqref{e:dedt} to find the radiated power.  In this simple case, 
we could also find the power directly from Eq.~\eqref{e:dotEdoth} for $\dot E$ in terms of $\dot h_+$ and $\dot h_\times$ -- the angle-average machinery leading to \eqref{e:dedt} 
is not needed.  

  Because $I_{xx}+I_{yy}=$ constant, we have $I_{ij} = \Ibar_{ij}+$ constant, and 
\footnote{For a general periodic source, one would take a time average over one cycle.  
In our case, however, the expression $\dddot \Ibar_{ij} \dddot\Ibar^{ij} $ is already time 
independent.} 
\beq 
\dddot \Ibar_{ij} \dddot\Ibar^{ij} 
= \dddot I_{xx}^2+\dddot I_{yy}^2+2\dddot I_{xy}^2  = 32(\mu a^2 \Omega^3)^2.  
\eeq 

The energy of the binary is
\beq
   E = - T = -\frac12\mu a^2\Omega^2, \qquad \mbox{or}\qquad E = \frac12 U_G = - \frac{\mu M}{2a}. 
\label{e:Ebinary}\eeq
From Eq.~\eqref{e:dedt} 
\beq
  \crv \frac{dE}{dt} = -\frac{32}5 \mu^2 a^4\Omega^6\cb.
\label{e:Edot1}\eeq

Because the frequency of the wave is directly measurable, while the 
binary separation is not, it is helpful to use Kepler's law, 
$\Omega^2 = M/a^3$ to replace $a$ by $M^{1/3} \Omega^{-2/3}$:
\beq
  \cblue \frac{dE}{dt} = -\frac{32}5 ({\cal M}\Omega)^{10/3},  \quad 
		{\cal M} \equiv \mbox{chirp mass} := \mu^{3/5}M^{2/5}\cb,
\label{e:Edot2}\eeq 
or, with $G$ and $c$ restored,
\[
  \frac{dE}{dt} = -\frac{32}5 \frac{G^{7/3}}{c^5} ({\cal M}\Omega)^{10/3}.
\]

To find an equation for the change in the orbital radius as the 
binary loses energy to gravitational waves, use the second 
version of $E$ in \eqref{e:Ebinary} to write
\[
  \frac{\dot E}E = -\frac{\dot a}a.   
\]
Then replace $\Omega^2$ by $M/a^3$ in $dE/dt$ of \eqref{e:Edot1}:
\be\cblue
\frac{dE}{dt} = -\frac{32}5 \frac{\mu^2 M^3}{a^5},
\ee  
giving
\begin{align}
  \frac{\dot a}a &= - \frac{64}5 \frac{\mu M^2}{a^4}  
		\Longrightarrow a^3\dot a =- \frac{64}5 \mu M^2 
\nonumber\\
	a^4  &= a_0^4-\frac{256}5 \mu M^2 t, 
\end{align}
with $a=a_0$ at $t=0$.  Alternatively, with $t=0$ at coalescence, 
\beq
   a\propto |t|^{1/4},\quad |t|\ \mbox{the time until coalescence}, 
\eeq 
and, from Kepler's law, $\Omega\propto |t|^{-3/8}$.\\

The first discovery of gravitational waves was indirect:\cite{wt81}\cite{tw82}
An initial measurement by Joel Weisberg and Joseph Taylor of the orbital decay of PSR B1913+16 (the Hulse-Taylor pulsar) followed by decades of further observation, agrees with the predicted rate of energy loss to 
gravitational waves by a binary system, with an accuracy of 0.3\%.    
For a brief summary see \href{https://aspbooks.org/custom/publications/paper/328-0025.html}{Weisberg-Taylor 2005}\cite{wt05}
\index{Hulse-Taylor pulsar} \index{pulsar!Hulse-Taylor}
\begin{figure}[H]
               \begin{center}
		\includegraphics[width=.5\textwidth]{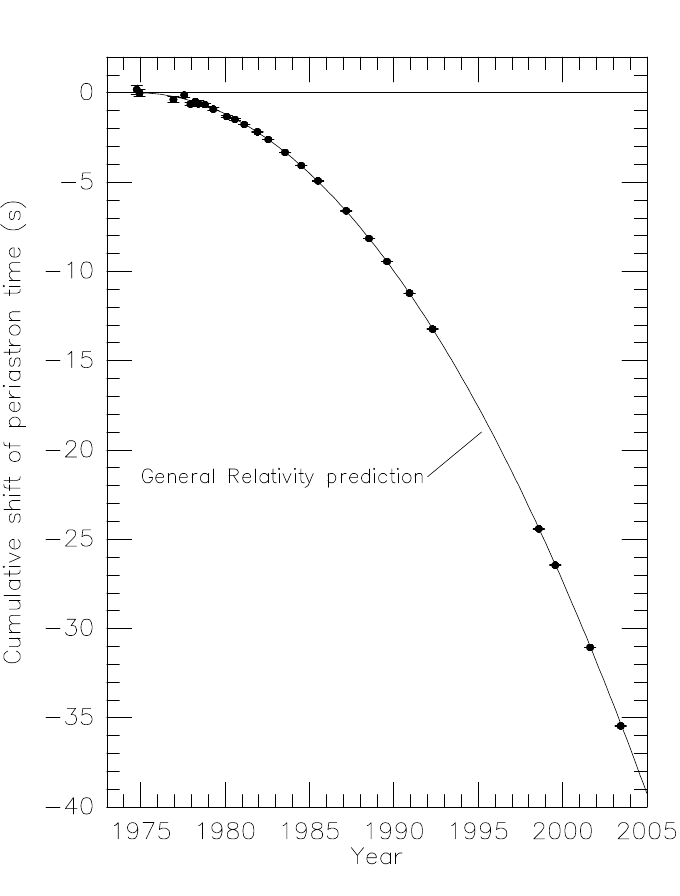}
		\end{center}
	\caption{Orbital decay of PSR B1913+16.  The data points have 
width generally smaller than the error bars. The parabola shows the 
orbital decay predicted by general relativity. From \cite{wt05}}
\label{f:orbital_decay}\end{figure}
\newpage

\benr \item \benalph\item Find the number of gravitons emitted per unit time by a spinning rod of length $\ell$ and negligible mass joining two point 
masses, each of mass $m$. Assume each graviton has energy $\hbar\omega$, 
with $\omega$ the frequency of classical gravitational waves.
\index{gravitons}

\item  Estimate this rate of emission for $m= 1\ \rm kg$, $\ell = 1\ \rm m$, 
and rotational frequency 500 rad/s. What is the probability that the system 
will radiate at all if it spins for a time $t=1/\omega$?  Weirdly, you 
should find Einstein's spinning rod rarely radiates gravitational 
waves!  
\een

\item Estimate the number of gravitons emitted per second by two equal-mass neutron stars, 
each of mass $m= 1.4 M_\odot$, when the  distance between their centers of mass is 20 $m$.

\item 
\benalph
\item In a principal-axis basis where $I_{ij}$ is diagonal, relate the eigenvalues of $\Ibar_{ij}$ to those of $I_{ij}$
\item  Using the relations 
\[ 
  r^2 Y_{2\pm 2} = \sqrt{\frac{15}{32\pi}} (x\pm iy)^2, \quad 
  r^2 Y_{21} = \mp\sqrt{\frac{15}{8\pi}} z(x+iy), \quad 
  r^2 Y_{20} = \sqrt{\frac5{16\pi}} (3z^2 - r^2), 
\]
relate the components of $\Ibar_{ij}$ along the complex basis 
$\{\hat e_{\pm} = \frac1{\sqrt2} (\hat x\pm i\hat y),\ \hat z\}$ to the quantities  
\[
  \Ibar_{2m} = \int \rho Y_{\ell m} dV.  
\]
\item Write $h_+$ and $h_\times$ in terms of $Y_{2m}$, with the $z$-axis along 
the direction from source to observer.    

\een     
\een 
\index{gravitational waves!binary inspiral|)}
\newpage

\section{Strain and detection of gravitational waves}
\index{gravitational waves!detection|(}\index{detection of gravitational waves}
\index{gravitational waves!strain}
Because of the difficulty in defining energy in a curved spacetime, parts of 
the physics community were uncertain whether gravitational waves carried energy 
from a system.  Pirani and days later (but apparently independently) 
Feynman gave equivalent arguments at the same 1957 Chapel Hill 
conference to resolve the issue.  Pirani had just published a paper 
using the equation of geodesic deviation \eqref{e:geodev},            
\[
\ddot S^\alpha  = R^\alpha {}_{\beta\gamma\delta}T^\beta T^\gamma S^\delta,
\]
to describe the relative acceleration of free particles as a gravitational wave passes.  
\footnote{
The question also led to a study of asymptotic flatness, with an isolated system modeled as an asymptotically 
flat spacetime.  The total mass of the spacetime can then be defined, and that mass decreases by the 
energy radiated to null infinity.  That is, along a sequence of asymptotically null slices 
(like the $u=$ constant slices of Schwarzschild), the mass between successive slices decreases by 
the energy radiated to null infinity between the two slices. 
On asymptotically {\sl spacelike} slices of the spacetime, the mass of the 
spacetime has a fixed, constant value.  The radiation is simply at successively larger distances from the source along a sequence of spacelike slices. See references on p.~\pageref{p:aflat} } 

Here is Pirani's argument, with the idea of energy absorption put in by Bondi.  
Feynman's is at the end of the same volume:\\

\noindent Felix Pirani, {\sl Measurement of Classical Gravitational Fields}, 
in \href{http://www.edition-open-sources.org/media/sources/5/Sources5.pdf}{\sl Report from the 1957 Chapel Hill Conference}\cite{chapel_hill57}, Chap. 14.\\

\hspace{1cm}\begin{minipage}[h]{6.2in}\leftskip 0.2in 

\index{principle of equivalence}
PIRANI: Because of the principle of equivalence, one cannot ascribe a direct physical interpretation to the gravitational field insofar as it is characterized by Christoffel symbols $\Gamma^\mu_{\nu\rho}$. One 
can, however, give an invariant interpretation to the variation of the gravitational field.  These 
variations are described by the Riemann tensor; therefore, measurements of the relative acceleration 
of neighboring free particles, which yield information about the variation of the field, will also yield 
information about the Riemann tensor.  

Now the relative motion of free particles is given by the equation of geodesic deviation 
\[
  \frac{\pa^2 \eta^\mu}{\pa\tau^2} + R^\mu_{\nu\rho\sigma} v^\nu\eta^\rho v^\sigma = 0 \ldots
\] 

BONDI: Can one construct in this way an absorber for gravitational energy by
       inserting a $d \eta/ d\tau$ term to learn what part of the Riemann tensor would
       be the energy producing one, because it is that part that we want to isolate
       to study gravitational waves?\\

PIRANI: I have not put in an absorption term, but I have put in a ``spring.''
             You can invent a system with such a term quite easily.

\end{minipage}
\vspace{5mm}

We will look at such a system, a damped oscillator driven by gravitational waves, in Eq.~\eqref{e:nongeodev1} below.  
\newpage
First, let's use the equation of geodesic deviation to get 
the standard picture of a circle of free particles stretched and 
compressed in orthogonal directions to become an oscillating ellipse.  We look at 
the Riemann tensor of a plane wave with wave vector $k^\a$ orthogonal to the plane of the 
circle of particles.  

Here's the Thorne-Blandford figure: 
\begin{figure}[H]
               \begin{center}
		\includegraphics[width=\textwidth]{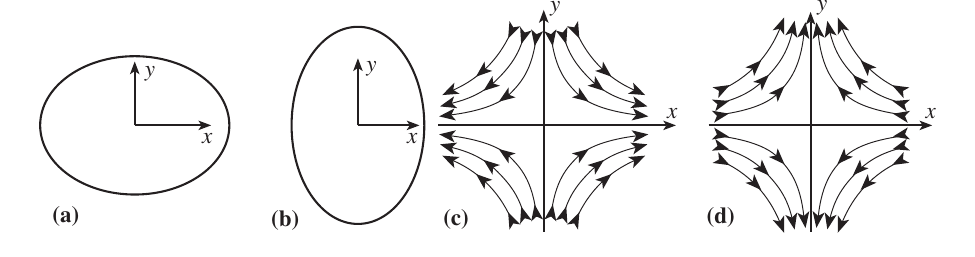}
		\end{center}
	\caption{Physical manifestations, in a particle's local Lorentz frame, of gravitational 
waves with polarization in the $x$-$y$ plane.  (a) Transverse deformation of an initially spherical cloud of test particles in a transverse plane at a phase of the wave when $h_+>0$.  (b) Deformation of the cloud when $h_+<0$. (c)-(d) Field lines representing the acceleration field 
$\delta\ddot{\bm x}$ when $\ddot h_+>0$ and $\ddot h_+<0$, respectively.}
\label{f:wave}\end{figure}

A plane wave with wavevector $k^\a = \omega(\widehat t^\a +\widehat z^a)$ and $+$ polarization in the 
$x$-$y$ plane has the form  
\be
   h_{\a\b} = (\widehat x_\a \widehat x_\b -\widehat y_\a \widehat y_\b) A\cos[\omega(t-z)] 
=  (\widehat x_\a \widehat x_\b - \widehat y_\a \widehat y_\b) h_+ ,
\ee
following the notation of Eq.\eqref{e:htt1}.
The linearized Riemann tensor \eqref{e:delta_riemann} is then given by 
\be
R^{(1)}_{\a\c\b\d} =\frac12(-k_\b k_\c h_{\a\d} - k_\a k_\d h_{\b\c} 
				+k_\a k_\b h_{\c\d} + k_\c k_\d h_{\a\b}).
\ee
In the geodesic deviation equation, we take $T^\a=\widehat t^\a$, the velocity of a particle prior to the 
arrival of the wave, and consider particles lying in the $z=0$ plane.  First consider 
two particles with separation vector $S^\a$ of the form   
\[
  S^\a = \ell\, \widehat x^\a + \xi^\a,
\]
where $\xi^\a = 0$ before the wave arrives.  The geodesic deviation 
equation then has the form 
\begin{align}
  \ddot \xi^\a &= R^\a{}_{\b\g\d} t^\b t^\g \widehat x^\d \ell + O(A\,\xi)
		=  \widehat x^\a R_{xttx} \ell + O(A\,\xi) 
		= -\widehat x^\a k_t^2 h_+ \ell  + O(A\,\xi)\nn\\
	      & =  -\widehat x^\a \frac12 \omega^2 \ell A \cos\omega t  + O(A\,\xi). 
\label{e:gw2}\end{align}
At linear order in the amplitude of the wave, the solution is 
\be
   \xi^\a = \frac12\ell h_+ \widehat x^\a 
   	  = \frac12 A\ell\cos\omega t \widehat x^\a, 
\label{e:gw}\ee
where we have chosen $t=0$ as a time of maximum displacement. 
For particles with displacement $\ell \widehat y^\a$ instead of 
$\ell \widehat x^\a$ prior to arrival of the wave, it is easy to see 
that the corresponding solution is 
\mbox{$\xi^\a = -\frac12\ell h_+ \widehat y^\a$}.  Then for a displacement 
of particles initially along a line at arbitrary angle $\phi$ from the 
$x$-axis, we have 
\be 
   \xi^\a(\phi) =  (\cos\phi\widehat x^\a -\sin\phi \widehat y^\a)\frac12\ell h_+, 
\label{e:gw1}\ee   
and 
\[
  S^\a(\phi) = (\cos\phi\widehat x^\a +\sin\phi \widehat y^\a)\ell  + \xi^\a(\phi).
\]  
Particles initially in a circle in the $z=0$ plane then comprise an 
oscillating ellipse, with particle separation $S^\a(\phi)$, as shown in 
Fig.\ref{f:wave}(a)-(b). Their relative acceleration $\ddot\xi^\a$ agrees 
with the quadrupole pattern of Fig.\ref{f:wave}(c)-(d).  

We can now return to Pirani's argument, with two masses joined by a 
spring of natural length $\ell$, frequency $\omega_0$, and damping coefficient 
$\gamma$. In flat space, with no gravitational waves, the relative 
acceleration of the particles with connecting vector 
$S^\a = \ell \widehat x^\a+ \xi^\a$ is given by 
\be
   \ddot\xi^\a = -\omega_0^2\xi^\a - 2\gamma \dot\xi^\a.  
\label{e:damped_oscillator}\ee
Intuitively, then, in the equation of geodesic deviation, one should add 
to the relative acceleration of the particles the forcing terms,  giving 
(at linear order in the gravitational wave amplitude) 
\be 
   \ddot\xi^\a = R^\a{}_{\b\g\d} t^\b t^\g x^\d \ell 
		-\omega_0^2\xi^\a - 2\gamma \dot\xi^\a.
\label{e:nongeodev1}\ee

  The geodesic deviation equation looks at the change in a separation vector $S^\a
$ that is dragged along by a family of timelike geodesics, and it is not immediately obvious that the derivation leads to Eq.\eqref{e:nongeodev1}.  But is not hard to 
see.  In the formal 
deviation before Eq.~\eqref{e:geodev}, only the last line uses the geodesic 
equation $T^\b\na_\b T^\a = 0$, and only the fact that $S^\a$ is tangent to a family of 
timelike curves with tangent $T^\a$ is used to reach the next-to-last line. This equation 
(with dummy indices renamed) is 
\begin{align}
\ddot S^\alpha  &= S^\g\nabla_\g \underbrace{(T^\beta\nabla_\b T^\alpha)}_{\mbox{now not $0$}}  
	+R^\alpha {}_{\beta\gamma\delta}T^\beta T^\gamma S^\delta ,\nonumber\\
\label{e:geodevp}\end{align}
with $t^\a$ the 4-velocity of the unperturbed mass - along a timelike Killing vector 
of the flat background spacetime.  
Take the spring to be a thin rod whose particles move along a family of 
timelike trajectories parameterized by proper time, this time with an 
external force acting.  Each particle's velocity $T^\a$ then satisfies 
$T^\b\na_\b T^\a = F^\a/m \equiv a^\a$.  Using this in Eq.~\eqref{e:geodevp} gives the relative acceleration as a function of particle position, 
\be
   \ddot S^\a = R^\alpha {}_{\beta\gamma\delta}T^\beta T^\gamma S^\delta +  S^\b\na_\b a^\a.
\label{e:nongeodev}\ee 
For the Pirani-Bondi-Feynman example of a damped 
spring consider two small masses joined by a massless rod, initially along the $x$-axis, again taking the 
rod to have natural length $\ell$, frequency $\omega_0^2$, and damping coefficient $\gamma$. The masses  
at the right and left ends of the rod have acceleration 
$T^\b\na_\b T^\a = \mp \frac12(\omega_0^2 \xi^\a +2\gamma\dot\xi^\a)$. 
Write $S^\a = S_+^\a - S_-^\a$, with $S_\pm^\a$ the separation vector of 
each mass from the center of mass.  Choose a coordinate $s$ for which 
$|s|$ is proper distance from the center of the rod.  The acceleration of a 
particle at $s$ then proportional to $s$: 
\[
	a^\a = -\frac {s}{\ell}(\omega_0^2 \xi^\a +2\gamma \dot \xi^\a),   
\]
The separation $|S_\pm^\a|$ of each mass from the center of mass is $\ell/2$ at zeroth 
order in $\xi^\a$, implying \\
$\dis S_\pm^\b\na_\b a^\a = \frac\ell2 \pa_s a^\a =\mp \frac12(\omega_0^2 \xi^\a +2\gamma \dot \xi^\a$). Then Eq.~\eqref{e:nongeodev} implies \eqref{e:nongeodev1} at linear order in $\xi^\a$.  
  
\benr
\item \benalph \item 
Check that the separation vector $S^\a$, with $\xi^\a$ given by the solution \eqref{e:gw1}, 
gives the stretching and contracting ellipse of Fig.\ref{f:wave}.  
\item Check that the acceleration vector conforms to parts (c)-(d) of the Figure.  
\item Find the form of $\xi^\a$ for the polarization $h_\times$ and sketch the corresponding 
figures.  
\een

\item {\sl Energy transferred by a gravitational wave to an oscillator.}  
\benalph\item For two particles connected by 
a spring, Eq.~\eqref{e:gw2} is changed by the additional terms of Eq.~\eqref{e:nongeodev1}.  Find the 
corresponding steady-state solution to this equation for a driven oscillator.  
\item Find the rate at which the oscillator absorbs energy from the gravitational wave. What is 
the time-average of this rate?  \\

Hint:  The rate at which the oscillator absorbs energy is the rate at which its energy would decrease if there were no gravitational wave.  
\een
\een

\noindent{\sl Interferometric detectors}\ (See, for example, the Creighton-Anderson text \cite{creightonanderson} on which this section is based.
)\index{interferometric detector}\\

If, when you think of optics, the first thing that comes to mind is the angular resolution $\lambda/d$, it seems impossible that an interferometer using light with wavelength of roughly $10^{-4}$ cm can detect a gravitational wave that changes the distance between mirrors by less than $10^{-15}$ cm.  Obtaining angular resolution 
better than $\lambda/d$ means measuring a difference in intensity corresponding to 
a fraction of the distance between dark fringes.  In an interferometer, one is again 
trying to measure the difference in intensity corresponding to a fraction of a fringe, 
but it turns out to be possible to find that difference to very high accuracy.  
How small a difference in intensity can be measured by a photodiode?  We begin with 
the usual diagram of a simple interferometer and then look at a limit on the accuracy with 
which it can measure a change in length.  
\begin{figure}[H]
               \begin{center}
		\includegraphics[width=0.6\textwidth]{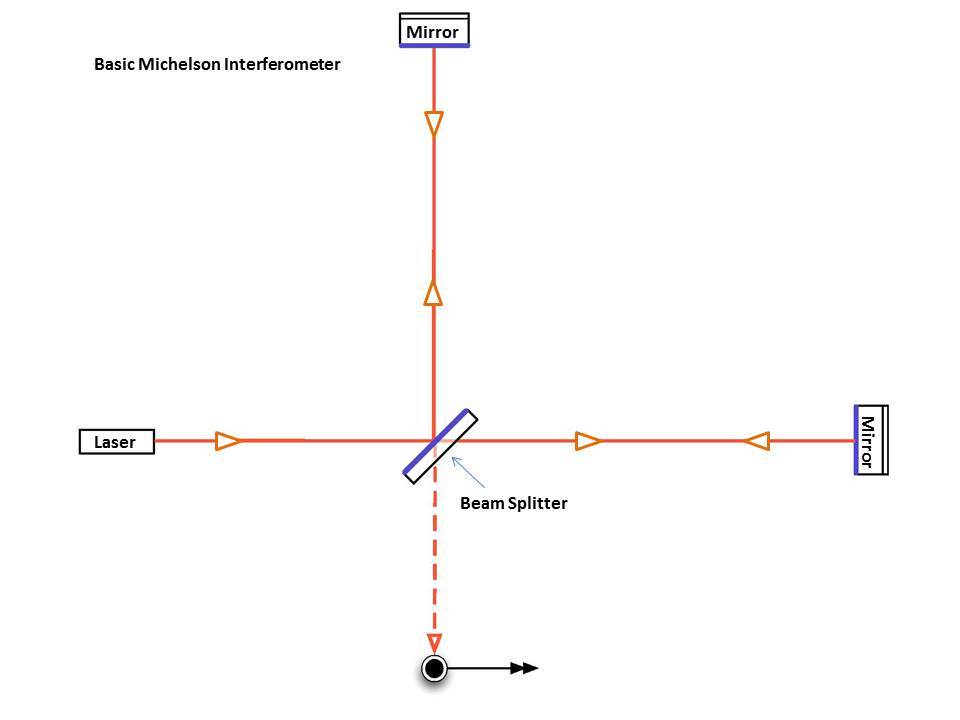}
		\end{center}
		\caption{Diagram from a LIGO \href{https://www.ligo.caltech.edu/page/what-is-interferometer}{webpage}}
\index{LIGO detector}
\label{f:interferometer}\end{figure}

Photons from the laser can traverse two alternative paths to the detector.  
Along the first path, light reflects upward from the beam splitter, 
then reflects off the top mirror, and finally passes through the beam splitter 
to the detector.  Along the second, light passes through the beam splitter, 
reflects off the mirror on the right, and then reflects off the beam splitter 
to the detector.  The number of photons reaching the detector depends on the 
phase difference between the two paths; by changing the arm lengths 
(the distances between mirrors and beam splitter), a gravitational wave changes 
the number of detected photons.  

Detector accuracy is limited by noise.  At frequencies above about 200 Hz 
(in the present 4 km detectors), the 
limit is set by {\sl shot noise}, random fluctuations in the flux of photons.
Roughly stated, a gravitational wave is detectable if the change $\delta N_{GW}$ 
in the number of photons reaching the detector is larger than the random 
fluctuation $\delta N_{\rm random}$.  
Let $N$ be the average number of photons that reach the detector in time $T$ 
in the absence of a gravitational wave. Random fluctuations in photon number 
change that by%
\footnote{The random-walk relation $\delta N_{\rm random}=\sqrt{N}$ holds for 
macroscopic incandescent and florescent light sources because a large number of 
independent contributing degrees of freedom yield a Poisson distribution of photons.  For a laser, it holds because the quantum state of the light is a coherent 
state (or a sum of coherent states of different phases), having the 
form $\exp(\alpha a^\dagger)\vert\,0\rangle$, with $a^\dagger$ the creation 
operator associated with a particular polarization and frequency and 
$\alpha$ a complex number. For a state of this form, the probability of $n$
photons is again given by the Poisson distribution 
\[
   P(n) = e^{-\langle n \rangle} \frac{\langle n\rangle^n}{n!}, 
\]   
where $n = a^\dagger a$ is the number operator and $\langle n \rangle = |\alpha|^2$ is its 
expectation value.  Finally, for a Poisson distribution with large $\langle n\rangle$, 
the rms deviation is 
$\sqrt{\langle (n - \langle n\rangle)^2\rangle } = \sqrt{\langle n\rangle}$.
\label{ft:laser}}
\be
  \delta N_{\rm random} \approx \sqrt N.
\label{e:nrandom}\ee

From Eq.~\eqref{e:gw1}, a wave of amplitude $h$ changes the distance $\ell$ 
between mirrors by a length $\delta \ell \sim h \ell$, implying a change of 
phase 
\be
   \delta \phi \sim h\frac \ell{\lambda_{\rm laser}}, 
\ee 
$\lambda_{\rm laser}$ the wavelength of the light. 
Now the amplitude of interfering beams with phase difference $\phi$ is proportional to 
\be 
  \sin(\omega_{\rm laser} t+\phi) + \sin(\omega_{\rm laser} t) = 2\sin(\omega_{\rm laser} t +\phi/2)\cos(\phi/2).
\ee
The time-averaged power $P$ reaching the detector is then proportional to 
$\cos^2\frac\phi2$.  Because the number of photons reaching the detector is 
proportional to $P$, 
\be 
   N \propto \cos^2\frac\phi2, \quad\mbox{implying }\ \ \delta N_{GW} \propto \delta\phi\ \frac d{d\phi} \cos^2\frac\phi2 = \delta\phi\ \sin\phi, 
\ee
taking its maximum value at $\phi=\pi/2$,\ $\sin\phi=1$.  Then, for a detector operating 
with unperturbed phase difference  $\pi/2$, we have $N(\phi) = 2 N \cos^2\frac\phi2$, 
and 
\be
   \delta N_{GW} = N\delta\phi \sim N h \frac\ell{\lambda_{\rm laser}}.
\label{e:ngw}\ee
From Eqs.\eqref{e:nrandom} and \eqref{e:ngw}, the detector can hear a wave when 
\[
  1\lesssim \frac{\delta N_{GW}}{\delta N_{\rm random}} 
  	\sim \sqrt N\ h \frac\ell{\lambda_{\rm laser}},
\] 
or
\be \crv
  	h\gtrsim \frac1{\sqrt N}\ \frac{\lambda_{\rm laser}}\ell\cb.
\label{e:hmax}\ee
 For a laser of power $P$, 
\be
  N = \frac{PT}{\hbar \omega_{\rm laser}} .
\label{e:N}\ee
 		
The laser power and wavelength of the \href{https://www.ligo.caltech.edu/page/laser}{LIGO laser} and interferometer arm length are 
 \\
\centerline{$P=20$ W, 
$\qquad \lambda_{\rm laser} = 1.064\times 10^{-4}$cm \ ($\omega_{\rm laser} = 1.8\times 10^{15}\rm s^{-1}$),\qquad 
$\ell = 4\,$ km. }
The binary black hole systems with individual masses $\gtrsim 30 M_\odot$ that have dominated the LIGO/Virgo observations have frequencies above 200 Hz for less than 10 cycles, implying an observation time $T\lesssim 10/f$.  
\index{LIGO/Virgo}
The number of photons in this time is, by Eq.~\eqref{e:N},
\be 
   N \approx \frac{(20\times 10^7 \mbox{erg/s})(10/200\,\rm s^{-1})}
			{\hbar(1.8\times 10^{15}\,\rm s^{-1})}
				=  5\times 10^{18}.
\ee 
If nothing further were done, the amplitude of an observable $300$ Hz wave 
($\lambda_{GW}= 10^8\rm cm$) would, by Eq.~\eqref{e:hmax}, be 
\be
  h \gtrsim \frac1{\sqrt{5\times 10^{18}}}\frac{10^{-4}\rm cm}{4\times 10^5 \rm cm}\\
 	\approx  10^{-19}, 
\ee
a few orders of magnitude larger than any observed source.    

Two of the improvements that dramatically enhance the sensitivity are   
\ben
\item[1.] The effective length of each arm is greatly increased by adding an extra mirror along the arm.  After entering the instrument via the beam splitter, the light in each arm bounces between its two mirrors about 300 times before leaving from the beam splitter and merging with the light from the other arm.  The result is an average photon path of about 1200 km and a decrease by a factor of $300$ in the smallest observable $h$. \footnote{This effective optical path length is deliberately kept below the wavelength corresponding to detectable frequencies.  A greater path length decreases sensitivity because the distance between mirrors changes in the time a photon traverses the path.}  (Each pair of mirrors with light bouncing between them constitutes a {\sl Fabry-Perot cavity}.) 
\begin{figure}[H]
               \begin{center}
		\includegraphics[width=0.6\textwidth]{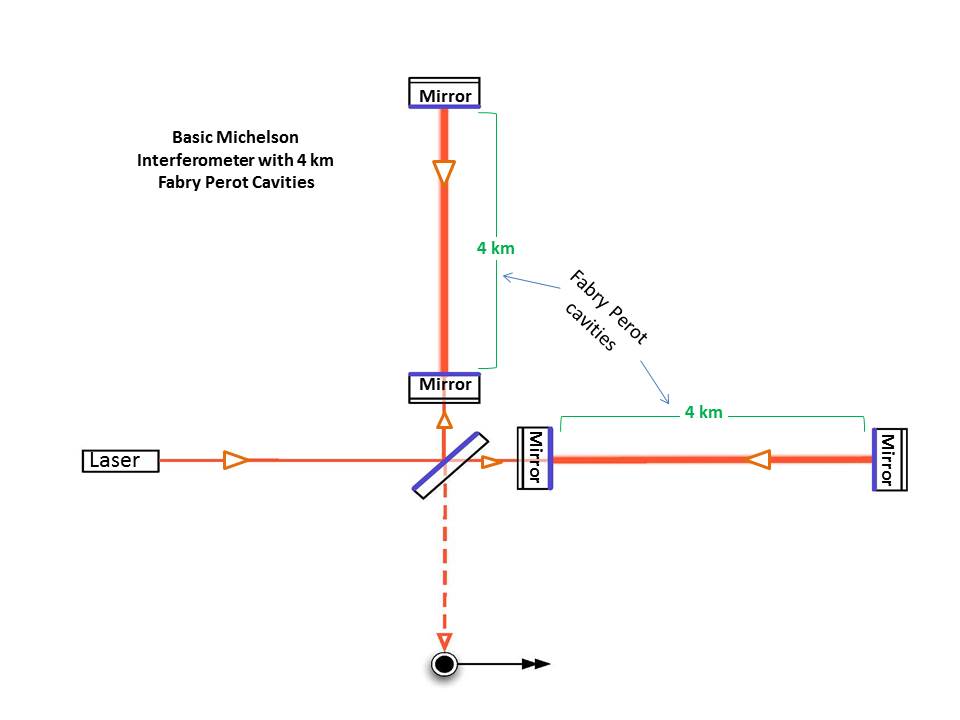}
		\end{center}
		\caption{Diagram from the same LIGO \href{https://www.ligo.caltech.edu/page/what-is-interferometer}{webpage} as that of Fig.~\ref{f:interferometer}}
\label{f:fpi}\end{figure}
\item[2.] A power recycling mirror, with high reflectivity, is placed between the laser 
and the interferometer.  The light that would reflect off the beam splitter in the 
direction of the laser is thus recycled, and the net power entering the intereferometer 
is enhanced by a factor of about 35.  
\een
Our estimate for the minimum observable amplitude then decreases by a factor of 300 from 
(1) and a factor of $\sqrt{35}$ from (2), giving 
\be
   h\gtrsim 6\times 10^{-23}.
\ee 
This value is now smaller than $h$ for the first observed source, GW150917.   

Below about 200 Hz, other sources of noise are larger than shot noise.  Sensitivity below about 100 Hz is limited by fluctuations in radiation pressure on the mirrors; like shot noise this 
is due to random fluctuations in the number of photons, this time in the number of photons 
hitting each mirror.  Below about 10 Hz, seismic noise is dominant, and between about 100 Hz 
and 200 Hz, thermal noise in the mirror coatings is important.  Shot noise and radiation-pressure noise are together called quantum noise because they both due to the Poisson 
distribution of photon number that describes the quantum state of laser light (see footnote \ref{ft:laser}).      
 
\benr\item\benalph
\item  Find the maximum value of $h_+$ or $h_\times$ for a binary neutron star system, with 
stars of mass $1.4 M_\odot$, at a distance $100$ Mpc.  You can use Eq.\eqref{e:Ixy} or the preceding equations for $I_{ij}$; assume that the maximum amplitude is reached just 
before the stars merge, with a distance of about $30$ km between their centers.  
\item Check that the maximum amplitude for a distance of $30$ Mpc agrees with that 
reported by LIGO/Virgo for GW170817.  \een\label{ex:170817}\een
\index{gravitational waves|)}\index{gravitational waves!detection|)}
\newpage
\noindent{\sl Gauge transformations as infinitesimal coordinate transformations}\\
\label{p:gtransf}
\index{infinitesimal coordinate transformation}\index{coordinates!infinitesimal coordinate transformation}\index{gauge transformation!as infinitesimal coordinate transformation}\index{passive transformation!infinitesimal coordinate transformation}

In Sect. \ref{s:gauge}, we defined a gauge transformation using an active transformation, an infinitesimal diffeo that drags along all tensor fields to give an physically equivalent system. One 
can equivalently keep the tensor fields fixed and make an infinitesimal coordinate 
transformation. This is the more common approach, and it is given here for 
reference. 

We consider a coordinate transformation from $x$ to $\bar x(x)$.  The equations in 
this section will be more transparent if the bar is on $g$ instead of on  
its indices. We will write the barred components of the tensor $g_{\alpha\beta}$ as  
\beq
   \bar g_{\mu\nu}(\bar x) 
	= \frac{\pa x^\sigma}{\pa\bar x^\mu}\frac{\pa x^\tau}{\pa\bar x^\nu}g_{\sigma\tau}(x).
\label{e:gbar}\eeq
Now write $x^\mu(\bar x) = \bar x^\mu + \xi^\mu$ with $\xi$ small.  To first order in $\xi$, we 
have   
\beq
   \bar g_{\mu\nu}(\bar x) = \bar g_{\mu\nu}(x-\xi) 
		= \bar g_{\mu\nu}(x)-\xi^\sigma\pa_\sigma g_{\mu\nu}(x).  
\label{e:gauge1}\eeq 
On the right side of \eqref{e:gbar}, we have
\[
  \frac{\pa x^\sigma}{\pa\bar x^\mu} = \frac{\pa\bar x^\sigma}{\pa\bar x^\mu} + \frac{\pa\xi^\sigma}{\pa\bar x^\mu}
	= \delta^\sigma_\mu +\frac{\pa\xi^\sigma}{\pa x^\mu}+ O(\xi^2), 
\]
giving 
\beq
  \frac{\pa x^\sigma}{\pa\bar x^\mu}\frac{\pa x^\tau}{\pa\bar x^\nu}g_{\sigma\tau}(x)
	 = g_{\mu\nu}(x) + \partial_\mu\xi^\sigma g_{\sigma\nu} + \partial_\nu\xi^\sigma g_{\mu\sigma}.
\label{e:gauge2}\eeq
Then, replacing the left and right sides of Eq.~\eqref{e:gbar} 
by the expressions in Eq. \eqref{e:gauge1} and \eqref{e:gauge2}, 
respectively, we obtain
\begin{align}
 \bar g_{\mu\nu}(x) 
	&= g_{\mu\nu}(x) + (\xi^\sigma\pa_\sigma g_{\mu\nu} +\partial_\mu\xi^\sigma g_{\sigma\nu} 			 + \partial_\nu\xi^\sigma g_{\mu\sigma}) 
\nonumber\\
	&= g_{\mu\nu}(x) + \Lie_{\bm\xi} g_{\mu\nu}(x).
\end{align} 
That is, to linear order in $\xi$ a change of coordinates changes the metric 
components by the components of its Lie derivative $\Lie_{\bm\xi} g_{\alpha\beta}$.  

Equivalently, given a family of coordinate transformations $\bar x(\lambda,x)$ 
with 
\beq
  	\bar x^\mu(0,x) = x^\mu, \qquad 
	\frac d{d\lambda}\left.\bar x^\mu(\lambda,x)\right|_{\lambda=0} = - \xi^\mu, 
\eeq
i.e., $x^\mu = \bar x^\mu + \lambda\xi^\mu + O(\lambda^2)$, \ \ \ we have 
\beq
  \frac d{d\lambda}\left.\bar g_{\mu\nu}(\lambda)\right|_{\lambda=0} = \Lie_{\bm\xi} g_{\mu\nu}.
\eeq
The family $\bar g_{\mu\nu}(\lambda)$ differs only by a coordinate transformation from 
$g_{\mu\nu}(\lambda)$, and at linear order in $\lambda$,  
\beq
  \delta \bar g_{\mu\nu} = \frac d{d\lambda}\left.\bar g_{\mu\nu}(\lambda)\right|_{\lambda=0} 
                     = h_{\mu\nu} + \Lie_{\bm\xi} g_{\mu\nu} , 
\eeq 
where
\beq
   h_{\mu\nu}= \frac d{d\lambda}\left. g_{\mu\nu}(\lambda)\right|_{\lambda=0} .
\eeq

More generally, if $S^{\mu\cdots\nu}_{\sigma\cdots\tau}$ are the components of any tensor, 
then its components in the chart $\bar x(\lambda)$ differ from the original components by 
$\Lie_{\bm\xi} S$ at linear order in $\lambda$: 
\beq
  \frac d{d\lambda}\left.\bar S^{\mu\cdots\nu}_{\sigma\cdots\tau}\right|_{\lambda=0} 
	= \Lie_{\bm\xi} S^{\mu\cdots\nu}_{\sigma\cdots\tau}
\eeq 
and a family of tensors, e.g., $g_{\alpha\beta}(\lambda,x),\ T_{\alpha\beta}(\lambda,x)$, 
has components in the family of charts $\bar x(\lambda,x)$ given at linear order in $\lambda$ by
\beq
  \bar h_{\mu\nu} = h_{\mu\nu} +\Lie_{\bm\xi} g_{\mu\nu}, \qquad 
  \bar \delta T_{\mu\nu} = \delta T_{\mu\nu} +\Lie_{\bm\xi} T_{\mu\nu}\,.
\label{e:gauge0}\eeq 
In particular, for a flat background, the perturbation  $h_{\alpha\beta}$ is physically 
equivalent to \\
\mbox{ $h_{\alpha\beta} + \Lie_{\bm\xi} \eta_{\alpha\beta} 
	= h_{\alpha\beta} + \nabla_\alpha\xi_\beta + \nabla_\beta\xi_\alpha$}, 
where $\nabla_\alpha$ is the flat covariant derivative, with components
$\partial_\mu$ in 
Cartesian coordinates. 

\newpage

\chapter{Cosmology from A to B} \label{c:cosmology}\index{cosmology|(}

\begin{verse}
{\sl \hspace{3mm} And more than a little arrogance is required for
creatures that evolved from quantum fluctuations and quark soup, that only exist for a
short time and are stuck on a small backwater outpost to think that they might be able to
understand the whole shebang}.
Michael Turner \cite{turner22} 
\end{verse}
This chapter starts by showing that the set of homogeneous, isotropic spatial geometries comprise the 
geometry of the 3-sphere, of flat space, and of the 3-dimensional hyperboloid.  The standard 
solutions to the field equations are derived, corresponding to radiation-dominated, matter-dominated, and vacuum-energy-dominated ($\Lambda$-dominated) stages in the universe's evolution; and the present composition of the universe -- vacuum energy, dark matter, baryonic matter, neutrinos and the cosmic microwave background (CMB)\index{cosmic microwave background} -- is used to find the times of transition between these stages. The cosmological metrics have a conformal Killing vector, and 
that is used to obtain the cosmological redshift. The chapter ends by 
showing how one uses gravitational-waves from binary inspiral to measure
the Hubble constant. 

For a summary of the standard $\Lambda$CDM (cosmological-constant-cold-dark-matter) cosmological model 
and the evidence for the present composition of the universe 
see, for example, Chaps.~17-19 of Hartle, 
Ryden's 2017 {\sl Introduction to Cosmology}\cite{ryden17}, or Mike Turner's recent 
\href{https://arxiv.org/pdf/2201.04741.pdf}{Road to Precision Cosmology}.\cite{turner22} 
Hartle and Ryden show how the numbers are derived. Turner gives a detailed description 
of the current state of cosmology, with a history of advances over the 
last century, and his summary also serves as an extensive guide to the literature.      

For most of universe's 13.8 billion year life, its density was dominated 
by the matter (ordinary and dark) that comprises its clusters of galaxies.  
These structures track the expansion of the universe, 
following timelike geodesics and so behaving like dust -- a pressureless fluid.  
We will see that the interior of a uniform-density collapsing ball of dust is 
identical to a matter-dominated, homogeneous isotropic universe, and we begin 
with that collapsing ball.   

\section{ Collapsing dust} \index{collapse!gravitational}\index{collapse, gravitational!collapse of ball of dust}\index{gravitational collapse!collapse of ball of dust}\index{gravitational collapse}

Consider a spherically symmetric ball of dust, falling from outer radius $R$. \\
Each dust particle follows a radial geodesic, with dust at the outer edge of the ball 
following a geodesic of the vacuum Schwarzschild geometry.  From Eq.~\eqref{rdot1}, 
the effective potential for a particle with zero angular momentum is 
\[
V_{\rm eff} = - \frac{M}{r}+ \frac{L^2}{2r^2} -\frac {ML^2}{r^3} = -\frac Mr.
\]
Then
\begin{align} 
\frac12\dot r^2& = e_N + \frac Mr \ \Longrightarrow\ 0 = e_N + \frac MR, \nonumber \\
\dot r^2 &= \frac {2M}r -\frac {2M}R \ .
\label{e:radial_geodesic}\end{align} 
Fall from $R=\infty$ at $\tau=-\infty$, reaching $r=0$ at $\tau=\tau_0$: 
\be
 \cblue r(\tau) = \left(\frac92M\right)^{1/3} (\tau_0-\tau)^{2/3} \cb \ 
\label{e:collapse}\ee
For a particle falling from at finite $R$, the solution (checked below) has 
the parametric form 
\bsube\begin{align}
r&= \frac R2(1+\cos\eta) \\
\tau&=\frac R2\sqrt{\frac R{2M}}\ (\eta+\sin\eta) 
\end{align}\label{e:cycloid}\esube
\index{cycloid}
The path $r(\tau)$ is half a cycloid, shown in the MTW figure below.  
\begin{figure}[H]
               \begin{center}
		\includegraphics[width=0.6\textwidth]{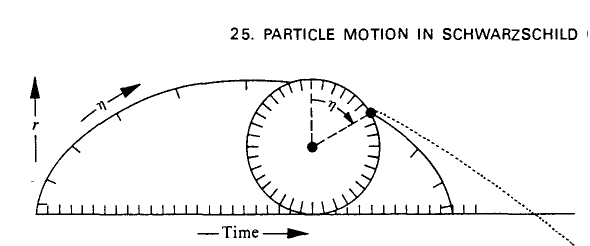}
		\end{center}
		\caption{Figure 25.3 of MTW}
\end{figure}

Check that $r(\tau)$ given by Eqs.~\eqref{e:cycloid} satisfies 
the geodesic equation \eqref{e:radial_geodesic}:  $(') = \frac d{d\eta}$  \\
\begin{align*}
r = \frac R2(1+\cos\eta) \quad\Longrightarrow\quad r' &= -\frac R2\sin\eta \\
\tau =\frac R2\sqrt{\frac R{2M}}(\eta+\sin\eta) \quad \Longrightarrow\quad 
 \tau' &= \frac R2\sqrt{\frac R{2M}} (1+\cos\eta) \\
\dot r^2 = \left(\frac{r'}{\tau'}\right)^2 
	= \frac{2M}R \frac{\sin^2\eta}{(1+\cos\eta)^2} 
	&= \frac{2M}R \frac{1-\cos\eta}{1+\cos\eta}.\\
\frac{2M}r - \frac{2M}R = \frac{2M}{\frac R2(1+\cos\eta)}  - \frac{2M}R 
	&= \frac{2M}R\frac{1-\cos\eta}{1+\cos\eta}. \qquad \Box
\end{align*}

We'll see that the interior of homogeneous collapsing dust is a homogeneous cosmology 
and that the interior particles obey the same geodesic equation, acting as if they 
see only the mass within their radius.  

\section{Spatially homogeneous, isotropic cosmology}
\index{cosmology!homogeneous isotropic|textbf}

The cosmic background radiation (CMB)\index{cosmic microwave background} is a gas of photons filling 
the universe.\index{cosmic microwave background}  Its extreme homogeneity and isotropy implies 
a similarly homogeneous and isotropic early universe, before cooling led to structure formation.  

Measurement of the CMB (CMB)\index{cosmic microwave background} by the the \href{https://articles.adsabs.harvard.edu/pdf/1992ApJ...396L...1S}{COBE}\cite{cobe92},
\href{https://iopscience.iop.org/article/10.1088/0067-0049/208/2/20/meta}{WMAP}\cite{WMAP13},
and \href{https://arxiv.org/abs/1807.06205}{Planck}\cite{planck20} satellites (as well as more recent
observations) show an early universe that was isotropic to about one part in $10^5$,
with temperature fluctuations in the CMB of about 80 $\mu$K about the 2.73 K average.
 That anisotropy was also the anisotropy of the matter at the 
time light decoupled from baryonic matter, about 300,000 years after the initial
Big Bang. Gravitational instability turned the small initial anisotropy of dark and baryonic matter 
into the large structures of the late universe -- 
superclusters comprising millions of galaxies, the {\sl filaments} that are their
walls, and the emptier {\sl voids} between the clusters.  The present universe looks
like a web of these structures, inhomogeneous on scales of $10^9$ ly.  At the 
$10^{10}$ ly scale of the visible universe itself, however, the distribution of 
matter and the geometry of space retain their homogeneous, isotropic character.\\

A 3-dimensional homogeneous, isotropic geometry looks identical at every point. 
The obvious example is flat space, whose group of isometries is the 
6-dimensional Euclidean group of translations and rotations.  We will see that  
there are only two other cases: The 3-spheres, the subspaces of Euclidean 4-dimensional 
space with $T^2 + x^2 + y^2 + z^2 = a^2$; and 3-dimensional hyperboloids, the 
subspaces of Minkowski space with $-t^2+x^2+y^2+z^2 =a^2$.  \\

\noindent{\bf Definitions}:\\
\index{homogeneous space}
A space is {\sl homogeneous} if, for any two points $P$ and $Q$, there is 
an isometry that maps $P$ to $Q$.  In flat space, the isometry is a translation.\footnote{Rotations also map any $P$ to any $Q$ in flat space, but it is translations that remain as symmetries of homogeneous but anisotropic metrics.  A simple example is the Kasner family of homogeneous, anisotropic spacetimes, 
whose metrics have the form  
$-dt^2+ e^{2p_1t} dx^2 + e^{2p_2t} dy^2+ e^{2p_3t} dz^2$. When $\sum p_i = \sum p_i^2 = 1$, these are vacuum solutions to the field equations.}
A space is {\sl isotropic}\index{isotropic space|textbf} if it has no preferred direction at any point:  
For any two unit vectors $u^a$ and $v^a$ at a point $P$ 
there is an isometry that keeps $P$ fixed and maps $u^a$ to $v^a$.  
In flat space, the isometry is a rotation fixing $P$. 

In our discussion of homogeneous, isotropic 3-dimensional spaces, we 
will write the Riemann tensor, Ricci tensor and Ricci scalar as $^3\!R^a{}_{bcd}$,
$^3\!R_{ab}$, and $^3\!R$.   
In a homogeneous space, any scalar constructed from the metric must have the same 
value at all points, so the Ricci scalar $^3\!R$ is constant, and we will 
show below that the homogeneous, isotropic spaces are determined by the value of $^3\!R$:\\ 
For $^3\!R>0$, the space is a 3-sphere \index{sphere!3-sphere}\\
For $^3\!R=0$, the space is flat; \\
For $^3\!R<0$, the space is a hyperboloid.\index{hyperboloid} \\ 

First we'll find the metrics on the unit 3-sphere and the unit hyperboloid. 
The 3-sphere of radius $a$ is the subspace of $\mathbb R^4$ given by 
\[
 a^2 =   T^2 + x^2 + y^2 + z^2 = T^2 + r^2.
\]
On the unit 3-sphere, define an angle $\chi$ from the $T$ axis to each point $P$ (analogous to the angle $\theta$ from the $z$-axis).  The 2-sphere at angle $\chi$ has radius $a\sin\chi$:  
\[
   T = \cos\chi, \quad r=\sin\chi; 
\]
with the usual polar coordinates ($x=r\sin\theta\cos\phi, y=r\sin\theta\sin\phi,z= r\cos\theta$), we have 
\begin{align}
  ds^2 & = dT^2 + dx^2 + dy^2 + dz^2 = dT^2 + dr^2 + r^2 d\Omega^2,  \nonumber\\
  &= d\chi^2 + \sin^2\chi d\Omega^2,
\end{align}
where $d\Omega^2$ is the metric on the unit 2-sphere, 
\[
    d\Omega^2 = d\theta^2 + \sin^2 \theta d\phi^2.  
\]
\begin{figure}[H]
               \begin{center}
		\includegraphics[width=0.5\textwidth]{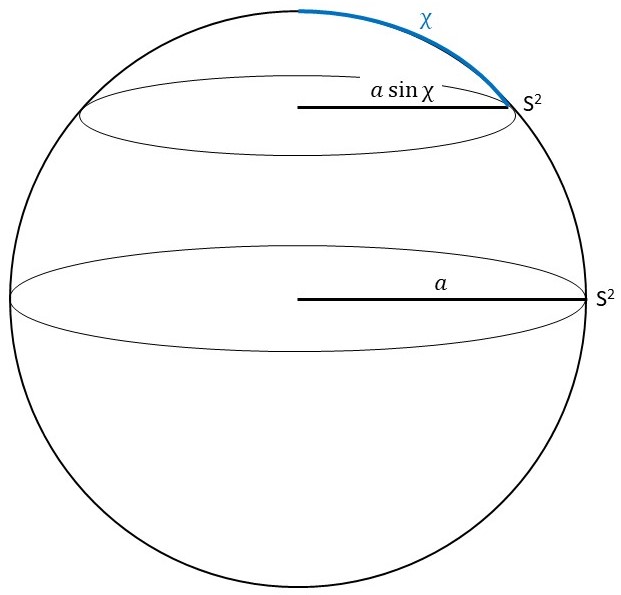}
		\end{center}
		\caption{The angle $\chi$ extends from the vertical $T$ axis.  Each $\chi=$ constant surface is a 2-sphere of radius $a\sin\chi$.  }
\end{figure}

Replacing $T$ by $it$ and taking $t$ to be real gives the corresponding hyperboloid:
\benr
\item \label{ex:hyperboloid} The unit spacelike hyperboloid in 4-dimensional Minkowski space, is the set of 
points at unit distance from the origin:  
\be
  \eta_{\mu\nu}x^\mu x^\nu =  -t^2 + x^2 + y^2 + z^2 = -1, \qquad t>0.    
\ee
\benalph\item Suppress a dimension (take z=0), and sketch the 2-D hyperboloid as a submanifold of $\mathbb R^3$.
\item Show that the 3-D hyperboloid is homogenous, that (1) any Lorentz transformation maps the hyperboloid to itself; and (2) for any points $P$ and $Q$ of the hyperboloid, there is a Lorentz transformation that takes $P$ to $Q$. 
\item Show that the metric on the hyperboloid is given by 
\[
   ds^2 = d\chi^2 + \sinh^2\chi d\Omega^2, 
\] 
where \vspace{-5mm}
\index{Lorentz transformation!symmetry of hyperboloid}

\begin{align*}
  x&=\sinh\chi \sin\theta\cos\phi\qquad y = \sinh\chi\sin\theta\sin\phi \\
  z&= \sinh\chi\cos\theta\qquad \qquad t = \cosh\chi
\end{align*}
Start with polar coordinates $t,r,\theta,\phi$ in 
Minkowski space, checking first that $-t^2+r^2 = -1$, with 
$r^2 = x^2+y^2+z^2$ and $t$, $x$, $y$ and $z$ in the form given above.\\
No need to rederive the metric on the unit two-sphere $S^2$.  
\een
\een

\noindent{\sl Claim}: The only homogeneous, isotropic, 3-dimensional spaces have the metrics of a sphere of radius $a$, the corresponding hyperboloid 
(the set of points in Minkowski space at a distance $a$ from the origin), 
and flat space.  For the sphere and hyperboloid, the curvature scalar has the values $\dis\pm \frac6{a^2}$.  
\index{cosmology!hyperbolic space|textbf}\index{cosmology!spherical space|textbf}\index{hyperbolic space|textbf}\index{sphere!3-sphere|textbf}
\be
  ds^2 = \begin{cases}
  	    a^2\left[ d\chi^2 + \sin^2\chi(d\theta^2 + \sin^2 \theta d\phi^2)\right],
  	    	 & ^3\!R=\dis\frac6{a^2}>0, \\
  	   \hspace{7mm} dx^2+ dy^2 + dz^2, & ^3\!R=0,\\
	   a^2\left[ d\chi^2 + \sinh^2\chi(d\theta^2 + \sin^2 \theta d\phi^2)\right], 
		& ^3\!R=\dis-\frac6{a^2} <0.
 \end{cases} 
\ee
Notice that we have said nothing about a spacetime metric or the 
field equations:  This is simply a description of all homogeneous 
isotropic 3-dimensional geometries.  

We need only show the result for the unit sphere and hyperboloid (and for $^3\!R=0$) 
because multiplying $g_{ab}$ by $a^2$ just multiplies the scalar curvature 
by $a^{-2}$.  This follows from dimensional analysis: All lengths change by 
the factor $a$ and $^3\!R$ has dimension $L^{-2}$. (Alternatively, 
because $g^{ab}\rightarrow a^{-2} g^{ab}$, there is no change in $\Gamma^i{}_{jk}$, 
so $^3\!R^a{}_{bcd}$ and $^3\!R_{ab}$ 
are unchanged, and $^3\!R = {}^3\!R_{ab}g^{ab} \rightarrow {}^3\!R/a^2$.) 

Because the space is isotropic, it is spherically symmetric about 
any point, and we already know from Sect.~\ref{s:spherical_symmetry} 
that a spherically symmetric space has a metric of the form 
\[
  ds^2 = f^2dr^2 + g^2 d\Omega^2,  
\] 
with $f$ and $g$ functions of $r$ only. Now define $\chi(r)$ by 
$d\chi = f dr$.  Then  
\be
   ds^2 = d\chi^2 + h^2(\chi) d\Omega^2.
\label{e:3g}\ee 
Isotropy implies that the component $^3\!R_{ab}u^a u^b$ of $R_{ab}$ along 
a unit vector $u^a$ is independent of the direction of $u^a$.
Then, for the unit vectors $\hat\chi^a$
, $\hat\theta^\a$, $\hat\phi^a$, 
\be
  ^3\!R_{\hat\chi\hat\chi} = {}^3\!R_{\hat\theta\hat\theta} 
  	= {}^3\!R_{\hat\phi\hat\phi} \ \Longrightarrow
{}^3\!R^\chi{}_\chi = {}^3\!R^\theta{}_\theta = {}^3\!R^\phi{}_\phi.
\ee

To compute Ricci tensor components we will use the form \eqref{e:Ricci_compute}, 
valid in any dimension:  
\[
R_{ij} = \partial_k \Gamma ^k {}_{i j }
			- \partial_i \partial _j \log \sqrt{-g} 
			+ \Gamma ^k{}_{i j }\partial _k \log \sqrt{-g} 
			- \Gamma^k{}_{\ell j }\Gamma^\ell {}_{k i} \ .
\] 
For a metric of the form \eqref{e:3g}, the only nonzero $\Gamma$'s are (up to index symmetry)
\bsube\begin{align}
  \Gamma^\phi{}_{\theta\phi} & = \cot\theta \ \ \quad\qquad
  \Gamma^\theta{}_{\phi\phi} = -\sin\theta\cos\theta \\ 
  \Gamma^\theta{}_{\chi\theta} &= \Gamma^\phi{}_{\chi\phi}=\frac{h'}h ,\quad
  \Gamma^\chi{}_{\theta\theta} = -hh'\ ,\quad \Gamma^\chi{}_{\phi\phi}= -hh'\sin^2\theta .
\end{align}\label{e:Gamma_3sphere}\esube
Using $\sqrt{^3\!g}=h^2\sin\theta$, we have 
\begin{align}
   ^3\!R_{\chi\chi} &= -\pa_\chi^2 \log\sqrt{^3\!g} - \left(\Gamma^\theta{}_{\chi\theta}\right)^2 - \left(\Gamma^\phi{}_{\chi\phi}\right)^2 
    = - 2\frac{h''}{h}{\color{gray} + 2\left(\frac{h'}h\right)^2 
       - 2\left(\frac{h'}h\right)^2} \nonumber\\
	 &= - 2\frac{h''}h \nonumber\\
  ^3\!R &=   {}^3\!R^\chi{}_\chi+  {}^3\!R^\theta{}_\theta +   {}^3\!R^\phi{}_\phi 
  	= 3\ {}^3\!R^\chi{}_\chi = -6 \frac{h''}h.
\end{align}

We now pick the value of the Ricci scalar that will give the metric 
on the sphere of radius $a=1$, namely $^3\!R = 6$.  As noted above, 
we can do this because we already know that the corresponding solution 
for any other positive Ricci scalar, $^3\!R=6/a^2$, is that metric 
multiplied by $a^2$.   
\be
   -6 \frac{h''}h = 6 \Longrightarrow h'' + h = 0,  \quad 
   	        h = A \sin(\chi + \eta),  
\ee
for some constants $A$ and $\eta$.  A choice of origin for $\chi$ sets 
$\eta$ to $0$, and we recover the metric of the unit sphere when $A=1$.  
When $A\neq 1$, the metric is not isotropic; $h$ does not satisfy 
$^3\!R^\theta{}_\theta={}^3\!R^\chi{}_\chi = 2$:
\begin{align}
   ^3\!R_{\theta\theta} 
     &= \pa_\chi\Gamma^\chi{}_{\theta\theta}-\pa_\theta^2 \log\sqrt{^3\!g}
     	+\Gamma^\chi{}_{\theta\theta}\pa_\chi\log\sqrt{^3\!g} 
     	-2\Gamma^\chi{}_{\theta\theta}\Gamma^\theta{}_{\chi\theta}
     	- \left( \Gamma^\phi{}_{\phi\theta}\right)^2 \nonumber\\
     &= -hh''-h'^2 +1.  \nonumber\\
^3\!R^\theta{}_\theta &= - \frac{h''}h -\left(\frac{h'}h\right)^2+\frac1{h^2}\ .
\end{align}
For $h=A\sin\chi$, we have  
$\dis ^3\!R^\theta{}_\theta = 1 - \cot^2\chi + \frac1{A^2}\csc^2\chi,$ whence
\be
	^3\!R^\theta{}_\theta = 2\ \Longrightarrow A=1.   
\ee
To summarize:  The unique homogeneous, isotropic 3-metrics with $^3\!R>0$ 
are the metrics of 3-spheres of radius $a$,  
\be
 \cblue ds^2 = a^2\left( d\chi^2 + \sin^2\chi d\Omega^2 \right),\cb
\label{e:g_3sphere}\ee
with
\be
  ^3\!R^a{}_b= \frac2{a^2} \delta^a_b \qquad ^3\!R = \frac6{a^2}.\cb
\label{e:ricci_3sphere}\ee
(Unique means unique up to a diffeo or, in passive language, a change of coordinates.)   

\benr\item Check that the same steps imply for the homogeneous, 
isotropic 3-metrics with $^3\!R<0$ the form 
\be
 ds^2 = a^2\left( d\chi^2 + \sinh^2\chi d\Omega^2 \right), 
\ee
with 
\be
  ^3\!R^a{}_b= -\frac2{a^2} \delta^a_b \qquad ^3\!R = -\frac6{a^2}.\cb
\label{e:ricci_3hyperboloid}\ee
\een 

On $\mathbb R^3$, the flat metric $dx^2+dy^2+dz^2$ is equivalent to the 
metric
\be
   ds^2 = a^2(dx^2+dy^2+dz^2),
\ee
for any nonzero constant $a$.  That is, the metrics are related by a 
diffeo mapping $(x,y,z)$ to $(ax,ay,az)$ (or, passively, by the corresponding 
coordinate transformation).  In spacetime, however, with 
$a$ a function of time, the 4-dimensional metric is not flat, and 
a time-dependent {\sl scale factor} $a$ describes an expanding or contracting 
universe. \\ 

Now that we know the form of the spatial metric, it is not difficult
to write the general form of the spacetime metric.  We first extend to 
spacetime our definitions of spatial homogeneity and isotropy.\\
{\bf Definition}:  A spacetime is homogeneous and isotropic if 
\ben\item[1.]the spacetime is the union of a family of disjoint spacelike 
slices (3-dimensional surfaces); 
\item[2.] for any two points $P$ and $Q$ on the same slice $\Sigma$, there 
is an isometry that maps $P$ to $Q$; 
\item[3.] for any two unit vectors $u^\a$ and $v^\a$ tangent to $\Sigma$ at a 
point $P$ there is an isometry that keeps $P$ fixed and maps $u^\a$ to $v^\a$.
\een
Corresponding to the three classes of homogeneous spaces with $R>0$, $R=0$, 
and $R<0$ are the three 6-dimensional isometry groups: the group $O(3)$ of 
rotations of the 3-sphere, the Euclidean group of rotations and translations 
of $\mathbb R^3$, and the Lorentz group (acting on the 3-dimensional hyperboloid).  

We can use the vector field $n^\a$ of unit normals to 
these surfaces to define a natural chart on the spacetime.  The integral curves 
of $n^\a$ are a family of timelike trajectories parameterized by 
proper time $\tau$.  This preferred family of trajectories is called 
the {\sl Hubble flow}.  

 Let $\Sigma_0$ be one of the homogeneous isotropic slices, and set 
$\tau=0$ on $\Sigma_0$. Spatial homogeneity means that the 4-metric 
$g_{\a\b}$ does not distinguish one point of $\Sigma_0$ from another.  
In particular, if one moves a proper time $\tau$ along the trajectories
through any two points $P$ and $Q$ of $\Sigma_0$, one reaches a 
slice with the same value $a(\tau)$ of the scale factor:  
Each homogeneous isotropic slice is the set of points at a fixed proper 
time $\tau$ from $\Sigma_0$. Then, with $\tau$ as our time coordinate, 
the scale factor is a function $a(\tau)$, and we can choose spacetime 
coordinates $\tau, \chi,\theta, \phi$ 
for which the spacetime metric is given by  
\be{\cblue
  ds^2 = - d\tau^2 + \begin{cases}
  	    a^2(\tau)\left( d\chi^2 + \sin^2\chi d\Omega^2\right), \\	
	   a^2(\tau)\left( d\chi^2+\chi^2 d\Omega^2\right)\quad = a^2(\tau)\left( dx^2+ dy^2 + dz^2 \right),\\
a^2(\tau)\left( d\chi^2 + \sinh^2\chi d\Omega^2\right) .
 \end{cases} }
\label{e:ghi}\ee
We do this using the 6-dimensional group of spacetime isometries.  
First choose natural coordinates on $\Sigma_0$. Next, 
use the trajectory through the origin on $\Sigma_0$ to define the 
origin $(\tau,0,0,0)$ on each slice $\Sigma_\tau$ (on each $\tau=$ constant surface).  Finally use the isometry $\psi$ that maps $(0,0,0)$ to 
$(0,\chi,\theta,\phi)$ to assign coordinates $(\tau,\chi,\theta,\phi)$ 
to the point $\psi(\tau,0,0,0)$. The spatial metric on each $\Sigma_\tau$ 
then has the form in \eqref{e:ghi}.  Finally, because an isometry maps a 
trajectory orthogonal to $\Sigma$ to a trajectory orthogonal to $\Sigma$, 
the lines of constant $\chi,\theta,\phi$ are the trajectories normal 
to the slices.  Their tangent vector $\bm\pa_\tau$ is therefore normal 
to $\Sigma$, implying $g_{\tau i}=0$. 

\index{metric!Friedmann,FLRW}\index{Lema\^itre}\index{Robertson-Walker metrics}
Metrics of the form \eqref{e:ghi} are variously called Friedmann, 
Robertson-Walker, FL, FRW, or FLRW metrics (the L for Lema\^itre).  
They were first found by Alexandr Friedmann in 1922 \href{https://ia803104.us.archive.org/30/items/aleksandr-friedmann.-on-the-curvature-of-space/Aleksandr%20Friedmann.%20On%20the%20Curvature%20of%20Space.pdf}{English translation} \cite{friedmann22g} 
for the spherical case and in 1924 for the hyperbolic case \cite{friedmann24}.     
Lema\^itre first pointed out that an expanding universe 
implies Hubble's law \href{https://articles.adsabs.harvard.edu/pdf/1931MNRAS..91..483L}{(1931 translation of 1927 article)}\cite{lemaitre31}. 
\footnote{After being ordained as a Catholic priest, Lema\^itre was allowed 
to study at MIT, where he got his PhD.  He apparently didn't know Friedmann's work, 
independently obtained the positive-curvature metric, and found Eq.~\eqref{e:taua} 
below for arbitrary amounts of pressure-free matter and radiation.  He made the connection to Hubble's law after visiting Hubble at Caltech and Slipher at Arizona.  
Slipher had preceded Hubble and Humason in discovering the redshifts of 
galaxies, but not the redshift-distance relation.  Robertson \cite{robertson33} 
looked at solutions with general $k,\Lambda$, $P$ and $\rho$.}
\index{redshift!redshift-distance relation}

 Another way to write the family of homogeneous isotropic metrics is 
at least as common in the literature as Eq.~\eqref{e:ghi}: 
Introducing a radial coordinate 
 $r$ by 
 \[
    r = \begin{cases}
  	    \sin\chi,  	    	 & ^3\!R>0, \\
  	    \chi = \sqrt{x^2+y^2+z^2}, & ^3\!R=0,\\
	   \sinh\chi,  
		& ^3\!R<0,
	\end{cases}
 \]
 immediately gives the metric in all three cases in the form
 \be
     ds^2 =-d\tau^2 + a^2\left[ \frac{dr^2}{1-kr^2} + r^2 d\Omega^2 \right],
 \label{e:gkr}\ee
 with $k=1, 0, -1$, for the spherical, flat, and hyperbolic cases, 
respectively.\\ 
\noindent{\sl Exercise}. Do the two-line calculation, replacing $r$ in 
Eq.\eqref{e:gkr} by $\sin\chi$ and $\sinh\chi$, to check this.   \\ 

\noindent{\sl\crv Finite or infinite?} 
\begin{verse}
{\sl \hspace{3mm} It is evident that from this point of view many 
assertions concerning space made by previous writers are no longer
correct (e.g. that infinity of space is a consequence of zero 
curvature) . . .}
\end{verse}
\vspace{-2mm}

\hspace{4mm}{Felix Klein, 1893 \href{https://babel.hathitrust.org/cgi/pt?id=wu.89041215435&seq=107}{\sl Researches in Non-Euclidean Geometry} p. 93.}\\
\vspace{-3mm}

The fact that flat $\mathbb R^3$ and the 3-dimensional hyperboloid are spatially infinite, while the 3-sphere is finite misleads people 
into thinking that non-positive spatial curvature implies an 
infinite universe.  Although Eq.~\eqref{e:ghi} gives all 
the homogeneous isotropic metrics, the metric does not determine 
the large-scale topology.\\ 
\noindent
\index{cosmology!finite flat, hyperbolic, and spherical spaces}
{\cblue Ten finite 3-manifolds 
allow flat metrics; a countably infinite 
collection of finite 3-dimensional manifolds allow the 
hyperboloidal metrics of Eq.~\eqref{e:ghi}; and a countably 
infinite collection of finite 3-dimensional manifolds have 
the spherical metrics of Eq.~\eqref{e:ghi}}.\\  More precise  
language replaces {\sl finite manifold} by {\sl compact manifold without boundary}.

The simplest example is the flat torus.  In 
2-dimensions, identifying the opposite edges of a flat 
rectangle gives a flat two-dimensional torus. This is 
the finite flat space of the video games Ms. PacMan and 
Asteroids, in which the animations identify the left and 
right edges and the top and bottom edges of the screen.  \vspace{-5mm}
\begin{figure}[H]
               \begin{center}
		\includegraphics[width=0.6\textwidth]{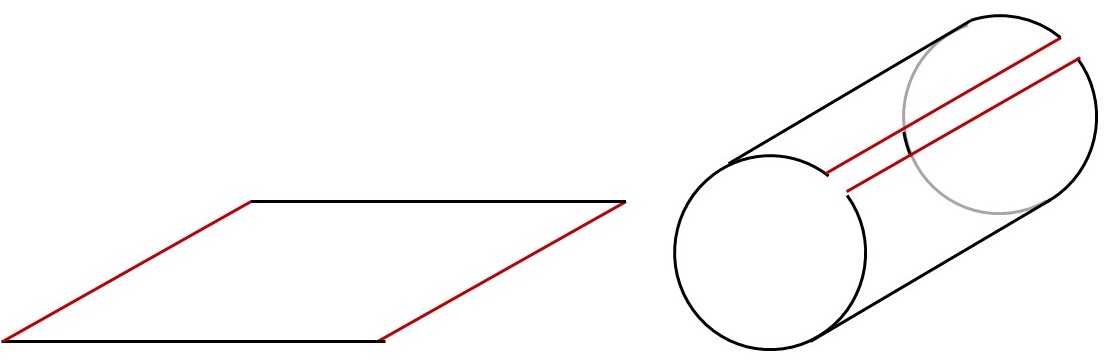}
		\end{center}
		\caption{A flat torus is constructed by identifying 
opposite edges of a flat sheet.  Identifying one pair of edges gives 
a flat cylinder, shown here embedded in flat $\mathbb R^3$.  Identifying 
the front and back circles of the cylinder then gives a flat torus. 
There is no similar embedding of the flat torus in $\mathbb R^3$.}
\end{figure}  \vspace{-4mm}
 
One similarly constructs a flat 3-torus by identifying 
opposite faces of a cube. \vspace{-5mm}
\begin{figure}[H]
               \begin{center}
		\includegraphics[width=0.42\textwidth]{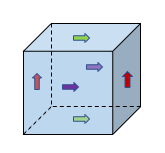}
		\end{center}
		\caption{A flat 3-torus is a cube with opposite faces identified by translation.  Figure from \href{https://sites.nova.edu/mjl/graphics/spaced-out/3-tori/}{Michael Laszlo}}
\end{figure}
You can regard the cube as the unit cell of a group of 
translations by a fixed distance $d$ in the $x$, $y$, 
and $z$ directions, the symmetry group of a simple crystal.
All finite 3-manifolds with locally homogeneous isotropic metrics are obtained in a similar way:  A discrete subgroup of the isometry group 
of $S^3$, $\mathbb R^3$ or $H^3$ (the hyperboloid) maps a cell into 
a set of copies of itself that tile the space.  And points of the 
cell boundary related by an isometry in the subgroup are identified.
This is Klein's construction. (The subgroup has to act freely 
for the construction to work: That is, no isometry in the subgroup 
except the identity can fix a point of space.) %
\footnote{More technically, each compact 3-manifold with locally homogeneous isotropic 
metric is a quotient space $M/G$ where $M$ is $S^3$, $\mathbb R^3$ or $H^3$ and $G$ 
is a finite subgroup of the isometry group, with $G$ acting freely on $M$.}

Several groups have studied the intriguing possibility that our universe might be much {\sl smaller} than that the visible universe, with 
the same galaxies seen repeatedly, as their light loops around 
the universe (already suggested by Friedmann in 1924).  The key observations, by the WMAP satellite, looked for pairs of identical circles in different directions 
in the CMB data.  The results were negative.  
See \href{https://arxiv.org/abs/1601.03884}{Luminet 2016}\cite{luminet16} for a summary of the circle method and references.  

For non-technical descriptions of finite hyperbolic spaces 
and spherical spaces, see 
\href{https://www.geometrygames.org/CurvedSpaces/index.html}{Jeffery Weeks' Curved Spaces},  
\href{https://static.scientificamerican.com/sciam/cache/file/DE663036-FFFB-4E81-B2F068C08A2BBBDA.pdf}{Thurston and Weeks 1984},\cite{tw84} and 
Ellis and Williams \cite{ew00} Sect. 7.7.

\section{The field equations and their solutions}  
\index{expanding universe}

For metrics of the form \eqref{e:ghi}, the Ricci tensor 
splits into a sum of two terms: The first, computed only from spatial derivatives, is the Ricci tensor of the spatial metric, already given by Eqs.~\eqref{e:ricci_3sphere} and \eqref{e:ricci_3hyperboloid} for the 
3-dimensional sphere and hyperboloid.  The second part involves the time 
derivative of the 3-metric along the unit normal $\na^\a \tau$ to the 
$\tau=$ constant surfaces, the {\sl extrinsic curvature}.   
If we write the metrics \eqref{e:ghi} in the form 
\index{normal to a spacelike hypersurface!relation to extrinsic curvature}
\[
  g_{\a\b} = -\na_\a\tau \na_\beta\tau + \gamma_{\a\b}, 
\]
then we can regard $\gamma_{\a\b}$ as the spatial metric: Its $\tau$ 
components vanish, $\gamma_{\a\b}\na^\b\tau = 0$, and its spatial components are those of the spatial metric, $\gamma_{ij} = g_{ij}$.  
The extrinsic curvature $K_{\a\b}$ is defined by
\index{extrinsic curvature} 
\be
   K_{\a\b} = \frac12\Lie_{\bm\na\dis \tau}\ \gamma_{\a\b},  
\label{e:Kalphabeta}\ee
whose nonzero components are 
\be
  K_{ij} = -\frac12 \pa_\tau \gamma_{ij} = -\frac{\dot a}a \gamma_{ij},   
\label{e:Kij}\ee
where $(^\cdot):=\frac d{d\tau}$.   

The initial-value formalism, detailed in Chap.~\ref{c:iv}, takes the 3-metric and its time derivative -- the extrinsic curvature-- 
as initial data for the vacuum field equations.   
The quick calculations here do not need that more elaborate 
study, but it is worth pointing out that the split of the curvature tensor holds for a generic spacetime and underlies the 
initial-value formalism.

\begin{figure}[H]
               \begin{center}
		\includegraphics[width=0.5\textwidth]{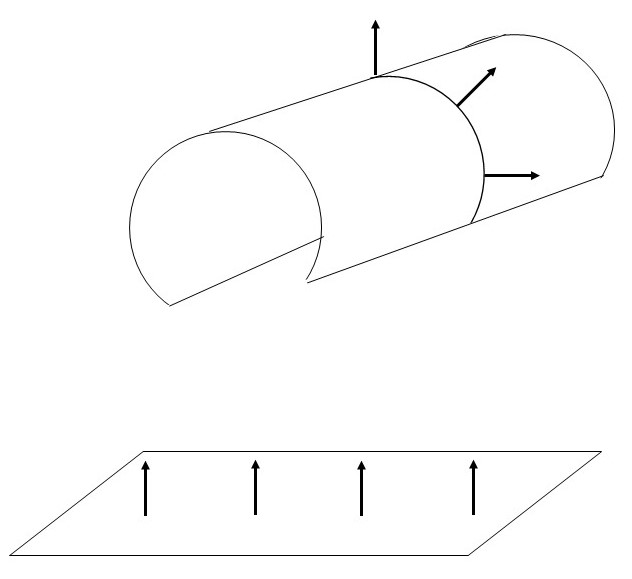}
		\end{center}
		\caption{Two embeddings of a flat surface in $\mathbb R^3$, with unit normals shown.  The extrinsic curvature measures the rate of change of the unit normal $n^a$ in directions tangent to the surface. 
Dragging the surface along the normal by a distance $ds$ gives a new surface whose 
fractional increase in volume, induced by the flat metric of $\mathbb R^3$, is 
given by $\na_a n^a\ ds$. }
\end{figure}     

\subsection{Spatially flat} 
 \index{cosmology!spatially flat|textbf}\index{metric!cosmological|(}
 
 We begin with the simplest case, a spatially flat metric. A nearly flat spatial metric is 
is also favored by current observations.  
For the spatially flat cosmology, with metric  
\[
     ds^2 = -d\tau^2 + a^2(\tau) (dx^2+dy^2 + dz^2),
\]
the intrinsic curvature vanishes, and the curvature tensor involves 
only the extrinsic curvature $K_{\a\b}$, and its time derivative, 
i.e., only $a, \dot a$ and $\ddot a$.  
The nonzero metric derivatives and $\Gamma$'s are   
\begin{align}
   \dot g_{xx} & = \dot g_{yy} = \dot g_{zz} = 2 a\dot a, \nonumber\\
   \Gamma^\tau{}_{xx} &= \Gamma^\tau{}_{yy} = \Gamma^\tau{}_{zz} 
   	=a\dot a\qquad 
   \Gamma^x{}_{x\tau} = \Gamma^y{}_{y\tau}=\Gamma^z{}_{z\tau}=\frac{\dot a}a\ .
\label{e:Gamma_sflat}\end{align}
{\sl Quick exercise}.  Use the definition \eqref{e:Kalphabeta} of $K_{\a\b}$ to check Eq.~\eqref{e:Kij} and the relation 
\[
	\Gamma^\tau{}_{ij} = -K_{ij}.
\]

Again using Eq.~\eqref{e:Ricci_compute} for components of $R_{\a\b}$ and 
writing $\log\sqrt{-g} = 3\log a$, we have
\bsube\begin{align}  
   R_{\tau\tau} &=  - \partial_\tau^2 \log \sqrt{-g} 
		 - 3(\Gamma^x{}_{x \tau})^2 = - 3\frac{\ddot a}a,\\
   R^\tau{}_\tau &= 3\frac{\ddot a}a ,\qquad 
   R^x{}_x = \frac{\ddot a}a + 2\frac{\dot a^2}{a^2},\qquad 
   R = R^\tau{}_\tau + 3R^x{}_x = 6\left(\frac{\ddot a}a +\frac{\dot a^2}{a^2}\right).
\label{e:ricci_sflat} \end{align}\esube
 
 Clusters of galaxies (including dark matter) move with the Hubble flow: with velocity $\bm u = \bm\pa_\tau$ orthogonal to the flat slices.  We approximate the large-scale average of the stress-energy tensor as a perfect fluid. Then $(u^\mu) = (1,0,0,0)$, and the two independent field equations are  
 \bsube\begin{alignat}{2} 
   G_{\tau\tau} &=8\pi T_{\tau\tau}:\cblue \hspace{3cm}&& \cblue
	3\frac{\dot a^2}{a^2} = 8\pi\rho \\
   R^\tau{}_\tau &= 8\pi (T^\tau{}_\tau- \frac12 T):
&&\cblue 3\frac{\ddot a}a  = -4\pi(\rho+3 P) .
\end{alignat}\esube

At times when dark energy is negligible and 
matter dominates, we can model the matter as dust: \\

\noindent {\cblue $P=0$ } 
\bsube\begin{align}
	3\frac{\dot a^2}{a^2} &= 8\pi\rho
\label{e:P=0a}\\
	3\frac{\ddot a}a &= -4\pi \rho.
\label{e:P=0b}\end{align}\esube
Conservation of mass for dust, 
$u_\a \na_\b T^{\a\b} = -\na_\b(\rho u^\b) = 0$  implies 
\index{conservation laws!mass}
\be
\rho a^3 = \mbox{ constant.}
\label{e:dustcons}\ee
Because of the contracted Bianchi identity, these three 
equations are redundant: Eq.~\eqref{e:dustcons} together with a $\tau$ derivative of 
($a^3\times $~\eqref{e:P=0a}) implies \eqref{e:P=0b}.\index{Bianchi identities!and independent field equations}  

Write the conserved mass inside a ball of radius $a$ as 
\be
 	\wt M := \frac43\pi \rho a^3 .
\ee
Eq.~\eqref{e:P=0a} then has the form
\[
\dot a^2 = \frac{2\wt M}a,
\]
with the expanding solution
\be
\cblue a(\tau) \cblue=   \left(\frac{9\wt M}2\right)^{1/3} \tau^{2/3} \cb.
\ee

The corresponding contracting solution is equivalent to the interior of a 
collapsing ball of dust with $E=1$ ($e_N=0$): The geometries and the dust 
trajectories are the same. For the ball of dust, we found  \\
\[
r(\tau) = \left(\frac92M\right)^{1/3} 
			(\tau_0-\tau)^{2/3}.
\]
To make the identification, first change collapsing to expanding ($-\tau\rightarrow \tau$) 
and take $\tau_0 = 0$.\\
Next choose spheres of the same area about $r=0$ and about $x=y=z=0$
to relate $r$ to $a$, 
\[
[\,r(\tau)\,]^2 = [\,a(\tau)\,]^2(x^2+y^2+z^2).
\] 
Finally, pick some value of $x^2+y^2+z^2$ --we'll arbitrarily pick 
$x^2+y^2+z^2=1$, to identify $\dis\wt M = \frac43\pi a^3\rho$ with $\dis M=\frac43\pi r^3\rho$. The 
interior of a uniform ball of collapsing dust is thus identical in its geometry and 
dust trajectories to the interior of a sphere of the same area in the cosmological solution.  \\

\subsection{Nonzero spatial curvature}\vspace{3mm}

\noindent{\sl\crv Positive spatial curvature}\\
\index{cosmology!positive spatial curvature|textbf}

We have already computed the intrinsic and extrinsic contributions 
to the Ricci tensor $R_{\a\b}$ of the metric 
\[
 ds^2 = -d\tau^2 + a^2(\tau) (d\chi^2 + \sin^2\chi d\Omega^2).  
\]
The intrinsic Ricci tensor is the Ricci tensor 
 \eqref{e:ricci_3sphere} of the 3-sphere of radius $a$.  The 
contribution to $R_{\a\b}$ from the extrinsic curvature is 
unchanged from the spatially flat case: It is just the Ricci 
tensor \eqref{e:ricci_sflat} of the spatially flat cosmology.  

To see that the Ricci tensor splits in this way, first note the split of 
$\log\sqrt{-g}$ into a term involving spatial coordinates and a 
term involving $\tau$:   
\[
	\log\sqrt{-g} = \log(\sin^2\chi\sin\theta) + 3\log a.
\]
Next, a quick check shows that the only nonzero $\Gamma$'s are 
the 3-sphere $\Gamma^i{}_{jk}$, from spatial derivatives of the metric, 
and the $\Gamma^\tau{}_{ij}, \Gamma^i{}_{j\tau}$ of the spatially flat 
cosmology, given by Eq.~\eqref{e:Gamma_sflat}.  Then, in 
Eq.~\eqref{e:Ricci_compute} for $R_{\mu\nu}$, the terms coming from 
the {\cred intrinsic curvature} (spatial derivatives of the 3-metric) and from the {\cblue extrinsic curvature} (derivatives of $a(\tau)$) separate, as claimed, and we have 
\begin{align} 
	R_{\a\b} &= \mbox{\cred Ricci tensor \eqref{e:ricci_3sphere}}
		    +  \mbox{\cblue Ricci tensor \eqref{e:ricci_sflat}} \nonumber\\
	R_{\chi\chi}&= {\cblue a\ddot a + 2\dot a^2} \cred+ 2\nonumber\\  
R^\chi_{\ \chi} &={\cblue\frac{\ddot a}a +2\frac{\dot a^2}{a^2}} + {\cred\frac 2{a^2}} =R^{\,\theta}_{\ \theta} = R^{\,\phi}_{\ \phi} \nonumber\\
R^\tau{}_\tau &= \cblue 3\frac{\ddot a}a \qquad 
	     R = 6\frac{\ddot a}{a} + 6\frac{\dot a^2}{a^2} + {\cred\frac 6{a^2}}
\label{e:Rmdom}\\
G_{\tau\tau} &= R_{\tau\tau} + \frac12 R ={\cblue 3\frac{\dot a^2}{a^2} }\cred +\frac3{a^2} 
\end{align} 

The field equations are now
\bsube\begin{alignat}{2}
  R^\tau{}_\tau &= 8\pi (T^\tau{}_\tau - \frac12 T):\qquad &&3\frac{\ddot a}a = -4\pi(\rho+3P) \\
  G_{\tau\tau}&= 8\pi T_{\tau\tau}: &&
  3\frac{\dot a^2}{a^2} +\frac3{a^2} = 8\pi\rho. 
\label{e:Gtautau}\end{alignat}\esube
They give the cycloidal behavior of the surface of a collapsing ball of dust 
with $E < 1\qquad (e_N <0$). \\

\noindent{\cblue $P=0$} \vspace{-10mm} 

\begin{align*}
\frac{\dot a^2}{a^2} +\frac1{a^2} &= \frac{8\pi}3 \rho = \frac{2\wt M}{a^3}, \qquad \wt M := \frac43\pi a^3 \rho\\
a(\eta) &= \wt M(1-\cos\eta)\\
\tau(\eta) &= \wt M(\eta -\sin\eta) 
\end{align*} 
Again identify the solution with that of collapsing dust by reversing time, 
$\eta\rightarrow \pi-\eta$, and identifying spheres of of the 
same area. 
 \newpage
  
\noindent{\sl \crv Negative spatial curvature}.\\
\index{cosmology!negative spatial curvature|textbf}

We have looked at the negative curvature analog of the unit 3-sphere 
in \ref{ex:hyperboloid}.  The analog of the 3-sphere of radius $a$ 
is the hyperboloid at proper time $a$ from the origin of Minkowski 
space, the spacelike 3-surface given by 
\[
   -t^2 + x^2+y^2 + z^2 = -a^2\ .
\] 
The Minkowski metric induces on the hyperboloid the positive-definite 
homogeneous, isotropic metric
\[
   ds^2 = a^2[\,d\chi^2 + \sinh^2\chi d\Omega^2\,], 
\] 
where \vspace{-15mm}

\begin{align*}
  x&=a\sinh\chi \sin\theta\cos\phi\qquad y = a\sinh\chi\sin\theta\sin\phi \\
  z&= a\sinh\chi\cos\theta\qquad \qquad t = a\cosh\chi\ .
\end{align*}
The only difference in $R_{\a\b}$ from that of the 3-sphere 
comes from changing $\sin\chi$ to $\sinh\chi$ in the metric,  
which gives a sign change in the spatial Ricci tensor:  
\begin{align*}
{}^3\!R_{\chi\chi} &= - \pa_\chi^2\log \sinh^2\chi - 2\coth^2\chi = -2\\
{}^3\!R^a{}_b &= -\frac2{a^2} \delta^a_b \qquad {}^3\!R = -\frac6{a^2}.   
\end{align*} 

 For the Ricci tensor of the corresponding spacetime metric, 
only the sign of intrinsic curvature changes.  The extrinsic 
curvature is that of the flat and positively curved cases, 
and the field equations are (with the sign change in red)\vspace{-3mm}
\bsube\begin{align}
  R^\tau{}_\tau &= 8\pi (T^\tau{}_\tau - \frac12 T):\qquad 3\frac{\ddot a}a = -4\pi(\rho+3P) \\
  G_{\tau\tau}&= 8\pi T_{\tau\tau}: \qquad\qquad
  3\frac{\dot a^2}{a^2} {\cred -\frac3{a^2}} = 8\pi\rho \ .
\label{e:Gtautau-}\end{align}\esube
{\cblue $P=0$} \vspace{-12mm} 

\begin{align*}
\frac{\dot a^2}{a^2} {\cred-\frac1{a^2} }&= \frac{8\pi}3 \rho = \frac{2\wt M}{a^3}\\
a(\eta) &= \wt M(\cosh\eta-1)\\
\tau(\eta) &= \wt M(\sinh\eta -\eta) 
\end{align*} 

These equations describe a universe that expands forever.  For the corresponding ball of dust, the velocity of each dust particle is greater than its escape velocity.
\newpage

\noindent
{\sl\crv Summary} \\
\label{p:matter_dominated}

Write \vspace{-3mm}
\be
	k:=0,\pm 1 \mbox{, \hspace{5mm} for flat space, positive curvature, 
			and negative curvature, respectively.}
\ee
The field equations for a homogeneous isotropic cosmology then have the form 
\bsube\begin{align}
\frac{\ddot a}a & = -\frac43\pi(\rho+3P) \\
\frac{\dot a^2}{a^2} +\frac k{a^2} &= \frac{8\pi}3 \rho.
\end{align}\esube

\noindent
$\cblue P=0$ \ (matter dominated cosmology -- the Friedmann solutions): \\
\index{cosmology!matter dominated|textbf}\index{matter-dominated universe}\index{Friedmann solutions}\index{FLRW solutions}\index{cosmology!Friedmann solutions}

$\dis\quad \wt M:=\frac43\pi a^3\rho$.  To have a common set of spherical coordinates for all values of $k$, we will write the 
flat spatial metric as $dx^2+dy^2+dz^2 = d\chi^2 + \chi^2 d\Omega^2$, where $\chi^2 = x^2+y^2+z^2$.  
  
\noindent Spatially flat, $k=0$:\vspace{-3mm}
\begin{align}
ds^2&= - d\tau^2 + a^2(\tau)(d\chi^2 + \chi^2 d\Omega^2) \nonumber \\
\cblue a(\tau) &\cblue =   \left(\frac{9\wt M}2\right)^{1/3} \tau^{2/3}\cb \ .
\label{e:mflat}\end{align}

\<Positive curvature, $k=1$:\vspace{-3mm}
\begin{align}
ds^2 &= -d\tau^2 
	+a^2[d\chi^2 +\sin^2\chi (d\theta^2 + \sin^2\!\theta\, d\phi^2)\ ]\nonumber\\
\cblue a(\eta) &\cblue = \wt M(1-\cos\eta)\quad \tau(\eta) = \wt M(\eta -\sin\eta)\cb  \ .
\label{e:m+}\end{align} 
Negative curvature, $k=-1$:\vspace{-3mm}
\begin{align}
ds^2 &= - d\tau^2 + a^2 [ d\chi^2 + \sinh^2\chi(d\theta^2 + \sin^2\theta d\phi^2)\ ]
\nonumber\\
 \cblue a(\eta) &\cblue = \wt M(\cosh\eta-1)\quad
 \tau(\eta) = \wt M(\sinh\eta -\eta) \cb \ .
\label{e:m-}\end{align}
Because the present universe is nearly spatially flat, Eq.~\eqref{e:mflat} is 
a good approximation for the matter dominated epoch;  it is the small-$\eta$ 
approximation of $a(\tau)$ in Eqs.\eqref{e:m+} and \eqref{e:m-}.  

For a spatially flat cosmology, multiplying the scale factor by a constant, $C$,
gives the same metric components as changing the coordinates from $(x,y,z)$ to 
$(\bar x,\bar y, \bar z)$ with $x=C \bar x,\, y= C\bar y,\, z= C\bar z$. By arbitrarily 
choosing a present value $a_0$ of the scale factor, say $a_0 = 1$ Mpc, one 
is assigning a 1~Mpc proper distance to a coordinate distance $\Delta x = 1$.  
The value of $a(\tau)$ at any other time is then the distance in Mpc associated 
with unit coordinate distance. With this assignment of $a_0$, two objects 
that are at present 1 Mpc apart and follow the Hubble flow are separated by a distance 
$a(\tau)$ in Mpc at any other time $\tau$. 

When the spatial curvature is nonzero, $a_0$ is not arbitrary 
because $a$ is related to the scalar curvature by $^3\!R = 6k/a^2$.  
But the present universe is too close to flat to measure $^3\!R$ or 
know its sign (given current uncertainty in the average density), 
so in practice one can set $k=0$, pick a unit of length, and take 
$a_0=1$. Equivalently, one simply uses the variable 
$\wt a: a/a_0$.
\\ 

The matter-dominated solutions with positive curvature recontract, 
the corresponding finite ball 
of dust expanding with less than the escape velocity. Solutions with negative 
curvature expand forever, and the corresponding dust ball expands faster  
than its escape velocity.  \vspace{-3mm}

\subsection{Solutions with radiation and a cosmological constant}  

\noindent{\sl\crv Radiation: photons, gravitons, and highly relativistic massive particles}  \vspace{3mm}
\index{cosmology!radiation dominated|textbf}

The pressure of gas of photons or highly relativistic particles satisfies 
$
  P = \frac13\rho  
$.
This follows from the general expression for the pressure of an ideal gas of particles with number density $n$, momentum $p$ and speed $v$, %
\footnote{The force needed to keep the matter on one side of a cut of area $A$ 
is the force on an area $A$ of a box that holds the gas.  Particles with $v_x>0$ 
hit the right wall of the box in a time $t$ if they are within distance $v_x t$ of the wall, hence within a volume $A v_x t$.  When they bounce off the wall, they transfer 
momentum $2p_x v_x$.  Because half the particles in the volume have $v_x>0$, 
the total momentum transferred is 
$  nAt\langle v_x p_x \rangle$.
Using $\langle v_x p_x \rangle = \frac13\langle v_x p_x+v_y p_y+v_zp_z \rangle 
= \frac13 \langle pv\rangle$ gives  
$
  P = \frac13 n\langle pv\rangle$.   } 
\be
  P = \frac13 n p v,
\ee
using $v=1$, $E=p$, $\rho = nE$.  (Alternatively, use the relation 
$T^\a{}_\a = -\rho + 3P$, together with the fact that the stress-energy 
tensor \eqref{e:Tem} of the electromagnetic field is tracefree.) 

The homogeneity and isotropy of the CMB defines both the preferred spacelike slices and 
the corresponding normal trajectories of observers that move with the Hubble flow.  
To these observers, the gas of photons is at rest. So, like the clusters of galaxies 
but with much greater accuracy, the CMB is a fluid that moves with the Hubble flow,  

 As the universe expands, 
the number of photons in a comoving volume $V$ stays constant, once the universe is 
sparse enough that the photons don't interact with matter.  The energy 
density $\rho_r$ of radiation is given by 
\index{energy density!radiation}
\be
   \rho_r = \frac NV\,\hbar\,\langle\omega\rangle 
\label{e:rho_r}\ee 
with $\langle\omega\rangle$ the average photon frequency and $N$ the number of 
photons in the volume $V$.  We show in Sect.~\ref{s:ckv-redshift} below that 
light traveling across the universe is redshifted, its wavelength increasing 
from $\lambda$ to $\wh\lambda$ in tandem with the the size of the universe: 
\be
   \wh\lambda = r\,\lambda , \mbox{ where $r$ is the ratio } r=\frac{\wh a}{a}.
\ee
A comoving volume changes from $V$ to $\wh V = r^3 V$, while the frequency of 
each photon changes from $\omega$ to $\wh \omega = \omega/r$.  
Eq.~\eqref{e:rho_r}, with $\wh N = N$, then implies that the density of radiation 
is proportional to $a^{-4}$: 
\be\cblue
   \rho_r a^4 = \mbox{ constant }.  \cb
\ee
\vspace{-6mm}

One additional comment before we use this relation to find solutions to the field 
equations for radiation:\index{blackbody radiation}\index{cosmic microwave background}   
The initial blackbody spectrum of the CMB is the result of thermal equilibrium 
of the radiation and matter prior to the decoupling of photons from baryonic matter 
as the universe expands and cools.  Implicit in many discussions of the CMB is 
that, as the universe expands, the photon gas retains a blackbody spectrum, with its temperature proportional to $1/a$.  
To see that this is true, recall that the 
blackbody spectrum has the following form for the energy $dE$ of photons in the frequency range $d\omega$ and in a box of volume $V$ (see, e.g.,
\href{https://www.feynmanlectures.caltech.edu/I_41.html}{Feynman v. I, Chap. 41}\cite{feynmanI}) 
\be
  d E(\omega) = V\,\frac{\hbar\omega^3 d\omega}{\pi^2 c^3} \frac1{e^{\hbar\omega/kT}-1} ,    
\ee
with corresponding number $dN$ of photons in this frequency range given by 
\be
   dN = \frac{dE}{\hbar\omega} = V\,\frac{\omega^2 d\omega}{\pi^2 c^3}\frac1{e^{\hbar\omega/kT}-1}.
\ee
At a later time, the same number of photons occupy a box of size $\wh V$. 
Using $V=\wh V/r^3$, $\omega = r\wh\omega$, we have 
\begin{align} 
   d\wh N = dN & = V\, \frac{\omega^2 d\omega}
   			 {\pi^2 c^3}\frac1{e^{\hbar\omega/kT}-1}  \nonumber \\
   &= \frac{\wh V}{\color{gray}r^3} \frac{{\color{gray}r^3}\wh\omega^2 d\wh\omega}
			{\pi^2 c^3} \frac1{e^{r\hbar\wh\omega/kT}-1} , 
\end{align}
a blackbody distribution at temperature 
\be
   \wh T = \frac Tr = T\frac{a}{\wh a}.
\ee

\noindent{\sl\crv Radiation dominated solutions} \\

Starting from 
$\dis \frac{\dot a^2}{a^2} +\frac k{a^2} = \frac{8\pi}3 \rho$, we easily 
obtain $a(\tau)$ for a radiation dominated universe with $k=\pm 1$.
 With $\dis C:= \frac{8\pi}3\rho a^4$, \vspace{-6mm}
\bsube \begin{alignat}{3}
 \cblue a &=\cblue  \sqrt C \left[1-\left(1-\frac \tau{\sqrt C}\right)^2\right]^{1/2}, \qquad &k&= 1\cb \\
 \cblue a& \cblue = (4C)^{1/4} \tau^{1/2}\ , \qquad &k&=0 \cb\\
 \cblue a &\cblue = \sqrt C \left[\left(1+\frac \tau{\sqrt C}\right)^2 -1\right]^{1/2},  & k&= -1
 \end{alignat}\label{e:arad} \esube

\noindent Check of solutions~\eqref{e:arad} for $k=1$ and $k=0$.\\

\noindent$k=1$:  \vspace{-5mm}
\[
\frac{\dot a^2}{a^2} +\frac k{a^2}  = \frac{8\pi}3 \rho = \frac C{a^4}.
\]
Writing $u=a^2$ gives
\begin{align*}
\dot u^2 + 4u &= 4 C \\
\int_0^u \frac{du}{\sqrt{C-u}}&= 2\tau \ \Longrightarrow \ 
-2\sqrt{C-u} +2\sqrt C = 2\tau\\
a^2 =u &= C - (\sqrt C-\tau)^2 \\
a &= \sqrt C \left[1-\left(1-\frac\tau{\sqrt C}\right)^2\right]^{1/2} .
\end{align*}

\noindent $k=0$: \vspace{-10mm} 
\begin{align*}
\frac{\dot a^2}{a^2} &= \frac C{a^4} \Longrightarrow a\dot a = \sqrt C \\
\frac12 a^2 &= \sqrt C\ \tau \Longrightarrow a=(4C)^{1/4} \sqrt\tau.
\end{align*}

\noindent Note: All of the radiative solutions \eqref{e:arad} 
have the same small-$\tau$ behavior
\be
	\cblue a=(4C)^{1/4} \tau^{1/2}, \quad\mbox{all } k,
		 \mbox{with }\tau\ll \sqrt C.\cb
\label{e:rflat}\ee
This is the relevant solution during the radiation-dominated epoch of the 
early universe.\\

\noindent{\sl\crv $\Lambda$ dominated solutions} \\
\index{cosmology!$\Lambda$-dominated|textbf}\index{cosmology!cosmological constant}\index{inflationary cosmology}

In the standard $\Lambda$CDM cosmology, a period of rapid inflation, mimicking 
a cosmological constant $\Lambda >0$, precedes the radiation-dominated epoch.  
It's easier to discuss the way this model accounts for successive epochs of 
our universe after we have the solution with nonzero $\Lambda$ in hand, so that 
discussion is relegated to Sect. ~\ref{s:parameters}.  

The great surprise in cosmology of the last quarter century was the discovery 
that a nonzero $\Lambda$ is not confined to the very early universe:  The  expansion of the universe is accelerating 
\href{https://arxiv.org/pdf/astro-ph/9805201.pdf}{Riess et al.'98}, 
\href{https://arxiv.org/pdf/astro-ph/9812133.pdf}{Perlmutter et al.99}.
\cite{riess98,perlmutter99}
The acceleration, apparent over the last 4 billion years, matches that of a 
cosmological constant, $\Lambda$ \eqref{e:Lambda}, with 
\be
   G_{\a\b} = 8\pi T_{\a\b} - \Lambda g_{\a\b}.
\label{e:GLambda}\ee 
Now this is exactly the field equation that would result from a stress-energy 
tensor $T_{\a\b} + T_{\Lambda\,\a\b}$, where 
\[
   T_{\Lambda\,\a\b} = -\frac1{8\pi}\Lambda g_{\a\b}.  
\]
From the relation, 
\[
	-\Lambda g_{\a\b} = \Lambda u_\a u_\b - \Lambda q_{\a\b},
\]
we have 
\be
T_{\Lambda\,\a\b} = \rho_\Lambda u_\a u_\b + P q_{\a\b}, \ \ \mbox{ where } \ \ 
   \rho_{\Lambda} = \frac1{8\pi}\Lambda,\quad P_\Lambda = - \frac1{8\pi}\Lambda.
\label{e:TLambda}\ee
In other words, a cosmological constant is equivalent to a homogeneous, 
isotropic fluid with stress-energy tensor, density and and pressure given by 
Eq.\eqref{e:TLambda}.  

\index{stress-energy tensor!of dark energy} 
$\rho_\Lambda$ is called {\sl dark energy} or {\sl vacuum energy}.  The latter 
term reflects the fact that a scalar field $\Phi$ in its ground state 
has a stress-energy tensor of exactly this form:  That is, if $\Phi$ has a 
nonzero value for the ground state of a potential $V(\Phi)$, the 
vacuum expectation value of its stress-energy tensor mimics the contribution 
to the field equation of a cosmological constant.  The discovery of 
a Higgs particle confirmed the existence of at least one fundamental scalar field. 
In the the standard model of particle physics, couplings to the stationary 
vacuum expectation values of these fields give mass to the other particles. 
\\

In $\Lambda$CDM cosmology, $\rho_\Lambda$ is about 70\% of the total 
density of the present universe, from observations of the universe's acceleration 
using type IA supernovae as standard candles, together with observations of the 
cosmic microwave background.  In the very early universe, its assumed 
value was much larger, leading to a de Sitter geometry, a homogeneous, 
isotropic solution to $G_{\a\b} = - \Lambda g_{\a\b}$. 

The pure $\Lambda$ solutions satisfy  
\be
\frac{\dot a^2}{a^2} + \frac k{a^2} = \frac13\Lambda , \quad k=0,\pm 1 \,.
\label{e:ELambda}\ee

\<{\sl\crv de Sitter space: $\Lambda >0$}.
\index{cosmology!de Sitter space|textbf}\index{de Sitter space}
\index{metric!deSitter}

The solutions are easy to find.\\
 
\noindent $\cblue k=0$
\begin{align}
\frac{\dot a^2}{a^2} &= \frac\Lambda3\ \Longrightarrow \ 
\frac d{d\tau}(\log a) = \sqrt{\frac\Lambda3} \nonumber\\
\cblue a(\tau)&\cblue = \alpha\, e^{\sqrt{\frac\Lambda3}\ \tau}\cb,\quad
	\alpha\mbox{ an arbitrary constant} \\
  ds^2 &  = -d\tau^2 + \alpha^2 e^{\sqrt{\frac43\Lambda}\ \tau}\ (dx^2+dy^2+dz^2).
\nonumber\end{align}
$\cblue k=1$ (\href{https://dwc.knaw.nl/DL/publications/PU00012455.pdf}{de Sitter}\cite{desitter17} found this first.) 
\begin{align}
\frac{\dot a^2}{a^2} +\frac1{a^2} &= \frac13\Lambda\nonumber\\
\cblue a(\tau) 
	&\cblue = \sqrt{\frac3\Lambda} \cosh\left(\sqrt{\frac\Lambda3} \ \tau\right)\\
ds^2 & = -d\tau^2 + a^2 \left[d\chi^2 + \sin^2\chi d\Omega^2\right].\nonumber
\end{align}
Notice that the scale factor $a$ has a minimum value in this solution.\\

\noindent Check: 
For $a = \sqrt{\frac3\Lambda}\cosh\left(\sqrt{\frac\Lambda3}\tau\right)$,  
\begin{align*}
\frac{\dot a}a &= \sqrt{\frac\Lambda3} \tanh \sqrt{\frac\Lambda3}\tau\\
\frac{\dot a^2}{a^2} +\frac1{a^2} 
	&= \frac\Lambda3\left( \tanh^2\sqrt{\frac\Lambda3}\tau 
		+ \sech^2\sqrt{\frac\Lambda3}\tau\right)  \\
&= \frac\Lambda3 \quad \Box
\end{align*}

$\cblue k=-1$
\begin{align}
\frac{\dot a^2}{a^2} -\frac1{a^2} &= \frac13\Lambda	\nonumber\\
\cblue a(\tau) 
	&\cblue = \sqrt{\frac3\Lambda} \sinh\left(\sqrt{\frac\Lambda3} \ \tau\right)\\
 ds^2 &  = -d\tau^2 + a^2 \left(d\chi^2 + \sinh^2\chi d\Omega^2\right) \nonumber
\end{align}

\< For completeness:\\
\<{\crv $\Lambda < 0$ anti-de Sitter} (also found by \href{https://dwc.knaw.nl/DL/publications/PU00012216.pdf}{de Sitter},\cite{desitter18} and published the next year). \\
\index{cosmology!anti-de Sitter space}\index{anti-de Sitter space}
\index{metric!anti-deSitter}

This is a spatially infinite universe that recontracts.\\
$\rho = \rho_\Lambda < 0$.  The field equation \eqref{e:ELambda} 
then implies $k=-1$.  
\begin{align*}
  a &= \sqrt{\frac3{|\Lambda|}} \sin\left(\sqrt{\frac{|\Lambda|}3}\ \tau\right)\\
  ds^2 &= -d\tau^2 + a^2\left( d\chi^2 + \sinh^2\chi d\Omega^2\right)\ .
\end{align*}

The way we have written the de Sitter solutions uses a slicing in which the universe 
is expanding or contracting. It hides a key feature:\\
  \centerline{\cblue The entire 4-dimensional de Sitter spacetime is invariant under the 
5-dimensional Lorentz group.}
The geometry at each point of the spacetime is identical to the geometry at every other point!  This maximal symmetry is implied by the Lorentz invariance of  Eq.~\eqref{e:deSsymmetric} below. 

\benr
\index{hyperbolic space!4-dimensional}
\item Consider the unit {\sl timelike} hyperboloid in 5-dimensional Minkowski space: 
\vspace{-3mm}
\be
  \eta_{\a\b}X^\a X^\b =  -T^2 + W^2 + X^2 + Y^2 + Z^2 = 1.  
\label{e:deSsymmetric}\ee
Show that the metric on the hyperboloid is given by 
\vspace{-3mm}
\[
   ds^2 = -d\tau^2 + \alpha^2 \cosh^2(\tau/\alpha) \left(d\chi^2 + \sin^2\chi d\Omega^2\right),
\] 
agreeing with the de Sitter metric for $\a^2 = 3/\Lambda$.    
where \vspace{-3mm}
\begin{align*}
  T &= \alpha\sinh(\tau/\alpha) \qquad W = \alpha\cosh(\tau/\a)\cos\chi\\
  X&=\a\cosh(\tau/\a) \sin\chi\sin\theta\cos\phi\qquad Y =\a \cosh(\tau/\a) \sin\chi\sin\theta\sin\phi \\
  Z&=\a \cosh(\tau/\a) \sin\chi\cos\theta.
\end{align*}
\een
\begin{figure}[H]
               \begin{center}
		\includegraphics[width=0.3\textwidth]{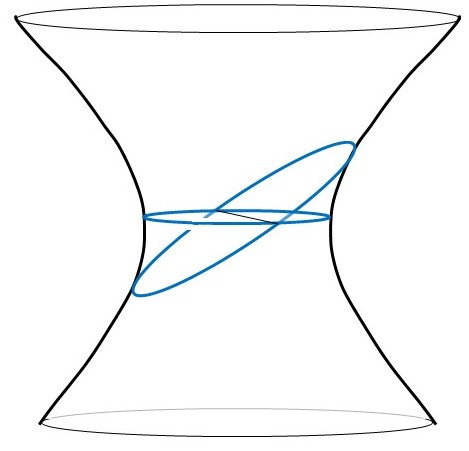}
		\end{center}
		\caption{The de Sitter spacetime as a submanifold of Minkowski space, shown here with one dimension suppressed. Boosting the horizontal blue circle gives the tilted blue circle with the same radius, each corresponding to a 2-sphere of minimal radius. An isometry (boost composed with rotation) maps any point of de Sitter space to any other point.}
\end{figure}
\index{metric!cosmological|)}

\index{collapse, gravitational!cosmological collapse}\index{gravitational collapse!cosmological collapse} \index{cosmology!recollapse}
\subsection{\texorpdfstring{$\bm H,\,\bm\Omega$}\,, recollapse or exponential expansion} 
\label{s:parameters}

In the matter dominated solutions of Eqs.~\eqref{e:mflat}-\eqref{e:m-}, we found that the 
criterion for recollapse was equivalent to requiring that the velocity of 
particles in the corresponding finite ball of dust was less than the escape 
velocity,
\be
   \dot a^2 < \frac{2 \wt M}{a} = \frac{8\pi}3 \rho a^2.  
\label{e:crit0}\ee 
\index{Hubble parameter $H$|textbf}\index{cosmology!Hubble parameter (Hubble constant)|textbf} 
The {\cblue Hubble parameter} (or ``Hubble constant'') $H$ is the fractional rate of expansion, 
\[
  \cblue H := \frac{\dot a}{a}. 
\] x
Written in terms of $H$, the criterion \eqref{e:crit0} is
\be
   H^2 < \frac{8\pi}3 \rho.  
\label{e:ineq}\ee
For the $P=0$ models, the inequality is satisfied when the spatial scalar 
curvature is positive, $^3\!R > 0$.  
For a universe with nonzero $P$ and $\Lambda$, we will see that 
the inequality \eqref{e:ineq} again governs the sign of the scalar curvature
(as is the case in the radiation-dominated and $\Lambda$-dominated solutions 
above). But for nonzero $\Lambda$, the link to collapse is gone:  
A positive constant value of $\Lambda$ prevents recollapse.  

The \mbox{$G_{\tau\tau}=8\pi T_{\tau\tau}$} equation has the form 
\be
  \frac{\dot a^2}{a^2} +\frac k{a^2} = \frac{8\pi}3 \rho,  
\label{e:friedman}\ee
independent of $P$, where the density $\rho$ is a sum
\be
  \rho = \rho_m + \rho_r + \rho_\Lambda
\ee
of contributions $\rho_m$ from pressure-free matter, $\rho_r$ from radiation 
(including relativistic particles) and from $\Lambda$.   
A { \cblue density parameter $\Omega$ } is defined as the ratio 
\index{density parameter, $\Omega$|textbf}\index{cosmology!density parameter, $\Omega$|textbf}
\be
  \cblue \Omega := \frac{8\pi}3 \frac\rho{H^2},  
\label{e:Omegadef}\ee
\< Critical values of $\rho$ and $\Omega$: \ \ $\cblue\Omega_c=1$, 
\index{critical density|textbf}\index{cosmology!critical density|textbf}
$\cblue\dis\rho_c=\frac3{8\pi}H^2$. 
Dividing Eq.~\eqref{e:friedman} by $H^2$ gives 
\be
	\Omega - 1 = \frac k{H^2 a^2}.  
\label{e:Omega-1}\ee
Then 
\begin{alignat*}{3} 
\Omega &>1 \Longrightarrow k = 1&\quad&\mbox{positive spatial curvature}\\
\Omega&=1 \Longrightarrow k =0& \quad&\mbox{zero spatial curvature}\\
\Omega&<1 \Longrightarrow k =-1& &\mbox{negative spatial curvature} .
\end{alignat*}

The present value $H_0$ of $H$ (uncertainties are discussed in Sect.~\ref{s:H0}) is 
\index{Hubble parameter $H$!present value $H_0$} \index{cosmology!$H_0$, present value of Hubble parameter}
\be
 \cblue  H_0 \approx 70\ \rm km/s/Mpc = 2.3\times 10^{-18}\rm s^{-1}\cb, 
\label{e:H0}\ee
implying for the critical density a value
\be
   \rho_c = \frac3{8\pi G} H_0^2 = 9\times 10^{-30} \rm g/cm^3.
\ee

The matter density $\rho_m$ of galaxies and clusters and the density $\rho_r$ of 
radiation decrease as $a$ increases, with $\rho a^2 \rightarrow 0$ as 
$a\rightarrow \infty$.  If $\rho$ behaves in this way, then Eq.~\eqref{e:friedman} in the form, 
\[
   \dot a^2 = -k + \frac{8\pi}3\rho a^2, 
\]  
again links the sign of the curvature to recollapse: 
For $k=1$, $\dot a$ goes through zero at some value $a_{\rm max}$ of the scale 
factor, and the universe recollapses;
for $k=0$ and $k=-1$, $\dot a$ is never zero and the universe expands forever.  

For a constant $\Lambda>0$, however, $\rho a^2$ increases with increasing $a$, 
once the matter and radiation contributions to $\rho$ are small.  In this 
$\Lambda$-dominated epoch $\Lambda$ not only prevents recollapse, it forces  
the expansion of the universe to accelerate, independent of the sign of the 
scalar curvature or the present value of $\Omega$.  \\

The $\Lambda$CDM model, with parameters set by observation 
is also called the {\sl Benchmark Model} or {\sl Standard Model} of cosmology;
\index{cosmology!benchmark model, standard model}\index{benchmark cosmological model}
\index{standard cosmological model}
in this model, the present universe is flat ($\Omega=1$), 
and parameters have the values%
\footnote{The value of each $\Omega$ depends on the present value of the 
Hubble parameter.   
\href{https://www.aanda.org/articles/aa/pdf/2020/09/aa33880-18.pdf}{Table 6 of 
the Planck collaboration's 2020 summary} gives $\Omega_m = \Omega_{m,\rm baryon}+\Omega_{m,\rm dark}$, with 
\begin{alignat*}{3}
&\Omega_{m,\rm baryon} h^2 && =0.02242\pm 0.00014\\
&\Omega_{m,\rm dark} h^2  &&=0.11933\pm0.00091 ,
\end{alignat*}
where $h:=H/(100\rm\ km/s/Mpc)$. The observed spatial flatness implies 
$\Omega_\Lambda = 1-\Omega_m$ with uncertainty of about 1\%.
}
\bsube
\begin{alignat}{5}
&\cblue\Omega_m &&\cblue=0.31 \qquad\quad&&\mbox{cold dark matter and baryonic matter} \\
&\cblue\Omega_{r} &&\cblue= 9.0\times 10^{-5} &&\mbox{photons and neutrinos} \\
&\cblue\Omega_\Lambda &&\cblue= 0.69&&
\end{alignat}\esube \vspace{-6mm}

\<(See, e.g., Ryden's 2017 {\sl Introduction to Cosmology}\cite{ryden17} or 
\href{http://carina.fcaglp.unlp.edu.ar/extragalactica/Bibliografia/Ryden_IntroCosmo.pdf}{Ryden's 2006 version}, or Turner's 
\href{https://arxiv.org/pdf/2201.04741.pdf}{Road to Precision Cosmology}.\cite{turner22})
Gravitational waves also contribute to $\rho_r$, but at a density small 
compared to that of photons and neutrinos in the Benchmark Model. 
The present value of $\Omega = \Omega_m+\Omega_r+\Omega_\Lambda$ is then 
surprisingly close to its critical value:
\be
  |\Omega -1|<0.01,  
\label{e:omega-1}\ee  
implying a universe whose large-scale spatial curvature is nearly zero.  

The evolution of the universe for a mix of pressure-free matter, radiation, and 
vacuum energy is governed by Eq.~\eqref{e:friedman} with 
$\rho = \rho_m + \rho_r+\rho_\Lambda$.     
Using $\rho_m \propto a^{-3}, \ \rho_r \propto a^{-4}$, $\rho_\Lambda=$ constant, 
we have
\[
  \frac{8\pi}3\rho = \Omega_m \left(\frac{a_0}a\right)^3 +\Omega_r \left(\frac{a_0}a\right)^4 + \Omega_\Lambda , 
\] 
where $\Omega_m,\Omega_r$ and $\Omega_\Lambda$ denote present values.  
Setting $\dis\wt a:= \frac a{a_0}$, $\wt \tau = H_0\tau$, we write the field equation in the form 
\be
   \left(\frac{d \wt a}{d\wt \tau}\right)^2 = \frac{\Omega_m}{\wt a} + \frac{\Omega_r}{\wt a^2} 
   			+\Omega_\Lambda \wt a^2,  
\label{e:atau}\ee
an energy-conservation equation with a negative effective potential that 
blows up as $-1/a^2$ for small $a$ in the radiation-dominated epoch and 
and blows up as $-a^2$ for large $a$ in the late $\Lambda$-dominated epoch.

Eq.~\eqref{e:atau} immediately gives  $\wt\tau = \wt\tau(\wt a)$ as the integral 
\be
 \cblue \wt\tau = \int_0^{\wt a} \frac{d\wt a}
  		{\sqrt{\Omega_m \wt a^{-1} +\Omega_r \wt a^{-2}+\Omega_\Lambda \wt a^2}}\cb.
\label{e:taua}\ee 
Without doing the integral, we can easily estimate the time at which 
the universe changed from matter-dominated to $\Lambda$-dominated 
and find the redshift before which it was radiation-dominated.  
\benr \item\phantom{x}

\benalph\item
Estimate $\wt a = a/a_0$ at the time the universe changed from matter-dominated 
to $\Lambda$-dominated; from $\wt a$, estimate how long ago that was.  
\item  
Estimate $\wt a = a/a_0$ at the time the universe changed from radiation-dominated 
to to matter-dominated.  Find the corresponding redshift, 
$z = a_0/a -1$.      
\een 
\een
{\sl Solution}. From Eq.~\eqref{e:H0}, 
\be
	H_0^{-1} = 4.4\times 10^{17}{\rm s} = 1.4\times 10^{10} \rm yr. 
\label{e:1/H0}\ee
a. Because $\Omega_r$ is negligible compared to $\Omega_m$ and 
$\Omega_\Lambda$, the change from matter- to $\Lambda$-dominated occurred when 
\begin{align}
   \frac{\Omega_m}{\wt a} &= \Omega_\Lambda \wt a^2 
   	\Longrightarrow \wt a_{m\mbox{-}\Lambda} 
   	= \left(\frac{\Omega_m}{\Omega_\Lambda}\right)^{1/3} = 0.77.
\label{e:arLambda}\\
   \frac d{d\tau}\ln(\wt a) &= H \approx H_0 \Longrightarrow \ln\frac{a_0}a \approx H_0 \Delta \tau\nonumber \\
   \Delta \tau &\approx \frac{0.27}{H_0} = 3.8\times 10^9 \rm yr.
\end{align} 
b. When radiation and matter had comparable densities, $\rho_\Lambda$ was 
negligible. Then
\begin{align}
   \frac{\Omega_m}{\wt a} &= \frac{\Omega_r}{\wt a^2} \Longrightarrow 
   \wt a_{r\mbox{-}m} = \frac{\Omega_r}{\Omega_m} = 2.9\times 10^{-4}.
\label{e:arm}\\
   z &= \frac{a_0}a-1 = 3.4\times 10^3. \nonumber
\end{align} 
The actual value is close:  $z=3.6\times 10^3$.\\

The value \eqref{e:1/H0} of $H_0^{-1}$ estimates the age of the universe by mechanically 
running $a$ back from its present value $a_0$ to $0$, using the constant 
value $H_0$ for $\dot a/a$:  $\tau_0 \approx a_0/\dot a_0 = H_0^{-1} = 14$ billion years,
remarkably close to the $13.8$ billion years of the Benchmark Model. \\    

A universe dominated at late times by a positive $\Lambda$ expands forever,  
but the real fate of the universe is unclear for many reasons. Here are a few:\\
 
\noindent 
If the value of $\Lambda$ is the vacuum energy density of a Higgs field, 
the potential governing the field may change, giving rise to a new value of 
$\Lambda$ that could be zero or possibly negative.\cite{rajantie18}\vspace{1mm} 
 
\noindent In the causal-set approach to quantum gravity,\index{causal set} 
$\Lambda$ 
emerges as a fundamental quantity that fluctuates and can again have 
any sign; that approach gives a value of $|\Lambda|$ of order the critical density 
at any time, in agreement with the observed present value and with a large 
early value needed for inflation. \cite{sorkin90,sorkin97,adgs04}\vspace{1mm}

\noindent The universe may be vastly larger than the visible universe 
(as is the case in the Benchmark Model of cosmology with rapid early inflation), 
and a universe much larger than the part that has inflated could look 
entirely different from the part of the universe we observe, leading to 
collapse on a timescale that is much longer 
than $1/H_0$ and that is not related to the local value of $\Lambda$.\\ 
\newpage

\section{Conformal time $\eta$, conformal Killing vector, and redshift}\index{cosmology!conformal time}\index{cosmology!cosmological redshift|textbf}\index{redshift!cosmological}
\label{s:ckv-redshift}
We have written the the spatially homogeneous isotropic metrics ~\eqref{e:ghi} in terms 
of the proper time $\tau$ of observers moving with the Hubble flow, 
\be
  ds^2 = -d\tau^2 + a^2 d\ell^2.
\ee
Because the scale factor is a function only of $\tau$, one can define 
a {\sl conformal time}\index{conformal time} $\eta$ by 
\[
\frac{d\eta}{d\tau} = \frac1a, \qquad \eta = \int^\tau \frac{d\tau'}{a(\tau')}. 
\] 
With $\eta$ replacing $\tau$ and the unit 3-metrics denoted by $d\ell^2$, 
the spacetime metrics take the form 
\be
  {\cblue ds^2 = a^2(\eta) (-d\eta^2 + d\ell^2)} ,
\ee 
where the spatial metric $d\ell^2$ is the 
the metric of the unit sphere, unit hyperboloid, or flat space.  

The time coordinate $\eta$ here is the same $\eta$ that appeared in 
our matter-dominated solutions with $k=\pm 1$:
\begin{align*}
ds^2 &= -d\tau^2 
	+a^2[d\chi^2 +\sin^2\chi (d\theta^2 + \sin^2\!\theta\, d\phi^2)\ ]\\
a(\eta) &= \wt M(1-\cos\eta)\\
\tau(\eta) &= \wt M(\eta -\sin\eta) 
\end{align*} 
and 
\begin{align*}
ds^2 &= - d\tau^2 + a^2 [ d\chi^2 + \sinh^2\chi(d\theta^2 + \sin^2\theta d\phi^2)\ ]\\
 a(\eta) &= \wt M(\cosh\eta-1)\\
 \tau(\eta) &= \wt M(\sinh\eta -\eta),
\end{align*} 
In these solutions, $\eta$ already satisfies
\begin{align*}
 d\tau &= a d\eta \\
 ds^2 &= a^2 (-d\eta^2 + d\ell^2).
\end{align*}

\noindent{\bf Definition}: Two metrics $g_{\a\b}$ and $\bar g_{\a\b}$ are {\cblue conformal} to each other if there is a nonzero scalar field $a$ for which 
$g_{\a\b} = a^2\bar g_{\a\b}$. The quantity $a^2$ is called a {\cblue conformal factor}. \\

\<{\sl\crv Conformal Killing vector} \vspace{3mm}
\index{conformal Killing vector|textbf}\index{Killing vector!conformal Killing vector|textbf}

If $\xi^\a$ is a Killing vector of $\bar g_{\a\b}$, then 
\[
\Lie_{\bm\xi} g_{\a\b} = \Lie_{\bm\xi} (a^2) \bar g_{\a\b} = \left(\frac 2a\Lie_{\bm\xi} a\right)\  g_{\a\b}.
\]
A vector field for which $\cblue \Lie_{\bm\xi} g_{\a\b} = f\ g_{\a\b}$ is a conformal Killing vector. \\

\noindent{\sl Claim}:  If $\xi^\a$ is a conformal Killing vector, $k_\a \xi^\a$ is conserved along null geodesics.  \\

\noindent{\sl Proof}: 
\begin{align*}
 k^\b \na_\b(k^\a\xi_\a) &= k^\a k^\b \na_{(\b}\xi_{\a)} = \frac12 k^\a k^\b f\  g_{\a\b} \\
 &= 0.
\end{align*}

\<{\sl\crv Cosmological redshift}\vspace{3mm}

For the metrics 
\[
    ds^2 = a^2(\eta) (-d\eta^2 + d\ell^2),
\]
$\eta^\a = \bm\pa_\eta$ is a conformal Killing vector because the only 
dependence on $\eta$ is in the conformal factor $a^2$.
Then $k_\eta = k_\a \eta^\a$ is conserved.  An observer moving with the 
Hubble flow has velocity \mbox{$\dis u^\a =\bm \pa_\tau = \frac1a\, \eta^\a$}. 
For light traveling from $A$ to $B$, the frequency measured by these observers 
is then 
\begin{align*}
\omega_A &= -k_\a u^\a (A) = -\frac1{a_A} k_\a\eta^\a \\
 \omega_B &= -\frac1{a_B} k_\a\eta^\a\\
 \cblue \frac{\omega_B}{\omega_A} &\cblue = \frac{a_A}{a_B}.
\end{align*}

For galaxies moving with the Hubble flow, this had better agree with 
the usual Doppler shift if the distance $d$ between galaxies is short enough 
that we can work in a local inertial frame and ignore the spacetime curvature.  
In particular, assume that during the light travel time $\tau_B-\tau_A$, 
the change in $d$ is small compared to $d$ -- that the relative velocity is 
nonrelativistic.  

To recover the nonrelativistic Doppler shift in this approximation, 
 we first show the following exact relation.\\ 
{\sl Claim}:  For galaxies whose relative motion tracks the expansion, 
\be
  v=Hd,
\label{e:H}\ee
where $d$ is the proper distance between the galaxies and $v$ 
is their relative velocity, the rate of change of their 
proper distance; here $d$, $v$ and $H$ are all evaluated at 
the same time $\tau$.\\

\noindent{\sl Proof}.  Given two galaxies, we can orient the coordinates 
so they have the same values of $\theta$ and $\phi$ and differ only in 
$\chi$, and we can take one galaxy to be at the origin.  At fixed $\tau$, 
\[
	ds = a\, d\chi, 
\]
for the 3-sphere, flat space, or the 3-hyperboloid.  
Galaxies following the Hubble 
flow stay at the same value of the spatial coordinates $\chi,\theta, \phi$, 
so the proper distance between the galaxies is just 
\[
   d = a\chi.
\]
The relative velocity, the rate of change of proper distance, is 
\[
   v := \frac{d a}{d\tau} \chi = \frac{\dot a}a\ a\chi = H d,
\]
as claimed.    \hspace{5mm}$\Box$
 
For nearby galaxies, the times $\tau_A$ and $\tau_B$ at which a light ray passes 
galaxies $A$ and $B$ are related by $  d\approx \tau_B-\tau_A $.  
Within this short-distance approximation, we have
\begin{align}
  a_B &= a_A + \dot a\ (\tau_B -\tau_A) = a_A + \dot a d\nonumber\\
  1+z &= \frac{a_B}{a_A} = 1 + \frac{\dot a}a d \nonumber\\
\cblue z & \cblue = Hd = v. 
\label{e:hubblelaw}\end{align}

  The Hubble parameter $H$ has dimension $1/T$, and $1/H$ is approximately the 
age of the universe. So the condition that Hubble's law \eqref{e:hubblelaw} hold 
is that the light-travel time between galaxies be short compared to the age of 
the universe or, equivalently, that the distance between galaxies be short 
compared to the size of the visible universe.  For distances of order $1/H$, 
there is no obvious way to identify the cosmological redshift with a 
relativistic Doppler shift.%
\footnote{As \href{https://blog.richmond.edu/physicsbunn/files/2009/03/doppler.pdf}{Bunn and Hogg} point out, you can think of light traveling past a set of observers that move with the Hubble flow, with the distance 
between successive observers small compared to $1/H$. Then the cosmological redshift 
between successive observers can be naturally viewed as a Doppler shift, 
so the long-distance cosmological redshift can in this way be regarded as a 
succession of local Doppler shifts. 
}  \vspace{3mm}

\< {\sl\crv Conformal invariance of the light cones}\index{conformal invariance}\vspace{3mm} 

Conformally related metrics $\bar g_{\a\b}$ and $g_{\a\b}$ have the same light-cone structure:  Whether a vector is null, spacelike, or timelike is unchanged when 
$g_{\a\b}$ is multiplied by a positive scalar $a^2$: 
\[
   g_{\a\b} v^a v^b \begin{cases} >0 \\ = 0 \\ < 0 \end{cases} \Longleftrightarrow 
   a^2 g_{\a\b} v^a v^b \begin{cases} >0 \\ = 0 \\ < 0 \end{cases}.
\]   
A much stronger statement is also true: \\
When there is no source (when $j^\a=0$), Maxwell's equations are 
conformally invariant.  That is, if a covariant tensor $F_{\a\b}$ satisfies 
Maxwell's equations for $g_{\a\b}$, it satisfies them for a conformally 
related metric $\bar g_{\a\b}$.  The short proof is given before
\ref{ex:maxconf}.\\

\< {\sl\crv The horizon problem and arguments for early inflation}\vspace{3mm}  
\index{cosmology!particle (cosmological) horizon|textbf}\index{cosmology!horizon problem}\index{horizon problem in cosmology}
\index{inflationary cosmology}

\index{spacetime diagram}
Conformal invariance means that the light cones of the cosmological metrics 
are the same as the light cones of the static metrics $-d\eta^2 + d\ell^2$.
In a spacetime diagram with coordinates $\eta, \chi$, the light cones 
are at 45$^\circ$, as in a Kruskal diagram.  The diagram shows  
the following difficulty with the observed homogeneity and isotropy of 
the universe, called the {\sl horizon problem}: 
CMB\index{cosmic microwave background} light reaching an observer from two opposite directions last interacted 
with matter at points $P$ and $Q$ whose past light cones are disjoint. 
The boundary of the past light cone of an observer at a point $P$ is 
called the {\sl particle horizon} or {\sl cosmological horizon}; 
it bounds the region from which signals can reach $P$.  
\index{horizon!paricle (cosmological) horizon|textbf}
\index{particle horizon|textbf}\index{cosmological horizon|textbf}
No information from the initial conditions near $a=0$ that determine the 
geometry and matter distribution at $P$ can reach $Q$.  So why is it that 
these unrelated regions have the same temperature?   

\newpage

Here's a more formal treatment.  
The past light cones of $P$ and $Q$ will be disjoint if the coordinate 
distance $2\chi_2$ between $P$ and $Q$ is larger than the maximum coordinate 
diameter $2\chi_1$ of their light cones, the diameter at $a=0$.  

\begin{figure}[H]
               \begin{center}
		\includegraphics[width=0.7\textwidth]{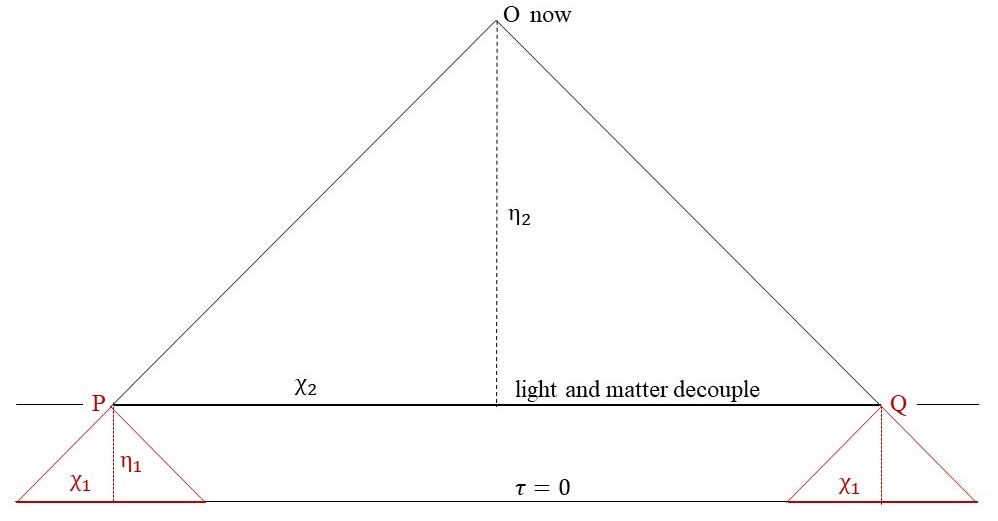}
		\end{center}
		\caption{CMB photons reaching an observer $O$ from opposite directions come from parts of the universe that were never in causal contact if the early universe was radiation dominated. The diagram shows the past light cones of the observer and of points $P$ and $Q$ at the time of decoupling, points at which the photons last interacted with matter before reaching $O$.  }
\end{figure}

\noindent{\sl Claim}: If the early universe is radiation dominated, 
$\dis \frac{\chi_1}{\chi_2} \ll 1$.\\

\noindent{\sl Check}. 
We look at null rays with $d\chi = d\eta$.  The distance $PQ$ 
is the diameter of the past light cone extending from an observer at 
time $\tau_0$ to points $P$ and $Q$ at the time $\tau_d$ of decoupling of 
light from matter.  To make the calculation easy, we underestimate 
the radius $\chi_2$ by restricting the expansion to the time of matter domination, 
starting from the transition from radiation to 
matter dominance (slightly after decoupling) and ending at the scale factor 
$a_{m\mbox{-}\Lambda}$ of the matter-$\Lambda$ transition.  
We'll use Eqs.~\eqref{e:arLambda} and \eqref{e:arm} for the dimensionless 
scale factors $\wt a_{r\mbox{-}m}$ and $\wt a_{m\mbox{-}\Lambda}$ at the two transition times.   
 
To find the coordinate radius $\chi_2$ of the past light cone from a point 
$O$ at the matter-$\Lambda$ transition time, use Eq.~\eqref{e:taua} 
for the matter-dominated epoch prior to the transition:  
\begin{align*}
d\chi &= d\eta = \frac{d\tau}{a} = \frac1{a_0H_0}\frac{d\wt\tau}{\wt a} 
	= \frac1{H_0 a_0}\frac{d\wt a}{\sqrt{\Omega_m\wt a}} \\
\chi_2 &= \int d\chi = \frac1{H_0 a_0\sqrt{\Omega_m}}
	\int_{\wt a_{r\mbox{-}m}}^{\wt a_{m\mbox{-}\Lambda}} 
	  \frac{d\wt a}{\sqrt{\wt a}}
      = \frac{2}{H_0 a_0\sqrt{\Omega_m}}
        ( \sqrt{\wt a_{m\mbox{-}\Lambda}} -\sqrt{\wt a_{r\mbox{-}m}}) \\
     & > \frac2{H_0 a_0}\sqrt{\frac{\wt a_{m\mbox{-}\Lambda}}{\Omega_m}}.
\end{align*}

The maximum coordinate radius $\chi_1$ of the past light cone of a point $P$ at 
decoupling is smaller than the maximum coordinate radius of the past light 
cone from a point at the the radiation-matter transition time: 
For radiation-dominance, Eq.~\eqref{e:taua} implies
\begin{align*}
d\chi &= \frac1{H_0 a_0} \frac{d\wt \tau}{\wt a} 
	= \frac1{H_0 a_0\sqrt{\Omega_r}} d\wt a \\
\chi_1 &< \frac1{H_0 a_0\sqrt{\Omega_r}}\int_0^{\wt a_{r\mbox{\tiny-}m}} d\wt a = \frac{\wt a_{r\mbox{\tiny-}m}}{H_0 a_0\sqrt{\Omega_r}}.\\
\frac{\chi_1}{\chi_2} 
   &<  \frac12\sqrt{\frac{\Omega_m}{\Omega_r}}\ 
   	\frac{\wt a_{r\mbox{-}m}}{\sqrt{\wt a_{m\mbox{-}\Lambda}}}
    = \frac12 \Omega_\Lambda^{1/6}\Omega_r^{1/2}\Omega_m^{-2/3} =0.01 \qquad \Box 
\end{align*}  

Because the past light cones of $P$ and $Q$ are disjoint if the very early 
universe is radiation-dominated, there is no reason why light reaching 
an observer from $P$ and $Q$ should have the same temperature.  
The reader may object to a calculation that extrapolates the the universe back to $a=0$ when the density is infinite and classical general relativity is 
no longer valid. The same problem is present, however, if we look at 
initial conditions near the Planck time, when fluctuations in the metric 
should be large and uncorrelated in the past light cones of $P$ and $Q$.  

Early inflation offers one solution to the mystery.  For a $\Lambda$-dominated 
early universe, the scale factor increases exponentially. Regions 
that are now far apart were close and in causal contact in the early 
universe.  In particular, for an early universe that begins to be 
$\Lambda$-dominated when $\wt a= \wt a_\Lambda$,
Eq.~\eqref{e:atau}, with $\Omega_\Lambda$ replaced by its early-universe value $\Omega_{\Lambda_E}$, gives
\begin{align}
d\chi &= \frac1{H_0 a_0} \frac{d\wt \tau}{\wt a}  
      = \frac1{H_0 a_0\sqrt{\Omega_{\Lambda_E}}}\frac{d\wt a}{\wt a^2}\nonumber\\
 \chi_1 &\approx \frac1{H_0a_0\sqrt{\Omega_{\Lambda_E}}}\ 
		\frac1{\wt a_{\Lambda}},     
\end{align}
where we have assumed that the value of $\wt a$ at the end of inflation is 
large compared to its starting value $\wt a_\Lambda$.  
Then $\chi_1$ is proportional to $1/\wt a_\Lambda$ and is much larger than $\chi_2$.  \\

A second argument for inflation is called the {\sl flatness problem}.
\index{flatness problem}\index{cosmology!flatness problem}  
We have seen in Eq.~\eqref{e:omega-1} that, measured by the difference $|1-\Omega|$, the average density, $\rho= \rho_m+\rho_r+ \rho_\Lambda$, is nearly equal to the critical density $\rho_c$.  That means that the present universe is nearly spatially flat, 
with a radius of spatial curvature much larger than the visible universe. This 
coincidence has a more extreme form in the early universe:  As the universe expands, the departure from critical density increases.  
Unless $\rho$ was excruciatingly close to critical density at early times, 
either $\Omega$ would now be much smaller than its observed value near 1 or the universe would have quickly collapsed, never surviving to its present time. 

To see why $|1-\Omega|$ is much smaller in the early universe, it is helpful to 
understand its meaning as a ratio of spatial (intrinsic) curvature, 
involving spatial derivatives of the metric, to extrinsic curvature, 
involving the time derivative $\dot a/a$.
The radius of spatial curvature is of order $a$, while the radius of extrinsic curvature 
is of order $a/\dot a \sim \tau$, for the power law behavior 
 $a\propto \tau^{2/3}$ and $a\propto \tau^{1/2}$ of the the matter-dominated and  
radiation dominated epochs (Eqs.\eqref{e:mflat} and \eqref{e:rflat}). 
Because the ratio $\tau/a$ increases as the universe expands, the ratio of 
spatial curvature to extrinsic curvature increases during the radiation- and matter-dominated epochs.    
 
Here's a more formal version:  
  From Eqs.~\eqref{e:Gtautau}, \eqref{e:Gtautau-}, and \eqref{e:Kij}, 
the ratio of contributions of spatial curvature and extrinsic curvature to $G_{\tau\tau}$ 
is 
\be 
   \frac12\frac{^3\!R}{K_{\a\b}K^{\a\b}} = \frac{k/a^2} {(\dot a/a)^2} = \frac k{\dot a^2} = \Omega-1. 
\ee
Then 
\[
  \Omega-1  \propto a , \ \mbox{matter-dominated}, \quad \Omega-1  \propto a^2,   \ \mbox{radiation-dominated}. 
\]
At the radiation-matter transition, $\wt a_{r\mbox{-}m} =2.9\times 10^{-4} $, implying 
$\Omega-1 \lesssim 3\times 10^{-6}$, with a vastly smaller value at earlier, radiation-dominated times.   

Early inflation gives an explanation for the extraordinary early-time flatness:  \\
For $a\propto e^{\sqrt\Lambda \tau}$, 
\[
   \Omega-1 = \frac k{\dot a^2} \propto e^{-2\sqrt\Lambda \tau},
\]
decreasing exponentially.  This behavior would, for example, allow spatial and temporal derivatives of the metric to be of the same order at Planck time, 
with a radius of curvature of order $1/\ell_{\rm Planck}^2$, consistent with 
quantum fluctuations of order unity on scales of order $\ell_{\rm Planck}$.

 \section{Using gravitational waves to measure $H_0$}
 \label{s:H0}
 \index{Hubble parameter, Hubble constant!measured by gravitational waves}
 \index{cosmology!gravitational-wave measurement of $H_0$}
 index

\noindent

Late-universe measurements of $H_0$ conflict with early-universe measurements from the cosmic microwave background at $z\approx 1100$. The early-universe measurements 
of the cosmic microwave background by the 
\href{https://iopscience.iop.org/article/10.1088/0067-0049/208/2/20/meta}{WMAP},
and \href{https://arxiv.org/abs/1807.06205}{Planck} satellites give
\be
  H_0 = 69.32\pm 0.80\ \rm km/s/Mpc \mbox{ and } H_0 = 67.66\pm 0.42 \rm km/s/Mpc, 
\ee
respectively.\cite{WMAP13,planck20} Here $H_0$ is not directly measured; its value relies on the $\Lambda$CDM model, whose six parameters fit with high precision the cosmological data.

In contrast, on the astrophysical side, the SH$_0$ES collaboration uses the brightness of type Ia supernovae (SNe Ia) as standard candles and finds the larger value 
\[
  H_0 = 74.03\pm 1.42\ \rm km/s/Mpc.
\]
This measurement rests on the Leavitt relation between period and luminosity of Cepheid variables, now found directly by parallax and by geometrical determination of distance to the Large Magellanic Cloud.  Independent ways to determine $H_0$ include water maser lines in imaged thin disks about galactic black holes, which give a value 73.9$\pm$ 3.0 km/s/Mpc; and measurements of time delays of different images from the same multiply-imaged quasar giving $73.3_{-1.8}^{+1.7}$. The {\sl Hubble tension}, the discrepancy between the early- and late-universe values, is now above 4$\sigma$. \footnote{However, a recent late-universe measurement by the Carnegie-Chicago Hubble Project calibrates Type Ia Sne by the brightness and luminosity of the tip of the red-giant sequence, finding $H_0 = 69.8 \pm 0.8({\rm statistical})\pm 1.7({\rm systematic})$ km/s/Mpc. } \index{Hubble tension}\index{cosmology!Hubble tension}  

The key question is whether the discrepancy indicates new physics beyond the $\Lambda$CDM model or is due to unknown systematic errors.  
In 1986, Schutz\cite{schutz86} pointed out that the binary-inspiral waveform 
could be used to give an independent measure of $H_0$ if one could identify a host galaxy 
with measured redshift.  Remarkably, the first observed inspiral waveform of a 
binary neutron star system gave exactly that measurement.\index{binary system!insprial}\index{inspiral}\index{gravitational waves!binary inspiral}  

The relation is derived as follows.
The distance $D$ to the binary is given in terms of three waveform observables, 
the frequency $\omega$, its first time derivative, $\dot\omega$ and its amplitude $h$.  
At lowest post-Newtonian order, three equations relate these quantities to $D$: 
\begin{enumerate} 
\item[I.] The radiation is quadrupole, and the frequency of quadrupole radiation
is related to the frequency $\Omega$ of the orbit by 
\index{frequency!gravitational wave}\index{gravitational wave!frequency}
\be
 \omega = 2\Omega.
\ee
\item[II.] For a circular orbit, the quadrupole form \eqref{e:hbinary} of 
 $h_+$ and $h_\times$ gives 
$\displaystyle 
   \bar h = \sqrt{\langle h^2 \rangle} \propto\frac1{D}\mu a^2 \Omega^2 ,
$ for the rms amplitude $\bar h$ of the wave, which we can write as  
\be
  \bar h \propto \frac ED.
\label{e:hED}\ee
where $E$ is the Newtonian energy of the binary.
\item[III.]
Finally, the rate $\dot E$ at which the system loses energy to gravitational waves 
is the energy flux at the distance $D$; from Eq.~\eqref{e:dotEdoth},   
\be 
   \dot E \propto \langle\dot h^2\rangle D^2 = \omega^2 \bar h^2 D^2. 
\label{e:Edot}\ee
\end{enumerate}  

In each case, the constant of proportionality depends only on $G$ and $c$: It is 
purely numerical in gravitational units. Then, with $k$ a constant of this form,
\be
  \frac{\dot\omega}\omega = \frac32 \frac{\dot E}E = \frac1k \frac{\omega^2 \bar h^2 D^2}{\bar h D},  
\label{e:dotomega}\ee
where Eqs.~\eqref{e:hED} and \eqref{e:Edot} were used in the last equality 
to write $E$ and $\dot E$, respectively, in terms of $h$ and $D$.  Thus 
\be
   \cblue D = k \frac{\dot\omega}{\omega^3 \bar h}. 
\label{e:D}\ee
The simplicity of this relation comes from that fact that $\dot E/E$ 
is independent of the two masses.   With the constants included and $\omega=2\pi f$, 
the relation has the form 
\be
  \crv D = 780 \frac{\dot f_{100}}{f_{100}^3 \bar h_{23}}\ \rm Mpc, 
\ee
where $f_{100} = f/(100\rm Hz)$, $\bar h_{23} = 10^{23}\bar h$, and $\dot f/f$ is given in 
s$^{-1}$.  \\

\noindent{\sl Cosmological correction} \vspace{2mm}

For a binary system 50 Mpc away, the relative velocity is about 
\[
   v = 70\mbox{ km/s/Mpc} \times 50 \mbox{ Mpc} = 3500 \mbox{\ km/s} 
\]
implying a redshift 
\[
   z \approx \frac{3500}{3\times 10^5} = 0.01,
\] 
too small to significantly alter the value current detectors assign to $H_0$.

For systems far enough away that the redshift is important, we need to clarify 
the distance $D$ of Eq.~\eqref{e:D} that is measured here.  
The equations came from an asymptotically
flat context, with $4\pi D^2$ the area of a distant sphere.  In a cosmological 
context, the amplitude $\bar h$ of the wave is still proportional to 
$D^{-1}$ if we define $D$ by demanding that $4\pi D^2$ be the area of a 
sphere. Spheres of coordinate radius $\chi$ have area 
\be
  A = 4\pi D^2 = 4\pi a^2 \begin{cases} \sin^2 \chi, & k = 1\\
  				\chi^2, & 	k=0\\
  			\sinh^2\chi, & k=-1,
  		\end{cases}
\label{e:As}\ee
and this relation defines $D$.  

Use subscripts $e$ and $o$ to denote 
quantities measured in the emitting system's host galaxy and by gravitational 
wave observatories here.  
An observer here sees a gravitational wave with redshifted frequency
\[
\omega_o = \frac{\omega_e}{1+z}\ , 
\] 
and the corresponding redshifted time interval, $d\tau_o = (1+z) d\tau_e$, gives
\[
   \dot \omega_o \equiv \frac{d\omega_o}{d\tau_o} = (1+z)^{-2} \dot\omega_e \equiv (1+z)^{-2} \frac{d\omega_e}{d\tau_e}  .
\]
Eq. \eqref{e:D}, correct in the emitter's galaxy, is   
\[
\bar h_e D_e = k \frac{\dot\omega_e}{\omega_e^3 }. 
\] 
\vspace{-5mm}

\noindent The relation $\bar h_o D_o = \bar h_e D_e$ then implies
\[
\bar h_o D_o = \frac1{1+z} k \frac{\dot\omega_o}{\omega_o^3}. 
\] 
\vspace{-5mm}

\noindent Dropping the subscript $o$, we have
\be
  D = \frac1{1+z} k \frac{\dot\omega}{\omega^3\bar h} .
\ee

The {\sl luminosity distance} $D_L$ is defined in terms of the observed energy flux  
$\dis \frac{L_o}{4\pi D_o^2} $ by 
\vspace{-3mm}

\be
  \frac{L_e}{4\pi D_L^2} :=\frac{L_o}{4\pi D_o^2}\ , 
\ee
with $L_e = \dot E_e$, the absolute luminosity of the source. 
Using $L_e = k \omega_e^2 \bar h_e^2 D_e^2 $,  
$\ \ L_o = k \omega_o^2 \bar h_o^2 D_o^2$ (same $k$) gives 
\vspace{-7mm}

\[
   L_o = \frac{\omega_0^2}{\omega_e^2} L_e = \frac1{(1+z)^2} L_e\,.
\]  
Then 
\[
  D_L = (1+z)D_o,   
\]
so 
\[
   \cblue D_L = k \frac{\dot\omega}{\omega^3\bar h}, 
\]
and this is what one measures by observing the waveform. \\       

For the first observed binary neutron star inspiral and coalescence, 
$GW170817$, the distance $D$, when found solely from the gravitational-wave observation, 
has the value $D = 43.8^{+2.9}_{-6.9}\ \rm Mpc$ .  The redshift, adjusted to account 
for the galaxy's peculiar velocity (its deviation from the Hubble flow), is 
$z = (1.01\pm 0.06)\times 10^{-2}$, corresponding to a Hubble-flow velocity 
$v=3.02\pm 17$ km/s, and giving  
\be
  H_0 = 70^{+12}_{-8}\rm km/s/Mpc. 
\ee
The central value is consistent with the current electromagnetic measurements, 
but we need to observe of order 40 systems with this accuracy to reduce the statistical error to 
a level that could discriminate between the values from the cosmic background and 
from local electromagnetic distance determinations.\\

  Although binary black hole inspirals are far more common than neutron star inspirals,  for them there has been no electromagnetic counterpart and so no redshift 
measurement. Their mass distribution, however, has characteristic features that 
can be used to disentangle $H_0$, $z$, and $D$ from observations of a large number of 
events. In particular,  there is a gap between the maximum-mass of neutron stars 
and the lightest black holes, with very few objects seen in the gap. 
(These gap objects may be small black holes formed by NS-NS coalescence.)   
And a expected rough cutoff in black-hole mass above about 50 $M_\odot$ is seen in the 
observed inspirals.  That expectation comes from an instability of high-mass 
stars tied to electron-positron pair production in their cores.%
\footnote{Stars with adiabatic index\index{adiabatic index} $\gamma < 4/3$ are unstable to radial 
perturbations.  Pair production reduces the radiation pressure in the hot 
core, and the collapse of the unstable core ends with a supernova that, in 
numerical simulations, entirely distrupts the star:   
The {\sl pair-instabilty supernova} (PISN) leaves no 
black-hole remnant. Above about 135 $M_\odot$, however, the supernova 
energy is not enough to blow the star apart, and pair instability leads 
to complete collapse to a black hole. \index{collapse, gravitational}\index{gravitational collapse}}  
See \href{https://journals.aps.org/rmp/abstract/10.1103/RevModPhys.74.1015}{Woosley et al. 2002}, 
\href{https://iopscience.iop.org/article/10.3847/1538-4357/ab1b41/pdf}{Woosley 2019}\cite{woosley19} 
and references therein. Using the known behavior of the waveforms with luminosity 
distance, one finds a best fit of $H_0$ and parameters characterizing the 
unkown true mass distribution.
See \href{https://arxiv.org/pdf/2202.08240.pdf}{Ezquiaga \& Holz 2022},\cite{eh22} 
Hernandez and Ray\cite{eh22,hr24} and references therein. 
One can also (or simultaneously) fit, as Schutz initially suggested, to the known 
spatial distribution of galaxies.  \\

  Messenger and Read \cite{mr12} noticed that, in NS-NS inspiral, the tidal terms 
add to the waveform an additional measurable parameter. This, in principle, allows 
one to measure both luminosity distance and redshift, thereby determining the 
Hubble constant from the waveform alone. Although well beyond the scope of the 
current detectors, it may be within reach of the next generation.

\newpage

\benr \item 
\label{ex:maxconf}
Here is a first proof of the assertion: \\
{\sl If a covariant tensor $F_{\a\b}$ satisfies 
Maxwell's equations for $g_{\a\b}$, it satisfies them for a conformally 
related metric $\bar g_{\a\b}$. }\\
1) $dF=0$ never heard of a metric.\\
2) The second Maxwell equation is $\overline\na_\b \bar F^{\a\b} :=0$, where 
$\bar F^{\a\b} := \bar g^{\a\g}\bar g^{\b\d} F_{\g\d}$.  Then 
\begin{align*}
 F^{\a\b} &= g^{\a\c} g^{\b\d} F_{\c\d} = a^{-4} \bar F^{\a\b} \\
 \na_\nu F^{\mu\nu} &= \frac1{\sqrt{-g}} \pa_\nu(\sqrt{-g} F^{\mu\nu}) 
 = \frac1{a^4 \sqrt{-\bar g}} 
	\pa_\nu({\cblue a^4 a^{-4}}\sqrt{-\bar g} \bar F^{\mu\nu}) \\
&=\frac1{a^4} \overline\na_\nu \bar F^{\mu\nu} = 0.\quad \Box
\end{align*}  
Write the second part of the proof as $d\,\overline{^*\!F} = d^*\!F = 0$,
where $^*F_{\a\b} := \frac12\ep_{\a\b}{}^{\c\d} F_{\c\d}$, using 
\mbox{$\bar\epsilon_{0123} = \sqrt{-\bar g}$} to check that  
$\bar\ep_{\a\b}{}^{\c\d} = \ep_{\a\b}{}^{\c\d}$.

 \item
 Because null geodesics are a geometric-optics limit of Maxwell's equations 
they must also be conformally invariant.  In particular, if the covariant 
vector $k_\a$ satisfies 
\be
 k^\b \na_\b k_\a=0, \ \mbox{ then } k^\b \overline \na_\b k_\a = 0 .
\label{e:confgeod} \ee
Check this equation, using the fact that the difference between 
two derivative operators is a tensor $\Gamma^\a{}_{\b\g}$, as in \ref{ex:nabla-partial}  
\[
  (\overline\na_\b - \na_\b) k_\a = -\Gamma^\g{}_{\a\b} k_\g, \qquad
  \Gamma^\g{}_{\a\b} = \frac12 \bar g^{\g\d}(\na_\a \bar g_{\b\d} 
  + \na_\b \bar g_{\a\d} - \na_\d \bar g_{\a\b}).
\]
Show that this is the form of $\Gamma^\g{}_{\a\b}$ for the derivative operators 
of any metrics $g_{\a\b}$ and $\bar g_{\a\b}$, and infer for the conformally 
related metrics the relation 
\[
   \Gamma^\g{}_{\a\b} 
   = \delta^\g_\a \na_\b \log a + \delta^\g_\b \na_\a \log a 
   -g_{\a\b} \na^\g \log a .
\]
Finally, verify Eq.~\eqref{e:confgeod}.
\een
 
More generally, an equation for a tensor field $T^{\cdots}_{\cdots}$ is said to be
conformally invariant\index{conformal invariance|textbf} if there is a power of $a$, 
called the conformal weight $s$,
for which the equation is satisfied when $g_{\a\b}$ is replaced by $a^2 g_{\a\b}$
and $T^{\cdots}_{\cdots}$ is replaced by $a^s T^{\cdots}_{\cdots}$.   In the examples
here, Maxwell's equations and the null-geodesic equation are conformally invariant
when $F_{\a\b}$ and $k_\a$ are assigned conformal weight $s=0$. Using $\bar g^{\a\b}$ 
to raise indices then gives the contravariant tensors 
$F^{\a\b}$ and $k^\a$ the conformal weights $-4$ and $-2$, respectively. 
(The contravariant vector field $k^\a$ is tangent to a null geodesic of $\bar g_{\a\b}$,
but it is $a^{-2}k^\a$ that is affinely parametrized with respect to $\bar g_{\a\b}$.) 
\index{cosmology|)}

\chapter{3+1 Split and Initial Value Equations}\label{c:iv}

\section{Notation}\index{notation}
A detailed table is given here in case of confusion from differences between notation
in the notes, in the text by Baumgarte \& Shapiro\cite{bsbook21} and in the text by Wald\cite{waldbook}. 

\begin{tabular}{*{4}{l}}
\\
\hline\hline
Item & These notes & Baumgarte \& Shapiro & Wald \\
[0.5ex]
\hline
Signature & $-+++$ & $-+++$& $-+++$\\
Indices $\alpha, \beta, \ldots$ & spacetime abstract& not used & spacetime concrete \\
Indices $\mu, \nu$ & spacetime concrete& not used & spacetime concrete\\
Indices $a,b, \ldots$ & space and $n$-dimensional abstract & spacetime & spacetime abstract\\
Indices $i,j, \ldots$ & space and $n$-dimensional concrete& space & not used \\
Spacetime & $M$ &  not named & $M$ \\
Spacelike hypersurface & $\Sigma_t$ &  not named & $\Sigma_t$ \\
3-dimensional surface & $\Sigma$ & not named & $\Sigma$ \\
Future pointing normal & $n^\a$ & $n^a$ & $n^a$ \\
Projection $\perp \bm n$ & $\g_\a^\b$ & $\g_a^b$ & $h_a^b$ \\
Spatial metric on $M$ & $\gamma_{\a\b}$ & $\gamma_{ab}$ & $h_{ab}$\\
Pullback from $M$ to $\Sigma$ & $\gamma^\alpha_a$ & $\gamma^a_i$ & not used \\
Extrinsic curvature & $K_{\a\b} = -\frac12\Lie_{\bm n} \gamma_{\a\b} $
		    & $K_{ab} = -\frac12\Lie_{\bm n} \gamma_{ab} $ 
		    & $K_{ab} = \frac12\Lie_{\bm n} h_{ab} $ \\
Riemann tensor on $M$ & $R_{\a\b\c\d}$ & $^{(4)}R_{abcd}$ & $R_{abcd}$ \\
Spatial Riemann on $M$ & $^3R_{\a\b\c\d}$ & $R_{ijkl}$ & $^{(3)}R_{abcd}$ \\
Lapse &$\a$ &$\a$ &$N$ \\
Shift &$\beta^\a$ & $\beta^a$ & $N^a$\\
$\bm \pa_t$ & $t^\a$ & $t^a$ & $t^a$ \\
Energy density $T_{\a\b}u^\a u^\b$ & $\rho$  & not used & $\rho$ \\
Energy density $T_{a\b}n^\a n^\b$ & $\rho_E$ & $\rho$ & not used \\
\hline\hline
\end{tabular}

\section{Introduction}
\label{Introduction}
\index{initial value problem|(}

Like the equations of ordinary mechanics, the equations governing classical fields -- 
scalar, electromagnetic, and gravitational -- allow one to  
predict the future (or retrodict the past) from initial data at a given time. 
For a system of particles, one specifies their initial positions $q^i$ and momenta $p_i$ 
(or velocities $\dot q^i$) 
and finds their time evolution from equations for $\dot q^i$ and $\dot p_i$ in terms 
of $q^i$ and $p_i$:  
\be
  \dot q^i = \frac{\pa H}{\pa p_i}, \qquad \dot p_i = \frac{\pa H}{\pa q^i},
\label{e:ham1}\ee
where $H(p,q)$ depends on $q$ and $p$, not on their time derivatives. For example, 
with $H=\frac12 g^{ij}p_i p_j + V(q)$, where $g^{ij}$ is constant,  
\be
  \dot q^i = p^i, \qquad \dot p_i = \frac{\pa V}{\pa q^i}.
\label{e:ham2}\ee
Here $p^i := g^{ij} p_j$.  Notice that, with $\dot q^i$ regarded as a vector, 
the momentum $p_i$ is naturally a dual vector (covariant vector).
\benr \item What is $(g^{ij})$ for a two particle system with masses $m$ and $M$?\\ Assume Cartesian coordinates.  
\item What is $(g^{ij})$ for one particle of mass $m$ in spherical coordinates? Find the equations of motion that generalize Eqs.~\eqref{e:ham2}.      
\een

\noindent
{\sl Scalar field} \\

We consider first a scalar field in flat space, a field satisfying the scalar wave 
equation 
\be 
  \raisebox{-1mm}{\text{\Large$\Box$}}\,\Phi 
		:= \nabla_\alpha \na^\alpha\Phi = 0.  
\ee 
We begin with a coordinate-based description of the initial value problem and 
then rephrase it in a way that can be can be carried over to curved spacetime. 
The coordinate description uses natural coordinates $\{t,x^i\}$ on Minkowski space. 
As in the case of a system of particles, one freely specifies the initial values 
of the field $\Phi$ and its conjugate momentum 
$p=\dot\Phi=\partial_t\Phi$ at a fixed time $t_0$, now at each point $(t_0,x^i)$.  
The time evolution of the field is given by 
\be
\dot\Phi = p, \qquad \dot p = \nabla^2 \Phi = \pa_i\pa^i\Phi.  
\label{e:sham}\ee

\benr \item With Hamiltonian $\dis H = \frac12\int d^3r [p^2 + (\nabla\Phi)^2]$,
show that 
\be
   \frac{\delta H}{\delta p(\bm r)} = p(\bm r), \qquad 
   \frac{\delta H}{\delta \Phi(\bm r)} = -\nabla^2\Phi(\bm r),
\ee
for $\delta\Phi$ vanishing as $\bm r\rightarrow \infty$. 
That is, Eqs.~\eqref{e:sham} are Hamilton's equations for the scalar field. 
\index{Hamiltonian!scalar field}   
\een

Here's the rephrasing, without introducing spatial coordinates.   
A choice $t$ of Minkowski time slices spacetime into a set of $t=$ constant 
surfaces, which we call $\Sigma_t$.  The future pointing unit normal to the 
surface is $n^\alpha=-\nabla^\alpha t$.  Using $t$ and the flat metric, we can 
decompose spacetime into a product $\mathbb R\times\Sigma$ of time $\mathbb R$ and space $\Sigma=\mathbb R^3$, with $\Sigma_t= \{t\}\times\Sigma$; here, if $\bf r$ 
is a point of $\Sigma$, the lines of fixed $\bm r$ are orthogonal to $\Sigma_t$, 
with tangent $n^\alpha = - \nabla^\alpha t$.  

We can similarly decompose the 4-dimensional gradient $\nabla_\alpha \Phi$ into the time derivative $\dot\Phi=n^\a \na\!_\a t$ and spatial gradient $\bm\nabla\Phi$. 
The spatial gradient $\bm\nabla\Phi$ is the gradient of $\Phi$ regarded as 
a function on the hypersurface $\Sigma$.     

One freely specifies 
the initial values of the field $\Phi$ and its conjugate momentum 
$p=n^\a\na\!_\a\Phi$ at a fixed time $t_0$ at each point 
$\bm r$ of $\Sigma$.    
The time evolution of the field is given by    
\be
\dot\Phi = p, \qquad \dot p = \nabla^2 \Phi.  
\label{e:sham1}\ee
We can then regard the time evolution as giving successive values of 
$\Phi(\bm r), p(\bm r)$ on the fixed space $\Sigma$. 

Equivalently, one can maintain a spacetime view, in which $\Phi$ and $p$ are 
functions on spacetime and the spatial gradient of $\Phi$ is the spatial 
projection of the 4-dimensional gradient:  One subtracts from the gradient
$\na\!_\a f$ of a function the part of the gradient along $\nabla_\alpha t$.  
That is, we define the spatial gradient $D_\alpha$ of a function $f$ on spacetime by 
\index{gradient!spatial gradient}
\be
   D_\alpha f = \nabla_\alpha f - \na\!_\a t\ n^\beta \nabla_\b f 
	      = (\delta_\a^\b - \nabla_\a t\, n^\b) \na\!_\b f.
\label{e:Da}\ee 
(We will from now on use $D$ instead of $\nabla$ for the spatial gradient, to distinguish 
it from the spacetime gradient.) 
We will repeatedly use this spatial projection.  Using $n_\alpha = - \nabla_\alpha t$,  
we write
\be
  \gamma_\a^\b := \delta_\a^\b + n_\a\, n^\b, \qquad 
		  D_\a f= \gamma_\a^\b \na\!_\b f.  
\ee 
One says that $\gamma_\a^\b$ projects vectors and dual vectors tangent to $\Sigma_t$ and 
orthogonal to its normal $n^\alpha$.    
In this spacetime description, one is specifying $\Phi$ and $\dot\Phi$ on a given 
$\Sigma_t$ and finding $\Phi$ everywhere else in spacetime from the scalar wave equation, 
$\raisebox{-1mm}{\text{\Large$\Box$}}\,\Phi = 0$.  

Notice that the nonzero part of $\gamma_{\alpha\beta}$ is the spatial part of the 
Minkowski metric, the Euclidean metric on each $\Sigma_t$.    
In the chart $\{t,x^i\}$, it has components 
\be
  \left[ \gamma_{ij} \right] 
		= \begin{bmatrix} 0 & 0\\	
				0 & \delta_{ij} 
		   \end{bmatrix} \qquad \mbox{(Minkowski space)}.
\ee 

In Wald's textbook, the projection operator is denoted 
by $h_\alpha^\beta$ instead of $\gamma_\alpha^\beta$.  These days, however, 
you are most likely to encounter the initial value equations in numerical relativity 
papers.  In this part of the recent literature, the spatial metric on $t=$ constant surfaces is generally written as $\gamma_{ab}$, with $\gamma_\alpha^\beta$ the projection operator. 
\vspace{3mm} 

We can extend the 3+1 split of vector fields to a 3+1 split of 
tensor fields: We decompose each index of a tensor into temporal and
spatial parts, parts along and orthogonal to $n^\alpha$, writing   
\be
  \delta_\a^\b = \gamma_\a^\b - n_\a n^\b. 
\ee
For example, for a two-index tensor, 
\begin{align}
  T_{\alpha\beta} &= \delta_\a^\d\, \delta_\b^\ep T_{\d\ep}
	 = (\gamma_\a^\d-n_\a n^\d)(\gamma_\b^\ep- n_\b n^\ep)T_{\d\ep} \nonumber\\
	&= \g_\a^\d\, \g_\b^\ep T_{\d\ep} -n_\a n^\d\g_\b^\ep T_{\d\ep} 
		-\g_\a^\d\, n_\b n^\ep T_{\d\ep} + n_\a n^\d n_\b n^\ep T_{\d\ep}
\label{e:t3+1}\end{align}

For the electromagnetic field, $F_{\a\b}$, this is equivalent 
to the decomposition we encountered in Sect.~\ref{s:em}.  
Here the magnetic field is the spatial part of $F_{\a\b}$; 
its natural form is a 2-index antisymmetric tensor, 
\index{magnetic field!from electromagnetic tensor $F_{\a\b}$}
\be  
  B_{\alpha\beta} = \g_\a^\d \g_\b^\ep F_{\d\ep}.
\ee
To define the corresponding vector, one needs to use the spatial antisymmetric tensor $\epsilon_{\a\b\c}$ of Eq.\eqref{e:3epsilon}, 
\[
  \epsilon^{\a\b\c} := \epsilon^{\d\a\b\c}n_\d .
\]
Then 
\be
  B^\a = \frac{1}{2} \epsilon^{\a\b\c}B_{\b\c}.
\label{e:Ba}\ee 
The projection of one index of $F_{\a\b}$ along $n^\a$ and one orthogonal to 
$n^\a$ gives the electric field: 
\[
 E_\a = \gamma_\a^\b n^\c F_{\b\c} = F_{\a\c}n^\c.
\] 
\index{electric field!from electromagnetic tensor $F_{\a\b}$}  
Because of the antisymmetry of $F_{\a\b}$, the time projection of both indices vanishes: $F_{\a\b}n^\a n^\b = 0$.    
The decomposition of $F_{\a\b}$ then has the form \eqref{e:FEB}, 
\begin{align*}
  F_{\a\b} &= n_\a E_\b - n_\b E_\a + \epsilon_{\a\b\c} B^\c. \tag{\ref{e:FEB}}
\end{align*} 
\noindent

\noindent{\sl Constraints and dynamical equations} \\

In the initial value problem for electromagnetism and gravity, an additional feature arises:  The equations 
are now analogous to those governing a system of particles with a constraint 
that is preserved by the time evolution-- for example, a set of particles 
constrained to move on a fixed surface.  \\ 

From Eq.~\eqref{e:Maxwell4}, the 3+1 split of the Maxwell equations for a free electromagnetic field in Minkowski space has the form 
\bsube\begin{align}
\quad \nabla_\beta  E^\beta  = 0 &\qquad \nabla_\beta  B^\beta =0\vspace{1mm}
\label{e:Maxconstraint}\\
-n^\beta \nabla_\beta  E^\alpha  +\epsilon^{\alpha\beta\gamma} \nabla_\beta  B_\gamma  = 0,
&\qquad n^\beta \nabla_\beta  B^\alpha  + \epsilon^{\alpha\beta\gamma} \nabla_\beta  E_\gamma  =0. 
 \label{e:Maxdynamical}\end{align}\esube
 The first two equations involve only spatial derivatives and can be written 
\be
  D_\b E^\b = 4\pi\rho_e, \qquad  D_\b B^\b = 0, 
\ee 
with $D_\a = \gamma_\a^\b \na_\b$ as in Eq.~\eqref{e:Da}, or  
as 3-dimensional divergences on the 3-surface $\Sigma$ 
\be
\bm\na\cdot \bm E= 0, \qquad \bm\na\cdot \bm B = 0.
\ee
They are {\sl constraint equations}, \index{constraint equations!electromagnetism}
\index{electromagnetism!initial value problem}
constraining the initial values of $E^\a$ and $B^\a$ that can be specified on 
$\Sigma$.  

The other two Maxwell equations \eqref{e:Maxdynamical} are the {\sl dynamical equations}, 
giving the time evolution of the initial data, with 3-dimensional form on $\Sigma$, 
\be
 \dot{\bm E} = \na\times \bm B\qquad  \dot{\bm B} = -\na\times \bm E.
\ee

\noindent{\sl Mathematical note}\\
We have introduced a one-one correspondence between vectors 
$v^a$ on $\Sigma$ and vectors $v^\a$ at each point of a given $\Sigma_t$ 
that are orthogonal to $n^\alpha$.  That correspondence 
is naturally described using the definitions of {\sl pullback} and {\sl push forward} (drag) given on p. \pageref{p:diffeo}.  The map $\psi:\Sigma\rightarrow \Sigma_t\subset M$ 
takes each point $\bm r$ of $\Sigma$ to the point $\psi(\bm r) = (t,\bm r)$ of 
$M=\mathbb R\times \Sigma$, and $\psi$ drags a vector $v^a$ at the point $\bm r$ to a vector $\psi^*v=v^\alpha$ at $(t,\bm r)$ of $M$.  Because $t$ is constant on $\Sigma_t$, $v^\alpha\nabla_\a t =0$. 
There is similarly a one-one correspondence between tensors $T^{a\ldots b}$ on $\Sigma$ and tensors $T^{\alpha\ldots\beta}$ at each point of a given $\Sigma_t$ that are orthogonal in each index to $n^\alpha$, 
\be
  T^{\alpha\ldots\beta} n_\alpha =0, \ \ \ldots , \ \ 
	T^{\alpha\ldots\beta}n_\beta = 0. 
\ee
That is, the map $\psi$ drags contravariant tensors on $\Sigma$ to spatial tensors on $M$, and $\psi$ pulls back covariant tensors on $M$ to tensors on 
$\Sigma$.  We will revisit this correspondence after introducing the 3+1 split in curved spacetime.

\section{The 3+1 Split in Curved Spacetime}
\index{initial value problem!3+1 split}

The initial value problem for fields in a curved spacetime is associated with 
a slicing of spacetime by a family of spacelike hypersurfaces. For detailed accounts
on the historical development of the subject, see for, example, \cite{Alcubierre2008,Bona2009,Baumgarte2021,Gourgoulhon2012}. The formalism we and 
much of the literature uses was introduced about sixty years ago by Arnowitt, 
Deser and Misner\cite{ADM62} (out of print and republished as \cite{ADM62a}). 

The Einstein equation (and the associated conservation of energy and momentum), 
conservation of baryons, and the equation of state of the fluid, 
determine the evolution of the geometry and the fluid variables from one hypersurface to another.  We will see that the 3+1 decomposition of the spacetime
is mirrored in a 3+1 decomposition of the equations and of the tensors 
whose time-evolution they describe. 
\index{hypersurface!slicing of spacetime}

\begin{figure}[H]
               \begin{center}
		\includegraphics[width=0.4\textwidth]{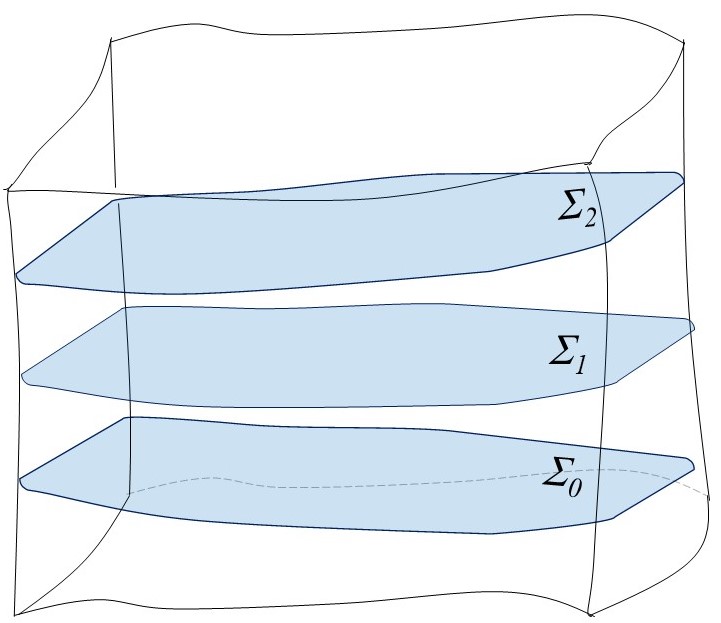}
		\end{center}
		\caption{Three of the hypersurfaces that slice a spacetime}
\end{figure}
A choice of time coordinate $t$ again gives a slicing by $t=$ constant surfaces 
$\Sigma_t$ whose normal $\nabla_\a t$ is everywhere timelike.   
The spacetimes we consider have the form ${\mathbb R}\times \Sigma$, 
with each hypersurface $\Sigma_t$ a copy $\{t\}\times \Sigma$ of $\Sigma$. 
\footnote{The mathematical literature calls a decomposition of an $n$-dimensional manifold $M$  into a set ${\mathbb R}\times \Sigma$ of $(n-1)$-dimensional manifolds a {\sl foliation}, with each copy of $\Sigma$ a {\sl leaf}; following the physics literature, we use the word {\sl slicing}, and each copy $\Sigma_t$ of $\Sigma$ is a slice. }
We call the future direction the direction in which $t$ increases: That is, a 
timelike path is future directed if $t$ increases along the path.  The $\ -+++$ metric signature implies that the future pointing contravariant unit vector 
normal to each slice $\Sigma_t$ is 
\be\cblue
  n^\a = -\frac{ \na^\a t}{\sqrt{-\na\!_\b t\na^\b t}} = -\alpha \na^\a t\cb, 
\index{normal to a spacelike hypersurface}
\label{e:normal}\ee
where 
\be
   \alpha :=  (-\na\!_\b t\na^\b t)^{-1/2} = |g^{tt}|^{-1/2},  
\label{e:alpha}\ee
in a chart of the form $\{t,x^i\}$. The components of $\nabla_\alpha t$ and 
$n_\alpha$ are then 
\be
   \nabla_\mu t = \delta_\mu^t, \qquad n_\mu =-\a \delta_\mu^t.  
\ee
The scalar $\alpha$ is called the {\em lapse}, because
$\alpha\, dt$ is the proper time elapsed in a normal direction between coordinate times $t$ and $t+dt$.  It is the proper time, measured normal to the hypersurfaces between 
slices $\Sigma_t$ and $\Sigma_{t+dt}$.  

\benr
\item Check that $\alpha\, dt$ is the proper time between slices by showing that, if the chart $\{t,x^i\}$ is chosen to have lines of constant 
$x^i$ perpendicular to $\Sigma_t$, then $g_{ti}=0$ and $\dis g_{tt} = \frac1{g^{tt}}$. 
Conclude that proper time elapsed normal to $\Sigma_t$ is $d\tau = \alpha dt$.   
\een

We now extend the 3+1 decomposition ${\mathbb R}\times \Sigma$ of spacetime 
into space and time to a decomposition of vectors and tensors.  Call a vector 
$v^\alpha$ a {\sl spatial} vector if it is tangent to a path $c(\lambda)$ that 
lies in a spacelike slice $\Sigma_t$; equivalently $v^\alpha n_\alpha = 0$.  
We say such a vector $v^\alpha$ is tangent to $\Sigma_t$.  If $v^\alpha$ 
is tangent to $\Sigma_t$, then $v^t=0$ in a chart $\{t,x^i$\}.  That is,  
if $\{t,x^i$\} are the coordinates of $c(0)$, then the coordinates of the path $c$ are 
$(x^\mu(\lambda) )= (t,x^i(\lambda))$, and the components of $v^\alpha$ are 
\[
  v^\mu = \left.\frac d{d\lambda} x^\mu(\lambda)\right|_{\lambda=0}, 
\]
implying
\be 
	  (v^\mu) = (0,v^i).
\ee  

We again introduce the projection operator  
\be\crv
\gamma^\a{}_\b := \delta^\a_\b + n^\a n_\b \cb.
\index{metric!spatial metric}\ee  
onto each $\Sigma_t$:  $\gamma^\a{}_\b$ kills vectors 
$v^\a= v n^\a$ normal to $\Sigma_t$,
and it leaves invariant vectors $v^\a$ tangent to $\Sigma_t$: 
\be
\gamma^\a{}_\b n^\b = 0, \qquad  \gamma^\a{}_\b v^\b = v^\a\ \mbox{for } v_\a n^\a=0.
\label{e:gamma_ab}\ee 

Any vector $v^\a$ (or any tensor index) can be decomposed into a spatial vector 
(one that is tangent to $\Sigma_t$) and a vector orthogonal to $\Sigma_t$:
\index{spatial vector!in 3+1 decomposition|textbf}
\be
  v^\a = \delta^\a_\b v^\b = (\gamma^\a{}_\b-n^a n_\b)v^\b = \gamma^\a{}_\b v^\b - n_\b v^\b n^\a .
\ee
The corresponding decomposition of the field equation \mbox{$\texttt E^{\a\b}:=G^{\a\b}-8\pi T^{\a\b}=0$},
is 
\index{Einstein field equation!three plus one@3+1 split|(}
\be
  \texttt E^{\a\b} 
=\gamma^\a{}_\c \gamma^\b{}_\d \texttt E^{\c\d} - 2 n_\d \texttt E^{\c\d}\gamma_\c{}^{(\a} n^{\b)} 
	+ n_\c n_\d \texttt E^{\c\d}n^\a n^\b .
\label{eq:e3+1}\ee
We will see that the projection $\gamma^\a{}_\c \gamma^\b{}_\d \texttt E^{\c\d}$ of $\texttt E^{\a\b}=0$ 
onto the hypersurface $\Sigma_t$ 
is a dynamical equation for the metric (involving its second time derivative),  while the remaining parts, 
$G_{\a\b}n^\b=8\pi T_{\a\b}n^\b$, of the field equation serve as constraints (equations involving only the values of the metric and its first time derivative on $\Sigma_t$). 
\index{constraint equations!gravitational}

The 3+1 decomposition of the metric is just 
\be 
   g_{\a\b} = \g_{\a\b} - n_\a n_\b, 
\ee 
the result of lowering the index $\a$ in Eq.~\eqref{e:gamma_ab}. 
In particular, $g_{\|\perp\,\alpha}=g_{\sigma\tau}\g_\a^\sigma n^\tau = \g_{\a\tau} n^\tau = 0$. 

The tensor $\g_{\a\b}$ can be regarded as the 3-metric on $\Sigma_t$, written 
with spacetime indices:  Acting on vectors tangent to $\Sigma_t$, it gives their dot 
product, and it kills vectors orthogonal to $\Sigma_t$.  More formally, at each time $t$, 
the spacetime metric $g_{\a\b}$ induces a metric $\gamma_{ab}$, on $\Sigma$. The 
map $\psi:\Sigma\rightarrow \Sigma_t$ pulls back $g_{\a\b}$ to 
$\g_{ab} = (\psi_* g)_{ab}$.  In a chart of the form $\{t,x^i\}$, the components 
of $\g_{ab}$ and the spatial components of $\g_{\a\b}$ and $g_{\a\b}$ coincide: 
\be
   \gamma_{ij} = g_{ij}.  
\ee
We use the same symbol $\gamma$ for the tensor $\gamma_{ab}$ on $\Sigma$ and 
$\gamma_{\a\b}$ on $\Sigma_t$: $\gamma_{\a\b}$ is the unique spacetime tensor 
orthogonal to $n^\a$ whose pullback is $\gamma_{ab}$. 

It will simplify covariant derivations of the initial value equations to stick with spacetime 
indices, instead of using both tensors on $\Sigma$ and tensors on $M$.  We do this 
as follows: \\
{\bf Definition}.  A tensor $T^{\a\cdots\b}{}_{\c\cdots\d}$ {\sl spatial} if it orthogonal 
in all its indices to $n_\a$:
\be
  T^{\a\cdots\b}{}_{\c\cdots\d}n_\a = 0, \ldots, T^{\a\cdots\b}{}_{\c\cdots\d} n^\d = 0.
\ee 
We then note:\\
{\cblue There is a one-one correspondence between spatial tensors $T^{\a\cdots\b}{}_{\c\cdots\d}$ 
at a point $(t,x)$ in $M$ and tensors $T^{a\cdots b}{}_{c\cdots d}$ at $x$ in $\Sigma$.}\\  
This is easy to show. We'll again use a chart $\{t,x^i\}$ on $M$ with corresponding chart 
$\{x^i\}$ on $\Sigma$.  We've already seen that a vector on $\Sigma$ with components $v^i$ is 
dragged to a vector on $M$ with components $(v^\mu) = (0,v^i)$, or 
\be
   v^\mu = \frac{\pa x^\mu}{\pa x^i} v^i = \delta^\mu_i v^i,  
\ee
A contravariant tensor  
on $\Sigma$ with components $T^{i\cdots j}$ is then dragged to the spatial tensor with vanishing $t$ 
components and with spatial components $T^{i\cdots j}$.  
\be 
   T^{\mu\cdots\nu} = \d^\mu_i\cdots \d^\nu_j T^{i\cdots j}.  
\ee  
Conversely, any contravariant spatial tensor $T^{\a\cdots\b}$ has vanishing $t$ components and 
corresponds to the unique tensor $T^{a\cdots b}$ on $\Sigma$ with components $T^{i\cdots j}$.  

While spatial {\sl vectors} are naturally defined as tangents to curves that lie 
in $\Sigma_t$ or vectors for which $v^\a \na_\a t = 0$, a 
{\sl spatial dual vector} $\sigma_\a$ uses the metric on $M$ for its definition.  
That is, one needs the metric to define the dot product 
$\sigma^\a \na_\a t= g^{\a\b} \sigma_\b \na_\a t = 0$. 
To define the tensor $T^{\a\cdots \b}{}_{\c\cdots \d}$ corresponding to $T^{a\cdots b}{}_{c\cdots b}$, 
one raises the indices of each tensor and requires $T^{\a\cdots \b \c\cdots \d}$ to be the unique spatial 
tensor corresponding to $T^{a\cdots b c\cdots d}$.  Then the spatial components of corresponding 
tensors--contravariant or covariant -- agree.  But be careful: Covariant $t$ components need not vanish.\\

\centerline{ $\sigma_\alpha$ spatial does {\sl not} imply $\sigma_t = 0$.}  
\benr
\item Let $\sigma_a$ be a dual vector at a point $x\in\Sigma$.  Show that the corresponding 
vector $\sigma_\a$ at $(t,x)$ has spatial components $\sigma_i$, but that $\sigma_t$ is nonzero if $g_{ti}$ is nonzero.  To avoid confusion in this exercise, write $\sigma_a$ for the 3-vector on $\Sigma$, 
$\widetilde\sigma_\alpha$ for the 
4-vector on $M$.  
\een 

Finally, note that the pullback of a tensor $T_{\a\cdots \b}$ on $M$ at $\{t,x^i\}$ 
is a tensor $T_{a\cdots b}$ at $\{x^i\}$ with the same spatial components $T_{i\cdots j}$:  
The pullback of $T_{\a\cdots\b}$ has components 
\be
   T_{i\cdots j} = \frac{\pa x^\mu}{\pa x^i} \cdots \frac{\pa x^\nu}{\pa x^j} T_{\mu\nu}
		=\d^\mu_i \cdots \d^\nu_j T_{\mu\cdots \nu}.
\ee
That means the pullback of a spatial covariant tensor on $M$ is the corresponding tensor on $\Sigma$.
Because $\gamma^\mu_i = \delta^\mu_i + n_\mu n_i$, and $n_i=0$ in a chart $(t,x^i)$, we have 
\[
   \gamma^\mu_i = \delta^\mu_i, \qquad (\Psi_* T)_{i\cdots j} = \g^\mu_i \cdots \g^\nu_j T_{\mu\cdots \nu},
\]
where $\Psi:\Sigma\rightarrow \Sigma_t \subset M$.  \newpage

\index{covariant derivative!of spatial metric}
We need two final definitions to cast the field equation in the standard 
initial-value form, the covariant derivative of $\gamma_{ab}$ and the extrinsic curvature of 
a hypersurface $\Sigma_t$. \\
{\sl Covariant derivative, $D_a, D_\a$} \\
Denote by $D_a$ the covariant derivative operator 
on $\Sigma$ associated at a time $t$ with the metric $\gamma_{ab}(t)$.   
If $v^\a$ is the spatial vector corresponding to the vector $v^a$ on $\Sigma$, we will see that 
its spatial covariant derivative is 
\be
   D_\b v^\a := \g_\b^\p \g^\a_\q \na\!_\p v^\q.   
\ee
\index{spatial covariant derivative, $D_\a$}
That is, with this definition, $D_\a v^\b$ is the unique spatial tensor corresponding to 
$D_a v^b$ on $\Sigma$. On a general spatial tensor $T^{\a\cdots\b}_{\c\cdots\d}$, 
\be
  D_\ep T^{\a\cdots\b}_{\c\cdots\d} 
   := \g_\ep^\lambda \g^\a_\mu\cdots\g^\b_\nu \g_\c^\sigma\cdots \g_\d^\tau 
	\na\!_\lambda T^{\mu\cdots\nu}_{\sigma\cdots\tau}.
\ee 
It is easy to see that $D_\alpha$ is a derivative operator, satisfying 
the conditions following Eq.~\ref{e:deriv} (linear, Leibnitz rule, torsion free). 
The covariant derivative $D_a$ on $\Sigma$ is the unique derivative operator 
for which $D_d\gamma_{ab} = 0$; that $D_\alpha$ corresponds to $D_a$ 
on $\Sigma$ follows from the fact that $D_\d \g_{\a\b} = 0$:  
\[
  D_\d \g_{\a\b} = \g_\d^\lambda \g_\a^\sigma \g_\b^\tau 
				\na\!_\lambda(g_{\sigma\tau}+ n_\sigma n_\tau)
	= \g_\d^\lambda \g_\a^\sigma \g_\b^\tau (0 + n_\sigma \nabla_\lambda n_\tau +  n_\tau \nabla_\lambda n_\sigma) = 0, 	 
\]
where we used $\g_\a^\b n_\b = 0$. 
We will generally adopt spacetime indices. \\ 

\noindent
{\sl Extrinsic curvature}\\
\index{curvature tensor!extrinsic curvature|textbf}
\index{Riemann tensor!extrinsic curvature|textbf}
Although one ordinarily chooses 
coordinates for which $t^\a$ is not normal to $\Sigma_t$, it is the time derivative 
normal to the surface that naturally enters the projections of the field equation, 
(\ref{e:kabevol},\ref{e:Hamconstraint},\ref{e:momconstraint}) below.  
This time derivative of the 3-metric is (up to a factor -1/2) 
the {\em extrinsic curvature} $K_{\a\b}$ of the hypersurface $\Sigma_t$:%
\footnote{There is no consistent convention in the literature for 
the sign of the extrinsic curvature of a spacelike hypersurface.  
Our convention agrees with that of MTW \cite{MTW} and has sign opposite to  
Wald's \cite{waldbook}.}
\be\crv
 K_{\a\b} := -\frac12 \Lie_{\bf n} \gamma_{\a\b}.
\index{extrinsic curvature|textbf}
\label{eq:kab0}\ee 
$K_{\a\b}$ is spatial, orthogonal to $n^\a$ on each index. This is immediate 
from $\Lie_{\bm n} n^\a = 0$ and $\g_{\a\b}n^\b=0$:  
\[
   K_{\a\b} n^\b = -\frac12 n^\b \Lie_{\bf n} \gamma_{\a\b}
		 = -\frac12\left[ \Lie_{\bf n} (n^\b\gamma_{\a\b}) 
			- (\Lie_{\bf n} n^\b)\gamma_{\a\b}\right]
		 = 0.
\] 
We denote by $K$ its contraction 
\be
K:=K_\a{}^\a = \gamma^{\a\b}K_{\a\b}.
\ee
The extrinsic curvature is thus the time derivative of the 3-metric  
for an observer with velocity $n^\alpha$; it describes the change in the geometry of 
$\Sigma_t$ as one moves a small proper distance orthogonal to the hypersurface. 

The {\sl intrinsic} curvature, the Riemann tensor of the a surface, depends only on the metric $\gamma_{ab}$ on that surface.  It does not care how the surface is embedded in a higher 
dimensional manifold.  The {\sl extrinsic} curvature looks at the embedding.  
We'll see below in Eq.~\eqref{e:Kab2} that $K_{\a\b}$ measures the rate at which the normal to 
the surface changes as one moves along the surface.  First an example.   \\
{\sl Example}.  Consider the flat cylinder obtained by gluing two opposite edges of a sheet of paper. Because the paper is not stretched or sheared, its intrinsic geometry is still flat.  
(Think of the universe of Ms. Pac Man or Asteroids with the top and bottom edges of 
the screen identified.  The screen is still flat after the identification.)  But 
the extrinsic curvature is not zero:  The flat Euclidean 3-metric is 
\[
  ds^2 = dz^2 + dr^2 + r^2 d\phi^2.  
\]
A cylinder of radius $r=a$ has the flat 2-metric 
\[
  d\sigma^2 = dz^2 + a^2 d\phi^2 = dz^2 + dx^2, \mbox{ with } x:= a\phi. 
\]
($\phi=0$ is identified with $\phi=2\pi$ and $x=0$ with $x=2\pi a$.) The unit normal 
to the cylinder is $n^a = \nabla^a r$, with components $n^i = \delta^i_r$.  Then, 
because $\pa_k n^i = 0$, we have 
$\Lie_{\bm n} g_{ij} = n^k\pa_k g_{ij} = 2r \na\!_i \phi \na\!_j \phi$, 
where, in our cylindrical coordinates, $\na\!_i\phi = \delta_i^\phi$.   
The extrinsic curvature of the $r=a$ cylinder is then
\be
   K_{ab} = \mp a \na\!_a \phi \na\!_b \phi,   
\ee 
the sign depending on which unit normal is chosen.   
\vspace{3mm}


The 3-metric and extrinsic curvature on an initial hypersurface $\Sigma_t$ will 
constitute our initial data.  They are constrained by the projections 
$\texttt E^{\a\b}n_\a n_\b=0$ and $\texttt E^{\d\b}\g^\a{}_\d n_\b=0$ of the Einstein equation, 
called, respectively, the {\it Hamiltonian constraint} and the {\sl momentum 
constraint}:
The terminology is related to the fact that 
\[
	\rho_E:=T^{\a\b}n_\a n_\b \quad\mbox{and}\quad j^\a:=-T^{\d\b}\g^\a{}_\d n_\b
\]
are, respectively, 
the densities of energy and momentum for an observer with velocity $n^\a$.
We will show that the constraint equations are\\
{\sl Momentum constraint}:
\index{momentum constraint}
\be\crv
  D_\b(K^{\a\b}-\gamma^{\a\b}K) =  8\pi j^\a; 
\label{e:momconstraint}\ee
{\sl Hamiltonian constraint}:\index{Hamiltonian constraint}
\be\crv
  \tr -K^{\a\b}K_{\a\b} + K^2 = 16\pi \rho_E.
\label{e:Hamconstraint}\ee

The time evolution of the initial data is a pair of equations for 
$\Lie_{\bm n} \gamma_{\a\b}$ and $\Lie_{\bm n} K_{\a\b}$.  The evolution equation for 
$\gamma_{\a\b}$, analogous to $\dot q = p$, is simply Eq.~\eqref{eq:kab0}, 
\be\cvi
   \Lie_{\bf n} \gamma_{\a\b} = -2 K_{\a\b}.  
\label{e:gabevol}\ee
The time evolution of the extrinsic curvature is given by 
the remaining, spatial, part of the Einstein equation, 
written using the Ricci tensor, namely 
\[
\gamma_\a{}^\sigma\gamma_\b{}^\tau [R_{\sigma\tau}  
		- 8\pi (T_{\sigma\tau}-\frac12 g_{\sigma\tau}T)]=0, \quad\mbox{with } T=T_\c{}^\c.
\]
It has the form 
\be\cvi
  \Lie_{\bm n} K_{\a\b} 
	= - \frac1\alpha \,D_\a D_\b \alpha - 2K_{\a\c}K_\b{}^\c + K K_{\a\b} +\tr_{\a\b} 
		-8\pi (\g_\a^\d \g_\b^\ep T_{\d\ep}-\frac12 \g_{\a\b} T).
\label{e:kabevol}\ee
Here $\tr_{\a\b}$ is the spatial Ricci tensor of the spatial metric $\gamma_{\a\b}$, and we will denote by 
\be
	\tr = \gamma^{\a\b}\ \tr_{\a\b}
\ee
the corresponding 3-dimensional Ricci scalar.    

Our main job in this section is to derive these equations: the momentum and 
Hamiltonian constraints and the evolution equation for $K_{\a\b}$.  
We begin by showing that the projections $n_\a G^{\a\b} n_\b$ and 
$\g^\a{}_\d G^{\d\b} n_\b$ of the Einstein tensor 
are given in terms of the 3-metric and extrinsic curvature by  
\be
 \cblue n_\a G^{\a\b} n_\b = {}^3\!R+K^2-K_{\a\b}K^{\a\b}, 
\ee
and 
\be 
    -\g^\a{}_\d G^{\d\b} n_\b = D_\b(K^{\a\b}-\gamma^{\a\b}K). 
\index{momentum constraint}\label{eq:momconstraint0}
\ee
Because $\gamma^\a_\b n^\b=0$, this last equation involves only the Ricci tensor: 
\be\cblue 
   - \g^\a{}_\d R^{\d\b} n_\b 
		= D_\b(K^{\a\b}-\gamma^{\a\b}K). 
\label{e:gc1}\ee
It is one of two {\sl Gauss-Codazzi equations}. \index{Gauss-Codazzi equations}\index{Codazzi, Gauss-Codazzi equations}\index{Riemann tensor!Gauss-Codazzi equations}\index{curvature tensor!Gauss-Codazzi equations} 
\index{initial value problem!Gauss-Codazzi equations}  

The second Gauss-Codazzi equation relates the Riemann tensor of the 3-metric 
$\gamma_{\a\b}$ to the Riemann tensor of the 4-metric, and its contraction 
has the form 
\be \cblue 
  {}^3\!R_{\a\b} 
	= \gamma_\a{}^\sigma\gamma_\b{}^\tau \gamma^{\d\ep} R_{\sigma\d\tau\ep}  
		+ K K_{\a\b} - K_{\a\c}K^{\c}{}_{\b} .
\ee

We now derive the Gauss-Codazzi equations for a hypersurface $\Sigma_t$.  We want to relate its 
3-dimensional Riemann tensor to the to 4-dimensional Riemann tensor of spacetime and the 
extrinsic curvature $K_{\a\b}$ that describes the way $\Sigma_t$ is embedded in $M$. Here 
are two derivations; the first uses an appropriate choice of chart, while  
the second is covariant and uses the projection operator $\g^\a_\b$.  \\

\noindent{\sl Derivation by a choice of chart}\\

The quickest derivation uses the fact that an equation involving tensors holds 
if holds in one chart. We choose a chart $(x^0,x^i)$ for which $x^0$ is proper time from $\Sigma_t$ along curves normal to 
$\Sigma_t$ and $x^i$ is constant along the curves, so that $n^i = n^\a\nabla_\a x^i = 0$.  Then  
\be
	g_{00} = -1,\quad g_{0i}=0, \qquad n^\mu = \delta^\mu_0, \quad n_\mu = - \delta_\mu^0,  
\ee
($g_{0i}=0$ because $\bm\partial_i$ lies in an $x^0=$ constant surface and is therefore 
orthogonal to $\bm\pa_0 = n^\a$), implying
\be 
   \gamma_{ij} = g_{ij} \mbox{ (true for any $\{x^i\}$ on $\Sigma$)}, \qquad \gamma^{ij} = g^{ij}
				\mbox{ (true because } n^i = 0),\qquad g^{0i}=0. 
\label{e:gamg}\ee 
Eqs.~\ref{e:gamg} imply that the 3-dimensional Christoffel 
symbols are the spatial parts of the 4-dimensional Christoffel symbols
\[
   ^3\Gamma^i{}_{jk} = \Gamma^i{}_{jk}  , 
\]  
and the 3-dimensional Riemann tensor on $\Sigma_t$ therefore has components 
\be
     ^3\!R^i{}_{jkl}
	= \pa_k\Gamma^i{}_{jl} - \pa_l\Gamma^i{}_{jk} 
	  + \Gamma^i{}_{m k}\Gamma^m{}_{jl} - \Gamma^i{}_{m l}\Gamma^m{}_{jk} \ . 
\ee 

Because $n^\mu=\delta^\mu_0$, the components of $K_{ab}$ are 
\begin{align*}
    K_{ij} = - \frac12 \Lie_{\bm n} g_{ij} = -\frac12\partial_0 g_{ij}.    
\end{align*} 
Then 
\begin{align*}
    \Gamma^0{}_{ij} =  - K_{ij}, \quad \Gamma^i{}_{0j} = - K^i{}_j.   
\end{align*} 
The spatial components of the 4-dimensional Riemann tensor at $P$ are now 
\begin{align*}
     R^i{}_{jkl} &=  \pa_k\Gamma^i{}_{jl} - \pa_l\Gamma^i{}_{jk}
				 +  \Gamma^i{}_{\mu k}\Gamma^\mu{}_{jl} - \Gamma^i{}_{\mu l}\Gamma^\mu{}_{jk}
				    \\
		 &=\hspace{8mm}  ^3\!R^i{}_{jkl} \hspace{10mm}
					  +  \Gamma^i{}_{0 k}\Gamma^0{}_{jl} - \Gamma^i{}_{0 l}\Gamma^0{}_{jk}
				    \\
		 &= {}^3\!R^i{}_{jkl} + K^i{}_k K_{jl} - K^i{}_l K_{jk} 
\end{align*} 
and these are the nonzero components of the second Gauss-Codazzi relation (Gauss's Theorema Egregium)  
\[
    {}^3\!R^\a{}_{\b\c\d} = \g^\a_\ep \g_\b^\zeta \g_\c^\sigma \g_\d^\tau R^\ep{}_{\zeta\sigma\tau} 
			  - K^\a{}_\c K_{\b\d} + K^\a{}_\d K_{\b\c}
\]
Equivalently, denoting by $R_{abcd}$ the pullback of $R_{\a\b\g\d}$ to $\Sigma_t$, we have 
\be 
    {}^3\!R_{abcd} = R_{abcd} - K_{ac} K_{bd} + K_{ad} K_{bc}.  
\ee

\benr \item Prove the first Gauss-Codazzi relation, 
\[ 
   n_\d R^\d{}_{cab} = D_a K_{bc} - D_b K_{ac},  
\]
where $n_\d R^\d{}_{cab}$ is the pullback of $n_\d R^\d{}_{\c\a\b}$ to $\Sigma_t$.  
It is simplest to note that one can pick the chart $\{x^i\}$ on $\Sigma_t$ to make 
$\Gamma^i{}_{jk}=0$ at a point $P$; then show that the relation holds at $P$.  
Because $P$ is arbitrary, it holds everywhere.   
\een

\noindent{\sl Covariant derivations}\\

\index{Christoffel symbol!relation to extrinsic curvature}
We turn now to covariant proofs of the relations, with no coordinates introduced.  In the 
coordinate-based derivation, because $n_\a$ was the timelike basis vector, the 
Christoffel symbol $\Gamma^0{}_{ij}$ had the meaning of the spatial part of $\na_\a n_\b$, 
so that is the form we will find for $K_{\a\b}$.   The derivations end with 
the paragraph containing Eq.~\eqref{e:gc3}.

The Gauss-Codacci equations are true in 
any dimension; our convention is to use Latin indices for arbitrary dimensions, 
and that may make the derivations easier to read and follow. The derivations work 
in the following way.  The Riemann tensor in any dimension $n$ is defined by the 
commutator of two covariant derivatives \eqref{rabcd}: 
\index{commutator!of covariant derivatives}\index{Riemann tensor!on hypersurface}\index{curvature tensor!on hypersurface}
\be 
   R_{abc}{}^d\xi_d = [\nabla_a, \nabla_b] \xi_c.
\ee
Then the Riemann tensor on an $(n-1)$-dimensional submanifold is given by 
\be
  ^{(n-1)}\!R_{abc}{}^d\xi_d = [D_a, D_b] \xi_c 
\ee 
where $\xi_c$ is tangent to the submanifold ($\xi_c n^c=0$). 
As in the earlier covariant sections, we work with $n$-dimensional tensors, 
instead of introducing additional indices for tensors on an $(n-1)$-dimensional submanifold, 
and this is useful in the covariant derivations.  
And, as before, 
$D_a\xi_b := \g_a^c\g_b^d \nabla_c \xi_d$.  In the derivations, we 
multiply covariant derivatives $\na_b$ by projection operators $\gamma_a^b$. 
We then bring the projection operators inside $\nabla$, using 
\[
\nabla_a \g_b^c= \nabla_a (\d_b^c+n_b n^c) = \nabla_a (n_b n^c).
\] 

All that remains is to compute the spatial projection of $\na_a n^b$.  It is given by  
\be
   K_{ab} = -\g_a{}^c\na\!_c n_b. 
\label{e:Kab2}\ee
To show this, use the definition \eqref{eq:kab0} and the 
relation $\Lie_{\bf n} g_{ab} = \na\!_a n_b + \na\!_b n_a$ (see, e.g., p. \pageref{e:lieg}).  
We have 
\begin{align*}
 \Lie_{\bf n} \gamma_{ab} &= \Lie_{\bf n} (g_{ab}+n_a n_b) 
			   = 2\na\!_{(a}n_{b)} + n_b\Lie_{\bf n} n_a+ n_a\Lie_{\bf n} n_b \Rightarrow \\
 K_{ab} &= -\frac12\gamma_a^c \gamma_b^d \Lie_{\bf n} \gamma_{cd} 
 	   = -\gamma_a^c \gamma_b^d \na\!_{(c} n_{d)},\quad \mbox{where we used } \g_a^b n_b = 0.
\end{align*} 
Now use $n_a=-\alpha\na\!_a t$ to show that $\gamma_a^c \gamma_b^d \na\!_c n_d$ is already a symmetric tensor: 
\[
 \gamma_a^c \gamma_b^d \na\!_{[c} n_{d]} 
	= -\gamma_a^c \gamma_b^d \na\!_{[c} (\a \na\!_{d]}t) 
        = -\gamma_a^c\gamma_b^d \a \na\!_{[c}\na\!_{d]}t = 0.
\]
Then $K_{ab}= -\gamma_a^c \gamma_b^d \na\!_c n_d$.  Finally, 
the second index of $\na\!_a n_b$ is already orthogonal to $n^b\ $  
($n^b\na\!_a n_b = 0$), so the second projection $\g_b^d$ is not needed, 
implying Eq.~\eqref{e:Kab2}.\hspace{3mm}\raisebox{-1mm}{\text{\Large$\Box$}}\\

\noindent
First Gauss-Codazzi equation: $\gamma^a_b R^{bc}n_c = -D_b(K^{ab} -\gamma^{ab}K) $. 
 (This is Wald's problem 10.4.) 

We start with the definition of the Riemann tensor, writing 
$[\na_b,\na_a] n_c = R_{bac}{}^e n_e$. 
We hit all indices with $\gamma$'s, contracting the indices $b$ and $c$ with $\gamma^{bc}$: 
\begin{align}
\gamma^{ad}\gamma^{bc} [\na_b,\na_d] n_c &= \gamma^{ad}\gamma^{bc} R_{bdc}{}^e n_e = \gamma^{ad}R_d{}^e n_e,
\quad\mbox{where we used } R_{bdc}{}^e n^c n_e =0. 
\label{e:lhs}\\
\mbox{LHS} &= (\gamma^{ad}\gamma^{bc}-\gamma^{ab}\gamma^{cd}) \na_b\na_d n_c 
		= \gamma^a_e \gamma^b_f\g^c_g (\gamma^{ed}\gamma^{fg}-\gamma^{ef}\gamma^{gd}) \na_b\na_d n_c
\nonumber\\
	   &= \gamma^a_e \gamma^b_f\g^c_g \na_b[(\gamma^{ed}\gamma^{fg}-\gamma^{ef}\gamma^{gd}) \na_d n_c]
	     -\gamma^a_e \gamma^b_f\g^c_g \na_b(\gamma^{ed}\gamma^{fg}-\gamma^{ef}\gamma^{gd}) \na_d n_c .
\label{e:gcterms}
\end{align}
The first term is what we want, and the second term vanishes:  Use the form \eqref{e:Kab2} for $K_{ab}$ and then 
the definition of $D_a$ to write the first term as 
\begin{align}
 	\gamma^a_e \gamma^b_f\g^c_g \na_b[(\gamma^{ed}\gamma^{fg}-\gamma^{ef}\gamma^{gd}) \na_d n_c] 
		&= \gamma^a_e \gamma^b_f\g^c_g \na_b(-K^e{}_c\g^{fg} + K^g{}_c\g^{ef} ) \nonumber\\
	   	&= -(D_b K^{ab} - D^a K). 
\label{e:rhs}\end{align} 
To get the last equality, we used $D_f\gamma^{fg}=0, \ D_f\gamma^{ef}=0$.

To show that the second term in \eqref{e:gcterms} vanishes, we again use 
Eq.~\eqref{e:Kab2} as well as the relations 
\be
  n^b\na\!_a n_b = 0, \quad \gamma_{ab}n^b=0, \quad
	\na\!_a\gamma_{bc} = \na\!_a(n_b n_c).    
\label{e:relns}\ee 
We have
\begin{align*}
  -\gamma^a_e \gamma^b_f\g^c_g \na_b(\gamma^{ed}\gamma^{fg}-\gamma^{ef}\gamma^{gd}) \na_d n_c
    &= -\gamma^a_e \gamma^b_f\g^c_g 
	[\na_b(n^en^d) \g^{fg} + \gamma^{ed}\na_b(n^fn^g) - \na_b(n^en^f) \g^{gd} - \gamma^{ef}\na_b(n^gn^d)]
	 \na_d n_c\\
    &= -\gamma^a_e \gamma^b_f\g^c_g 
	[\na_b(n^e)n^d \g^{fg}   - \gamma^{ef}\na_b(n^g)n^d]\na_d n_c \\
    &=  K^{ca}n^d \na\!_d n_c - K^{ac}+= n^d\na\!_d n_c
     = 0, 
\end{align*}

Then the right side of \eqref{e:lhs} is equal to the expression \eqref{e:rhs}, giving the 
first Gauss-Codazzi relation (in its contracted form), Eq.~\eqref{e:gc1}. \hspace{3mm}
\raisebox{-1mm}{\text{\Large$\Box$}}\\ 

The second Gauss-Codazzi relation relates the Riemann tensor $R_{abcd}$ 
on an $n$-dimensional manifold with metric $g_{ab}$ to the Riemann tensor
$^{(n-1)}\!R_{abcd}$ on an $(n-1)$-dimensional hypersurface:  
\be\crv
  ^{(n-1)}\!R_{abc}{}^d 
	= \g_a^l \g_b^m \g_c^n \g^d_p R_{lmn}{}^p + K_a{}^dK_{bc}-K_{ac}K_b{}^d.
\label{e:gc2}\ee
To prove this, we again write the Riemann tensor in terms 
of the commutator,  
\be
  ^{(n-1)}\!R_{abc}{}^d\xi_d = [D_a, D_b] \xi_c 
\ee 
this time with $\xi_c$ a vector tangent to the submanifold ($\xi_c n^c=0$). 
We have 
\[
 D_a D_b \xi_c =\g_a^l \g_b^m \g_c^n \na\!_l (D_m \xi_n) 
	= \g_a^l \g_b^m \g_c^n \na\!_l (\g_m^r \g_n^s \na\!_r\xi_s) \ .
\]
In the derivation, we once again use Eqs.~\eqref{e:Kab2} and \eqref{e:relns}. In particular, each time $\nabla_.$ hits $n_.$
 we get $K_{..}\ \ $. We also need 
 \mbox{$n^b\na\!_a\xi_b = -\xi_b \na\!_a n^b +\na\!_a(\xi_b n^b) = -\xi_b \na\!_a n^b$}.
 
We have
\begin{align*}  
  D_a D_b \xi_c &= \g
_a^l \g_b^m \g_c^n\left[\na\!_l(n_m n^r) \g_n^s \na\!_r\xi_s 
	  + \g_m^r\na\!_l(n_n n^s) \na\!_r\xi_s 
		+ \g_m^r \g_n^s\na\!_l\na\!_r\xi_s \right] \\
 &= - K_{ab} n^r \g_c^s\na\!_r \xi_s -\gamma^r_b K_{ac}n^s\nabla_r \xi_s 
	+ \g_a^l \g_b^m \g_c^n \na\!_l \na\!_m\xi_n\ .
\end{align*}
Because the first term is symmetric in $a$ and $b$, its contribution to 
$[D_a, D_b] \xi_c$ vanishes.  The second term is 
\[
  \gamma^r_b K_{ac} \xi_s \na\!_r n^s = - K_{ac} K_b{}^d\xi_d. 
\]
Finally, antisymmetrizing on $a$ and $b$ gives the $n$-dimensional Riemann tensor in the last term: 
\be
[D_a, D_b] \xi_c = (-K_{ac} K_b{}^d+K_{bc} K_a{}^d)\xi_d 
			+ \g_a^l \g_b^m \g_c^n \g_p^d R_{lmn}{}^p\xi_d, 
\ee
where we have used $\xi_a = \gamma_a^b\xi_b$ to get the fourth $\gamma$ in the last term. 
Because this equation holds for all spatial $\xi_d$, we recover the second 
Gauss-Codazzi relation \eqref{e:gc2}.   

Contracting Eq.~\eqref{e:gc2} on $b$ and $d$ and then changing the index $c$ back to $b$, 
gives us the contracted identity, 
\be\cblue
   ^{(n-1)}\!R_{ab} = \gamma_a^c \g_b^d \g^{ef} R_{cedf} + K_a{}^c K_{bc} - K_{ab} K, 
\label{e:gc3}\ee
where $K:=K_c{}^c$.\  \  \raisebox{-1mm}{\text{\Large$\Box$}}

We now return to four dimensions and the Einstein equation.  
From the first Gauss-Codazzi equation~\eqref{e:gc1} we can immediately write the momentum constraint 
in terms of $\g_{\a\b}$ and $K_{\a\b}$, recovering 
Eq.~\eqref{e:momconstraint}: 
\[
  D_\b(K^{\a\b}-\gamma^{\a\b}K) = -8\pi \g^\a{}_\b T^{\b\c} n_\c. 
\]

To obtain the Hamiltonian constraint, we need another contraction of the second Gauss-Codazzi equation \eqref{e:gc3}, namely 
\be
\tr = \g^{\a\b}\, \tr_{\a\b} = \g^{\a\b}\g^{\c\d}R_{\a\c\b\d} +K^{\a\b}K_{\a\b} - K^2
\label{e:gc5}\ee
The first term on the right is 
\begin{align}
(g^{\a\b} + n^\a n^\b)(g^{\c\d}+n^\c n^\d) R_{\a\c\b\d} 
	&= R + 2n^\a n^\b R_{\a\b} = 2 G_{\a\b}n^\a n^\b 
\end{align}
whence
\[ 
  2 G_{\a\b}n^\a n^\b = \tr -K^{\a\b}K_{\a\b} + K^2, 
\] 
and the Hamiltonian constraint, Eq.~\eqref{e:Hamconstraint}, is
\[
  \tr -K^{\a\b}K_{\a\b} + K^2 = 16\pi T^{\a\b}n_\a n_\b,
\]
as claimed.

The derivation of the evolution equation \eqref{e:gabevol} for $K_{\a\b}$ 
again involves Gauss-Codazzi and needs only one additional relation, 
the derivative of $n^\a$ along itself:  
\be \cblue
  n^\b\na\!_\b n_\a = D_\a \ln\alpha.  \cb
\label{e:nbna1}\ee
To see this, use $n_\a = -\alpha \na\!_\a t$ to write 
\be
  n^\b\na\!_\b\, n_\a = - n^\b \na\!_\b\, \a\,  \na\!_\a t - \a n^\b \na\!_\b\na\!_\a t.  
\label{e:nbna2}\ee
From the definition ~\eqref{e:alpha} of $\alpha$, we have 
\[
  \na\!_\a \alpha = \alpha^{\!3}\ \na^\b t \na\!_\b\na\!_\a t = - \a^2 n^\b\ \na\!_\b\na\!_\a t.
\]
Using this in the second term of Eq.~\eqref{e:nbna2} and replacing $-\na\!_\a t$ by $\a^{-1}n_\a$ 
in the first term gives 
\[
  n^\b\na\!_\b n_\a = \a^{-1} (n_\a n^\b \na\!_\b \a + \na\!_\a \a) 
		    = \a^{-1} \g^\b_\a \nas_\b\, \a, 
\]  
implying Eq.~\eqref{e:nbna1}.

We can now compute $\Lie_{\bm n} K_{\a\b}$, noting first that it is spatial:\\
\centerline{
 $n^\b \Lie_{\bm n} K_{\a\b} = \Lie_{\bm n} (n^\b K_{\a\b}) - (\Lie_{\bm n} n^\b)K_{\a\b}
				=0-0=0$.} 
Then 
\[
    \Lie_{\bm n} K_{\a\b} 
	= -\g_\a^\sigma \g_\b^\tau \Lie_{\bm n}(\g_\sigma^\d \na_\d n_\tau)
	 = {\color{red}-\g_\a^\sigma \g_\b^\tau \Lie_{\bm n}(\g_\sigma^\d) \na_\d n_\tau} 
	   {\cblue -\g_\a^\sigma \g_\b^\tau \g_\sigma^\d\Lie_{\bm n} \na_\d n_\tau }
\]
In the red term on the right, because $\Lie_{\bm n} n^\d=0$, 
\[
\Lie_{\bm n}(\g_\sigma^\d)   = \Lie_{\bm n}(n_\sigma) n^\d
	= (n^\g \na_\g n_\sigma + n_\g \na_\sigma n^\g)n^\d = n^\g (\na_\g n_\sigma) n^\d = D_\sigma\ln\a\, n^\d,  
\]
and the red term becomes 
\[
{\color{red}-\g_\a^\sigma \g_\b^\tau \Lie_{\bm n}(\g_\sigma^\d) \na_\d n_\tau }
	= -D_\a\ln\alpha\, D_\b\ln\alpha.
\]
The blue term is \vspace{-2mm}
\begin{align}
    {\cblue -\g_\a^\sigma \g_\b^\tau \Lie_{\bm n} \na_\sigma n_\tau} 
	&= - \g_\a^\sigma \g_\b^\tau \left( n^\g\nas_\g \nas_\sigma n_\tau
		+ \nas_\g n_\tau \nas_\sigma n^\g + \nas_\sigma n_\g \nas_\tau n^\g \right)\nonumber\\
	&= -\g_\a^\sigma \g_\b^\tau n^\g\nas_\g \nas_\sigma n_\tau 
		- K_{\c\b} K_\a{}^\c -K_{\a\c}K_\b{}^\c\nonumber\\
\mbox{1st term}	
	&=-\g_\a^\sigma \g_\b^\tau n^\g 
	   \left(  [\nas_\g,\nas_\sigma] n_\tau +\nas_\sigma \nas_\g n_\tau\right)
\nonumber\\
	& = -\g_\a^\sigma \g_\b^\tau 
	   \left[ n^\g R_{\g\sigma\tau}{}^\d n_\d +\nas_\sigma(n^\g \nas_\g n_\tau) 
		  - \nas_\sigma n^\g \nas_\g n_\tau \right]
\nonumber\\
& =  R_{\a\g\b\d}n^\g n^\d  - D_\a D_\b\ln \alpha + K_\a{}^\c K_{\b\c}
\end{align} 
Then
\[
{\cblue -\g_\a^\sigma \g_\b^\tau \Lie_{\bm n} \na_\sigma n_\tau}  
	=  R_{\a\g\b\d}n^\g n^\d - D_\a D_\b\ln \alpha - K_\a{}^\c K_{\b\c} \nonumber
\]
Combining red and blue terms, we have 
\be
\Lie_{\bm n} K_{\a\b} = R_{\a\g\b\d}n^\g n^\d - \frac1\alpha\,D_\a D_\b \alpha - K_\a{}^\c K_{\b\c}.   
\label{e:LieK}\ee
Finally, we use the contracted Gauss-Codazzi equation \eqref{e:gc3} and the remaining part of 
the Einstein equation to write the Riemann-tensor term as follows: 
\begin{align*}
   R_{\a\g\b\d}n^\g n^\d &= \g_\a^\p \g_\b^\q n^\c n^\d R_{\c\p\d\q}
			 = \g_\a^\p \g_\b^\q (\g^{\c\d}-g^{\c\d})R_{\c\p\d\q}\\
 \g_\a^\p \g_\b^\q \g^{\c\d}R_{\c\p\d\q}
			&= \tr_{\a\b} -K_{\a\c}K_\b{}^\c +K K_{\a\b},
				 \quad\mbox{by \eqref{e:gc3}};\\
  -\g_\a^\p \g_\b^\q g^{\c\d}R_{\c\p\d\q}
			&= -\g_\a^\p \g_\b^\q R_{\p\q} 
			 = -8\pi \g_\a^\p \g_\b^\q(T_{\p\q}-\frac12 g_{\p\q} T)
			= -8\pi (\g_\a^\p \g_\b^\q T_{\p\q}-\frac12 \g_{\a\b} T)\\
\Longrightarrow\ \  R_{\a\g\b\d}n^\g n^\d &=  \tr_{\a\b} -K_{\a\c}K_\b{}^\c +K K_{\a\b}
			-8\pi (\g_\a^\d \g_\b^\ep T_{\d\ep}-\frac12 \g_{\a\b} T).  
\end{align*}

Using this equation to replace $R_{\a\g\b\d}n^\g n^\d$ in Eq.~\eqref{e:LieK}, we 
obtain the second evolution equation \eqref{e:kabevol}.\  \  \raisebox{-1mm}{\text{\Large$\Box$}}

This completes our derivations of the initial value equations.  To make contact 
with the numerical relativity literature, two additional pieces are needed.  
Because derivatives are computed in a general timelike direction $t^\a$ that is 
not along $n^\alpha$, we need to express $\Lie_{\bm n}$ in terms of $\Lie_{\bm t}$.  
And we need explicit forms of the projections of the stress-energy tensor for a fluid.

\section{Initial value equations in the NR literature.} 

Because of the dragging of frames, the timelike 
Killing vector $t^\a$ of a rotating star or a rotating black hole is not in 
general orthogonal to the surfaces $\Sigma_t$. Not surprisingly, 
in the more general spacetimes that describe oscillating stars and binary 
systems, convenient choices of coordinates $\{t,x^i\}$ again have a time direction 
$t^\a\equiv \bm\pa_t$ that is not along the normal.  That is, a chart $\{t,x^i\}$
picks out the direction of time $\bm \pa_t$ along which the coordinates $x^i$ are 
fixed and only $t$ changes. We write $t^\alpha$ for $\bm\pa_t$.  Then, in the chart  
$\{t,x^i\}$, 
\bsube\be
	t^\mu = \delta_t^\mu, 
\label{e:tmu}\ee
\be
\nabla_\mu t = \delta_\mu^t.
\ee\esube

The vector field $t^\alpha$ is the sum of orthogonal components, $\alpha n^\alpha $
along the unit normal $n^\alpha$ and a vector $\beta^\alpha$ tangent to the slices: 
\be 
   t^\alpha = \alpha n^\alpha + \beta^\alpha,\ \mbox{where }\beta^\alpha n_\alpha = 0.
\label{eq:ta}\ee 
\begin{figure}[H]
               \begin{center}
		\includegraphics[width=0.4\textwidth]{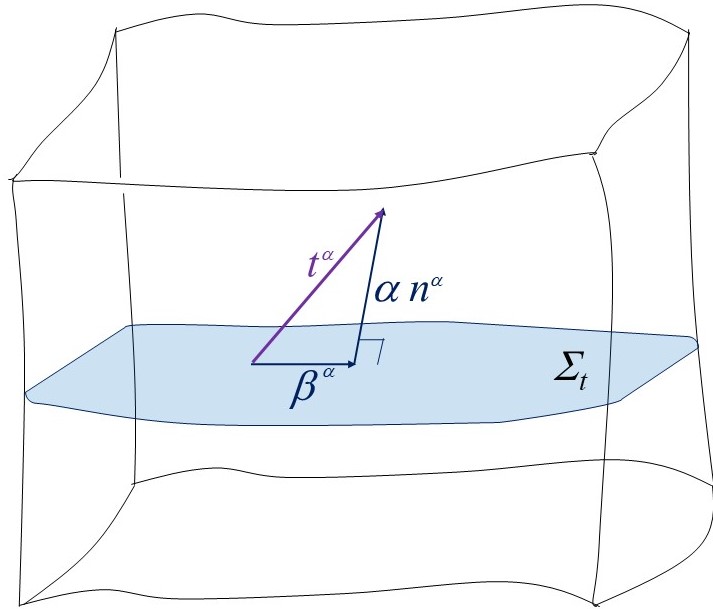}
		\end{center}
		\caption{The decomposition of $t^\alpha$ into lapse and shift.}
\end{figure}
As before, the component $\alpha$ of $t^\a$ along the unit normal is called the 
{\em lapse}, because $\alpha\, dt$ is the proper time elapsed in a normal direction 
between coordinate times $t$ and $t+dt$.  The spatial vector $\beta^\a$ is called the 
{\em shift}, because $\beta^\a\, dt$ is the spatial shift in the position from a 
path along the normal $n^\a$ to the actual path along $t^\alpha$.

One regards the choice of the vector $t^\a$ as the direction in which the time-evolution 
proceeds, noting that the time derivative of a function $f$, defined by
\be
 \dot f := \Lie_{\bm t} f = t^\alpha \nabla_\alpha f,
\label{eq:dott}\ee 
is the partial derivative $\dot f = \partial_t f$ in the chart $\{t,x^i\}$. 
For any covariant spatial tensor $S_{\a\cdots\b}$, 
\be
  \Lie_{\bm t} S_{\a\cdots\b} 
		= (\alpha\Lie_{\bm n} + \Lie_{\bm \b}) S_{\a\cdots\b}.  
\label{e:Liet}\ee
\benr
\item Use the fact that $S_{\a\cdots\b}$ is spatial to check this relation.  Show 
that it is not true for a contravariant tensor by checking that it does not hold 
for a contravariant spatial vector $v^\a$, and show 
\[
  \Lie_{\bm t} v^\b = (\alpha\Lie_{\bm n} + \Lie_{\bm \b}) v^\b + n^\b v^\g\na_\g \alpha.
\]
\een
Because $t^\a = \bm\pa_t$ has components $\delta^\mu_t$ in the chart $\{t,x^i\}$.  
the Lie derivative of any tensor $T^{\a\cdots\b}{}_{\c\cdots\d}$, 
has, in this chart, components
\be
  \Lie_{\bm t}T^{\mu\cdots\nu}{}_{\kappa\cdots\lambda} = \partial_t T^{\mu\cdots\nu}{}_{\kappa\cdots\lambda}.  
\ee 
More generally, for any nonzero vector field $\xi^a$ on a manifold, if one chooses the 
parameter distance along its integral curves as a coordinate, say $x^1$, then in a 
chart $x^1, x^2, \cdots, x^n$, we have $\xi^\mu = \delta_1^\mu$, implying 
\[
  \Lie_{\bm\xi} T^{i\cdots j}{}_{k\cdots l} = \pa_1 T^{i\cdots j}{}_{k\cdots l}.
\]   

In particular, the derivatives 
$\Lie_{\bm t} \g_{\a\b}$ and $\Lie_{\bm t}K_{\a\b}$ have spatial components 
\be
  \Lie_{\bm t} \g_{ij} = \pa_t \g_{ij}, \qquad \Lie_{\bm t} K_{ij} = \partial_t K_{ij},\qquad 
\ee 
implying the pullbacks of $\g_{\a\b}$ by the family of maps $\Sigma \rightarrow \Sigma_t$ 
is the time-dependent tensor $g_{ab}(t)$ on $\Sigma$, and the family of pullbacks of 
$\Lie_{\bm t} \g_{\a\b}$ is $\pa_t g_{ab}(t)$. \\

The spacetime metric at a time $t$ can be written in terms of the lapse, the shift, and 
$\gamma_{\alpha\beta}$.  A small coordinate displacement $(\Delta t, \Delta x^i)$ gives 
a change in proper time $\alpha\Delta t$ perpendicular to $\Sigma_t$ and a net spatial 
coordinate change $\Delta x^i + \beta^i \Delta t$ tangent to $\Sigma_t$. The proper length 
of the displacement is then given by 
\[ 
   \Delta s^2 = - \alpha^2 \Delta t^2 
		+ \gamma_{ij}(\Delta x^i + \beta^i\Delta t)(\Delta x^j +\beta^j \Delta t)
\]
implying 
\be
ds^2 = -\alpha^2 dt^2 +\gamma_{ij}(dx^i+\beta^i dt)(dx^j+\beta^j dt).
\label{e:gmunu}
\ee
More formally, in the coordinates $\{t,x^i\}$, the 3+1 decomposition of the 4-metric,  
\mbox{$g_{\alpha\beta} =  \gamma_{\alpha\beta} - n_\alpha n_\beta$}, has, from 
Eq.~(\ref{eq:ta}) and the relations $\beta^t=0$ and $n_i=0$, components
\be 
  g_{tt} = t_\a t^\a=-\a^2+\b_i \b^i, \qquad g_{ij}=\gamma_{ij},\qquad
g_{ti} = g_{\mu i} t^\mu =\alpha n_i + \beta_i  = \beta_i,
\ee
or 
\[
  [g_{\mu\nu}] = \begin{bmatrix} -\alpha^2 + \b_i\b^i & \beta_i\\
				 \beta_i & \gamma_{ij} 
		 \end{bmatrix}.
\]

The corresponding decomposition of the volume element is given by 
\be
\cblue\sqrt{|g|}= \alpha\sqrt\gamma\cb,
\ee
where $\c={\rm det}(\g_{ij})$.  This follows from the fact that the  
determinant $\gamma$ is the cofactor of $g_{tt}
$ in the matrix $g_{\mu\nu}$, so 
$\displaystyle   g^{tt} = \frac\gamma g $. Because $g^{tt} = - \a^{-2}$, we have
\be 
   \ g=-\a^2\gamma, \qquad \sqrt{|g|} = \alpha\sqrt\gamma.    
\ee 
The inverse metric has components
\bea
g^{tt} &=& -\frac{1}{\a^2}, \qquad
g^{ti} = \frac{\b^i}{\a^2}, \qquad 
g^{ij} = \g^{ij}-\frac{\b^i \b^j}{\a^2},
\label{eq:contrametric}
\eea\label{eq:g3+1}
or
\be
   [g^{\mu\nu}] 
         = \begin{bmatrix} -\alpha^{-2} & \beta^i\alpha^{-2}\\
			\b^i \alpha^{-2}& \gamma^{ij} -\beta^i\beta^j \alpha^{-2}
	   \end{bmatrix}.
\ee
Equivalently,%
\footnote{This is a shorthand for $g^{\mu\nu}\bm\partial_\mu\otimes\bm\partial_\nu 
	= -\frac1{\alpha^2}(\bm\partial_t-\beta^i\bm\partial_i)\otimes(\bm\partial_t-\beta^i\bm\partial_i)
	  +\gamma^{ij}\bm\pa_i\otimes\bm\pa_j$.
}
\be
  g^{\mu\nu}\partial_\mu\partial_\nu 
	= -\frac1{\alpha^2}(\partial_t-\beta^i\partial_i)^2+\gamma^{ij}\pa_i\pa_j.
\label{e:ginv}\ee  

The next two exercises give a quick way to invert the metric $g_{\alpha\beta}$, 
showing why a metric written in the form \eqref{e:gmunu} has an inverse of the 
form \eqref{e:ginv}.
\benr
\item Let $\omega^\mu$ and $e_\mu$ be dual bases, i.e., 
$\omega^\mu(e_\nu)= \delta^\mu_\nu$.  
Show that a metric $g_{\mu\nu}\omega^\mu\omega^\nu$ has inverse 
$g^{\mu\nu} e_\mu e_\nu$, where, as usual, $[g^{\mu\nu}]$ is the matrix inverse of 
$[g_{\mu\nu}]$.
\item Invert the metric \eqref{e:gmunu} to obtain \eqref{e:ginv} by showing that 
the basis $\omega^0 =dt$, $\omega^i = dx^i+\beta^i dt$ has dual 
$e_0 = \partial_t - \beta^i\partial_i$, $e_i = \partial_i$ and using the 
previous exercise.
\een

Finally, for those of you familiar with wedge products: \vspace{-2mm}
\benr
\item\label{ex:det}
Show that a metric of the form $g_{\mu\nu}dx^\mu dx^\nu 
		= \bar g_{\mu\nu}\omega^\mu\omega^\nu$ 
has determinant satisfying \\
$\sqrt g\, dx^0\wedge dx^1\wedge dx^1\wedge dx^3 
		= \sqrt{\bar g}\, \omega^0\wedge \omega^1\wedge\omega^2\wedge\omega^3$. 
Compute the expression on the right for the metric \eqref{e:gmunu}, 
with the basis $\omega^\mu$ of the last problem to show 
$\sqrt{g} = \alpha\sqrt{\gamma}$.   
\een

Using Eq. (\ref{e:Liet}), we can write the dynamical equations \eqref{e:gabevol} and 
\eqref{e:kabevol} for $\gamma_{\a\b}$ and $K_{\a\b}$ in the forms
\be
  \dot\gamma_{\a\b} = - 2\alpha K_{\a\b} +  \Lie_{\bm \b}\g_{\a\b}, 
\label{e:gabevol1}\ee
\be
  \dot K_{\a\b} 
	=   \alpha\left(\tr_{\a\b} - 2K_{\a\c}K_\b{}^\c + K K_{\a\b} \right) 
	  - \,D_\a D_\b \alpha -8\pi\alpha(\g_\a^\d \g_\b^\ep T_{\d\ep}-\frac12 \g_{\a\b} T)
	  + \Lie_{\bm \b}K_{\a\b}. 
\label{e:kabevol1}\ee

The system (\ref{e:gabevol1}), (\ref{e:kabevol1}) describes the 
evolution of the spatial tensors $\g_{\a\b}$ 
and $K_{\a\b}$.  Because the equations do not contain time derivatives of
the lapse function $\alpha$ or of the shift vector $\beta^\a$, these 
metric functions are not dynamical variables. One can regard the four degrees 
of gauge freedom associated with the choice of coordinates $(t, x^i)$ as 
the freedom to choose $\alpha$ and $\beta^\a$.  From this point of view, 
once $\alpha$ and $\beta^\a$ are prescribed, initial data for the geometry 
consist of initial values $\gamma_{\a\b}(0)$ and $K_{\a\b}(0)$ of the 
metric and extrinsic curvature satisfying the constraint equations 
\ref{e:momconstraint}), (\ref{e:Hamconstraint}.

For black hole spacetimes, there is no matter, and the 3+1 decomposition of 
the Einstein equation, the set \ref{e:momconstraint})-\eqref{e:kabevol} \  \ 
(or (\ref{e:momconstraint}),(\ref{e:Hamconstraint},(\ref{e:gabevol1}), (\ref{e:kabevol1})), 
determines the spacetime metric given initial data and a choice of gauge.  
For spacetimes with matter, one must add equations determining the 
evolution of the matter. \\
\newpage

\noindent{\sl Matter}

 As discussed in Sect.~\ref{s:neutron stars}, one can accurately model stationary neutron stars, and also oscillating neutron stars and  
neutron stars in binary inspiral, as perfect fluids, whose stress-energy 
tensor has the form \eqref{e:Tab_fluid}, 
\be
  T^{\a\b} = \rho u^\a u^\b + P q^{\a\b},
\ee 
where
\be
q_\a^\b:= \d_\a^\b + u_\a u^\b
\ee
is the projection operator orthogonal to $u^\a$. 
Here $\rho = T_{\a\b} u^\a u^\b$ is the energy density seen by a family of {\sl comoving observers}, observers with velocity $u^\alpha$.  
\index{energy density!of comoving observer}\index{density!energy density}\index{energy!energy density of comoving observer}

The evolution of the matter variables $u^\a$, $\rho$, $P$ is given by the equation, 
\be
  \na_\b T^{\a\b}=0.  
\ee 
For analytic problems, the equation is ordinarily decomposed into parts along and orthogonal to the velocity $u^\alpha$ 
(see Sects.~\ref{s:perfect_fluid}) and \ref{s:em_fluid_cst}, p.~\pageref{p:perfect fluid}), 
giving equations expressing conservation of 
energy, \index{conservation laws!energy}\index{conservation laws!momentum}
\be
   \nabla_\beta (\rho u^\beta ) = -P\nabla_\b u^\b,
\label{e:econs}\ee
and momentum (the relativistic Euler equation)     
\be (\rho +P)u^\beta\nabla_\beta u^\alpha =-q^{\alpha\beta}\nabla_\beta P.
\label{e:relEuler}\ee
\index{Euler equation!relativistic}
The first equation gives the time derivative $u^\a \na_\a \rho$ of the energy density; the 
second gives the time derivative of the velocity, and an equation of state of the form $P=P(\rho)$ completes the system.  

For numerical work, one generally uses the 3+1 decomposition along and orthogonal to $n_\a$.  
The energy and momentum densities and spatial stress tensor seen by an observer with velocity 
$n^\a$ are 
\bsube\begin{align}
\rho_E = T_{\a\b}n^\a n^\b, \qquad j^\a = -\g^\a_\d n_\b T^{\d\b}, 
\qquad S^{\a\b} =  \g^\a_\d \g^\b_\ep T^{\d\ep}, 
\end{align}\esube
\index{energy density!observer with velocity $n^\a$} 
with corresponding decomposition
\be
  T^{\a\b} = \rho_E n^\a n^\b + j^\a n^\b + n^\a j^\b + S^{\a\b}.
\ee 
Before writing the decomposition of $\na_\b T^{\a\b}=0$, it is helpful to look at the 
simpler conservation law $\na_\a J^\a =0$, where $J^\a$ is a conserved current:  Examples 
are the currents associated with conservation of baryons and conservation of charge. 
Decompose the baryon current $J^\a$ into baryon density and spatial current ${\rm j}^\a$,
\index{conservation laws!baryons}
\be
   J^\a = \rho_0 n^\a + {\rm j}^\a, \quad \mbox{with}\quad \rho_0 = -J^\a n_\a, \quad 
		{\rm j}^\a = \gamma^\a_\b J^\b,  
\ee 
(Note that $\rho_0$ here is the baryon density seen by an observer with velocity $n^\a$, 
not that of a comoving observer.)\,  Next, 
the contraction of Eq.~\eqref{e:Kab2} gives
\be
   K = -\gamma^{\a\b}\na_\a n_\b = -\na_\b n^\b.   
\ee
Then 
\begin{align*}
  \na_\b J^\b &= \na_\b (\rho_0 n^\b + {\rm j}^\b) = n^\b\na_\b \rho_0 + \rho_0 \na_\b n^\b 
		+ (\gamma^{\a\b} - n^\a n^\b) \na_\a {\rm j}_\b \\
	      &= n^\b\na_\b \rho_0 - K\rho_0 + D_\b {\rm j}^\b + {\rm j}^\b D_\b\ln\a,
\end{align*}
where we use $n^\b \na_\a {\rm j}_\b = - {\rm j}_\b \na_\a n^\b$ and $n^\a\na_\a n_\b = \na_\b \ln \a$ 
to get the last term.  Equivalently, 
\be
  \Lie_{\bm t} \rho_0 = \Lie_{\bm \b} \rho_0 + \a K\rho_0 - D_\b(\a {\rm j}^\b).  
\ee

We can now follow essentially the same pattern in writing  
the projections of $\na_\b T^{\a\b}=0$ along and orthogonal to $n^\a$.  
These equations for the evolution of energy density and momentum density are 
\bsube
\begin{align}
 \Lie_{\bm t} \rho_E &= \Lie_{\bm\beta} \rho_E +\alpha K\rho_E -\frac1\alpha D_\b(\alpha^2 j^\b) 
		  +\alpha K_{\a\b}S^{\a\b},
\label{e:econs3+1}\\
 \Lie_{\bm t} j_\a &= \Lie_{\bm\beta} j_\a +\a K j_\a 
		- D_\b(\alpha S_\a{}^\b) -\rho_E D_\a \alpha.
\label{e:euler3+1}
\end{align}\esube

\benr
\item Derive Eq.~\eqref{e:econs3+1}.  
\een

Here's the derivation of Eq.~\eqref{e:euler3+1}: \\
The time derivative of $j_\a$ comes from the blue term below where 
$n^\b$ hits $\na_\b$.
\begin{align*}
  \g_\a^\d \na_\b T^\b{}_\d &=\g_\a^\d \na_\b (\rho_E n^\b n_\d + j^\b n_\d + j_\d n^\b+S^\b{}_\d)\\
	&= \rho_E n^\b\na_\b n_\a + j^\b \g_\a^\d \na_\b n_\d + j_\a \na_\b n^\b 
		+ {\cblue \g_\a^\d n^\b\na_\b j_\d} + \g_\a^\d \na_\b S^\b{}_\d \\
	&= \rho_E D_\a \ln\a -\cancel{ K_\a{}^\b j_\b }{-15}- K j_\a 
		+\g_\a^\d( \Lie_{\bm n} j_\d -\cancel{j_\b \na_\d n^\b}{-15}) 
		+ \g_\a^\d(\g^{\b\ep}-n^\b n^\ep) \na_\b S_{\d\ep}\\
	&= \rho_E D_\a \ln\a - K j_\a + \g_\a^\d \Lie_{\bm n} j_\d  
		+ D_\b S_\a{}^\b + \g_\a^\d n^\b \na_\b n^\ep S_{\d\ep}\\
	&= \rho_E D_\a \ln\a -K j_\a + \g_\a^\b \Lie_{\bm n} j_\b  
		+ D_\b S_\a{}^\b + S_\a{}^\b \na_\b\ln\alpha.
\end{align*}

The equation of state and local thermodynamics of the fluid involve the 
pressure and comoving energy density $\rho$, so one needs the relations between 
$\rho_E, j^\a$ and $\rho, u^\a$.  These involve the 3+1 decomposition of 
$u^\a$, 
\be
 u^\a = \Upsilon(n^\a+ v^\a), \mbox{ where } \Upsilon = - u^\a n_\a, \quad 
				\qquad v_\a = \Upsilon^{-1}\g_{\a\b} u^\b.
\ee
Then $v^\a$ is the 3-velocity of an observer with 4-velocity $u^\a$, and 
$u^\a u_\a = -1$ implies 
\[
   \Upsilon = \frac1{\sqrt{1-v^2}}.  
\]
We easily find 
\bsube
\begin{align}
   \rho_E &= \frac{\rho + Pv^2}{1-v^2}, \qquad j^\a = (\rho+P)  \frac{v^\a}{1-v^2},
\label{e:rhoE_j}\\
   S_{\a\b} &= j_\a v_\b-P\g_{\a\b}.
\end{align}\esube
In numerical evolutions that use variables $j^\a$ and $\rho_E$, one must numerically 
invert Eqs.~\eqref{e:rhoE_j}, using the equation of state $P=P(\rho)$, to obtain the 
{\sl primitive} fluid variables $\rho$, $P$ and $u^\a$.

\newpage
    
\noindent
{\sl 3-dimensional form of the equations} 

By sticking to 4-dimensional indices we simplified the covariant derivation of the 3+1 form of the 
field equation.   To write the four equations, \eqref{e:momconstraint}, \eqref{e:Hamconstraint}, 
\eqref{e:gabevol1}, \eqref{e:kabevol1}, using 3-dimensional indices is simple.  
Because every spatial tensor $T^{\a\cdots\b}{}_{\c\cdots\d}$ 
at a time $t$ (at a point of $\Sigma_t\subset M$) corresponds to a unique tensor 
$T^{a\cdots b}_{c\cdots d}$ on $\Sigma$, we can write the constraint equations 
as 
\index{constraints!gravitational|textbf}\index{momentum constraint|textbf}\index{Hamiltonian constraint|textbf}
\index{initial value problem!Hamiltonian constraint}
\index{initial value problem!Momentum constraint} 
\begin{align}
D_b(K^{ab}-\gamma^{ab}K) &= 8\pi j^a, \\
  \tr -K^{ab}K_{ab} + K^2 &= 16\pi \rho_E.
\end{align} 
The components of the momentum constraint in a chart $\{x^i\}$ on $\Sigma$ are the same 
as its nonzero components in the chart $\{t,x^i\}$ on $M$, 
\[
  D_j(K^{ij}-\gamma^{ij}K) = 8\pi j^i.  
\]
Similarly, using Eq.~\eqref{e:Liet} to replace $\Lie_{\bm t}$ on $M$ by $\partial_t$ on $\Sigma$, 
we obtain the dynamical equations on $\Sigma$ corresponding to \eqref{e:gabevol1} and \eqref{e:kabevol1} 
on $M$: 
\bsube
\begin{align}
   \partial_t\gamma_{ab} &= - 2\alpha K_{ab} +  \Lie_{\bm \b}\g_{ab}, 
\label{e:gabevola}\\
  \pa_t K_{ab} 
	&=   \alpha\left(\tr_{ab} - 2K_{ac}K_b{}^c + K K_{ab} \right) 
	  - \,D_a D_b \alpha -8\pi\alpha(T_{ab}-\frac12 \g_{ab} T)
	  + \Lie_{\bm \b}K_{ab}. 
\label{e:Kabevola}\end{align} \esube
\index{initial value problem!dynamical equations}

We have already seen that a spatial tensor $K_{\a\b}$  
corresponds to a tensor $K_{ab}$ whose components in a chart $\{t,x^i\}$ 
are the spatial components $K_{ij}$ of $K_{\a\b}$.  
The components of the 3+1 Einstein equations on  
$\Sigma$ in the chart $x^i$ are then just the spatial components of the corresponding equations on $M$ in the chart $t,x^i$.  \\

\noindent
{\sl Electromagnetism} \\

The equations of electromagnetism closely resemble 
those of gravity in harmonic coordinates (see below), when the tensor 
$F_{\a\b}$ is written in terms of a vector potential, 
\be
  F_{\a\b} = \na_\a A_\b - \na_\b A_\a.  
\ee 
Then the first Maxwell equation $\na_{[\a} F_{\b\c]} = 0$ is automatically satisfied, 
\[
   \nabla_{[\a} \nabla_\b A_{\c]} = 0,
\] 
because partial derivatives commute and the antisymmetry gets rid of Christoffel symbols \\  
(or, use $d d\omega = 0$ for any form $\omega$, here with $\omega=A$).  

The remaining equation, $\na_\b F^{\a\b} = 4\pi j^\a$, takes the form of the wave equation for 
the vector potential if we use the Lorenz gauge condition: 
\be
   \na_\a A^\a = 0.  
\ee
(This was introduced by Hedvig Lorenz, not the famous Hendrik Lorentz.)
Then 
\[
   \na_\b \na^\a A^\b - \na_\b \na^\b A^\a = 4\pi j^\a.
\]
Writing 
\be
   \na_\b \na^\a A^\b = [\na_\b, \na^\a] A^\b + \underbrace{\na^\a \na_\b A^\b}_{0} = R_{\a\b}A^\b, 
\ee
we obtain a wave equation for $A^\a$ on a curved background spacetime:    
\be
  \na_\b \na^\b A^\a  - R^\a{}_\b A^\b = -4\pi j^\a.  
\label{e:boxA}\ee

Once a vector potential is introduced, the constraint $\nabla_\a B^\a=0$ is automatically satisfied.  In its place is the gauge condition, here a relation between $\dot A^0$ and $A^i$ 
that again must be satisfied by the initial data, now the values of $A^\a$ and $\dot A^\a$ on $\Sigma_t$.  

It is the 2nd derivatives that determine the character of the dynamical equation \eqref{e:boxA}: As in flat space, they 
have the form of a wave equation, a hyperbolic equation that gives a unique solution for initial 
data satisfying the gauge condition.  \\ 

\noindent
{\sl Gravity in harmonic coordinates} \\  
\index{initial value problem!in harmonic coordinates}

\index{coordinates!harmonic}
The harmonic coordinate condition for gravity is a nonlinear analog of the Lorenz gauge.  \index{coordinates!harmonic}\index{harmonic coordinates}
It has the form, 
\be
\partial_\nu(\sqrt{-g} g^{\mu\nu}) = 0,  
\ee
and is equivalent to requiring that each coordinate $x^\mu$ satisfy the scalar wave equation:  
\be
  \nabla_\a \na^\a (x^\mu) = 0.  
\ee
Note that each $x^\mu$ is a scalar.
This condition was first used by Fock and deDonder.  Fock called it {\sl harmonic}, presumably because in the mathematics literature a harmonic function is a solution to $\nabla_a\na^a f=0$ for a Riemannian 
metric (signature $+\cdots + $). \\
The condition is also equivalent to 
\be
	g^{\mu\nu}\Gamma^\lambda{}_{\mu\nu} = 0,
\ee
and this relation lets us write $R_{\mu\nu} =0$ as a nonlinear wave equation, an equation of the form 
\be
    \pa_\lambda \pa^\lambda g_{\mu\nu} + F(g,\pa g) = 0,
\ee
where $F$ involves only the metric components and their first derivatives.  

To show all of this, we begin by checking the equivalence of our three forms of the harmonic condition. 
From \ref{p:div} or Eq.~\eqref{e:Gammajij}, the divergence of a vector field $v^\alpha$ can be written as 
\be
   \na_\nu v^\nu = \frac1{\sqrt{-g}} \pa_\nu(\sqrt{-g}\, v^\nu). 
\label{e:div}\ee

For each $\mu$,  $\na^\a x^\mu$ is a vector, and we have
\begin{align*}
    \nabla_\nu \na^\nu (x^\mu)  &= \frac1{\sqrt{-g}}\pa_\nu(\sqrt{-g}\,\pa^\nu x^\mu) 
				 = \frac1{\sqrt{-g}}\pa_\nu(\sqrt{-g}\,g^{\nu\lambda}\pa_\lambda x^\mu)\\
				&= \frac1{\sqrt{-g}}\pa_\nu(\sqrt{-g}\,g^{\mu\nu}).
\end{align*} 
For the third form, write 
\begin{align*}
  g^{\mu\nu}\Gamma^\lambda{}_{\mu\nu} 
	&= \frac12 g^{\mu\nu}g^{\lambda\sigma}
		(\pa_\mu g_{\nu\sigma} + \pa_\nu g_{\mu\sigma} - \pa_\sigma g_{\mu\nu} ) 
	= g^{\mu\nu}g^{\lambda\sigma}\pa_\mu g_{\nu\sigma} 
		- \frac12  g^{\mu\nu}g^{\lambda\sigma}\pa_\sigma g_{\mu\nu} \\
	&= - \pa_\mu g^{\mu\lambda} - \frac1{\sqrt{-g}}\frac12 \pa^\lambda \sqrt{-g}
	 = - \frac1{\sqrt{-g}} \pa_\mu(g^{\mu\lambda}\sqrt{-g}).
\end{align*}
In the last line we used, for a matrix $M$, 
$\partial_\mu M^{-1} = - M^{-1}(\partial_\mu M) M^{-1}$, implying\\ 
\be
	\pa_\lambda g^{\mu\nu} = - g^{\mu\sigma}g^{\nu\tau}\pa_\lambda g_{\sigma\tau}.
\label{e:dgup}\ee
Check: Write
\mbox{ $0=\pa_\mu (M^{-1} M) = \pa_\mu(M^{-1}) M + M^{-1}\pa_\mu M$},  and multiply on 
the right by $M^{-1}$.  

From the usual coordinate expression \eqref{e:rijklup} for the Riemann tensor, 
\begin{align}
   R^\mu{}_{\sigma\nu\tau} &= \partial_\nu\Gamma^\mu{}_{\sigma\tau} - \pa_\tau\Gamma^\mu{}_{\sigma\nu}
				+ \Gamma^\mu{}_{\lambda\nu}\Gamma^\lambda{}_{\sigma\tau} 
				- \Gamma^\mu{}_{\lambda\tau}\Gamma^\lambda{}_{\sigma\nu}\nonumber\\ 
			   &= \partial_\nu\Gamma^\mu{}_{\sigma\tau} - \pa_\tau\Gamma^\mu{}_{\sigma\nu}
				+ F(g,\pa g), 
\end{align} 
where $F(g,\pa g)$ here and in later equations will mean some function of $g_{\mu\nu}$ and 
$\partial_\lambda g_{\mu\nu}$, but not the same function each time the symbol is used. 
Contracting on $\sigma$ and $\tau$ and using our gauge condition in the form 
$\Gamma^\mu{}_{\sigma\tau} g^{\sigma\tau} = 0$, we have 
\[
  R^\mu{}_\nu = -\pa^\sigma\Gamma^\mu{}_{\sigma\nu} + F(g,\pa g).  
\]
Now lowering the index $\mu$ and noting that 
$g_{\mu\lambda}\pa^\sigma \Gamma^\lambda{}_{\sigma\nu} = \pa^\sigma \Gamma_{\mu\sigma\nu} + F(g,\pa g)$, 
we have 
\be 
   R_{\mu\nu} = -\pa^\sigma\Gamma_{\mu\sigma\nu} + F(g,\pa g).  
\ee  
Because $R_{\mu\nu}$ is a symmetric tensor, we can write this as  
\be 
   R_{\mu\nu} = -\frac12\pa^\sigma(\Gamma_{\mu\sigma\nu} +\Gamma_{\nu\sigma\mu}) + F(g,\pa g).  
\ee
Finally, in the sum of the two $\Gamma$'s, the derivatives with respect to $\mu$ and $\nu$ cancel, 
leaving 
\[
\Gamma_{\mu\sigma\nu} +\Gamma_{\nu\sigma\mu} =  \partial_\sigma g_{\mu\nu},
\]
whence 
\be
   R_{\mu\nu} = -\frac12 \pa^\sigma\pa_\sigma g_{\mu\nu} + F(g,\pa g), 
\ee
as claimed.  

\benr\item
\label{ex:harmonic}
\benalph
\item
Use this relation and Eqs. \eqref{e:dgup} and \eqref{e:Gammajij} for $\pa_\lambda g^{\mu\nu}$ and $\pa_\lambda \log\sqrt{-g}$, show  
\be
   G^{\mu\nu} = \frac12 \frac1{\sqrt{-g}}\pa^\lambda\pa_\lambda (g^{\mu\nu}\sqrt{-g}) + F(g,\pa g).
\label{e:Gmunu}\ee
\item Now write $g_{\a\b} = \eta_{\a\b} + h_{\a\b}$ and show that, to $O(h_{..})$, 
$g^{\a\b} = \eta^{\a\b} - h^{\a\b}$, $\sqrt{-g} = 1+\frac12 h$, and 
\[
	g^{\a\b}\sqrt{-g} = - (h^{\a\b} -\frac12 \eta^{\a\b} h), 
\]
where $h^{\a\b} = \eta^{\a\c} \eta^{\b\d} h_{\c\d}$ and $h:=h_\a{}^\a$.
Show that the linearized harmonic condition is then
\be
   \na_\b (h^{\a\b} -\frac12 \eta^{\a\b} h) = 0,
\ee 
with $\na_\a$ the flat space derivative operator.   
\een
\een
This gauge condition on a metric perturbation $h_{\a\b}$, introduced earlier in Eq.~\eqref{e:dedonder}, 
was first used by Hilbert and then Einstein.  It is referred to in the literature 
by at least seven different names:   
the Hilbert, Einstein, Fock, deDonder, Lorenz, Lorentz (mistakenly), and harmonic gauge. 
Now $F(g,\pa g)$ is quadratic in $\pa g$ (this follows from the fact that every term has 
two derivatives and they can't both hit the same $g_{..}$), so at linear order in $h_{..}$, 
$F(g,\pa g)$ vanishes and Eq.~\eqref{e:Gmunu} has the form  
\be
   \delta G^{\a\b} = -\frac12  \na_\c \na^\c \bar h^{\a\b}, \quad\mbox{where }  
			\bar h_{\a\b} :=  h_{\a\b} -\frac12 \eta_{\a\b} h.
\ee
The field equation linearized about flat space is then 
\be
   \na_\c \na^\c \bar h^{\a\b} = -16\pi \delta T^{\a\b}.
\ee
\vspace{1cm}

\index{initial value problem|)}
%
%

\chapter{Newman-Penrose (NP) formalism and the Teukolsky equation}
\label{c:np}
\section{NP formalism}
\index{Newman-Penrose formalism|(}\index{NP formalism|(}
The Newman-Penrose paper\cite{np62} uses a signature $+ - -\, -\,$, and this convention is followed 
by the papers by Teukolsky\cite{teukolsky72,teukolsky73}, 
and by Chandrasekhar's detailed derivation of the Teukolsky equation in 
{\sl Mathematical Theory of Black Holes}\cite{chandra83}.  
Because of that we will switch (in this chapter only) to the $-2$ signature.  
The definition of the Riemann tensor 
$R^a{}_{bcd}$ is unchanged, and the conventions for the Ricci tensor and scalar are 
still $R_{ab} = R^c{}_{acb}$, $R=R^a{}_a$.  

\begin{tabular}{*{5}{l}}
\\
\hline\hline
Item & These notes & Teukolsky & Chandrasekhar & NP \\
[0.5ex]
\hline
Signature & $+---$ & $+---$& $+---$ &$ +---$\\
Coordinate indices & $\mu, \nu, \ldots$ & $\mu, \nu, \ldots$ & $i, j, \ldots$
						& $\mu, \nu, \ldots$ \\
Tetrad indices & $m,n, p, \ldots$ & not used  & $(a), (b), \ldots$
						& $\bf m, n, p, \ldots$\\
Numbered tetrad vectors & $e_m{}^\a$ & not used & $e_{(a)}{}^i$ &$ z_{\bf m}{}^\mu$\\
$r^2 + a^2\cos^2\theta$ & $\rho^2$ & $\Sigma$ &$\rho^2$ & not used\\
$r^2-2Mr+a^2$ & $\Delta$ & $\Delta$ & $\Delta$ & not used \\
$n^\mu\pa_\mu$ &\,\DDelta& $\Delta$ & \DDelta & $\Delta$ \\
Christoffel symbol & $\Gamma^m{}_{np}$ & not used & $\gamma^{(a)}{}_{(b)(c)}$ 
			& $-\gamma^{\bf m}{}_{\bf np}$ \\
$\Gamma_{314}=\barm^\b\na_\b\ell_\a m^\a$  & $\varrho $& $\rho$ & $\bar\rho$ & $\rho$ \\ 

\hline\hline
\end{tabular}

\newpage

A radiation field far from its source and the Kerr family of geometries have something in common: a preferred null direction.  In the case of radiation, the direction is a 
radial null vector $\ell_\a$ whose asymptotic form is $\ell_\a = \na_\a u = \na_\a(t-r)$. 
An outgoing radiation field has asymptotic form proportional to 
$\dis\frac{f(\ell_\mu x^\mu)}r = \frac{f(u)}r$.  The corresponding contravariant vector, 
$\ell^\alpha$ is along $\widehat t^\a+\widehat r^\alpha$.
The polarization of the field is perpendicular to the null direction, so an 
electromagnetic field in a Lorenz gauge has 
\[
   A_\a = \ep_\a \frac{f(u)}r+O(r^{-2}), \qquad F_{\a\b} 
		= (\ell_\a \ep_\b - \ell_\b\ep_\a)\frac{f'(u)}r + O(r^{-2}),  
\]
with $\epsilon_\a$ in the $\widehat\theta$-$\widehat\phi$ plane.  
A weak gravitational wave is described by a metric perturbation 
\[
   h_{\a\b} = \ep_{\a\b}\frac{f(u)}r + O(r^{-2})  
\]
with corresponding Riemann tensor given by Eq.~\eqref{e:RP1}.    
\begin{align}
 R_{\alpha\gamma\beta\delta} 
	    &=  \frac{1}{2} (\pa_\beta \pa_\gamma h_{\alpha\delta} 
		+ \pa_\alpha \pa_\delta h_{\beta\gamma} - \pa_\alpha \pa_\beta h_{\gamma\delta} 
		- \pa_\gamma \pa_\delta h_{\alpha\beta})\nn\\
	    &= \frac12 \frac{f''}r(\ell_\b \ell_\c \epsilon_{\a\d} + \ell_\alpha \ell_\delta \ep_{\beta\gamma} - \ell_\alpha \ell_\beta \ep_{\gamma\delta} 
		- \ell_\gamma \ell_\delta \ep_{\alpha\beta}) + O(r^{-2})   	
\label{e:outgoing}\end{align} 
The null vector $\ell_\a$ is called a {\sl principal null direction} of the geometry. 
The NP formalism is designed to exploit symmetries of the geometry tied to principal 
null directions. Although their 1962 paper emphasized the asymptotic behavior of 
gravitational waves, Newman and Penrose knew the vacuum Schwarzschild geometry 
has two principal null directions, tangent to ingoing and outgoing 
radial null geodesics $n_\a$ and $\ell_\a$, with each term in the Riemann tensor proportional to $\ell_\a$ and $n_\a$. 

The NP formalism completes the basis (tetrad) by using complex combinations $\bm m$ and 
$\bm{\barm}$ of unit vectors 
orthogonal to the null vectors $n^\a$ and $\ell^\a$.  Like the 
combinations $x\pm iy$ that you've encountered in spherical harmonics, $\bm m$ and 
$\bm{\barm}$  change by a phase under rotations about the $z$-axis. For null vectors 
\[
\ell^\a = \frac1{\sqrt 2} (\widehat t^\a + \widehat z^\a), \qquad 
n^\a =  \frac1{\sqrt 2} (\widehat t^\a - \widehat z^\a),
\] 
a corresponding choice of $\bm m$ and $\overline{\bm m}$ is 
\[
   m^\a = \frac1{\sqrt 2} (\widehat x^\a + i \widehat y^\a), \qquad 
\barm^\a =  \frac1{\sqrt 2} (\widehat x^\a -i \widehat y^\a),
\] 
Then $m^\a$ and $\barm^\a$ are complex null vectors, and the tetrad satisfies 
\be 
   \ell_\a n^\a = 1, \qquad m_\a \barm^\a = -1, \qquad \mbox{all other dot products vanishing}.
\ee

An NP tetrad is then a null tetrad $\{\bm e_m\} = \{e_m{}^\a\}$, $m=1-4$, 
\be
  e_1{}^\a = \ell^\a, \ e_2{}^\a = n^\a, \ e_3{}^\a = m^\a, \ e_4{}^\a = \barm^\a, \ 
\ee 
for which the spacetime metric has components $g_{mn}  = e_m{}^\a e_{n\,\a}$ given by 
\be
    [g_{mn}] = [\eta_{mn}] = \begin{bmatrix}    
				  0 & 1 & 0 & 0 \\
				  1 & 0 & 0 & 0 \\
				  0 & 0 & 0 & -1 \\
				  0 & 0 & -1 & 0
	  	              \end{bmatrix}
\label{e:etamn}\ee   
Then $g^{\a\b}$ has the same components  
\be
   [\eta^{mn}] = \begin{bmatrix}    
				  0 & 1 & 0 & 0 \\
				  1 & 0 & 0 & 0 \\
				  0 & 0 & 0 & -1 \\
				  0 & 0 & -1 & 0
	  	              \end{bmatrix}
\ee 
and 
\be 
  g_{\a\b} = \eta^{mn}e_{m\a} e_{n\b} = \ell_\a n_\b + \ell_\b n_\a - m_\a \barm_\beta - m_\b \barm_\a .
\ee 

\noindent{\sl Examples}:  \\
For the Schwarzschild geometry, with
\bsube\begin{align}  
  ds^2 & = \left(1-\frac{2M}r\right) dt^2 -\left(1-\frac{2M}r\right)^{-1} dr^2 
	 - r^2 d\theta^2 - r^2\sin^2\theta d\phi^2 \\
	&= \frac\Delta{r^2} dt^2 -\frac{r^2}\Delta dr^2 
	 - r^2 d\theta^2 - r^2\sin^2\theta d\phi^2, \quad \Delta := r^2-2Mr,
\end{align}\esube 
an orthonormal basis is 
\be
   	\widehat t^\a = \frac1{\sqrt{1-\frac{2M}r}} \bm \pa_t, 
\quad 	\widehat r^\a = \sqrt{1-\frac{2M}r}\bm\pa_r,\quad 
   	\widehat\theta^\a = \frac1r\bm\pa_\theta, \quad 
	\widehat\phi^\a = \frac1{r\sin\theta} \bm\pa_\phi. 
\ee
We could take 
\be
	\ell^\a = \frac1{\sqrt2}(\widehat t^\a + \widehat r^\a),\quad 
	   n^\a = \frac1{\sqrt2}(\widehat t^\a - \widehat r^\a),\quad
	   m^\a = \frac1{\sqrt2}(\widehat \theta^\a +i\widehat \phi^a). 
\ee
with contravariant components
\be
  \ell^\mu = \frac1{\sqrt2}\left(\frac r{\sqrt\Delta},\frac{\sqrt\Delta}r,0,0\right),\quad
     n^\mu = \frac1{\sqrt2} \left(\frac r{\sqrt\Delta},-\frac{\sqrt\Delta}r,0,0\right),\quad
     m^\mu = \frac1{\sqrt2\ r} \left(0,0,1,\frac i{\sin\theta}\right) 
\label{e:sk}\ee
Notice that multiplying $\ell^\a$ by any function $f$ and dividing 
$n_\a$ by $f$ gives another null tetrad with the same null directions 
and the same metric components $\eta_{mn}$.  Kinnersley, 
one of Kip Thorne's first students, used this freedom to get rid of the 
square roots in $\bm \ell$ and $\bm n$, making the commutators and the Christoffel 
symbols (Ricci rotation coefficients) simpler, at the cost of 
breaking the symmetry associated with time reversal.  The Kinnersley 
tetrad is 
\be
	(\ell^\mu) = (r^2/\Delta, 1,0,0), \quad
	(n^\mu) = \frac12(1,-\Delta/r^2, 0 ,0), \quad
        (m^\mu) = \frac1{\sqrt2\ r} (0,0,1,\frac i{\sin\theta}) 
\label{e:kins}\ee
 
The Kerr metric has the factored form 
\be
  ds^2 = \frac{\Delta }{\rho^2}~(dt-a\sin^2\theta d\phi)^2 
 	-\frac{\sin^2 \theta }{\rho^2} [(r^2+a^2)d\phi - a dt]^2
	- \frac{\rho ^2}{\Delta } dr^2 - \rho ^2 d\theta ^2,
\ee
where 
\be
  \Delta = r^2 - 2Mr+a^2, \qquad \rho^2 = r^2 + a^2\cos^2\theta.
\label{e:Delta_rho}\ee
An orthonormal covariant (dual) basis is 
\[
  \qquad \Omega^0 = \frac{\sqrt\Delta}\rho (dt-a\sin^2\theta d\phi),
  \quad \Omega^1 = \frac{\rho}{\sqrt\Delta }dr, \qquad 
	\Omega^2 = \rho d\theta,\qquad  
	\Omega^3=\frac{\sin\theta}\rho[(r^2+a^2)d\phi - a dt],\qquad
\]  
and the corresponding contravariant basis to which it is dual is 
\be
   \bm E_0 =\frac1{\rho\sqrt\Delta}[(r^2+a^2) \pa_t + a\pa_\phi],\quad
  \bm E_3=\frac1{\rho\sin\theta} (\pa_\phi + a\sin^2\theta\pa_t), \qquad
   \bm E_1=\frac{\sqrt\Delta}\rho \pa_r, \qquad \bm E_2=\frac1\rho \pa_\theta.   
\ee
The associated Kinnersley tetrad is 
\begin{align*}
   \bm e_1 = \bm\ell &= \frac\rho{\sqrt\Delta}(\bm E_0+\bm E_1) \\
   \bm e_2 = \bm n &= \frac12\frac{\sqrt\Delta}\rho(\bm E_0 - \bm E_1) \\
   \bm e_3 = \bm m &= \frac\rho{\sqrt2\ \tilde\rho}\left(\bm E_2+ i\bm E_3\right)\\
   \bm e_4 = \barbm &= \frac\rho{\sqrt2\ \tilde\rho^*}\left(\bm E_2- i\bm E_3\right)
\end{align*} 
where 
\be
  \tilde\rho := r+ia\cos\theta
\ee
is a kind of complex square root of $\rho^2$: 
$\ \tilde\rho\tilde\rho^* = \rho^2$.  Introducing it gets rid of the square 
root, \mbox{$\rho=\sqrt{r^2+a^2\cos^2\theta}$}, while keeping $m^\a \barm_\a = -1$:    
\bsube\begin{align}
   (e_1{}^\mu) = (\ell^\mu) &= \left(\frac{r^2+a^2}\Delta, 1, 0, \frac a\Delta\right), \\
    (e_2{}^\mu) =  (n^\mu) &= \frac12\left(\frac{r^2+a^2}{\rho^2}, -\frac{\Delta}{\rho^2}, 0, 
		\frac a{\rho^2} \right),\\
   (e_3{}^\mu) =  (m^\mu) &= \frac1{\sqrt2\ \tilde\rho}\left(ia\sin\theta, 0, 1, \frac i{\sin\theta}\right).
\end{align}\label{e:kinnersley} \esube
When $a=0$, we recover the Kinnersley tetrad for Schwarzschild \eqref{e:sk}. \\

\noindent{\sl Weyl tensor}

The tracefree part of the Riemann tensor is called the Weyl tensor and written 
$C_{abcd}$.  In a vacuum spacetime the two coincide.  In general the decomposition of 
the Riemann tensor into its tracefree part and parts involving the Ricci tensor and 
Ricci scalar is 
\be
   R_{abcd} = C_{abcd} 
		+ \frac12 (g_{ac}R_{bd} + g_{bd}R_{ac} - g_{ad}R_{bc} - g_{bc}R_{ad})
		- \frac16 (g_{ac}g_{bd} - g_{ad}g_{bc})R.
\ee
Up to the constant factor 1/2, the Ricci tensor part is the only combination of 
$g_{ab}$ and $R_{ab}$ with Riemann tensor index symmetries (and linear in $R_{ab}$); 
and the Ricci scalar part is the only combination of $R$ and $g_{ab}$  
with those symmetries (linear in $R$). All that remains is to take the trace of both sides to get the factors $1/2$ and $1/6$.  We will be working with vacuum spacetimes, and for those $R_{abcd} = C_{abcd}$.  

\benr \item 
Show that, for an $n$-dimensional spacetime, the decomposition of the Riemann tensor is given by
\[
   R_{abcd} = C_{abcd} 
		+ \frac1{n-2} (g_{ac}R_{bd} + g_{bd}R_{ac} - g_{ad}R_{bc} - g_{bc}R_{ad})
		- \frac1{(n-1)(n-2)} (g_{ac}g_{bd} - g_{ad}g_{bc})R. 
\]   
\een

The Teukolsky equation is an equation for a single tetrad component of the Weyl tensor.  
The NP formalism gives names to each independent Christoffel symbol (Ricci rotation coefficient)  and to each independent tetrad component of the Weyl tensor.  We begin 
by noting that the Weyl tensor has 10 independent real components and that these 
correspond to 5 independent complex components along the NP tetrad.  

First: The Riemann tensor in $n$ dimensions has 
$\dis \frac{n^2(n^2-1)}{12}$ independent components. \\
This is Prob. 3b of Wald, and if you 
are already familiar with it, skip to the end of this paragraph.  
The index symmetries of $R_{ijkl}$ are 
\[
  R_{ijkl} = R_{[ij][kl]}, \quad R_{i[jkl]} = 0, \quad R_{ijkl} = R_{klij}.
\]
The last of these is implied by the others, so we do not 
include it in our counting.  
Because the index pair $ij$ is antisymmetric, the independent 
pairs have $i<j$, and there are $n(n-1)/2$ such pairs. There are similarly $n(n-1)/2$ pairs $kl$ with $k<l$.  The number of pairs $(ij)(kl)$ is then $n^2(n-1)^2/4$.  These are not all independent
because we have the relations $R_{i[jkl]}$ = 0. Here we can choose $i$ in $n$ different ways; 
and for $[jkl]$ we have $n$ objects taken 3 at a time or 
$\dis\begin{pmatrix}n\\3\end{pmatrix}=\frac{n(n-1)(n-2)}6$. 
The number of relations is then $\dis n^2(n-1)(n-2)/6$, and each relation reduces the number 
of independent components by one, leaving 
\[
   \frac{n^2(n-1)^2}4-\frac{n^2(n-1)(n-2)}6 = \frac{n^2(n^2-1)}{12} 
\] 
independent components of the Riemann tensor in $n$ dimensions. 

To find the number of independent components of $C_{abcd}$, we subtract the number 
of independent components of $R_{ab}$. Here the only symmetry is $R_{ab} = R_{ba}$, so 
there are $n(n+1)/2$ independent pairs $(ab)$.  Then the Weyl tensor has \vspace{-2mm}
\be 
   \frac{n^2(n^2-1)}{12} - \frac{n(n+1)}2 = \frac{(n-3)n(n+1)(n+2)}{12} 
\ee 
independent components, 10 in four dimensions. In fewer than four dimensions, the Weyl tensor \mbox{vanishes}. \\ 

\noindent {\sl Spin weight}\\ 
A rotation of the tetrad in the plane spanned by $\bm m$ and $\bm \barm$ has the 
form 
\be
     \bm m\rightarrow e^{i\eta} \bm m, \quad   \barbm\rightarrow e^{-i\eta} \barbm
\ee
\benr\item Two-line problem:  Write $\bm m = \widehat{\bm x} +i\widehat{\bm y}$ and check that a clockwise rotation of the basis by $\eta$ in the $\bm{\widehat x}$-$\bm{\widehat y}$ plane gives the above equation.   
\een 
Then a component of a tensor with $p$ indices along $\bm m$ and $q$ indices along 
$\barbm$ (and remaining indices along $\bm n$ and $\bm \ell$) changes under a 
tetrad rotation by the factor $e^{i(p-q)\eta}$, and is said to have {\sl spin weight} 
$p-q$.  This is, of course, analogous to an eigenvalue of $L_z$ in quantum 
mechanics: A wavefunction $\psi$ with eigenvalue $m\hbar$ changes under 
a rotation by $\eta$ about the $z$-axis by the factor $e^{im\eta}$.   

In particular, NP order the five independent complex components of the Weyl tensor by their spin weight, from 2 to $-2$: 
\bsube
\begin{align}
\Psi_0 &= - C_{1313} = - C_{\a\b\g\d} \ell^\a m^\b \ell^\g m^\d\\
\Psi_1 &= - C_{1213} = - C_{\a\b\g\d} \ell^\a n^\b \ell^\g m^\d\\
\Psi_2 &= - C_{1342} = - C_{\a\b\g\d} \ell^\a m^\b \barm^\g n^\d\\
\Psi_3 &= - C_{1242} = - C_{\a\b\g\d} \ell^\a n^\b \barm^\g n^\d\\
\Psi_4 &= - C_{2424} = - C_{\a\b\g\d} n^\a \barm^\b n^\g \barm^\d\ .
\end{align}\label{e:Psi}\esube
It would have made more sense to write these as $\Psi_2, \Psi_1, \cdots ,\Psi_{-2}$, 
and Press and Teukolsky tried that, but there were too many other papers 
with the NP convention, and the change didn't stick.

For Schwarzschild and Kerr the only nonzero component of the Weyl tensor 
is $\Psi_2$, the spin-weight zero part.  Geometries with this character 
are Petrov type $D$.  They have the two principal null directions 
$\ell_\alpha$ and $n_\alpha$.  

\benr\item NP components of $F_{\a\b}$.  
\benalph
\item
How many independent components does the electromagnetic tensor $F_{\a\b}$ have? 
\item In analogy with the components of the Weyl tensor above, write these 
components of $F_{\a\b}$ as complex components with definite spin weight, 
taking as the spin-weight 0 component $F_{\a\b} (\ell^\a n^\b + \barm^\a m^\b)$.    
\item Using the flat space tetrad 
\[
\bm\ell = \frac1{\sqrt2}(\widehat{\bm t} +\widehat{\bm z}), \quad \bm n = \frac1{\sqrt2}(\widehat{\bm t} -\widehat{\bm z}), \quad \bm m = \frac1{\sqrt2}(\widehat{\bm x} +i\widehat{\bm y}), \quad 
\barbm = \frac1{\sqrt2}(\widehat{\bm x} - i\widehat{\bm y}),
\] 
write each of these complex components in terms of the $t,x,y,z$ components of 
$\bm E$ and $\bm B$.  
\een
\een

We have already seen in Eq.~\eqref{e:outgoing} that outgoing waves have, at leading 
order in $1/r$, a Weyl tensor that is quadratic in $\ell_\a$ and is constructed 
from $\ell_\a$ and a polarization tensor in the $\bm m$-$\barbm$ plane.  
From $\ell_\a \ell^\a = \ell_\a m^\a = 0$, it immediately follows that 
the only nonvanishing $\Psi_i$ is $\Psi_4$.  Similarly, purely ingoing radiation 
has $\Psi_0$ as its leading asymptotic term.  And it is $\Psi_0$ and $\Psi_4$ 
that appear in the decoupled wave equations that describe perturbations of 
Schwarzschild and Kerr spacetimes.  

\benr\item  Writing the asymptotic form of $\bm m$ as $\frac1{\sqrt2}(\widehat{\bm x} +i\widehat{\bm y})$ (corresponding to a chart near an observer with $z$ in the direction of propagation 
of the wave), show that the two polarizations of the wave are given by the real 
and imaginary parts of $\Psi_4$. ($+$ and $\times$ polarizations are proportional 
to $h_{xx} - h_{yy}$ and $h_{xy}$, respectively, for a given choice of $x$ and $y$.)
Use Eq.~\eqref{e:outgoing}.     
\een

\noindent {\sl Christoffel symbols (Ricci rotation coefficients)}  

We looked in Section \ref{s:g_nabla} at the Christoffel symbols \eqref{gamc} 
and Riemann tensor components 
for a basis field $\{ \bm e_n\}$ in which the components $g_{mn}$ are constant. 
As we saw, the quantity $\Gamma^m{}_{n\a} v^\a$ is the derivative of the 
family of rotations or Lorentz transformations that relate a basis parallel 
transported along $v^\a$ to the original basis field.  That's why $\Gamma^m{}_{np}$ 
is called a rotation coefficient and why $\Gamma_{mnp}$, defined by Eq.~\eqref{e:Gamma_up}, 
is antisymmetric in $m$ and $n$.  Chandrasekhar's convention is 
\[
   \gamma^m{}_{np} = \Gamma^m{}_{np}, 
\]
while NP take the opposite sign for $\gamma^{\bf m}{}_{\bf np}$.  Chandra also uses 
$(a), (b), \cdots$ for concrete tetrad indices, instead of the $\bf m,n,p,q$ of 
NP.
\index{Ricci rotation coefficients}\index{Lorentz transformation!and Christoffel symbols (Ricci rotation coefficients)} 
\index{Christoffel symbol!for an orthonormal basis}

The NP naming of independent symbols is 
\begin{align}
   \kappa &= \Gamma_{311} &\varrho &= \Gamma_{314} & 
				\ep &= \frac12(\Gamma_{211}+\Gamma_{341}) \nn\\
   \sigma &= \Gamma_{313} & \mu &= \Gamma_{243} &
				\gamma &= \frac12(\Gamma_{212}+\Gamma_{342})\nn\\
   \lambda &= \Gamma_{244} & \tau &= \Gamma_{312} &
				\alpha &= \frac12(\Gamma_{214}+\Gamma_{344})\nn\\
   \nu &= \Gamma_{242} & \pi &= \Gamma_{241} &
				\beta &= \frac12(\Gamma_{213}+\Gamma_{343})\nn\\
\Gamma_{211} = \ep+\bar\ep &\quad  \Gamma_{341} = \ep - \bar\ep & \Gamma_{212} =\gamma+\bar\gamma 
&\quad\Gamma_{342} = \gamma-\bar\gamma & \Gamma_{214} =\alpha +\bar\beta 
& \quad\Gamma_{344} = \alpha - \bar\beta.
\end{align}

Each of these has a fixed spin weight. The quantities $\Gamma_{21\alpha}$ and 
$\Gamma_{34\alpha}$ are respectively real and imaginary, and the third column lists 
the four tetrad components of the complex vector $\Gamma_{21\alpha} + \Gamma_{34\alpha}$. \\
Quick question:  Why is $\Gamma_{34\alpha}$ a purely imaginary dual vector?     

The signs are the same for Chandra and NP if one writes the NP 
symbols directly as derivatives of tetrad vectors: That is, 
\[
  \kappa = e_{3\alpha} \nabla_1 e_1{}^\alpha  = m_\a\ell^\b\na_\b \ell^\a, \qquad
  \pi =  e_{2\alpha} \nabla_1 e_4{}^\alpha  = n_\a\ell^\b\na_\b \barm^\a 
					    = - \barm^\a\ell^\b\na_\b n_\a. 
\] 
for everyone. \\

\noindent{\sl Symbols for derivatives}  

A final set of individual symbols introduced by NP are the names of the 
basis vectors regarded as derivatives (as we have defined vectors):
\be
  \ell^\mu\pa_\mu = D, \quad n^\mu\pa_\mu =\DDelta\ ,
\quad m^\mu\pa_\mu = \delta, \quad \barm^\mu\pa_\mu = \bar\delta .   
\ee  
\index{Newman-Penrose formalism|)}\index{NP formalism|)}

\section{Bardeen-Press and Teukolsky equations}

The presentation here illustrates how you might find your way to the 
Teukolsky equation starting from the wave equation for the perturbed Weyl 
tensor on a flat background and then 
moving on to Schwarzschild (the Bardeen-Press equation) and finally to Kerr.  
For a concise derivation of the Teukolsky equation using the NP formalism, 
you can skip to the section \ref{s:teukolsky} on Kerr.   Everything uses
the equations in the NP paper that express the Weyl tensor components, the 
commutators of basis vectors, and the Bianchi identities in terms of 
Christoffel symbols (spin coefficients).

\subsection{Flat space}
This second section of the notes will eventually follow NP equations for 
commutators and for the Weyl tensor in terms of the NP symbols of the last section.  
For now, to see where we're going, it will be helpful to look at the wave 
equation in its 
simplest context -- flat space with basis constructed from a $t,x,y,z$ chart 
(instead of spherical coordinates).    

We start with the scalar wave equation with partial derivatives along the null 
basis vectors, 
\begin{align*}
   \bm e_1 = \bm\ell &= \widehat{\bm t} + \widehat{\bm z}  \\
   \bm e_2 = \bm n &= \frac12(\widehat{\bm t} - \widehat{\bm z}) \\
   \bm e_3 = \bm m &= \frac1{\sqrt2}\left(\widehat{\bm x} + i \widehat{\bm y}\right)\\
   \bm e_4 = \barbm &= \frac1{\sqrt2}\left(\widehat{\bm x} - i \widehat{\bm y}\right).
\end{align*} 
Then 
\[ 
  \na_\a\na^\a\Phi = \eta^{mn}\pa_m\pa_n\Phi = (\pa_1 \pa_2 +\pa_2\pa_1 -\pa_3\pa_4 - \pa_4\pa_3)\Phi = 2(\pa_1\pa_2 -\pa_3\pa_4)\Phi,    
\] 
and the scalar wave equation is 
\be
   (\pa_1\pa_2 - \pa_3\pa_4)\Phi = 0.  
\ee
In NP notation, this is 
\be
  (D\DDelta - \delta\overline\delta )\Phi = 0.
\label{e:scalarwave}\ee

\benr\item \benalph\item Two-line exercise:  Replace the null basis vectors by their expressions in terms of the Cartesian basis to recover the usual form of the scalar wave equation.
\item Another two-line exercise in flat space.  Show that Maxwell's equations,
$\na_{[\a} F_{\b\g]}=0$ and $\na_\b F^{\a\b}=0$, imply $\na_\g \na^\g F_{\a\b}=0$.  (Hint: Take the divergence 
of the first Maxwell equation and then use the second.)    
\een
\een

The wave equations satisfied by $\Psi_0$ and $\Psi_4$ are easily derived as a perturbation 
of flat space. In curved space, nonzero Christoffel symbols mix together Weyl tensor 
parts with different spin weights and leave only these extreme spin components with their own decoupled equations.  We'll find the equation for $\Psi_0 = -C_{1313}$.  Begin with the Bianchi identity, 
\index{Bianchi identities!in Newman-Penrose formalism}
\be
   \nabla_{[\a} C_{\b\g]\d\ep} = 0,
\label{e:bianchi1}\ee  
writing the two nontrivial components that involve $C_{1313} = -\Psi_0$.  These are 
\be
   \nabla_{[4} C_{13]13} = 0 \mbox{ and } \nabla_{[2} C_{13]13} = 0.  
\label{e:bianchi2}\ee
Next, we'll use the Riemann-tensor symmetries and the fact that $C_{. . . .}$ is 
tracefree to show 
\be
  C_{4113} = 0, \quad C_{3213} = 0, \qquad C_{3413} = - C_{1213} = \Psi_1.  
\label{e:crelns}\ee 
Proof:  The Riemann-tensor symmetries imply $C_{4113} = C_{3114}$, and 
$C_{1112} = 0 =  C_{2111} $.  Then 
\begin{align*}
   C_{4113} & = \frac12(\, C_{4113}+C_{3114} - C_{1112} - C_{2111} \,)\\
	    &= - \frac12 \eta^{mn} C_{m11n} = 0, 
\end{align*} 
where the last equality is the statement that $C_{\a\b\g\d}$ is tracefree. 
The same steps, using $\eta^{mn} C_{3mn3} = 0$, show $C_{3213} = 0$.
Finally, again using the tracefree condition, we have  
\begin{align*}
   0 &= \eta^{mn} C_{1mn3} = C_{1123} + C_{1213} - C_{1343} - C_{1433} 
			  =C_{1213} - C_{1343} \Longrightarrow \\
    \Psi_1 &: = - C_{1213} = -C_{1343} = C_{3413} . 
\end{align*} 

Now we're ready to use the Bianchi identity: Eqs.~\eqref{e:bianchi2} and 
\eqref{e:crelns} give 
\be 
    \na_4 C_{1313} + \na_1 C_{3413} = 0, \qquad \na_2 C_{1313} + \na_3 C_{2113} = 0.    
\label{e:psi0psi1}\ee
Now linearize about flat space and 
write $\psi_i$ for the first-order change in $\Psi_i$.  Because 
the Weyl tensor and all Christoffel symbols vanish in the background spacetime 
for a Cartesian basis, we have, for example, $\na_4 C_{1313} = -\bar\delta\psi_0$, and 
the equations take the form, 
\[
   -\bar\delta\psi_0 + D\psi_1 = 0, \qquad -\DDelta \psi_0 + \delta\psi_1 = 0.  
\] 
To eliminate $\psi_1$, act with $\delta$ on the first equation and with $D$ on the second   and subtract to obtain
\be
    (D \DDelta -\delta\bar\delta)\psi_0 = 0.  
\ee
That is, $\psi_0$ satisfies the scalar wave equation \eqref{e:scalarwave}.\\

\noindent{\sl Flat space, spherical coordinates}.  \\
We look next at the key equations \eqref{e:psi0psi1} for $\Psi_0$ in the Kinnersley tetrad associated with spherical coordinates but still in flat space:  
\bsube\begin{align}
   \bm e_1 = \bm\ell^\cdot &= \pa_t +\pa_r, \\
    \bm e_2 = \bm n^\cdot &= \frac12\left(\pa_t-\pa_r\right),\\
   \bm e_3 =  \bm m^\cdot 
		&= \frac1{\sqrt2\ r}\left(\pa_\theta+\frac i{\sin\theta}\pa_\phi\right)\\
   \bm e_4 = \barbm^{\,\cdot} 
\end{align}\label{e:e_spherical} \esube
The independent nonzero commutators are 
\bsube\begin{align}
 [\bm e_1,\bm e_3] 
	&= \left[\pa_r, 
		\frac1{\sqrt2\ r}\left(\pa_\theta+\frac i{\sin\theta}\pa_\phi\right)\right]  
	 = -\frac1{\sqrt2\ r^2}\left(\pa_\theta+\frac i{\sin\theta}\pa_\phi\right)\\
	& = -\frac1r \bm e_3 ,\\
 [\bm e_1,\bm e_4] &= -\frac1r \bm e_4,  \\
 [\bm e_2,\bm e_3] &= \frac1{2r}\bm e_3,\\
 [\bm e_2,\bm e_4] &= \frac1{2r}\bm e_4,\\
 [\bm e_3,\bm e_4]&= \frac{\cot\theta}{\sqrt2\ r}(\bm e_3 - \bm e_4)
\end{align}\label{e:kflat} \esube
By inspection, the corresponding nonzero coefficients $c^m{}_{np}$ (and then with lowered indices)  are
\begin{align}
  c^3{}_{13}&=c^4{}_{14} = -\frac1r, \quad c^3{}_{23}=c^4{}_{24} = \frac1{2r}, \quad
 c^3{}_{34}=- c^4{}_{34} = \frac{\cot\theta}{\sqrt2\ r},\nn\\
  c_{413}&=c_{314} = \frac1r, \qquad c_{423}=c_{324} = -\frac1{2r}, \qquad
 c_{434}=- c_{334} = -\frac{\cot\theta}{\sqrt2\ r}, 
\label{e:commflat}\end{align}
together with those obtained by exchanging the last two indices: $c_{pmn} = - c_{pnm}$.  
\benr \item Check the last four commutators of Eqs.~\eqref{e:kflat}.
\een
In NP notation for derivative operators, Eqs.~\eqref{e:kflat} are
\be
[D,\delta]  = -\frac1r \delta, \ \ [D,\bar\delta] = -\frac1r \bar\delta,\ \ 
[\DDelta,\delta] = -\frac1{2r}\delta,\ \ [\DDelta,\bar\delta]= -\frac1{2r}\bar\delta,\ \ 
[\delta,\bar\delta] = -\frac{\cot\theta}{\sqrt2\ r}(\delta-\bar\delta).
\label{e:kflatnp}\ee

The nonzero Christoffel symbols and corresponding spin coefficients are now obtained 
from Eq.~\eqref{gamc}
\be
 \Gamma_{mnp} = \frac12(c_{pmn}-c_{npm}-c_{mnp}).
\ee
Notice that the nonzero $c_{mnp}$ have at most one index in $\{1,2\}$ and the first 
index is $3$ or $4$. Because $\Gamma_{341}$ and $\Gamma_{342}$ are imaginary while all 
$c_{mnp}$ are real $\Gamma_{341} = \Gamma_{342}=0 $.   We are left with    
\be
   \Gamma_{134} = \frac1r = -\varrho, \quad 
   \Gamma_{234} = \Gamma_{243}  = -\frac1{2r} = \mu, \quad 
   \Gamma_{343}  = \frac1{\sqrt2}\frac{\cot\theta}r = \beta-\bar\alpha, 
\quad  \Gamma_{434}= \frac1{\sqrt2}\frac{\cot\theta}r  = \alpha-\bar\beta, 
\label{e:lG}\ee
together with those obtained by exchanging the first two indices: $\Gamma_{mnp}= -\Gamma_{nmp}$.
Notice also that taking the complex conjugate exchanges $3$ and $4$: 
$\Gamma_{243} = \overline \Gamma_{234}$, and in this flat-space example, the $\Gamma$'s are real.  

For reference in computing the Bianchi identities, here are the corresponding raised 
$\Gamma$'s and the nonzero spin coefficients:  
\begin{align}
      \Gamma^1{}_{34} &= \Gamma^1{}_{43}  = \mu, \quad 
	\Gamma^2{}_{34} = \Gamma^2{}_{43} =-\varrho , \quad 
	\Gamma^3{}_{33} = \Gamma^4{}_{44} = 2\beta = - 2\alpha  , \nn\\
   \Gamma^3{}_{13} &= \Gamma^4{}_{14}  = -\varrho, \quad \Gamma^4{}_{24} = \Gamma^3{}_{23} =\mu,
	\quad \Gamma^3{}_{34} = \Gamma^4{}_{43} = -2\beta = 2\alpha .
\label{e:Gamma}\\
\alpha &= -\frac1{2\sqrt2}\frac{\cot\theta}r ,\quad \beta = \frac1{2\sqrt2}\frac{\cot\theta}r, 
\qquad \varrho = -\frac1r, \quad \mu = -\frac1{2r}.
\label{e:spinflat} \end{align}
We'll see later that Schwarzschild is only slightly different.  

For the two components of the Bianchi identities, we'll need to show that the component 
$\nabla_\a C_{4113} \equiv \nabla_\ep C_{\a\b\g\d}\barm^\a \ell^\b \ell^\c m^\d$ is 
zero.  We already showed that $C_{4113}=0$, but the covariant derivative involves other 
components of $C_{\a\b\c\d}$ multiplied by Christoffel symbols.  The proof, however, 
is essentially identical if we keep the basis vectors outside the derivative operator   
\begin{align*}
      \nabla_\ep C_{\a\b\c\d}\barm^\a \ell^\b \ell^\g m^\d 
   &=  \nabla_\ep C_{\d\c\b\a}\barm^\a \ell^\b \ell^\c m^\d 
    = \nabla_\ep C_{\a\b\c\d} m^\a \ell^\b \ell^\c \barm^\d \\
   &= \frac12\na_\ep C_{\a\b\c\d}\ell^\b \ell^\c 
		(\barm^\a m^\d + m^\a\barm^\d - \ell^\a n^\d - \ell^\d n^\a) 
    = \frac12\na_\ep C_{\a\b\c\d}\ell^\b \ell^\c (-g^{\a\d}) \\
   &= 0.  
\end{align*}   
The equivalent steps show $\na_\a C_{3213} = 0$.  Set off and numbered for future reference:   
\be
    \na_\a C_{4113} \equiv \nabla_\a C_{\b\g\d\ep}\barm^\b \ell^\c \ell^\d m^\ep =0, \quad
     \na_\a C_{3213}\equiv \nabla_\a C_{\b\g\d\ep} m^\b n^\c \ell^\d m^\ep =0.   
\ee 

\noindent{\sl Naming of parts} 

Again for reference, here's a list of the Weyl tensor components in terms of $\Psi_i$ ordered top to bottom and then left to right.  
Those in blue are, up to exchanging antisymmetric indices, the definitions \eqref{e:Psi}. 
\begin{align*}
C_{1212} &= -(\Psi_2+\bar\Psi_2) & \cblue C_{1313} &\cblue = -\Psi_0 & C_{1414} &= -\bar\Psi_0 & C_{2323} &= -\bar\Psi_4 & C_{3434} &= -(\Psi_2+\bar\Psi_2)\\
\cblue C_{1213} &\cblue = -\Psi_1 & C_{1314} &= 0 & C_{1423} &= \bar\Psi_2 & C_{2324} &= 0 && \\
C_{1214} &= -\bar\Psi_1 & C_{1323} &= 0 & C_{1424} &= 0 & C_{2334} &= \bar\Psi_3 &&\\
C_{1223} &= \bar\Psi_3 &\cblue C_{1324} &= \cblue \Psi_2 & C_{1434} &= -\bar\Psi_1 & C_{2424} &= -\Psi_4 &&\\
\cblue C_{1224} &\cblue = \Psi_3 & C_{1334} &= \Psi_1 &&& \cblue C_{2434} &\cblue = -\Psi_3&&\\
C_{1234} &= \Psi_2 -\bar\Psi_2 &&&&&&
\end{align*} 
Every term comes from the definition via either complex conjugation or the tracefree condition
or both.  For example, to get $C_{1212}$, start from $C_{1342}=-\Psi_2$. Weyl tracefree implies 
\[
 0=\eta^{mn}C_{1mn2} = \underbrace{C_{1122}}_0+C_{1212} - C_{1342}-C_{1432} \Longrightarrow 
   C_{1212} = C_{1342}+\overline C_{1342} = - (\Psi_2+\overline\Psi_2).\vspace{-3mm}
\]
All vanishing components also come from the tracefree condition, as in the 
lines following Eq.~\eqref{e:crelns}. 

  Up to sign everything is determined by spin-weight and: \\
More $\ell$'s than $n$'s $\Rightarrow\Psi_0$, $\Psi_1$; \\
more $n$'s than $\ell$'s $\Rightarrow \Psi_3$, $\Psi_4$;\\  
if the component has the same number of $\ell$'s and $n$'s, it involves $\Psi_2$ and if 
real or imaginary is twice the real or imaginary part of $\Psi_2$.      

The two Bianchi identities involving $\psi_0$, \eqref{e:psi0psi1}, now involve the 
nonzero $\Gamma$'s in Eq.~\eqref{e:Gamma}.  \\
First identity:\vspace{-2mm}
\begin{align}
  \na_4 C_{1313} & = - \bar\delta \psi_0 -2\Gamma^m{}_{14} C_{m313} 
					- 2\Gamma^m{}_{34} C_{1m13}\nn\\
		 &= - \bar\delta \psi_0 -2\Gamma^4{}_{14} C_{4313}- 2\Gamma^3{}_{34} C_{1313}\nn\\
  		 &= - \bar\delta \psi_0 -4\varrho \psi_1 + 4\alpha\psi_0 \nn\\
  \na_1 C_{3413} &= D\psi_1 \quad\mbox{ because no $\Gamma^m{}_{n1}$ is nonzero}\nn\\
	\cblue0 & = 3\na_{[4} C_{13]13} \cblue= -(\bar\delta -4\alpha)\psi_0 +(D- 4\varrho) \psi_1.\vspace{-2mm}
\label{e:b1}\end{align}  
Second identity: \vspace{-3mm}
\begin{align}
  \na_2 C_{1313} & = - \DDelta \psi_0 \quad\mbox{ because no $\Gamma^m{}_{n2}$ is nonzero}\nn\\
  \na_3 C_{2113} &=  \delta \psi_1 -\Gamma^m{}_{23} C_{m113} - \Gamma^m{}_{13} C_{2m13}
			- \Gamma^m{}_{13} C_{21m3} - \Gamma^m{}_{33} C_{211m}\nn\\
		 &=  \delta \psi_1 -\Gamma^3{}_{23} C_{3113} 
				  - \Gamma^3{}_{13} \underbrace{C_{2313}}_0
			- \Gamma^3{}_{13} \underbrace{C_{2133}}_0 - \Gamma^3{}_{33} C_{2113}\nn\\
		 &=  \delta \psi_1 -\mu \psi_0 - 2\beta\psi_1 \\
	\cblue0 &= 3 \na_{[2} C_{13]13} \cblue=- (\DDelta +\mu)\psi_0 +(\delta-2\beta)\psi_1
\label{e:b2}\end{align}
Wave equation:  Again we need to eliminate $\psi_1$, but this time $\delta-2\beta$ 
doesn't commute with $D-4\varrho$.  This is easy to rectify because the operator\vspace{-2mm}
\[
r(\delta-2\beta) = \frac1{\sqrt2}\left(\pa_\theta + \frac i{\sin\theta}\pa_\phi -2\cot\theta\right) 
\]
is independent of $r$. So we multiply Eq.~\ref{e:b2} by $r$ before applying $D-4\varrho$: Using 
\\ $[D-4\rho,\ r(\delta -4\beta)]=0$, we write 
\[
    \frac1r \left(D+\frac4r\right) (r \times\mbox{ Eq.~\eqref{e:b2}}) - (\delta- 2\beta)(\mbox{ Eq.~\eqref{e:b1}}) =0, 
\] 
and obtain our decoupled equation for $\psi_0$: 
\be
  \frac1r(D-4\varrho) [r(\DDelta +\mu)\psi_0] - (\delta-2\beta)(\bar\delta -4 \alpha)\psi_0 = 0,
\label{e:bpflat1}\ee 
or 
\be
  \left[\left(D+\frac5r\right)\left(\DDelta + \frac1{2r}\right) 
    - \left(\delta-\frac1{\sqrt 2}\frac{\cot\theta}r\right)\left(\bar\delta+\sqrt2\frac{\cot\theta}r\right)\right]\psi_0
  = 0.
\label{e:bpflat2}\ee

In generalizing the derivation of our wave equation \eqref{e:bpflat1} for $\psi_0$ to Kerr, 
it will be helpful to write the commutation relation we used to eliminate $\psi_1$ in terms 
of spin coefficients and operators $D$ and $\delta$.  Replacing $r$ by $\dis-\frac1\varrho$, we have 
\be
  \crv\left[\left(D-5\varrho\right)\left(\DDelta +\mu\right) 
    - \left(\delta-2\beta\right)\left(\bar\delta-4\alpha\right)\right]\psi_0 = 0\cb.
\label{e:bpflat3}\ee  
\benr\item  Fill in the steps to obtain Eq.~\eqref{e:bpflat2} and then write it in terms 
of $\pa_t, \pa_r, \pa_\theta, \pa_\phi$, multiplying by $2r^2$ to get rid of the factors 
of $1/(2r^2)$.
\item Check that NP Eqs. (4.2e), and the second equation in NP (4.4) are, respectively 
\be
   D\beta = \rho\beta,\qquad [D,\delta] = \rho\delta, 
\ee 
and use these to go from Eqs.~\eqref{e:b1} and \eqref{e:b2} to the wave equation
in the form \eqref{e:bpflat3}.    
\label{ex:bp2}\een

\subsection{Schwarzschild}
\index{Bardeen-Press equation|(}  
We start with the Christoffel symbols and Weyl tensor for Schwarzschild.  Because the 
$\bm m$-$\barbm$ part of the metric is unchanged from flat space, only the commutators 
involving $D$ and $\DDelta$ are different.  From Eq.~\eqref{e:kins}, we have 
\be
\bm e_1 = D = \frac{r^2}\Delta \pa_t+\pa_r, \quad
\bm e_2 = \bm n = \DDelta = \frac12\left(\pa_t - \frac{\Delta}{r^2} \pa_r\right), \quad
\bm e_3 = \bm m = \delta = \frac1{\sqrt2 \, r}\left(\pa_\theta+\frac i{\sin\theta}\pa_\phi\right),
\label{e:sbasis}\ee
and 
\[
 \pa_t = \frac12\frac\Delta{r^2} \bm e_1 + \bm e_2, \quad 
 \pa_r = \frac12\bm e_1 -\frac{r^2}\Delta \bm e_2.
\]
Then
\begin{align*}
  [\bm e_1,\bm e_2] & = -\frac12\pa_r\left(1-2M/r\right)\pa_r 
		+ \frac12\left(1-2M/r\right)\pa_r\left(1-2M/r\right)^{-1}\pa_t\\
	    & = -\frac M{r^2}\pa_r -\frac12\left(1-2M/r\right)^{-1}\frac{2M}{r^2}\pa_t\\
	    & = -\frac M{r^2}\left(\frac12\bm e_1 -\frac{r^2}\Delta \bm e_2\right) 
		- \frac M{\Delta}\left(\frac12\frac\Delta{r^2} \bm e_1 + \bm e_2\right)\\
	    & = -\frac M{r^2}\bm e_1.\\
  [\bm e_1,\bm e_3] &= \pa_r \left(\frac1r\right) r\bm e_3 = -\frac1r \bm e_3, 
		\mbox{( This one didn't change)}\\
  [\bm e_2,\bm e_3] &= -\frac12\frac\Delta{r^2} \pa_r\left(\frac1r\right) r\bm e_3 = \frac12\frac\Delta{r^3}  \bm e_3
\end{align*} 
The flat-space $c^m{}_{np}$ and $c_{mnp}$ of Eq.~\eqref{e:commflat}, with changed coefficients in blue, 
are 
\begin{align*}
 {\cblue c^1{}_{12} = -\frac M{r^2}}\quad c^3{}_{13}&=c^4{}_{14} = -\frac1r, \quad 
 {\cblue c^3{}_{23}=c^4{}_{24} = \frac\Delta{2r^3}}, \quad
 c^3{}_{34}=- c^4{}_{34} = \frac{\cot\theta}{\sqrt2\ r},\nn\\
 {\cblue c_{212} = -\frac M{r^2}} \quad c_{413}&=c_{314} = \frac1r, \qquad 
 {\cblue c_{423}=c_{324} = -\frac\Delta{2r^3}}, \quad
 c_{434}=- c_{334} = -\frac{\cot\theta}{\sqrt2\ r}. 
\end{align*}

With $c_{212}$ nonzero, we have a new nonzero $\Gamma$, and Eq.~\eqref{e:lG} becomes 
\be
   {\cblue\Gamma_{212} = \frac M{r^2}}\quad \Gamma_{134} = \frac1r = -\varrho, \quad 
{\cblue\Gamma_{234} = \Gamma_{243}  = -\frac\Delta{2r^3} = \mu}, \quad 
   \Gamma_{343}  = \frac1{\sqrt2}\frac{\cot\theta}r  = \alpha-\bar\beta, 
\label{e:lGsch}\ee

\begin{align}
 {\cblue\Gamma^1{}_{12} = \g+\bar\g}\quad      \Gamma^1{}_{34} &= \Gamma^1{}_{43}  = \mu, \quad 
	\Gamma^2{}_{34} = \Gamma^2{}_{43} =-\varrho , \quad 
	\Gamma^3{}_{33} = \Gamma^4{}_{44} = 2\beta = - 2\alpha  , \nn\\
 \Gamma^3{}_{13} &= \Gamma^4{}_{14}  = -\varrho, \quad \Gamma^4{}_{24} = \Gamma^3{}_{23} =\mu,
	\quad \Gamma^3{}_{34} = \Gamma^4{}_{43} = -2\beta = 2\alpha .
\label{e:Gamma1}\\
{\cblue \gamma= \frac M{2r^2}},\quad  \alpha &= -\frac1{2\sqrt2}\frac{\cot\theta}r ,\quad \beta = \frac1{2\sqrt2}\frac{\cot\theta}r, 
\qquad \varrho = -\frac1r, \quad {\cblue \mu = -\frac\Delta{2r^3}}.
 \label{e:sspin}\end{align}

We next show that the Schwarzschild Weyl tensor is type D: Only $\Psi_2$ is nonzero.  
Instead of writing out the equation for each component of the Riemann tensor in terms of 
the $\Gamma$'s, let's take advantage of the fact that Newman and Penrose have already done this 
for us:  Eqs. (4.2) list the expressions for components of the Weyl and Ricci tensors in terms 
of the spin coefficients.  The Ricci tensor components are given as $\Phi_{mn}$, so for us 
$\Phi_{mn}=0$ in these equations.  The equations for the relevant components of the 
Weyl tensor are (4.2b) for $\Psi_0$, (4.2c) for $\Psi_1$ and (4.2f) for $\Psi_2$.  Because of the $\ell\leftrightarrow n$, $m\leftrightarrow\barm$ symmetry, we need not check that $\Psi_3$ and $\Psi_4$ vanish. 

First (4.2b) for $\Psi_0 = -C_{1313}= C^4{}_{131}$. This one time we'll check the NP equation 
from scratch.  There's no subtlety. Begin with the Riemann tensor written in terms 
of $\Gamma$'s and then write the NP symbol for each $\Gamma$ (listed on the last page of these notes).  In the second line below, we omit terms that vanish by 
$\Gamma_{ijk}=-\Gamma_{jik}$, (like $\G1{23}$).  For a general basis, the 
components of the Riemann tensor are given by Eq.~\eqref{rijkl1}:     
\begin{align*}
   R^i{}_{jkl} & = e_k \Gamma^i{}_{jl} - e_l \Gamma^i{}_{jk} 
	+ \Gamma^i{}_{mk} \Gamma^m{}_{jl} - \Gamma^i{}_{ml} \Gamma^m{}_{jk}  
	+2\Gamma^i{}_{jm}\Gamma^m{}_{[kl]}\\
  C^4{}_{131}&= e_3\G4{11}-e_1\G4{13} + \G{4}{m3}\G{m}{11}
			-\G4{m1}\G{m}{13} +2 \G{4}{1m}\G{m}{[31]}\\
	    &= -\delta\kappa + D \sigma 
		+ (\G{4}{13}\G{1}{11}+\G{4}{43}\G{4}{11})  
			-(\G4{11}\G{1}{13}+\G4{41}\G{4}{13}) 
		+2( \G{4}{11}\G{1}{[31]}+\G{4}{12}\G{2}{[31]})\\
&= -\delta\kappa + D \sigma + [-\sigma(\ep+\bar\ep) + (\bar\a -\b))(-\kappa) ] 
			-[(-\kappa)(-\sigma) +(\bar\ep-\ep)(-\sigma)] \\  
	&\qquad\qquad +[(-\kappa)\bar\pi +(-\tau)(-\kappa) +(-\sigma)(\ep-\bar\ep) 
	  -(-\sigma)(-\bar\varrho) - (-\varrho)(-\sigma)] \quad \Longrightarrow 
\end{align*}
\be
D\sigma-\delta\kappa = (\varrho+\bar\rho+3\ep-\bar\ep)\sigma 
			-(\tau-\bar\pi+\bar\a+3\b)\kappa +\Psi_0   \tag{NP 4.2b}
\label{4.2b}\ee 
 
For Schwarzschild, only $\a,\b,\g,\mu$ and $\varrho$ are nonzero. 
Because $\sigma$ and $\kappa$ vanish, we have
\be
\Psi_0 = 0.  
\ee
The next Weyl-tensor equation we need is (4.2c) for $\Psi_1 = -C_{1213}=-C^2{}_{312}$:
\be 
  D\tau-\Delta\kappa = (\tau+\bar\pi)\varrho + (\bar\tau+\pi)\sigma
			+(\ep-\bar\ep)\tau - (3\g+\bar\g)\kappa+\Psi_1 \tag{NP\ 4.2c}
\label{4.2c}\ee  
In this case, because $\tau$, $\pi$, $\sigma$, and $\kappa$ vanish, we have 
\be
  \Psi_1 = 0.  
\ee
The symmetry of Schwarzschild under $t\rightarrow -t$ and under $\phi\rightarrow -\phi$ implies a symmetry of the NP equations under $\ell\leftrightarrow n$ and $m\leftrightarrow \bar m$, giving $\Psi_3=\Psi_4 = 0$, and leaving only $\Psi_2$ 
nonzero as claimed.  
To find its value, we use a third Weyl-tensor equation, (4.2f): The only nonzero 
spin coefficient in that equation is $\gamma$ and the calculation is simple:  
\be
  \Psi_2 = D\g  = \pa_r\left(\frac M{2r^2}\right) =-\frac M{r^3}.  
\ee

\newpage
We now turn to the wave equation for Schwarzschild, again looking
at the two components of the Bianchi identity that gave us the 
wave equation for $\psi_0$.  The Bianchi identity components are NP (4.5). 
$\na_{[4}C_{13]13}$ is the first of these equations, 
\index{Bianchi identities!in Newman-Penrose formalism}
\be
   (D - 2\ep -4\varrho)\Psi_1 -(\bar\delta -4\alpha +\pi)\Psi_0 = -3\kappa\Psi_2,
\label{e:bs1}\ee
and the fifth is $\na_{[2}C_{13]13}$, 
\be
   (\DDelta -4\gamma +\mu)\Psi_0 - (\delta-2\beta-4\tau)\Psi_1  =3\sigma\Psi_2.
\label{e:bs2}\ee
We need to linearize the equations about the Schwarzschild background.  
That is, we consider a family of metrics, 
$g_{\a\b}(\lambda) = g_{\a\b}(0) + \lambda \dot g_{\a\b} + O(\lambda^2)$, 
with $g_{ab}(0)$ the unperturbed spacetime, and we write
\be
  \dot g_{\a\b} := \left.\frac d{d\lambda}g_{\a\b}(\lambda)\right|_{\lambda=0}. 
\ee
We'll write the unperturbed quantity, like $\Psi_2(0)$ or the spin-coefficient 
$\alpha(0)$ as simply $\Psi_2$ and $\alpha$.  We'll again use $\psi_i:=\dot\Psi_i$, 
and will just write $\dot\sigma$, $\dot\kappa$ for the perturbed spin coefficients.

Then, because $\ep, \kappa, \mu$, $\Psi_0$ and $\Psi_1$ vanish at $\lambda=0$,  $\dis \left.\frac d{d\lambda}\mbox{Eq.~\eqref{e:bs1}}(\lambda)\right|_{\lambda=0} $
is 
\be\cblue
 (D-4\varrho)\psi_1 -(\bar\delta -4\alpha)\psi_0 = -3\dot\kappa\Psi_2.
\label{e:lbs1}\ee
Similarly, the linearized equation \eqref{e:bs2} is 
\be\cblue
   (\DDelta -4\gamma +\mu)\psi_0 - (\delta-2\beta)\psi_1  =3\dot\sigma\Psi_2.
\label{e:lbs2}\ee
Again we need to eliminate $\psi_1$, and again just multiplying \eqref{e:lbs2} by 
$r$ gets rid of the $r$ dependence in $(\delta-2\beta)$, allowing it to commute 
with $D$.  But now there's a problem.  The right sides of the two equations 
involve $\dot\kappa$ and $\dot\sigma$, so we don't succeed in getting a 
decoupled equation for $\psi_0$.  

We are saved by a miracle that will look more miraculous for Kerr.  
The equation for $\psi_0$ in terms of the $\Gamma$'s is the NP equation \eqref{4.2b}
that we derived above, and its linearized version is
\be
   (D-\varrho-\bar\varrho)\dot\sigma - (\delta - \bar\alpha - 3\beta)\dot \kappa 
	= \psi_0, 
\ee 
or, using $\varrho = \bar\varrho,\ \bar\alpha = -\beta$ for Schwarzschild), 
\be\cblue
   (D-2\varrho)\dot\sigma - (\delta - 2\beta)\dot \kappa 
	= \psi_0. 
\ee 
We will simply follow our path to the equation in flat space, leaving 
terms involving $\dot\sigma$ and $\dot\kappa$ on the right side 
of the wave equation. And those terms will be the above expression for $\psi_0$.    

As in Exercise 8, we use NP equations 
(4.2e) and (4.4); and we also use the second Bianchi identity in (4.5), 
all for the unperturbed Schwarzschild spacetime:  
\be \cblue
  D\beta = \varrho \beta, \qquad [D,\delta] = \varrho\delta, \qquad 
	D\Psi_2 = 3\varrho\Psi_2.  
\label{e:relns1}\ee
We just follow the steps for flat space, writing 
$(\delta-2\beta) [\mbox{Eq.~\eqref{e:lbs1}}] + (D-5\varrho)[\mbox{Eq.\eqref{e:lbs2}}]$ 
\begin{align}
  (\delta -2\beta)(D-4\varrho)\psi_1& -(\delta-2\beta)(\bar\delta-4\alpha)\psi_0\nn \\
    + (D-5\varrho) (\DDelta-4\gamma+\mu)\psi_0& - (D-5\varrho)(\delta-2\beta)\psi_1 
  = -3\Psi_2(\delta-2\beta)\dot\kappa + 3(D-5\varrho)(\Psi_2\dot\sigma).
\label{e:prebp}\end{align}
The commutation relation that eliminates $\psi_1$ is unchanged from the flat-space case:  Using the first two relations in Eq.~\eqref{e:relns1}, we have
\begin{align*}
  [\delta-2\beta,D-4\varrho] &= [\delta,D] +2D\beta = -\varrho(D-2\beta) \\
\Longrightarrow\ \ (\delta-2\beta)(D-4\varrho)\psi_1 &= (D-5\varrho)(D-2\beta)\psi_1,  
\end{align*}
and the terms involving $\psi_1$ cancel exactly as before.  
On the right side, the term involving $\dot \sigma$ is 
\begin{align*}
 3(D-5\varrho)(\Psi_2\dot\sigma)  
			&= 3[(D\Psi_2)\dot\sigma + \Psi_2(D-5\varrho)\dot\sigma] 
			 = 3[3\varrho\Psi_2\dot\sigma + \Psi_2(D-5\varrho)\dot\sigma]\\
			&= 3\Psi_2 (D-2\varrho)\dot\sigma.
\end{align*}
Finally, Eq.~\eqref{e:prebp} becomes 
\[
[(D-5\varrho) (\DDelta-4\gamma+\mu)-(\delta-2\beta)(\bar\delta-4\alpha)]\psi_0 
  = 3\Psi_2[-(\delta-2\beta)\dot\kappa + (D-2\varrho) \dot\sigma) = 3\Psi_2\psi_0,  
\]
or 
\be\cvi
  [(D-5\varrho) (\DDelta-4\gamma+\mu)-3\Psi_2 -(\delta-2\beta)(\bar\delta-4\alpha)]\ \psi_0 = 0.  
\label{e:bp}\ee
This is the Bardeen-Press equation\cite{bp73}. Multiplying by $r^2$ (or $\varrho^{-2}$) gives an equation with all radial dependence and derivatives in the first terms and all angular dependence and derivatives in the last: The separation of variables is essentially already here. \\

\noindent{\sl Explicit equation}: \\
Multiply by $-2r^2$ and bring the $r^2$ 
inside $(D-5\rho)$, changing it to $(D-3\rho) = (D+3/r)$.  Then, using 
Eqs.~\eqref{e:sbasis} for the derivative operators and Eqs.~\eqref{e:sspin} for 
the spin coefficients, we obtain the explicit form 
\index{Bardeen-Press equation|textbf}
\begin{align*}
\left\{\phantom{\frac12}\right.\hspace{-4mm}
    &\left(\frac{r^2}\Delta\pa_t + \pa_r +\frac3r\right)
      \left[\Delta\left(-\frac{r^2}\Delta \pa_t+ \pa_r 
				+\frac1r + \frac{4M}\Delta\right)\right] \nn\\
    &\left.+ \left(\pa_\theta+\frac i{\sin\theta} \pa_\phi - \cot\theta\right)
     \left(\pa_\theta-\frac i{\sin\theta} \pa_\phi +2\cot\theta\right)
   \right\}\psi_0 = 0.	 
\end{align*} 
\vspace{1mm}

\noindent In our derivation, the input was the set of six equations in 
blue on the last page: \\
\indent two perturbed Bianchi identities, \\
\indent the equation for $\psi_1$ in terms of Christoffel symbols (spin coefficients), \\
\indent the vanishing of $\Psi_1$ in Schwarzschild (giving $D\beta=\varrho\beta$), \\
\indent the commutator $[D,\delta]=[\bm e_1,\bm e_3]$ in Schwarzschild, \\
\indent a Bianchi identity for $\Psi_2$ in Schwarzschild.\\
We also used the fact that all spin coefficients are real. We'll follow the 
same path for Kerr, with a couple of additional relations associated with 
$\theta$ derivatives that vanish for Schwarzschild and with spin coefficients 
that are complex.   \\        
\index{Bardeen-Press equation|)}
\newpage
\subsection{Kerr}
\label{s:teukolsky}
\index{Teukolsky equation|(}
Although the explicit forms of the spin coefficients is given below, 
with a sample calculation, they are not used to derive the Teukolsky 
equation in the form~\eqref{e:teuk} below.  You need only the NP 
equations and the fact that the other spin coefficients vanish.  
The vanishing of $\kappa, \sigma, \lambda, \nu, \tau$ is automatic for 
a type D Weyl tensor: Five components of the Bianchi identities-- 
i.e., five of NP Eqs. (4.5)-- are simply
\[
3\kappa\Psi_2 = 0,\quad -3\lambda\Psi_2=0,\quad 3\sigma\Psi_2=0,\quad -3\tau\Psi_2=0, \quad 3\nu\Psi_2=0.
\] 
The choice of Kinnersley tetrad also makes $\epsilon=0$ and $\alpha=\pi-\beta$. 
The derivation of Eq.~\eqref{e:teuk} using only this information, begins 
after Eq.~\eqref{e:rho} below.  
  
From Eq.~\eqref{e:kinnersley}, we have 

\bsube\begin{align}
   \bm e_1 = \bm\ell^\cdot 
	&= \frac1\Delta [(r^2+a^2)\pa_t +\Delta\pa_r + a\pa_\phi], 
\label{e:lkin}\\
    \bm e_2 = \bm n^\cdot
	&= \frac1{2\rho^2}[(r^2+a^2)\pa_t -\Delta\pa_r+a\pa_\phi],\\
   \bm e_3 = \bm m^\cdot 
	&= \frac1{\sqrt2\ (r+ia\cos\theta)}
	   \left(ia\sin\theta\pa_t +\pa_\theta+ \frac i{\sin\theta}\pa_\phi\right)
\end{align}\label{e:e_m} \esube
 
In calculating the commutation coefficients $c^p{}_{mn}$, you'll need to 
invert these expressions -- to invert $\bm e_m = e_m{}^\mu\pa_\mu$ in order to 
express $\pa_\mu$ in terms of $\bm e_m$.  The easy way to do this is to 
use the dual basis $\omega^m$.  That is, because 
$\omega^m(\bm e_n) = \omega^m{}_\mu e_n{}^\mu = \delta^m_n$, \ $\omega^m{}_\mu$ 
is the inverse of $e_n{}^\mu$: $\ \ \omega^m{}_\mu e_m{}^\nu = \delta_\mu^\nu$.  Then
\[
   \partial_\mu = \omega^m{}_\mu \bm e_m.  
\]
  From the orthonormal dual basis vectors listed after 
Eq.~\eqref{e:Delta_rho}, we quickly see   

\[
 \omega^1 = \frac{\sqrt\Delta}{2\rho}(\Omega^0+\Omega^1), \quad
\omega^2 = \frac\rho{\sqrt\Delta}(\Omega^0 - \Omega^1), \qquad
\omega^3 = \frac{r+ia\cos\theta}{\sqrt2\,\rho}(\Omega^2-i\Omega^3)
\] 
giving
\bsube
\begin{align}
 \omega^1 &= \frac\Delta{2\rho^2}(dt-a\sin^2\theta d\phi) + \frac12 dr, \\
  \omega^2 &=dt-a\sin^2\theta d\phi - \frac{\rho^2}\Delta dr, 
  \label{e:omega2}\\
  \omega^3 &= \frac1{\sqrt2(r-ia\cos\theta)}\left[ 
		ia\sin\theta\, dt +\rho^2 d\theta 
		- (r^2+a^2)\, i\sin\theta d\phi
		\right].
\label{e:omega3}
\end{align}\esube

Here are the nonzero spin coefficients, followed by a sample calculation of one 
of them: 
\begin{align}
  \varrho &= - \frac1{r-ia\cos\theta} \quad 
     \beta = \frac1{2\sqrt2\,(r+ia\cos\theta)} \cot\theta, \quad
     \pi = \frac{ia}{2\sqrt2\,(r-ia\cos\theta)^2} \sin\theta,\quad
     \alpha  = \pi - \bar\beta, \nn\\ 
     \mu &= -\frac\Delta{2\rho^2(r-ia\cos\theta)}, \quad 
     \gamma = \mu+\frac{r-M}{2\rho^2},\quad \tau = -\frac{ia}{2\rho^2} \sin\theta.
\end{align}

Calculation of $\varrho = \Gamma_{314} = \frac12(c_{431}-c_{143}-c_{314})$: \\
\bsube\begin{align}
c_{143} &= c^2{}_{43} = -\omega^2([\bm e_3,\bm e_4]),\\
c_{431} &= -c^3{}_{31} = \omega^3([\bm e_1,\bm e_3]),\quad 
c_{314} = -c_{341} = \overline c_{431}. 
\end{align}\esube 
\newpage

Next, use Eqs.~\eqref{e:e_m} for $\bm e_m$ to write  
\begin{align*}
[\bm e_3,\bm e_4] 
	&= \frac1{\sqrt2(r+ia\cos\theta)} \pa_\theta
  	  \left[\frac1{\sqrt2(r-ia\cos\theta)}	
	      \left(-ia\sin\theta\pa_t +\pa_\theta-\frac i{\sin\theta}\pa_\phi\right)
	  \right] - \mbox{c.c.} \\
	&= -\frac{\pa_\theta(r-ia\cos\theta)}{\sqrt2\rho^2} \bm e_4 
	   +\frac{\pa_\theta(r+ia\cos\theta)}{\sqrt2\rho^2} \bm e_3 
	   + \frac2{2\rho^2}\left(-ia\cos\theta\pa_t 
	                          +\frac {i\cot\theta}{\sin\theta}\pa_\phi\right)\\
	&=-\frac{ia\sin\theta}{\sqrt2\rho^2}(\bm e_3+\bm e_4) 
	   +\frac i{\rho^2}\cot\theta
	   		\left(-a\sin\theta\pa_t + \frac1{\sin\theta}\pa_\phi\right),\\
[\bm e_1,\bm e_3] &= - 2\frac{\pa_r(r+ia\cos\theta)}{r+ia\cos\theta}\bm e_3 
	   = -\frac1{r+ia\cos\theta}\bm e_3 
\end{align*}
Using Eqs.~\eqref{e:omega2} and \eqref{e:omega3}, we have
\begin{align*}
c_{143} &=  -\omega^2([\bm e_3,\bm e_4]) 
	= -\frac i{\rho^2}\cot\theta(-a\sin\theta + a\sin\theta)
	= 2\frac{ia\cos\theta}{\rho^2}\\
c_{413} &= \omega^3([\bm e_1,\bm e_3]) = -\frac1{r+ia\cos\theta},  	
\end{align*} 
and 
\be
  \varrho = \Gamma_{314} 
	= \frac12\left(-\frac{2r}{\rho^2} -2\frac{ia\cos\theta}{\rho^2} \right) 
	= -\frac{2r}{\rho^2}.
\label{e:rho}\ee

We'll use the following equations from NP, most of which we've already used: 
Starting from Eq.~\eqref{4.2b} for $\Psi_0$ in terms of spin coefficients, they are
\begin{align*}
  (D-\varrho-\bar\varrho)\sigma - (\delta-\tau-2\beta)\kappa &=\Psi_0,\hspace{8.7cm}\mbox{(4.2b)} \nn\\ 
 (D-4\varrho)\Psi_1 - (\bar\delta-4\alpha+\pi)\Psi_0 &= - 3\kappa\Psi_2, \tag{4.5-first}\\
 (\DDelta - 4\g +\mu)\Psi_0 - (\delta -2\beta-4\tau)\Psi_1 &= -3\sigma\Psi_2 \tag{4.5-fifth}\\
 D\Psi_2 &= 3\varrho\Psi_2 \tag{4.5-second}\\
 \delta\Psi_2 &= 3\tau\Psi_2 \tag{4.5-sixth}\\
 D\tau &= \varrho \tau + \varrho\bar\pi +\Psi_1 \tag{4.2c}\\
 D\beta&=\bar\varrho\beta \tag{4.2e}\\
 \delta\varrho &= (\bar\alpha +\beta)\varrho + (\varrho - \bar\varrho)\tau \tag{4.2k} \\
[D,\delta] = \bar\varrho\delta \tag{4.4} 
\end{align*} 

In the first paragraph, we used the vanishing of $\Psi_i$, $i\neq 2$ to infer 
the vanishing of a set of spin coefficients.  Conversely, if one uses the 
explicit tetrad to show $\kappa = \sigma  = 0 $, then, from the NP 
equation for $\Psi_0$ in terms of spin coefficients and \eqref{4.2b}, we immediately have $\Psi_0=0$ for the background 
spacetime. Eq.~\eqref{4.2c} implies $\Psi_1 = 0$, after one checks that 
$ D\tau = (\tau+\bar\pi)$. 
\footnote{The relations $\kappa=\sigma = 0 \Longleftrightarrow \Psi_0=\Psi_1=0$ 
and $\nu=\lambda = 0 \Longleftrightarrow \Psi_3=\Psi_4=0$ are two instances of 
the Goldberg-Sachs theorem.  The vanishing of $\kappa$ and $\sigma$ is the 
statement that $\ell^\a$ generates a congruence of shear-free null geodesics and 
the theorem states that such a congruence exists if and only if the Weyl 
tensor is algebraically special, $\Psi_0=\Psi_1 = 0$. Similarly, $\nu=\lambda = 0$ means that $n^\alpha$ generates a second family of shear-free null geodesics.} 

 We use the remaining equations in their perturbed forms to get the Teukolsky equation.  
As usual, we start with the Bianchi identities for $\psi_0$, 
\begin{align}
(D-4\varrho)\psi_1 - (\delta -4\alpha+\pi)\psi_0 &= -3\dot\kappa\Psi_2, \label{e:kappa1}\\
(\DDelta - 4\gamma+\mu) \psi_0 - (\delta -4\tau- 2\beta)\psi)_1 &= -3\dot\sigma\Psi_2.
\label{e:sigma}\end{align}   
From the right sides, we infer the form of the operators that should act on each 
equation to eliminate $\dot\sigma$ and $\dot\tau$ and hope that they will also 
eliminate $\psi_1$:  From \eqref{4.2b}, we need $(D-\varrho-\bar\varrho)$ to act on
$\dot\sigma$, and $(\delta-\tau-2\beta)$ to act on $\dot\kappa$.  By (4.5-second) and 
(4.5-sixth), we have 
\begin{align*}
  (D-4\varrho - \bar\varrho)(\Psi_2\dot\sigma) &= \Psi_2(D-\varrho - \bar\varrho)\dot\sigma, \\
  (\delta - 4\tau-2\beta) (\Psi_2\dot\kappa) &= \Psi_2(\delta - \tau - 2\beta)\dot\kappa,
\end{align*} 
So we use $(D-4\varrho - \bar\varrho)$[Eq.\eqref{e:sigma}$ - (\delta-4\tau-2\beta)$[Eq.~\eqref{e:kappa1}]: 
\begin {align*}
(D-4\varrho - \bar\varrho)&[(\DDelta - 4\gamma+\mu) \psi_0 - (\delta -4\tau- 2\beta)\psi)_1] \\
- (\delta-4\tau-2\beta)&[(D-4\varrho)\psi_1 - (\delta -4\alpha+\pi)\psi_0] = -3\Psi_2 \psi_0.
\end{align*}
Then what remains is to show 
\[ 
[(D-4\varrho - \bar\varrho)(\delta -4\tau- 2\beta) -(\delta-4\tau-2\beta)(D-4\varrho)]\psi_1 = 0. 
\] 
The operator on the left side is 
\be 
  -\bar\varrho(\delta-4\tau-2\beta) + [D-4\varrho, \delta-4\tau - 2\beta].
\label{e:op}\ee 
Expanding the commutator, and using the last set of NP equations above, we have 
\begin{align*}
   [D-4\varrho, \delta-4\tau - 2\beta] &=[D,\delta] -4 D\tau-2D\beta+4\delta\varrho \\
&= \bar\varrho\delta - 4\tau\varrho - 4\bar\pi\varrho - 2\bar\varrho \beta +4(\bar\alpha+\beta)\varrho + 4(\varrho-\bar\varrho)\tau \\
&= \bar\varrho(\delta-4\tau-2\beta), 
\end{align*}
canceling the first term in \eqref{e:op} ! (Use $\bar\pi = \bar\alpha +\beta$ to get the 
last equality.) 

The Teukolsky equation is then 
\be\crv
\left[\,(D-4\varrho - \bar\varrho)(\DDelta - 4\gamma+\mu)+3\Psi_2 
			- (\delta-4\tau-2\beta)(\delta -4\alpha+\pi)\,\right]\psi_0 = 0 \cb.
\label{e:teuk}\ee 
 \index{Teukolsky equation|(}\index{Teukolsky equation|textbf} 
\newpage
  
\section{Appendix to Chapter \ref{c:np}}

\noindent{\sl Naming of parts} 

\[
  \ell^\mu\pa_\mu = D, \quad n^\mu\pa_\mu =\DDelta\ ,
\quad m^\mu\pa_\mu = \delta, \quad \barm^\mu\pa_\mu = \bar\delta .   
\]  

\begin{align*}
   \kappa &= \Gamma_{311} &\varrho &= \Gamma_{314} & 
				\ep &= \frac12(\Gamma_{211}+\Gamma_{341}) \nn\\
   \sigma &= \Gamma_{313} & \mu &= \Gamma_{243} &
				\gamma &= \frac12(\Gamma_{212}+\Gamma_{342})\nn\\
   \lambda &= \Gamma_{244} & \tau &= \Gamma_{312} &
				\alpha &= \frac12(\Gamma_{214}+\Gamma_{344})\nn\\
   \nu &= \Gamma_{242} & \pi &= \Gamma_{241} &
				\beta &= \frac12(\Gamma_{213}+\Gamma_{343})\nn\\
\Gamma_{211} = \ep+\bar\ep &\quad  \Gamma_{341} = \ep - \bar\ep & \Gamma_{212} =\gamma+\bar\gamma 
&\quad\Gamma_{342} = \gamma-\bar\gamma & \Gamma_{214} =\alpha +\bar\beta 
& \quad\Gamma_{344} = \alpha - \bar\beta.
\end{align*}

\begin{align*}
\G1{11} &=\ep+\bar\ep &\G2{21}&=-(\ep+\bar\ep) &\G3{11}  &= -\bar\kappa 	&\G4{11}  &= -\kappa&\\
\G1{12} &=\g+\bar\g &  \G2{22} &=-(\g+\bar\g)	&\G3{12} &= -\bar\tau   	& \G4{12}  &= -\tau& \\
\G1{13} &= \bar\a+\b & \G2{23} &= -(\bar\a+\b)	&\G3{13} &= -\bar\varrho    	& \G4{13}  &= -\sigma&\\
\G1{14} &= \a+\bar\b & \G2{24} &= -(\a+\bar\b)	&\G3{14} &= -\bar\sigma		&\G4{14} &=-\varrho&\\
\G1{31} &=\bar\pi   & \G2{31}  &=-\kappa    	&\G3{21} &= \pi          	& \G4{21}& =\bar\pi&\\
\G1{32} &=\bar\nu   & \G2{32}  &= -\tau 	&\G3{22} &=  \nu 			&\G4{22}& = \bar\nu&\\
\G1{33} &=\bar\lambda&\G2{33}  &= -\sigma	&\G3{23} &= \mu 		&\G4{23}&=\bar\lambda \\
\G1{34} &= \bar\mu  & \G2{34}  &=-\varrho	&\G3{24} &=\lambda 		&\G4{24}&=\bar\mu\\
\G1{41} &= \pi       &\G2{41}  &=-\bar\kappa	&\G3{31} &= \ep-\bar\ep		&\G4{41}&=\bar\ep-\ep\\
\G1{42} &= \nu       &\G2{42}  &= -\bar\tau	&\G3{32} &=\g-\bar\g		&\G4{42}&=\bar\g-\g\\
\G1{43} &= \mu       &\G2{43}  &=-\bar\varrho	&\G3{33} &=\b-\bar\a		&\G4{43}&=\bar\a-\b\\
\G1{44} &= \lambda   &\G2{44}  &=-\bar\sigma	&\G3{34} &=\a-\bar\b		&\G4{44}&=\bar\b-\a\end{align*}

\begin{align*}
C_{1212} &= -(\Psi_2+\bar\Psi_2) & \cblue C_{1313} &\cblue = -\Psi_0 & C_{1414} &= -\bar\Psi_0 & C_{2323} &= -\bar\Psi_4 & C_{3434} &= -(\Psi_2+\bar\Psi_2)\\
\cblue C_{1213} &\cblue = -\Psi_1 & C_{1314} &= 0 & C_{1423} &= \bar\Psi_2 & C_{2324} &= 0 && \\
C_{1214} &= -\bar\Psi_1 & C_{1323} &= 0 & C_{1424} &= 0 & C_{2334} &= \bar\Psi_3 &&\\
C_{1223} &= \bar\Psi_3 &\cblue C_{1324} &= \cblue \Psi_2 & C_{1434} &= -\bar\Psi_1 & C_{2424} &= -\Psi_4 &&\\
\cblue C_{1224} &\cblue = \Psi_3 & C_{1334} &= \Psi_1 &&& \cblue C_{2434} &\cblue = -\Psi_3&&\\
C_{1234} &= \Psi_2 -\bar\Psi_2 &&&&&&
\end{align*} 

\chapter{Black Hole Thermodynamics}
\label{laws}
\index{thermodynamics!black hole}
\index{entropy!black hole}

We're back to  $-+++$ and are roughly following the Bardeen-Carter-Hawking paper, 
{\sl The Four Laws of Black Hole Mechanics}, Comm. Math. Phys. {\bf 31}, 161-170 (1973).   These analogs of four laws of thermodynamics are presented below 
in the order 
\begin{enumerate}
\item[] 2nd law:  Area increase theorem, corresponding to entropy increase. 

\item[] 0th law:  The surface gravity $\kappa$ of a black hole is constant, 
corresponding to the constant temperature of a system in thermodynamic equilibrium.

\item[] 3rd law: One cannot reduce $\kappa$ to zero, corresponding to the inability to 
reach $T=0$ in classical thermodynamics.  

\item[]1st law:  $\delta M = \frac1{8\pi}\kappa\delta A + \Omega_H \delta J$.
 \end{enumerate}   

\section{The 2nd Law:  The Area of the Event Horizon Can Never Decrease}
\index{area increase theorem} 

Recall that the event horizon, $H$, is the boundary of the past of
future null infinity.  From any point outside $H$ there is a future
directed timelike line that reaches infinity---an inextendible timelike
line without future endpoint that remains forever outside $H$.  By
considering a family of such lines that begin from points $P_1, P_2,
\cdots$ converging to $P$ on $H$, one acquires at each $P$ in $H$ a
future directed null line $l$ that is inextendible  and has no
future endpoint.  
\begin{figure}[H]
               \begin{center}
		\includegraphics[width=.6\textwidth]{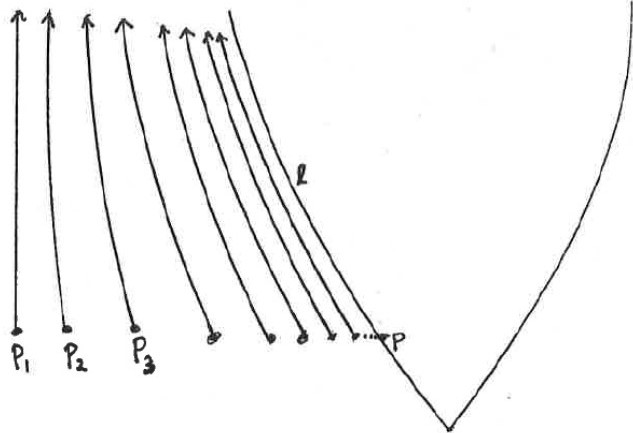}
		\end{center}
\caption{Similar to figures in a Geroch-Horowitz paper\cite{gh79}}
\end{figure}
\hspace{-10mm}

\noindent These null lines are called the null generators of $H$.\\

Let $\ell^\alpha$ be a nonvanishing vectorfield that is tangent
on $H$ to the null generators.  

\noindent {\sl Claim}:  The null generators of a null hypersurface (e.g.,
$H$) are geodesics.\\

\noindent {\bf Proof}:  Call the null hypersurface $H$.  Let $u$ be a scalar 
that is constant on $H$ and with $\nabla_\alpha u \not= 0$.  Note first that 
$\nabla_\alpha u$ is null.  This follows from the fact that the tangent space 
at a point $P$ of $H$ is spanned by a null vector $\ell^\a$ and two orthogonal 
spacelike vectors.  
Then $\ell^\a$ is orthogonal to every vector in $H$; and any vector 
orthogonal to all other vectors in $H$ must be along $\ell^\a$ (other vectors 
are not null and so not orthogonal to themselves). So  
$\ell_\alpha=f\nabla_\alpha u$, some scalar $f$.  We have
\begin{eqnarray*}
\ell^\b\n_\b \ell^\a &=& \ell^\b\n_\b (f\n^\a u) 
	= (\ell^\b\n_\b f)\n^\a u + f\ell^\b\n_\b \n^\a u\\
&=& \frac{\ell^\b\n_\b f}{f} \ell^\a + f^2 \n^\b u\n^\a \n_\b u 
= \frac{\ell^\b\n_\b f}{f} \ell^\a 
+ f^2 \frac12 \n^\a\underbrace{ (\n^\b u\n_\b u)}_0 
\\
&=& \frac{\ell^\b\n_\b f}{f} \ell^\a.
\end{eqnarray*}
Thus $\ell^\a$ generates a geodesic congruence.\hspace{4mm} $\Box$

When we get to stationary, axisymmetric black holes, we will see that 
$\ell^\alpha$ is along a Killing vector, and we can set it equal to 
the Killing vector if we do not use affine parametrization.  Following 
Bardeen-Carter-Hawking, we write 
\be 
	\ell^\b\nabla_\b\ell^\a = \kappa\ell^\a. 
\label{e:kappa}\ee
This $\kappa$ is {\sl not} the NP spin coefficient, 
which we will denote by $\kappa_{NP}$ on the one occasion when we 
need it.  

We can now state the area increase law.  Let $S$ be a spacelike slice
of $H$ orthogonal to $\ell^\a$; and let $\psi_\la$ be the family of
diffeos generated by $\ell^\a$, so $\psi_\la$ moves you a parameter
distance $\la$ along the null generators of $H$.  Let $S_\la = \psi_\la
(S)$ and let $A_\la$ be the area of the black hole with boundary
$S_\lambda$: $\ A_\la = \int\limits_{S_{\la}} \ep_{\a\b}dS^{\a\b}$.

\noindent {\bf 2nd Law}: 
\be \frac{d}{d\la} A_\la \geq 0 . \label{e:dAdlambda}\ee

This theorem, due to Hawking, follows quickly from the {\em Raychaudhuri equation}, 
the general relativistic form of the {\em optical scalar equation}.  This equation describes 
the cross section of a bundle of nearby null rays -- called a {\em congruence} -- 
and it is applied to the generators of the horizon.  Let $\ell^\alpha$ be a null 
vector field tangent to a family of affinely parametrized null geodesics.  
(For this proof we use the affinely parametrized $\ell^\a$.) 
A circular cross section of the family distorts as it evolves, and the distortion 
can be regarded as a sum of three parts, a change in its area, a rotation of the 
circle and a shearing of the circle, taking the circle to an ellipse.  In terms of 
$\ell^\alpha$ this describes a decomposition of $\nabla_\alpha \ell_\beta$ or, more 
accurately, of the projection of $\nabla_\alpha \ell_\beta$ onto the cross section 
orthogonal to $\ell^\alpha$.  

Let $n^\a$ be a null vector with $\ell^\a n_\a =-1$.
Let $e_{\a\b}$ be the projection $\perp$ to $\ell^\a$ and $n^\a$:
\be 
	e_{\a\b} = g_{\a\b} + \ell_\a n_\b + \ell_\b n_\a 
		= m_\a \overline m_\b + \overline m_\a m_\b.
\label{2}\ee
We can regard $e_{\a\b}$ as the Euclidean 2-metric on a cross section of the 
congruence of null geodesics. 

The divergence, shear and rotation of the congruence are defined by 
\bea
e_\a^\c e_\b^\de \n_\c\ell_\de &=& \sigma_{\a\b} +\frac12 e_{\a\b}\theta + \omega_{\a\b}, \nonumber \\
\theta &=& e^{\a\b}\n_\a \ell_\b,\qquad 
\sigma_{\a\b} = e_\a^\c e_\b^\de\n_{(\c}\ell_{\de)}-\frac12 e_{\a\b}\theta,\qquad
\omega_{\a\b} = e_\a^\c e_\b^\de\n_{[\c}\ell_{\de]}
\eea

Because $\ell_\a = f\n_\a u$, $\omega_{\a\b}$ vanishes:
\begin{eqnarray*}
\omega_{\a\b} &=& e_{[\a}{}^\g e_{\b ]}{}^\de \n_\g\ell_\de\\
&=& e_{[\a}{}^\g e_{\b]}{}^\de \n_\g f\, \n_\de u\\
&=& 0, \hspace{3mm} {\mbox{using }} e_\a{}^\b\n_\b u=0 .
\end{eqnarray*}

Then, using 
$\ell^\b\na_\b \ell_\a = \kappa \ell_\a,\ \ell^\b \na_\a \ell_\b = 0$, we have 
\begin{eqnarray}
\n_\a\ell_\b &=& (e_\a{}^\g - \ell_\a n^\g -\ell^\g n_\a )(e_\b{}^\de -\ell_\b
n^\de - \ell^\de n_\b )\n_\g\ell_\de \nonumber\\
&=& e_\a{}^\g e_\b^\de\n_\g\ell_\de + \ell_\a\ell_\b n^\g n^\de\n_\g\ell_\de
- e_\a^\g \ell_\b n^\de \n_\g \ell_\de - e_\b^\de \ell_\a n^\g \n_\g \ell_\de - \kappa n_\a\ell_\beta.
\nonumber\\
&=& \sigma_{\a\b} + \frac{1}{2} e_{\a\b}\theta +
	\ell_\a\ell_\b n^\g n^\de \n_\g\ell_\de 
 - e_\a^\g \ell_\b n^\de \n_\g \ell_\de - e_\b^\de \ell_\a n^\g \n_\g \ell_\de  - \kappa n_\a\ell_\beta.
\label{e:grad-ell}\end{eqnarray}
        
\noindent {\bf Proof of 2nd Law}:  We have\\
\[
	\frac{d}{d\la} A_\la = \int_\Sigma {\mbox {\mbox \pounds}}_\ell \ep_{\a\b} dS^{\a\b},
\]
where $\epsilon_{\a\b} = \ep_{\a\b\gamma\delta}\ell^\gamma n^\delta$.
We will, for this proof, choose $\ell^\alpha$ to be affinely parametrized 
($\kappa=0$),
and it will be helpful to extend $\ell^\alpha$ to a null vector field on a  
neighborhood of $H$. Then 
\[
 \Lie_\ell \ell_\alpha = \ell^\beta\nabla_\beta l_\alpha 
		+ \ell_\beta\nabla_\alpha\ell^\beta = 0 
		+ \frac12\nabla_\alpha(\ell_\beta\ell^\beta) = 0.
\]
Using 
\[
  \Lie_\ell\epsilon_{\a\b\g\de} =
\n\cdot\ell\, \epsilon_{\a\b\g\de},\ \ \Lie_\ell \ell^\g =0, \ \ 
\ell_\de \Lie_\ell n^\de  = - n^\delta \Lie_\ell \ell_\delta = 0,
\]
we obtain

\[
\Lie_\ell\ep_{\a\b}\ dS^{\a\b} = \n_\g\ell^\g\ \ep_{\a\b} dS^{\a\b}.
\]
Let us now use (\ref{e:grad-ell}) to obtain an equation for $\ell^\b\n_\b\theta =
\dot\theta$ --- we will show that $\dot\theta \leq 0$.
\[ \n_\g\n_\de\ell_\a - \n_\de\n_\g\ell_\a = R_{\a\b\g\de}\ell^\b\]
\be \Rightarrow \ell^\de e^{\a\g} (\n_\g\n_\de\ell_\a - \n_\de\n_\g\ell_\a) =
R_{\a\b}\ell^\a\ell^\b\label{4}\ee
\begin{eqnarray}
LHS &=& (\ell^\a e^{\beta \epsilon} -\ell^\epsilon 
e^{\a\b})\nabla_\epsilon\nabla_\a\ell_\b\nonumber\\
&=& (\ell^\a e^{\b\ep} - \ell^\epsilon e^{\a\b})\n_\ep
(\sigma_{\a\b} + \frac{1}{2} e_{\a\b}\theta +\ell_\a\ell_\b\, n^\g n^\de
\n_\a\ell_\de
- e_\a^\g \ell_\b n^\de \n_\g \ell_\de - e_\b^\de \ell_\a n^\g \n_\g 
\ell_\de)\ \ \ \ 
\label{5ins}
\end{eqnarray}
The last three terms in parentheses all die by virtue of the relations
\[ e^{\a\b}\ell_\b = 0, \hspace{3mm} \ell^\b\nabla_\b\ell_\a =-0, 
\hspace{3mm} \ell^\b\nabla_\a\ell_\b =0, \hspace{3mm} \ell^\a\ell_\a =0,\]
and we are left with
\be LHS = \ell^\a e^{\b\gamma}\nabla_\gamma (\sigma_{\a\b} + \frac{1}{2} 
e_{\a\b}\theta ) - \ell^\gamma e^{\a\b}\nabla_\gamma (\sigma_{\a\b} + 
\frac{1}{2} e_{\a\b}\theta ) . \label{6ins}\ee
Considering each term in turn, we have
\begin{eqnarray}
\ell^\a e^{\b\gamma}\nabla_\gamma (\sigma_{\a\b} + \frac{1}{2} 
e_{\a\b}\theta ) 
&=& e^{\b\gamma}\nabla_\gamma \underbrace{[\ell^\a (\sigma_{\a\b} + 
\frac{1}{2} e_{\a\b}\theta )]}_0\mbox{} -e^{\b\gamma}\nabla_\gamma\ell^\a\ (\sigma_{\a\b} + \frac{1}{2} 
e_{\a\b}\theta )\nonumber\\
&=& -e^{\b\gamma}\nabla_\gamma\ell_\delta\, e^{\a\delta}\, (\sigma_{\a\b} + 
\frac{1}{2} e_{\a\b}\theta )\nonumber\\
&=& -(\sigma^{\a\b} + \frac{1}{2} e^{\a\b}\theta )(\sigma_{\a\b} 
+\frac{1}{2} e_{\a\b}\theta )\nonumber\\
&=& -\sigma^{\a\b}\sigma_{\a\b} - \frac{1}{2} \theta^2 \label{7ins}
\end{eqnarray}

\begin{eqnarray}
-\ell^\gamma e^{\a\b}\nabla_\gamma (\sigma_{\a\b} + \frac{1}{2} 
e_{\a\b}\theta ) &=& -\ell^\gamma\nabla_\gamma [e^{\a\b}(\sigma_{\a\b} + 
\frac{1}{2} e_{\a\b}\theta )]
\mbox{} + \ell^\gamma\nabla_\gamma e^{\a\b} (\sigma_{\a\b} + \frac{1}{2} 
e_{\a\b}\theta )\nonumber\\
&=& -\ell^\gamma\nabla_\gamma\theta + \ell^\gamma\nabla_\gamma (\ell^\a 
n^\b +\ell^\b n^\a )(\sigma_{\a\b} + \frac{1}{2} e_{\a\b}\theta )\nonumber\\
&=& -\ell^\gamma\nabla_\gamma\theta . \label{8ins}
\end{eqnarray}
 From Eqs.\  (\ref{4}), (\ref{6ins}), (\ref{7ins}), and (\ref{8ins}), we 
obtain the Raychaudhuri equation,
\[
 -\sigma^{\a\b}\sigma_{\a\b} - \frac{1}{2} \theta^2-\dot\theta =
R_{\b\a}\ell^\b\ell^\de
\]
or
\be \cblue
\dot\theta =-\frac{1}{2} \theta^2 -\sigma^{\a\b}\sigma_{\a\b} -R_{\a\b}\ell^\b\ell^\a .
\label{ray} \ee
Then the energy condition
\[ R_{\a\b}\ell^\a\ell^\b \geq 0, \hspace{2mm} {\mbox{all null $\ell^\a$}} ,\]
implies
\be \dot\theta \leq -\frac{1}{2} \theta^2 . \label{7}\ee

	This last equation is the key to the area theorem.  If $\theta =
\theta_0<0$ anywhere on $H$, then (\ref{7}) implies
\begin{align}
\theta^{-2} \frac{d\theta}{d\lambda} \leq&-\frac{1}{2} \Longrightarrow
\int_0^\lambda \theta^{-2} \frac{d\theta}{d\lambda} d\lambda \leq
-\frac{1}{2} \lambda\nonumber\\
&\mbox{\hspace{17mm}} -\frac{1}{\theta} + \frac{1}{\theta_0} \leq -
\frac{\lambda}{2}\nonumber\\
&\Rightarrow \theta \leq \frac{2}{\lambda - \left(
-\frac{2}{\theta_0}\right)}\label{8}
\end{align}
Eq.~(\ref{8}) implies $\theta \rightarrow -\infty$ within an affine
parameter length $\lambda < \left| \frac{2}{\theta_0}\right|$ from the point
where $\theta = \theta_0$.

	This means that two future-directed neighboring null generators 
of $H$ cross in finite affine parameter length---at some point $P$ of $H$.  
But then their tangents will be {\em two} null vectors at $P$:  $\ell^\a$ and
$\hat\ell^\a$ (!), and a linear combination will be a timelike vector on
$H$, contradicting the fact that $H$ is null.  Therefore $\theta \geq 0$ on
$H$ and $\frac{dA}{d\lambda} \geq 0$. \hspace{3mm} $\Box$

\subsection{Area of a Kerr black hole and irreducible mass}
The area increase theorem gives an upper limit on the energy that can be extracted from 
a black hole, by, for example, the Penrose process or in the inspiral and merger of 
two black holes.  In the Penrose process, the area of the final black hole must be larger 
than that of the initial black hole; in inspiral and merger, the final black hole must 
have an area larger than the sum of the initial areas.  (The horizon of a spacetime 
of two merging black holes is a 
continuous 3-surface; an early spacelike slice is the disjoint union of two spheres, 
and each late slice is a single sphere.)   

  In the Kerr geometry, the horizon area $A$ is the area of a $\theta,\phi$ sphere with $\Delta =0$. Although the Boyer-Lindquist form \eqref{e:kerr1} of the metric gives the correct 
induced metric on the horizon 2-spheres spanned by $\theta$ and $\phi$, these are 
surfaces of constant $t$ and $\phi$, and $t$ is undefined on the horizon.  
So we should use the metric in Kerr coordinates \eqref{e:kmetric}, 
\index{Kerr spacetime!Kerr coordinates}\index{Kerr coordinates}
\index{coordinates!Kerr coordinates}
setting $r=r_+$, $\Delta=0$, and $v=$ constant, to read off the metric on a slice of 
the future horizon:  
\be
   ds^2 = \frac{\sin^2\theta}{\rho^2} (r_+^2+a^2) d\phi^2 + \rho^2 d\theta^2, \qquad
   \sqrt{^2g} = (r_+^2 + a^2) \sin\theta \ .    
\ee
Then, writing $A=\int \sqrt{^2g}\ d\theta d\phi$ and using $r_+^2 + a^2 = 2Mr_+\ $, we have
\be\cblue
   A = 8\pi M r_+ = 8\pi M (M+\sqrt{M^2 - a^2}).    
\label{e:Akerr}\ee  

The maximum energy that can be extracted is the difference in mass between this rotating 
black hole and the smallest-mass black hole with an area this large, namely a Schwarzschild 
black hole, with  
\be
  A = 16\pi M_{\rm final}^2 .  
\ee  
\index{energy!maximum extractable from black hole}\index{mass!irreducible mass of black hole}
\index{irreducible mass of black hole}
The {\sl irreducible mass} of the original black hole is this smallest final mass, 
\be
   M_{\rm irreducible} = \sqrt{\frac A{16\pi}} =  \frac M{\sqrt 2} \sqrt{1+\sqrt{1-(a/M)^2}}.
\label{e:mirred}\ee
Then the largest amount of available rotational energy is $M-M_{\rm irreducible}$ for 
an initial maximally rotating black hole; from Eq.~\eqref{e:mirred}  $a<M$ 
implies $\dis\phantom{\frac12}\!\! M_{\rm irreducible} > 1/\sqrt2$,  and the extracted energy is limited by 
\be
   \frac{\Delta M}M < 1 - \frac1{\sqrt2} \approx 0.29.  
\ee

It is easy to check that the limit for collision of two Schwarzschild black holes 
with negligible initial kinetic energy is again 
\be
   \frac{\Delta M}{M_{\rm total}} < 1 - \frac1{\sqrt2},
\ee
much larger than the $\sim 0.05 M_{\rm total}$ actually radiated in binary inspiral 
(see, e.g., Eq.~\eqref{e:gwemri} or 
\href{https://arxiv.org/pdf/1606.01210.pdf}{GW150914}).\cite{gw150914}  
\benr\item A 2-line exercise:  Show for equal-mass Kerr black holes with 
opposite spins, area increase imposes only the weaker limit 
\be
    \frac{\Delta M}{M_{\rm total}} < \frac12. 
\ee
\een 

\noindent{\sl Stationary, axisymmetric horizons}\\

We look now at equilibrium configurations---stationary, axisymmetric
black holes, with Killing vectors $t^\a$ and $\phi^\a$.  
Because the spacetime is stationary and axisymmetric, rotations and time-translations map
the horizon to itself. It follows that $t^\alpha$ and $\phi^\alpha$ are tangent to the 
horizon $H$. This is because $\phi^\alpha$ is tangent to the circular 
trajectories of rotated points, and $t^\alpha$ is tangent to the trajectories of time-translated points.

We have seen (beginning on pp. \pageref{s:horizon}) that the horizon is the set of points 
where the determinant of the 2-metric in the $\bm t$-$\bm \phi$ plane 
vanishes $g_{tt} g_{\phi\phi} - g_{t\phi}^2 = \bm{t \cdot t ~\phi \cdot \phi} - (\bm{t \cdot \phi})^2 = 0$.   

Frame dragging is measured by $\omega = -\bm{t\cdot\phi}/\bm{\phi\cdot\phi}$, 
the angular velocity of freely falling observers with zero angular momentum.\index{angular momentum!zero angular momentum observer}  And, 
as shown in Eqs.~\eqref{e:omega1}, the vanishing of the 2-metric determinant 
implies that it is the vector field $t^\alpha + \omega \phi ^\alpha $ that is null on the horizon.  That is, the generators of the horizon are the paths of zero-angular-momentum photons.
  
\benr \item The tangent space at each point of spacetime is a copy of Minkowski space. 
Show that the metric in a null hyperplane in Minkowski space has vanishing determinant. 
Pick your favorite null hyperplane and show that 
the induced metric on it has vanishing determinant.  This means that 
the 3-metric is degenerate, mapping some vector to $0$.  Show that 
$^3g_{\a\b}\ell^\b = 0$, where $\ell^\alpha$ is a null vector in 
the null plane.  Finally, defining $n^\alpha$ as the null vector orthogonal to the 
horizon for which $n^\alpha \ell_\a = -1$, show that $^3g_{\a\b} = e_{\a\b}:= g_{\a\b} + \ell_\a n_\b + \ell_\b n_\a$.    
\een  

\section{Mass and angular momentum of stationary axisymmetric black-hole spacetimes} 
\index{angular momentum!total angular momentum of spacetime}

  The asymptotic metric of rotating stars and black holes has the form that we 
 saw for Kerr black holes, agreeing with Schwarzschild at order $1/r$  and with Kerr  
 to order $1/r^2$ in an asymptotically Cartesian chart. 
In particular, the mass appears in the asymptotic form of $t^\a t_\a$, 
 \be
    g_{tt} = - (1-2M/r) + O(r^{-2}); 
 \ee
and the angular momentum appears in the leading term in $t^\a\phi_\a$:  As in 
\eqref{e:asymp}, 
\be
    g_{t\phi} = -2\frac Jr\sin^2\theta \ [1 + O(r^{-1})],  
 \ee 
corresponding to the form ~\ref{e:asymp2}, 
$g_{ti} = 2\epsilon_{ijk} \frac{x^j J^k}{r^3} +O(r^{-3})$, in the associated Cartesian chart. 
 
   We will see that the asymptotic behavior is tied to expressions for mass and angular momentum 
 as integrals at infinity involving the timelike and rotational Killing vectors, respectively.  
 Gauss's theorem then relates the asymptotic integrals to volume integrals over the matter.  
 (Underlying all this is the general relation between symmetries and conserved quantities, 
 discussed in the Noether part of Sect.~\ref{s:action_noether}.) 
We begin with surface integral expressions for $M$ and $J$, introduced by Komar, 
showing 
\index{conservation laws!Komar mass and angular momentum}\index{Komar mass}\index{angular momentum!Komar}\index{mass!Komar mass}     
\be\cblue
M = -\frac1{4\pi}\int_\infty \na^\a t^\b  dS_{\a\b} , \qquad 
   J = \frac1{8\pi} \int_\infty \na^\a \phi^\b  dS_{\a\b} . 
\label{e:mkjk}\ee
When $\pa\Sigma$ is a $t= $ constant, $r= $ constant surface, 
$dS_{\mu\nu} = \sqrt{-g} \na_{[\mu} t\ \na_{\nu]}r d\theta d\phi$. 
 
For our asymptotic calculations, we'll use a chart in which, as in Boyer-Lindquist coordinates 
for Kerr, 
\bea
 ds^2 = &-&\left[1-\frac{2M}r + O(r^{-2})\right]\, dt^2 
 	- \frac{4J}r \sin^2\theta\, dt\, d\phi [1+ O(r^{-1})]
 	+ r^2\sin^2\theta d\phi^2 [1+O(r^{-2})] \nonumber\\ 
 	&+& \left[1+\frac{2M}r+O(r^{-2})\right] dr^2
 	+ r^2 d\theta^2 [1+O(r^{-2})] .
\label{asympt}\eea 
Then, recalling $t_\mu = g_{\mu t}, \ \phi_\mu = g_{\mu\phi}$, we have  
\begin{align}
 \int_\infty \na^\a t^\b  dS_{\a\b} 
	&= \lim_{r\rightarrow\infty}\int g^{tt} g^{rr} 
			\frac12(\pa_t t_r - \pa_r t_t)r^2 d\Omega
	= -\frac12\lim_{r\rightarrow\infty} \int\pa_r(1-2M/r) r^2 d\Omega \nn\\
	&= -4\pi M. 
\label{e:mk0}\end{align}  
Similarly, 
\begin{align}
 \int_\infty \na^\a \phi^\b  dS_{\a\b} 
   &= \lim_{r\rightarrow\infty}\int g^{t\mu} g^{r\nu}\pa_{[\mu} \phi_{\nu]}r^2 d\Omega 
	=  \lim_{r\rightarrow\infty}\int \frac12[g^{tt} g^{rr} (- \pa_r g_{t\phi})
					+ g^{t\phi} g^{rr} (- \pa_r g_{\phi\phi})]r^2 d\Omega\nn\\
   &= \lim_{r\rightarrow\infty}\int \frac12\left[\pa_r\left(-\frac{2J}{r}\sin^2\theta\right) 
					    +\frac{2J}{r^3}\pa_r\left(r^2\sin^2\theta\right)\right]d\Omega\nn\\
   & = 8\pi J.
\label{e:jk0}\end{align} 
 
Now, from the identity \eqref{e:kv} for a Killing vector $\xi^\a$, 
\be
  \na_\b\na^\a \xi^\b = R^\a{}_\b \xi^\b , 
\label{e:kdr}\ee
Gauss's theorem gives the integral $\int_\Sigma \na_\b\na^\a \xi^\b dS_\alpha$ 
over a 3-dimensional hypersurface in terms of the integral over its boundary, 
\be
   \int_\Sigma \na_\b\na^\a \xi^\b dS_\a 
   		 = \int_{\pa\Sigma} \na^\a \xi^\b  dS_{\a\b}.
\ee
(See Sect.~\ref{s:gauss} of the integration appendix for a proof of Gauss's law in this form.)  
 
For a rotating star (or, more generally, a stationary, axisymmetric, asymptotically flat orientable spacetime with no interior boundary), the only boundary is at infinity, and 
Gauss's theorem relates $M$ and $J$, given as integrals at infinity, to volume integrals 
over the matter:
\bsube\begin{align}
   M & = -\frac1{4\pi}\int_\infty \na^\a t^\b  dS_{\a\b} 
   	= -\frac1{4\pi} \int_\Sigma  R^\a{}_\b t^\b dS_\a, 
 \label{e:mk}\\
   J &= \frac1{8\pi} \int_\infty \na^\a \phi^\b  dS_{\a\b} 
   	= \frac1{8\pi}  \int_\Sigma R^\a{}_\b \phi^\b dS_\a.  
\label{e:jk}\end{align}\label{e:komar}\esube 

\benr\item  Using the field equation in the form 
$R_{\a\b} = 8\pi(T_{\a\b} -\frac12 g_{\a\b}T)$, 
check that these expressions have the Newtonian limits 
$M=\int \rho dV$, $J=\int \rho \varpi^2\Omega dV$, for a rotating fluid with density 
$\rho$ and velocity $v=\varpi\omega$, with $\varpi$ the cylindrical radius.  
\een
 
Eqs.~\eqref{e:komar} show that a stationary, asymptotically flat, orientable vacuum spacetime with no interior boundary and a single asymptotic region has vanishing mass and angular momentum: It is flat space.  

For a black hole spacetime, the boundary $\pa\Sigma$ is the disjoint union of the 
sphere at infinity and a slice $S$ of the horizon.  Gauss's theorem now relates 
the surface integrals $M$ and $J$ at infinity to the surface integrals over the black hole:  
\bsube\begin{align}\cblue
   M & \cblue= -\frac1{4\pi}\int_S \na^\a t^\b  dS_{\a\b} 
   	 -\frac1{4\pi} \int_\Sigma  R^\a{}_\b t^\b dS_\a \equiv M_H + M_{\rm matter},\cb 
   \label{e:mk1} \\
   J &\cblue= \frac1{8\pi} \int_S \na^\a \phi^\b  dS_{\a\b} 
   	+ \frac1{8\pi}  \int_\Sigma R^\a{}_\b \phi^\b dS_\a \equiv J_H + J_{\rm matter}.  
\cb\label{e:jk1}\end{align}\label{e:komar1}\esube \cb
Here, the surface integrals over the horizon, $M_H$ and $J_H$, can be regarded as the 
horizon's mass and angular momentum;\index{angular momentum!of black hole}  the volume integrals $M_{\rm matter}$ and $J_{\rm matter}$ 
are the contributions of the matter to the total $M$ and $J$.  Note, however,
that the contribution of the matter to the total mass includes its gravitational 
binding energy. \index{binding energy!gravitational}\index{energy!gravitational binding energy}\footnote{One {\sl defines} the gravitational binding energy $W$ indirectly 
by identifying a rotational kinetic energy,\\ 
$T_{\rm rotational} = \frac12\int_\Sigma\Omega dJ$ and proper mass 
$M_P= \int\rho u^\a dS_\a$, and writing $W:=M-M_H-M_P-T_{\rm rotational}$.
Here, as in Eq.~\eqref{deltam1}, $dJ= T^\a{}_\b \phi^\b dS_\a$ for a rotating star. }


\section{Two Constant Scalars on \texorpdfstring{$H$}.}

To summarize:  The horizon $H$ is the surface on which $t^{[\a}\phi^{\b ]} $ is null:
\be 0 = 2 t^{[\a}\phi^{\b ]} t_\a\phi_\b = t\cdot
t\; \phi\cdot\phi -(t\cdot\phi )^2; \label{9}\ee
and the linear combination $t^\a+\Omega_H\phi^\a$ is null, where
\be \Omega_H \equiv\omega_H = -\left.\frac{t\cdot\phi}{\phi\cdot\phi}\right|_H
.\label{10}\ee
The null generators of the horizon then have tangent 
$\ell^\a = (t^\a + \Omega_H\phi^\a)f$, where $f$ is any nonzero scalar. 
If one chooses $f=1$, then  
\be \ell^\a = t^\a + \Omega_H\phi^\a .\label{11}\ee
By setting the coefficient of $t^\a$ to $1$, one sacrifices the
affine parametrization of $\ell^\a$.  Away from the horizon, we 
will take $\ell^\a$ to be tangent to outgoing null rays and denote 
by  $k^\a$ the Killing vector 
\be
   k^\a = t^\a + \Omega_H\phi^\a \mbox{ \ \ everywhere}, 
\ee 
coinciding with $\ell^\a$ only on the horizon: 
\footnote{Bardeen-Carter-Hawking paper\cite{bch73} use $\ell^\alpha$ to mean the Killing vector, so for them, $\ell^\a$ is null only on the horizon.  We keep 
$\ell^\alpha$ null to conform to NP formalism, which we'll use in the calculations.}

	$\Omega_H$ is the angular velocity of $H$ in the sense that a point
moving along the generators of $H$ (along the integral curves of
$\ell^\a$) has angular velocity $\Omega_H$ as seen by an observer at
infinity; well, a point on the horizon can't be seen from infinity, so 
$\Omega_H$ is the limiting angular velocity as seen from infinity of particle trajectories 
just outside the horizon.   The amazing thing is that the horizon rotates as a rigid
body:  $\Omega_H$ is constant.  For Kerr black holes, this is already implied by the explicit 
form \eqref{e:omegaH}, namely $\Omega_H = a/2Mr_+$.  For distorted black holes -- e.g., stationary 
axisymmetric black holes surrounded by an accretion disk, the result is still valid:  \\

\noindent $\bm {\Omega_H}$ {\bf is constant on }$\bm H$.\\

\noindent {\bf Proof}. 
We already know that the component of $\nabla \Omega_H$ along 
$\ell^\alpha$ vanishes; that is, since $t^\alpha$ and $\phi^\alpha$ are 
Killing vectors, $t^\a\nabla_\a \Omega_H = 0 = \phi^\a\nabla_\a\Omega_H$, 
implying
\[ 
	\ell^\a \nabla_\a\Omega_H = 0. 
\]
Our job is to show that $e_\a^\b\nabla_\b \Omega_H = 0$, 
or, equivalently, $m^\a \na_\a \Omega_H = 0$. 
We begin by using the same argument to write
\[ 
	\dot\theta \equiv \ell^\a \nabla_\a\theta = 0. 
\]
When the null energy condition is satisfied, all terms on the RHS of the 
Raychaudhuri equation, (\ref{ray}), are negative, and $\dot\theta = 0$ then 
implies that each term must vanish,
\be 
	\sigma_{\a\b} = 0,\ \theta = 0, \ R_{\a\b} l^\a l^\b = 0. 
\label{zero}\ee
(Because the Raychaudhuri equation was written for an affinely parametrized 
$\ell^\alpha$, we have shown only that $\sigma_{\a\b}$ and $\theta$ 
vanish for an affinely parametrized $\ell^\alpha$; but it is easy to 
check that they then vanish for any parametrization of $\ell^\alpha$.) 
\benr\item Check this.
\een
 
  We next use the vanishing of the shear and divergence on $H$:  
\be
  \sigma_{\a\b} = 0 = \theta \ \Rightarrow 
    0 =  m^\a m^\b \nabla_\a \ell_\b = m^\a m^\b \nabla_\a k_\b\quad
			\mbox{(this is the NP } \sigma).
\label{kill}\ee
Using $\Lie_{\bm t} g_{\a\b} = 0 = \Lie_{\bm\phi} g_{\a\b}$, we have
\bea 
 \Lie_{\bm k} g_{\a\b} &=& 2\na_{(\a}k_{\b)} = 2\na_{(\a}\Omega_H \phi_{\b)}
\eea
Contracting with $m^\a m^\b$, already symmetric in the indices ${}^{\a\b}$, gives
\[ 
 m^\a m^\b \phi_\a \nabla_\b\Omega_H = 0.  
\]
Now $m_\a \phi^\a$ is nonzero; otherwise $\phi^\a$ would be orthogonal to $m^\a$ and $\barm^\a$ , implying it would be parallel to $\ell^a$.  This is  
impossible because the trajectories of $\phi^\a$ are circles, 
closed curves with finite 
affine parameter length, contradicting the infinite affine parameter length of the 
horizon's generators.  (It would also violate causality - photons would follow 
closed curves, looping backward in time.)  Thus 
\be
	m^\b \nabla_\b\Omega_H = 0.\qquad \Box
\ee 

$\Omega_H$ constant on the horizon means that 
$\ell^\alpha = t^\alpha + \Omega_H \phi^\a$ is a Killing vector on the horizon.  
A horizon whose null generator is a Killing vector is called a {\em Killing horizon}; 
so we have just shown that the horizon of a stationary, axisymmetric spacetime is 
a Killing horizon.  

Note that $\ell^\alpha$ is a Killing vector on the horizon  
for the parametrization of the null generators that gives $\ell^\a$ the form 
$t^\alpha + \Omega_H \phi^\alpha$. As mentioned above, with this choice, 
$\ell^\a$ is not affinely parametrized: 
\[ 
	\ell^\b\nabla_\b\ell^\a = \kappa\ell^\a, 
\]
with $\kappa$ nonzero.  $\kappa$ is called the ``surface gravity,'' and 
it measures the acceleration of $\ell^\a$.  It will play the role of 
temperature. (This is {\sl not} the NP spin-coefficient labeled $\kappa$; 
it is the real part of the NP $\epsilon$ in a tetrad for which $\ell$ 
is not affinely parametrized.)  Once the QFTCST calculation is done, 
the identification is $\dis T= \frac\kappa{2\pi}$, entropy $\dis S = \frac A4$, 
in units with $G=c=\hbar=k_B=1$, where $k_B$ is Boltzmann's constant.  \\
(Keeping the constants and taking $\kappa$ to have dimension of acceleration, $L/T^2$, 
we have \\
$\dis k_B T= \kappa\hbar /(2\pi c),\quad S = k_BA/(4\ell_P^2)$, with 
\mbox{$\ell_P=\sqrt{G\hbar/c^3}=$ Planck length}. )
\index{entropy!black hole entropy|textbf}\index{black hole!entropy|textbf}
\\

\benr\item To prove the $0^{th}$ law, that $\kappa$ is constant on the horizon, 
we'll need the following identity for a Killing vector $\xi^\a$:    
\be
  \na_\a \na_\b \xi_\gamma = -R_{\b\gamma\a\delta} \xi^\delta.  
\label{e:kv}\ee
Prove this, starting from the definition $[\na_\b,\na_\g]\xi_\a = R_{\b\g\a\d}\xi^\d$.  \\
Hint: Use $\na_\a \xi_\b + \na_\b\xi_\a = 0$ to replace one of the terms, 
write down two cyclic permutations, and add the three equations with one minus 
sign to get the relation.  (You'll also need $R_{[\a\b\gamma]\delta} = 0$.)   
\label{e:killing}\een 

\noindent  $\bm{0^{th}}$ {\bf Law}:  
$\bm \kappa$ {\bf is constant on} $\bm H$  (for vacuum or for matter satisfying the 
dominant energy condition).\\
This time, we'll first give the general proof and then explicitly find $\kappa$ for Kerr.  

\noindent {\bf Proof}. 
Again, to prove $\kappa$ constant on $H$, it suffices to show
\be m^\b \nabla_\b \kappa = 0.\label{e:0th}\ee
The definition (\ref{e:kappa}) of $\kappa$ is equivalent to
\be 
	\kappa =-n_\a\ell^\b\nabla_\b\ell^\a .
\label{19}\ee
We'll use the vanishing of shear,  divergence and twist of $\ell^\a$ in
Eq.~\eqref{e:grad-ell} for $\na_\a \ell_\b$:    
\be
  \na_\a \ell_\b = \ell_\a\ell_\b n^\g n^\de \n_\g\ell_\de 
                   - e_\a^\g \ell_\b n^\de \n_\g \ell_\de 
                   - e_\b^\de \ell_\a n^\g \n_\g \ell_\de - \kappa n_\a \ell_\b.
\ee
In particular, we'll need the result of dotting this with $m^\a$. Only the 
second term on the right survives:    
\be
  m^\a \na_\a \ell_\b = -\ell_\b m^\g n^\de \na_\g \ell_\de,
\label{e:mdl}\ee
implying that the NP spin coefficients $\kappa_{NP}$ and $\varrho$, as 
well as $\sigma$,  vanish on $H$.%
\footnote{From the fact that $\ell^\a$ is a Killing vector and Eq.~\eqref{e:mdl}, we have 
(with our $-+++$ convention) \\
$\kappa_{NP}:=-\ell^\b m^\a\na_\b\ell_\a = \ell^\b m^\a\na_\a\ell_\b 
	     =  \ell^\b(-\ell_\b m^\g n^\de \na_\g \ell_\de) = 0$ and  
$\varrho:=-m^\b \barm^\a \na_\a \ell_\b = -m^\b (-\ell_\b \barm^\g n^d\na_\b \ell_\d) = 0$. 
}
(With the Kinnersley tetrad, $\ell^\alpha$ blows 
up on the horizon.  Our $\ell^\alpha$ is $l^\alpha_{\rm Kinnersley}\Delta$  
evaluated on the horizon where $\Delta=0$. The rescaling gives $\varrho = 0$, 
although it is nonzero on $H$ for Kinnersley.)   
  
Now on to $\na \kappa$:  
\begin{align*}
-m^\de \nabla_\de \kappa &= m^\de \nabla_\de (n^\g\ell^\b\nabla_\b\ell_\g )\\
	&= m^\de\nabla_\de n^\g\ \ell^\b\nabla_\b\ell_\g 
	   + m^\de\nabla_\de \ell^\b\ n^\g\nabla_\b\ell_\g 
	   +m^\de\ell^\b\nabla_\de \nabla_\b\ell_\g n^\g 
\end{align*}
Use \eqref{e:mdl} in 2nd term and the Killing vector identity \eqref{e:kv} in the 3rd: 
\begin{align*}
   -m^\de \nabla_\de \kappa 
	&= m^\de \na_\de n^\g \kappa\ell_\g 
	   -\ell^\b m^\de n^\ep\na_\de \ell_\ep\ n^\g \nabla_\b \ell_\g 	   + R_{\b\g\de\ep} m^\b \ell^\g n^\de \ell^\ep  	
\end{align*}
\be
{\mbox{2nd term}} =  -m^\de n^\ep \nabla_\de \ell_\ep
\underbrace{n^\g\ell^\b\nabla_\b\ell_\g}_{-\kappa } = \kappa m^\de n^\ep \nabla_\de \ell_\ep
	=- {\mbox{ 1st term}}. 
\ee
Then only the 3rd term remains:  
\begin{align*}
 -m^\de\nabla_\de \kappa &= m^\de\ell^\b R^\epsilon{}_{\de\b\g}\ell_\epsilon n^\g 
		= R_{\a\b\g\de} \ell^\a m^\b \ell^\g n^\de  \label{21} \\
	&= \left[C_{\a\b\g\de} 
		+\frac12(g_{\a\g}R_{\b\de} +g_{\b\de} R_{\a\g} 
			-g_{\b\g}R_{\a\de} -g_{\a\de}R_{\b\g} )\right]
		\ell^\a m^\b \ell^\g n^\de \\
	&= -\Psi_1 + R_{\b\g}m^\b\ell^\g  \quad\mbox{(coefficient of $R$ vanishes)}.
\end{align*}
Now $\Psi_1$ vanishes:  Because $\sigma = 0 = \kappa_{NP}$, NP (4.2k) implies $\Psi_0=0$ 
as usual.  With \mbox{$D\Psi_1 = \ell^\a\na_\a \Psi_1 = 0$} and $\varrho=0$, there is only one remaining term in NP (4.5, first equation), which now has the form 
\[
  2\epsilon \Psi_1 = 0.
\]  
But $\epsilon = \frac12(\Gamma_{121} + \Gamma_{431})$, with nonzero real part 
$-\frac12\ell^\b\na_\b \ell_\a n^\a = \kappa$. 
So $\Psi_1 = 0$.  

Finally, we show $R_{\b\g}e_\a{}^\b\ell^\g = 0$. We have already seen that 
$R_{\b\g}\ell^\b\ell^\g = 0$.  The {\em dominant energy condition } is the condition that 
$J^\a := T^\a{}_\b\ell^\b$ is nonspacelike. Now 
\bea T_{\b\g}\ell^\b\ell^\g = J_\b\ell^\b = 0 &\Rightarrow & J_\b = f\ell_\b + e_\b{}^\g J_\g 
 \nonumber \\
  &\Rightarrow& J_\b J^\b = e^{\b\g}J_\b J_\g > 0 {\mbox{ unless }}
e^{\b\g}J_\g = 0.
\eea
Then $J^\a$ nonspacelike implies
\begin{eqnarray}
0 &=& e^{\a\b}J_\b = \frac1{8\pi}e^{\a\b} (R_{\b\g} - \frac{1}{2} 
\underbrace{ g_{\b\g}R}_0)\ell^\g 
\nonumber\\
&=& e^{\a\b} R_{\b\g} \ell^\g . \label{24}
\end{eqnarray}
$\Rightarrow \kappa$ constant on $H$. \hspace{10mm} $\Box$\\

\noindent{\sl Surface gravity $\kappa$ for Kerr} \\

To find $\kappa$ for a Kerr black hole, 
we'll evaluate $\ell^\b \na_\b \ell^\a$ on the 
horizon, with $\ell^\a$ the null generator normalized to agree with the Killing vector 
$k^\a= t^\a + \Omega_H \phi^\a$ on the horizon. 
We use the fact that the generator $\ell^\a$ of the horizon is along 
a principal null direction, agreeing with the affinely parametrized null vector $\ell_{\rm affine}^\a$ of Eq.~\eqref{e:laffine} up to an overall factor $f$:  
\be
    \ell^\a|_H = \lim_{r\rightarrow r_+} (f \ell_{\rm affine}^\a).  
\ee
where $\bm \ell_{\rm affine}$ satisfies the affinely parametrized geodesic equation, 
\be
  \ell_{\rm affine}^\b \na_\b \ell_{\rm affine}^\a = 0.   
\ee 
Although $\ell_{\rm affine}^\a$ is not defined on $H$, 
$\ell^\a$ is smooth everywhere, so its value on $H$ is the limit 
as $r\rightarrow r_+$ of its value for $r>r_+$.  Then 
\be
   \left. \kappa \ell^\a\right|_H= \lim_{r\rightarrow r_+}  \ell^\b \na_\b (f\ell_{\rm affine}^\a) 
   				= \lim_{r\rightarrow r_+}  (\ell^\b \na_\b f) \ell_{\rm affine}^\a. 
\label{e:kappaf}\ee

To find $f$, write $\ell_{\rm affine}^\a$ in terms of the basis vectors of the Kerr coordinates 
$v,r,\widetilde\phi, \theta$: From Eqs.\eqref{e:laffine} we have 
\be
   \pa_t = \pa_v, \quad \pa_\phi = \pa_{\widetilde\phi}, \quad 
   \partial_r \mbox{(with $t,\phi$ fixed)} 
	=  \frac{r^2+a^2}\Delta\partial_v +\partial_r\mbox{(with $v,\widetilde\phi$ fixed)} 
	   + \frac a\Delta\partial_{\widetilde\phi}\ . 
\ee  
(the last equality is \eqref{e:partialr} with $\partial_v$ and $\partial_{\widetilde\phi}$ 
replacing $\pa_t$ and $\pa_\phi$).  In particular, in a Kerr basis, the generator  
of the horizon is 
\be
   \bm \ell|_H = \pa_v + \Omega_H \pa_{\widetilde\phi}.  
\ee  
From Eq.~\eqref{e:lkin}, in the Boyer-Lindquist coordinate basis, 
$\dis\bm \ell_{\rm affine} = \frac{r^2+a^2} \Delta\pa_t + \pa_r + \frac a\Delta \pa_\phi$, 
so in the Kerr basis, 
\begin{align}
  \bm \ell_{\rm affine} 
  	&= \frac{r^2+a^2}\Delta \pa_v 
  	  +\left( \frac{r^2+a^2}\Delta\pa_v + \pa_r+\frac a\Delta\pa_{\widetilde\phi}\right) 
  	  + \frac a\Delta \pa_{\widetilde\phi} 
 	\nonumber\\
  	 & = 2~\frac{r^2+a^2}\Delta\left(\pa_v +\frac\Delta{2(r^2+a^2)}\pa_r 
  	 			+ \frac a{r^2+a^2}\pa_{\widetilde\phi}\right).
\end{align}
Then, 
\be
  \lim_{r\rightarrow r_+} \left[\frac\Delta{2(r^2+a^2)} \bm\ell_{\rm affine}\right] 
        = \pa_v + \frac a{r_+^2+a^2} \pa_{\wt \phi}
        = \pa_v + \Omega_H \pa_{\wt \phi}  
  	=\bm\ell|_H.
 \ee 
 Thus $\dis f = \frac\Delta{2(r^2+a^2)}$, 
 \be
   \bm \ell = \pa_v + \frac\Delta{2(r^2+a^2)}\pa_r + \frac a{r^2+a^2}\pa_{\widetilde\phi} , 
 \ee 
and Eq.~\eqref{e:kappaf} gives 
\begin{align}
  \kappa \ell^\a|_H 
      &=\lim_{r\rightarrow r_+} 
          \left. \ell_{\rm affine}^\a~ \frac\Delta{2(r^2+a^2)} ~ \pa_r ~\frac{\Delta}{2(r^2+a^2)} 
          \right|_H
      = \frac{r_+-M}{r_+^2 + a^2} \ell^\a\,, \nonumber\\
  \crv
   \kappa &\crv = \frac{r_+ -M}{r_+^2+a^2} = \frac{\sqrt{M^2 - a^2}}{2Mr_+}.\cb
\label{e:kappakerr} \end{align}

\noindent  $\bm{3^{rd}}$ {\bf Law}:  $\kappa$ cannot be reduced to zero.  \\

The third law of thermodynamics is the statement: {\sl It is impossible by
any process, no matter how idealized, to reduce the temperature to 
zero  in a finite sequence of operations.} \\
Its black hole analog is: \\
 {\sl It is impossible by
any process, no matter how idealized, to reduce $\kappa$ to 
zero  in a finite sequence of operations.}   

As one spins up a Kerr black hole to its maximum value, $a=M$ (an {\sl extremal} black hole), its surface gravity drops to zero. 
Reaching or surpassing $a=M$ means creating a naked singularity, 
and the cosmic censorship hypothesis \index{cosmic censorship} is the statement 
this is impossible.  
As mentioned in our previous discussion of Kerr black holes,\pageref{p:censor} there is no formal proof of cosmic censorship, but all calculations that attempt 
to add enough angular momentum to a black hole by sending in particles or fields 
end up adding enough mass to keep $\kappa$ positive, as long as the matter satisfies 
a positive energy condition.  An extremal black hole has no trapped surface, and 
Israel proves a \href{https://journals.aps.org/prl/abstract/10.1103/PhysRevLett.57.397}{version of the third law}\cite{israel86} that prevents a spacetime with trapped 
surfaces from evolving to one with no trapped surface.

\section{The 1st Law}
\index{conservation laws!first law of black hole thermodynamics|(}
\index{first law of thermodynamics!for black holes|(}
The first law of thermodynamics gives the difference in energy 
between nearby equilibrium configurations.
We will begin with the first law for vacuum, asymptotically flat, 
stationary black hole spacetimes.  The Kerr family of geometries are 
the only such spacetimes, so the first calculation uses explicit expressions 
for the energy (mass) of Kerr black holes.  

We then present a more general derivation for stationary axisymmetric 
black holes surrounded by matter (e.g., an accretion disk) following BCH, 
the original Bardeen-Carter-Hawking 
\href{https://projecteuclid.org/search?term=four+laws+of+black+hole+mechanics}{paper}\cite{bch73}, 
and filling in steps.  They use $\ell^\alpha$ to mean the Killing vector 
\be
  \ell^\alpha_{\rm BCH} = t^\a + \Omega_H \phi^\a, 
\ee
so $\ell^\a_{\rm BCH}$ is null only on the horizon. To avoid confusion, these notes 
will denote the Killing vector by 
\be
   k^\a = t^\a + \Omega_H\phi^\a, 
\ee
with $\ell^\a$ a null vector for which $\ell^\a = k^\a$ only on $H$.  

Consider a family of stationary axisymmetric black holes with masses,
angular momenta, horizon areas, and horizon angular velocities 
$M(\lambda), J(\lambda), A(\lambda), \Omega_H(\lambda)$.  
The first law of thermodynamics  expresses the change in the mass (energy) 
along such a family of equilibria in terms of the thermodynamic variables 
that describe the system.  For the black hole, these will be entropy and 
angular momentum, with the entropy identified with the black hole's area
(more precisely, after the BH temperature was discovered, with $A/8\pi$): 
With $\delta$ denoting $d/d\lambda$ at $\lambda = 0$, it has the form   
\be\crv
   \delta M = \frac1{8\pi} \kappa \delta A + \Omega_H\delta J, 
\label{e:first1}\ee
analogous to 
\be
  \delta M = T\delta S +\Omega \delta J,   
\label{e:first0}\ee
for a uniformly rotating system.  For the unperturbed quantities, 
the black-hole relation is 
\be \crv
   M = \frac1{4\pi} \kappa A + 2\Omega_H J, 
\label{e:smarr}\ee
reminiscent of the $M=TS - PV$ thermodynamic relation.

\subsection{First law for Kerr black holes} 

To obtain Eq.~\eqref{e:first1} for Kerr black holes, we first need to find the 
surface gravity $\kappa$, for a black hole of mass $M$ and angular momentum $J$.  
We already know $A$ and $\Omega_H$ from Eqs.~\eqref{e:Akerr}, ~\eqref{e:omegaH} and \eqref{e:rpm}, 
\be 
   A = 8\pi M r_+, \quad \Omega_H = \frac a{2M r_+}, \mbox{ where } r_+ = M+\sqrt{M^2-a^2}.
\ee   

We can now easily check the first law \eqref{e:first1} and the corresponding Smarr 
relation \eqref{e:smarr}.  We'll start with the Smarr relation.  Using Eqs.\eqref{e:omegaH}, \eqref{e:Akerr} and \eqref{e:kappakerr}, for $A,\Omega_H$ 
and $\kappa$, together with $J=aM$, we have 
\be
   \frac A{4\pi} \kappa + 2\Omega_H J 
   	= \frac{8\pi M r_+}{4\pi} \ \frac{r_+-M}{2Mr_+} + 2 \frac a{2Mr_+} aM 
   	= \frac{r_+^2 -Mr_+ + a^2}{r_+} = M.  
\ee 
The last equality used $r_+^2 + a^2 = 2Mr_+$ once again.  

To obtain the first law, we express the Smarr relation in terms of of the variables 
$M$, $A$ and $J$.  Writing $\dis r_+=\frac A{8\pi M}$, we have 
\be
\kappa = \frac{r_+-M}{2Mr_+}= \frac1{2M} - \frac{4\pi M}A, \quad
\Omega_H = \frac J{2M^2 r_+} = \frac{4\pi J}{AM}, \\
\label{e:kappaomega}\ee
and Eq.~\eqref{e:smarr} takes the form
\[
   M = \frac A{8\pi M} - M + \frac{8\pi J^2}{AM}, 
\]
or
\be
   M^2 = \frac{A}{16\pi} + \frac{4\pi J^2}A. 
\ee
Then  
\be 
  2 M\delta M = \left(\frac 1{16\pi} -\frac{4\pi J^2}{A^2}\right)\delta A 
  		+ \frac{8\pi J}A\delta J.  
\ee
Finally, divide by $2M$, replace $4\pi J^2/A$ by $M^2 - A/(16\pi)$, and then 
use Eq.~\eqref{e:kappaomega} in reverse to restore $\kappa$ and $\Omega_H$, 
to get the first law.  
\[
   \delta M = \frac{\kappa}{8\pi}\delta A + 2\Omega_H \delta J.  \quad \Box
\tag{e:first1}\]

\subsection{First law for relativistic stars and matter outside a black hole}
\label{s:starlaw}\index{first law of thermodynamics!for relativistic stars}

We show in next section, Sect.~\ref{s:firstg} below, that the first law differs from 
its vacuum form, \eqref{e:first1}, only by a contribution $\delta M_{\rm matter}$ from the 
matter.  For an axisymmetric disk outside a black hole and for a rotating star with no black 
hole, modeled in each case by a perfect fluid, the contribution of the matter to change in 
total energy is a consequence of the local form of the first law:  The change in mass 
is given in terms of the change in the entropy $S$, baryon number $N$.  
The change in the energy of a fluid element, measured by an comoving observer 
(an observer whose velocity $u^\a$ is the velocity of the fluid element), is 
$
  dE = TdS +\mu dN.
$
The corresponding change $dM$ in the mass of the spacetime involves a redshift 
factor and a term $\Omega dJ$, with $\Omega$ the star's angular velocity (which may depend on position), 
and $d J$ the angular momentum of the fluid element.\index{angular momentum!of a fluid element}    

The derivation in this section is drawn from previous discussions by Thorne \cite{rssd}, Zel'dovich 
and Novikov \cite{zeldovich} (with a section on injection energy by Thorne), and by 
\href{https://link.springer.com/article/10.1007/s10714-009-0920-9}{Carter}\cite{carterlh}.  The text below is largely taken from Friedman and Stergioulas \cite{fsbook}.   
\vskip0.4cm
\noindent{\it Injection energy: First law deduced by dropping a group of baryons from infinity} \index{energy!injection energy}\index{injection energy}

 Consider a stationary, axisymmetric star (or fluid outside a black hole) with fluid velocity \\
\be
u^\a = u^t (t^\a + \Omega\phi^\a).  
\label{e:vfluid}\ee
The contents of a box at infinity are to be injected into the star at a point $P$. 
\footnote{To preserve axisymmetry, we can take the box to be an axisymmetric ring of fluid to be 
inserted at an axisymmetric circle in the star.}
The box holds a collection of baryons having the same composition
as the matter at the point $P$. The collection has baryon number $\delta N$ and entropy $\delta S$; 
the box, including its contents, has energy $\delta M_1$ and angular momentum $\delta J$.
In order that the matter can be injected in a thermodynamically reversible process, 
the entropy per baryon $s$ in the box is to be that of the star at the point $P$, 
so that $\delta S = s\delta N$ is the entropy of the $\delta N$ baryons.  
Denoting by
$p_{1\alpha}$ the initial four-momentum of the box, we have 
\be
	\delta M_1 = - p_{1\alpha} t^\alpha, \qquad \delta J = p_{1\alpha} \phi^\alpha.
\label{dm}\ee
Imagine the box falling freely to the point $P$ of the star. An observer at rest with respect 
to the fluid has velocity $u^\a(P)$ given by Eq.~\eqref{e:vfluid}, and when the box reaches $P$, 
the observer measures an energy 
\bea
 \delta E_1 &=& - p_{1\alpha} u^\alpha = -u^t(p_{1\alpha} t^\alpha 
		+ \Omega p_{1\alpha} \phi^\alpha)
\nonumber\\
	 &=& u^t(\delta M_1 - \Omega \delta J).
\eea
Now, following Thorne, we 
suppose the observer catches the box at $P$ and reversibly injects its contents into 
the fluid, imparting to the fluid an energy (as measured by the observer)
\be
	\delta E = T\delta S + \mu \delta N. 
\ee 

Because the initial entropy per baryon was already $s$, not all of the 
available energy is used:   Our active observer uses the remaining energy to throw
the empty box back up to infinity, on a trajectory with zero angular momentum, so 
that the angular momentum $\delta J$ is retained by the fluid.  
The returning box then has momentum $p_{2\alpha}$ with vanishing angular 
momentum, $p_{2\a}\phi^\a = 0$.  Its locally measured energy is  
$\delta E_2 = \delta E_1 - \delta E = p_{2\alpha} u^\alpha = p_{2\a}t^\a u^t$.  Because its free trajectory 
conserves $p_{2\alpha} t^\alpha$, the box reaches infinity with redshifted energy 
$\dis\delta M_2 = p_{2\alpha}t^\alpha = \frac{\delta E_2}{u^t}$.
The change in mass of the star is then related to the change in 
baryon number, entropy, and angular momentum of the fluid by 
\bea 
  \delta M &=& \delta M_1 - \delta M_2 = \frac1{u^t}(\delta E_1-\delta E_2)+\Omega\delta J
	= \frac{\delta E}{u^t} + \Omega\delta J
\nonumber\\
	&=& \frac \mu{u^t}\delta N + \frac T{u^t}\delta S 
			+ \Omega \delta J.
\eea \label{law1matter}

The first law for generic black holes surrounded by an axisymmetric configuration of fluid 
or for a relativistic star has the form 
\bea 
	\crv \delta M 
	&=&\crv \frac1{8\pi} \kappa \delta A + \Omega_H\delta J  
		+\int_\Sigma \left(  \frac \mu{u^t}\delta dN + \frac T{u^t}\delta dS 
			+ \Omega \delta dJ\right).
\label{deltam1}  
\eea 
where $dJ = T^\a_\b \phi^\b dS_\a$, $dN = nu^\a dS_\a$, and $dS = s dN$.  

Although the reversible injection of a ring of matter preserves the 
local thermodynamic equilibrium, the star itself is no longer in gravitational 
equilibrium after the process described in our derivation of the 
first law.  But the ignored global correction to the mass is second order in 
the perturbation:  Readjusting to the new equilibrium is a perturbation that 
preserves entropy, baryon number and angular momentum, and an equilibrium 
configuration is an extremum of the mass for perturbations that conserve entropy, 
baryon number and angular momentum.  

\subsection{First law for generic stationary axisymmetric black holes.} 
\label{s:firstg}
We show here that, as noted above, the vacuum terms in the first law have the same form 
for perturbations of a stationary axisymmetric black hole spacetime with matter.  
 We first show that the surface terms at the horizon and at 
infinity again give $\frac{\kappa}{8\pi}\delta A + 2\Omega \delta J$ as the black-hole 
contribution to $\delta M$. For a perfect fluid, the matter contribution is again 
given by Eq.~\eqref{deltam1}.  Here we leave the contribution from the matter in its 
form for an arbitrary stress-energy tensor.  

We start by generalizing the Smarr relation \eqref{e:smarr}.
We can assume that the Killing vectors $t^\a$ and $\phi^\a$ are the same for all $\lambda$: 
This uses our freedom to make an arbitrary diffeo or, in coordinate language, to 
to say that in a fixed chart, each metric $g_{\a\b}(\lambda)$ has components independent of 
$t$ and $\phi$.  Consistent with this, we can also require that the set of points 
comprising the horizon is the same for all metrics $g_{\a\b}(\lambda)$.%
\footnote{If the new and old horizons do not coincide, one can always make 
a $t$- and $\phi$-independent diffeo that maps the new horizon to the old 
one.  Keeping $t^\a$ and $\phi^\a$ fixed was implicit in our derivation of the 
first law for Kerr, and fixing the horizon as well is a sensible choice for proving the first law in general.  
Although this way of describing the family of black holes is natural for proving 
the first law, it is not the way one would ordinarily describe changes in a black 
hole that alter its orientation:  Galactic black holes that accrete matter by 
swallowing stars change from one stationary axisymmetric configuration to another, 
and the changes in area, mass, and total angular momentum are related by the first law. 
But to describe the time-dependent accretion or the change in spin direction, 
one would typically pick a chart in which the coordinate direction of the rotation 
axis changes.}    

From the Killing vector identity \eqref{e:kdr}, we have
\be
  \na_\b\na^\a k^\b = R^\a{}_\b k^\b,
\ee
vanishing for our vacuum spacetimes.  Gauss's theorem now gives the integral 
$\int_\Sigma \na_\b\na^\a k^\b dS_\alpha$ 
over a 3-dimensional hypersurface in terms of the integral over its boundary, 
\be
   \int_\Sigma \na_\b\na^\a k^\b dS_\a 
   		 = \int_{\pa\Sigma} \na^\a k^\b  dS_{\a\b}.
\ee

When $\pa\Sigma$ is a $t= $ constant, $r= $ constant surface, 
$dS_{\mu\nu} = \sqrt{-g} \na_{[\mu} t\ \na_{\nu]} r d\theta d\phi$.   
  With $k^\a$ replaced by $t^\a$ 
or $\phi^\a$, the identity shows that a stationary, asymptotically flat, orientable vacuum spacetime with no interior boundary and a single asymptotic region has vanishing mass and angular momentum: It is flat space. 
In our case, denoting by $S$ the 2-dimensional slice of the horizon, 
$S:=\Sigma\cap H$, we have 
\be
   \int_\infty \na^\a k^\b  dS_{\a\b} 
   	= \int_S \na^\a k^\b  dS_{\a\b} + \int_\Sigma R^\a_\b k^\b dS_\a ,
\label{e:kkomar}\ee 
where we have used Eq.~\eqref{e:mk1}.
From Eqs.~\eqref{e:mkjk}, we have 
\be
  \frac1{4\pi}\int_\infty \na^\a k^\b dS_{\a\b} = -M+2\Omega_H J.  
\label{e:kkomar1}\ee

Now to the integral over the horizon: We can take our slice $S$ of $H$ to be 
orthogonal to $\ell^\a$ and $n^\a$, spanned by $m^\a$ and $\barm^\a$.  
On $H$,  $k^\alpha = \ell^\alpha$, and the area element of $S$ is  
\be \cblue
	dS_{\alpha\beta} = \frac12(\ell_\alpha n_\beta - \ell_\beta n_\alpha)dA,
\label{abh}\ee
where the orientation is chosen in the same way as at the sphere at infinity:
$dS_{tr} = \epsilon_{tr\theta\phi}d\theta d\phi$ for $t$ increasing to the 
future and $r$ increasing in the usual way for a radial coordinate. 
\benr\item Quick exercise:  Check this, starting with the area element in the 
form $\omega^0{}_{[\alpha} \omega^1{}_{\beta]}dA$, for an orthonormal basis $\{\bm\omega^\mu\}$
with $\bm\omega^0$ and $\bm\omega^1$ normal to $\bm m$.  Then for some constant $a$, 
 $\ell_\alpha = a(\omega^0{}_\a - \omega^1{}_\b)$, $\displaystyle n_\alpha = \frac1{2a}(\omega^0{}_\a + \omega^1{}_\b)$. 
\een

Next, using the Killing equation, 
$\nabla^\alpha k^\beta = \nabla^{[\alpha} k^{\beta]}$, we have on $S$ 
\be
  \nabla^\alpha k^\beta\frac12(k_\alpha n_\beta - k_\beta n_\alpha) 
  	= k^\a\nabla_\a k^\b n_\b = \ell^\a\nabla_\a \ell^\b n_\b = -\kappa,
\ee
implying
\be\cblue
 \frac1{4\pi} \int_S \na^\a k^\b  dS_{\a\b} = - \frac{1}{4\pi}\kappa A. 
\label{45}\ee
With this expression for the surface integral at the horizon and Eq.~\eqref{e:kkomar1} for 
the integral at infinity, Eq.~\eqref{e:kkomar} gives the generalized Smarr relation  	
\be\crv
   M = \frac{1}{4\pi}\kappa A + 2\Omega_H J -\frac1{4\pi} \int_\Sigma R^\a_\b k^\b dS_\a ,
\label{e:gsmarr}\ee
with the last term on the right the contribution from the matter.

   To obtain the first law, we will use this relation, in the perturbed form
\begin{align}\cblue
   \delta M &= \cblue\frac{1}{4\pi}\delta(\kappa A) + 2\delta(\Omega_H J) +\delta_1  M_{\rm matter},
\label{e:dm1} \\ 
   \delta_1  M_{\rm matter} &= -\frac1{4\pi} \int_\Sigma R^\a_\b k^\b dS_\a , 
\label{e:d1mmatter}\end{align} 
together with a relation 
\be\cblue
  \delta M = -\frac1{4\pi}\delta\kappa\ A - 2\delta\Omega_H\ J +\delta_2 M_{\rm matter}, 
\label{e:dm2}\ee
that comes from varying the Hilbert action, 
$\displaystyle I = \frac1{16\pi} \int R\sqrt{-g}d^4x$.   
Adding these two equations immediately gives the first law.  

The variation of the action is given in Sect.~\ref{s:action} below.      
To avoid factors of 1/2 and conform to BCH, we'll look at the change in 
$\displaystyle \frac{1}{8\pi}R \sqrt{-g}$  
from an arbitrary perturbation $h_{\a\b}$ of the metric. From Eq.\eqref{drsqrtg2}, 
we have
\be
\frac1{\sqrt{-g}}\delta\left(R\sqrt{-g}\right) 
	= - \Gabu h_{\a\b} 
	  +\nabla_\alpha A^\alpha, 
\label{e:dR}\ee
where
\be\color{blue} 
	 A^\alpha = \nabla_\b h^{\a\b} - \na^\a h.   
\label{thetag}\ee

One ordinarily obtains the field equation by varying the action, defined as an 
integral over a region of spacetime between an initial and final hypersurface.  
Here, because we are looking at stationary solutions, we can omit the integral 
over time and write the action as an integral over $\Sigma$: Eq.~\eqref{e:dR} gives  
\be
  \frac1{8\pi} \int_\Sigma \delta (R\sqrt{-g}) d^3x  
		= -\frac1{8\pi} \int_\Sigma \Gabu h_{\a\b} \sqrt{-g}\, d^3x
		  + \frac1{8\pi} \int_\Sigma \na_\a A^\a \sqrt{-g}\, d^3x  .
\label{e:drsqrtg}\ee
We now show that the second integral on the right is the sum of surface terms at infinity 
and at the horizon, having the form 
\begin{align}
 \frac1{8\pi} \int_\Sigma \na_\a A^\a \sqrt{-g}\, d^3x  &= \mbox{surface integral at infinity} -
 \mbox{surface integral at horizon} \nonumber\\
 &= \delta M + \frac1{4\pi}\delta\kappa\,A + \delta\Omega_H\ J_H,
\end{align} 
where the minus sign in the first equality corresponds to using a normal that is radially outward at 
both the horizon and at spatial infinity.   
In a vacuum spacetime, only the surface terms are nonzero, and the change in mass 
is given by 
$\dis \delta M = -\frac1{4\pi}\delta\kappa\,A -\delta\Omega_H\ J$. 
When matter is present, the remaining terms in Eq.~\eqref{e:drsqrtg} 
are the contributions from the matter to $\delta M$ that we have 
called $\de_2 M_{\rm matter}$ in Eq.~\eqref{e:dm2}
\be
\de_2 M_{\rm matter} =  \frac1{8\pi} \int_\Sigma \delta (R\sqrt{-g}) d^3x  
			+\frac1{8\pi} \int_\Sigma \Gabu h_{\a\b} \sqrt{-g}\, d^3x.
\label{e:d2mmatter}\ee 
  \\   

\noindent{\sl Surface integrals}. 

Choose an axisymmetric scalar $t$ for which $\Sigma$ is a $t$ = constant surface 
and with $t^\a\nabla_\a t = 1$. (For Kerr or Schwarzschild, a natural choice is a scalar agreeing 
with $v$ near the future horizon and with the usual coordinate $t$ near infinity.)   
In a chart $\{t, x^i\}$, \\
$dS_\a = \nabla_\a t {\sqrt{-g}}\, d^3x$, and $t^\a dS_\a = {\sqrt{-g}}\, d^3x$.  
Then
\be \de\int_\Sigma R t^\a dS_\a = \de\int_\Sigma R {\sqrt{-g}}\, d^3x 
	=  -\int_\Sigma G_{\b\g} h^{\b\g}t^\a dS_\a +\int_\Sigma \nabla_\b A^\b t^\a dS_\a . 
\label{47}
\ee

 Because $h_{\a\b}$ and $g_{\a\b}$ are time independent, 
we have 
\[
  \sqrt{-g} \na_\a A^\a = \pa_\mu(A^\mu\sqrt{-g}) = \pa_i(A^i\sqrt{-g}).
\] 
The surface term at infinity from $\n_\a A^\a$ in Eq.~\eqref{47} can be written as  
\begin{align}
\int_\infty A^\a dS_\a & =\lim_{r\rightarrow\infty}\int A^\a\na_\a r\  r^2 d\Omega 
		= \lim_{r\rightarrow\infty}\int [\pa_\mu(r^2 h^{r\mu}) -r^2\pa^r h] d\Omega\nn\\
	&= \lim_{r\rightarrow\infty}\int [\pa_r(r^2 h^{rr}) -r^2\pa_r h] d\Omega.\nn 
\end{align}
For our family of asymptotic metrics \eqref{asympt}, to order $1/r$, 
\[
 h^{rr} = h_{rr} = \frac{2\delta M}r, \qquad 
h^t{}_t+h_r^r = - h_{tt} + h_{rr} = 0,
\]
and we have   
\be\color{blue}
  \frac1{8\pi}\int_\infty A^\a dS_\a = \delta M\color{black}.  
\ee

Finally, to compute the surface integral at the horizon, where what we understand is 
$dS_{\a\b}$ in terms of the two normals $n_\a$ and $\ell_\a$, it will be helpful to 
rewrite the surface term.  Start with 
\begin{eqnarray*}
\nabla_\b A^\b t^\a &=& \nabla_\b (A^\b t^\a - A^\a t^\b ) -
\underbrace{A^\b\nabla_\b t^\a +t^\b\nabla_\b  A^\a}_{{\mbox {\mbox \pounds}}_{\bm t}A^\a =0} -
\underbrace{A^\a\nabla_\b t^\b}_{0}\\
&=& \nabla_\b (A^\b t^\a - A^\a t^\b ) .
\end{eqnarray*}
Then the divergence term in the varied action~\eqref{47} can be written as
\begin{align}
  \int_\Sigma \nabla_\b A^\b t^\a dS_\a 
	&= \int_\Sigma \nabla_\b (A^\b t^\a -A^\a t^\b)dS_\a \nonumber\\
	&= - \int_{\partial \Sigma} (A^\b t^\a -A^\a t^\b )dS_{\a\b} , 
\label{48}\end{align}
and the surface term at the horizon is 
\begin{align}\color{blue}
 \int_S (A^\b t^\a -A^\a t^\b )dS_{\a\b} 
	&= \int_S (A^\b t^\a -A^\a t^\b )\frac{1}{2} (\ell_\a n_\b-\ell_\b n_\a )dA 
	= \int_S A^\b t^\a (\ell_\a n_\b -\ell_\b n_\a )dA  \nonumber\\
	& =\frac1{8\pi}\int_S (\nabla_\g h^{\b\g} - \nabla^\b h)
	   (\ell^\a -\underbrace{\Omega_H\phi^\a}_{0}) 
		\begin{array}[t]{l} (\ell_\a n_\b -\ell_\b n_\a)dA \\ 
		           \bm\phi\cdot\bm\ell = \bm\phi\cdot \bm n=0\end{array}\nonumber\\
	&\color{blue} =\frac1{8\pi}\int_S \ell^\alpha\nabla^\beta h_{\alpha\beta} dA.
\label{e:dhl}\end{align}
where we have used $\ell\cdot\ell = 0,\; \ell\cdot n =-1$, and $ \ell^\a\nabla_\a h=0$ in the last equality. \\

We need to relate the integrand to $\delta\kappa$, and we compute $\delta\kappa$ as 
follows. From the definition of $\kappa$, we have
\be
 \de\kappa = -\de (\ell^\b n^\a\nabla_\b\ell_\a ) 
		=-\de\ell^\b n^\a\nabla_\b\ell_\a
		 - \de n^\a \ell^\b \nabla_\b\ell_\a 
		- \ell^\b n^\a\de(\nabla_\b \ell_\a).  
\ee
Now on the horizon, 
\be
\delta\ell^\a = \delta k^\a = \delta\Omega_H \phi^\alpha,
\label{e:deltaka}\ee
because $\de t^\a = 0 =\de\phi^\a$.  
In the third term, we can replace $\ell_\a$ by its value $k_\a$ on the horizon, 
because the only derivative is along the horizon.  Then Killing's 
equation gives \\
 $\delta (\nabla_\b \ell_\a ) = \delta (\nabla_{[\b}k_{\a]}) 
	= \nabla_{[\b}\delta k_{\a]}$, 
\begin{align}
\delta\kappa &= -n^\a\de\Omega_H \phi^\b\nabla_\b\ell_\a 
		- \de n^\a \ell^\b \nabla_\b\ell_\a 
		-  \ell^\b n^\a\nabla_{[\b} \delta k_{\a]} . 
\label{e:dk}\end{align}
On to the second term on the right:  
Because the horizon is unchanged in our gauge, $\ell_\alpha(\lambda)$ is parallel
to the null normal $\na_\a u$ to $H$, implying $\delta \ell_\alpha$ is parallel to $\na_\a u$: 
$\ \delta \ell_\alpha = f\ell_\alpha$, some
function $f$ on $H$, and we can write
\begin{align}
-\delta n^\alpha \ell^\beta\nabla_\b \ell_\a 
	&= -\delta n^\alpha\kappa \ell_\alpha 
	 = \kappa  n^\alpha \delta \ell_\alpha = -\kappa  f\nonumber\\
	&= f n^\alpha \ell_\beta\nabla^\b \ell_\a 
	 =n^\alpha\ell_\beta\nabla_\b (f \ell_\a),\quad
	 	\mbox{using $\ell^\b\na_b f = k^\b\na_b f = 0$ on $H$}, \nonumber \\
 	&= \ell^\b n^\a\nabla_\b\delta \ell_\a = \ell^\b n^\a\nabla_\b\delta k_\a . 
\label{e:nldl}\end{align}
The second and third terms on the right of \eqref{e:dk} now combine to give
\[ 
\frac{1}{2} (\ell^\alpha n^\beta + \ell^\beta n^\alpha )\nabla_\alpha\delta k_\beta.
\]
The shear and divergence, $\sigma_{\alpha\beta}(\lambda)$ and $\theta(\lambda)$ vanish, so 
$0= \delta (m^\a \bar m^b \na_\a k_b)$;  because the position of the horizon is unchanged, $\delta m^\a$ 
is a linear combination of $\ell^\a, m^\a$ and $\bar m^\a$, implying $\delta m^\a \bar m^\b\na_\a k_\b = 0$.  Then 
$m^\a \bar m^\b \na_\a \d k_\b =0$ on $H$, and we have 
\begin{align} 
\frac{1}{2} (\ell^\alpha n^\beta + \ell^\beta n^\alpha )\nabla_\alpha\delta k_\beta
     &= -\frac{1}{2} (-\ell^\alpha n^\beta - \ell^\beta n^\alpha 
			+m^\alpha \barm^\beta +m^\b \barm^\a)\nabla_\alpha\delta k_\beta 
	=-\frac{1}{2} \nabla^\alpha\delta k_\alpha\nonumber\\
	&\color{blue}= - \frac{1}{2} \ell^\alpha\nabla^\beta h_{\alpha\beta} \color{black}.\label{49}
\end{align}
In the last equality, we have used Eq.~\eqref{e:deltaka}, 
$\de k_\a = \de (g_{\a\b}k^\b) = h_{\a\b}k^\b +\Omega_H \phi_\beta$, and $\na^\beta\phi_\beta=0$.  
Then
\be {\color{blue} 
   \delta\kappa }= - \frac{1}{2} \ell^\alpha\nabla^\beta h_{\alpha\beta} 
		   -n^\a\de\Omega_H \phi^\b\nabla_\b\ell_\a
     \color{blue}= -\frac{1}{2} \ell^\b\nabla^\a h_{\a\b} 
		   - \de\Omega_H n^\a\ell^\b\nabla_\b\phi_\a ,
\label{44}\ee
this time using $\Lie_{\bm\phi}\ell_\a=0$. 

Using this relation to replacing $\frac{1}{2} \ell^\alpha\nabla^\beta h_{\alpha\beta}$ in Eq.~\eqref{e:dhl} gives for the surface integral over the horizon
\be {\color{blue}\int_S (A^\b t^\a -A^\a t^\b )dS_{\a\b}} =- \frac{1}{4\pi}\int_S \delta\kappa dA - 
	\delta\Omega_H\ \frac{1}{4\pi}\int_S  \nabla^\a\phi^\b \ell_\a n_\b dA
								\nonumber\\
={\cblue-\frac{1}{4\pi} \delta \kappa A - 2\delta\Omega_H J_H},
\label{50}
\ee
where, in the last line, we used the antisymmetry of $\nabla^\a\phi^\b$.
Thus the variation \eqref{e:drsqrtg} of $\int_\Sigma d^3xR\sqrt{-g}/8\pi $ yields 
\be\cblue
\delta M = -\frac1{4\pi}\delta\kappa\,A -2\delta\Omega_H\ J_H + \delta_2 M_{\rm matter}\color{black},  
\ee
as claimed.  

Adding this to the relation $\delta M = \delta(\kappa A/4\pi +2\Omega_H J_H) + \de_1 M_{\rm matter}$, Eq.~\eqref{e:dm1}, and dividing by 2, gives the first law, 
\[ \crv
   \delta M = \frac1{8\pi} \kappa \delta A + \Omega_H\delta J_H + \de M_{\rm matter}\color{black},
\tag{\ref{e:first1}}
\]
with 
\[ 
  \de M_{\rm matter} = \frac12\left(\de_1 M_{\rm matter}+\de_2 M_{\rm matter}\right).
\]
In Eq.~\eqref{e:d2mmatter} for $\delta_2 M_{\rm matter}$, we can write $\sqrt{-g}d^3x= t^\alpha dS_\alpha = k^\alpha dS_\alpha$, because $\phi^\alpha$ lies in $\Sigma$ and is therefore orthogonal to $dS_\alpha$. Using Eq.~\eqref{e:d1mmatter} for $\delta_1 M_{\rm matter}$, we then have
\begin{align} 
  \de M_{\rm matter} &= -\frac1{8\pi}\delta \int G^\a_\b k^\b dS_\a +\frac1{16\pi} \int G^{\b\g} h_{\b\g} k^\a dS_\a
  \nonumber\\
  		    &= -\delta \int T^\a_\b k^\b dS_\a +\frac12 \int T^{\b\g} h_{\b\g} k^\a dS_\a.
\end{align} 

\section{Gauss}\label{s:gauss1}

For an antisymmetric tensor $F^{\a\b}$, 
$\nabla_\nu F^{\mu\nu} = (-g)^{-1/2}\pa_\nu [(-g)^{1/2}F^{\mu\nu}]$.  This is 
the same identity that gives Gauss's law for a vector $A^\nu$; the extra Christoffel 
symbol in the derivative vanishes because of the antisymmetry of $F$.  
Consider a $t=$constant surface
$\Sigma$ bounded by an $r=$ constant surface $\pa\Sigma$. (For any smooth 3-dimensional 
submanifold with the topology of a ball bounded by a topological sphere, just choose coordinates with $t$ constant on the ball and $r$ constant on the boundary.) 
Write $dS_\nu=\na_\nu t \sqrt{-g}d^3 x $ and 
$dS_{\mu\nu} = \sqrt{-g} \na_{[\mu} t\ \na_{\nu]}rd^2 x$. Then 
\begin{align*}
  \int_\Sigma \na_\b F^{\a\b} dS_\a 
		&= \int_\Sigma \pa_\nu (F^{\mu\nu} \sqrt{-g})\na_\mu t d^3x 
			= \int_\Sigma F^{\mu\nu} \na_\mu t \na_\nu r \sqrt{-g}d^2 x\\
		&= \int_{\pa\Sigma} F^{tr} \sqrt{-g} d\theta d\phi\\
		&= \int_{\pa\Sigma} F^{\a\b} dS_{\a\b}.  
\end{align*}
\index{charge!as surface integral of $F^{\a\b}$}\index{conservation laws!charge}
If $n_\a$ and $\tilde n_\a$ are two orthogonal unit normals spanning the tangent space to 
$\pa\Sigma$,\\
 $dS_{\a\b} = n_{[\a} \hat n_{\b]} dA$.  
For $F^{\mu\nu}$ the electromagnetic field, this gives the electric charge: 
\[
  4\pi j^\a =  \na_\b F^{\a\b} \Longrightarrow \frac1{4\pi }\int_\Sigma j^\a dS_\a 
			=\frac1{4\pi }\int_{\pa\Sigma} F^{tr} r^2\sin\theta d\theta d\phi \\
			= \frac1{4\pi }\int E^r r^2 d\Omega =  Q.
\] 

\section{First Law in a Hamiltonian Framework} 
\subsection{The gravitational Hamiltonian} 
\label{s:H_G}
The first law for black-holes is simplest to state (but not simplest to prove) 
when written in terms of the Hamiltonian for gravity.  A summary is given 
in \href{https://arxiv.org/pdf/gr-qc/9305022.pdf}{Wald '92}, based on work by Wald and Sudarsky. \cite{sw92,wald93}  We'll go over the 
statement of it first, then verify the claims. 

Write the expressions for the Hamiltonian and momentum constraints in the form 
\index{constraints!gravitational|textbf}\index{momentum constraint|textbf}\index{Hamiltonian constraint|textbf}
\index{initial value problem!Hamiltonian constraint}
\index{initial value problem!Momentum constraint} 
\be
  {\cal H} = (\tr -K^{ab}K_{ab} + K^2)\sqrt{\g}, \qquad 
  {\cal H}_a = 2D^b(K_{ab}-\g_{ab}K)\sqrt{\g}.  
\ee
We will show: A Hamiltonian governing the spacetime geometry is 
\be
  H_G = \int d^3 x  [\alpha{\cal H}+\beta^a {\cal H}_a].
\label{e:hg}\ee
the momentum conjugate to $\g_{ab}$ is  
\be
  \pi^{ab} = -(K^{ab} -\gamma^{ab} K)\sqrt{\gamma}.
\label{eq:piab0}\ee 
Varing the Hamiltonian gives the equations of motion, 
\be
  \dot\gamma_{ab} = \frac{\delta H_G}{\delta \pi^{ab}}, \qquad 
  \dot\pi^{ab} = -\frac{\delta H_G}{\delta \g_{ab}},
\label{e:hameqs}\ee
where $(^\cdot)$ is the time derivative associated with $t^\a = \alpha n^\a + \beta^\a$, 
and where $\alpha$, $\beta^a$, $h_{ab}$ and $\pi^{ab}$ vanish rapidly enough at infinity that there are no surface terms. \footnote{
\index{Hamiltonian!gravitational} 
For asymptotically flat spacetimes, the Hamiltonian $H_G$ is not the true Hamiltonian, because the surface terms are nonzero for generic perturbations that preserve asymptotic 
flatness. Moreover, the value of Hamiltonian for solutions to the field equation should be the 
mass (energy) $M$ of the spacetime, but $H_G$ vanishes.  The two problems are related: 
Adding a divergence that gets rid of second derivatives in $H_G$ removes the surface term, and 
the revised Hamiltonian is equal to $M$ for solutions to the field equation (equivalently, 
for initial data $g_{ab}, \pi^{ab}$ satisfying the constraint equations). A similar problem 
arises for the Hilbert Lagrangian density $R\sqrt{|g|}$, and a first-order Lagrangian is presented in Sect. \ref{s:firstorderaction} below.
}

  The first law involves the opposite case, with 
$t^\a$ the timelike Killing vector and $\beta^\a = \phi^a$ the rotational Killing vector 
of a stationary, axisymmetric vacuum black hole, and of a perturbation to 
a nearby  stationary, axisymmetric black hole.  In this case $\dot\gamma_{ab} =0$ and
$\dot \pi_{ab} = 0$; the field equation is satisfied, so 
\[ 
 \frac{\delta H_G}{\delta \pi^{ab}} = 0, \qquad 
  \frac{\delta H_G}{\delta \g^{ab}} =0; 
\]  
and the constraint equations are satisfied by the perturbed and unperturbed 
solutions, so $\delta{\cal H}=0$ and $\delta{\cal H}_a=0$.  But the surface terms are nonzero and we are left with 
\[\crv
  \mbox{surface term at $\infty$} = \mbox{surface term at horizon}, 
\] 
or 
\be\crv
  -\delta M + \Omega_H \delta J = -\kappa \frac{\delta A}{8\pi}.
\ee

To check all this, we will find the gravitational Hamiltonian, show that the surface term at 
infinity is $-\delta M + \Omega_H \delta J$, and then use our original derivation of the first law to infer that the surface term at the horizon is $-\kappa \delta A/8\pi$.  
We first add to the Lagrangian density ${}^4\!R\sqrt{|g|}$ 
a total divergence that replaces the second time derivatives in ${}^4\!R$ by a kinetic term 
quadratic in first time derivatives.  From the 
contracted Gauss-Codazzi equation (1.70) in the initial value notes, we have
\be
\tr = \g^{\a\b}\, \tr_{\a\b} = \g^{\a\b}\g^{\c\d}\,{}^4\!R_{\a\c\b\d} +K^{\a\b}K_{\a\b} - K^2
\label{e:gc4}\ee
which we can rewrite as 
\be
  {}^4\!R - 2\nabla_\a (n^\c\nabla_\c n^\a -n^\a \na_\c n^\c) 
	= {}^3\!R + K_{\a\b}K^{\a\b} - K^2.    
\ee 
That is, the right side differs from $^4\!R$ by a total divergence, so the Lagrangian  
density,
\be
  \Lag  = (K_{ab}K^{ab} - K^2+\tr)\a\sqrt{\gamma},
\label{tildel}\ee 
gives the field equations for variations that vanish fast enough at infinity.
\index{Lagrangian!gravitational}

\index{conjugate momentum, $\gamma_{ab}$|textbf}\index{momentum!momentum conjugate to $\gamma_{ab}$}
Because $\tr$ involves only spatial derivatives of $\gamma_{ab}$, 
the time derivative $\dot\gamma_{ab}$ occurs only in the 
terms involving the extrinsic curvature $K_{ab}$, so the momentum 
conjugate to $\gamma_{ab}$ with respect to $\Lag$ is given by  
\index{extrinsic curvature!relation to conjugate momentum}  
\be
  \pi^{ab} := \frac{\partial{\Lag}}{\partial \dot\gamma_{ab}}
	= \frac{\partial{\Lag}}{\partial K_{ab}}\frac{\partial K_{ab}}{\partial\dot\g_{ab}} 
	= -(K^{ab} - \g^{ab}K)\sqrt{\g}.
\label{eq:piab}
\ee 
where we have used Eq.~\eqref{e:gabevol1}, namely 
\be
 K_{ab} = -\frac1{2\alpha}\left(\dot \g_{ab} -  \Lie_{\bm \b}\g_{ab}\right).
\tag{\ref{e:gabevol1}}
\label{e:gabevol1a}
\ee 
In the canonical formulation of general relativity, 
$\gamma_{ab}, \pi^{ab},\alpha$ and $\beta^a$  
are regarded as independent gravitational field variables,   
the arguments of the geometrical Hamiltonian
\be
H_G=\int d^4x\left(\pi^{ab}\dot \g_{ab} - \Lag\right).
\ee
To write $H_G$ as a function of the four variables 
$\gamma_{ab}, \pi^{ab},\alpha$ and $\beta^a$, we first invert Eq.~(\ref{eq:piab}) to write $K^{ab}$ in terms 
of $\pi^{ab}$: The trace of Eq.~(\ref{eq:piab}) gives $\pi = 2K\sqrt{\gamma}$, 
where $\pi := \gamma_{ab}\pi^{ab}$. From (\ref{eq:piab}) itself we then have 
\be
K_{ab} = - \frac1{\sqrt{\g}} \left(\pi_{ab} -\frac12\gamma_{ab}\pi \right),
\label{eq:kpi}\ee
and the Lagrangian density \eqref{tildel} has the form 
\be
{\Lag} 
  = \alpha \left[\sqrt\g \ \tr + \frac1{\sqrt{\g}} \left(\pi^{ab}\pi_{ab} 
			  - \frac {1}{2}  \pi^{2} \right) \right], 
\ee
Next use \eqref{eq:kpi} and \eqref{e:gabevola} to write 
$\dot\g_{ab}$ in terms of $\pi_{ab}$:  
\be
\dot\gamma_{ab}
 = \frac2{\sqrt{\g}}\a \left(\pi_{ab}-\frac12\g_{ab}\pi \right)+D_a\beta_b+D_b\b_a.
\label{eq:gammaab}\ee 
The Hamiltonian density now takes the form
\bea
\pi^{ab}\dot \g_{ab} -  {\Lag}
   &=& \frac\a{\sqrt\g} \left (\pi^{ab}\pi_{ab}- \frac {1}{2}  \pi^{2}\right )
		-\a\sqrt\g\, \tr + 2\pi^{ab}D_a\beta_b\\
   &=& \alpha{\cal H} + \beta^a{\cal H}_{a} + D_a(2\pi^{ab}\beta_b),  
 \label{eq:calh}\eea
where the scalar density 
\be
{\cal H} :=\frac1{\sqrt{\g}} \left(\pi ^{ab} \pi_{ab} - \frac {1}{2} \pi^{2} \right)
		-\sqrt\g \ \tr,
\label{eq:hamg}\ee
and the vector density
\be
 {\cal H}_{a} := -2D_b \pi^b_{\ a},
\label{eq:momga}\ee
are the Hamiltonian and momentum constraints, written in terms of $\pi_{ab}$.
\index{constraints!gravitational}\index{momentum constraint}\index{Hamiltonian constraint}

 To relate geometrical 
variables to mass and angular momentum, we need to multiply the geometrical Hamiltonian $H_G$ 
by $c^4/(16\pi G) = 1/16\pi$ to get the physical Hamiltonian    
\mbox{$\dis H:=\frac1{16\pi}\int d^3x (\pi^{ab}\dot \g_{ab} -  {\Lag})$}. 
The gravitational action $\dis I = \frac1{16\pi}\int d^3x {\Lag}$ 
can now be written in the form 
\be
 I =\frac1{16\pi} \int d^4x  (\pi^{a b}\dot\gamma_{ab}
			-\alpha{\cal H}-\beta^a {\cal H}_a),
\ee
and the gravitational Hamiltonian is 
\be
  H = \frac1{16\pi}  \int d^3 x  [\alpha{\cal H}+\beta^a {\cal H}_a].
\label{e:hg0}\ee 

Because $\beta^a$ and $\alpha$ do not appear in $\cal H$ or ${\cal H}_a$, defined as  
functions of $\pi^{ab}$ and $\gamma_{ab}$,   
the variational derivatives $\delta I/\delta\alpha$ and  $\delta I/\delta\beta^a$ 
are the constraint equations, the vacuum constraints 
\be
  {\cal H} = 0, \qquad {\cal H}_{a}=0,
\ee 
and the dynamical Einstein equations are given by \ref{e:hameqs}  
for variations vanishing rapidly enough at spatial infinity and at the horizon that 
there are no surface terms.  \\  

\subsection{Surface terms and the first law}

 In the first law of black hole thermodynamics, our interest is in 
the opposite case, in surface terms for variations 
associated with nearby solutions for which $\dot\gamma_{ab}=0=\dot\pi^{ab}$. 
As described at the beginning of this section, because solutions to the Einstein field equation satisfy ${\cal H} = 0$ and
${\cal H}_a = 0$, the Hamiltonian $H$ also vanishes, implying  
the variation $\delta H = 0$. Now $\delta H$ has the form 
\be
   \delta H = \int d^3 x \left[
   		  \frac{\delta H}{\delta \gamma_{ab}} \delta \gamma_{ab}
   		+ \frac{\delta H}{\delta \pi^{ab}}\delta \pi_{ab}\right] + 
   		  \mbox{surface terms at infinity and at the horizon}.
\ee 
Because the solutions we are looking at 
are time independent, we have
\be
   \frac{\delta H}{\delta \gamma_{ab}} = 0 =  \frac{\delta H}{\delta \pi^{ab}},
\ee 
whence  
\be
  0 = \delta H = \mbox{surface terms at infinity and at the horizon}.
\ee

We consider an evolution along $k^\a = t^\a + \Omega_H\phi^\a$ with 
$\Omega_H$ fixed.   In this case, 
\[
    H = \frac1{16\pi} \int d^3 x  [\alpha{\cal H}+\Omega_H\phi^a {\cal H}_a], 
\]
\index{Hamiltonian!gravitational}
and in
\[ 
 \alpha{\cal H} = \frac\a{\sqrt{\g}} \left(\pi ^{ab} \pi_{ab} 
			- \frac {1}{2} \pi^{2} \right) -\a\sqrt\g \ \tr,
\]
the first term (the part quadratic in $\pi^{ab}$) has no derivatives of $\gamma_{ab}$ or 
$\pi^{ab}$, so its variation gives no surface terms.  As a result, in 
\[
  \delta(\alpha{\cal H}) = \delta\left[\frac\a{\sqrt{\g}} \left(\pi ^{ab} \pi_{ab} 
			- \frac {1}{2} \pi^{2} \right) -\a\sqrt\g \ \tr\right], 
\]
the surface term comes only from $\delta\left(-\a\sqrt\g \ \tr\right) =\delta\left(-\sqrt{|g|} \ \tr\right)$.  From 
Eq.~\eqref{eq:dr1}, we have
\be
\delta (-\sqrt{|g|}\ \tr) 
	= \left[ (R^{ab}-\frac12 g^{ab}R)h_{ab} -D_a A^a]\right]\sqrt{|g|} \,, 
\label{drsqrtg}\ee
where 
\be
  A^a = D_b h^{ab} - D^a \,^3h,  \mbox{ with }  ^3h=\gamma^{ab}h_{ab}.  
\ee
The surface term comes from the divergence $-D_a A^a$.  
Here, at order $1/r$ only $h_{rr} =2\delta M/r$ is nonzero.  Then $^3h = h_{rr}+O(r^{-2})$,
$\ h^{rr}=h_{rr}+O(r^{-2})$, and the first surface term is 
\be\cblue
  -\frac1{16\pi} \int_\infty A^a dS_a = -\frac1{16\pi} \int_\infty
		   \left[\frac1{r^2}\pa_r(r^2 h_{rr}) - \pa^r h_{rr}) \right]r^2 d\Omega 
		= -\delta M.   
\label{e:sm}\ee
\index{mass!ADM}\index{ADM mass}
This is the ADM expression for the mass, involving only the spatial metric; it is named 
after Arnowitt, Deser, and Misner, who introduced it in their famous 3+1 paper \cite{ADM62}. 

A second surface term comes from $\Omega_H\delta(\phi^a{\cal H}_a)$. In fact, 
$\int_\Sigma\phi^a{\cal H}_a$ is itself a sum of surface terms at infinity and at the 
horizon, each proportional to the angular momentum.  We convert $\delta(\phi^a{\cal H}_a)$ to the form Eq.~\eqref{e:jk} of $J$ as follows:
With $n_\a$ and $\hat r_\a$ the unit normals to $\Sigma$, we have 
\[
\nabla^\a \phi^\b\ dS_{\a\b}
	= \nabla^\a \phi^\b \ \hat r_{[\a} n_{\b]} dS 
	= \nabla^\a \phi^\b \ \hat r_\a\,n_\b \, dS,
\]
using in the last equality the antisymmetry of $\na^\a\phi^\b$.
Then $\phi^\b n_\b =0$ implies
\bea
 \na^\a\phi^\b \hat r_\a n_\b 
	&=& -\phi^\b\hat r^\a\na_\a n_\b  = K_{\a\b}\phi^\b\hat r^\a 
	 = -\frac1{\sqrt{\g}}\pi^{ab} \phi_a\hat r_b, \nn\\
\cblue 8\pi J &=& \int_\infty \nabla^\a \phi^\b dS_{\a\b}
	=- \int_\infty\pi^{ab} \phi_a\hat r_n\, dS 
	\cblue = -\int_\infty\pi^{ab} \phi_a dS_b.
\eea
\index{angular momentum!as a surface integral}
We can now find our second surface term as follows: From
\begin{align*}
\frac1{16\pi} \delta\int_\Sigma \Omega_H\phi^a{\cal H}_a 
		&= -2\Omega_H\delta\int_\Sigma (\phi_a D_b \pi^{ab}) d^3 x
		 = -2\Omega_H\delta\int_\Sigma D_b(\pi^{ab}\phi_a) d^3 x, 
\end{align*}
the surface term at infinity is 
\be\cblue
   \frac1{16\pi}\left(-2\Omega_H\delta\int_\infty \pi^{ab}\phi_a dS_b\right)
		 = \Omega_H \delta J.
\label{e:sj}\ee  

The sum of the two surface terms at infinity, \eqref{e:sm} and \eqref{e:sj},  is thus 
\be\crv
   \mbox{surface term at infinity} = -\delta M + \Omega_H \delta J, 
\ee   
as claimed.  Our proof of the first law implies for the sum of terms at the horizon,
\be\crv
   \mbox{surface term at the horizon } =\ -\kappa\frac{\delta A}{8\pi}. 
\ee 

A direct calculation of the surface term at the horizon would give an alternative 
proof of the first law, 
\be  
   \delta M =\Omega_H \delta J + \kappa\frac{\delta A}{8\pi},
\ee  
but the calculation, first done by Sudarsky and Wald,\cite{sw92} involves theorems on black holes whose proofs are beyond the scope of these notes.  

\chapter{Action, Noether, Surface term}\index{conservation laws!Noether's theorem}
\index{action!gravitational|(}
\label{s:action_noether}

The first section in this chapter, showing that $R\sqrt{|g|}$ is a Lagrangian
density for the Einstein equation, is a natural part of a second 
semester GR course.  The subsequent sections are less standard:  
They use Noether's theorem to introduce conserved quantities in GR, following Katz\cite{katz85}, Chrusciel\cite{chrusciel85}, and Sorkin\cite{sorkin88}.  
To do so, one uses an action that involves only first derivatives of the metric, 
or, equivalently an action that differs from $\int R\sqrt{|g|} d^4x$ by a 
surface term involving the extrinsic curvature $K$\cite{york72}. 
The revised action is needed because $\int R\sqrt{|g|} d^4x$ is not 
a true action:  Variations of the action for which $\delta g_{\a\b}$ vanishes 
on an initial and final surface-- and is well-behaved asymptotically-- 
do not yield the field equations. Instead, they 
leave a divergence whose integral is a nonvanishing surface term.     
These last sections may be helpful to readers using work by Hojman, 
Kuchar \& Teitleboim\cite{hkt76} relating surface terms to conserved quantities, 
or work by York\cite{york72} and later Gibbons \& Hawking\cite{gh77} using the first-order action 
written as the Hilbert action with an extrinsic curvature surface term.  
\index{conservation laws!first law of black hole thermodynamics|)}
\index{first law of thermodynamics!for black holes|)}     

\section{Variation of the gravitational (Hilbert) action}
\label{s:action}\index{action!Hilbert (Einstein-Hilbert) action} 
\index{Einstein field equation!from action}\index{field equation!from action}
\index{Hilbert action\textbf}\index{Einstein-Hilbert action\textbf}
\index{Lagrangian!gravitational}

\noindent{\sl Varying $\int d^nx R\sqrt{|g|}$ in any dimension.}\\  

The Hilbert (or Einstein-Hilbert) action is \href{https://echo-old.mpiwg-berlin.mpg.de/ECHOdocuView?url=/permanent/echo/einstein/MHTT6W63/index.meta&pn=2}{Hilbert 1915} 
\be
  I= \int \Lag d^4x = \frac1{16\pi}\int R\sqrt{|g|} d^n x.    
\ee
Finding the field equations by varying this action is a quick calculation once 
one realizes that the variation of the Ricci tensor $\delta R_{ab}$ is a pure divergence
\be
   \delta R_{ab} = \na_c(B^c{}_{ab}),
\label{e:dRB}\ee 
some tensor $B^c{}_{ab}$.

We'll do the easy part first, getting the field equations assuming 
$\delta R_{ab}$ has the form \eqref{e:dRB}, and then do the harder part, 
finding $\delta R_{ab}$ and showing it 
contributes only a divergence $\nabla_a A^a$ to the variation of $\Lag$.  
For the first law for black holes, the Ricci tensor vanishes, and it is $A^a$ 
that is the important term.  

Here's the first, quick, calculation:  
\bea
\delta R &=& \delta (R_{ab}g^{ab})
	= - R_{ab}h^{ab}+ g^{ab}\delta R_{ab} 
\nonumber\\
&=& -R_{ab}h^{ab} + \nabla_c A^c,
\label{eq:dr}\eea
where $A^c = B^{cb}{}_b$.
Eq.~(\ref{eq:dr}) and 
\[
   \delta\sqrt{|g|} = \frac12 \sqrt{|g|} \, h 
\]
immediately give
\be \cblue
\frac1{\sqrt{|g|}}\delta (R\sqrt{|g|}) = -(R^{ab}-\frac12 g^{ab}R)h_{ab} +\nabla_a A^a\color{black},
\label{drsqrtg2}\ee
as claimed.  

Now we find $A^a$.  We start by varying the Riemann tensor, using the commutator 
\be 
  R^a_{\ \ bcd}v^b=[\nabla_c,\nabla_d] v^a ,
\ee 
for any vector $v^a$. For fixed $v^a$, we need to compute
\be
\delta R^a_{\ \ bcd}v^b =\d[\nabla_c\nabla_d v^a - \nabla_d\nabla_c v^a].
\label{driem1}\ee
The perturbed covariant derivative operator, $\delta \nabla_{a}$, acts as a tensor%
\footnote{Although the Christoffel symbols $\Gamma^\mu{}_{\nu\lambda}$ do not naturally constitute 
a tensor, the difference between the Christoffel symbols of two different 
covariant derivative operators do so. To linear order in the difference, this is implied by the 
explicit covariance of Eq.~(\ref{eq:dgamma}).  
From p. \pageref{p:tensorfield} a tensor field is a map of vector fields that is 
linear under multiplication by scalar fields, and the difference between two covariant 
derivative operators, $\tilde\nabla_a -\nabla_a$ is linear in this sense: \\
$(\tilde\nabla_a -\nabla_a)(f\xi^b) = f(\tilde\nabla_a -\nabla_a)\xi^b $.
This was \ref{ex:nabla-partial}}%
$\ \delta \Gamma^c_{\ ba}$; on a vector $v^a$, its action is
\be 
  \delta\nabla_c v^a = \delta\Gamma^a_{\ bc}v^b, 
\ee
where  
\be
 \delta \Gamma^a_{\ bc}
	= \frac12\left(\nabla_b h^a_{\ c}+\nabla_c h^a_{\ b} - \nabla^a h_{bc}\right).
\label{eq:dgamma}
\ee
Eq.~(\ref{eq:dgamma}) can be derived from the relation
\be
0= \delta (\nabla_a g_{bc})
	= \nabla_a h_{bc}-\delta\Gamma^d_{\ ba} g_{dc}-\delta\Gamma^d_{\ ca} g_{bd},
\ee
via the linear combination 
\be
0=\frac12 g^{ad}[\de(\nabla_b g_{cd})+\de(\nabla_c g_{bd})-\de(\nabla_d g_{bc})].
\ee

To compute $\delta R_{abc}{}^{d}$, we write
\bea
\delta(\nabla_{c}\nabla_{d}v^a) 
&=& -\delta\Gamma^{b}{}_{cd}\nabla_b v^a
    + \delta\Gamma^a{}_{bc}\nabla_d v^b    
	+\nabla_c\de(\nabla_{d}v^a)
\nonumber\\
&=& -\delta\Gamma^{b}{}_{cd}\nabla_b v^a
    + \delta\Gamma^a{}_{bc}\nabla_d v^b  
    + \delta\Gamma^a{}_{bd}\nabla_c v^b \nonumber\\
&\quad& +\nabla_c(\delta\Gamma^a{}_{bd})v^b.
\eea
Because the first three terms on the right side are, together, 
symmetric in $c$ and $d$, Eq.~(\ref{driem1}) implies, for all $v^a$,
\[
\delta R^a_{\ \ bcd}v^b =
(\nabla_c \delta\Gamma^a{}_{bd}
- \nabla_d \delta\Gamma^a{}_{bc})v^b,
\]
whence
\be
\delta R^a_{\ \ bcd} 
 = \nabla_c \delta\Gamma^a{}_{bd}- \nabla_d \delta\Gamma^a{}_{bc}.
\ee
The perturbation of $R_{ab}$ follows immediately:
\be
\delta R_{ab} = \delta R^c_{\ \ acb} 
= \nabla_c\delta\Gamma^c{}_{ab}-\nabla_b\delta\Gamma^c{}_{ca};
\label{eq:drab}\ee
and we have
\bea
\delta R &=& \delta (R_{ab}g^{ab})
	= - R_{ab}h^{ab}+ g^{ab}\delta R_{ab} 
\nonumber\\
&=& -R_{ab}h^{ab} + \nabla_a A^a,
\label{eq:dr1}\eea
where 
\begin{align}
A^a &= (g^{bc}\delta\Gamma^a{}_{bc}-g^{ac}\delta\Gamma^b{}_{bc})
	= (g^{ac}g^{bd} -g^{ab}g^{cd}) 
          \na_b h_{cd} \quad\mbox{ or }\nn\\
\color{blue} A^a 	&{\color{blue}=\na_b h^{ab} -\na^a h.}
\label{e:thetagrav}\end{align}

To summarize: The variation \eqref{drsqrtg2} has the form
\be \crv
\frac1{\sqrt{|g|}}\delta (R\sqrt{|g|}) = -G^{ab}h_{ab} +\nabla_a\na_b (h^{ab} -\na^a h) \color{black}.
\label{drsqrtg3}\ee

\vspace{2cm}
The remaining sections of this chapter are related to conserved quantities.
A generic spacetime has no Killing vectors and no natural definition of 
energy-momentum conservation.   
\index{conservation laws!no energy-momentum conservation in generic spacetime}
\index{conservation laws!asymptotic}
In an asymptotically flat spacetime, 
however, one can recover the Poincar\'e group as a symmetry 
group at spatial infinity. (See, for example, \href{https://journals.aps.org/pr/pdf/10.1103/PhysRev.122.997}{ADM}\cite{adm61},\href{https://aip.scitation.org/doi/pdf/10.1063/1.1666094}{Geroch}\cite{geroch72}, \href{https://aip.scitation.org/doi/pdf/10.1063/1.523863}{Ashtekar-Hansen}\cite{ah78},\href{https://aip.scitation.org/doi/pdf/10.1063/1.528208}{Chrusciel}\cite{chrusciel89},\href{https://aip.scitation.org/doi/10.1063/1.523698}{Sommers}\cite{sommers78}). 
At null infinity (in the limit as one moves away along outward null geodesics), 
one can again recover a unique group of asymptotic translations and a corresponding 
conserved energy-momentum of the spacetime, with the mass of the spacetime decreasing 
by the energy radiated in gravitational (or electromagnetic) waves between successive 
null hypersurfaces (\href{https://www.jstor.org/stable/2414436?origin=ads#metadata_info_tab_contents}{Bondi,van der Burg, Metzner}\cite{bms62}, \href{https://journals.aps.org/pr/pdf/10.1103/PhysRev.128.2851}{Sachs}\cite{sachs62},\href{https://www.jstor.org/stable/2415306?origin=ads#metadata_info_tab_contents}{Penrose}\cite{penrose65}).  
But the presence of radiation at null infinity leads to 
 weaker conditions on asymptotic fall-off than at spatial infinity because radiation 
lies within the future light cone of its source.  The weaker asymptotic conditions 
at null infinity lead to a larger group of asymptotic symmetries 
(the Bondi-Metzner-Sachs or BMS group), in which one cannot uniquely define angular 
momentum: The asymptotic symmetry group includes a family of Poincar\'e subgroups, 
with no preferred choice.  The conserved angular momentum in the 
following discussion, is then appropriate for spatial infinity or for the null infinity 
of a stationary spacetime, or for an arbitrary choice of an asymptotic Lorentz 
subgroup of the symmetry group at null infinity for a generic asymptotically flat 
spacetime. \label{p:aflat}

\section{First-order action}
\label{s:firstorderaction}
\index{action!first-order action} \index{Einstein field equation!from first-order action}\index{field equation!from first-order action}

The presentation in this and the next section is tied to work by Katz\cite{katz85}, 
Chrusciel\cite{chrusciel85}, and Sorkin\cite{sorkin88}; their papers follow earlier 
work by Hojman, Kuchar \& Teitleboim\cite{hkt76} that associates surface 
terms with conserved quantities.  See also Trautman \cite{trautman62,trautman67} for 
careful treatments of Noether's theorem.
    
If a function $I(\phi)$ on a space of fields, $\{\phi\}$, is an action for the field equation $E(\phi)=0$, 
then solutions to $E=0$ are extrema of $I$ for perturbations of the fields that vanish on an initial and 
final hypersurface.    
Perturbations $h_{\a\b}=\delta g_{\a\b}$ from one asymptotically flat solution to another do {\sl not} 
extremize the Hilbert action because the surface terms at spatial infinity are nonzero.  
This is related to the presence of second derivatives of the metric in $R$ and to 
the fact that the metric itself is nonzero at infinity.  One obtains a true action 
for asymptotically flat spacetimes by adding a surface term, $-\int_\infty B^a dS_a$, whose variation 
cancels the surface term $\int_\infty A^\a dS_\a$.  Equivalently, one adds to the Lagrange density the term $-\na_\a B^a\sqrt{-g}$.  

In Sect.~\ref{s:H_G} we added to the Lagrangian density a total divergence that removed the second time derivatives, obtaining a Lagrangian density,  
\eqref{tildel}, quadratic in first time derivatives of the metric.   Here we remove all second derivatives and obtain a first-order action, and action 
involving only the metric and its first derivatives.     
In the next section, Sect.~\eqref{s:surfaceK}, we show that the surface term is proportional to $\int K dS$, with 
$K$ the extrinsic curvature of the boundary of the region of integration, and this is the way 
in which York \cite{york72} and Gibbons \& Hawking\cite{gh77} write the first-order action.  

Denote by $\dflat_\a$ the covariant derivative operator of the flat metric 
$\eta_{\a\b}$ (introduced previously in \ref{ex:christoffel}\hspace{-1mm}{\color{white}.} and \ref{ex:nabla-partial}\hspace{-1mm}{\color{white}.}) and let $\xi^\a$ be a Killing vector of $\eta_{\a\b}$.  
Denote by $C^\a{}_{\b\c}$ the tensorial form of the Christoffel symbol 
$\Gamma^\lambda{}_{\mu\nu}$, 
\be
  C^\a{}_{\b\c} := \frac12 g^{\a\de} (\dflat_\b g_{\c\de} + \dflat_\c g_{\b\de}- \dflat_\de g_{\b\c} ).
\ee
We cancel the second derivatives in $R\sqrt{-g}$ by subtracting the divergence $\dflat_\a \cB^\a$, where 
\bsube\begin{align}
   \cB^\a &= B^\a\sqrt{-g}, \mbox{ with }
   \label{e:Ba0}\\
   \cblue B^\a &:= g^{\b\c}C^\a{}_{\b\c} - g^{\a\b}C^\c{}_{\c\b} 
\label{e:Bb}\\
	&\cblue = \left( g^{\a\c} g^{\b\de} - g^{\a\b} g^{\c\de} \right) \dflat_\b g_{\g\de}.
\label{e:Bc}\end{align}\label{e:B}\esube
From Eq.~\eqref{e:thetagrav},
\be
\delta B^\a = A^\a, \mbox{  at leading order in } r^{-1}.  
\ee
Then, in the variation of the action $\int {\cal L} d^4x$, with 
\be
{\cal L} := R\sqrt{-g} - \pa_a {\cal B}^\a, 
\label{e:lag1a}\ee
the surface terms at spatial infinity vanish. 

We now show that $\scrL$  is quadratic in  $C$,   
\be\crv
  I=\int \scrL d^4x, \qquad \scrL  :=  (C_{\a\b\c} C^{\b\a\c} - C^{\a\b}{}_\b C^\g{}_{\g\a})\sqrt{-g}.
\label{e:lag1b}\ee
\index{Lagrangian!gravitational, first-order}
This is the $\Gamma\Gamma$ part of $R\sqrt{-g}$, but with the opposite sign.

Here's the calculation:  
From Eq.~\eqref{e:rijklup}, the usual form of $R$ in terms of $\Gamma$ is  
\be
	R = g^{\g\de}(\dflat_\a C^\a{}_{\c\de} - \dflat_\de C^\a{}_{\c\a})
			+ C^{\a\b}{}_\b C^\c{}_{\c\a}  - C_{\a\b\c} C^{\b\a\c}\ .
\ee
Using Eqs.~\eqref{e:Ba0},\eqref{e:Bb}, we have
\[
\dflat_\a {\cal B}^\a =  \dflat_\a(g^{\b\c}\sqrt{-g}) C^\a{}_{\b\c}
		    - \dflat_\a(g^{\a\b}\sqrt{-g}) C^\c{}_{\c\b} 
			+ g^{\g\de}(\dflat_\a C^\a{}_{\c\de} -\dflat_\de C^\a{}_{\c\a})\sqrt{-g}.
			\]
We replace derivatives of the metric by $C$'s, using 
$0 = \na_\a g_{\b\g} = \dflat_\a g_{\b\g} - C^\d{}_{\b\a}g_{\d\g} - C^\d{}_{\a\g}g_{\b\d}$:  
\begin{align*}    
 \dflat_\a(g^{\b\c}\sqrt{-g}) 
  	&= (-g^{\b\sigma} g^{\c\tau} \dflat_\a g_{\sigma\tau}  
           + \frac12g^{\b\c} g^{\sigma\tau}\dflat_\a g_{\sigma\tau})\sqrt{-g} 
         = (- C^\b{}_\a{}^\c - C^\c{}_\a{}^\b + g^{\b\c}C^\de{}_{\a\de})\sqrt{-g}\\
 \dflat_\a(g^{\a\b}\sqrt{-g}) 
        &= -C^{\beta\a}{}_\a\sqrt{-g}.
\end{align*}
Then
\begin{align*}
 \dflat_\a(g^{\b\c}\sqrt{-g}) C^\a{}_{\b\c} 
	&= (-2C_{\a\b\c}C^{\b\a\c} + C^{\a\b}{}_\b C^\c{}_{\c\a})\sqrt{-g}\\
    -\dflat_\a(g^{\a\b}\sqrt{-g}) C^\c{}_{\c\b} & =  C^{\b\a}{}_\a C^\c{}_{\c\b}\sqrt{-g}\\
 \dflat_\a {\cal B}^\a 
	       &= \left[- 2C_{\a\b\c} C^{\b\a\c} + 2 C^{\a\b}{}_\b C^\c{}_{\c\a} 
			+ g^{\g\de}(\dflat_\a C^\a{}_{\c\de} -\dflat_\de C^\a{}_{\c\a})
		   \right] \sqrt{-g}\\ 
R\sqrt{-g} - \dflat_\a {\cal B}^\a 
	&=  \left( C_{\a\b\c} C^{\b\a\c} - C^{\a\b}{}_\b C^\c{}_{\c\a}\right)\sqrt{-g}
 \,.\ \ \Box 
\end{align*} 

\section{Noether}
\label{s:noether}\index{conservation laws!Noether's theorem|textbf}\index{Noether's theorem|textbf}\index{symmetry!relation to conserved current}
\index{symmetry!Noether's theorem|textbf}

Noether's theorem relates symmetries to conserved currents.  It can be stated and 
quickly derived as follows.  Let $\scrL$ be a Lagrangian density for a field or
set of fields $\Phi^I$.  (For us, $\Phi^I = g_{\a\b}$.)  Varying 
the action $I=\int_M\mathscr L d^4x$ gives the field equations $E_I=0$ satisfied by $\phi^I$; 
that means   
\be
   \delta  \scrL  = E_I\delta\Phi^I + \na_\a {\cal A}^\a(\Phi,\delta\Phi).  
\ee
Because ${\cal A}^\a$ is a density, its divergence is independent of derivative 
operator and could equally well have been written $\pa_\a {\cal A}^\a$

The symmetries of Noether's theorem are maps of the space $\{\Phi\}$ of fields on $M$, 
maps from $\{\Phi\}$ to $\{\Phi\}$, that  change $\scrL$ by a total divergence 
whether or not $\Phi$ satisfies the field equations.  
Consider a smooth family $\Psi_\lambda:\{\Phi\}\rightarrow\{\Phi\}$ of such maps. 
Denote by $\widehat \de$ the corresponding 
``infinitesimal symmetry." That is, $\widehat\de \Phi$ is the derivative of $\Phi$ 
along the family of maps:  
$\dis \widehat\de \Phi 
	= \left.\frac d{d\lambda}\Psi_\lambda[\Phi]\right|_{\lambda = 0}$.
Then $\widehat \de  \scrL$ is a total divergence, 
\be
  \widehat \de  \scrL  = \na_\a \Theta^\a.  
\ee

When the field equations are satisfied, taking as $\delta\Phi$ the symmetry $\widehat\delta\Phi$ 
gives a conserved current
\index{symmetry!relation to conserved current} 
\be\crv
	{\cal J}^a = {\cal A}^\a(\Phi,\widehat\delta\Phi) - \Theta^\a\cb:
\ee
\be
  0 =  \de \scrL  - \de \scrL  = \de \scrL  - \widehat\de \scrL  
	=\na_\a {\cal J}^\a, 
\ee
and the associated conserved quantity is the {\sl Noether charge}, the integral of ${\cal J}^\a$ 
over a spacelike hypersurface $\Sigma$,  
\be\crv
  Q = \int_\Sigma {\cal J}^\a d\sigma_\a\cb, 
\ee
where $d\sigma_\a = \eta_{\a\b\c\d}dS^{\b\c\d}$, with $\eta_{\a\b\c\d}$ the totally antisymmetric tensor density of weight $-1$ with $\eta_{0123} = 1$ (for any basis).%
\footnote{
Like $\delta_\a^\b$ the density $\eta_{\a\b\c\d}$ is {\sl natural}, invariant under diffeos, 
and hence with vanishing Lie derivatives: $\Lie_{\bm\xi} \eta_{\a\b\c\d} = 0$, for any $\xi$.  
It does not need a metric for its definition, but once $g_{\a\b}$ is defined, 
\[
    \eta_{\a\b\c\d} = \epsilon_{\a\b\c\d}/\sqrt{-g}. 
\]
} 
For $\Sigma$ a $t=$ constant surface, $d\sigma_\a = \na_\a t\ d^3x$.\\

\noindent
{\sl Example}: A massive scalar field on a 
spacetime with metric $g_{\a\b}$ has Lagrange density 
\[
   \scrL  = - \frac12\left( \na^\a\Phi \na_\a \Phi + m^2 \Phi^2\right)\sqrt{-g}.
\] 
A Killing vector $\xi^\a$ of the metric $g_{\a\b}$ generates a family of diffeos for which 
$\delta_\xi\Phi = \Lie_{\bm\xi}\Phi$ and 
\[
   \delta \scrL  =  \Lie_{\bm\xi}  \scrL  = \na_\a ( \scrL \xi^\a).  
\]
The diffeos generated by $\xi^\a$ are thus symmetries, and $\Theta^\a =  \scrL \xi^\a$.\\
  We find ${\cal A}^\a$ by varying $\scrL$ for a general $\delta\Phi$, writing 
\[
  \delta\scrL = -(\na^\a\Phi \na_\a\delta\Phi +m^2\Phi\de\Phi) \sqrt{-g} 
		= (\na_\a\na^\a -m^2)\Phi\,\de\Phi \sqrt{-g} - \na_\a(\na^\a\Phi\de\Phi\sqrt{-g}).
\]
Then ${\cal A}^\a = - \na^\a\Phi\de\Phi\sqrt{-g}$.  For $\delta\Phi = \Lie_{\bm\xi}\Phi$, we have
\begin{align*}
  {\cal A}^\a &= - \na^\a\Phi \Lie_{\bm\xi}\Phi\sqrt{-g}, \qquad \Theta^\a = \xi^\a\cal L \\
  {\cal J}^\a &= - \na^\a\Phi \Lie_{\bm\xi}\Phi\sqrt{-g}-\Theta^\a 
	 =  \left[- \na^\a\Phi \Lie_{\bm\xi}\Phi 
		+\frac12 \xi^\a \left(\na^\b\Phi \na_\b \Phi + m^2 \Phi^2\right)\right]\sqrt{-g}\,.
\end{align*}
The corresponding Noether charge is then 
\be
   Q = \int {\cal J}^t d^3x 
     = \int \left[- \na^t\Phi \Lie_{\bm\xi}\Phi 
	  +\frac12 \xi^t \left(\na^\b\Phi \na_\b \Phi + m^2 \Phi^2\right)\right]\sqrt{-g}\,d^3x.
\ee
If the $t=$ constant hypersurface is invariant under rotations or spatial translations, 
|$\xi^\a\na_\a t = \xi^t=0$, and 
\be
  Q =  -\int \pa^t\Phi \Lie_{\bm\xi} \Phi \sqrt{-g}\,d^3x = \int\pi \Lie_{\bm\xi}\Phi \, d^3x, 
\label{e:Qscalar}\ee 
with $\pi = -\pa^t\Phi\sqrt{-g}$ the momentum density conjugate to $\Phi$, equal to $\pa_t\Phi\sqrt{-g}$ in 
Minkowski space. 
\index{conservation laws!relation to Killing vector}   
Thus if the metric is invariant under translations along $\bm\pa_x$ or is invariant under rotations 
generated by $\bm\pa_\phi$, 
the scalar field's linear momentum $P_x$ or angular momentum $J$ will be conserved, and they are given by 
\be
  P_x =  \int\pi \pa_x\Phi \, d^3x, \qquad J =  \int\pi \pa_\phi\Phi\, d^3x.
\ee
For a time-independent metric, the field's energy is conserved.  Because $\xi^\a = t^\a$ and $\xi^t = t^\a\na_\a t = 1$, we get 
$\pi\Lie_{\bm\xi}\Phi -\mathscr L=\pi\dot\Phi-\mathscr L$ instead of $\pi\dot\Phi$ in the integrand of Eq.~\eqref{e:Qscalar}:  
\be
   E = \int (\pi\dot\Phi -\mathscr L)d^3x.  
\ee
In particular, in Minkowski space, 
\be
  E = \int \frac12\left[ \dot\Phi^2 +\bm\na\Phi\cdot\bm\na\Phi +m^2\Phi^2\right]\sqrt{-g}\,d^3x.
\ee
\index{energy!scalar field}
\vspace{5mm}

\noindent{\sl Gravity}\\
 We now turn to the conserved quantities associated with the first-order gravitational Lagrangian 
\eqref{e:lag1a}, \eqref{e:lag1b},  
\be\cblue
    \scrL  = R\sqrt{|g|} + \na_\a {\cal B^\a}\cb.  
\label{e:lagc}\ee
We will find that, for a Killing vector of the flat metric $\eta_{\a\b}$, the associated conserved 
quantity can be written as a surface integral at infinity%
\footnote{Do not be confused by the cancellation of surface integrals at infinity 
from $\cal A$ and $\delta\cal B$ in varying the action and the lack of cancellation in the 
contributions to $Q$ from $\cal A$ and $\delta\cal B$.  The surface integral at infinity from varying the action involves $\int_\infty \cal A$, 
whereas the contribution from $\cal A$ to $Q$ comes from 
$\int_\Sigma {\cal A}$.  We will see that ${\cal A}(\Lie_{\bm\xi} g)$ is a divergence 
$\na\cdot\na\xi$, allowing us to write $\int_\Sigma {\cal A} = \int_\infty \na\xi$. Similarly, 
the contribution from $\cal B$ to $\delta I$ involves $\int_\infty\delta\cal B$, whereas 
the contribution to $Q$ from $\delta{\cal B} = \Lie_{\bm\xi}\cal B$ involves 
$\int_\Sigma \de{\cal B} = \int_\infty {\cal B}\xi $.   
}
of the form
\be
   Q = 2\int_\infty \left( \na^{[\a} \xi^{\b]} + B^{[\a}\xi^{\b]}\right) dS_{\a\b}, 
\ee
or, with $\scrL$ replaced by the correctly normalized Lagrangian density 
$\dis\frac1{16\pi} \scrL$, 
\be\crv
   Q = \frac1{8\pi}\int_\infty \left( \na^{[\a} \xi^{\b]} + B^{[\a}\xi^{\b]}\right) dS_{\a\b}\cb. 
\ee
  
Now $\scrL$ is constructed from $g_{\a\b}$ and the derivative operator $\dflat_\alpha$ of the flat metric $\eta_{\a\b}$, so dragging both $g_{\a\b}$ and $\eta_{\a\b}$ by a diffeo just drags 
$\scrL$ by the diffeo.  For a Killing vector $\xi^\a$ of $\eta_{\a\b}$, dragging  
$g_{\a\b}$ alone by the diffeo associated with $\xi^\a$ is equivalent to dragging both 
$g_{\a\b}$ and $\eta_{\a\b}$ by the diffeo, because the diffeo doesn't change $\eta_{\a\b}$.  
Then the change in $\scrL$ from a symmetry perturbation $h_{\a\b} = \Lie_{\bm\xi} g_{\a\b}$ is 
\be
  \widehat\de \scrL = \Lie_{\bm\xi}\scrL = \na_\a(\xi^\a\scrL),   
\ee
so
\be
  \Theta^\a = \xi^\a\scrL.
\label{e:Theta}\ee

Call ${\cal A}^\a$ the divergence associated with $\delta\scrL$ for an arbitrary $h_{\a\b}$,  
\[
  \de\scrL = -G^{\a\b} h_{\a\b} + \na_\a {\cal A}^\a.   
\]
Using the form \eqref{e:lagc} of $\scrL$ and Eq.\eqref{e:thetagrav},  we have 
\begin{align*}
  \de\scrL &= \de(R\sqrt{-g}) - \na_\a \de{\cal B}^\a 
	    =[- G^{\a\b}h_{\a\b} + \na_\a\na_\b (h^{\a\b}-\na^\a h)]\sqrt{-g}  
		+ \na_\a \de{\cal B}^\a,\\
	{\cal A}^\a &= (\na_\b h^{\a\b}-\na^\a h)\sqrt{-g}  +  \de{\cal B}^\a\,.
\end{align*}

To find the conserved current, we need to evaluate ${\cal A}^\a$ for 
$h_{\a\b} = \Lie_{\bm\xi} g_{\a\b}$. Its first term is 
\be
  [\,\na_\b(\na^\a\xi^\b +\na^\b\xi^\a) -2 \na^\a\na_\b \xi^\b\,]\sqrt{-g} 
	 = [\, \na_\b(\na^\b\xi^\a - \na^\a\xi^\b ) + 2 R^\a{}_\b \xi^\b\,]\sqrt{-g}.
\ee
Second term:  
From its definition \eqref{e:B},  ${\cal B}^\a$ is constructed from $\eta_{\a\b}$ and $g_{\a\b}$, so $\delta {\cal B}^\a = \Lie_{\bm\xi} {\cal B}^\a$. 
The Lie derivative of a vector density can be written as 
\[ 
  \Lie_{\bm\xi}{\cal B}^\a 
	= \na_\b({\cal B}^\a\xi^\b - {\cal B}^\b\xi^\a) + \xi^\a \na_\b {\cal B}^\b, 
\]
and replacing $\na_\b{\cal B}^\b$ in the last term by $\scrL -R\sqrt{-g}$ gives
\be
  \delta {\cal B}^\a = \na_\b({\cal B}^\a\xi^\b - {\cal B}^\b\xi^\a) 
				+ \xi^\a  (\scrL -R\sqrt{-g})   \,.
\ee
Then 
\be
 {\cal A}^\a = \na_\b[(\na^\b\xi^\a - \na^\a\xi^\b )\sqrt{-g} 
		+ {\cal B}^\a\xi^\b - {\cal B}^\b\xi^\a] 
				+ (\xi^\a \scrL +2\xi^\b G^\a{}_\b)\sqrt{-g}).
\label{e:A}\ee
With ${\cal B}^\a$ written as $B^\a\sqrt{-g}$, Eqs.~\eqref{e:Theta} and \eqref{e:A} give 
for the conserved current \\ ${\cal J}^\a=\Theta^\a - {\cal A}^\a$ the form 
\be
{\cal J}^\a = \na_\b(\na^\b\xi^\a - \na^\a\xi^\b 
		+ B^\a\xi^\b - B^\b\xi^\a)\sqrt{-g} 
				 +2\xi^\b G^\a{}_\b\sqrt{-g}).
\ee 
When $G_{\a\b} = 0$, the corresponding conserved quantity is given by  
\begin{align}
  Q &= \int_\Sigma \na_\b\left(\na^\b\xi^\a - \na^\a\xi^\b 
		+ B^\a\xi^\b - B^\b\xi^\a\right) dS_\a \nonumber\\
	&=\int_\infty \left(\na^\a\xi^\b - \na^\b\xi^\a 
		+ B^\a\xi^\b - B^\b\xi^\a\right) dS_{\a\b}, 
\end{align} 
or, with the normalized action, 
\be
 Q = \frac1{8\pi} \int_\infty \left(\na^\a\xi^\b+ B^{[\a}\xi^{\b]}\right) dS_{\a\b}, 
\ee
as claimed, with $B^\a$ given by Eq.~\eqref{e:Bc}. 
The first term, involving $\na^{\a}\xi^{\b}$ gives the Komar expression for 
a conserved quantity associated with $\xi^\a$. It is the only contribution 
when $\xi^\a$ is a rotational Killing vector, giving for $\bm\xi = \bm\phi$, 
\index{angular momentum!asymptotic}
\be
   Q[\bm\phi] = J.  
\ee
\index{mass!Komar mass}\index{Komar mass!relation to Noether mass}
For $\xi^\a = t^\a$ (i.e., an asymptotically timelike Killing vector with asymptotic norm 1), 
\be 
   Q[\bm t] = -M.
\ee 
\index{energy!mass of asymptotically flat spacetime}\index{energy!Noether's theorem}\index{gravitational mass!as Noether charge}\index{mass!gravitational mass}
The term involving $B^\a$ contributes to the mass an amount equal to the mass from the Komar 
expression.  This is why, when one writes the mass in terms of the Komar 
expression alone, one needs a coefficient $1/4\pi$, twice the factor of $1/8\pi$ in the 
expression for linear and angular momentum.  (The sign difference, $-1/4\pi$ instead of $1/4\pi$, 
is the usual Lorentz signature difference in sign in $\delta I = -E\delta t + p\delta x = p_\mu \de x^\mu$.)\\

\section{ Surface term and extrinsic curvature \texorpdfstring{$K$}. }
\label{s:surfaceK}
We now write the first-order action \eqref{e:lag1b} as the sum of the Hilbert action and a surface integral of the boundary's extrinsic curvature.    

The surface term that arises from varying the Hilbert action 
involves derivatives of the metric on the boundary $\pa \Omega$ of a
4-dimensional region $\Omega$ of integration.  As \href{https://journals.aps.org/prl/pdf/10.1103/PhysRevLett.28.1082}{York}\cite{york72} and later 
\href{https://journals.aps.org/prd/pdf/10.1103/PhysRevD.15.2752}{Gibbons \& Hawking}\cite{gh77} 
pointed out, for variations of the metric with 
\be 
   h_{\a\b}=0 \ \mbox{on }\ \pa \Omega,
\ee 
the surface term $\int_{\pa \Omega} A^\a dS_\a$ is proportional to the integral of $\d K$.  
The reason is simple: Because $h_{\a\b}$ vanishes on the boundary, so do its 
tangential derivatives.  All that remains is the normal derivative, and the 
scalar built from the normal derivative of the metric is $K$.  

More formally, on $\pa \Omega$, $h_{\a\b}=0$ implies 
\bsube
\begin{align}
\delta n_\a &= 0,\quad \delta n^\a = 0, \quad \delta\gamma_{\a\b}=0,\quad 
		\delta\gamma^\a_\b = 0,
 \\
  \g^{\a\b}\na_\b h_{\g\de}|_{\pa\Omega}& =0,\quad \g^{\a\b}\na_\b \de n_\c|_{\pa\Omega} = 0\ \mbox{ (tangential derivatives vanish). }
\label{e:tang}\end{align}
\label{e:zero}\esube
For a spacelike part of $\pa \Omega$, from Eq.~\eqref{e:Kab2}, the trace of 
the extrinsic curvature is  
\[
      K = - \g^\a_\b\na_\a n^\b. 
\]
Using Eqs.~\eqref{e:zero}, we have
\begin{align*}
   \d K|_{\Omega} &= - \g^{\a\b}(\d\na_\a) n_\b = \g^{\a\b} \d \Gamma_{\g\a\b}n^\g \\
  	&= \g^{\a\b} \frac12(\na_\a h_{\b\c} + \na_\b h_{\a\c} -\na_\c h_{\a\b})n^\g\\
  	&= -\frac12 \g^{\a\b} n^\c\na_\c h_{\a\b}, \quad \mbox{by Eq.\eqref{e:tang}.}   
\end{align*}

The integrand of the surface term has the form
\begin{align*}
     A^\a n_\a &= n^\a g^{\b\c}(\na_\b h_{\a\c} - \na_\a h_{\b\c}) 
		= n^\a (\g^{\b\c}-n^\b n^\c) (\na_\b h_{\a\c} - \na_\a h_{\b\c})\\
		&= n^\a \g^{\b\c} (\na_\b h_{\a\c} - \na_\a h_{\b\c})\ \ \mbox{ (by $\a$-$\b$ antisymmetry})\\
		&= -n^\a \g^{\b\c}\na_\a h_{\b\c}   \ \ \mbox{(tangential derivatives vanish})\\
		&= 2\delta K.
\end{align*}

For a timelike part of $\pa \Omega$, with the sign convention,
\be
  K = \g^{\a\b}\frac12\Lie_{\bm n} \g_{\a\b} \quad \pa M \mbox{ timelike},
\ee
we have
\be
  A^\a n_\a = - 2\de K\quad \pa M \mbox{ timelike}.
\ee
Then 
\begin{align*} \cblue
   \de\int R\sqrt{|g|} d^4x 
	&= - \int G^{\a\b}h_{\a\b} \sqrt{|g|} d^4x +\int_{\pa M} A^\a n_\a dS \\
	&\cblue= - \int G^{\a\b}h_{\a\b} \sqrt{|g|} d^4x +2\int_{\pa M} e\,\de K dS,
\end{align*} 
where 
\[
  e = 	\begin{cases}
	 \  \ 1, & \pa \Omega\ \text{spacelike}\\
	     -1 & \pa \Omega\ \text{timelike}
	\end{cases}.    
\]
Thus, by adding to the Hilbert action a surface term involving the 
trace of the extrinsic curvature,
\be\crv
   \hat I := \frac1{16\pi} \left[\int_\Omega R\sqrt{|g|} d^4x - 2\int_{\pa \Omega}e\, K dS\right]\cb, 
\ee 
we obtain an action that enforces the field equations for metric perturbations 
vanishing on $\partial\Omega$: 
\be\cblue
  \de \hat I = - \frac1{16\pi}\int_\Omega G^{\a\b}h_{\a\b} \sqrt{|g|} d^4x\cb.
\ee
\index{action!gravitational|)}

\bibliography{grg_notes}

\appendix
\setcounter{equation}{0}
\setcounter{chapter}{0}

\chapter{Integration, Forms and Densities}
\label{appendix}
\index{integration!on manifolds|(} 

\section{Forms and densities}
\label{s:forms}
\index{exterior derivative|textbf}\index{forms, differential|textbf}
\index{differential forms|textbf}

The tensor $\epsilon_{a\cdots b}$ and the quantity $\sqrt{|g|}$ that appear in the 
alternative ways of writing an integral are, respectively, an example of a form and 
a scalar density. As we will see, there is a duality between forms and densities that 
underlies a equivalence between Stokes's theorem and Gauss's theorem.  We define forms 
and densities, present the duality that relates them, and go on to the corresponding 
duality relating the integral theorems.
\index{duality!between forms and densities}

\vskip0.4cm
\noindent
{\em Forms.}\\
\noindent{\bf Definition}. A {\em p-form} $\sigma_{a\cdots b}$ is an antisymmetric, covariant 
tensor  with $p$ indices. \\
In particular, a scalar $f$ is 
a 0-form, a covariant vector $A_a$ is a 1-form, and an antisymmetric 2-index tensor 
$F_{ab}$ is a 2-form.  

\noindent{\bf Definition}. The exterior derivative $d\sigma$ of a $p$-form $\sigma$
is the $p+1$ form 
\be 
 (d\sigma)_{ab\cdots c} = (p+1)\nabla_{[a}\sigma_{b\dots c]}.
\label{e:exteriord}\ee
\index{exterior derivative|textbf}
The factor $p+1$ is the number of independent ways of distributing the $p+1$ indices 
between $\nabla$ and $\sigma$.  The antisymmetry implies that $d\sigma$ is independent 
of the derivative operator; in any chart it has components
\be
(d\sigma)_{ij\cdots k} = (p+1)\partial_{[i}\sigma_{j\dots k]}.
\ee 
Antisymmetry and the commutativity of partial derivatives imply for any form $\sigma$
\be
    d^2\sigma = 0.
\label{d2}\ee

Lie derivatives and exterior derivatives commute, 
\be 
	\Lie_{\bf v} d\sigma = d\Lie_{\bf v}\sigma,
\label{lied}\ee
and the two derivatives satisfy the Cartan identity,
\be
	\Lie_{\bf v}\sigma = v\cdot(d\sigma) + d(v\cdot \sigma), 
\mbox{where } (v\cdot\sigma)_{a\cdots b} := v^c\sigma_{ca\cdots b}.
\label{liesigma}\ee
Each relation can be proved by induction on $p$ (the number of indices 
of $\sigma$).  For an $n$-form in $n$ dimensions, the second relation 
can be written as
\be
	\Lie_{\bf v}\sigma_{a\cdots b} = \nabla_c(\sigma_{a\cdots b}v^c),
\ee
with special case
\be
   \Lie_{\bf v}\epsilon_{a\cdots b} = \nabla_c(\epsilon_{a\cdots b}v^c)=\epsilon_{a\cdots b}\nabla_c v^c.
\ee

In $n$-dimensions, any nonzero $n$-form is functionally proportional to any other, because 
each has only one independent component.  In particular, because any $n$-form
$\sigma$ is given by $\sigma_{a\cdots b} = f\epsilon_{a\cdots b}$ for some scalar 
$f$, the integral of an $n$-form is well-defined, given in any coordinate system by 
\[
\int_\Omega \sigma_{a\cdots b}dS^{a\cdots b} = \int \sigma_{1\cdots n} dx^1\cdots dx^n.
\]
(Again the integral is well defined because a change of coordinates multiplies 
the value of a $p$-form by the Jacobian of the transformation.)\\ 

\noindent{\em Densities.}
\index{density! scalar and tensor|textbf}
\index{scalar density|textbf}\index{vector density|textbf}\index{tensor density|textbf}

A {\em scalar density} $\mathfrak f$ (of weight 1), by definition, transforms under 
a change of coordinates in the same way as one component of an $n$-form in 
$n$ dimensions:  
${\mathfrak f}\rightarrow  \left|\frac{\partial x}{\partial x'}\right|\mathfrak f$.  
Just as one can write any $n$-form as a multiple of $\epsilon_{a\cdots b}$ 
(once one is given a metric), one can write any scalar density as a scalar multiple 
of $\sqrt{|g|}$: 
\be 
	{\mathfrak f} = f\sqrt{|g|}, 
\ee
for some scalar $f$ (namely ${\mathfrak f}/\sqrt{|g|}$).  
One can analogously introduce vector and tensor densities by transformation laws that 
differ from those of vectors and tensors by the Jacobian of the transformation: 
The change of the components of a vector density under a coordinate transformation
is given by  
$
 {\mathfrak j}^i \rightarrow \left|\frac{\partial x}{\partial x'}\right|\,\frac{\partial x'^{i}}{\partial x^k}\,{\mathfrak j}^k.
$  
Again one can write any vector density in the form ${\mathfrak j}^a = j^a \sqrt{|g|}$, with 
$j^a$ a vector field.  

The Lie derivative of a scalar or tensor density can be deduced from this fact 
as follows 
(a metric-free derivation uses the geometric definition of Lie derivative and 
the fact that the action of a diffeo on a density differs from its action on 
a tensor by the inverse Jacobian of the diffeo):
We have 
\[
\Lie_{\bf v} \sqrt{|g|} = \frac{\partial \sqrt{|g|}}{\partial g_{ab}}\Lie_{\bf v} g_{ab}
 = \frac12 \frac1{\sqrt{|g|}}\frac{\partial g}{\partial g_{ab}}\Lie_{\bf v} g_{ab}.
\]
Now the coefficient of the component $g_{ij}$ in the determinant $g$ is the 
cofactor $\Delta^{ij}$ of $g_{ij}$, and the inverse metric is given by 
$\dis g^{ij}=\frac{\Delta^{ij}}g$. Thus 
$\dis \frac{\partial g}{\partial g_{ij}} = \Delta^{ij} = gg^{ij}$, and we have 
\be 
      \Lie_{\bf v} \sqrt{|g|} = \frac12\sqrt{|g|}g^{ab}\Lie_{\bf v} g_{ab} = \sqrt{|g|}\nabla_a v^a
	= \nabla_a(\sqrt{|g|}v^a).
\label{lieg}\ee 
Finally, Eqs.~(\ref{lief}) and (\ref{lieg}) imply that the Lie derivative of a 
general scalar density $\mathfrak f$ is 
\be
   \Lie_{\bf v} {\mathfrak f} = \nabla_a({\mathfrak f} v^a). 
\label{lieff}\ee 

\vskip0.4cm
\noindent{\em Duality.}
A duality between $p$-forms $\sigma_{a_1\cdots a_p}$ and antisymmetric tensor densities 
${\cal A}^{a_1\cdots a_q}$, with $q=n-p$ indices ($n$ the dimension of the space) is 
given by
\be 
   {\cal A}^{a_1\cdots a_q} 
	= \frac1{p!} \sqrt{|g|}\epsilon^{a_1\cdots a_q b_1\cdots b_p}\sigma_{b_1\cdots b_p}.
\ee 
The inverse relation is 
$\dis\sigma_{a_1\cdots a_p}
=\pm \frac1{q!}|g|^{-1/2} \epsilon_{a_1\cdots a_p b_1\cdots b_q}{\cal A}^{b_1\cdots b_q}$, 
where the sign is positive for a positive definite metric, negative for a metric with 
Lorentz signature.

The divergence $\nabla\cdot {\cal A}$, of an antisymmetric density, 
\be
 (\nabla\cdot {\cal A})^{b\cdots c} := \nabla_a {\cal A}^{ab\cdots c}, 
\ee
is dual 
to the exterior derivative $d\sigma$ of the form $\sigma$ dual to $\cal A$, and it is 
independent of the choice of $\nabla$, having components
\be
   \nabla_k {\cal A}^{ki\cdots j} = \partial_k {\cal A}^{ki\cdots j}.
\ee   
In particular, the equation for the divergence of a vector density ${\cal A}^a$,
\be
	\nabla_a{\cal A}^a = \partial_i {\cal A}^i,
\label{eq:divdensity}\ee
is equivalent (once one has a metric) to the familiar relation 
$\nabla_a A^a = \frac1{\sqrt{|g|}}\partial_i(\sqrt{|g|}A^i)$.

The relations (\ref{d2}), (\ref{lied}), and (\ref{liesigma}) are dual to the relations
\bea
\nabla\cdot(\nabla\cdot {\cal A}) &=& 0, \\
\Lie_{\bf v} \nabla\cdot {\cal A} &=& \nabla\cdot\Lie_{\bf v} {\cal A},
\label{liediv}\\
\Lie_{\bf v} {\cal A} &=& \nabla\cdot (v\wedge {\cal A}) + v\wedge \nabla\cdot{\cal A}, 
\eea
where $(v\wedge{\cal A})^{ab\cdots c} := (q+1) v^{[a}{\cal A}^{b\cdots c]}$. 

On an $n$-dimensional manifold, integrals are naturally defined for $n$-forms or for 
scalar densities, which can be said to be {\em dual} to their corresponding $n$-forms.
That is, the integrals 
\be
  \int \omega_{a\cdots b} dS^{a\cdots b} = \int \omega_{1\cdots n} dx^1\cdots dx^n 
\quad \mbox{ and } \quad
  \int {\mathfrak f} dx^1\cdots dx^n
\ee   
are well defined because under a change of coordinates, each integrand is multiplied by 
the Jacobian of the transformation.  The mathematical literature adopts an index-free 
notation in which the integral of an $n$-form $\omega$ over an $n$-dimensional region $\Omega$ 
is written $\dis \int_\Omega \omega$.

\section{Stokes's theorem and Gauss's theorem for antisymmetric tensors}
\label{sec:gauss-stokes}

The simplest version of Stokes's theorem is its 2-dimensional form, 
namely Green's theorem:
\[
\int_S (\partial_x A_y - \partial_y A_x)dx\,dy = \int_c (A_x dx + A_y dy), 
\] 
where $c$ is a curve bounding the 2-surface $S$.  The theorem involves the integral over a 
2-surface of the antisymmetric tensor $\nabla_a A_b - \nabla_b A_a$.  In three dimensions, 
the tensor is dual to the curl of $\bf A$: $(\dis \nabla\times {\bf A})^a = \epsilon^{abc}\nabla_b A_c$; 
and Stokes's generalization of Green's theorem can be written in either the form 
\[ 
\int_S (\bm{ \nabla\times A})\cdot d{\bf S} = \int_c {\bf A\cdot dl}
\] 
or in terms of the antisymmetric tensor $\nabla_a A_b - \nabla_b A_a$
\be 
  \int_S (\nabla_a A_b - \nabla_b A_a) dS^{ab} = \int_c A_a dl^a,
\ee
where, for an antisymmetric tensor $F_{ab}$,  $\dis F_{ab} dS^{ab}$ means 
$F_{12} dx^1 dx^2 + F_{23} dx^2 dx^3 + F_{31}dx^3 dx^1$.  Written in this form, the
theorem is already correct in a curved spacetime. The reason is that the antisymmetric 
derivative $\nabla_a A_b - \nabla_b A_a$ has in curved space the same form it has in flat 
space: As we have seen, its components in any coordinate system are just $\partial_i A_j -\partial_j A_i$. 

As a result, the flat-space proof of Green's theorem and Stokes's theorem, based on 
the Fundamental Theorem of calculus ($\int_a^b f'(x) dx = f(b)-f(a)$), 
holds in curved space as well: Let $S$ be a coordinate square in a surface of constant 
$x^3$.  Then \\
\[
\int_S (\partial_1 A_2 -\partial_2 A_1) dx^1 dx^2 = \int_c (A_1 dx^1 + A_2 dx^2),
\] 
with the boundary of the square traversed counterclockwise as seen from above the square.  

Gauss's theorem, 
\be
\int_V \nabla_a A^a d^3x = \int_S A^a dS_a, 
\ee
with $S$ a surface bounding the volume $V$ again has a simple generalization to curved space.  Although the divergence $\nabla_a A^a$ does not have components 
$\partial_i A^i$, the divergence of 
${\cal A}^a:=A^a \sqrt{g}$ does: $\nabla_a {\cal A}^a = \partial_i {\cal A}^i$.
As we saw in Sect.~\ref{s:gauss0}, the proof of Gauss's theorem follows, as in flat space, from an integration over coordinate cubes using the fundamental theorem of calculus for the integral over each coordinate:
With $d^nV = \sqrt{|g|} d^n x$, and $\Omega$ an $n$-dimensional region 
with boundary $\pa\Omega$, 
\be
   \int_\Omega \na_a A^a d^n V = \int_{\pa\Omega} A^a dS_a.  \tag{\ref{e:gauss0}}
\ee

\subsection{Integrals on a submanifold}

A metric $g_{\alpha\beta}$ on spacetime induces a 3-metric $^3g_{ab}$ on a 3-dimensional hypersurface $V$, 
its pullback to $V$.   
Given the 3-metric $g_{ab}$, the 3-form $\epsilon_{abc}$ on $V$ is uniquely defined 
up to sign.  When one chooses coordinates $\{x^0,x^1, x^2,x^3\}$ on a spacetime, 
one ordinarily chooses the tensor $\epsilon_{\alpha\beta\gamma\delta}$ for which the coordinate system has positive orientation, for which $\epsilon_{0123} = \sqrt{|g|}$, 
not $-\sqrt{|g|}$, and one calls the direction of increasing $x^0$ the future.
When $\epsilon_{0123}>0$ and when $x^0$ is constant on $V$ and increasing to the 
future, we will say the corresponding 
positive orientation $V$ is the one for which $\epsilon_{123}=\sqrt{^3g}$.

In discussing hypersurfaces in spacetime it is convenient to avoid the distinction
between tensors on $V$ and tensors on $M$.   
In the same way that a vector $v^a$ on $V$ is uniquely identified with a vector 
$v^\alpha$ on $M$ for which $v^\alpha n_\alpha = 0$, we can identify the 3-metric 
$^3g_{ab}$ and 3-form $\epsilon_{abc}$ with tensors
$^3g_{\alpha\beta}$ and $\epsilon_{\alpha\beta\gamma}$ on $M$ that are orthogonal
to $n_\alpha$ and whose spatial indices coincide with those of $^3g_{ab}$ and 
$\epsilon_{abc}$.  
\be
	^3g_{\alpha\beta} = g_{\alpha\beta} + n_\alpha n_\beta, 
\qquad
	\epsilon_{\alpha\beta\gamma} = n^\delta\epsilon_{\delta\alpha\beta\gamma},
\label{gep}\ee
where $n^\alpha$ is the future pointing unit normal to $V$, namely  
\be
 n_\mu  = \frac{-\delta^0_\mu}{\left( -g^{00}\right)^{1/2}} .
\ee
We then have $n^0= \sqrt{-g^{00}}>0 $, and as we show next, the spatial components of $\epsilon_{\alpha\beta\gamma}$ are $\epsilon_{ijk} = \sqrt{^3g}$.  

It follows immediately from Eq.~(\ref{gep}) that $^3g_{\alpha\beta}$ satisfies 
the defining conditions $^3g_{\alpha\beta}n^\beta = 0$ and 
$^3g_{ab}u^a v^b = {}^3g_{\alpha\beta}u^\alpha v^\beta = g_{\alpha\beta}u^\alpha v^\beta$, 
for $u^\alpha$ and $v^\alpha$ vectors on $V$ (equivalently $^3g_{\alpha\beta}$ 
has spatial components $g_{ij}$). 

The spatial components of $\epsilon_{\alpha\beta\gamma}$ are 
\[ 
 \epsilon_{\mu ijk} n^\mu   
 = \left(-g^{00}\right)^{1/2} \epsilon_{0 ijk},
\]
or $\dis \epsilon_{123} = {\sqrt{gg^{00}}}$, implying
\be 
 \epsilon_{123} = \sqrt{^3g}.
\ee
{\bf Warning}:  Although contravariant tensors $T^{\alpha\cdots\beta}$ are 
tensors on $V$ if and only if their $0$th components all vanish, the 0th 
components of covariant tensors $T_{\alpha\cdots\beta}$ on $V$ will not generally 
vanish, because in general $g_{0i}\neq 0$.  The information about the 
3-dimensional objects is carried by the spatial components, and the spatial 
components of the corresponding 3-dimensional and 4-dimensional tensors coincide.\\  

One writes
\be \int dV = \int_V \epsilon_{\alpha\beta\gamma} dS^{\alpha\beta\gamma},  \ee
for the volume of $V$, where one thinks of $dS^{\alpha\beta\gamma}$ as antisymmetric, 
with nonzero components 
$\dis \pm\frac1{3!} dx^idx^jdx^k$, with the sign depending on whether $i,j,k$ is 
an even or odd permutation of $1,2,3$. 
In coordinates $\{x^0,x^1, x^2,x^3\}$ with $V$ an $x^0 =$ constant surface,
\[ 
 \int dV = \int {\sqrt{^3g}}\, d^3x = \int {\sqrt{gg^{00}}}\, dx^1dx^2dx^3.
\]
For any 3-form, 
\[
 \int \omega_{\alpha\beta\gamma}  dS^{\alpha\beta\gamma}  
 	= \int\omega_{123}dx^1 dx^3 dx^3.
\]

	If $S$ is a 2-dimensional surface with unit normals $n^\alpha $ and
$m^\alpha $, and we pick the order of the two normals correctly, its area is 
given in terms of $\epsilon_{\alpha\beta} =
\epsilon_{\alpha\beta\gamma\delta}  n^\alpha  m^\beta $ by 
\be
 \int_S dS
= \int_S \epsilon_{\alpha\beta}dS^{\alpha\beta} .
\ee
In a chart $x^0, x^1, x^2, x^3$, with $S$ a surface of constant $x^0$ and $x^1$,
we have
\[
\epsilon_{23} = {\sqrt{^2g}} 
= \left[ g_{22}g_{33}-(g_{23})^2\right]^{1/2},
\]
and $\int dS = \int {\sqrt{^2g}} dx^2dx^3$.  \\

\index{flux!magnetic}
\noindent {\bf Example}.  The magnetic flux through a surface $S$
of radius $R$ in $t, r, \theta , \phi$ coordinates is
\[ 
 \int_S F_{ij} dS^{ij} 
		= \int_S F_{\theta\phi}d\theta d\phi .
\]

Finally, the integral of a scalar along a curve $c$ is 
$f = \int_c fdl$, with $dl$ proper length along $l$; and the corresponding 
line integral of a 1-form along $c$ is 
\[
 \int_c  A_\alpha  dl^\alpha  
 	= \int_c A_\alpha t^\alpha dl,
\] 
where $t^\alpha $ is the unit tangent to $c$.

\noindent{\em Summary of Notation}
\begin{eqnarray*}
 \int fd{}^4V &=& \int f\epsilon_{\alpha\beta\gamma\delta} dS^{\alpha\beta\gamma\delta}  
 = \int f{\sqrt{|g|}}~d^4x\\
\int \bm\omega &=&
\int\omega_{\alpha\beta\gamma\delta}  dS^{\alpha\beta\gamma\delta} 
=\int \omega_{0123}d^4x\\
\int fdV &=& \int f\epsilon_{\alpha\beta\gamma}  dS^{\alpha\beta\gamma}
    =  \int f {\sqrt{^3g}}~ d^3x
    = \int_{x^0\ {\text{constant}}} f{\sqrt{gg^{00}}}\ d^3x\\
\int\bm\omega &=&\int\omega_{\alpha\beta\gamma} dS^{\alpha\beta\gamma}
= \int_{x^0 \ {\text{constant}}} \omega_{123}dx^1dx^2dx^3\\
\int fdS &=& \int f\epsilon_{\alpha\beta} dS^{\alpha\beta}
= \int f {\sqrt{^2g}}~d^2x 
= \int_{\scriptstyle{x^0,x^1\ {\rm constant} }} 
 f\left[g_{22}g_{33}-(g_{23})^2\right]^{1/2} d^2x\\
\int\bm\omega &=& \int\omega_{\alpha\beta}
dS^{\alpha\beta} 
=\int_{\scriptstyle{x^0,x^1\ {\rm constant} }}\omega_{23}dx^2dx^3\\
\int_c fdl 
 &=& \int_{x^0,x^1,x^2\ \scriptstyle{\rm constant}}   f{\sqrt{g_{33}}}~dx^3 \\
 \int_c \bm\omega &=& \int_c  \omega_\alpha  dl^\alpha  
=\int_{\scriptstyle{x^0,x^1,x^2\ {\rm constant} }} \omega_3 dx^3.
\end{eqnarray*}

\noindent {\bf Example}:  The integral form of baryon mass
conservation $\nabla_\alpha  (\rho u^\alpha  )=0$ is
\begin{eqnarray*}
0 &=& \int_\Omega \nabla_\alpha  (\rho u^\alpha  )d{}^4V 
  = \int_{\partial\Omega} \rho u^\alpha dS_\alpha\\
&=& \int_{V_2} \rho u^\alpha  dS_\alpha  +
\int_{V_1} \rho u^\alpha  dS_\alpha .
\end{eqnarray*}  
Here the fluid is taken to have finite spatial extent, and the spacetime 
region $\Omega$ is bounded by the initial and final spacelike hypersurfaces
$V_1$ and $V_2$.  In a coordinate system for which $V_1$ and $V_2$ are 
surfaces of constant $t$, with $t$ increasing to the future, we have 
$dS_\mu = \nabla_\mu t \sqrt{|g|}d^3x = \delta_\mu^t \sqrt{|g|}d^3x$ on $V_2$, 
$dS_\mu = - \delta_\mu^t \sqrt{|g|}d^3x$ on $V_1$, and 
\be
\int_\Omega \nabla_\alpha  (\rho u^\alpha  )d{}^4V 
	= \int_{V_2} \rho u^t \sqrt{|g|} d^3x - \int_{V_1} \rho u^t \sqrt{|g|} d^3x. 
\ee 
If, on a slicing of spacetime one chooses on each hypersurface $V$ a surface element 
$dS_\alpha$ along $+\nabla_\alpha t$, the conservation law is then  
\be
M_0 = \int_V \rho u^\alpha  dS_\alpha  = \mbox{ constant}.
\ee

Note that the fact that one can write the conserved quantity associated with a current $j^\alpha$ in the form,   
\[ 
	\int_V j^\alpha  dS_\alpha  = \int_V j^t {\sqrt{|g|}}~d^3x ,
\]
means that there is no need to introduce $n_\alpha $ and ${\sqrt{^3g}}$ 
to evaluate the integral.  This fact is {\em essential} if one is evaluating an 
integral $\int j^\alpha dS_\alpha$ over a null surface, where there is no unit 
normal.  The flux of energy or of baryons across the horizon of a Schwarzschild 
black hole, for 
example, can be computed in Eddington-Finkelstein or Kruskal coordinates: 
In ingoing Eddington-Finkelstein coordinates $v,r,\theta,\phi$, 
the horizon is a surface of constant $r$, and we have
\[
 \int j^\alpha dS_\alpha = \int j^r \sqrt{|g|}\ dv d\theta d\phi.
\]
\label{p:bcons}
\subsubsection{Generalized divergence theorem}
\index{Gauss's theorem}\index{integration!Gauss's theorem}\index{divergence theorem}\index{conservation laws!Gauss's theorem}
\label{s:gauss} 

The key to Gauss's theorem is the fact that the divergence of a 
vector density has the form  
$ \partial_i \left( {\sqrt{|g|}}~A^i \right)$ or $\partial_i {\cal A}^i$.
This is true of any $q$-index antisymmetric tensor density, 
${\cal A}^{a\cdots b} = A^{a\cdots b}\sqrt{|g|}$, and  
an analogous theorem holds.  Because the text uses only the case of a two-index 
antisymmetric tensor, the electromagnetic field $F^{\alpha\beta}$, 
and because the way one extends the proof will be clear, we will give the 
generalization in detail for this case.  The theorem now relates an integral over 
an $n-1$-dimensional submanifold $S$ of $M$ to an integral over its $n-2$-dimensional
boundary $\partial S$:
\be 
  \int_S \nabla_b A^{ab} dS_a = \int_{\partial S} A^{ab} dS_{ab},
\label{gauss2}\ee  
where the sign of $dS_a$ and the meaning of $dS_{ab}$ are defined as follows.

Let $x^1, \ldots, x^n$ be a positively oriented chart on a subset of $M$ 
for which $x^1$ is constant on $S$, and with $x^2$ constant 
on $\partial S$ and increasing outward.  Choose the sign of $dS_a$ by 
$dS_a = \epsilon_{ab\cdots c}dS^{b\cdots c}$ or, equivalently, by 
requiring $dS_a = \nabla_a x^1 \sqrt{|g|} dx^2\cdots dx^n$.  The volume 
element $dS_{ab}$ is similarly chosen to satisfy, in our oriented chart,  
\be
   dS_{ab} = \nabla_{[a} x^1 \nabla_{b]}x^2 \sqrt{|g|}dx^3\cdots dx^n.  
\label{eq:dsab}\ee

Then in the coordinates' domain, 
\be
  \int_S \nabla_b A^{ab} dS_a
	=\int_S \nabla_b (A^{ab}\nabla_{a}x^1\sqrt{|g|})
	  dx^2\cdots dx^n.
\ee 
But the last integrand is just the divergence of the vector density 
$\widetilde{\cal A}^b 
 = A^{ab}\nabla_a x^1\sqrt{|g|}$, 
and we have already proved Gauss's law for this case: 
\be
  \int_S \nabla_b {\tilde A}^b \sqrt{|g|} dx^2\cdots dx^n 
 =  \int_{\partial S} {\tilde A}^b d{\tilde S}_b,
\label{atilde}\ee 
where, in each chart, 
$d{\tilde S}_b = \nabla_b x^2 \sqrt{|g|} dx^3\cdots dx^n$.  
Finally, ${\tilde A}^b dS_b = A^{ab} dS_{ab}$, whence Eq.~(\ref{atilde}) 
is identical to the generalized divergence theorem, Eq.~(\ref{gauss2}). 

\index{flux!electric}\index{charge!electric}
\noindent {\bf Example} (Electric charge).  Let $V$ be a ball containing a
charge $Q$.  The 4-dimensional form of Gauss's law 
relates the charge $Q=\int_V j^\alpha dS_\alpha $  in $V$ to the electric flux 
$\int F^{\alpha\beta}dS_{\alpha\beta}$ through the 2-dimensional surface of $V$: 
\be
Q=\int_V j^\alpha dS_\alpha  = \frac{1}{4\pi} \int_V \nabla_\beta F^{\alpha\beta}dS_\alpha  = \frac{1}{4\pi}
\int_{\pa V} F^{\alpha\beta}dS_{\alpha\beta}.
\ee

Pick positively oriented coordinates $t,r,\theta,\phi$ 
for which $V$ is a $t=$constant surface and  $\partial V$ an $r=$ constant surface. Then
\be   
 Q = \frac{1}{4\pi} \int F^{tr}\sqrt{|g|}d\theta d\phi.
\ee
Flat space:
\[
F^{tr} = E^r = \frac{Q}{r^2} \Longrightarrow \hspace{5mm} 
\int_V j^\alpha dS_\alpha =
\frac{1}{4\pi} \int \left(\frac{Q}{r^2}\right) (r^2\sin\theta )d\theta
d\phi = \frac{Q}{4\pi} \int d\Omega = Q .
\]
\newpage	

For a $q$-index antisymmetric tensor $A^{a\cdots bc}$, the generalized divergence 
theorem takes the 
form \\
$\int_S \nabla_c A^{a\cdots bc}dS_{a\cdots b} 
= \int_{\partial S} A^{a\cdots bc}dS_{a\cdots bc}$. \\

\subsection{Stokes's theorem}
\index{Stokes's theorem}\index{integration!Stokes's theorem}

The divergence $\nabla_a A^a$ of a vector on an $n$-dimensional manifold is dual to the 
exterior derivative $(d\omega)_{ab\cdots c} = n \nabla_{[a}\omega_{b\cdots c]}$ of the  $(n-1)$-form $\omega$ dual to $A^a$.
That is, with 
\be
\omega_{b\cdots c} := A^a \epsilon_{ab\cdots c},
\label{oa}\ee
we have 
\be
 \epsilon_{ab\cdots c} \nabla_d A^d = (d\omega)_{ab\cdots c}.
\label{doda}\ee    
Because both sides of the equation are $n$-forms, one need only check one component:
$n \nabla_{[1}\omega_{2\cdots n]}= \nabla_1 (A^d\epsilon_{d2\cdots n})-\nabla_2 (A^d\epsilon_{d13\cdots n})-\cdots-\nabla_n (A^d\epsilon_{d2\cdots n-1\,1})
 = \epsilon_{12\cdots n} \nabla_d A^d$. 
The corresponding dual of the generalized divergence theorem above is  \\
{\bf Stokes's Theorem}.  Let $\omega$ be an $n-1$-form on an $n$-dimensional 
manifold $S$ with boundary $\partial S$.  Then
\bsube 
\bea
\int_S d\omega &=& \int_{\partial S}\omega.\hspace{30mm}\\
\mbox{In index notation,}\hspace{38mm}&& \nonumber\\ 
\int_S (d\omega)_{ab\cdots c}dS^{ab\cdots c} &=& \int_{\partial S} \omega_{b\cdots c} dS^{b\cdots c}.\hspace{30mm}
\eea\label{stokes}\esube

The theorem implicitly assumes an orientation for $\partial S$ obtained 
from that of $S$ by requiring that, if $x^1,\cdots, x^n$ is a positively 
oriented chart on $S$ with $\partial S$ a surface of constant $x^1$ and 
$x^1$ increasing outward, then $x^2,\cdots, x^n$ is a positively oriented 
chart for $\partial S$.    

\noindent {\bf Proof}.  This dual of Gauss's theorem  follows quickly from 
Eq.~(\ref{e:gauss}), in the form
\be
 \int_S\nabla_d A^d \epsilon_{ab\cdots c}dS^{ab\cdots c}
= \int_{\partial S} A^a dS_a. 
\ee
Define $A^a$ by Eq.~(\ref{oa}), and note that, with the orientation chosen above, 
$dS_a = \epsilon_{ab\cdots c}dS^{b\cdots c}$.  Then 
\be
\int_S\nabla_d A^d \epsilon_{ab\cdots c}dS^{ab\cdots c}
= \int_{\partial S} A^a \epsilon_{ab\cdots c}dS^{b\cdots c},
\ee
and, from Eqs.~(\ref{oa}) and (\ref{doda}), the result follows:
\be  
\int_S (d\omega)_{ab\cdots c} dS^{ab\cdots c} 
= \int_{\partial S} \omega_{b\cdots c} dS^{b\cdots c}.\qquad \Box
\ee

\noindent
{\bf Example: Stokes's theorem in three dimensions}.\\  
Let $A_a $ be a 3-vector.  $(dA)_{ab} =
\nabla_a A_b -\nabla_b A_a $
\[ \int_S (\nabla_a A_b -\nabla_b A_a)dS^{ab} = \int_c  A_a dl^a ,\]
$c$ the curve bounding $S$.   As noted at the beginning of this section, 
this is equivalent to the usual form of Stokes's theorem in 
vector calculus:
\[ 
\int_S \vec\nabla\times\vec A\cdot d\vec S 
= \int_S \left( \epsilon^{abc} \nabla_b A_c \right) \epsilon_{ade}dS^{de} 
= \int_S(\nabla_a A_b -\nabla_b A_a )dS^{ab} = \int_c  \vec A\cdot d\vec l .
\]
\subsection{Diffeomorphism invariance}
\index{diffeomorphism!invariance of integrals}\index{integration!diffeomorphism invariance}
\index{active transformation!diffeomorphism}

The usual invariance of an integral under a coordinate transformation 
has as its active equivalent the invariance of an integral under a diffeo.
Let $\omega_{a\ldots b}$ be an $n$-form on an $n$-dimensional volume $V$.  
With $\psi(V)$ the image of $V$ and $\psi\omega_{a\ldots b}$ the 
dragged $n$-form, the invariance relation is 
\be
  \int_{\psi(V)} \psi\omega = \int_V \omega.  
\label{eq:diffeoinv}\ee
As in the case of coordinate transformations, invariance under diffeos follows from the 
fact that the components of $\omega_{a\ldots b}$ change by a 
Jacobian. Intuitively, the invariance follows from the equivalence of diffeo-related 
tensors on diffeo-related domains.  

\index{conservation laws!circulation, vorticity}\index{circulation}\index{vorticity}
We'll now use a corollary for a family of diffeos, $\psi_\lambda$, 
to obtain a relation between conservation of vorticity and conservation of circulation:\\
Let $\xi^a$ be the generator of the family $\psi_\lambda$.  
Then 
\be
    \frac d{d\lambda} \int_{\psi_\lambda(V)} \omega = \int_V\Lie_{\bm\xi}\omega. 
\label{eq:dlie}\ee
The proof is immediate from Eq.~(\ref{eq:liepsi}), starting from Eq.~(\ref{eq:diffeoinv}) in 
the form \\
$\dis\int_{\psi_\lambda(V)} \omega = \int_V \psi_{-\lambda}\omega$: \ \ \ 
$\dis
     \frac d{d\lambda} \int_{\psi_\lambda(V)} \omega 
  =\int_V \frac d{d\lambda} \psi_{-\lambda}\omega 
    = \int_V\Lie_{\bm\xi}\omega .\\
$

The equations of motion for a fluid imply that the flow conserves 
the vorticity, 
\be
\omega_{\a\b} :=\nabla_\a (hu_\b) - \na_\b (hu_a), 
\ee
where $h = \frac{\rho+p}\rho_0$, is the fluid's enthalpy ($\rho_0$ is the 
baryon mass density).  The differential conservation law is 
\be
{\cal L}_{\bf u} \omega_{\alpha\beta} = 0.
\label{lieom}
\ee
Then, denoting by $\psi_\tau$ the family of diffeos describing the flow of a fluid 
with velocity $u^\a$, we have 
\begin{eqnarray}
0 = \int_S {\cal L}_{\bf u} \omega_{\alpha\beta} dS^{\alpha\beta} &=&
\int_c
{\cal L}_{\bf u}
(h\,u_\alpha) d l^\alpha \nonumber\\
&=& \frac d{d\tau} \int_{c_\tau} h\, u_\alpha\,
d l^\alpha,
\label{vortiii}
\end{eqnarray}
where Stokes's theorem was used to obtain the second equality and Eq.~(\ref{eq:dlie}) 
 was used in the last equality.  That is, the line integral,
\be
\int_{c_\tau} h\,u_\alpha\, d l^\alpha=\int_{c_\tau} \frac{\epsilon +p}\rho\;u_\alpha\,
d l^\alpha, 
\label{vortiv}
\ee
(the {\it circulation} of the flow along a closed curve)\index{circulation|textbf} 
is independent of $\tau$, conserved by the fluid flow.  In the Newtonian limit, 
the conserved circulation is just
\be 
\int_{c_t} v_a\, d l^a;
\ee
that is, the line integral of the fluid velocity along a curve embedded in (moving with) the 
fluid is constant.  The corresponding differential law, the Newtonian limit of Eq.~(\ref{lieom}), is 
\be
  (\pa_t+\Lie_{\bf v}) \nabla_{[a}v_{b]} = 0.
\ee
\index{integration!on manifolds|)}
\newpage

\section{Cartan calculus} \index{Cartan calculus}
\label{s:cartan} 
This section uses the first three equations in Appendix \ref{s:forms}, defining forms 
and the exterior derivative but nothing else from that section.  

\noindent{\bf Definition}.  The {\em wedge product} $\sigma \wedge \tau$ 
of a $p$-form $\sigma_{a\dots b}$ and a $q$-form $\tau_{c\dots d}$ 
is the $(p+q)$-form 
\be 
(\sigma \wedge \tau)_{a\dots bc\dots d} := 
\frac{(p+q)!}{p!q!}\sigma_{[a\dots b}\tau_{c\dots d]}.
\label{e:wedge}\ee
The factor 
$\dis \left(\begin{array}{cc}p+q\\ q\end{array} \right) = \frac{(p+q)!}{p!q!}$ 
gives a number of terms equal to the number of independent ways of assigning
$p$ indices to $\sigma$ and $q$ indices to $\tau$.\\

\noindent{\bf Examples}. If $\sigma$ and $\tau$ are 1-forms, 
\be 
	(\sigma \wedge \tau)_{ab} = \sigma_a \tau_b - \sigma_b \tau_a;
\ee
equivalently, 
\[ 
	(\sigma \wedge \tau)(u,v) = \sigma(u)\tau(v) - \sigma(v) \tau(u).
\]
If $\sigma$ is a 1-form and $\tau$ is a 2-form, 
\be 
 (\sigma \wedge \tau)_{abc} 
 	= \sigma_a \tau_{bc} + \sigma_b \tau_{ca} + \sigma_c \tau_{ab}.
\ee
\subsection{The Riemann Tensor in a General basis}

Cartan's formalism provides an efficient way to calculate the
Riemann tensor, picking out a minimal set of connection
coefficients to be evaluated.  One uses an orthonormal basis, and
we will begin by finding the components of the Riemann tensor with
respect to an arbitrary basis, $e_i$. Starting from the definition
\be
	\frac{1}{2}R^a_{\ bcd}v^{b}=\nabla_{[c}\nabla_{d]}v^a,
\label{riem}\ee
we have,
\[
	\nabla_{[c}\nabla_{d]}e_i^a=\frac{1}{2}R^a_{\ bcd}e_i^b.
\]

Then
\beaa
\na_k\na_l\, e_j^a = e_k^{\ c} e_l^{\ d}\nabla_c\nabla_d e_j^{\ a}
&=&
e_k^{\ c}\nabla_c(e_l^{\ d}\nabla_d e_j^{\ a})
	-(e_k^{\ c}\nabla_c e_l^{\ d})\nabla_d e_j^{\ a}\\
&=&\nabla_k(\Gamma^m_{\ jl}e_m^{\ a})-\Gamma^m_{\ lk}e_m^{\ d}\nabla_d e_j^{\ a}\\
&=&e_k\Gamma^m_{\ jl}\ e_m^{\ a}+\Gamma^m_{\ jl}\Gamma^i_{\ mk}e_i^{\ a}
		        -\Gamma^m_{\ lk}\Gamma^i_{\ jm}e_i^{\ a}.
\eeaa

Finally, antisymmetrizing on $k$ and $l$, we have 
\be\cblue
R^i_{\ jkl} =  ([\na_k,\na_l]e_j)^i
	=e_k\Gamma^i_{\ jl}-e_l\Gamma^i_{\ jk}+\Gamma^i_{\ mk}\Gamma^m_{\ jl}
		 -\Gamma^i_{\ ml}\Gamma^m_{\ jk}+2\Gamma^i_{\ jm}\Gamma^m_{\ [kl]}
\index{Riemann tensor!components in general basis}
\index{curvature tensor!components in general basis}
\label{rijkl1}\ee
Here $e_k f$ means the derivative of a scalar $f$ along $\bm e_k$: That is, $e_k f \equiv e_k{}^a\na_\a f$.  
Eq.~\eqref{rijkl1} agrees with the usual expression for the coordinate
components when $e_i=\partial_i$, for then $\Gamma^m_{\ [kl]}=0$.\\

\subsection{ Cartan's structure equations}\index{Cartan structure equations}

A definition of the exterior derivative of a 1-form that does
not make use of a covariant derivative operator is 
\be
  d\sigma(u,v) = u \cdot d[\sigma(v)] - v\cdot d[\sigma(u)] - \sigma([u,v]),
\label{dsig1}\ee 
where $df$ means the gradient of the scalar $f$; in the equation, 
$\sigma(v)$ and $\sigma(u)$ are, of course, scalars. (A similar 
equation can be written down for the exterior derivative of a $p$-form.)  
Here's a check that the two equations for $d\sigma$ are equivalent:
\beaa 
 d\sigma(u,v) &=& (\nabla_a\sigma_b - \nabla_b \sigma_a)u^a v^b
 	 =  u^a\nabla_a(\sigma_b v^b) - v^b\nabla_b( \sigma_a u^a) 
 	  - u^a(\nabla_a v^b) \sigma_b + v^b(\nabla_b u^a)\sigma_a \\
 	&=& u\cdot d[\sigma(v)] - v\cdot d[\sigma(u)] - \sigma([u,v]).
\eeaa
Calculations repeatedly use the identity, for any form $\omega$ and scalar $f$, 
\be
   d(f\, \omega) = df\wedge \omega + f\,d\omega.  
\ee
This follows immediately from the Leibnitz rule and our initial definitions of $d$ and $\wedge$,
\eqref{e:exteriord} and \eqref{e:wedge}:
\[
	d(f\, \omega)_{ab\cdots c}  = (p+1)\na_{[a} (f\omega_{b\cdots c]}) 
	= (p+1)\na_{[a} f\ \omega_{b\cdots c]} +f (p+1) \na_{[a} \omega_{b\cdots c]}
	= (df\wedge\omega + f d\omega)_{ab\cdots c}
\]
In particular, for two scalars $f$ and $g$, because $d^2 g=0$, the identity is 
\be
  d(f dg) = df\wedge dg.  
\label{e:dfg}\ee

The connection coefficients (Christoffel symbols) will be calculated by taking exterior
derivatives of the dual basis vectors $\omega^i$.  Because
$\omega^i(e_j)=\delta^i_j$ and the components $\delta^i_j$ are constants,
Eq. (\ref{dsig1}) gives
\be
(d\omega^i)_{jk}=d\omega^i(e_j , e_k)=-\omega^i([e_j ,
e_k])=-c^i_{\ jk}.
\label{domi}\ee
Then $\Gamma$ will be obtained below from $d\omega$ using Eq.~\eqref{gamc}, namely\\
\be 
\Gamma_{ijk} = \frac{1}{2} (\, c_{kij} - c_{jki} - c_{ijk}\,).
\label{e:gammac}\ee

Cartan defines connection l-forms
\be
	\omega^i_{\ j}=\Gamma^i_{\ jk}\omega^k, 
\label{e:omGam}
\ee
\index{Christoffel symbol!relation to connection 1-forms}\index{connection 1-form}\index{connection coefficient}
or $(\omega^i_{\ j})_k=\Gamma^i_{\ jk}$.
Equivalently,
\be
	\nabla_a e_j^{\ b}=\omega^i_{\ ja}\,e_i^{\ b}.
\ee
Then for an orthonormal basis, or any basis in which the metric
has constant components,
\be 
	\omega_{ij}=-\omega_{ji}.
\ee 
Check: 
$\omega_{ija} = e_{ib}\nabla_a\ e_j^{\ b}=\nabla_a(g_{ij})-e_j^{\ b}\nabla_a e_{ib} 
			   =-e_{jb}\nabla_a e_i^b = - \omega_{jia}.$\\
Note that $\nabla_a(g_{ij})$ here is the gradient of the scalar $g_{ij}$, {\sl not} the $ij$ component 
of the covariant derivative of the metric. It vanishes because $g_{ij}$ is a constant. \\ 

Cartan's first structure equation relates 
$\omega^i_{\ j}$ to $d\omega^i$:
\be \crv
d\omega^i=-\omega^i_{\ j}\wedge\omega^j\cb .
\label{c1}\ee
Proof:  Using Eqs.~(\ref{domi}), \eqref{gasym}, and then $(\omega^i)_j = \delta^i_j$, we have
\[
(d\omega^i)_{\ kl}=-c^i_{\ kl}=\Gamma^i_{\ kl}-\Gamma^i_{\ lk}
	=(\omega^i_{\ j})_l(\omega^j)_k-(\omega^i_{\ j})_k(\omega^j)_l
	=-(\omega^i_{\ j}\wedge\omega^j)_{kl}.
\]
Eq.~\eqref{e:gammac} is essentially the 
inverse of this last equation, allowing one to write $\omega_{ij}$ in terms of 
$d\omega^i$.  After rearranging the $c$'s and using Eq.~(\ref{domi}) to replace $c_{ijk}$ by 
$-(d\omega_i)_{jk}\equiv - d\omega_{i.jk}$, Eq.~\eqref{e:gammac} becomes  
\be\crv
\omega_{ij.k}=\frac{1}{2}(d\omega_{i.jk}+d\omega_{j.ki}-d\omega_{k.ij})\cb,
\label{c1a}\ee
Here $\omega_{ij.k}$ means $(\omega_{ij})_k$.  Once one has computed the 
connection coefficients using (\ref{c1}) or (\ref{c1a}), the Riemann tensor is found from 
Cartan's second structure equation, namely
\be \crv
	{\cal R}^i_{\ j}=d\omega^i_{\ j} + \omega^i_{\ k}\wedge\omega^k_{\ j}\cb. 
\label{c2}\ee
where ${\cal R}^i_{\ j}$ is a 2-form defined by ${\cal R}^i_{\ j}:=
R^i_{\ jab}$.

Proof of (\ref{c2}):  We must check that (\ref{c2}) agrees with (\ref{rijkl1}).
Now from Eq.~\eqref{dsig1}, the first term on the right of \eqref{c2} is 
\begin{align*}
 d\omega^i_{\ j}(e_k,e_l)
 	&= e_k [\omega^i{}_j (e_l)] - e_l[\omega^i{}_j (e_k)] - \omega^i{}_j ([e_k,e_l])\\
        &= e_k\Gamma^i_{\ jl} -e_l\Gamma^i_{\ jk} - \omega^i{}_{j.m} c^m{}_{kl}\\
        &= e_k\Gamma^i_{\ jl}-e_l\Gamma^i_{\ jk} +2\Gamma^i_{\ jm}\Gamma^m_{\ [kl]},
\end{align*}
where we used Eq.~\eqref{gasym} in the last line.  The second term in \eqref{c2} is 
\[
\omega^i_{\ m}\wedge\omega^m_{\ j}(e_k,e_l)= \Gamma^i_{\ mk}\Gamma^m_{\ jl}
				    -\Gamma^i_{\ ml}\Gamma^m_{\ jk}.
\]

Thus (\ref{c2}) implies \beaa
R^i_{\ jkl}&=&(d\omega^i_{\ j}+\omega^i_{\ m}\wedge\omega^m_{\ j})(e_k,e_l)\\
&=&  e_k\Gamma^i_{\ jl}-e_l\Gamma^i_{\ jk}+\Gamma^i_{\ mk}\Gamma^m_{\ jl}
	-\Gamma^i_{\ ml}\Gamma^m_{\ jk}+2\Gamma^i_{\ jm}\Gamma^m_{\ [kl]},
\eeaa
in agreement with (\ref{rijkl1}).\\

\noindent {\em Exercise}.  Write down the independent components of the
Rieman tensor in an orthonormal basis for the 2-sphere with metric
\[
ds^2=a^2(d\theta^2+\sin^2\theta \;d\phi^2)=a^2d\Omega^2,
\]
using Cartan calculus.  Find the Ricci scalar.\\

\noindent
{\sl Exercise with solution}. Compute the independent components of the Riemann tensor for 
a time-independent spherically symmetric (Schwarzschild) metric of the form 
\be
  ds^2 = -e^{2\Phi} dt^2 + e^{2\lambda} dr^2 + r^2 (d\theta^2+\sin^2\theta \;d\phi^2), 
\ee
with $\Phi$ and $\lambda$ functions of $r$.  \\

\noindent
Solution on next page.\\

\noindent {\em Exercise}.    Repeat the calculation of the first exercise above for the 3-sphere:
$\dis
ds^2=a^2(d\chi^2 +\sin^2\chi \;d\Omega^2)
$,
and check that $R_{abcd}=a^{-2}(g_{ac}g_{bd}-g_{ad}g_{bc}).$\\

\noindent {\em Exercise}.    Find the independent components of the Riemann
and Ricci tensor in orthonormal bases for the FRW spacetimes
with metrics
\begin{align*}
ds^2 &=-dt^2+a(t)^2(d\chi^2+\sin^2\chi\;d\Omega^2),\qquad\mbox{and }\\
ds^2&=-dt^2+a(t)^2(dx^2+dy^2+dz^2).
\end{align*}
\\

\noindent{\sl Solution to Schwarzschild exercise}. \\
When terms in the solution below are written in gray, it means they vanish.  \\
Spherical symmetry and the fact that the basis on the sphere is orthonormal means that Riemann, 
Ricci, and Einstein tensor 
components are invariant under the index replacement $2\leftrightarrow 3$:  
$R_{1212} = R_{1313}$.  
\def\om{\omega} 
\def\lm{\lambda} 
\begin{align*} 
   \omega^0 &= e^\Phi dt, \quad \omega^1 = e^\lambda dr,\quad \om^2= r\,d\theta, 
   \quad \om^3 = r\sin\theta\ d\phi.\\
   d\om^0 &= \Phi' e^\Phi dr\wedge dt = -\Phi'e^{-\lm}\om^0\wedge\om^1,
    \quad d\om^1 =  0 \quad \mbox{That is}, d [f(r)dr] = f'(r) dr\wedge dr = 0.\\
    d\om^2 &= dr\wedge d\theta = \frac1r e^{-\lm} \om^1\wedge\om^2 \\
    d\om^3 &= d(r\sin\theta)\wedge d\phi = \sin\theta \,dr\wedge d\phi + r d(\sin\theta)\wedge d\phi
    	    = \frac{e^{-\lm}}r \omega^1\wedge\om^3 + \frac{\cot\theta}r e^{-\lm} \om^2\wedge\om^3
\end{align*}
Notice now that $(d\omega^\lambda)_{\mu\nu}$ is nonzero only if the set of three indices is 
$001$, $212$, $313$ or $323$, with $\lambda$ the first index.  That means $\Gamma$ is nonzero 
only for these four sets of indices.  
\begin{align*}
   \om_{01} &= \Gamma_{010} \omega^0 
   	= \frac12(d\omega_{0.10} + {\color{gray}d\omega_{1.00} }-d\omega_{0.01} )\om^0
	= -\Phi' e^{-\lm} \om^0 = - \Phi'e^{\Phi-\lm}dt \quad  
    (\Gamma_{011} = \Gamma_{012} = \Gamma_{013} = 0).  \\
  \om_{02} &= \om_{03}=0 \quad 
		\mbox{(No nonvanishing } d\om_{\lambda.\mu\nu} \mbox{ has indices $02$ or $03$}). \\
  \om_{12} &= \Gamma_{122}\om^2 = \frac12 ({\color{gray} d\om_{1.22}}+d\om_{2.21}-d\om_{2.12})\om^3
  	    = -\frac1r e^{-\lm} \om^2 = - e^{-\lm} d\theta \\
  \om_{13} &= -\frac1r e^{-\lm} \om^3 = -e^\lm\sin\theta d\phi\\
  \om_{23} &= \Gamma_{233}\om^3 = \frac12 ({\color{gray} d\om_{2.33}}+d\om_{3.32}-d\om_{3.23})\om^3
  	    = -\frac1r\cot\theta\ \om^3 = -\cos\theta\,d\phi.\\
 {\cal R}^0{}_1 &= d\om^0{}_1 + {\color{gray}\om^0{}_\mu \wedge \om^\mu{}_1} = d(\Phi'e^{\Phi-\lm} dt) 
 		= -[\Phi'' + \Phi'(\Phi'-\lm')] e^{-2\lm}\om^0\wedge\om^1\\
 {\cal R}^0{}_2 &={\color{gray} d\om^0{}_2} +\om^0{}_\mu \wedge \om^\mu{}_2 = \om^0{}_1 \wedge \om^1{}_2
		= \left(\Phi'e^{-\lm}\om^0\right)\wedge\left(-\frac1r e^{-\lm} \om^2\right) 
		= -\frac1r \Phi'e^{-2\lm} \om^0\wedge\om^2\\
{\cal R}^0{}_3 &= -\frac1r \Phi' e^{-2\lm} \om^0\wedge\om^3 \quad \mbox{by spherical symmetry}\\
{\cal R}^1{}_2 &= d\om^1{}_2 +{\color{gray}\om^0{}_\mu \wedge \om^\mu{}_2} = d(-e^{-\lm}d\theta)  		 	= \frac1r\lm' e^{-2\lm}	\om^1\wedge\om^2 \\
{\cal R}^1{}_3 &= \frac1r\lm' e^{-2\lm}	\om^1\wedge\om^3 \quad \mbox{by spherical symmetry}\\
{\cal R}^2{}_3 &= d\omega^2{}_3 + \omega^2{}_1\wedge\om^1{}_3 
		= d(-\cos\theta d\phi) + (\frac1r e^{-\lm}\om^2)\wedge(-\frac1r e^{-\lm}\om^3) 
		= \frac1{r^2}\left(1-e^{-2\lm}\right)\om^2\wedge\om^3.\\
R_{0101} &= \left[\Phi'' + \Phi'(\Phi'-\lm')\right] e^{-2\lm} \qquad 
	 R_{2323}= \frac1{r^2} \left(1-e^{-2\lm}\right) \\
R_{0202} &= R_{0303} = \frac1r\Phi' e^{-2\lm} \qquad R_{1212} = R_{1313}=\frac1r\lm' e^{-2\lm}.
\end{align*}

\chapter{Constants (cgs and Gravitational Units)}\index{constants (physical constants)}\index{physical constants}
\vspace{-8mm}

See the \href{https://physics.nist.gov/cuu/Constants/index.html}{NIST constants} 
website for error bars and current values of fundamental physical constants\\

\noindent{\bf Fundamental Constants}
\vspace{-2mm}

\begin{tabbing}
Gravitational Constant\hspace{20mm} \=
$G=6.674\,30 \times10^{-8}{\rm {\rm cm}}^3/{\rm g}\,{\rm s}^2=1$
\\
Speed of light\>
$c= 299\,792\,458\times10^{10}{\rm {\rm cm}}/{\rm sec} =1$\\
Useful combinations 
\> $ G/c^2=0.7425\times10^{-28}{\rm cm}/{\rm g}=1.476\,64\ {\rm km}/M_\odot=1$\\
\> $c^5/G=3.629\times10^{59}{\rm erg}/{\rm s}=2.030\times10^5M_\odot
        c^2/{\rm s}=1$ (emission
factor)\\
\>$ G/c=2.226\times10^{-18}{\rm cm}^2{\rm Hz/ g}=1$ (receptor factor)\\
\>$ c^2/G^{\frac{1}{2}} =
3.479\times10^{24}{\rm gauss}\,{\rm cm}=3.479\times10^{24}{\rm statvolt}=1$
\\
Planck's reduced constant \>
$\hbar= 1.054\,571\,817. . .\times10^{-27}{\rm g}\,{\rm cm}^2/{\rm s} =2.6123\times10^{-66}{\rm cm}^2$\\

Planck length \> $(\hbar
G/c^3)^{\frac{1}{2}}=1.616\times10^{-33}{\rm cm}$ \\

Planck time  \> 
$(\hbar G/c^5)^{\frac12}=5.391\times10^{-44}{\rm s}$\\

Planck mass  \> 
$(\hbar c/G)^{\frac{1}{2}}=2.176 \times10^{-5}{\rm g}$\\

Planck density   \>
$(c^5\hbar G^2) =5.157\times10^{93}{\rm g}/{\rm cm}^3$\\

Quantum of charge
 \> $e=4.803\,25\times10^{-10}({\rm g}\,{\rm cm}^3/{\rm s}^2)^{\frac{1}{2}}
         =1.381\times10^{-34}{\rm cm}$\\

$1/\alpha$ \> $\hbar
c/e^2=137.0360$\\

Electron rest mass  \> $m_e
=9.109\,56\times10^{-28}{\rm g}=
8.1873\times10^{-7}{\rm erg}=0.511\,004\,{\rm MeV} $\\
\>\hspace{7mm}$ = 6.764\times10^{-56}{\rm cm}$\\

Proton rest mass
 \>
$m_p=1.672\,622\times10^{-24}{\rm g}
        =1.503\,28\times10^{13}{\rm erg}=0.938\,32\, {\rm GeV}$ \\
\>\hspace{7mm}$ =1.2419\times10^{-52}{\rm cm}$\\

Bohr radius
\> $a_0=\hbar^2/m_e e^2=0.529\,1772\times10^{-8}{\rm cm}$\\

Reduced Compton wavelength \>
$\lambda_e=\hbar/m_ec=3.861\,59\times
10^{-11}{\rm cm}$\\

Classical electron radius
\> $r_0=e^2/m_ec^2=2.817\,940\times10^{-13}{\rm cm}$\\

Atomic en{\rm erg}y unit \>$ e^2/a_0=m_e e^4/
\hbar^2=4.359\,83\times10^{-11}{\rm erg}=27.2116\mbox{ eV}
        =3.157\,86\times10^5 {\rm K}$\\
        \>\hspace{7mm}$=3.602\times
10^{-60}{\rm cm}$
\\

{\bf Conversion Factors}
\\
Distance\> $
1\;{\rm pc}=3.0856\times10^{18}{\rm cm}$\\
\>$\mbox{ly}=0.94605\times10^{18}{\rm cm}$\\
\>$ 1\;{\rm AU}=1.495\,985\times10^{13}{\rm cm}$
\\
\\
Time\>
$1\;{\rm yr}=3.1556926\times10^7{\rm s}$\\
\>$\mbox{1 day}=86\,400.{\rm s};\,1\;\mbox{sidereal day}=86\,164.091\;{\rm s}$
\\
\\
Mass, en{\rm erg}y, temperature\>
$1\;{\rm eV}=1.16048\times10^4{\rm K}=1.602192\times10^{-12}{\rm erg}$\\
\>$\qquad =1.782\,68\times10^{-33}{\rm g}=1.324\times10^{-61}{\rm cm}$
\end{tabbing}
\newpage

\noindent{\bf Electromagnetic}

\begin{tabbing}

Blackbody radiation\\
energy density $=aT^4,$\hspace{35mm}  \=
$a=7.5647\times10^{-15}{\rm erg}\;{\rm cm}^{-3}{\rm K}^{-4}$\\
emittance $\sigma T^4$,\>
$\sigma=\frac{1}{4}ac=5.6696\times10^{-5}{\rm erg}\;{\rm cm}^{-2}{\rm s}^{-1}{\rm K}^{-4}$\\
peak of $E(\lambda)$ spectrum\>
$\lambda_{\rm max}T=0.289\,79\mbox{ cm K}$\\
\\
Bohr magneton \hspace{45mm}  \= $e\hbar/2m_e c = 9.274\times 10^{-21}\mbox{erg/G} = 1.65\times10^{-12}\mbox{cm}^2$\\
Thomson cross section,\>
$\sigma_T=(8\pi/3)r_0^2=0.665\,24\times10^{-24}{\rm cm}^2$\\
\\
\end{tabbing}
{\bf The Universe}

\begin{tabbing}
Solar mass\hspace{55mm}\= $M_\odot=1.988\,475±0.000\,092\times 10^{33}{\rm g}
 = 1.476\,64 \times 10^5{\rm cm}$
\\ \> $\qquad =4.9255\;\mu {\rm s}$\\

Solar radius\>
$R_\odot=6.9570\times10^{10}{\rm cm}$\\

Solar luminosity\>
$L_\odot=3.828\times10^{33}{\rm erg}/{\rm s}=(1.07\times10^{-26})c^5/G$\\

Earth-Sun distance\>
$1\;{\rm AU}=(1.495\,985\pm
0.000\,005)\times10^{13}{\rm cm}=499.007\;{\rm s}$\\

Earth mass\>
$M_\oplus=5.9722\times10^{27}{\rm g}=0.4438\,{\rm cm}$\\

Earth mean radius\>
$R_\oplus=6.378\,14\times10^8{\rm cm}$\\
 
Hubble parameter (see Sect.~\ref{s:H0})\>
$H_0=71\pm 4\,{\rm km}\;{\rm s}^{-1}{\rm Mpc}^{-1}=(14\times 10^9{\rm yr})^{-1}
	=(1.3\times 10^{28}{\rm cm})^{-1}$
\index{Hubble parameter $H$!present value $H_0$}
\\
Critical density $\rho_c=3 H_0^2/8\pi G$\> $\rho_c=0.95\times 10^{-29}\mbox{g/cm}^3 
= 7.0\times 10^{-58} \mbox{cm}^{-2}$ \\
\> $\quad =1.88\times10^{-29} h^2 \mbox{g/cm}^3,\mbox{where } h=H_0/100\ {\rm km}\;{\rm s}^{-1}{\rm Mpc}^{-1}$

\end{tabbing}

\printindex

\end{document}